\documentclass[msmallroyalvopaper,oldfontcommands,openright]{memoir}

\setlrmarginsandblock{12mm}{12mm}{*}
\setulmarginsandblock{23mm}{20mm}{*}
\checkandfixthelayout

\newif\ifgreyprint
\greyprintfalse
\ifgreyprint\newcommand\greyprint{-grey}\else\newcommand\greyprint{}\fi
 
\usepackage{latexsym,xspace,calc,amsthm}
\usepackage{amsmath,amssymb} %
\usepackage{nicefrac} %

\usepackage{lmodern}
\usepackage{microtype}
\usepackage{tgtermes}
\usepackage{fix-cm}

\usepackage{bm} %

\usepackage{cancel}

\usepackage{savesym}
\savesymbol{subcaption}
\usepackage{subcaption}

\usepackage{graphbox} %
\newdimen\proofrulebreadth \proofrulebreadth=.05em
\newdimen\proofdotseparation \proofdotseparation=1.25ex
\newdimen\proofrulebaseline \proofrulebaseline=2ex
\newcount\proofdotnumber \proofdotnumber=3
\let\then\relax
\def\hfi{\hskip0pt plus.0001fil}
\mathchardef\squigto="3A3B
\newif\ifinsideprooftree\insideprooftreefalse
\newif\ifonleftofproofrule\onleftofproofrulefalse
\newif\ifproofdots\proofdotsfalse
\newif\ifdoubleproof\doubleprooffalse
\let\wereinproofbit\relax
\newdimen\shortenproofleft
\newdimen\shortenproofright
\newdimen\proofbelowshift
\newbox\proofabove
\newbox\proofbelow
\newbox\proofrulename
\def\shiftproofbelow{\let\next\relax\afterassignment\setshiftproofbelow\dimen0 }
\def\shiftproofbelowneg{\def\next{\multiply\dimen0 by-1 }%
\afterassignment\setshiftproofbelow\dimen0 }
\def\setshiftproofbelow{\next\proofbelowshift=\dimen0 }
\def\setproofrulebreadth{\proofrulebreadth}

\def\prooftree{
\ifnum  \lastpenalty=1
\then   \unpenalty
\else   \onleftofproofrulefalse
\fi
\ifonleftofproofrule
\else   \ifinsideprooftree
        \then   \hskip.5em plus1fil
        \fi
\fi
\bgroup
\setbox\proofbelow=\hbox{}\setbox\proofrulename=\hbox{}%
\let\justifies\proofover\let\leadsto\proofoverdots\let\Justifies\proofoverdbl
\let\using\proofusing\let\[\prooftree
\ifinsideprooftree\let\]\endprooftree\fi
\proofdotsfalse\doubleprooffalse
\let\thickness\setproofrulebreadth
\let\shiftright\shiftproofbelow \let\shift\shiftproofbelow
\let\shiftleft\shiftproofbelowneg
\let\ifwasinsideprooftree\ifinsideprooftree
\insideprooftreetrue
\setbox\proofabove=\hbox\bgroup$\displaystyle 
\let\wereinproofbit\prooftree
\shortenproofleft=0pt \shortenproofright=0pt \proofbelowshift=0pt
\onleftofproofruletrue\penalty1
}

\def\eproofbit{
\ifx    \wereinproofbit\prooftree
\then   \ifcase \lastpenalty
        \then   \shortenproofright=0pt  
        \or     \unpenalty\hfil         
        \or     \unpenalty\unskip       
        \else   \shortenproofright=0pt  
        \fi
\fi
\global\dimen0=\shortenproofleft
\global\dimen1=\shortenproofright
\global\dimen2=\proofrulebreadth
\global\dimen3=\proofbelowshift
\global\dimen4=\proofdotseparation
\global\count255=\proofdotnumber
$\egroup  
\shortenproofleft=\dimen0
\shortenproofright=\dimen1
\proofrulebreadth=\dimen2
\proofbelowshift=\dimen3
\proofdotseparation=\dimen4
\proofdotnumber=\count255
}

\def\proofover{
\eproofbit 
\setbox\proofbelow=\hbox\bgroup 
\let\wereinproofbit\proofover
$\displaystyle
}%
\def\proofoverdbl{
\eproofbit 
\doubleprooftrue
\setbox\proofbelow=\hbox\bgroup 
\let\wereinproofbit\proofoverdbl
$\displaystyle
}%
\def\proofoverdots{
\eproofbit 
\proofdotstrue
\setbox\proofbelow=\hbox\bgroup 
\let\wereinproofbit\proofoverdots
$\displaystyle
}%
\def\proofusing{
\eproofbit 
\setbox\proofrulename=\hbox\bgroup 
\let\wereinproofbit\proofusing
\kern0.3em$
}

\def\endprooftree{
\eproofbit 
  \dimen5 =0pt
\dimen0=\wd\proofabove \advance\dimen0-\shortenproofleft
\advance\dimen0-\shortenproofright
\dimen1=.5\dimen0 \advance\dimen1-.5\wd\proofbelow
\dimen4=\dimen1
\advance\dimen1\proofbelowshift \advance\dimen4-\proofbelowshift
\ifdim  \dimen1<0pt
\then   \advance\shortenproofleft\dimen1
        \advance\dimen0-\dimen1
        \dimen1=0pt
        \ifdim  \shortenproofleft<0pt
        \then   \setbox\proofabove=\hbox{%
                        \kern-\shortenproofleft\unhbox\proofabove}%
                \shortenproofleft=0pt
        \fi
\fi
\ifdim  \dimen4<0pt
\then   \advance\shortenproofright\dimen4
        \advance\dimen0-\dimen4
        \dimen4=0pt
\fi
\ifdim  \shortenproofright<\wd\proofrulename
\then   \shortenproofright=\wd\proofrulename
\fi
\dimen2=\shortenproofleft \advance\dimen2 by\dimen1
\dimen3=\shortenproofright\advance\dimen3 by\dimen4
\ifproofdots
\then
        \dimen6=\shortenproofleft \advance\dimen6 .5\dimen0
        \setbox1=\vbox to\proofdotseparation{\vss\hbox{$\cdot$}\vss}%
        \setbox0=\hbox{%
                \advance\dimen6-.5\wd1
                \kern\dimen6
                $\vcenter to\proofdotnumber\proofdotseparation
                        {\leaders\box1\vfill}$%
                \unhbox\proofrulename}%
\else   \dimen6=\fontdimen22\the\textfont2 
        \dimen7=\dimen6
        \advance\dimen6by.5\proofrulebreadth
        \advance\dimen7by-.5\proofrulebreadth
        \setbox0=\hbox{%
                \kern\shortenproofleft
                \ifdoubleproof
                \then   \hbox to\dimen0{%
                        $\mathsurround0pt\mathord=\mkern-6mu%
                        \cleaders\hbox{$\mkern-2mu=\mkern-2mu$}\hfill
                        \mkern-6mu\mathord=$}%
                \else   \vrule height\dimen6 depth-\dimen7 width\dimen0
                \fi
                \unhbox\proofrulename}%
        \ht0=\dimen6 \dp0=-\dimen7
\fi
\let\doll\relax
\ifwasinsideprooftree
\then   \let\VBOX\vbox
\else   \ifmmode\else$\let\doll=$\fi
        \let\VBOX\vcenter
\fi
\VBOX   {\baselineskip\proofrulebaseline \lineskip.2ex
        \expandafter\lineskiplimit\ifproofdots0ex\else-0.6ex\fi
        \hbox   spread\dimen5   {\hfi\unhbox\proofabove\hfi}%
        \hbox{\box0}%
        \hbox   {\kern\dimen2 \box\proofbelow}}\doll%
\global\dimen2=\dimen2
\global\dimen3=\dimen3
\egroup 
\ifonleftofproofrule
\then   \shortenproofleft=\dimen2
\fi
\shortenproofright=\dimen3
\onleftofproofrulefalse
\ifinsideprooftree
\then   \hskip.5em plus 1fil \penalty2
\fi
}

\usepackage{tikz}
\usepackage{tikz-cd}
\usetikzlibrary{cd,decorations.markings}
\tikzset{
    dharrow/.style={
        <->,
        postaction={decorate,-},
        }
}
\tikzset{
    dhdashedarrow/.style={
        <->,
        dashed,
        postaction={decorate,-},
        }
    }
\tikzset{
    lrharpoonarrow/.style={
        <[harpoon]->[harpoon],
        postaction={decorate,-},
        }
}
\tikzset{
    lrharpoondashedarrow/.style={
        <[harpoon]->[harpoon],
        dashed, %
        postaction={decorate,-},
        }
}
\usetikzlibrary{arrows}
\usetikzlibrary{arrows.meta} 
\usetikzlibrary{calc}
\usetikzlibrary{positioning}
\usetikzlibrary{snakes,automata,chains}
\usetikzlibrary{graphs}

\usepackage{binarytree}

\usepackage{amssymb,stmaryrd,amsmath}
\usepackage{mdwlist} %
\usepackage{float}   %
\usepackage{centernot} %

\PassOptionsToPackage{hyphens}{url}  %
\ifgreyprint
\usepackage[allcolors=black,colorlinks]{hyperref}
\else
\usepackage[colorlinks]{hyperref}
\fi
\usepackage{relsize}

\usepackage{breakurl}  %

\newtheoremstyle{jamiestyle}%
  {4pt}%
  {0pt}%
  {\it}%
  {0pt}%
  {\sc}%
  {.}%
  { }%
  {}%
\theoremstyle{jamiestyle}
\newtheorem{thrm}{Theorem}[section]
\newtheorem{prop}[thrm]{Proposition}
\newtheorem{lemm}[thrm]{Lemma}
\newtheorem{corr}[thrm]{Corollary}

\newtheoremstyle{jamienfstyle}%
  {4pt}%
  {0pt}%
  {\normalfont}%
  {0pt}%
  {\sc}%
  {.}%
  { }%
  {}%
\theoremstyle{jamienfstyle}
\newtheorem{nttn}[thrm]{Notation}
\newtheorem{defn}[thrm]{Definition}
\newtheorem{xmpl}[thrm]{Example}
\newtheorem{rmrk}[thrm]{Remark}

\usepackage{color}
\definecolor{mygreen}{rgb}{0,0.6,0}
\definecolor{mygray}{rgb}{0.5,0.5,0.5}
\definecolor{mymauve}{rgb}{0.58,0,0.82}

\usepackage{listings}
 
\definecolor{gray}{RGB}{128, 128, 128}
\definecolor{lightgray}{RGB}{200, 200, 200}
\definecolor{cyan}{RGB}{0, 255, 255}
\definecolor{blue}{RGB}{0, 0, 255}
\definecolor{red}{RGB}{255, 0, 0}
\definecolor{pink}{RGB}{255, 128, 128}
\definecolor{green}{RGB}{0, 128, 0}
\definecolor{lightyellow}{RGB}{255, 255, 200}
\definecolor{purple}{RGB}{128, 0, 128}

\lstdefinestyle{all}
    {basicstyle=\ttfamily\scriptsize,
     keywordstyle=\color{blue}\ttfamily\scriptsize,
     commentstyle=\color{green}\ttfamily\scriptsize,
     stringstyle=\color{red}\ttfamily\scriptsize}

\lstdefinelanguage{hask}{%
    frame=none,
    xleftmargin=2pt,
    belowcaptionskip=\bigskipamount,
    captionpos=b,
    tabsize=2,
    numbers=left,
    numberstyle=\tiny\color{gray},
    emphstyle={\bf},
	morecomment=[s][\color{green}]{\{-}{-\}},
    stringstyle=\mdseries\rmfamily,
    commentstyle=\color{green},
    keywords={},
    keywords=[1]{case, of, data, if, then, else, where, let, in, do},
    keywords=[2]{Chip, Config, CurrencySymbol, TokenName, PubKeyHash, Integer, Value, State, Action, Text, Maybe, Void, TxConstraints,  Contract},
    keywords=[3]{HasNative},
    keywords=[4]{=>},
    keywords=[5]{Just, Nothing, MkChip, MkConfig, SetPrice, Buy},
    keywordstyle=[1]\mdseries\sffamily\color{red},
    keywordstyle=[2]\mdseries\sffamily\color{blue},
    keywordstyle=[3]\mdseries\sffamily\color{green},
    keywordstyle=[4]\mdseries\sffamily,
    keywordstyle=[5]\mdseries\sffamily\color{purple},
    columns=flexible,
    basicstyle=\small\sffamily,
    showstringspaces=false,
    breaklines=false,
    showspaces=false,
    escapeinside={--}{\^^M},escapebegin={\color{green}--},escapeend={},
    literate= {+}{{$+$}}1 {/}{{$/$}}1 {*}{{$*$}}1 {=}{{$=$}}1
              {>}{{$>$}}1 {<}{{$<$}}1 {\\}{{$\lambda$}}1
              {\\\\}{{\char`\\\char`\\}}1
              {->}{{$\rightarrow$}}2 {>=}{{$\geq$}}2 {<-}{{$\leftarrow$}}2
              {<=}{{$\leq$}}2 {=>}{{$\Rightarrow$}}2
              {\ .}{{$\circ$}}2 {\ .\ }{{$\circ$}}2
              {>>}{{>>}}2 {>>=}{{>>=}}2
              {|}{{$\mid$}}1
              {\_}{{\underline{\hspace{2mm}}}}2
}

\lstdefinelanguage{solidity}{%
    frame=none,
    xleftmargin=2pt,
    belowcaptionskip=\bigskipamount,
    captionpos=b,
    tabsize=2,
    numbers=left,
    numberstyle=\tiny\color{gray},
    emphstyle={\bf},
	morecomment=[s][\color{green}]{\{-}{-\}},
    stringstyle=\mdseries\rmfamily,
    commentstyle=\color{green},
    keywords={},
    keywords=[1]{pragma, solidity, contract, event, constructor, require, function, return, emit},
    keywords=[2]{address, uint, mapping},
    keywords=[3]{public, payable, external, view, returns},
    keywords=[4]{=>, +=, -=, =, <=, ==},
    keywords=[5]{msg, sender, transfer, value},
    keywordstyle=[1]\mdseries\sffamily\color{red},
    keywordstyle=[2]\mdseries\sffamily\color{blue},
    keywordstyle=[3]\mdseries\sffamily\color{green},
    keywordstyle=[4]\mdseries\sffamily,
    keywordstyle=[5]\mdseries\sffamily\color{purple},
    columns=flexible,
    basicstyle=\small\sffamily,
    showstringspaces=false,
    breaklines=false,
    showspaces=false,
    escapeinside={--}{\^^M},escapebegin={\color{green}--},escapeend={},
    literate= {+}{{$+$}}1 {/}{{$/$}}1 {*}{{$*$}}1 {=}{{$=$}}1
              {>}{{$>$}}1 {<}{{$<$}}1 {\\}{{$\lambda$}}1
              {\\\\}{{\char`\\\char`\\}}1
              {->}{{$\rightarrow$}}2 {>=}{{$\geq$}}2 {<-}{{$\leftarrow$}}2
              {<=}{{$\leq$}}2 {=>}{{$\Rightarrow$}}2
              {\ .}{{$\circ$}}2 {\ .\ }{{$\circ$}}2
              {>>}{{>>}}2 {>>=}{{>>=}}2
              {|}{{$\mid$}}1
              {\_}{{\underline{\hspace{2mm}}}}2
}

\newcommand*\cleartorightpage{%
  \clearpage
  \ifodd\value{page}\hbox{}\newpage\thispagestyle{empty}\hbox{}\newpage\else \hbox{}\newpage\fi
}

\newcommand\jamiepart[1]{\cleartorecto\part{#1}}
\newcommand\jamiesection[1]{\cleartorecto\chapter{#1}\thispagestyle{headings}}
\newcommand\jamiesubsection[1]{\section{#1}}
\newcommand\jamiesubsubsection[1]{\subsection{#1}}

\usepackage{xcolor,fix-cm}
\ifgreyprint\definecolor{numbercolor}{gray}{0.4}\else\definecolor{numbercolor}{rgb}{0.7,0,0}\fi
\definecolor{chaptertitlecolor}{gray}{0.2}
\newif\ifchapternonum
\makechapterstyle{jenor}{

}
\chapterstyle{jenor}

\setsecnumformat{\color{numbercolor}\csname the#1\endcsname\quad}

\newcommand\flanks{\ltimes}

\newcommand\declaresoundness[2]{The case of \rulefont{#1} with $#2$.}
\newcommand\declaresoundnessshort[1]{The case of \rulefont{#1}.}

\makeatletter
\newcommand\hpn[2][]{%
  \ext@arrow 9999{\hpnfill@}{#1}{#2}}
\newcommand\hpnfill@{%
  \arrowfill@\leftharpoonup\relbar\rightharpoondown}
\makeatother

\newcommand\id{\f{id}}

\NewCommandCopy{\oldin}{\in}
\renewcommand\in{{{\hspace{1pt}{\oldin}\hspace{1pt}}}}
\NewCommandCopy{\oldnotin}{\notin}
\renewcommand\notin{{{\hspace{1pt}{\oldnotin}\hspace{1pt}}}}
\newcommand\compactin{\in} %

\NewCommandCopy{\oldsetminus}{\setminus}
\renewcommand\setminus{{{\hspace{1pt}{\oldsetminus}\hspace{1pt}}}}

\newcommand\THREE{{\mathbf 3}}

\newcommand\binaryconnectives{\ensuremath{\{\tand,\tor,\tnotor,\tlatticeiff,\timp,\tiff\}}\xspace} %
\newcommand\unaryconnectives{\ensuremath{\{\tneg,\modT,\modTB,\modB\}}\xspace}

\newcommand\tvT{{\mathbf t}}
\newcommand\tvF{{\mathbf f}}
\newcommand\tvB{{\mathbf b}}

\newcommand\tvsTB{{\mathit{tb}}}
\newcommand\tvsFB{{\mathit{fb}}}
\newcommand\tvsTT{{\mathit{tt}}}
\newcommand\tvsFF{{\mathit{ff}}}
\newcommand\atag{\mathtt{a}}
\newcommand\btag{\mathtt{b}}

\newcommand\xor{\mathbin{\mathsf{\small xor}}}
\newcommand\modT{{\tf T}}
\newcommand\modTB{{\tf T\hspace{-2.5pt}\tf B}}
\newcommand\modB{\tf B} %

\newcommand\witno{\prec}
\newcommand\atopen{T}
\newcommand\afilter{F}
\newcommand\apoint{P}
\newcommand\avaluation{f} %
\newcommand\indicator[1]{\delta_{#1}}

\newcommand\rulefont[1]{\ensuremath{{\mathrm{\bf (#1)}}}}
\newcommand\leftopeninterval[1]{(#1]}
\newcommand\rightopeninterval[1]{[#1)}
\newcommand\openinterval[1]{(#1)}

\newcommand\opens{\tf{Open}}
\newcommand\regularOpens{\tf{Open}_{\f{reg}}}
\newcommand\topens{\tf{Topen}}
\newcommand\closed{\tf{Closed}}
\newcommand\regularClosed{\tf{Closed}_{\f{reg}}}

\newcommand\closure[1]{|{#1}|}

\newcommand{\dotarrow}{%
   \mathrel{\ooalign{\hss\raise.85ex\hbox{\scalebox{1.25}{\normalfont .}}%
   \kern0.35ex\hss\cr$\rightarrow$}}}

\newcommand\ttplus[1]{{\text{\texttt{+}}\mathtt{#1}}}
\newcommand\ttminus[1]{{\text{\texttt{-}}\mathtt{#1}}}

\usepackage{letltxmacro}%
\newcounter{fnmarkcntr}\newcounter{fntextcntr}
\makeatletter
\renewcommand{\footnotemark}{%
   \@ifnextchar[\@xfootnotemark
     {\stepcounter{fnmarkcntr}%
      \refstepcounter{footnote}\label{footnotemark\thefnmarkcntr}%
      \protected@xdef\@thefnmark{\thefootnote}%
      \@footnotemark}}
\makeatother
\LetLtxMacro{\oldfootnotetext}{\footnotetext}%
\renewcommand{\footnotetext}[1]{%
  \stepcounter{fntextcntr}%
  \oldfootnotetext[\ref{footnotemark\thefntextcntr}]{#1}%
}

\newcommand\onlineref[2]{\url{#1} (permalink: \url{#2})}
\newcommand\footnoteref[2]{\footnote{See \onlineref{#1}{#2}.}}

\newcommand\overlaps{{\rlap{$>$}\,{<}}}
\newcommand\notbetween{\mathbin{\cancel{\between}}}
\newcommand\notintertwinedwith{\mathrel{\notbetween}}

\newcommand\notintersectswith{\notbetween}

\newcommand\intertwined[1]{#1_{\between}}
\newcommand\intertwinedX[1]{#1_{\intertwinedwithX}}
\newcommand\intertwinedC[1]{#1_{\intertwinedwithC}}
\newcommand\intertwinedwith{\mathrel{\between}}
\newcommand\intertwinedwithX{\mathrel{\between^{\hspace{-1.5pt}\modT}}}
\newcommand\intertwinedwithC{\mathrel{\between^{\hspace{-1pt}\texttt{=}}}}
\newcommand\topind{\mathrel{\mathring{=}}} %
\newcommand\intertwinedwithwitness{\mathbin{\compressthis{\between}_{\hspace{-0pt}\scalebox{0.55}{$\witness$}}}}

\newcommand\leqk{\leq_{\hspace{-.7pt}\intertwinedwith}}
\newcommand\geqk{\geq_{\hspace{-.7pt}\intertwinedwith}}
\newcommand\cw{\leftrightarrow}  %
\newcommand\cti{\leq}  %
\newcommand\nbhd[0]{\f{nbhd}}
\newcommand\interior[0]{\f{interior}}
\newcommand\kiss[0]{\f{kiss}}
\newcommand\community[0]{\f{K}}
\newcommand\framecommunity[0]{\f{k}}
\newcommand\cast[1]{#1^{\ast c}}
\newcommand\cclo[1]{#1^c}
\newcommand\kernel[0]{\f{ker}}

\newcommand\witness[0]{{w\hspace{-1.8pt}f}}
\newcommand\barwitness[0]{{\bar{w}\hspace{-1.8pt}f}}
\newcommand\blocking[0]{\f{B}}
\newcommand\thyAxW{\tf{Ax}(\hspace{-0pt}{\witness})}
\newcommand\thyAxExW{\tf{AxEx}(\hspace{-0pt}{\witness})}

\newcommand\blocks[1]{\mathbin{\compressthislight{!}_{\hspace{-1pt}#1}}}
\newcommand\nblocks[1]{\mathbin{{\compressthislight{\cancel!}_{\hspace{-1pt}#1}}}}

\newcommand\enabledby[1]{\mathbin{{?}_{\hspace{-1pt}#1}}}
\newcommand\enables[1]{\mathbin{{?}_{\hspace{-1pt}#1}}}
\newcommand\notenable[1]{\mathbin{{\cancel?}_{\hspace{-1pt}#1}}}

\newcommand\Kmod[1]{{\f{\forall\hspace{-2pt}Val}}_{\hspace{-0pt}#1}\,}
\newcommand\Emod[1]{{\f{\exists\hspace{-2pt}Val}}_{\hspace{-0pt}#1}\,}

\makeatletter
\newcommand\@deffont[2][]{{\bfseries #2}\index{#1}}
\newcommand\deffont{\@dblarg\@deffont}
\makeatother
\newcommand\powerset{\f{pow}}

\newcommand\finpow{\f{fin}}

\newcommand\f[1]{\mathit{#1}}
\newcommand\tf[1]{\mathsf{#1}}
\newcommand\ns[1]{\bm{\mathsf{#1}}}

\newcommand\liff{\Longleftrightarrow}
\newcommand\limp{\Longrightarrow}

\newcommand\ssm{{{:}\text{=}}}
\DeclareMathSymbol{\shortminus}{\mathbin}{AMSa}{"39}
\newcommand\minus{{\shortminus}}
\newcommand\plus{{+}}
\newcommand\Forall[1]{\forall #1.}
\newcommand\Exists[1]{\exists #1.}

\newcommand\cent{\vdash}
\newcommand\ncent{\not\vdash}
\newcommand\ment{\vDash}
\newcommand\mentX{\ment^{{}^{\hspace{-3pt}\tf{X}}}}
\newcommand\nment{\nvDash}

\newcommand\boundary{\f{boundary}}
\newcommand\lmodel{[\hspace{-0.2em}[}
\newcommand\rmodel{]\hspace{-0.2em}]}
\newcommand\model[1]{{\lmodel #1 \rmodel}}

\newcommand\atmclosure[1]{{\langle #1 \rangle}_P}
\newcommand\mone{{\text{-}1}}
\newcommand\fv{\f{fv}}

\makeatletter
\DeclareRobustCommand{\barcent}{\mathbin{\mathpalette\barcent@@\relax}}
\newcommand{\barcent@@}[2]{%
  \vbox{\offinterlineskip
    \sbox\z@{$\m@th#1\cent$}%
    \ialign{%
      \hfil##\hfil\cr
      $\m@th#1{}_{\minus}\kern-\scriptspace$\cr
      \noalign{\kern-.3\ht\z@}
      \box\z@\cr
    }%
  }%
}
\makeatother

\makeatletter
\def\pmb@#1#2{\setbox8\hbox{$\m@th#1{#2}$}%
  \setboxz@h{$\m@th#1\mkern-.1mu$}\pmbraise@\wdz@
  \binrel@{#2}%
  \dimen@-\wd8 %
  \binrel@@{%
    \mkern-.1mu\copy8 %
    \kern\dimen@\mkern-.2mu\copy8 %
    \kern\dimen@\mkern-.3mu\copy8 %
    \kern\dimen@\mkern-.4mu\copy8 %
    \kern\dimen@\mkern.1mu\copy8 %
    \kern\dimen@\mkern.2mu\copy8 %
    \kern\dimen@\mkern.3mu\copy8 %
    \kern\dimen@\mkern.0mu\raise\pmbraise@\copy8 %
    \kern\dimen@\mkern.4mu\box8 %
           }%
}
\makeatother

\newcommand\limitat[1]{\textstyle\lim_{#1}}
\newcommand\compressthislight[1]{{\hspace{.8pt}\raisebox{.5pt}{\scalebox{.85}{$#1$}}\hspace{.2pt}}}
\newcommand\compressthis[1]{\pmb{\compressthislight{#1}}}

\newcommand\tneg{{\pmb\neg}}

\newcommand\ttop{{\pmb\top}}
\newcommand\tbot{{\pmb\bot}}
\newcommand\teq{{\pmb{\text{=}}}}
\newcommand\tand{{\pmb\wedge}}

\newcommand\tor{{\pmb\vee}}
\newcommand\timp{{\pmb\Rightarrow}}

\newcommand\tnotor{{\compressthis{\supset}}}
\newcommand\tlatticeiff{{\compressthis{\equiv}}}
\newcommand\tiff{\compressthis{\Leftrightarrow}} %

\newcommand\tall{{\compressthis{\forall}}}
\newcommand\texi{{\compressthis{\exists}}}

\makeatletter
\newcommand{\circlearrow}{}%
\DeclareRobustCommand{\circlearrow}{%
  \mathrel{\vphantom{\shortrightarrow}\mathpalette\circle@arrow\relax}%
}
\newcommand{\circle@arrow}[2]{%
  \m@th
  \ooalign{%
    \hidewidth$#1\circ\mkern1mu$\hidewidth\cr
    $#1\longrightarrow$\cr}%
}
\makeatother

\makeatletter
\newcommand*\bigcdot{\mathpalette\bigcdot@{.5}}
\newcommand*\bigcdot@[2]{\mathbin{\vcenter{\hbox{\scalebox{#2}{$\m@th#1\bullet$}}}}}
\makeatother

\usepackage{datetime}
\yyyymmdddate

\makeatletter
\newlength\drop
\newcommand*{\titleGM}{%
\thispagestyle{empty}
\begingroup%
\drop = 0.1\textheight
\vspace*{\baselineskip}
\vfill
\hbox{%
  \hspace*{0.2\textwidth}%
  \rule{1.5mm}{\dimexpr\textheight-28pt\relax}%
  \hspace*{0.05\textwidth}%
  \parbox[b]{0.75\textwidth}{%
    \vbox{%
      \vspace{\drop}
      {\Huge\bfseries\raggedright\@title\par}\vskip2.37\baselineskip
      {\Large\bfseries\@author\par}
      \vspace{0.05\textheight}
\ \\
\includegraphics[width=0.28\columnwidth]{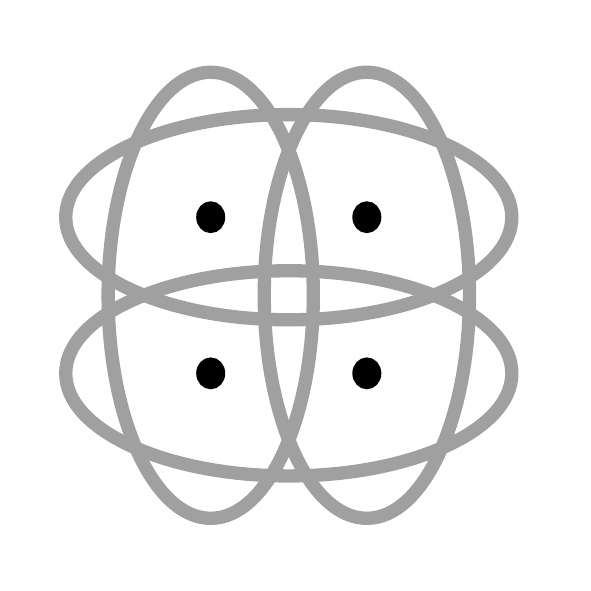}
      \vspace{0.3\textheight}
    }%
  }%
}%
\vfill
\null
\endgroup}
\makeatother

\AtBeginDocument{\addtocontents{toc}{\protect\thispagestyle{headings}}} 
\AtBeginDocument{\addtocontents{lof}{\protect\thispagestyle{headings}}} 

\title{Semitopology: decentralised collaborative action via topology, algebra, and logic} %

\author{Murdoch J. Gabbay} 

\makeindex

\begin{document}

\titleGM 
\thispagestyle{empty}

\pagenumbering{roman}
\setcounter{page}{4}

\tableofcontents

\clearpage
\begin{center}
    \thispagestyle{empty}
    \vspace*{\fill}
To my wife and son. 
Maths is delightful, and so are you.
    \vspace*{\fill}
\end{center}
\clearpage

\cleartorecto

\pagenumbering{arabic}
\setcounter{page}{1}

\jamiesection{Introduction}
\label{sect.intro}

\jamiesubsection{What is `decentralised collaborative action'?}
\label{subsect.what.is}

This book reports on the outcome of a mathematical investigation into what I will call \emph{decentralised collaborative action}, motivated by recent developments in \emph{decentralised permissionless heterogeneous computing systems}.
Let's unpack the jargon:
\begin{itemize*}
\item
A system is \emph{decentralised} when it is distributed over several machines, and the system as a whole is not centrally controlled.

Most blockchain systems and peer-to-peer networks are decentralised in the sense we intend, or at least they are supposed to be.
They are distributed over multiple participants, and no single entity controls the system. 
The internet is also (mostly) decentralised, at least in principle.\footnote{The internet was designed to be an information network that would be resilient to nuclear attack.  It did this by being `centrifugal'; emphasising node-to-node actions instead of centre-to-centre actions.  See~\cite{ryan:hisidf}, summarised by Ars Technica~\cite{ars-technica:howabg}.} 
\item
A system is \emph{permissionless} (or \emph{unpermissioned}) when participants can leave and join the system at any time.

Nature is naturally permissionless (living things do not need permission to be born or die).
National voting systems \emph{are} permissioned (because citizens require certification from the government to be allowed to vote).
\item
A system is \emph{heterogeneous} when participants may be following different rules.\footnote{By `different rules' we include the situation where an algorithm (such as a consensus algorithm) is agreed between participants but a critical parameter may vary substantively across them, e.g. imagine a blockchain in which some participants require a ${>}2/3$ majority to act, and others require just a ${>}1/2$ majority.}

Ethereum and Tezos are decentralised and permissionless, but they are not heterogeneous in the sense we intend.
If you are running a Tezos or Ethereum node, then you are not forced to follow the rules, but if you do not then by definition you are not following the rules.

In contrast, consider the combination of Tezos and Ethereum as a single system connected by a \emph{blockchain bridge}.\footnoteref{https://ethereum.org/en/bridges/}{https://web.archive.org/web/20240324090911/https://ethereum.org/en/bridges/}
This is heterogeneous, because Tezos nodes and Ethereum nodes have different rules and different consensus mechanisms.
A Tezos node is not a bad node just because it is not following the rules of Ethereum, and vice-versa, but because of the blockchain bridge, they can be considered to be operating within a single (heterogeneous) combined system.
\end{itemize*}
So a decentralised heterogeneous permissionless system consists of \emph{some} participants communicating to do \emph{something}, with no \emph{a priori} restrictions on who, what, or how.

If you are a blockchain engineer then this scenario --- with its weak well-behavedness assumptions that do not even assume all participants share a common ruleset --- might seem a terrible idea which we should not allow, because it admits %
crazy networks with bad behaviour.
But here the generality is a feature, not a bug:
\begin{enumerate*}
\item
As mathematicians, we \emph{want} to admit general models, including bad ones, so that we can formalise their good and bad behaviour\footnote{\dots which will vary by application; e.g. sometimes all participants should play by the same rules, but in the case of a blockchain bridge we specifically want to \emph{admit} different rules.} and express conditions to include or exclude it.
\item
Surprisingly, it will turn out that there is still a lot that we can say even about the general case, and we shall see that much useful structure will emerge \emph{even from very weak assumptions} (a detailed summary is listed in Subsection~\ref{subsect.map}).
\end{enumerate*} 
So granted that the generality of decentralised collaborative action is a feature, not a bug; but how should we approach this mathematical generality?
The key is to look at how groups of participants can \emph{progress their local state}.
To see this we need a little more discussion.

In a decentralised system, a participant must store state locally --- if there were a global source of truth for state then whoever controls that truth would \emph{de facto} control the system --- 
and communicate with other participants to decide on how their local states evolve.
There must be \emph{some} rules about how this state should be updated, even if these rules may differ across participants in the system, and even though the rules may not always be followed. 
It turns out that one common feature of decentralised heterogeneous permissionless systems is a notion of what I will call an \emph{actionable coalition},
by which I intend
\begin{quote}
\emph{a set of participants who are legally entitled (but not obliged) to collaborate to update their local state (possibly but not necessarily in identical ways).}
\end{quote}
We will consider three examples:
\begin{enumerate*}
\item
Ethereum.

Ethereum's consensus protocol is proof-of-stake, so an actionable coalition on Ethereum is any group of participants who hold a majority stake of tokens (this is a bit of a simplification, but it will do).
\item
Ethereum and Tezos with a blockchain bridge between them.

Tezos's consensus protocol is also proof-of-stake.
An actionable coalition in this system is either an actionable coalition of Ethereum, or one of Tezos, \emph{or} the sets union of an actionable coalition from each, along with the bridging node (again, a simplification, but it will do).\footnote{Typically, participants can update their state if they held a majority of the stake at some time in the past (e.g. two weeks ago) --- the idea being that all participants have reached agreement on, and learned, the state of the network two weeks in the past, so this can be treated as immutable common knowledge without undermining the decentralised nature of the system in the present~\cite[Subsection~3.2.1, final paragraph]{tezos:whitepaper}.}
\item
A Tango dance evening where men will only dance with women and vice-versa.\footnote{Many dancers can lead as well as follow, but for the sake of the mathematics we will simplify.}

An actionable coalition is any set containing equal numbers of male leads and female followers.
\end{enumerate*}
If an actionable coalition can communicate to agree on a set of local state updates, e.g. if the Tango lead leads a move and the Tango follower chooses to follow it, then the participants in this coalition are entitled to update their local states accordingly.
Note that local state updates need not be literally identical across participants; they just need to be mutually agreed upon and then actioned.

Some important notes:
\begin{enumerate*}
\item
The actionable coalition can progress \emph{without} consulting the rest of the system.
\item
Being in an actionable coalition does not imply control.
This set describes a potential legal collaboration, but participants can choose what actionable coalition to work with, if any, and they can also choose not to follow the rules.\footnote{If you put your elbow into your dance partner's eye, or simply deliver a poor lead or a poor follow, then the other dancer might stop dancing with you or turn you down if you ask for another dance.
But neither of you are \emph{compelled} to dance with one another, and if you do, you are not \emph{compelled} to dance well.}
\item 
If $O$ is an actionable coalition for $p\in O$, and $p'\in O$ is another participant in $O$, then $O$ is also an actionable coalition for $p'$.
Note that this makes actionable coalitions look a bit like open sets in a topology.
\end{enumerate*}
So we can now introduce our first mathematical abstraction: we identify participants as \emph{points}, and we let let \emph{open sets} be \emph{actionable coalitions}.
An actionable coalition is a \emph{coalition of participants with the capacity to act}.
They are not obliged to act, and if they do their action need not be identical across all participants, but the potential exists for this set to collaborate to progress their states.
\begin{enumerate*}
\item
With reference to our couples dance example:
an example of an actionable coalition that is not minimal is a set containing two male leads and two female followers.
There are two ways for the participants to pair off to collaborate (i.e. dance).
\item
With reference to our bridged blockchain example: an example of a set that contains an actionable coalition but is not one itself is an actionable coalition from Ethereum, along with the bridging node.
The Ethereum coalition on its own is actionable, but the bridging node cannot take any action without also collaborating with an actionable coalition from Tezos.
\end{enumerate*} 

To get a flavour of our mathematical results, consider a fundamental problem in any decentralised system: ensuring that its participants remain in agreement, for some suitable sense of `agree'.
To take a simple example from blockchain: if we reach a situation where half of the nodes say that we have paid for a service, and the other half say that we have not --- then \emph{everyone} has a problem, because the system has become incoherent and it is not clear how the system as a whole can restore coherence and progress.\footnote{coherent (adj.) 1550s, ``harmonious;'' 1570s, ``sticking together,'' also ``connected, consistent'' (of speech, thought, etc.), from French cohérent (16c.), from Latin cohaerentem (nominative cohaerens), present participle of cohaerere ``cohere,'' from assimilated form of com ``together'' (see co-) + haerere ``to adhere, stick'' (etymologyonline: \url{https://www.etymonline.com/word/coherent}).}
This phenomenon is called \emph{forking}, and blockchain designers really want to avoid it!

We will call our mathematical abstraction of agreement, \emph{antiseparation}.
In a little more detail, antiseparation properties are coherence properties that are guaranteed to hold of a decentralised system 
\emph{just} by analysing the structure of its actionable coalitions.
It turns out that we can get surprisingly detailed information about agreement/antiseparation properties, even working from quite weak and abstract mathematical assumptions on the actionable coalitions.

We emphasise this point: sometimes we can predict important macro properties of a system's behaviour without knowing anything about its specifics, so long as we have certain good properties on its actionable coalitions.

Let us start by considering a simple situation where participants are trying to agree on a binary consensus problem: whether to announce a single value `true' or `false'.
Continuing the theme of simplicity, assume some finite nonempty set of participants $\mathbb E$ and let their actionable coalitions be just any set of participants that forms a majority (so it contains strictly more than half of the set of all participants).
Now suppose that the participants in some actionable coalition $O\subseteq\mathbb E$ have communicated and have agreed on `true'.
Because they form an actionable coalition, they are entitled to act and to announce `true', and so they do.
They have now all committed to this state update and they cannot change their minds.

So: can this system fork?
Consider some participant $p\not\oldin O$.
If $p$ wants to make progress, is must also agree on `true', because all of its actionable coalitions intersect with $O$ and so contain at least one participant that has committed to `true' and cannot change its mind.
This does not mean that $p$ has to agree on `true'; it could choose not to progress, or it could break the rules.
But, by definition if $p$ does want to progress legally, then the decision has been made and it must eventually go along with the majority.
Thus, we have proved that any progress that is made by one participant within the rules (\dots must be shared with some actionable coalition of that participant, and since all such coalitions intersect it \dots) must eventually be followed any other participant that also progresses.
Thus forking is impossible.

The reader may already be familiar with this example, but note that this antiseparation property comes simply \emph{from the structure of the actionable coalitions}.
There is no need to consider the protocol, or even how values are interpreted.
It turns out that antiseparation-style behaviour is common, and arises even if we do not require actionable coalitions that are simple majorities.
For example, let participants be $\mathbb Z=\{0,1,\minus 1,2,\minus 2,\dots\}$ and let actionable coalitions be generated by sets of three consecutive numbers starting at an even number $\{2i,2i\plus 1,2i\plus 2\}$, and suppose again that we are trying to agree on `true' or `false'.
Note that in contrast to the previous example, actionable coalitions need not intersect.
Yet, the moment one triplet of participants commits to `true', the rest of the system is obliged to eventually agree, if all participants play by the rules.
Now this example system is not necessarily particularly safe or desirable in practice, because we can imagine that $\{0,1,2\}$ agree on `true', and $\{4,5,6\}$ acting independently but in good faith agree on `false', and then $3$ cannot legally progress, because within $\{2,3,4\}$, $2$ has announced `true' and $4$ has announced `false' and $3$ cannot agree with both.
But, we know that \emph{if} all participants do legally progress, then they announce the same value.
So this example illustrates how antiseparation can arise even when actionable coalitions are rather small.\footnote{See also Remark~\ref{rmrk.transitive.comment}.}

The two examples above are quite different.
In one, all actionable coalitions intersect, and in the other they mostly do not.
This suggests that a `general mathematics of (anti)separation' is possible, based on the study of actionable coalitions.
In a nutshell, that mathematical story is what we will develop.

\jamiesubsection{What is a semitopology?}

So at a high level, what do we have?
\begin{enumerate*}
\item
There is a notion of what we can call an \emph{actionable coalition} (or just: \emph{open set}).
This is a set $O\subseteq\ns P$ of participants with the capability, though not the obligation, to act collaboratively to advance (= update / transition) the local state of the elements in $O$, possibly but not necessarily in the same way for every $p\in O$. 
\item
$\varnothing$ is trivially an actionable coalition.
Also we assume that $\ns P$ is actionable, effectively assuming that every point is a member of at least one actionable coalition.
\item
A sets union of actionable coalitions, is an actionable coalition.
\end{enumerate*}
This leads us to the definition of a semitopology.

\begin{nttn}
\label{nttn.powerset}
Suppose $\ns P$ is a set.
Write $\powerset(\ns P)$ for the powerset of $\ns P$ (the set of subsets of $\ns P$); there will be more on this in Notation~\ref{nttn.finpow}.
\end{nttn}

\begin{defn}
\label{defn.semitopology}
A \deffont{semitopological space}, or \deffont{semitopology} for short, consists of a pair $(\ns P, \opens(\ns P))$ of 
\begin{itemize*}
\item
a (possibly empty) set $\ns P$ of \deffont{points}, and 
\item
a set $\opens(\ns P)\subseteq\powerset(\ns P)$ of \deffont[open sets $\opens$]{open sets}, 
\end{itemize*}
such that:
\begin{enumerate*}
\item\label{semitopology.empty.and.universe}
$\varnothing\in\opens(\ns P)$ and $\ns P\in\opens(\ns P)$.
\item\label{semitopology.unions}
If $X\subseteq\opens(\ns P)$ then $\bigcup X\in\opens(\ns P)$.\footnote{There is a little overlap between this clause and the first one: if $X=\varnothing$ then by convention $\bigcup X=\varnothing$.  Thus, $\varnothing\in\opens(\ns P)$ follows from both clause~1 and clause~2.  If desired, the reader can just remove the condition $\varnothing\in\opens(\ns P)$ from clause~1, and no harm would come of it.} 
\end{enumerate*}
We may write $\opens(\ns P)$ just as $\opens$, if $\ns P$ is irrelevant or understood, and we may write $\opens_{\neq\varnothing}$ for the set of nonempty open sets.
\end{defn}

The reader will recognise a semitopology as being like a \emph{topology} on $\ns P$ \cite{engelking:gent,willard:gent}, but without the condition that the intersection of two open sets necessarily be an open set.
This reflects the fact that the intersection of two actionable coalitions need not itself be an actionable coalition.

Armed with this simple definition and bearing in mind its modern relevance as noted above, we introduce and survey semitopologies and their properties.
There is an emphasis (though not an exclusive one) on studying decentralised collaborative actions, which (broadly speaking) amounts to studying antiseparation properties of points, and how this interacts with topological continuity of functions out of semitopologies. 
The details of what this means are unpacked below. 

We will proceed in three parts, by which we hope to give a comprehensive overview of our approach to decentralised collaborative action: 
\begin{quote}
point-set semitopologies, algebra, and logic.
\end{quote}
The point-set semitopologies are most pertinent to concrete models (i.e. real networks), the algebra is most pertinent to clarifying an abstract (algebraic) view of what are the essential structures in play, and the logic is pertinent to specifying properties and --- because logic is a portal to computation --- computing/checking these properties.

\begin{rmrk}
Traditional notions of consensus and voting can be understood in a semitopological framework.
For instance, a committee may make a decision by two-thirds majority vote; the set of all 2/3 majorities of some $\ns P$ is a semitopology (note that it is not a topology).
Also, concrete algorithms to attain consensus often use a notion of \emph{quorum}~\cite{lamport_part-time_1998,lamport:byzgp} as a set of participants whose unanimous adoption of a value guarantees that other (typically all other) participants will eventually also adopt this value.
Social choice theorists have a similar notion called a \emph{winning coalition}~\cite[Item~5, page~40]{riker:thepc}.
If the reader has a background in logic then they may be reminded of a whole field of \emph{generalised quantifiers} (a good survey is in~\cite{sep-generalized-quantifiers}).

The reader should just note that these examples have a synchronous, centralised flavour. 

For instance: a vote in the typical democratic sense is a synchronous, global operation (unless the result is disputed): votes are cast, collected, and then everyone gets together --- e.g. in a vote counting hall --- to count the votes and agree on who won and so certify the outcome.\footnote{I have seen this happen; votes being tallied up while under supervision by representatives of all parties on the ballot.  It is a moving sight.  But it is not decentralised.}
This is certainly a collaborative action, but it is centralised.

Our semitopological framework adds to the above by allowing us to study \emph{decentralised} collaborative action, which can progress by local state updates on actionable coalitions (which certainly do not need to be simple majorities), and they can so act without \emph{necessarily} having to synchronise step-by-synchronous-step on global state updates.
\end{rmrk}

\begin{rmrk}
\label{rmrk.collaboration}
Let us take a moment to give a high-level motivation for semitopologies, in the style of (very abstract) justifications that have been given for topologies. 
A classic text on topology~\cite{vickers:topvl} justifies topology as follows: 
\begin{enumerate*}
\item
Logically, open sets model \emph{affirmations}:\footnote{\emph{Affirmation:}\ Something declared to be true; a positive statement or judgment.  \onlineref{https://www.wordnik.com/words/affirmation}{https://web.archive.org/web/20230608073651/https://www.wordnik.com/words/affirmation}.} an open set $O$ corresponds to an affirmation (of $O$)~\cite[page~10]{vickers:topvl}.\footnote{\emph{Side note:}\ I sometimes get asked here why in topology, open sets are closed under \emph{arbitrary} unions but only \emph{finite} intersections.  Surely an infinite collection of positive affirmations is still a positive affirmation?  No, not quite: consider the affirmation `I am a nonempty open neighbourhood of $\pi$ in $\mathbb Q$'.  This is closed under finite intersections, but not infinite intersections.}
\item
Computationally, open sets model \emph{semidecidable properties}.  See the first page of the preface in~\cite{vickers:topvl}.
\end{enumerate*}
The notion of an actionable coalition justifies semitopologies in similar terms. 
\begin{enumerate*}
\item
Logically, open sets model \emph{collaborative} affirmations. 
An open set corresponds to an affirmation (by collaboration between the elements of $O$). 
\item
Computationally, open sets model \emph{actionable properties}, this being a (semidecidable) property that furthermore \emph{enough participants agree on that it can be acted on}.
\end{enumerate*}
Alternatively, we could just read this document in the spirit of pure mathematics --- I am sympathetic to this view --- in which case a test for whether semitopologies are an interesting definition is just whether we get interestingly more structure out of it than we put in to the definition. 
We shall see that this is indeed the case.
\end{rmrk}

\jamiesubsection{Who should read this document?}

\begin{enumerate}
\item
\emph{Practitioners,} especially those using quorum systems and fail-prone systems, looking for a mathematical framework that subsumes what they are already doing, puts it in a broader context, creates a common language to speak with one another and with mathematicians, and suggests new engineering options.
\item 
\emph{Theoreticians,} looking to use semitopologies to construct the next generation of advanced decentralised computer systems --- especially where the maths makes it easier to reuse existing tools whose applicability would not otherwise be clear (e.g. using SAT solvers to compute semitopological antiseparation properties from Part~\ref{part.1} using the logic in Part~\ref{part.3}).
\item
\emph{Pure mathematicians,} who might be delighted to discover a new topology-adjacent field and might see it as a fresh research opportunity.\footnote{E.g. \emph{algebraic semitopology} does not exist yet, nor does \emph{semitopological epistemological logic}; but perhaps somebody might read this work and be inspired to create them.}

Pure mathematicians can also learn a lot from this material regarding what things are important and interesting to look at, and what distinctions make a difference in practice; I know that I have.
\item
\emph{Mathematicians looking to get into practical systems.}

Real systems can be messy because they have to accommodate a messy reality.
Semitopologies provide a useful abstraction that can help understand what is going on at a high level.
\end{enumerate} 

\jamiesubsection{Why did I write it?}

\begin{enumerate*}
\item
Numerous authors have recently studied designing systems where participants have different opinions on who is part of the system or on who is trustworthy or not~\cite{Alpos2024,sheff_heterogeneous_2021,cachin_quorum_2023,li_quorum_2023,bezerra_relaxed_2022,garcia2018federated,lokhafa:fassgp,losa:stecbi,florian_sum_2022,li_open_2023}.
These systems go by names such as \emph{(permissionless) fail-prone systems} and \emph{(heterogeneous) quorum systems} (more discussion, with more references, is in Subsection~\ref{subsect.related.work}).

Most of these systems are (or to be more precise: they directly give rise to) semitopologies, and we would argue that the literature above is, in fact, \emph{rediscovering topology through semitopology}, but they did not know it. 
This document makes the connection to classical mathematics explicit, and builds on it in interesting ways. 
\item
Like topology, semitopology is not a single theorem; it is a method and a research topic. 
As such, the topic needs a text to describe its current scope, and show how the parts of the theory fit together.\footnote{\dots if only so that others can build on this and do even more, and even better!} 

For instance, the algebraic theory is motivated by the point-set semitopologies; and conversely, to fully understand what point-set semitopologies really are, we need the algebraic theory.
Similarly for the logic. 
Here, we have the space to tell a proper story arc. 
\item
By design, this document is interdisciplinary, speaking to two communities: 
\begin{itemize*}
\item
\emph{pure mathematicians}, who probably know lots about point-set topology but maybe know less about modern consensus or blockchain systems, and 
\item
\emph{practitioners}, who conversely are very familiar with the engineering, and may have already reinvented (semi)topologies but not realise it, but may be unfamiliar with the terminology and methods of pure mathematics.\footnote{This is not a hypothetical.}
\end{itemize*}
I intend this maths to be accessible and useful to both types of reader, and hope thereby to make a constructive difference to the development of both the theory and practice of decentralised systems.
\end{enumerate*}

\jamiesubsection{Map of the work}
\label{subsect.map}

As already mentioned, this work is in three parts: 
\begin{enumerate*}
\item
We consider point-set semitopologies in Sections~\ref{sect.semitopology} to~\ref{sect.dense}.
\item
We then take an algebraic, point-free, categorical approach by studying semiframes in Sections~\ref{sect.semiframes} to~\ref{sect.graphs}. 
\item
Finally, we consider logic over semitopologies in Sections~\ref{sect.three} to~\ref{sect.extremal.valuations}. 
\end{enumerate*}
In a nutshell, we will study the topology, algebra, and logic of semiframes, as follows:
\begin{enumerate}
\item
Section~\ref{sect.intro} is the Introduction.  You Are Here.
\end{enumerate}
\subsubsection*{\sc Semitopologies}
\begin{enumerate}
\setcounter{enumi}{1}
\item
In Section~\ref{sect.semitopology} we show how \textbf{continuity corresponds to local agreement} (Definition~\ref{defn.semitopology} and Lemma~\ref{lemm.open.lc}).
\item
In Section~\ref{sect.transitive.sets} we discuss \textbf{transitive sets}, \textbf{topens}, and \textbf{intertwined points}.
These are all different views on the anti-separation well-behavedness properties that will interest us. 
Most of Section~\ref{sect.transitive.sets} is concerned with showing how these different views relate and in what senses they are equivalent (e.g. Proposition~\ref{prop.cc.char}).
Transitive sets are guaranteed to be in agreement (in a sense made precise in Theorem~\ref{thrm.correlated} and Corollary~\ref{corr.correlated.intersect}), and we take a first step to understanding the fine structure of semitopologies by proving that every semitopology partitions into topen sets (Theorem~\ref{thrm.topen.partition}), plus other kinds of points which we classify in the next Section.
\item
In Section~\ref{sect.regular.points} we start to classify points in more detail, introducing notions of \textbf{regularity} for points in Definition~\ref{defn.tn}.
This is the start of a classification of good properties for points in semitopologies, including: regular, weakly regular, indirectly regular, quasiregular, unconflicted, hypertransitive, and more.
\item
In Section~\ref{sect.closed.sets} we study \textbf{closed sets}, and in particular the interaction between intertwined points, topens, and closures.
Typical results are Proposition~\ref{prop.intertwined.as.closure} and Theorem~\ref{thrm.up.down.char} which characterise sets of intertwined points as minimal closures.
The significance to consensus is discussed in Remarks~\ref{rmrk.fundamental.consensus} and~\ref{rmrk.why.top.closure}.
\item
In Section~\ref{sect.unconflicted.point} we study \textbf{unconflicted} and \textbf{hypertransitive} points, leading to two useful characterisations of regularity in Theorems~\ref{thrm.r=wr+uc} and~\ref{thrm.regular=qr+sc}.
\item
In Section~\ref{sect.product} we consider \textbf{product semitopologies}.
These are defined just as for topologies (Definition~\ref{defn.product.semitopology}) but we study how the semitopological properties we have considered above --- like being intertwined, topen, regular, conflicted, and so forth --- interact with taking products.
This is also useful for building large complex counterexamples out of smaller simpler ones (examples in Corollary~\ref{corr.conflicted.and.not.wr} or Theorem~\ref{thrm.nitpicked}).
\item
In Section~\ref{sect.witness} we construct a novel theory of computationally tractable semitopologies, based on \textbf{witness functions} (Definition~\ref{defn.witnessed.set}(\ref{witness.function})).
We call semitopologies generated by witness functions \emph{witness semitopologies}.
These display excellent algorithmic behaviour (Remarks~\ref{rmrk.computing.open.sets} and~\ref{rmrk.computing.closed.sets}) and we note deep reasons why this is so by showing that witness functions correspond to Horn clause theories, and that open and closed sets in the witness semitopology are related to answer sets to those theories; see Subsection~\ref{subsect.declarative.witness}.
\item
In Section~\ref{sect.cc.cb} we introduce \textbf{(strongly) chain-complete} semitopologies.
We argue in Remark~\ref{rmrk.plausible.abstraction} that these have properties making them a suitable abstraction of finite semitopologies --- finite semitopologies are of particular interest because these are the ones that we can build.
We study their properties and prove a key result that witness semitopologies are chain-complete (Theorem~\ref{thrm.lim.O.open}), even if they are infinite.\footnote{We discuss why infinite semitopologies matter, even in a world of finite implementations, in Remark~\ref{rmrk.infinite}.  Note also that in a real system there may be hostile participants who report an unbounded space of `phantom' points, either for denial-of-service or to create `extra voters'.  So even a system that is physically finite may present itself as infinite.}
\item
A key property in a strongly chain-complete semitopology is that the poset of open sets is \emph{atomic}, i.e. minimal nonempty open sets always exist.
In Section~\ref{sect.kernels} we study \textbf{kernels} --- unions of atomic transitive open sets --- especially in strongly chain-complete semitopologies where atoms are guaranteed to exist. 
We will see that the kernel dictates behaviour in a sense we make formal (see discussion in Remark~\ref{rmrk.arrow}).
\item
In Section~\ref{sect.dense} we study notions of \textbf{dense subset of} from topology and see that this splits into two notions: \emph{weakly dense in} and \emph{strongly dense in} (Definition~\ref{defn.dense}).
Transitivity turns out to be closely related to denseness (Proposition~\ref{prop.most.general}).
We prove a continuous extension result and show that this leads naturally back to the notion of regular point and topen set which we developed to begin with (Remark~\ref{rmrk.top.ce}).
\end{enumerate}
\subsubsection*{\sc Semiframes}
\begin{enumerate}
\setcounter{enumi}{11}
\item
In Section~\ref{sect.semiframes} we introduce \textbf{semiframes}.
These are the algebraic version of semitopologies, and they are to semitopologies as frames are to topologies.
We discover that semiframes are not just join-semilattices; \textbf{semiframes are \emph{compatible} semilattices}, which include a \emph{compatibility relation} $\ast$ to abstract the property of sets intersection $\between$ (see Remark~\ref{rmrk.amazing}).
\item
In Section~\ref{sect.semifilters.and.points} we introduce \textbf{semifilters}.
These play a similar role as filters do in topologies, except that semifilters have a \emph{compatibility condition} instead of closure under finite meets.
We develop the notion of abstract points (completely prime semifilters), and show how to build a semitopology out of the abstract points of a semiframe. 
\item
In Section~\ref{sect.spatial.and.sober} we introduce \textbf{sober semitopologies} and \textbf{spatial semiframes}.
The reader familiar with categorical duality will know these conditions.
Some of the details are significantly different (see for instance the discussion in Subsection~\ref{subsect.sober.top.contrast}) but at a high level 
these conditions work in the proofs just as they do for the topological duality.
\item
In Section~\ref{sect.duality} we consider the \textbf{duality} between suitable categories of (sober) semitopologies and (spatial) semiframes.
\item
In Section~\ref{sect.closer.look.at.semifilters} we \textbf{dualise the well-behavedness conditions} from Section~\ref{sect.transitive.sets} to algebraic versions.
The correspondence is good (Proposition~\ref{prop.regular.match.up}) but also imperfect in some interesting ways (Remark~\ref{rmrk.summary.of.sc}).
\item
In Section~\ref{sect.graphs} we briefly consider alternative \textbf{graph-based representations} of semitopologies.
\end{enumerate}
\subsubsection*{\sc Logic}
\begin{enumerate}
\setcounter{enumi}{17}
\item
In Section~\ref{sect.three} we introduce a \textbf{three-valued modal logic} to describe properties of semitopologies.
Logic is closely related to computation, so what we can describe we can --- by passing it to a solver or a prover --- also compute.
\item
In Section~\ref{sect.axiomatisations} we use our logic to start \textbf{axiomatising}.
In particular, in Definition~\ref{defn.Ax} we carry out the most basic task and write down axioms in our logic that correspond to continuity.
\item 
In Section~\ref{sect.logical.intertwinedwith} we \textbf{axiomatise the regularity properties} that we have already investigated. 
\item
In Section~\ref{sect.logic.programming} we look at computing in more detail.
We show how to \textbf{convert a witness function into a logical theory} suitable for passing to a SAT solver, we investigate notions of Horn clause programming (both for two-valued and three-valued logic), and we show in Theorem~\ref{thrm.NP.complete} that \textbf{determining whether two points are intertwined is NP-complete} in general.
\item 
In Section~\ref{sect.extremal.valuations} we open up a fresh line of inquiry and consider \textbf{extremal valuations}, which roughly speaking correspond to final states of a system in which every participant who could return a definite value, has returned a definite value.
This turns up to open a new design space which we put in the context of the maths thus far. 
\end{enumerate}

\subsubsection*{\sc Conclusions}
\begin{enumerate}
\setcounter{enumi}{22}
\item
In Section~\ref{sect.conclusions} we conclude and discuss related and future work.
\end{enumerate}

\begin{rmrk}
Algebraic topology has been applied to the solvability of distributed-computing tasks in various computational models (e.g. the impossibility of wait-free $k$-set consensus using read-write registers and the Asynchronous Computability Theorem~\cite{herlihy_asynchronous_1993,borowsky_generalized_1993,saks_wait-free_1993}; see~\cite{herlihy_distributed_2013} for a survey).
Semitopology is not topology, and this work is not about algebraic topology applied to the solvability of distributed-computing tasks!

We are interested in the mathematics of actionable coalitions, as made precise by point-set semitopologies; their antiseparation properties; and the implications to partially continuous functions on of them.
If we discuss distributed systems, it is by way of providing motivating examples or noting applicability.
\end{rmrk}

\jamiepart{Point-set semitopologies}
\label{part.1}
\jamiesection{Semitopology}
\label{sect.semitopology}

\jamiesubsection{Definitions, examples, and some discussion}

\jamiesubsubsection{Definitions}

Recall from Definition~\ref{defn.semitopology} the definition of a semitopology.

\begin{rmrk}
\label{rmrk.two.ways.to.think}
\leavevmode
\begin{enumerate*}
\item
As a sets structure, a semitopology on $\ns P$ is like a \emph{topology} on $\ns P$, but without the condition that the intersection of two open sets be an open set.
\item
As a lattice structure, a semitopology on $\ns P$ is a 
bounded complete join-subsemilattice of $\powerset(\ns P)$.\footnote{\emph{Bounded} means closed under empty intersections and unions, i.e. containing the empty and the full set of points.  \emph{Complete} means closed under arbitrary (possibly empty, possibly infinite) sets unions.
The reader may know that a complete lattice is also co-complete: if we have all joins, then we also have all meets.  However, note that there is no reason for the meets in $\opens$ to coincide with the meets in $\powerset(\ns P)$, i.e. for them to be sets intersections.  

Also, note that this does not mean that semitopologies are `just' bounded complete join-subsemilattices.  They are in fact \emph{compatible} bounded complete join-semilattices.  See Section~\ref{sect.semiframes}.
}
\item
Every semitopology $(\ns P,\opens)$ induces two natural topological completions: the least topology that contains $\opens$, and the greatest topology contained in $\opens$.
But there is more to semitopologies than just their topological completions, because:
\begin{enumerate*}
\item
We are explicitly interested in situations where intersections of open sets need \emph{not} be open.
\item
Completing to a topology loses information.
For example: the `many', `all-but-one', and `more-than-one' semitopologies in Example~\ref{xmpl.semitopologies} express three distinct notions of quorum, yet if $\ns P$ is infinite then for all three, the least topology containing them is the discrete semitopology (Definition~\ref{defn.value.assignment}(\ref{item.discrete.semitopology})), and the greatest topology that they contain is the trivial topology $\{\varnothing,\ns P\}$ (Example~\ref{xmpl.semitopologies}(\ref{item.trivial.topology})).
See also the overview in Subsection~\ref{subsect.vs}. 
\end{enumerate*}
\end{enumerate*}
\end{rmrk}

\begin{figure}
\vspace{-1em}
\centering
\includegraphics[align=c,width=0.4\columnwidth,trim={50 120 50 120},clip]{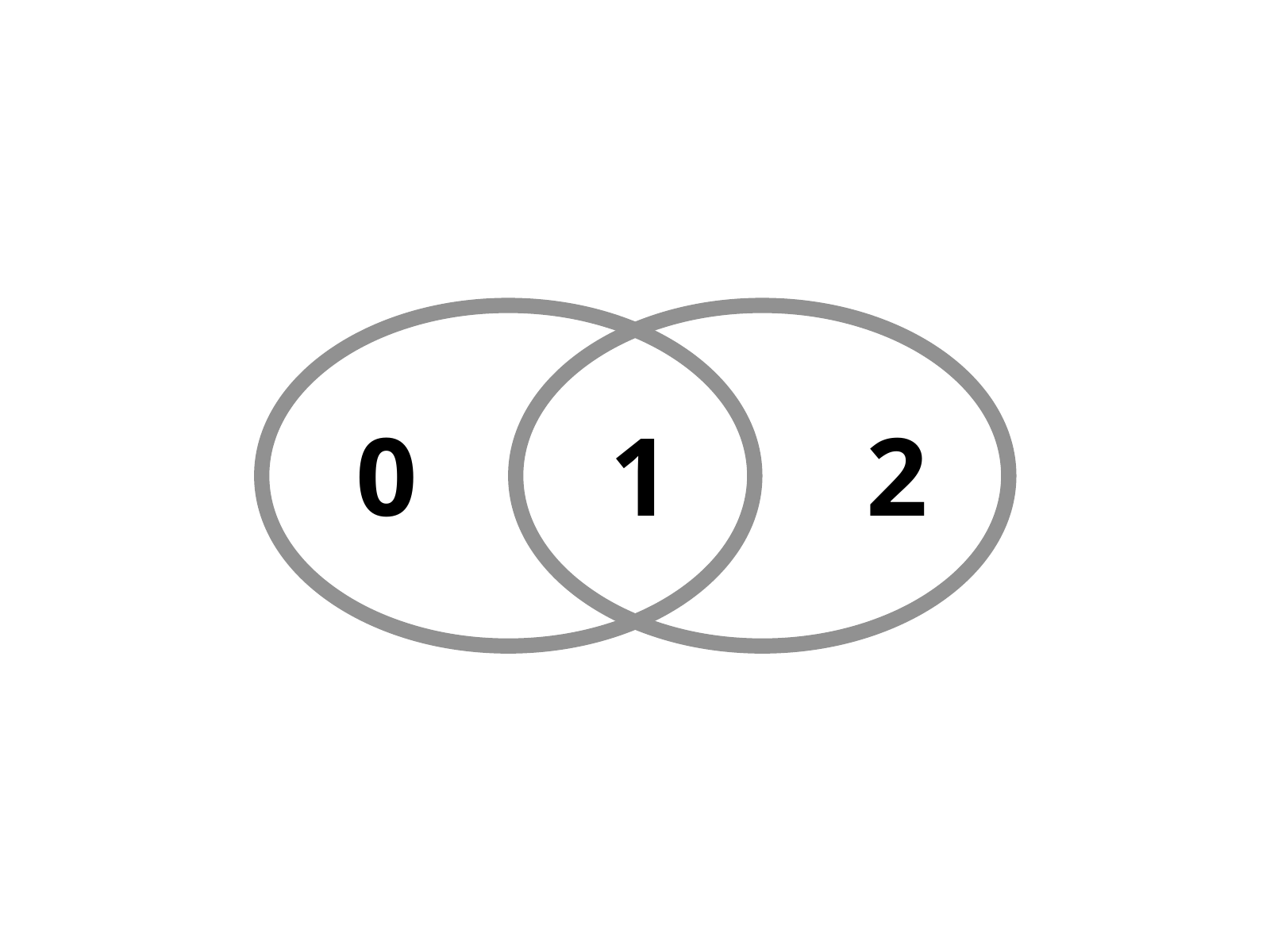}
\vspace{-1em}
\caption{An example of a point with two minimal open neighbourhoods (Lemma~\ref{lemm.two.min})}
\label{fig.two.min}
\end{figure}

Semitopologies are not topologies.
We take a moment to spell out one concrete difference:
\begin{lemm}
\label{lemm.two.min}
In topologies, if a point $p$ has a minimal open neighbourhood then it is least (= unique minimal).
In semitopologies, a point may have multiple distinct minimal open neighbourhoods.\footnote{We study minimal open neighbourhoods in detail, starting from Definition~\ref{defn.open.covers}.}
\end{lemm}
\begin{proof}
To see that in a topology every minimal open neighbourhood is least, just note that if $p\in A$ and $p\in B$ then $p\in A\cap B$.
So if $A$ and $B$ are two minimal open neighbourhoods then $A\cap B$ is contained in both and by minimality is equal to both.

To see that in a semitopology a minimal open neighbourhood need not be least, it suffices to provide an example.
Consider $(\ns P,\opens)$ defined as follows, as illustrated in Figure~\ref{fig.two.min}:
\begin{itemize*}
\item
$\ns P=\{0,1,2\}$
\item
$\opens = \bigl\{ \varnothing,\ \{0,1\},\ \{1,2\},\ \{0,1,2\} \bigr\}$
\end{itemize*}
Note that $1$ has two minimal open neighbourhoods: $\{0,1\}$ and $\{1,2\}$. 
\end{proof}

\jamiesubsubsection{Examples}

As standard, we can make any set $\tf{Val}$ into a semitopology (indeed, it is also a topology) just by letting open sets be the powerset: 
\begin{defn}
\label{defn.value.assignment}
\leavevmode
\begin{enumerate*}
\item\label{item.discrete.semitopology}
Call $(\ns P,\powerset(\ns P))$ the \deffont{discrete semitopology on $\ns P$}.
 
We may call a set with the discrete semitopology a \deffont{semitopology of values}, and when we do we will usually call it $\tf{Val}$.
We may identify $\tf{Val}$-the-set and $\tf{Val}$-the-discrete-semitopology; meaning will always be clear.
\item\label{item.value.assignment}
When $(\ns P,\opens)$ is a semitopology and $\tf{Val}$ is a semitopology of values, we may call a function $f:\ns P\to\tf{Val}$ a \deffont[value assignment $f:\ns P\to\tf{Val}$]{value assignment}.

Note that a value just assigns values to points, and in particular we do not assume \emph{a priori} that it is continuous, where continuity is defined just as for topologies (see Definition~\ref{defn.continuity}).
\end{enumerate*} 
\end{defn}

\begin{xmpl}
\label{xmpl.semitopologies}
We consider further examples of semitopologies:
\begin{enumerate}
\item
Every topology is also a semitopology; intersections of open sets are allowed to be open in a semitopology, they are just not constrained to be open.
In particular, the discrete topology is also a discrete semitopology (Definition~\ref{defn.value.assignment}(\ref{item.discrete.semitopology})).
\item
The \deffont{initial semitopology} $(\varnothing,\{\varnothing\})$ and the \deffont{final semitopology} $(\{\ast\},\{\varnothing,\{\ast\}\})$ are semitopologies. 
\item\label{item.boolean.discrete}
An important discrete semitopological space is 
$$
\mathbb B=\{\bot,\top\}
\quad\text{with the discrete semitopology}\quad
\opens(\mathbb B)=\{\varnothing, \{\bot\},\{\top\},\{\bot,\top\}\}.
$$
We may silently treat $\mathbb B$ as a (discrete) semitopological space henceforth.
\item\label{item.trivial.topology}
Take $\ns P$ to be any nonempty set.
Let the \deffont[trivial semitopology]{trivial semitopology} (this is also a topology) on $\ns P$ have 
$$
\opens =\{\varnothing, \ns P\}.
$$
So (as usual) there are only two open sets: the one containing nothing, and the one containing every point.\footnote{According to Wikipedia, this space is also called \emph{indiscrete}, \emph{anti-discrete}, \emph{concrete}, and \emph{codiscrete} (\url{https://en.wikipedia.org/wiki/Trivial_topology}).}

The only nonempty open is $\ns P$ itself, reflecting a notion of actionable coalition that requires unanimous agreement. 
\item
Suppose $\ns P$ is a set and $\mathcal F\subseteq\powerset(\ns P)$ is nonempty and up-closed (so if $P\in\mathcal F$ and $P\subseteq P'\subseteq\ns P$ then $P'\in\mathcal F$, then $(\ns P,\mathcal F)$ is a semitopology.
This is not necessarily a topology, because we do not insist that $\mathcal F$ is a filter (i.e. is closed under intersections).

We give four sub-examples for different choices of $\mathcal P\subseteq\powerset(\ns P)$.
Partly this is to illustrate how varying $\mathcal F$ can encode different systems, but also, these variations can have substantially different behavior, as Lemmas~\ref{lemm.all-but-one} and~\ref{lemm.more-than-one} will illustrate later.
\begin{enumerate}
\item\label{item.supermajority}
Take $\ns P$ to be any finite nonempty set.
Let the \deffont{supermajority semitopology} have 
$$
\opens =\{\varnothing\}\cup\{O\subseteq\ns P \mid \f{cardinality}(O)> \nicefrac{2}{3}*\f{cardinality}(\ns P)\}.
$$
So $O$ is open when it contains more than two-thirds of the points.

Two-thirds is a common threshold used for making progress in consensus and voting algorithms.\footnote{In the context of consensus algorithms, we usually take a strict inequality $\f{cardinality}(O)> \nicefrac{2}{3}*\f{cardinality}(\ns P)$ --- to guarantee that the intersection of any three nonempty open sets is nonempty.
However, in the context of voting (as in e.g. the rules for the US Senate to amend the Constitution) the majority may be taken to have the form $\f{cardinality}(O)\geq \nicefrac{2}{3}*\f{cardinality}(\ns P)$ --- to guarantee that votes \emph{for} outnumber votes \emph{against}, by at least two-to-one.} 
\item
Take $\ns P$ to be any nonempty set.
Let the \deffont{many semitopology} have
$$
\opens = \{\varnothing\}\cup\{O\subseteq\ns P \mid \f{cardinality}(O)=\f{cardinality}(\ns P)\} .
$$
For example, if $\ns P=\mathbb N$ then open sets include $\f{evens}=\{2*n \mid n\in\mathbb N\}$ and $\f{odds}=\{2*n\plus 1 \mid n\in\mathbb N\}$.

Its notion of open set captures an idea that an actionable coalition is a set that may not be all of $\ns P$, but does at least biject with it.
\item\label{item.counterexample.X-x}
Take $\ns P$ to be any nonempty set.
Let the \deffont{all-but-one semitopology} have
$$
\opens = \{\varnothing,\ \ns P\}\cup\{\ns P\setminus \{p\}\mid p\in\ns P\} .
$$
This semitopology is not a topology.
See also Lemma~\ref{lemm.all-but-one}.

The notion of actionable coalition here is that there may be at most one objector (but not two).
\item\label{item.counterexample.more-than-one}
Take $\ns P$ to be any set with cardinality at least $2$.
Let the \deffont{more-than-one semitopology} have
$$
\opens = \{\varnothing\}\cup\{O\subseteq\ns P \mid \f{cardinality}(O) \geq 2\} .
$$
This semitopology is not a topology.
See also Lemma~\ref{lemm.more-than-one}.

This notion of actionable coalition reflects a security principle in banking and accounting (and elsewhere) of \emph{separation of duties}, that functional responsibilities be separated such that at least two people are required to complete an action --- so that errors (or worse) cannot be made without being discovered by another person.
\end{enumerate}
\item
Take $\ns P=\mathbb R$ (the set of real numbers) and let open sets be generated by intervals of the form $\rightopeninterval{0,r}$ or $\leftopeninterval{\minus r,0}$ for any strictly positive real number $r>0$.

This semitopology is not a topology, since (for example) $\leftopeninterval{1,0}$ and $\rightopeninterval{0,1}$ are open, but their intersection $\{0\}$ is not open.
\item\label{item.quorum.system}
In~\cite{naor:loacaq} a notion of \emph{quorum system} is discussed, defined as any collection of pairwise intersecting sets.
Quorum systems are a field of study in their own right, especially in the theory of concrete consensus algorithms.

Every quorum system gives rise naturally to a semitopology, just by closing under arbitrary unions.
We obtain what we will call an \emph{intertwined space} (Notation~\ref{nttn.intertwined.space}; a semitopology all of whose nonempty open sets intersect)\footnote{A topologist would call this a \emph{hyperconnected space}, but be careful! There are multiple such notions in semitopologies, so intuitions need not transfer over.  See the discussion in Subsection~\ref{subsection.topens.in.topologies}.}
and this is also clearly related to the \emph{compatibility condition} in the notions of \emph{semifilter} and \emph{abstract point} in Section~\ref{sect.semifilters.and.points}.

Going in the other direction is interesting for a different reason, that it is slightly less canonical: of course every intertwined space is already a quorum system; but (for the finite case) we can also map to the set of all open covers of all points (Definition~\ref{defn.open.covers}(\ref{item.open.cover}); in the notation of that Definition, we would write this as $\bigcup_{p\in\ns P}\{O\in\opens \mid O\gtrdot p\}$).

To give one specific example of a quorum system from~\cite{naor:loacaq}, consider $n\times n$ grid of cells with quorums being sets consisting of any full row and a full column; note that any two quorums must intersect in at least two points.
We obtain a semitopology just by closing under arbitrary unions.
\end{enumerate}
\end{xmpl}

\begin{rmrk}[Logical models of semitopologies]

\noindent One class of examples of semitopologies deserves its own discussion.
Consider an arbitrary logical system with predicates $\tf{Pred}$ and entailment relation $\cent$.\footnote{A validity relation $\ment$ would also work.}
Call $\Phi\subseteq\tf{Pred}$ \deffont[deductively closed (set of predicates)]{deductively closed} when $\Phi\cent\phi$ implies $\phi\in\Phi$.
Then take 
\begin{itemize*}
\item
$\ns P=\tf{Pred}$, and 
\item
let $O\in\opens$ be $\tf{Pred}$ or the complement to a deductively closed set $\Phi$, so $O=\tf{Pred}\setminus\Phi$.
\end{itemize*}
Note that an arbitrary union of open sets is open (because an arbitrary intersection of deductively closed sets is deductively closed), but an intersection of open sets need not be open (because the union of deductively closed sets need not be deductively closed).
This is a semitopology.
This example will be important to us and we will return to it in Subsection~\ref{subsect.declarative.witness}. 
\end{rmrk}

\jamiesubsubsection{Why the name `semitopologies', and other discussion}

\begin{rmrk}[Why the name `semitopologies']
\label{rmrk.why.name.semitopologies}
When we give a name `semitopologies' to things that are like topologies but without intersections, this is a riff on 
\begin{itemize*}
\item
`semilattices', for things that are like lattices with joins but without meets (or vice-versa), and 
\item
`semigroups', for things that are like groups but without inverses.
\end{itemize*}
But, this terminology also reflects a real mathematical connection, because semitopologies \emph{are} semilattices \emph{are} semigroups, in standard ways which we take a moment to spell out: 
\begin{itemize*}
\item
A semitopology $(\ns P,\opens)$ is a bounded join subsemilattice of the powerset $\powerset(\ns P)$, by taking the join $\tor$ to be sets union $\cup$ and the bounds $\bot$ and $\top$ to be $\varnothing$ and $\ns P$ respectively. 
\item
A semilattice is an idempotent commutative monoid, which is an idempotent commutative semigroup with an identity, by taking the multiplication $\circ$ to be $\tor$ and the identity element to be $\bot$ ($\top$ becomes what is called a \emph{zero} or \emph{absorbing} element, such that $\top\circ x=\top$ always).
\end{itemize*} 
\end{rmrk}

\begin{figure}
\centering
\includegraphics[align=c,width=0.4\columnwidth,trim={50 0 50 0},clip]{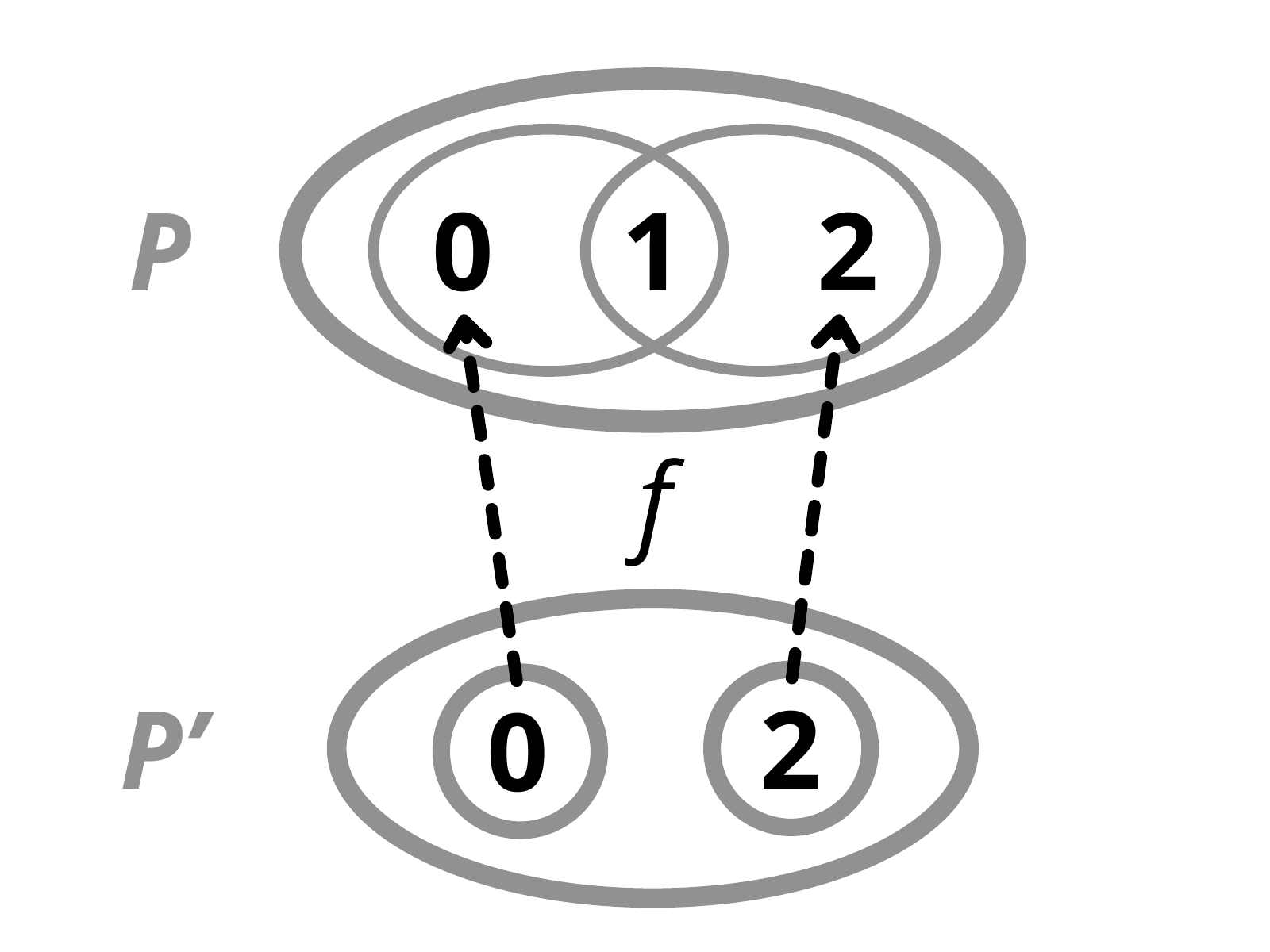}
\caption{Two nonidentical semitopologies (Remark~\ref{rmrk.PtoP})}
\label{fig.PtoP}
\end{figure}

\begin{rmrk}[Semitopologies are not \emph{just} semilattices]
\label{rmrk.PtoP}
We noted in Remark~\ref{rmrk.why.name.semitopologies} that every semitopology is a semilattice.
This is true, but the reader should not read this statement as reductive: semitopologies are not \emph{just} semilattices. 

To see why, consider the following two simple semitopologies, as illustrated in Figure~\ref{fig.PtoP}:
\begin{enumerate*}
\item
$(\ns P,\opens)$ where $\ns P=\{0,1,2\}$ and $\opens=\bigl\{\varnothing,\{0,1\},\{1,2\},\{0,1,2\}\bigr\}$.
\item
$(\ns P',\opens')$ where $\ns P=\{0,2\}$ and $\opens'=\bigl\{\varnothing,\{0\},\{2\},\{0,2\}\bigr\}$.
\end{enumerate*}
Note that the semilattices of open sets $\opens$ and $\opens'$ are isomorphic --- so, when viewed as semilattices these two semitopologies are the same (up to isomorphism).

However, $(\ns P,\opens)$ is not the same semitopology as $(\ns P',\opens')$.
There is more than one way to see this, but perhaps the simplest indication is that for every continuous $f:(\ns P,\opens)\to(\ns P',\opens')$, there is no continuous map $g:(\ns P',\opens')\to(\ns P,\opens)$ such that $g\circ f$ is the identity (we will define continuity in a moment in Definition~\ref{defn.continuity}(\ref{item.continuous.function}) but it is just as for topologies, so we take the liberty of using it here).
There are a limited number of possibilities for $f$ and $g$, and we can just enumerate them and check:
\begin{itemize*}
\item
If $f(0)=0$ and $f(2)=2$ and $g(1)=0$, then $g^\mone(\{2\})=\{2\}\not\oldin\opens$, and if $g(1)=1$ then $g^\mone(\{0\})=\{0\}\not\oldin\opens$. 
\item
If $f(0)=0$ and $f(2)=1$ and $g(1)=0$, then $g^\mone(\{2\})=\{1\}\not\oldin\opens$, and if $g(1)=2$ then $g^\mone(\{0\})=\{0\}\not\oldin\opens$. 
\item
Other possibilities are no harder.
\end{itemize*}
A similar observation holds for \emph{topologies}: for example, if we write $(\mathbb Q,\opens_{\mathbb Q})$ for the rational numbers with their usual open set topology, and $(\mathbb R,\opens_{\mathbb R})$ for the real numbers with their usual open set topology, then their topologies are isomorphic as lattices, with one direction of the isomorphism given just by $O\in \opens_{\mathbb R}$ maps to $O\cap \mathbb Q\oldin\opens_{\mathbb Q}$. 
This counterexample works for semitopologies too since every topology is also a semitopology.

However, we would still argue that the counterexample in Figure~\ref{fig.PtoP} is inherently stronger; not just because it is smaller (two and three points instead of countably and uncountably many) but also because --- while we can recover $\mathbb R$ from $\mathbb Q$ in a natural and canonical way by forming a completion --- the upper semitopology in Figure~\ref{fig.PtoP} is not \emph{a priori} canonically derived from the lower one.
The two semitopologies in Figure~\ref{fig.PtoP} seem to be distinct in some structural way, yet they still corresponding to the same semilattice, so we see that there is other structure here, which is not reflected by the pure semilattice derived from their open sets. 
\end{rmrk}

\begin{rmrk}[`Stronger' does not necessarily equal `better']
We conclude with some easy predictions about the theory of semitopologies, made just from general mathematical principles.
Fewer axioms means: 
\begin{enumerate*}
\item
\emph{more} models, 
\item
\emph{finer discrimination} between definitions, and 
\item
(because there are more models) \emph{more counterexamples}.
\end{enumerate*}
So we can expect a theory with the look-and-feel of topology, but with new models, new distinctions between definitions that in topology may be equivalent, and some new definitions, theorems, and counterexamples. 
 
Note that fewer axioms does not necessarily mean fewer interesting properties.
On the contrary: if we can make finer distinctions, there may also be more interesting things to prove; and assumptions can become \emph{more} impactful in a weaker system, because they may exclude more models than would have been the case with more powerful axioms.

For example consider semigroup theory and group theory: every group is a semigroup, but both groups and semigroups have their own distinct character, literature, and applications. 
To take this to an extreme, consider the \emph{terminal} theory, which has just one first-order axiom: $\Exists{x}\Forall{y}x=y$.
This `subsumes' groups, lattices, graphs, and much besides, in the sense that every model of the terminal theory \emph{is} a group, a lattice, and a graph, in a natural way.  
But models of this theory are so restricted (just the singleton model with one element) that there is not much left to say about it. 
Additional assumptions add nothing of value, because there was only one element to begin with!
\end{rmrk}

\begin{rmrk}[Counting semitopologies]
The number of semitopologies\index{number of semitopologies}\index{counting semitopologies} with $n$ points follows OEIS sequence~A102894 (\url{https://oeis.org/A102894}): the \emph{``number of families of subsets of $\{1, \dots, n\}$ that are closed under intersection and contain both the universe and the empty set.''}
(The sequence counts closed sets rather than open sets, but these are in bijection.)
\end{rmrk}

\jamiesubsection{Continuity, and its interpretation}
\label{subsect.continuity}

The topological notion of continuity works fine in semitopologies, and the fact that there are no surprises is a feature. 
In Remark~\ref{rmrk.continuity=consensus} we explain how these notions matter to us:

\begin{defn}
\label{defn.continuity}
We import standard topological notions of inverse image and continuity:
\begin{enumerate}
\item
Suppose $\ns P$ and $\ns P'$ are any sets and $f:\ns P\to\ns P'$ is a function.
Suppose $O'\subseteq\ns P'$.
Then write $f^\mone(O')$ for the \deffont[inverse image $f^\mone(O')$]{inverse image} or \deffont[preimage $f^\mone(O')$]{preimage} of $O'$, defined by
$$
f^\mone(O')=\{p{\in}\ns P \mid f(p)\in O'\} . 
$$
\item\label{item.continuous.function}
Suppose $(\ns P,\opens)$ and $(\ns P',\opens')$ are semitopological spaces (Definition~\ref{defn.semitopology}).
Call a function $f:\ns P\to\ns P'$ \deffont[continuous function]{continuous} when the inverse image of an open set is open.
In symbols:
$$
\Forall{O'\in\opens'} f^\mone(O')\oldin\opens .
$$
\item\label{item.continuous.function.at.p}
Call a function $f:\ns P\to\ns P'$ \deffont[continuous function at a point]{continuous at $p\in\ns P$} when
$$
\Forall{O'{\in}\opens'}f(p)\in O'\limp \Exists{O_{p,O'}{\in}\opens}p\in O_{p,O'}\land O_{p,O'}\subseteq f^\mone(O') .
$$
In words: $f$ is continuous at $p$ when the inverse image of every open neighbourhood of $f(p)$ contains an open neighbourhood of $p$.
\item
Call a function $f:\ns P\to\ns P'$ \deffont[continuous function on a set]{continuous on $P\subseteq\ns P$} when $f$ is continuous at every $p\in P$.
\end{enumerate}
\end{defn}

\begin{lemm}
\label{lemm.alternative.cont}
Suppose $(\ns P,\opens)$ and $(\ns P',\opens')$ are semitopological spaces (Definition~\ref{defn.semitopology}) and suppose $f:\ns P\to\ns P'$ is a function.
Then the following are equivalent:
\begin{enumerate*}
\item
$f$ is continuous (Definition~\ref{defn.continuity}(\ref{item.continuous.function})).
\item
$f$ is continuous at every $p\in\ns P$ (Definition~\ref{defn.continuity}(\ref{item.continuous.function.at.p})).
\end{enumerate*}
\end{lemm}
\begin{proof}
The top-down implication is immediate, taking $O=f^\mone(O')$.

For the bottom-up implication, given $p$ and an open neighbourhood $O'\ni f(p)$, we write
$$
O=\bigcup\{O_{p,O'}\in\opens \mid p\in\ns P,\ f(p)\in O'\}.
$$
Above, $O_{p,O'}$ is the open neighbourhood of $p$ in the preimage of $O'$, which we know exists by Definition~\ref{defn.continuity}(\ref{item.continuous.function.at.p}).

It is routine to check that $O= f^\mone(O')$, and since this is a union of open sets, it is open. 
\end{proof}

\begin{defn}
\label{defn.locally.constant}
Suppose that:
\begin{itemize*}
\item
$(\ns P,\opens)$ is a semitopology and 
\item
$\tf{Val}$ is a semitopology of values (Definition~\ref{defn.value.assignment}(\ref{item.discrete.semitopology})) and 
\item
$f:\ns P\to \tf{Val}$ is a value assignment (Definition~\ref{defn.value.assignment}(\ref{item.value.assignment}); an assignment of a value to each element in $\ns P$).
\end{itemize*}
Then:
\begin{enumerate*}
\item
Call $f$ \deffont[locally constant at a point]{locally constant at $p\in\ns P$} when there exists $p\in O_p\in\opens$ such that 
$$
\Forall{p'{\in}O_p}f(p)=f(p').
$$
So $f$ is locally constant at $p$ when it is constant on some open neighbourhood $O_p$ of $p$.
\item
Call $f$ \deffont[locally constant on a set]{locally constant} when it is locally constant at every $p\in\ns P$.
\end{enumerate*} 
\end{defn}

\begin{lemm}
\label{lemm.open.lc}
Suppose $(\ns P,\opens)$ is a semitopology and $\tf{Val}$ is a semitopology of values and $f:\ns P\to\tf{Val}$ is a value assignment.
Then the following are equivalent:
\begin{itemize*}
\item
$f$ is locally constant / locally constant at $p\in\ns P$ (Definition~\ref{defn.locally.constant}).
\item
$f$ is continuous / continuous at $p\in\ns P$ (Definition~\ref{defn.continuity}). 
\end{itemize*}
\end{lemm}
\begin{proof}
This is just by pushing around definitions, but we spell it out:
\begin{itemize}
\item
Suppose $f$ is continuous, consider $p\in\ns P$, and write $v=f(p)$.
By our assumptions we know that $f^\mone(v)$ is open, and $p\in f^\mone(v)$.
This is an open neighbourhood $O_p$ on which $f$ is constant, so we are done.
\item
Suppose $f$ is locally constant, consider $p\in\ns P$, and write $v=f(p)$.
By assumption we can find $p\in O_p\in\opens$ on which $f$ is constant, so that $O_p\subseteq f^\mone(v)$.
\qedhere\end{itemize}
\end{proof}

\begin{rmrk}[Continuity = agreement]
\label{rmrk.continuity=consensus}
Lemma~\ref{lemm.open.lc} tells us that
we can view the problem of attaining agreement across an actionable coalition (as discussed in Subsection~\ref{subsect.what.is}) as being the same thing as computing a value assignment that is continuous on that coalition (and possibly elsewhere).

To see why, consider a semitopology $(\ns P, \opens)$ and following the intuitions discussed in Subsection~\ref{subsect.what.is} view points $p\in \ns P$ as \emph{participants}; and view open neighbourhoods $p\in O\in\opens$ as \deffont{actionable coalitions} that include $p$.
Then to say ``$f$ is a value assignment that is continuous at $p$'' is to say that:
\begin{itemize*}
\item
$f$ assigns a value or belief to $p\in\ns P$, and
\item
$p$ is part of a (by Lemma~\ref{lemm.open.lc} continuity) set of peers that agrees with $p$ and (being open) can progress to act on this agreement.
\end{itemize*}
Conceptually and mathematically this reduces the general question 
\begin{quote}
\emph{How can we model collaborative action?} 
\end{quote}
(which, to be fair, has more than one possible answer!) to a more specific research question
\begin{quote}
\emph{Understand continuous value assignments on semitopologies}.
\end{quote}
We then devote ourselves to elaborating (some of) a body of mathematics that we can pull out of this idea.
\end{rmrk}

\jamiesubsection{Neighbourhoods of a point}

Definition~\ref{defn.open.neighbourhood} is a standard notion from topology, and Lemma~\ref{lemm.open.is.open} is a (standard) characterisation of openness, which will be useful later: 

\begin{defn}
\label{defn.open.neighbourhood}
Suppose $(\ns P,\opens)$ is a semitopology and $p\in\ns P$ and $O\in\opens$.
Then call $O$ an \deffont{open neighbourhood} of $p$ when $p\in O$.

In other words: an open set is (by definition) an \emph{open neighbourhood} precisely for the points that it contains.
\end{defn}

\begin{lemm}
\label{lemm.open.is.open}
Suppose $(\ns P,\opens)$ is a semitopology and suppose $P\subseteq\ns P$ is any set of points.
Then the following are equivalent:
\begin{itemize*}
\item
$P\in\opens$.
\item
Every point $p$ in $P$ has an open neighbourhood in $P$. 
\end{itemize*}
In symbols we can write:
$$
\Forall{p{\in}P}\Exists{O{\in}\opens}(p\in O\land O\subseteq P)
\quad\text{if and only if}\quad
P\in\opens
$$
\end{lemm}
\begin{proof}
If $P$ is open then $P$ itself is an open neighbourhood for every point that it contains. 

Conversely, if every $p\in P$ contains some open neighbourhood $p\in O_p \subseteq P$ then $P=\bigcup\{O_p\mid p\in P\}$ and this is open by condition~\ref{semitopology.unions} of Definition~\ref{defn.semitopology}.
\end{proof}

\begin{rmrk}
An initial inspiration for modelling collaborative action using semitopologies, came from noting that the standard topological property described above in Lemma~\ref{lemm.open.is.open}, corresponds to the \emph{quorum sharing} property in \cite[Property~1]{losa:stecbi}; the connection to topological ideas had not been noticed in~\cite{losa:stecbi}.
\end{rmrk}

\jamiesection{Transitive sets \& topens}
\label{sect.transitive.sets}

\jamiesubsection{Some background on sets intersection}

Some notation will be convenient:
\begin{nttn}
\label{nttn.between}
Suppose $X$, $Y$, and $Z$ are sets.
\begin{enumerate*}
\item\label{item.between}
Write 
$$
X\between Y
\quad\text{when}\quad 
X\cap Y\neq\varnothing.
$$
When $X\between Y$ holds then we say (as standard) that $X$ and $Y$ \deffont[intersecting sets $X\between Y$]{intersect}.\index{$X\between Y$ (intersection of sets)}
\item
We may chain the $\between$ notation, writing for example 
$$
X\between Y\between Z
\quad\text{for}\quad
X\between Y\ \land \  Y\between Z
$$
\item
We may write $X\notbetween Y$ for $\neg(X\between Y)$, thus $X\notbetween Y$ when $X\cap Y=\varnothing$.
\end{enumerate*}
\end{nttn}

\begin{rmrk}
\emph{Note on design in Notation~\ref{nttn.between}:}
It is uncontroversial that if $X\neq\varnothing$ and $Y\neq\varnothing$ then $X\between Y$ should hold precisely when $X\cap Y\neq\varnothing$ --- but there is an edge case! 
What truth-value should $X\between Y$ return when $X$ or $Y$ is empty?
\begin{enumerate*}
\item
It might be nice if $X\subseteq Y$ would imply $X\between Y$.
This argues for setting 
$$
(X=\varnothing\lor Y=\varnothing)\limp X\between Y .
$$
\item
It might be nice if $X\between Y$ were monotone on both arguments (i.e. if $X\between Y$ and $X\subseteq X'$ then $X'\between Y$).
This argues for setting 
$$
(X=\varnothing\lor Y=\varnothing)\limp X\notbetween Y .
$$
\item
It might be nice if $X\between X$ always --- after all, should a set \emph{not} intersect itself? --- and this argues for setting 
$$
\varnothing\between\varnothing ,
$$ 
even if we also set $\varnothing\notbetween Y$ for nonempty $Y$. 
\end{enumerate*}
All three choices are defensible, and they are consistent with the following nice property:
$$
X\between Y \limp (X\between X \lor Y\between Y) . 
$$
We choose the second --- if $X$ or $Y$ is empty then $X\notbetween Y$ --- because it gives the simplest definition that $X\between Y$ precisely when $X\cap Y\neq\varnothing$.
\end{rmrk}

We list some elementary properties of $\between$ from Notation~\ref{nttn.between}(\ref{item.between}):
\begin{lemm}
\label{lemm.between.elementary}
\leavevmode
\begin{enumerate*}
\item\label{item.between.nonempty}
$X\between X$ if and only if $X\neq\varnothing$.
\item\label{item.between.symmetric}
$X\between Y$ if and only if $Y\between X$.
\item\label{between.elementary.either.or}
$X\between (Y\cup Z)$ if and only if $(X\between Y) \lor (X\between Z)$.
\item\label{between.subset}
If $X\subseteq X'$ and $X\neq\varnothing$ then $X\between X'$.
\item\label{between.monotone}
Suppose $X\between Y$.
Then $X\subseteq X'$ implies $X'\between Y$, and $Y\subseteq Y'$ implies $X\between Y'$. 
\item\label{between.nonempty}
If $X\between Y$ then $X\neq\varnothing$ and $Y\neq\varnothing$.
\end{enumerate*}
\end{lemm}
\begin{proof}
By facts of sets intersection.
\end{proof}

\jamiesubsection{Transitive open sets and value assignments}

\begin{defn}
\label{defn.transitive}
Suppose $(\ns P,\opens)$ is a semitopology.
Suppose $\atopen\subseteq\ns P$ is any set of points.
\begin{enumerate*}
\item\label{transitive.transitive}
Call $\atopen$ \deffont[transitive set]{transitive} when 
$$
\Forall{O,O'{\in}\opens} O\between \atopen \between O' \limp O\between O'. 
$$
\item\label{transitive.cc}
Call $\atopen$ \deffont[topen set]{topen} when $\atopen$ is nonempty transitive and open.\footnote{%
The empty set is trivially transitive and open, so it would make sense to admit it as a (degenerate) topen.  However, it turns out that we mostly need the notion of `topen' to refer to certain kinds of neighbourhoods of points (we will call them \emph{communities}; see Definition~\ref{defn.tn}).  It is therefore convenient to exclude the empty set from being topen, because while it is the neighbourhood of every point that it contains, it is not a neighbourhood of any point.} 

We may write 
$$
\topens=\{ \atopen\in\opens_{\neq\varnothing} \mid \atopen\text{ is transitive}\} .
$$
\item\label{transitive.max.cc}
Call $S$ a \deffont[maximal topen set]{maximal topen} when $S$ is a topen that is not a subset of any strictly larger topen.\footnote{`Transitive open' $\to$ `topen', like `closed and open' $\to$ `clopen'.

For convenient reference, note that related notions of \emph{strong} transitivity and topen are in Definition~\ref{defn.strongly.transitive}.}
\end{enumerate*}
\end{defn}

Theorem~\ref{thrm.correlated} clarifies why transitivity is interesting: continuous value assignments are constant --- if we think of points as participants, `constant function' here means `in agreement' --- across transitive sets.
\begin{thrm}
\label{thrm.correlated}
Suppose that:
\begin{itemize*}
\item
$(\ns P,\opens)$ is a semitopology.
\item
$\tf{Val}$ is a semitopology of values (a nonempty set with the discrete semitopology; see Definition~\ref{defn.value.assignment}(\ref{item.discrete.semitopology})). 
\item
$f:\ns P\to\tf{Val}$ is a value assignment (Definition~\ref{defn.value.assignment}(\ref{item.value.assignment})). 
\item
$T\subseteq\ns P$ is a transitive set (Definition~\ref{defn.transitive}) --- in particular this will hold if $\atopen$ is topen --- and $p,p'\in T$.
\end{itemize*} 
Then:
\begin{enumerate*}
\item\label{item.correlated.1}
If $f$ is continuous at $p$ and $p'$ then $f(p)=f(p')$.
\item\label{item.correlated.2}
As a corollary, if $f$ is continuous on $\atopen$, then $f$ is constant on $\atopen$.
\end{enumerate*}
In words we can say: 
\begin{quote}
Continuous value assignments are constant across transitive sets.
\end{quote}
\end{thrm}
\begin{proof}
Part~\ref{item.correlated.2} follows from part~\ref{item.correlated.1} since if $f(p)=f(p')$ for \emph{any} $p,p'\in T$, then by definition $f$ is constant on $\atopen$.
So we now just need to prove part~\ref{item.correlated.1} of this result.

Consider $p,p'\in T$.
By continuity on $\atopen$, there exist open neighbourhoods $p\in O\subseteq f^\mone(f(p))$ and $p'\in O'\subseteq f^\mone(f(p'))$.
By construction $O\between \atopen \between O'$ (because $p\in O\cap T$ and $p'\in T\cap O'$).
By transitivity of $\atopen$ it follows that $O\between O'$. 
Thus, there exists $p''\in O\cap O'$, and by construction $f(p) = f(p'') = f(p')$.
\end{proof}

Corollary~\ref{corr.correlated.intersect} is an easy and useful consequence of Theorem~\ref{thrm.correlated}:
\begin{corr}
\label{corr.correlated.intersect}
Suppose that:
\begin{itemize*}
\item
$(\ns P,\opens)$ is a semitopology. 
\item
$f:\ns P\to \tf{Val}$ is a value assignment to some set of values $\tf{Val}$ (Definition~\ref{defn.value.assignment}). 
\item
$f$ is continuous on topen sets $\atopen, \atopen'\in\topens$.
\end{itemize*}
Then 
$$
\atopen\between \atopen'
\quad\text{implies}\quad 
\Forall{p\in\atopen,p'\in\atopen'} f(p)=f(p').
$$
\end{corr}
\begin{proof}
By Theorem~\ref{thrm.correlated} $f$ is constant on $\atopen$ and $\atopen'$.
We assumed that $\atopen$ and $\atopen'$ intersect, and the result follows.
\end{proof}

A converse to Theorem~\ref{thrm.correlated} also holds:
\begin{prop}
\label{prop.correlated.converse}
Suppose that:
\begin{itemize*}
\item
$(\ns P,\opens)$ is a semitopology.
\item
$\tf{Val}$ is a semitopology of values with at least two elements (to exclude a degenerate case that no functions exist, or they exist but there is only one because there is only one value to map to).
\item
$T\subseteq\ns P$ is any set. 
\end{itemize*} 
Then 
\begin{itemize*}
\item
\emph{if} for every $p,p'\in T$ and every value assignment $f:\ns P\to\tf{Val}$, $f$ continuous at $p$ and $p'$ implies $f(p)=f(p')$, 
\item
\emph{then} $\atopen$ is transitive.
\end{itemize*}
\end{prop}
\begin{proof}
We prove the contrapositive. 
Suppose $\atopen$ is not transitive, so there exist $O,O'\in\opens$ such that $O\between \atopen\between O'$ and yet $O\cap O'=\varnothing$.
We choose two distinct values $v\neq v'\in\tf{Val}$ and define $f$ to map any point in $O$ to $v$ and any point in $\ns P\setminus O$ to $v'$.

Choose some $p\in O$ and $p'\in O'$.
It does not matter which, and some such $p$ and $p'$ exist, because $O$ and $O'$ are nonempty by Lemma~\ref{lemm.between.elementary}(\ref{between.nonempty}), since $O\between\atopen$ and $O'\between\atopen$).

We note that $f(p)=v$ and $f(p')=v'$ and $f$ is continuous at $p\in O$ and $p'\in O'\subseteq\ns P\setminus O$, yet $f(p)\neq f(p')$.
\end{proof}

We can sum up what Theorem~\ref{thrm.correlated} and Proposition~\ref{prop.correlated.converse} mean, as follows:
\begin{rmrk}
\label{rmrk.transitive.correlated}
Suppose $(\ns P,\opens)$ is a semitopology and $\tf{Val}$ is a semitopology of values with at least two elements.
Say that a value assignment $f:\ns P\to\tf{Val}$ \deffont[splits (value assignment splits a set)]{splits} a set $T\subseteq\ns P$ when there exist $p,p'\in T$ such that $f$ is continuous at $p$ and $p'$ and $f(p)\neq f(p')$. 
Then Theorem~\ref{thrm.correlated} and Proposition~\ref{prop.correlated.converse} together say in words that: 
\begin{quote}
$T\subseteq\ns P$ is transitive if and only if it cannot be split by a value assignment that is continuous on $T$. 
\end{quote}
Intuitively, transitive sets characterise areas of guaranteed agreement.

This reminds us of a basic result in topology about \emph{connected spaces}~\cite[Chapter~8, section~26]{willard:gent}.
Call a topological space $(\ns T,\opens)$ \deffont[disconnected (semi)topology]{disconnected} when there exist open sets $O,O'\in\opens$ such that $O\cap O'=\varnothing$ (in our notation: $O\notbetween O'$) and $O\cup O'=\ns T$; otherwise call $(\ns T,\opens)$ \deffont[connected (semi)topology]{connected}.
Then $(\ns T,\opens)$ is disconnected if and only if (in our terminology above) it can be split by a value assignment. 
Theorem~\ref{thrm.correlated} and Proposition~\ref{prop.correlated.converse} are not identical to that result, but they are in the same spirit. 
\end{rmrk}

\begin{rmrk}
\label{rmrk.transitive.comment}
The notion of transitive set gives us enough to comment on the two examples in Subsection~\ref{subsect.what.is}.
Recall that we considered:
\begin{enumerate*}
\item
A nonempty finite set $\mathbb E$ with open sets $\opens(\mathbb E)$ (`actionable coalitions') being majority subsets $O\subseteq\mathbb E$.
\item
Integers $\mathbb Z$ with open sets $\opens(\mathbb Z)$ generated by triplets $\{2i,2i\plus 1,2i\plus 2\}$.
\end{enumerate*}
The reader can check that in $(\mathbb E,\opens(\mathbb E))$ \emph{every} set is transitive, because every pair of nonempty open sets intersect; thus, no $T\subseteq\mathbb E$ can be split by a value assignment that is continuous on $T$. 
In contrast, the reader can check that in $(\mathbb Z,\opens(\mathbb Z))$, most sets are not transitive, including (for example) $\{0,4\}$. 
This lack of transitivity reflects an intuitive observation we made in Subsection~\ref{subsect.what.is} that our second example was `not necessarily particularly safe or desirable in practice'; in our more technical language, we can now note that there exists a value assignment that splits $\{0,4\}$, yet is continuous at $0$ and $4$.
What $(\mathbb Z,\opens(\mathbb Z))$ does satisfy is the weaker (but still useful!) safety property that any continuous value assignment that is continuous everywhere, is constant (corresponding to our informal observation that ``\emph{if} all participants do legally progress, then they announce the same value'').\footnote{We can be more precise if we like: e.g. $T$ cannot be split by a value assignment that is continuous on a contiguous segment of $\mathbb Z$ that includes $T$.  Continuity on all of $\mathbb Z$ is one sufficient condition for this, which corresponds (in the language of consensus) to assuming that all participants are correct.  But we digress.}
This reflects a useful intuition, that the topological notion of `continuity at a point', corresponds to an intuition of $p$ as a participant `following the rules'.
\end{rmrk}

\jamiesubsection{Examples and discussion of transitive sets and topens}

We may routinely order sets by subset inclusion; including open sets, topens, closed sets, and so on, and we may talk about maximal, minimal, greatest, and least elements.
We include the (standard) definition for reference: 
\begin{nttn}
\label{nttn.min.max}
Suppose $(\ns P,\leq)$ is a poset.
Then:
\begin{enumerate*}
\item
Call $p\in\ns P$ \deffont[maximal element (in poset)]{maximal} when $\Forall{p'}p{\leq}p'\limp p'=p$ and \deffont[minimal element (in poset)]{minimal} when $\Forall{p'}p'{\leq}p\limp p'=p$.
\item
Call $p\in\ns P$ \deffont[greatest element (in poset)]{greatest} when $\Forall{p}p'\leq p$ and \deffont[least element (in poset)]{least} when $\Forall{p'}p\leq p'$.
\end{enumerate*}
\end{nttn}

\begin{xmpl}[Examples of transitive sets]
\label{xmpl.singleton.transitive}
\leavevmode
\begin{enumerate*}
\item\label{item.singleton.transitive}
$\{p\}$ is transitive, for any single point $p\in\ns P$. 
\item
The empty set $\varnothing$ is (trivially) transitive.
It is not topen because we insist in Definition~\ref{defn.transitive}(\ref{transitive.cc}) that topens are nonempty.
\item
Call a set $P\subseteq\ns P$ \emph{topologically indistinguishable} when (using Notation~\ref{nttn.between}) for every open set $O$, 
$$
P\between O\liff P\subseteq O .
$$ 
It is easy to check that if $P$ is topologically indistinguishable, then it is transitive.
\end{enumerate*} 
\end{xmpl}

\begin{figure}
\vspace{-1em}
\centering
\includegraphics[align=c,width=0.4\columnwidth,trim={50 60 50 120},clip]{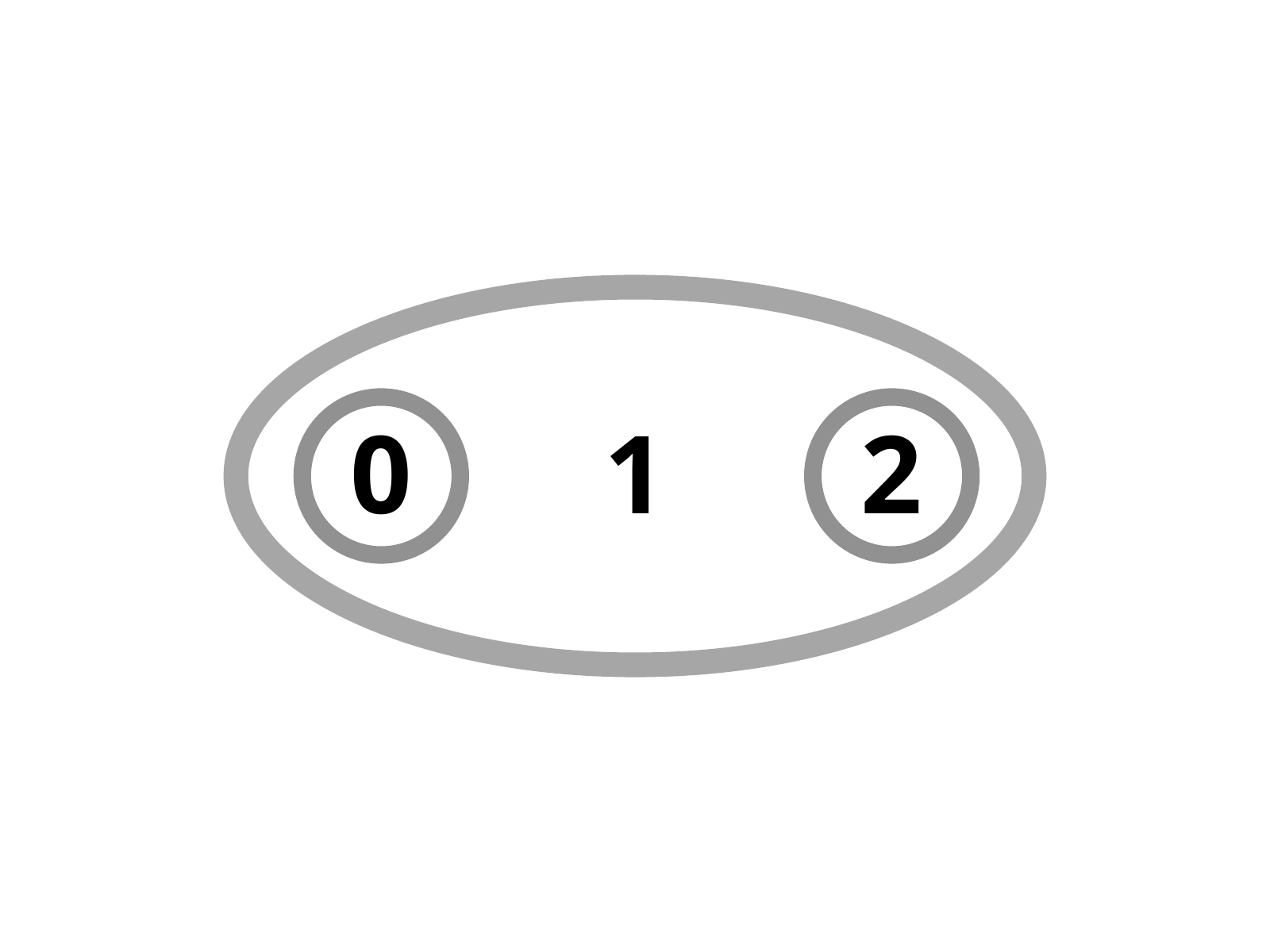}
\includegraphics[align=c,width=0.43\columnwidth,trim={50 60 50 220},clip]{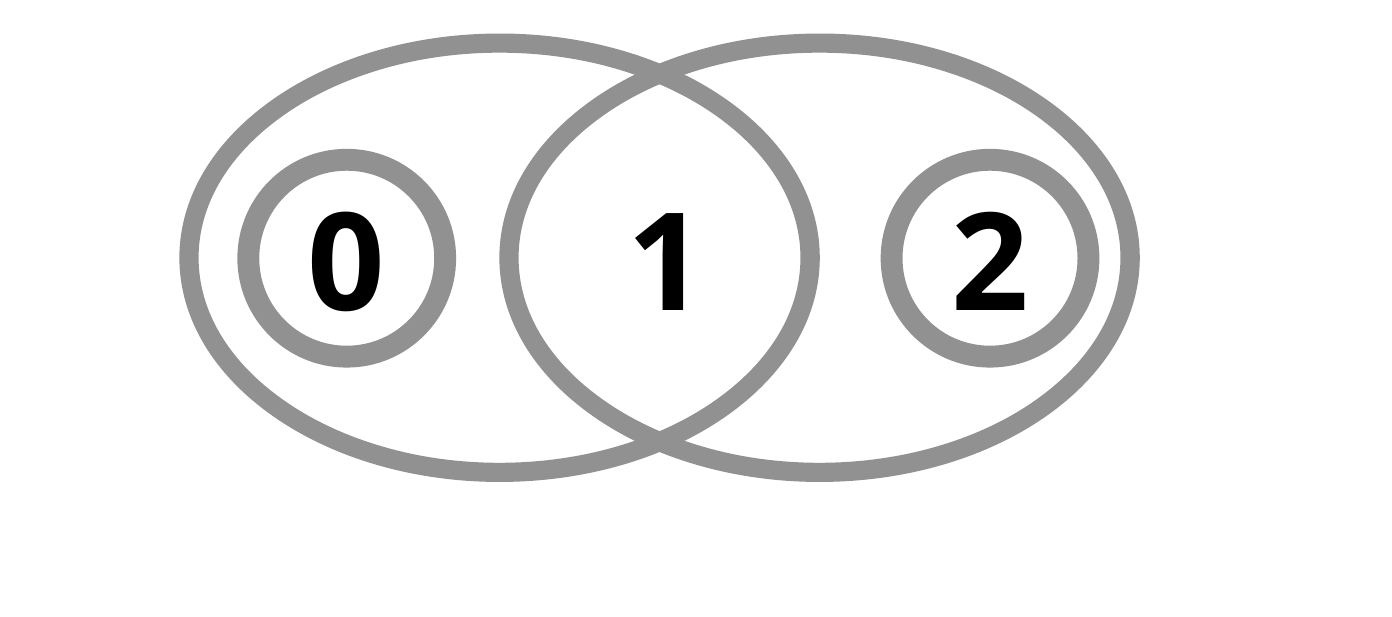}
\\
\includegraphics[align=c,width=0.35\columnwidth,trim={20 20 20 20},clip]{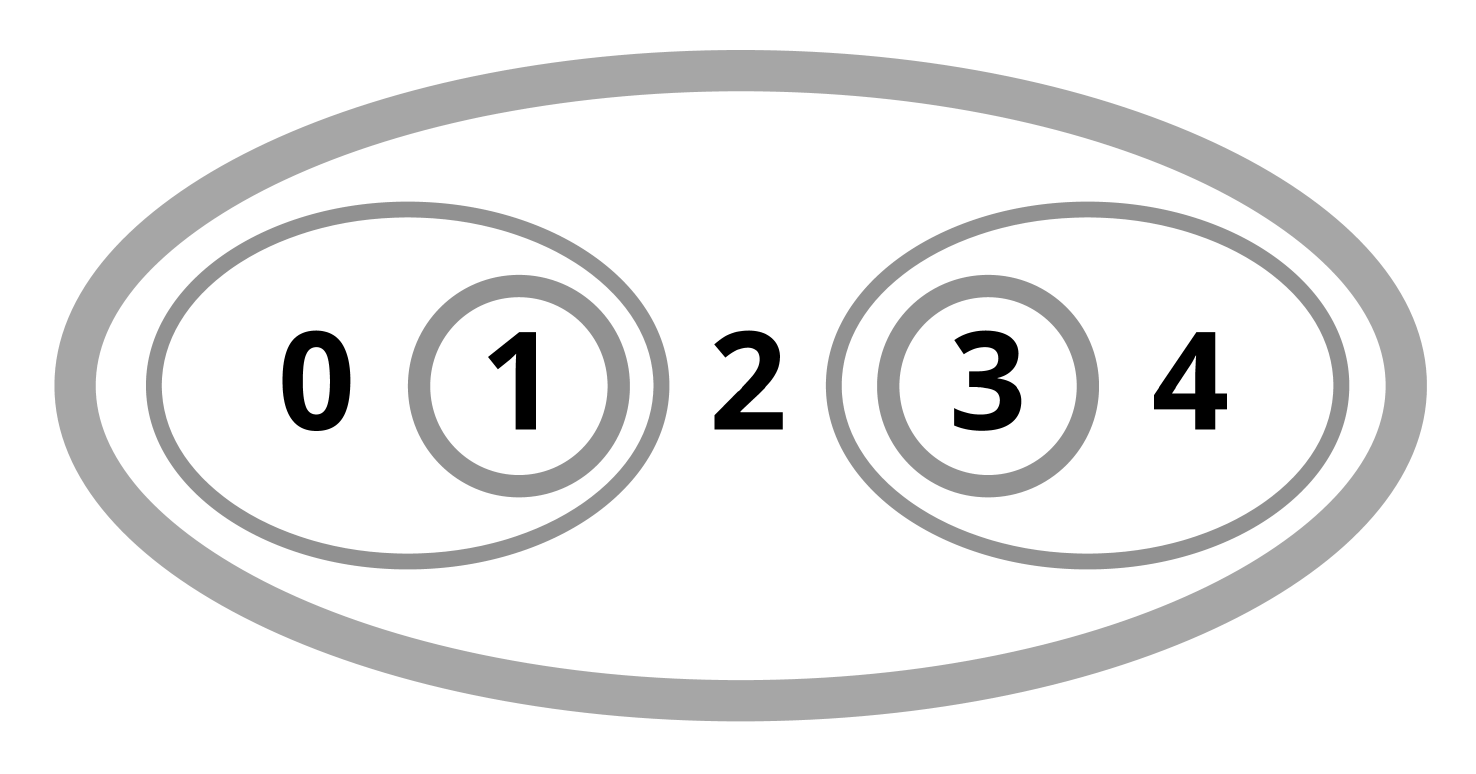}
\quad\  
\includegraphics[align=c,width=0.35\columnwidth,trim={50 20 50 20},clip]{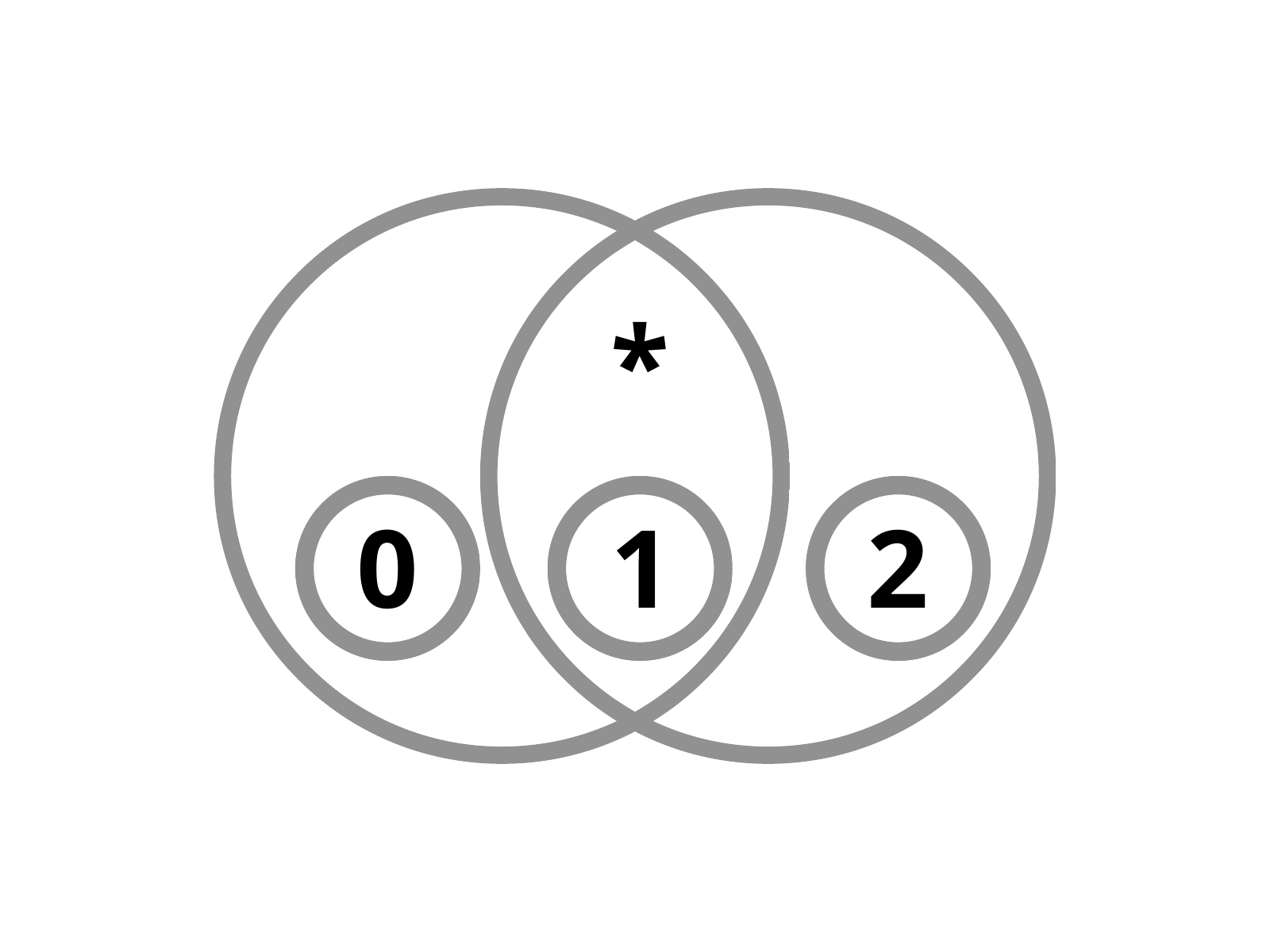}

\begin{flushleft}
\noindent\emph{Here and elsewhere, we might omit open sets that are unions of open sets that are illustrated.  
For example, we explicitly draw the universal open set in the left-hand diagrams above, but not in the right-hand diagrams above.
Meaning is clear and we get cleaner diagrams.
}
\end{flushleft}
\caption{Examples of topens (Example~\ref{xmpl.cc})}
\label{fig.012}
\end{figure}

\begin{xmpl}[Examples of topens]
\label{xmpl.cc}
\leavevmode
\begin{enumerate*}
\item\label{item.cc.two.regular}
Take $\ns P=\{0, 1, 2\}$, with open sets $\varnothing$, $\ns P$, $\{0\}$, and $\{2\}$. 
This has two maximal topens $\{0\}$ and $\{2\}$  as illustrated in Figure~\ref{fig.012} (top-left diagram). 
\item\label{item.cc.two.regular.b}
Take $\ns P=\{0, 1, 2\}$, with open sets $\varnothing$, $\ns P$, $\{0\}$, $\{0, 1\}$, $\{2\}$, $\{1,2\}$, and $\{0,2\}$. 
This has two maximal topens $\{0\}$ and $\{2\}$, as illustrated in Figure~\ref{fig.012} (top-right diagram). 
\item\label{item.xmpl.cc.3}
Take $\ns P=\{0,1,2,3,4\}$, with open sets generated by $\{0, 1\}$, $\{1\}$, $\{3\}$, and $\{3,4\}$.
This has two maximal topens $\{0,1\}$ and $\{3,4\}$, as illustrated in Figure~\ref{fig.012} (lower-left diagram). 
\item\label{item.xmpl.cc.4}
Take $\ns P=\{0,1,2,\ast\}$, with open sets generated by $\{0\}$, $\{1\}$, $\{2\}$, $\{0, 1,\ast\}$, and $\{1,2,\ast\}$.
This has three maximal topens $\{0\}$, $\{1\}$, and $\{2\}$, as illustrated in Figure~\ref{fig.012} (lower-right diagram). 
\item
Take the all-but-one semitopology from Example~\ref{xmpl.semitopologies}(\ref{item.counterexample.X-x}) on $\mathbb N$: so $\ns P=\mathbb N$ with opens $\varnothing$, $\mathbb N$, and $\mathbb N\setminus \{x\}$ for every $x\in\mathbb N$.
This has a single maximal topen $\mathbb N$.
\item
The semitopology in Figure~\ref{fig.square.diagram} has no topen sets at all ($\varnothing$ is transitive and open, but by definition in Definition~\ref{defn.transitive}(\ref{transitive.cc}) topens have to be nonempty).
\end{enumerate*}
\end{xmpl}

\begin{rmrk}[Discussion]
We take a moment for a high-level discussion of where we are going.

The semiopologies in Example~\ref{xmpl.cc} invite us to ask what makes these examples different (especially parts~\ref{item.cc.two.regular} and~\ref{item.cc.two.regular.b}).
Clearly they are not equal, but that is a superficial answer in the sense that it is valid just in the world of sets, and it ignores semitopological structure.

For comparison: if we ask what makes $0$ and $1$ different in $\mathbb N$, we could just to say that $0\neq 1$, but this ignores what makes them different \emph{as numbers}.
For more insight, we could note that $0$ is the additive unit whereas $1$ is the multiplicative unit of $\mathbb N$ as a semiring; or that $0$ is a least element and $1$ is the unique atom of $\mathbb N$ as a well-founded poset; or that $1$ is the successor of $0$ of $\mathbb N$ as a well-founded inductive structure. 
Each of these answers gives us more understanding, not only into $0$ and $1$ but also into the structures that can be given to $\mathbb N$ itself. 

So we can ask:
\begin{quote}
\emph{What semitopological property or properties on points can identify the essential nature of the differences between the semitopologies in Example~\ref{xmpl.cc}?}
\end{quote}
There would be some truth to saying that the rest of our investigation is devoted to developing and understanding answers to this question!
In particular, we will shortly define the set of \emph{intertwined points} $\intertwined{p}$ in Definition~\ref{defn.intertwined.points}.
Example~\ref{xmpl.how.different?} will note that $\intertwined{1}=\{0,1,2\}$ in Example~\ref{xmpl.cc}(\ref{item.cc.two.regular}), whereas $\intertwined{1}=\{1\}$ in Example~\ref{xmpl.cc}(\ref{item.cc.two.regular.b}), and $\intertwined{x}=\mathbb N$ for every $x$ in Example~\ref{xmpl.cc}(\ref{item.xmpl.cc.3}).
\end{rmrk}

\jamiesubsection{Closure properties of transitive sets}
\label{subsect.closure.properties.of.tt}

\begin{rmrk}
Transitive sets have some nice closure properties which we treat in this Subsection --- here we mean `closure' in the sense of ``the set of transitive sets is closed under various operations'', and not in the topological sense of `closed sets'.

Topens --- nonempty transitive \emph{open} sets --- will have even better closure properties, which emanate from the requirement in Lemma~\ref{lemm.transitive.transitive} that at least one of the transitive sets $\atopen$ or $\atopen'$ is open. 
See Subsection~\ref{subsect.closure.properties.of.cc}.
\end{rmrk}

\begin{lemm}
\label{lemm.transitive.subset}
Suppose $(\ns P,\opens)$ is a semitopology and $\atopen\subseteq \ns P$. 
Then:
\begin{enumerate*}
\item\label{item.transitive.subset.1}
If $\atopen$ is transitive and $\atopen'\subseteq \atopen$, then $\atopen'$ is transitive.
\item\label{item.transitive.subset.2}
If $\atopen$ is topen and $\varnothing\neq \atopen'\subseteq \atopen$ is nonempty and open, then $\atopen'$ is topen.
\end{enumerate*}
\end{lemm}
\begin{proof}
\leavevmode
\begin{enumerate}
\item
By Definition~\ref{defn.transitive} it suffices to consider open sets $O$ and $O'$ such that $O\between \atopen'\between O'$, and prove that $O\between O'$.
But this is simple: by Lemma~\ref{lemm.between.elementary}(\ref{between.monotone}) $O\between \atopen\between O'$, so $O\between O'$ follows by transitivity of $\atopen$. 
\item
Direct from part~\ref{item.transitive.subset.1} of this result and Definition~\ref{defn.transitive}(\ref{transitive.cc}).
\qedhere\end{enumerate}
\end{proof}

\begin{lemm}
\label{lemm.transitive.transitive}
Suppose that:
\begin{itemize*}
\item
$(\ns P,\opens)$ is a semitopology.
\item
$\atopen,\atopen'\subseteq\ns P$ are transitive.
\item
At least one of $\atopen$ and $\atopen'$ is open.
\end{itemize*}
Then:
\begin{enumerate*}
\item\label{item.transitive.transitive.1} 
$\Forall{O,O'\in\opens}O\between \atopen \between \atopen'\between O' \limp O\between O'$. 
\item\label{item.transitive.transitive.2} 
If $\atopen\between \atopen'$ then $\atopen\cup \atopen'$ is transitive.
\end{enumerate*}
\end{lemm}
\begin{proof}
\leavevmode
\begin{enumerate}
\item
We simplify using Definition~\ref{defn.transitive} and our assumption that one of $\atopen$ and $\atopen'$ is open.
We consider the case that $\atopen'$ is open: 
$$
\begin{array}{r@{\ }l@{\qquad}l}
O\between \atopen\between \atopen'\between O'
\limp&
O\between \atopen' \between O'
&\text{$\atopen$ transitive, $\atopen'$ open}
\\
\limp&
O\between O'
&\text{$\atopen'$ transitive}.
\end{array}
$$
The argument for when $\atopen$ is open, is precisely similar.
\item
Suppose $O\between \atopen\cup \atopen'\between O'$.
By Lemma~\ref{lemm.between.elementary}(\ref{between.elementary.either.or}) (at least) one of the following four possibilities must hold:
$$
O\between \atopen\land \atopen\between O',
\quad
O\between \atopen'\land \atopen\between O',
\quad
O\between \atopen\land \atopen'\between O',
\quad\text{or}\quad
O\between \atopen'\land \atopen'\between O' .
$$
If $O\between \atopen\ \land\ \atopen'\between O'$ then by part~\ref{item.transitive.transitive.1} of this result we have $O\between O'$ as required. 
The other possibilities are no harder.
\qedhere\end{enumerate}
\end{proof}

\begin{defn}[Ascending/descending chain]\leavevmode
\label{defn.ascending.chains}
A \deffont[chain of sets]{chain} of sets $\mathcal X$ is a collection of sets that is totally ordered by subset inclusion $\subseteq$.\footnote{A total order is reflexive, transitive, antisymmetric, and total.}

We may call a chain \deffont[ascending chain of sets]{ascending} or \deffont[descending chain of sets]{descending} if we want to emphasise that we are thinking of the sets as `going up' or `going down'.
\end{defn}

\begin{lemm}
\label{lemm.cac.transitive}
Suppose $(\ns P,\opens)$ is a semitopology and suppose $\mathcal \atopen$ is a chain of transitive sets (Definition~\ref{defn.ascending.chains}).
Then $\bigcup\mathcal \atopen$ is a transitive set.
\end{lemm}
\begin{proof}
Suppose $O\between \bigcup\mathcal \atopen\between O'$.
Then there exist $\atopen,\atopen'\in\mathcal\atopen$ such that $O\between \atopen$ and $\atopen'\between O'$.
But $\mathcal\atopen$ is totally ordered, so either $\atopen\subseteq\atopen'$ or $\atopen\supseteq\atopen'$.
In the former case it follows that $O\between \atopen'\between O'$ so that $O\between O'$ by transitivity of $\atopen'$; the latter case is precisely similar. 
\end{proof}

\jamiesubsection{Closure properties of topens}
\label{subsect.closure.properties.of.cc}

Definition~\ref{defn.connected.set} will be useful in Proposition~\ref{prop.cc.unions}(\ref{item.clique.of.topens}): 
\begin{defn}
\label{defn.connected.set}
Suppose $(\ns P,\opens)$ is a semitopology.
Call a set of nonempty open sets $\mathcal O\subseteq\opens_{\neq\varnothing}$ a \deffont[clique of sets]{clique} when its elements pairwise intersect.\footnote{%
We call this a \emph{clique}, because if we form the \emph{intersection graph} with nodes elements of $\mathcal O$ and with an (undirected) edge between $O$ and $O'$ when $O\between O'$, then $\mathcal O$ is a clique precisely when its intersection graph is indeed a clique.
See also Definition~\ref{defn.tangled}.
We will return to intersection graphs in Subsection~\ref{subsect.intersection.graph}.  
}
In symbols: 
$$
\mathcal O\subseteq\opens\ \text{is a clique}
\quad\text{when}\quad
\Forall{O,O'\in\mathcal O}O\between O'.
$$
Note that if $\mathcal O$ is a clique then every $O\in\mathcal O$ is nonempty, since if $O=\varnothing$ then $O\notbetween O$ by Lemma~\ref{lemm.between.elementary}(\ref{item.between.nonempty}).
\end{defn}

\begin{prop}
\label{prop.cc.unions}
Suppose $(\ns P,\opens)$ is a semitopology.
Then:
\begin{enumerate*}
\item\label{item.intersecting.pair.of.topens}
If $\atopen$ and $\atopen'$ are an intersecting pair of topens (i.e. $\atopen\between \atopen'$), then $\atopen\cup \atopen'$ is topen. 
\item\label{item.clique.of.topens}
If $\mathcal \atopen$ is a clique of topens (Definition~\ref{defn.connected.set}), then $\bigcup\mathcal \atopen$ is topen. 
\item\label{item.chain.of.topens}
If $\mathcal \atopen$ is a nonempty ascending chain of topens then $\bigcup\mathcal \atopen$ is topen.
\end{enumerate*}
\end{prop}
\begin{proof}
\leavevmode
\begin{enumerate}
\item
$\atopen\cup \atopen'$ is open because by Definition~\ref{defn.semitopology}(\ref{semitopology.unions}) open sets are closed under arbitrary unions, and by Lemma~\ref{lemm.transitive.transitive}(\ref{item.transitive.transitive.2}) $\atopen\cup \atopen'$ is transitive.
\item
$\bigcup\mathcal \atopen$ is open by Definition~\ref{defn.semitopology}(\ref{semitopology.unions}).
Also, if $O\between\bigcup\mathcal \atopen\between O'$ then there exist $\atopen,\atopen'\in\mathcal \atopen$ such that $O\between \atopen$ and $\atopen'\between O'$.
We assumed $\atopen\between \atopen'$, so by Lemma~\ref{lemm.transitive.transitive}(\ref{item.transitive.transitive.1}) (since $\atopen$ and $\atopen'$ are open) we have $O\between O'$ as required. 
\item
Any chain is pairwise intersecting.  We use part~\ref{item.clique.of.topens} of this result.\footnote{We could also use Lemma~\ref{lemm.cac.transitive}.  The chain needs to be nonempty because $\bigcup\varnothing=\varnothing$ and this is open but not topen (= nonempty, transitive, and open).  The reader might ask why Lemma~\ref{lemm.cac.transitive} was not derived directly from Lemma~\ref{lemm.transitive.transitive}(\ref{item.transitive.transitive.2}); this is because (interestingly) Lemma~\ref{lemm.cac.transitive} does not require openness.}
\qedhere
\end{enumerate}
\end{proof}

\begin{corr}
\label{corr.max.cc}
Suppose $(\ns P,\opens)$ is a semitopology.
Then every topen $\atopen$ is contained in a unique maximal topen.
\end{corr}
\begin{proof}
Consider $\mathcal \atopen$ defined by
$$
\mathcal \atopen = \{\atopen\cup \atopen' \mid \atopen'\text{ topen}\land \atopen\between \atopen'\} .
$$
By Proposition~\ref{prop.cc.unions}(\ref{item.intersecting.pair.of.topens}) this is a set of topens.
By construction they all contain $\atopen$, and by our assumption that $\atopen\neq\varnothing$ they pairwise intersect (since they all contain $\atopen$, at least), so by Proposition~\ref{prop.cc.unions}(\ref{item.clique.of.topens}) $\bigcup\mathcal \atopen$ is topen.
It is easy to check that this is the unique maximal transitive open set that contains $\atopen$. 
\end{proof}

\begin{thrm}
\label{thrm.topen.partition}
Suppose $(\ns P,\opens)$ is a semitopology.
Then any $P\subseteq \ns P$, and in particular $\ns P$ itself, can be partitioned into:
\begin{itemize*}
\item
Some disjoint collection of maximal topens.
\item
A set of other points, which are not contained in any topen.
\end{itemize*}
\end{thrm}
\begin{proof}
Routine from Corollary~\ref{corr.max.cc}.
\end{proof}

\begin{rmrk}
\label{rmrk.forward}
\label{rmrk.partition}
It may be useful to put Theorem~\ref{thrm.topen.partition} in the context of the terminology, results, and examples that will follow below. 
We will have Definition~\ref{defn.tn}(\ref{item.regular.point}\&\ref{item.irregular.point}) and Theorem~\ref{thrm.max.cc.char}.
These will allow us to call a point $p$ contained in some maximal topen $\atopen$ \emph{regular}, to call the maximal topen $\atopen$ of a regular point its \emph{community}, and a point that is not contained in any topen \emph{irregular}.
Then Theorem~\ref{thrm.topen.partition} says that a semitopology $\ns P$ can be partitioned into:
\begin{itemize*}
\item
Disjoint maximal communities of regular points which, in a sense made formal in Theorem~\ref{thrm.correlated}, are a coalition acting together --- and
\item
a set of irregular points, which are in no community and so are not members of any coalition.
\end{itemize*} 
We give examples in Example~\ref{xmpl.cc} and Figure~\ref{fig.012}, and we will see more elaborate examples below (see in particular the collection in Example~\ref{xmpl.two.topen.examples}). 

In the special case that the entire space consists of a single topen community, there are no irregular points and all participants are guaranteed to agree, where algorithms succeed.
For the application of a single blockchain trying to arrive at consensus, this discussion tells us that we want it to consist of a single topen.
\end{rmrk}

\jamiesubsection{Intertwined points} 
\label{subsect.intertwined.points}

\jamiesubsubsection{The basic definition, and some lemmas}

\begin{defn}
\label{defn.intertwined.points}
Suppose $(\ns P,\opens)$ is a semitopology and $p,p'\in\ns P$.
\begin{enumerate*}
\item\label{item.p.intertwinedwith.p'}
Call $p$ and $p'$ \deffont[intertwined (two points $p\intertwinedwith p'$)]{intertwined} when $\{p,p'\}$ is transitive.\index{$p\intertwinedwith p'$ (two intertwined points)}
Unpacking Definition~\ref{defn.transitive} this means:
$$
\Forall{O,O'{\in}\opens} (p\in O\land p'\in O') \limp O\between O' .
$$ 
By a mild abuse of notation, write 
$$
p\intertwinedwith p' \quad \text{when}\quad \text{$p$ and $p'$ are intertwined}.
$$
\item\label{intertwined.defn}
Define $\intertwined{p}$\index{intertwined of $p$ ($\intertwined{p}$)}\index{$\intertwined{p}$ (points intertwined with a point $p$)} (read `intertwined of $p$') to be the set of points intertwined with $p$.
In symbols: 
$$
\intertwined{p}=\{p'\in\ns P \mid p\intertwinedwith p'\} .
$$
\end{enumerate*}
\end{defn}

\begin{xmpl}
\label{xmpl.how.different?}
We return to the examples in Example~\ref{xmpl.cc}.  
There we note that:
\begin{enumerate*}
\item
$\intertwined{1}=\{0,1,2\}$ and $\intertwined{0}=\{0,1\}$ and $\intertwined{2}=\{1,2\}$.
\item
$\intertwined{1}=\{1\}$ and $\intertwined{0}=\{0\}$ and $\intertwined{2}=\{2\}$.
\item
$\intertwined{0}=\intertwined{1}=\{0,1,2\}$ and $\intertwined{3}=\intertwined{4}=\{2,3,4\}$ and $\intertwined{2}=\ns P$.
\item
$\intertwined{0}=\{0\}$ and $\intertwined{1}=\intertwined{\ast}=\{1,\ast\}$ and $\intertwined{2}=\{2\}$. 
\item
$\intertwined{x}=\ns P$ for every $x$. 
\item
$\intertwined{x}=\{x\}$ for every $x$. 
\end{enumerate*}
\end{xmpl}

Here is one reason to care about intertwined points; a value assignment is constant on a pair of intertwined points, where it is continuous:
\begin{lemm}
\label{lemm.intertwined.correlated}
Suppose $\tf{Val}$ is a semitopology of values and $f:\ns P\to\tf{Val}$ is a value assignment (Definition~\ref{defn.value.assignment})
and $p,p'\in\ns P$ and $p\between p'$.
Then if $f$ is continuous at $p$ and $p'$, then $f(p)=f(p')$.
\end{lemm}
\begin{proof}
$\{p,p'\}$ is transitive by Definition~\ref{defn.intertwined.points}(\ref{item.p.intertwinedwith.p'}).
we use Theorem~\ref{thrm.correlated}.
\end{proof}

We might suppose that being intertwined is transitive.
Lemma~\ref{lemm.intertwined.not.transitive} shows that this is not necessarily the case (the case when $\between$ \emph{is} transitive at $p$ is an important well-behavedness property, which we will call being \emph{unconflicted}; see Subsection~\ref{subsect.reg.tra.int} and Definition~\ref{defn.conflicted}):
\begin{lemm}
\label{lemm.intertwined.not.transitive}
Suppose $(\ns P,\opens)$ is a semitopology.
Then:
\begin{enumerate*}
\item
The `is intertwined' relation $\between$ is reflexive and symmetric. 
\item
$\between$ is not necessarily transitive.
That is: $p'\intertwinedwith p\intertwinedwith p''$ does not necessarily imply $p'\intertwinedwith p''$.
\end{enumerate*}
\end{lemm}
\begin{proof}
Reflexivity and symmetry are clear from Definition~\ref{defn.intertwined.points}(\ref{item.p.intertwinedwith.p'}) and Lemma~\ref{lemm.between.elementary}(\ref{between.elementary.either.or}).

To show that transitivity need not hold, it suffices to provide a counterexample.
The semitopology from Example~\ref{xmpl.cc}(\ref{item.cc.two.regular}) (illustrated in Figure~\ref{fig.012}, top-left diagram) will do.
Take 
$$
\ns P=\{0,1,2\}
\quad\text{and}\quad
\opens=\{\varnothing,\ns P,\{0\},\{2\}\}.
$$
Then 
$$
0\between 1
\ \ \text{and}\ \ 1\between 2,
\quad\text{but}\quad
\neg(0\between 2).
$$
\end{proof}

We conclude with an easy observation:
\begin{nttn}
\label{nttn.intertwined.space}
Suppose $(\ns P,\opens)$ is a semitopology.
Call $\ns P$ \deffont[intertwined (a set $\ns P$)]{intertwined} when 
$$
\Forall{p,p'\in\ns P}p\intertwinedwith p'.
$$
In words: $\ns P$ is intertwined when all of its points are pairwise intertwined.
\end{nttn}

Lemma~\ref{lemm.intertwined.space} will be useful later, notably for Lemma~\ref{lemm.intertwined.space.regular}:
\begin{lemm}
\label{lemm.intertwined.space}
Suppose $(\ns P,\opens)$ is a semitopology.
Then the following conditions are equivalent:
\begin{enumerate*}
\item\label{item.intertwined.space.P}
$\ns P$ is an intertwined space.
\item\label{item.intertwined.space.P.transitive}
$\ns P$ is a transitive set in the sense of Definition~\ref{defn.transitive}(\ref{transitive.transitive}).
\item
All nonempty open sets intersect.
\item
Every nonempty open set is topen.
\end{enumerate*}
\end{lemm}
\begin{proof}
Routine by unpacking the definitions.
\end{proof}

\begin{rmrk}
A topologist would call an intertwined space \emph{hyperconnected} (see Definition~\ref{defn.tangled} and the following discussion).
This is also --- modulo closing under arbitrary unions --- what an expert in the classical theory of consensus might call a \emph{quorum system}~\cite{naor:loacaq}.
\end{rmrk}

\jamiesubsubsection{Pointwise characterisation of transitive sets}

\begin{lemm}
\label{lemm.three.transitive}
Suppose $(\ns P,\opens)$ is a semitopology and $\atopen\subseteq\ns P$.
Then the following are equivalent:
\begin{enumerate*}
\item\label{item.three.transitive.1}
$\atopen$ is transitive.
\item\label{item.three.transitive.2}
$p\intertwinedwith p'$ (meaning by Definition~\ref{defn.intertwined.points} that $\{p,p'\}$ is transitive) 
for every $p,p'\in \atopen$.
\end{enumerate*}
\end{lemm}
\begin{proof}
Suppose $\atopen$ is transitive.
Then by Lemma~\ref{lemm.transitive.subset}(\ref{item.transitive.subset.1}), $\{p,p'\}$ is transitive for every $p,p'\in \atopen$.

Suppose $\{p,p'\}$ is transitive for every $p,p'\in \atopen$.
Consider open sets $O$ and $O'$ such that $O\between \atopen\between O'$. 
Choose $p\in O\cap \atopen$ and $p'\in O\cap \atopen'$.
By construction $\{p,p'\}\subseteq \atopen$ so this is transitive.
It follows that $O\between O'$ as required.
\end{proof}

The special case of Lemma~\ref{lemm.three.transitive} where $\atopen$ is an open set will be particularly useful:
\begin{prop}
\label{prop.cc.char}
Suppose $(\ns P,\opens)$ is a semitopology and $\atopen\subseteq\ns P$.
Then the following are equivalent:
\begin{enumerate*}
\item
$\atopen$ is topen.
\item
$\atopen\in\opens_{\neq\varnothing}$ and $\Forall{p,p'{\in}\atopen}p\intertwinedwith p'$.
\end{enumerate*}
In words we can say:
\begin{quote}
A topen is a nonempty open set of intertwined points.
\end{quote}
\end{prop}
\begin{proof}
By Definition~\ref{defn.transitive}(\ref{transitive.cc}), $\atopen$ is topen when it is nonempty, open, and transitive. 
By Lemma~\ref{lemm.three.transitive} this last condition is equivalent to $p\intertwinedwith p'$ for every $p,p'\in \atopen$. 
\end{proof}

\begin{rmrk}[Intertwined as `non-Hausdorff']
\label{rmrk.not.hausdorff}
\leavevmode
\\
\noindent Recall that we call a topological space $(\ns P,\opens)$ \deffont[Hausdorff space]{Hausdorff} (or \deffont[$T_2$ space (Hausdorff condition)]{$T_2$}) when any two points can be separated by pairwise disjoint open sets.
Using the $\between$ symbol from Notation~\ref{nttn.between}, we rephrase the Hausdorff condition as
$$
\Forall{p,p'}p\neq p'\limp \Exists{O,O'}(p\in O\land p'\in O'\land \neg (O\between O')) , 
$$
we can simplify to 
$$
\Forall{p,p'}p\neq p'\limp p\notintertwinedwith p' ,
$$
and thus we simplify the Hausdorff condition just to
\begin{equation}
\label{eq.hausdorff}
\Forall{p}\intertwined{p}=\{p\}.
\end{equation}
Note how distinct $p$ and $p'$ being intertwined is the \emph{opposite} of being Hausdorff: $p\intertwinedwith p'$ when $p'\in\intertwined{p}$, and they \emph{cannot} be separated by pairwise disjoint open sets.
Thus the assertion $p\intertwinedwith p'$ in Proposition~\ref{prop.cc.char} is a negation to the Hausdorff property:
$$
\Exists{p}\intertwined{p}\neq\{p\} .
$$
This is useful because for semitopologies as applied to consensus, 
\begin{itemize*}
\item
being Hausdorff means that the space is separated (which is probably a bad thing, if we are looking for a system with lots of points in consensus), whereas 
\item
being full of intertwined points means 
by Theorem~\ref{thrm.correlated} that the system will (where algorithms succeed) be full of points whose value assignment agrees (which is a good thing).
\end{itemize*}
In the blockchain literature, we say that a blockchain \emph{forks} when it partitions into two sets of participants with incompatible beliefs about the state of the system.
In this light, we can view Theorem~\ref{thrm.correlated} as a result making precise sufficient conditions to ensure that this does not happen. 
\end{rmrk}

\jamiesubsection{Strong topens: topens that are also subspaces}

\jamiesubsubsection{Definition and main result}

Let us take stock and recall that:
\begin{itemize*}
\item
$\atopen$ is \emph{topen} when it is a nonempty open transitive set (Definition~\ref{defn.transitive}).
\item
$\atopen$ is \emph{transitive} when $O\between \atopen \between O'$ implies $O\between O'$ for all $O,O'\in\tf{Opens}$ (Definition~\ref{defn.transitive}). 
\item
$O\between O'$ means that $O\cap O'\neq\varnothing$ (Notation~\ref{nttn.between}). 
\end{itemize*}
But, note above that if $\atopen$ is topen and $O\between \atopen\between O'$ then $O\cap O'$ need not intersect \emph{inside $\atopen$}.
It could be that $O$ and $O'$ intersect outside of $\atopen$ (an example is in the proof Lemma~\ref{lemm.cc.subspaces} below).

Definition~\ref{defn.subspace} spells out a standard topological construction in the language of semitopologies:
\begin{defn}[Subspaces]
\label{defn.subspace}
Suppose $(\ns P,\opens)$ is a semitopology and suppose $\atopen\subseteq\ns P$ is a set of points.
Write $(\atopen,\opens\cap \atopen)$ for the semitopology such that:
\begin{itemize*}
\item
The points are $\atopen$.
\item
The open sets have the form $O\cap \atopen$ for $O\in\opens$.
\end{itemize*}
We say that $(\atopen, \opens\cap \atopen)$ is $\atopen$ with the \deffont{semitopology induced by $(\ns P,\opens)$}.

We may call $(\atopen,\opens\cap \atopen)$ a \deffont{subspace} of $(\ns P,\opens)$, and if the open sets are understood then we may omit mention of them and just write:
\begin{quote}
A subset $\atopen\subseteq\ns P$ is naturally a \deffont{(semitopological) subspace} of $\ns P$.
\end{quote}
\end{defn}

\begin{figure}
\vspace{-1em}
\centering
\subcaptionbox{A topen that is not strong (Lemma~\ref{lemm.cc.subspaces})}{\includegraphics[width=0.4\columnwidth,trim={50 0 50 20},clip]{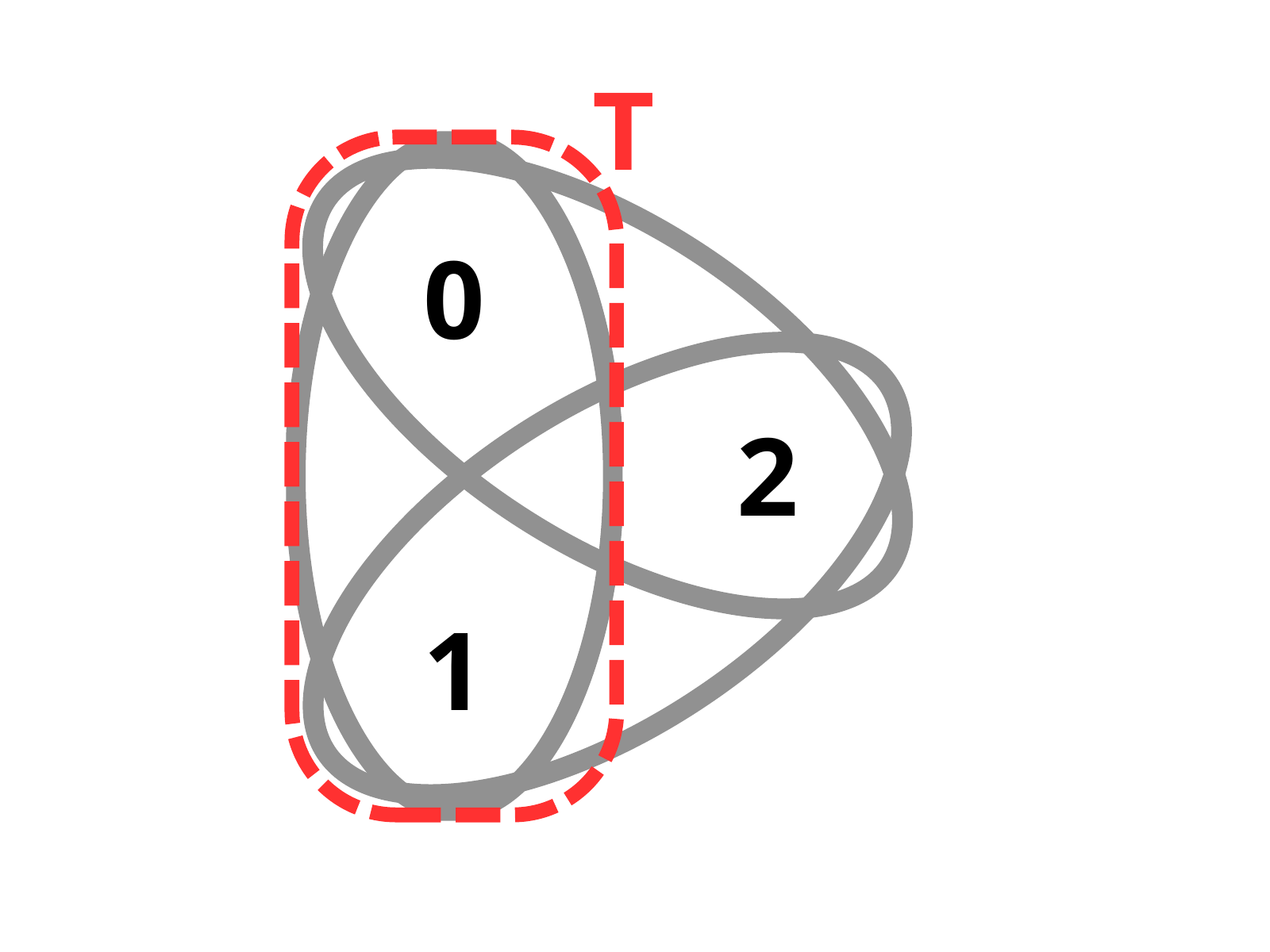}}
\qquad
\subcaptionbox{A transitive set that is not strongly transitive (Lemma~\ref{lemm.strong.is.stronger}(\ref{item.strong.is.stronger.2}))}{\includegraphics[width=0.5\columnwidth,trim={50 30 50 30},clip]{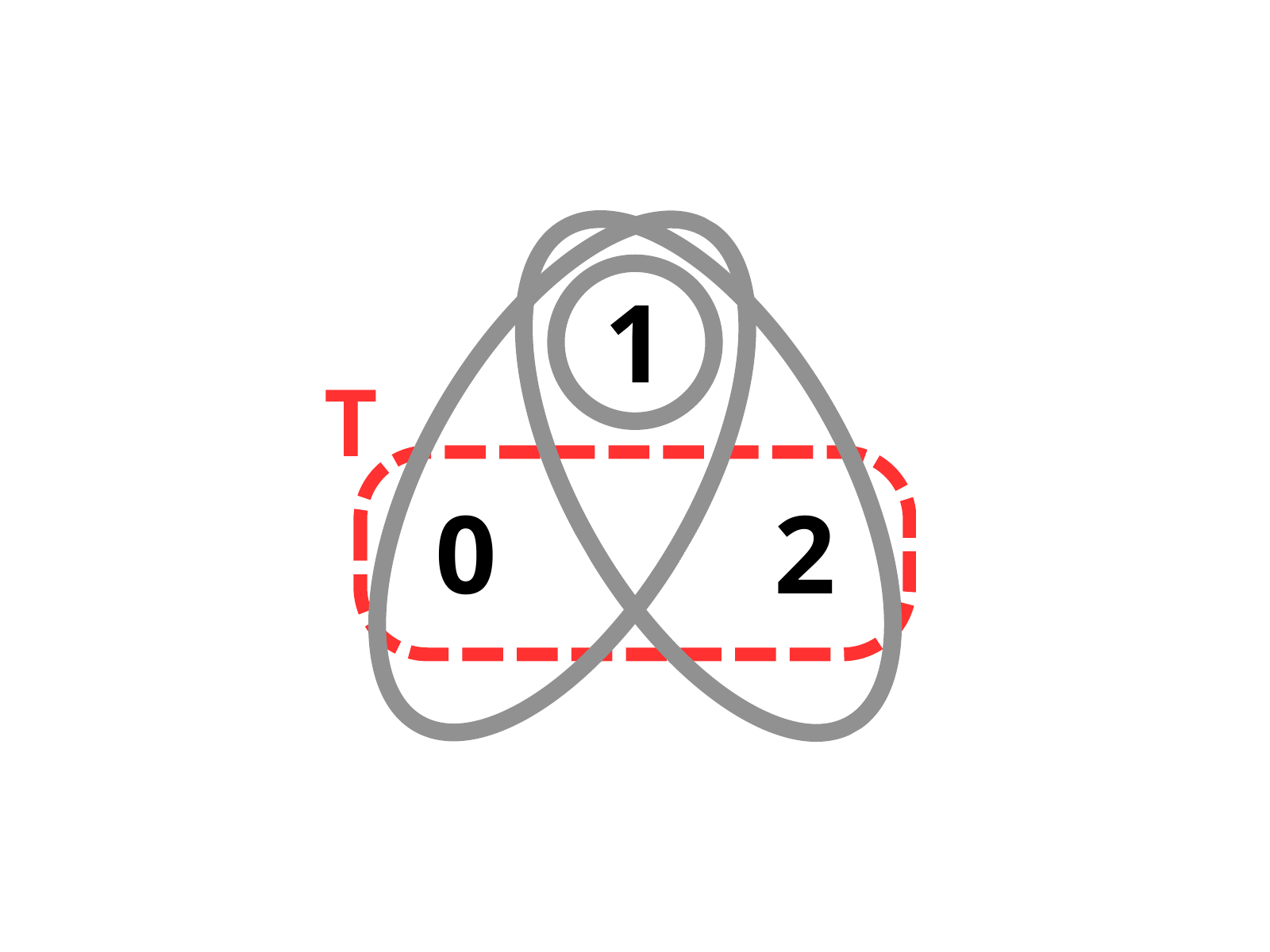}}
\caption{Two counterexamples for (strong) transitivity}
\label{fig.not-strong-topen}
\end{figure}

\begin{lemm}
\label{lemm.cc.subspaces}
The property of being a (maximal) topen is not necessarily closed under taking subspaces.
\end{lemm}
\begin{proof}
It suffices to exhibit a semitopology $(\ns P,\opens)$ and a subset $\atopen\subseteq\ns P$ such that $\atopen$ is topen in $(\ns P,\opens)$ but $\atopen$ is not topen in $(\atopen,\opens\cap \atopen)$.
We set:
$$
\ns P=\{0, 1, 2\}
\qquad
\opens=\{\varnothing,\ \{0, 2\},\ \{1, 2\},\ \{0,1\},\ \ns P\}
\qquad
\atopen=\{0,1\}
$$
as illustrated in Figure~\ref{fig.not-strong-topen} (left-hand diagram).
Now:
\begin{itemize*}
\item
$\atopen$ is topen in $(\ns P,\opens)$, because every open neighbourhood of $0$ --- that is $\{0,2\}$, $\{0,1\}$, and $\ns P$ --- intersects with every open neighbourhood of $1$ --- that is $\{1,2\}$, $\{0,1\}$, and $\ns P$.
\item
$\atopen$ is not topen in $(\atopen,\opens\cap \atopen)$, because $\{0\}$ is an open neighbourhood of $0$ and $\{1\}$ is an open neighbourhood of $1$ and these do not intersect.
\qedhere\end{itemize*}
\end{proof}

Lemma~\ref{lemm.cc.subspaces} motivates the following definitions:

\begin{defn}
\label{defn.betweenY}
Suppose $X$, $Y$, and $Z$ are sets.
Write $X\between_Y Z$, and say that $X$ and $Z$ \deffont[meet in $Y$ ($X\between_Y Z$)]{meet}\index{$X\between_Y Z$ ($X$ and $Z$ intersect in $Y$)} or \deffont[intersect in $Y$ ($X\between_Y Z$)]{intersect in $Y$}, when $(X\cap Y)\between (Z\cap Y)$.
\end{defn}

\begin{lemm}
\label{lemm.betweenY.basic.sets}
Suppose $X$, $Y$, and $Z$ are sets.
Then:
\begin{enumerate*}
\item\label{item.betweenY.basic.sets.1}
The following are equivalent:
$$
X\cap Y\cap Z\neq\varnothing 
\quad\liff\quad
X\between_Y Z
\quad\liff\quad
Y\between_X Z
\quad\liff\quad
X\between_Z Y .
$$
\item\label{item.betweenY.basic.sets.2}
$X\between_Y Y$ if and only if $X \between Y$.
\item\label{item.betweenY.basic.sets.3}
If $X\between_Y Z$ then $X\between Z$.
\end{enumerate*}
\end{lemm}
\begin{proof}
From Definition~\ref{defn.betweenY}, by elementary sets calculations.
\end{proof}

\begin{defn}
\label{defn.strongly.transitive}
Suppose $(\ns P, \opens)$ is a semitopology and recall from Definition~\ref{defn.transitive} the notions of \emph{transitive set} and \emph{topen}.
\begin{enumerate*}
\item\label{item.strongly.transitive}
Call $\atopen\subseteq\ns P$ \deffont[strongly transitive set]{strongly transitive} when
$$
\Forall{O,O'{\in}\opens} O\between \atopen \between O' \limp O\between_\atopen O' . 
$$
\item\label{strong.transitive.cc}
Call $\atopen$ a \deffont{strong topen}\index{strongly topen set} when $\atopen$ is nonempty open and strongly transitive, 
\end{enumerate*}
\end{defn}

\begin{lemm}
\label{lemm.strong.is.stronger}
Suppose $(\ns P, \opens)$ is a semitopology and $\atopen\subseteq\ns P$.
Then:
\begin{enumerate*}
\item\label{item.strong.is.stronger.1}
If $\atopen$ is strongly transitive then it is transitive.
\item\label{item.strong.is.stronger.2}
The reverse implication need not hold (even if $(\ns P,\opens)$ is a topology): it is possible for $\atopen$ to be transitive but not strongly transitive.
\end{enumerate*} 
\end{lemm}
\begin{proof}
We consider each part in turn:
\begin{enumerate}
\item
Suppose $\atopen$ is strongly transitive and suppose $O\between\atopen\between O'$.
By Lemma~\ref{lemm.betweenY.basic.sets}(\ref{item.betweenY.basic.sets.2}) $O\between_\atopen \atopen \between_\atopen O'$.
By strong transitivity $O\between_\atopen O'$.
By Lemma~\ref{lemm.betweenY.basic.sets}(\ref{item.betweenY.basic.sets.3}) $O\between O'$.
Thus $\atopen$ is transitive.
\item
It suffices to provide a counterexample.
This is illustrated in Figure~\ref{fig.not-strong-topen} (right-hand diagram).
We set:
\begin{itemize*}
\item
$\ns P = \{0,1,2\}$, and
\item
$\opens= \{\varnothing,\ \{1\},\ \{0,1\},\ \{1,2\},\ \{0,1,2\}\}$.
\item
We set $\atopen=\{0,2\}$.
\end{itemize*}
We note that $(\ns P,\opens)$ is a topology, and it is easy to check that $\atopen$ is transitive --- we just note that $\{0,1\}\between\atopen\between\{1,2\}$ and $\{0,1\}\between\{1,2\}$.
However, $\atopen$ is not strongly transitive, because $\{0,1\}\cap\{1,2\}=\{1\}\not\subseteq\atopen$.
\qedhere\end{enumerate}
\end{proof}

\begin{prop}
Suppose $(\ns P,\opens)$ is a semitopology and suppose $\atopen\in\opens$.
Then the following are equivalent:
\begin{enumerate*}
\item
$\atopen$ is a strong topen.
\item
$\atopen$ is a topen in $(\atopen,\opens\cap \atopen)$ (Definition~\ref{defn.subspace}).
\end{enumerate*} 
\end{prop}
\begin{proof}
Suppose $\atopen$ is a strong topen; thus $\atopen$ is nonempty, open, and strongly transitive in $(\ns P,\opens)$.
Then by construction $\atopen$ is open in $(\atopen,\opens\cap \atopen)$, and the strong transitivity property of Definition~\ref{defn.strongly.transitive} asserts precisely that $\atopen$ is transitive as a subset of $(\atopen,\opens\cap \atopen)$.

Now suppose $\atopen$ is a topen in $(\atopen,\opens\cap \atopen)$; thus $\atopen$ is nonempty, open, and transitive in $(\atopen,\opens\cap \atopen)$.
Then $\atopen$ is nonempty and by assumption above $\atopen\in\opens$.\footnote{It does not follow from $\atopen$ being open in $(\atopen,\opens\cap \atopen)$ that $\atopen$ is open in $(\ns P,\opens)$, which is why we included an assumption that this holds in the statement of the result.}
Now suppose $O,O'\in\opens$ and $O\between \atopen\between O'$.
Then by Lemma~\ref{lemm.betweenY.basic.sets}(\ref{item.betweenY.basic.sets.2}) $O \between_\atopen \atopen\between_\atopen O'$, so by transitivity of $\atopen$ in $(\atopen,\opens\cap \atopen)$ also $O\between_\atopen O'$, and thus by Lemma~\ref{lemm.betweenY.basic.sets}(\ref{item.betweenY.basic.sets.3}) also $O\between O'$. 
\end{proof}

\jamiesubsubsection{Connection to lattice theory}

There is a notion from order-theory of a \emph{join-irreducible} element (see for example in \cite[Definition~2.42]{priestley:intlo}), and a dual notion of \emph{meet-irreducible} element:
\begin{defn}
Call an element $s$ in a lattice $\mathcal L$ 
\begin{itemize*}
\item
\deffont[join-irreducible element]{join-irreducible} when $s$ is not a bottom element, and $s$ is not a join of two strictly smaller elements: if $x\vee y=s$ then $x=s$, or $y=s$, and
\item
\deffont[meet-irreducible element]{meet-irreducible} when $s$ is not a top element, and $s$ is not a meet of two strictly greater elements: if $x\wedge y=s$ then $x=s$ or $y=s$. 
\end{itemize*}
This definition is typically given for lattices, but it makes just as much sense for semilattices as well.
\end{defn}

\begin{xmpl}
\label{xmpl.meet-irreducible}
\leavevmode
\begin{enumerate*}
\item
Consider the lattice of finite (possibly empty) subsets of $\mathbb N$, with $\mathbb N$ adjoined as a top element.
Then $\mathbb N$ is join-irreducible; $\mathbb N\subseteq\mathbb N$ is not a bottom element, and if $x\cup y=\mathbb N$ then either $x=\mathbb N$ or $y=\mathbb N$.
\item\label{item.final.N}
Consider $\mathbb N$ with the \deffont{final segment semitopology} such that opens are either $\varnothing$ or sets $n_\geq = \{n'\in\mathbb N \mid n'\geq n\}$.

Then $\varnothing$ is meet-irreducible; $\varnothing$ is not a top element, and if $x\cap y=\varnothing$ then either $x=\varnothing$ or $y=\varnothing$.
\item
Consider the integers with the lattice structure in which meet is minimum and join is maximum.
Then every element is join- and meet-irreducible; if $x\vee y=z$ then $x=z$ or $y=z$, and similarly for $x\wedge y$. 
\end{enumerate*}
\end{xmpl}

We spell out how this is related to our notions of transitivity from Definitions~\ref{defn.transitive} and~\ref{defn.strongly.transitive}:
\begin{lemm}
\label{lemm.meet-irreducible}
Suppose $(\ns P,\opens)$ is a semitopology and $\atopen\subseteq\ns P$.
Then: 
\begin{enumerate*}
\item\label{item.meet-irreducible.1}
$\atopen$ is strongly transitive if and only if $\varnothing$ is meet-irreducible in $(\atopen,\opens\cap \atopen)$ (Definition~\ref{defn.subspace}). 
\item
$\atopen$ is transitive if $\varnothing$ is meet-irreducible in $(\atopen,\opens\cap \atopen)$.
\item
If $\atopen$ is transitive it does not necessarily follow that $\varnothing$ is meet-irreducible in $(\atopen,\opens\cap \atopen)$.
\end{enumerate*}
\end{lemm}
\begin{proof}
We reason as follows: 
\begin{enumerate}
\item
$\varnothing$ is meet-irreducible in $(\atopen,\opens\cap \atopen)$ means that $(O\cap \atopen)\cap (O'\cap \atopen)=\varnothing$ implies $O\cap \atopen=\varnothing$ or $O\cap \atopen'=\varnothing$.

$\atopen$ is strongly transitive when (taking the contrapositive in Definition~\ref{defn.strongly.transitive}(\ref{item.strongly.transitive})) $(O\cap \atopen)\cap (\atopen\cap O')=\varnothing$ implies $O\cap \atopen=\varnothing$ or $\atopen\cap O'=\varnothing$.

That these conditions are equivalent follows by straightforward sets manipulations. 
\item
We can use part~\ref{item.meet-irreducible.1} of this result and Lemma~\ref{lemm.strong.is.stronger}(\ref{item.strong.is.stronger.1}), or give a direct argument by sets calculations: if $O\cap O'=\varnothing$ then $(O\cap \atopen)\cap (\atopen\cap O')=\varnothing$ and by meet-irreducibility $O\cap \atopen=\varnothing$ or $\atopen\cap O'=\varnothing$ as required.
\item
Figure~\ref{fig.not-strong-topen} (left-hand diagram) provides a counterexample, taking $\atopen=\{0,1\}$ and $O=\{0,2\}$ and $O'=\{1,2\}$.
Then $(O\cap \atopen)\cap (\atopen\cap O')=\varnothing$ but it is not the case that $O\cap \atopen=\varnothing$ or $O'\cap \atopen=\varnothing$.
\qedhere\end{enumerate}
\end{proof}

\begin{rmrk}
\label{rmrk.imperfect}
The proof of Lemma~\ref{lemm.meet-irreducible} not hard, but the result is interesting for what it says, and also for what it does not say:
\begin{enumerate}
\item
The notion of being a strong topen maps naturally to something in order theory; namely that $\varnothing$ is meet-irreducible in the induced poset $\{O\cap \atopen\mid O\in\opens\}$ which is the set of open sets of the subspace $(\atopen,\opens\cap \atopen)$ of $(\ns P,\opens)$.
\item
However, this mapping is imperfect: the poset is not a lattice, and it is also not a sub-poset of $\opens$ --- even if $\atopen$ is topen.
If $\opens$ were a topology and closed under intersections then we would have a lattice --- but it is precisely the point of difference between semitopologies vs. topologies that open sets need not be closed under intersections. 
\item
Being transitive does not correspond to meet-irreducibility; there is an implication in one direction, but certainly not in the other. 
\end{enumerate}
So, Lemma~\ref{lemm.meet-irreducible} says that (strong) transitivity has a flavour of meet-irreducibility, but in a way that also illustrates --- as did Proposition~\ref{prop.max.topen.min.closed}(\ref{item.max.topen.min.closed.2}) --- how semitopologies are different, because they are not closed under intersections, and have their own behaviour.

See also the characterisation of strong transitivity in Lemma~\ref{lemm.strongly.dense.strongly.transitive} and the surrounding discussion.
\end{rmrk}

\jamiesubsubsection{Topens in topologies}
\label{subsection.topens.in.topologies}

We conclude by briefly looking at what `being topen' means if our semitopology is actually a topology.
We recall a standard definition from topology:
\begin{defn}
\label{defn.tangled}
Suppose $(\ns P,\opens)$ is a semitopology.
Call $\atopen\subseteq\ns P$ \deffont[hyperconnected set]{hyperconnected} when all nonempty open subsets of $\atopen$ intersect.\footnote{Calling this \emph{hyperconnected} is a slight but natural generalisation of the usual definition: in topology, `hyperconnected' is typically used to refer to an entire space rather than a subset of it.  In the case that $\atopen=\ns P$, our definition specialises to the usual one.}
In symbols: 
$$
\Forall{O,O'\in\opens_{\neq\varnothing}} O,O'\subseteq\atopen \limp O\between O' .
$$
\end{defn}

\begin{lemm}
\label{lemm.tran.neosi}
Suppose $(\ns P,\opens)$ is a semitopology.
Then if $\atopen\subseteq\ns P$ is transitive then it is hyperconnected.
\end{lemm}
\begin{proof}
Suppose $\varnothing\neq O,O'\subseteq\atopen$.
Then $O\between\atopen\between O'$ and by transitivity $O\between O'$ as required.
\end{proof}

What is arguably particularly interesting about Lemma~\ref{lemm.tran.neosi} is that its reverse implication does \emph{not} hold, and in quite a strong sense: 
\begin{lemm}
\label{lemm.tran.no.neosi}
Suppose $(\ns P,\opens)$ is a semitopology and $\atopen\subseteq\ns P$. 
Then:
\begin{enumerate*}
\item
$\atopen$ can be hyperconnected but not transitive, even if $(\ns P,\opens)$ is a topology (not just a semitopology).
\item
$\atopen$ can be hyperconnected but not transitive, even if $\atopen$ is an open set.
\end{enumerate*}
\end{lemm}
\begin{proof}
It suffices to provide counterexamples:
\begin{enumerate}
\item
Consider the semitopology illustrated in the lower-left diagram in Figure~\ref{fig.012} (which is a topology), and set $\atopen=\{0,4\}$.
This has no nonempty open subsets so it is trivially hyperconnected.
However, $\atopen$ is not transitive because $\{0,1\}\between \atopen \between \{3,4\}$ yet $\{0,1\}\notbetween\{3,4\}$.
\item
Consider the semitopology illustrated in the top-right diagram in Figure~\ref{fig.012}, and set $\atopen=\{0,1\}$.
This has two nonempty open subsets, $\{0\}$ and $\{0,1\}$, so it is hyperconnected.
However, $\atopen$ is not transitive, because $\{0\}\between \atopen \between \{1,2\}$ yet $\{0\}\notbetween\{1,2\}$.
\qedhere\end{enumerate}
\end{proof}

We know from Lemma~\ref{lemm.strong.is.stronger}(\ref{item.strong.is.stronger.2}) that `transitive' does not imply `strongly transitive' for an arbitrary subset $\atopen\subseteq\ns P$, even in a topology.
When read together with Lemmas~\ref{lemm.tran.neosi} and~\ref{lemm.tran.no.neosi}, this invites the question of what happens when 
\begin{itemize*}
\item
$(\ns P,\opens)$ is a topology, and \emph{also} 
\item
$\atopen$ is an open set.
\end{itemize*}
In this natural special case, strong transitivity, transitivity, and being hyperconnected, all become equivalent: 
\begin{lemm}
\label{lemm.transitive.topology}
Suppose $(\ns P,\opens)$ is a topology and suppose $\atopen\in\opens$ is an open set.
Then the following are equivalent:
\begin{itemize*}
\item
$\atopen$ is a strong topen (Definition~\ref{defn.strongly.transitive}(\ref{strong.transitive.cc})).
\item
$\atopen$ is a topen.
\item
$\atopen$ is hyperconnected.
\end{itemize*}
\end{lemm}
\begin{proof}
We assumed $\atopen$ is open, so the equivalence above can also be thought of as 
\begin{quote}
strongly transitive $\liff$ transitive $\liff$ all nonempty open subsets intersect.
\end{quote}
We prove a chain of implications:
\begin{itemize}
\item
If $\atopen$ is a strong topen then it is a topen by Lemma~\ref{lemm.strong.is.stronger}(\ref{item.strong.is.stronger.1}).
\item
If $\atopen$ is a topen then we use Lemma~\ref{lemm.tran.neosi}.
\item
Suppose $\atopen$ is hyperconnected, so every pair of nonempty open subsets of $\atopen$ intersect; and 
suppose $O,O'\in\opens_{\neq\varnothing}$ and $O\between\atopen\between O'$.
Then also $(O\cap\atopen) \between \atopen \between (O'\cap\atopen)$.
Now $O\cap\atopen$ and $O'\cap\atopen$ are open: because $\atopen$ is open; and $\ns P$ is a topology (not just a semitopology), so intersections of open sets are open.
By transitivity of $\atopen$ we have $O\cap\atopen\between O'\cap\atopen$.
Since $O$ and $O'$ were arbitrary, $\atopen$ is strongly transitive.
\qedhere\end{itemize} 
\end{proof}

\jamiesection{Interiors, communities \& regular points}
\label{sect.regular.points}

\jamiesubsection{Community of a (regular) point}

Definition~\ref{defn.interior} is standard:
\begin{defn}[Open interior]
\label{defn.interior}
Suppose $(\ns P,\opens)$ is a semitopology and $P\subseteq\ns P$.
Define $\interior(P)$ the \deffont{(open) interior of $P$}\index{$\interior(P)$ (open interior)} by
$$
\interior(P)=\bigcup\{ O\in\opens \mid O\subseteq P\} .
$$
\end{defn}

\begin{lemm}
\label{lemm.interior.open}
Suppose $(\ns P,\opens)$ is a semitopology and $P\subseteq\ns P$.
Then $\interior(P)$ from Definition~\ref{defn.interior} is the greatest open subset of $P$.
\end{lemm}
\begin{proof}
Routine by the construction in Definition~\ref{defn.interior} and closure of open sets under unions (Definition~\ref{defn.semitopology}(\ref{semitopology.unions})).
\end{proof}

\begin{corr}
\label{corr.interior.monotone}
Suppose $(\ns P,\opens)$ is a semitopology and $P,P'\subseteq\ns P$.
Then if $P\subseteq P'$ then $\interior(P)\subseteq\interior(P')$.
\end{corr}
\begin{proof}
Routine using Lemma~\ref{lemm.interior.open}.
\end{proof}

\begin{defn}[Community of a point, and regularity]
\label{defn.tn}
Suppose $(\ns P,\opens)$ is a semitopology and $p\in\ns P$.
Then:
\begin{enumerate*}
\item\label{item.tn}
Define $\community(p)$ the \deffont[community of $p$ ($\community(p)$)]{community of $p$}\index{$\community(p)$ (community of a point)} by 
$$
\community(p)=\interior(\intertwined{p}) .
$$
\item\label{item.community.P}
Extend $\community$ to subsets $P\subseteq\ns P$ by taking a sets union:
$$
\community(P) = \bigcup\{\community(p) \mid p\in P\} .
$$
\item\label{item.regular.point}
Call $p$ a \deffont{regular point} when its community is a topen neighbourhood of $p$.
In symbols:
$$
p\text{ is regular}\quad\text{when}\quad p\in\community(p)\in\topens .
$$
\item\label{item.weakly.regular.point}
Call $p$ a \deffont{weakly regular point} when its community is an open (but not necessarily topen) neighbourhood of $p$.
In symbols:
$$
p\text{ is weakly regular}\quad\text{when}\quad p\in\community(p)\in\opens .
$$
\item\label{item.quasiregular.point}
Call $p$ a \deffont{quasiregular point} when its community is nonempty.
In symbols:
$$
p\text{ is quasiregular}\quad\text{when}\quad \varnothing\neq\community(p)\in\opens .
$$
\item\label{item.irregular.point}
If $p$ is not regular then we may call it an \deffont{irregular point}, or just say that it is not regular.
\item\label{item.regular.S}
If $P\subseteq\ns P$ and every $p\in P$ is regular/weakly regular/quasiregular/irregular then we may call $P$ a \deffont{regular/weakly regular/quasiregular/irregular set} respectively (see also Definition~\ref{defn.conflicted}(\ref{item.unconflicted})).
\qedhere\end{enumerate*}
\end{defn}

\begin{rmrk}
\label{rmrk.r.wr.qr}
Lemmas~\ref{lemm.wr.r} and~\ref{lemm.wr.r.no} give an overview of the relationships between the properties in Definition~\ref{defn.tn}.
For reference, two more regularity-flavoured conditions will appear later: \emph{indirect regularity} in Definition~\ref{defn.indirectly.regular}, and an \emph{MCN} property is mentioned in Remark~\ref{rmrk.mcn}.
See also Remark~\ref{rmrk.linear.regularity}.
\end{rmrk}

\begin{lemm}
\label{lemm.wr.r}
Suppose $(\ns P,\opens)$ is a semitopology and $p\in\ns P$.
Then:
\begin{enumerate*}
\item\label{item.r.implies.wr}
If $p$ is regular, then $p$ is weakly regular.
\item\label{item.wr.implies.qr}
If $p$ is weakly regular, then $p$ is quasiregular.
\end{enumerate*}
\end{lemm}
\begin{proof}
We consider each part in turn:
\begin{enumerate}
\item
If $p$ is regular then by Definition~\ref{defn.tn}(\ref{item.regular.point}) $p\in\community(p)\in\topens$, so certainly $p\in\community(p)$ and by Definition~\ref{defn.tn}(\ref{item.weakly.regular.point}) $p$ is weakly regular.
\item
If $p$ is weakly regular then by Definition~\ref{defn.tn}(\ref{item.weakly.regular.point}) $p\in\community(p)\in\opens$, so certainly $\community(p)\neq\varnothing$ and by Definition~\ref{defn.tn}(\ref{item.quasiregular.point}) $p$ is quasiregular.
\qedhere
\end{enumerate}
\end{proof}

\begin{xmpl}
\label{xmpl.wr}
\leavevmode
\begin{enumerate*}
\item
In Figure~\ref{fig.not-strong-topen} (left-hand diagram),\ $0$, $1$, and $2$ are three intertwined points and the entire space $\{0,1,2\}$ consists of a single topen set.
It follows that $0$, $1$, and $2$ are all regular and their community is $\{0,1,2\}$.
\item\label{item.wr.2}
In Figure~\ref{fig.012} (top-left diagram),\ $0$ and $2$ are regular and $1$ is weakly regular but not regular ($1\in\community(1)=\{0,1,2\}$ but $\{0,1,2\}$ is not topen). 
\item\label{item.qr.2}
In Figure~\ref{fig.012} (lower-right diagram),\ $0$, $1$, and $2$ are regular and $\ast$ is quasiregular ($\community(\ast)=\{1\}$).
\item
In Figure~\ref{fig.012} (top-right diagram),\ $0$ and $2$ are regular and $1$ is neither regular, weakly regular, nor quasiregular ($\community(1)=\varnothing$).
\item
In a semitopology of values $(\tf{Val},\powerset(\tf{Val}))$ (Definition~\ref{defn.value.assignment}) every value $v\in\tf{Val}$ is regular, weakly regular, and unconflicted.
\item\label{item.wr.6}
In $\mathbb R$ with its usual topology (which is also a semitopology), every point is unconflicted because the topology is Hausdorff and by Equation~\ref{eq.hausdorff} in Remark~\ref{rmrk.not.hausdorff} this means precisely that $\intertwined{p}=\{p\}$ so $p$ is intertwined just with itself.
Furthermore $p$ is not (quasi/weakly)regular, because $\community(p)=\interior(\intertwined{p})=\varnothing$.
\end{enumerate*} 
\end{xmpl}

\begin{lemm}
\label{lemm.wr.r.no}
Suppose $(\ns P,\opens)$ is a semitopology and $p\in\ns P$.
Then:
\begin{enumerate*}
\item\label{item.wr.r.not.quasiregular}
$p$ might not be quasiregular (i.e. $\community(p)=\varnothing$); thus by Lemma~\ref{lemm.wr.r} it is also not weakly regular and not regular.
\item\label{item.wr.r.no.converse.1}
$p$ might be quasiregular but not weakly regular (i.e. $\community(p)\neq\varnothing$ but $p\notin\community(p)$); and 
\item\label{item.wr.r.no.converse.2}
$p$ might be weakly regular but not regular (i.e. $p\in\community(p)\notin\topens$). 
\end{enumerate*}
\end{lemm}
\begin{proof}
We consider each part in turn:
\begin{enumerate}
\item
Point $0\in\mathbb R$ in Example~\ref{xmpl.wr}(\ref{item.wr.6}) is not quasiregular.
\item
Point $1$ in Example~\ref{xmpl.wr}(\ref{item.wr.2}) (illustrated in Figure~\ref{fig.012}, top-left diagram) is weakly regular ($\community(1)=\{0,1,2\}$) but not regular ($\community(1)$ is open but not topen).
\item
Point $\ast$ in Example~\ref{xmpl.wr}(\ref{item.qr.2}) (illustrated in Figure~\ref{fig.012}, lower-right diagram) is quasiregular ($\community(\ast)=\{1\}$ is nonempty but does not contain $\ast$).
\qedhere
\end{enumerate}
\end{proof}

\begin{lemm}
\label{lemm.intertwined.space.regular}
Suppose $(\ns P,\opens)$ is a semitopology.
Then:
\begin{enumerate*}
\item\label{item.intertwined.space.regular.1}
If all nonempty open sets intersect then $(\ns P,\opens)$ is regular (meaning that every $p\in\ns P$ is regular).
\item\label{item.intertwined.space.regular.2}
The reverse implication need not hold: it is possible for $(\ns P,\opens)$ to be regular but not all open sets intersect (cf. Corollary~\ref{corr.topen.partition.char}).
\end{enumerate*}
\end{lemm}
\begin{proof}
We consider each part in turn:
\begin{enumerate}
\item
By Lemma~\ref{lemm.intertwined.space}(\ref{item.intertwined.space.P.transitive}) $\ns P\in\topens$ (since it is transitive and open).
By Lemma~\ref{lemm.intertwined.space}(\ref{item.intertwined.space.P}) $\intertwined{p}=\ns P$ for every $p\in\ns P$, thus $\community(p)=\interior(\intertwined{p})=\ns P$.
Thus $p\in\community(p)\in\topens$ for every $p\in\ns P$, so $\ns P$ is regular.
\item
It suffices to provide a counterexample.
We take any discrete semitopology with at least two elements; e.g. $(\{0,1\},\powerset(\{0,1\}))$.
Then $\{0\}\notintersectswith\{1\}$, but by Corollary~\ref{corr.when.singleton.topen} $0$ and $1$ are both regular.
\qedhere
\end{enumerate}
\end{proof}

\begin{xmpl}
When we started looking at semitopologies we gave some examples in Example~\ref{xmpl.semitopologies}.
These may seem quite elementary now, but we run through them commenting on which spaces are regular, weakly regular, or quasiregular:
\begin{itemize*}
\item
Any discrete semitopology is regular; topen neighbourhoods are just the singleton sets.
\item
The initial semitopology is regular: it has no topen neighbourhoods, but also no points.
The final semitopology is regular: it has one topen neighbourhood, containing one point.
The trivial topology is regular; it has a single topen neighbourhood that is $\ns P$ itself. 
\item
The supermajority semitopology is regular.
It has one topen neighbourhood containing all of $\ns P$.
\item
The many semitopology is regular if $\ns P$ is finite (because it is equal to the trivial semitopology), and not even quasiregular if $\ns P$ is infinite, because (for infinite $\ns P$) $\intertwined{p}=\varnothing$ for every point.
For example, if $\ns P=\mathbb N$ and $p$ is even and $p'$ is odd, then $\f{evens}=\{2*n \mid n\in\mathbb N\}$ and $\f{odds}=\{2*n\plus 1 \mid n\in\mathbb N\}$ are disjoint open neighbourhoods of $p$ and $p'$ respectively.
\item
The all-but-one semitopology is regular for $\ns P$ having cardinality of $3$ or more, since all points are intertwined so there is a single topen neighbourhood which is the whole space.
If $\ns P$ has cardinality $2$ or $1$ then we have a discrete semitopology (on two points or one point) and these too are regular, with two or one topen neighbourhoods. 
\item
The more-than-one semitopology is not even quasiregular for $\ns P$ having cardinality of $4$ or more.
If $\ns P$ has cardinality $3$ then we get the left-hand topology in Figure~\ref{fig.not-strong-topen}, which is regular.
If $\ns P$ has cardinality $2$ then we get the trivial semitopology, which is regular. 
\item
Take $\ns P=\mathbb R$ (the set of real numbers) and let open sets be generated by intervals of the form $\rightopeninterval{0,r}$ or $\leftopeninterval{\minus r,0}$ for any strictly positive real number $r>0$.
The reader can check that this semitopology is regular.
\item
Any quorum system induces an intertwined semitopology, as outlined in Example~\ref{xmpl.semitopologies}(\ref{item.quorum.system}).
By Lemmas~\ref{lemm.intertwined.space.regular}(\ref{item.intertwined.space.regular.1}) and~\ref{lemm.intertwined.space} this is a regular semitopology, and every nonempty open set is a topen neighbourhood.
\end{itemize*}
\end{xmpl}

\begin{rmrk}
We pause to recap:
\leavevmode
\begin{enumerate}
\item
$\community(p)$ always exists and always is open.
It may or may not be empty, may or may not be topen, and may or may not contain $p$.
\item
When $p\in\community(p)\in\topens$ we call $p$ `regular', which suggests that non-regular behaviour --- $p\notin\community(p)$ and/or $\community(p)\notin\topens$, or even $\community(p)=\varnothing$ --- is `bad behaviour', and being regular `good behaviour'.

But what is this good behaviour that regularity implies? 
Theorem~\ref{thrm.correlated} (continuous value assignments are constant on topens) tells us that a regular $p$ is surrounded by a topen neighbourhood of points $\community(p)=\interior(\intertwined{p})$ that must agree with it under continuous value assignments.
Using our terminology \emph{community} and \emph{regular}, we can say that \emph{the community of a regular $p$ shares its values}.
\item
We can sum up the above intuitively as follows: 
\begin{enumerate*}
\item
We care about transitivity because it implies agreement.
\item
We care about being open, because it implies actionability. 
\item
Thus, a regular point is interesting because it is a participant in a maximal topen neighbourhood and therefore can \emph{i)} come to agreement and \emph{ii)} take action on that agreement. 
\end{enumerate*}
\item
The question then arises how the community of $p$ can be (semi)topologically characterised.
We will explore, notably in Theorem~\ref{thrm.max.cc.char}, Proposition~\ref{prop.views.of.regularity}, and Theorem~\ref{thrm.up.down.char}; see also Remark~\ref{rmrk.arc}.
\end{enumerate}
\end{rmrk} 

\jamiesubsection{Further exploration of (quasi-/weak) regularity and topen sets}

\begin{rmrk}
\label{rmrk.T0-T2}
Recall three common separation axioms from topology:
\begin{enumerate*}
\item
$T_0$: if $p_1\neq p_2$ then there exists some $O\in\opens$ such that $(p_1\in O)\xor (p_2\in O)$, where $\xor$ denotes \emph{exclusive or}.
\item
$T_1$: if $p_1\neq p_2$ then there exist $O_1,O_2\in\opens$ such that $p_i\in O_j \liff i=j$ for $i,j\in\{1,2\}$.
\item
$T_2$, or the \emph{Hausdorff condition}: if $p_1\neq p_2$ then there exist $O_1,O_2\in\opens$ such that $p_i\in O_j \liff i=j$ for $i,j\in\{1,2\}$, and $O_1\cap O_2=\varnothing$.
Cf. the discussion in Remark~\ref{rmrk.not.hausdorff}.
\end{enumerate*}
Even the weakest of the well-behavedness property for semitopologies that we consider in Definition~\ref{defn.tn} --- quasiregularity --- is in some sense strongly opposed to the space being Hausdorff/$T_2$ (though not to being $T_1$), as Lemma~\ref{lemm.quasiregular.hausdorff} makes formal.
\end{rmrk}

\begin{lemm}
\label{lemm.quasiregular.hausdorff}
\leavevmode
\begin{enumerate*}
\item
Every quasiregular Hausdorff semitopology is discrete.

In more detail: if $(\ns P,\opens)$ is a semitopology that is quasiregular (Definition~\ref{defn.tn}(\ref{item.quasiregular.point})) and Hausdorff (equation~\ref{eq.hausdorff} in Remark~\ref{rmrk.not.hausdorff}), then $\opens=\powerset(\ns P)$. 
\item
There exists a (quasi)regular $T_1$ semitopology that is not discrete.
\end{enumerate*} 
\end{lemm}
\begin{proof}
We consider each part in turn:
\begin{enumerate}
\item
By the Hausdorff property, $\intertwined{p}=\{p\}$.
By the quasiregularity property, $\community(p)\neq\varnothing$.
It follows that $\community(p)=\{p\}$.
But by construction in Definition~\ref{defn.tn}(\ref{item.tn}), $\community(p)$ is an open interior.
Thus $\{p\}\in\opens$.
The result follows.
\item
It suffices to provide an example.
We use the left-hand semitopology in Figure~\ref{fig.not-strong-topen}.
Thus $\ns P=\{0,1,2\}$ and $\opens$ is generated by $\{0,1\}$, $\{1,2\}$, and $\{2,0\}$.
All nonempty open sets intersect, so by Lemma~\ref{lemm.intertwined.space.regular}(\ref{item.intertwined.space.regular.1}) $\ns P$ is regular.
It is also $T_1$ (Remark~\ref{rmrk.T0-T2}).
\qedhere\end{enumerate}
\end{proof}
 
Lemma~\ref{lemm.two.intertwined} confirms in a different way that regularity (Definition~\ref{defn.tn}(\ref{item.regular.point})) is non-trivially distinct from weak regularity and quasiregularity:
\begin{lemm}
\label{lemm.two.intertwined}
Suppose $(\ns P,\opens)$ is a semitopology and $p\in\ns P$.
Then:
\begin{enumerate*}
\item\label{item.two.intertwined.1}
$\community(p)\in\opens$.
\item\label{item.two.intertwined.2}
$\community(p)$ is not necessarily topen; equivalently $\community(p)$ is not necessarily transitive.
(More on this later in Subsection~\ref{subsect.irregular}.)
\end{enumerate*}
\end{lemm}
\begin{proof}
$\community(p)$ is open by construction in Definition~\ref{defn.tn}(\ref{item.tn}), since it is an open interior.

For part~\ref{item.two.intertwined.2}, it suffices to provide a counterexample.
We consider the semitopology from Example~\ref{xmpl.cc}(\ref{item.cc.two.regular}) (illustrated in Figure~\ref{fig.012}, top-left diagram). 
We calculate that $\community(1)=\{0,1,2\}$ so that $\community(1)$ is an open neighbourhood of $1$ --- but it is not transitive, and thus not topen, since $\{0\}\cap\{2\}=\varnothing$.

Further checking reveals that $\{0\}$ and $\{2\}$ are two maximal topens within $\community(1)$. 
\end{proof}

So what is $\community(p)$?
We start by characterising $\community(p)$ as the \emph{greatest} topen neighbourhood of $p$, if this exists:
\begin{lemm}
\label{lemm.intertwined.is.the.greatest}
\label{lemm.max.cc.intertwined}
Suppose $(\ns P,\opens)$ is a semitopology and recall from Definition~\ref{defn.tn}(\ref{item.regular.point}) that $p$ is regular when $\community(p)$ is a topen neighbourhood of $p$.
\begin{enumerate*}
\item\label{item.intertwined.is.the.greatest.1}
If $\community(p)$ is a topen neighbourhood of $p$ (i.e. if $p$ is regular) then $\community(p)$ is a maximal topen.
\item\label{item.intertwined.is.the.greatest.2}
If $p\in \atopen\in\topens$ is a maximal topen neighbourhood of $p$ then $\atopen=\community(p)$.
\end{enumerate*}
\end{lemm}
\begin{proof}
\leavevmode
\begin{enumerate}
\item
Since $p$ is regular, by definition, $\community(p)$ is topen and is a neighbourhood of $p$.
It remains to show that $\community(p)$ is a maximal topen.

Suppose $\atopen$ is a topen neighbourhood of $p$; we wish to prove $\atopen\subseteq \community(p)=\interior(\intertwined{p})$.
Since $\atopen$ is open it would suffice to show that $\atopen\subseteq\intertwined{p}$.
By Proposition~\ref{prop.cc.char} $p\intertwinedwith p'$ for every $p'\in \atopen$, and it follows immediately that $\atopen\subseteq\intertwined{p}$.
\item
Suppose $\atopen$ is a maximal topen neighbourhood of $p$.

First, note that $\atopen$ is open, and by Proposition~\ref{prop.cc.char} $\atopen\subseteq\intertwined{p}$, so $\atopen\subseteq\community(p)$.

By assumption $p\in\atopen\cap\community(p)$ and both are topen so by Proposition~\ref{prop.cc.unions}(\ref{item.intersecting.pair.of.topens}) $\atopen\cup\community(p)$ is topen, and by maximality $\community(p)\subseteq\atopen$.
\qedhere\end{enumerate}
\end{proof}

\begin{rmrk}
\label{rmrk.how.regularity}
We can use Lemma~\ref{lemm.max.cc.intertwined} to characterise regularity in five equivalent ways: see Theorem~\ref{thrm.max.cc.char} and Corollary~\ref{corr.regular.is.regular}.
Other characterisations will follow but will require additional machinery to state (the notion of \emph{closed neighbourhood}; see Definition~\ref{defn.cn}).
See Corollary~\ref{corr.corr.pKp} and Theorem~\ref{thrm.up.down.char}.
\end{rmrk}

\begin{thrm}
\label{thrm.max.cc.char}
Suppose $(\ns P,\opens)$ is a semitopology and $p\in \ns P$.
Then the following are equivalent:
\begin{enumerate*}
\item\label{char.p.regular}
$p$ is regular, or in full: $p\in\community(p)\in\tf{Topen}$.
\item\label{char.Kp.greatest.topen}
$\community(p)$ is the greatest topen neighbourhood of $p$.
\item\label{char.Kp.max.topen}
$\community(p)$ is a maximal topen neighbourhood of $p$.
\item\label{char.max.topen}
$p$ has a maximal topen neighbourhood. 
\item\label{char.some.topen}
$p$ has some topen neighbourhood.
\end{enumerate*}
\end{thrm}
\begin{proof}
We prove a cycle of implications:
\begin{enumerate}
\item
If $\community(p)$ is a topen neighbourhood of $p$ then it is maximal by Lemma~\ref{lemm.intertwined.is.the.greatest}(\ref{item.intertwined.is.the.greatest.1}).
Furthermore this maximal topen neighbourhood of $p$ is necessarily greatest, since if we have two maximal topen neighbourhoods of $p$ then their union is a larger topen neighbourhood of $p$ by Proposition~\ref{prop.cc.unions}(\ref{item.intersecting.pair.of.topens}) (union of intersecting topens is topen).
\item
If $\intertwined{p}$ is the greatest topen neighbourhood of $p$, then certainly it is a maximal topen neighbourhood of $p$.
\item
If $\intertwined{p}$ is a maximal topen neighbourhood of $p$, then certainly $p$ has a maximal topen neighbourhood.
\item
If $p$ has a maximal topen neighbourhood then certainly $p$ has a topen neighbourhood.
\item
Suppose $p$ has a topen neighbourhood $\atopen$.
By Corollary~\ref{corr.max.cc} we may assume without loss of generality that $\atopen$ is a maximal topen.
We use Lemma~\ref{lemm.max.cc.intertwined}(\ref{item.intertwined.is.the.greatest.2}).
\qedhere\end{enumerate}
\end{proof}

Theorem~\ref{thrm.max.cc.char} has numerous corollaries:
\begin{corr}
\label{corr.when.singleton.topen}
Suppose $(\ns P,\opens)$ is a semitopology and $p\in\ns P$ and $\{p\}\in\opens$.
Then $p$ is regular. 
\end{corr}
\begin{proof}
We noted in Example~\ref{xmpl.singleton.transitive}(\ref{item.singleton.transitive}) that a singleton $\{p\}$ is always transitive, so if $\{p\}$ is also open, then it is topen, so that $p$ has a topen neighbourhood and by Theorem~\ref{thrm.max.cc.char}(\ref{char.some.topen}) $p$ is topen.\footnote{%
It does not follow from $p\in\{p\}\in\topens$ that $\community(p)=\{p\}$: consider $\ns P=\{0,1\}$ and $\opens=\{\varnothing,\{0\},\{0,1\}\}$ and $p=0$; then $\{p\}\in\topens$ yet $\community(p)=\{0,1\}$.}
\end{proof}

\begin{corr}
\label{corr.regular.is.regular}
Suppose $(\ns P,\opens)$ is a semitopology and $p\in\ns P$.
Then the following are equivalent:
\begin{enumerate*}
\item
$p$ is regular.
\item
$p$ is weakly regular and $\community(p)=\community(p')$ for every $p'\in\community(p)$.
\end{enumerate*} 
\end{corr}
\begin{proof}
We prove two implications, using Theorem~\ref{thrm.max.cc.char}:
\begin{itemize}
\item
Suppose $p$ is regular.
By Lemma~\ref{lemm.wr.r}(\ref{item.r.implies.wr}) $p$ is weakly regular.
Now consider $p'\in\community(p)$.
By Theorem~\ref{thrm.max.cc.char} $\community(p)$ is topen, so it is a topen neighbourhood of $p'$. 
By Theorem~\ref{thrm.max.cc.char} $\community(p')$ is a greatest topen neighbourhood of $p'$. 
But by Theorem~\ref{thrm.max.cc.char} $\community(p)$ is also a greatest topen neighbourhood of $p$, and $\community(p)\between\community(p')$ since they both contain $p'$.
By Proposition~\ref{prop.cc.unions}(\ref{item.intersecting.pair.of.topens}) and maximality, they are equal.
\item
Suppose $p$ is weakly regular and suppose $\community(p)=\community(p')$ for every $p'\in\community(p)$, and consider $p',p''\in\community(p)$.
Then $p'\intertwinedwith p''$ holds, since $p''\in\community(p')=\community(p)$.
By Proposition~\ref{prop.cc.char} $\community(p)$ is topen, and by weak regularity $p\in\community(p)$, so by Theorem~\ref{thrm.max.cc.char} $p$ is regular as required. 
\qedhere\end{itemize}
\end{proof}

\begin{rmrk}
With regards to Corollary~\ref{corr.regular.is.regular}, it might be useful to look at Example~\ref{xmpl.cc}(\ref{item.cc.two.regular.b}) and Figure~\ref{fig.012} (top-right diagram).
In that example the point $1$ is \emph{not} regular, and its community $\{0,1,2\}$ is not a community for $0$ or $2$.
\end{rmrk}

\begin{corr}
\label{corr.p.p'.regular.community}
Suppose $(\ns P,\opens)$ is a semitopology and $p,p'\in\ns P$.
Then if $p$ is regular and $p'\in\community(p)$ then $p'$ is regular and has the same community.
\end{corr}
\begin{proof}
Suppose $p$ is regular --- so by Definition~\ref{defn.tn}(\ref{item.regular.point}) $p\in\community(p)\in\topens$ --- and suppose $p'\in\community(p)$.
Then by Corollary~\ref{corr.regular.is.regular} $\community(p)=\community(p')$, so $p'\in\community(p')\in\topens$ and by Theorem~\ref{thrm.max.cc.char} $p'$ is regular. 
\end{proof}

\begin{corr}
\label{corr.max.topen.char}
Suppose $(\ns P,\opens)$ is a semitopology. 
Then the following are equivalent for $\atopen\subseteq\ns P$:
\begin{itemize*}
\item
$\atopen$ is a maximal topen.
\item
$\atopen\neq\varnothing$ and $\atopen=\community(p)$ for every $p\in \atopen$.
\end{itemize*}
\end{corr}
\begin{proof}
If $\atopen$ is a maximal topen and $p\in\atopen$ then $\atopen$ is a maximal topen neighbourhood of $p$.
By Theorem~\ref{thrm.max.cc.char}(\ref{char.Kp.greatest.topen}\&\ref{char.some.topen}) $\atopen=\community(p)$.

If $\atopen\neq\varnothing$ and $\atopen=\community(p)$ for every $p\in\atopen$,
then $\community(p)=\community(p')$ for every $p'\in\community(p)$ and by Corollary~\ref{corr.regular.is.regular} $p$ is regular, so that by
Definition~\ref{defn.tn}(\ref{item.regular.point}) $\atopen=\community(p)\in\topens$ as required. 
\end{proof}

\jamiesubsection{Intersection and partition properties of regular spaces}
\label{subsect.topen.partitions}

Proposition~\ref{prop.topen.intersect.subset} is useful for consensus in practice.
Suppose we are a regular point $q$ and we have reached consensus with some topen neighbourhood $O\ni q$.
Suppose further that our topen neighbourhood $O$ intersects with the maximal topen neighbourhood $\community(p)$ of some other regular point $p$.
Then Proposition~\ref{prop.topen.intersect.subset} tells us that we were inside $\community(p)$ all along.
See also Remark~\ref{rmrk.gradecast}.
\begin{prop}
\label{prop.topen.intersect.subset}
Suppose $(\ns P,\opens)$ is a semitopology and $p\in\ns P$ is regular and $O\in\topens$ is topen.
Then 
$$
O\between\community(p)
\quad\text{if and only if}\quad
O\subseteq\community(p).
$$
\end{prop}
\begin{proof} 
The right-to-left implication is immediate from Notation~\ref{nttn.between}(\ref{item.between}), given that 
topens are nonempty by Definition~\ref{defn.transitive}(\ref{transitive.cc}).

For the left-to-right implication, suppose $O\between\community(p)$.
By Theorem~\ref{thrm.max.cc.char} $\community(p)$ is a maximal topen, and by Proposition~\ref{prop.cc.unions}(\ref{item.intersecting.pair.of.topens}) $O\cup\community(p)$ is topen.
Then $O\subseteq\community(p)$ follows by maximality.
\end{proof}

\begin{prop}
\label{prop.community.partition}
Suppose $(\ns P,\opens)$ is a semitopology and suppose $p,p'\in\ns P$ are regular.
Then
$$
\community(p)\between\community(p')
\quad\liff\quad
\community(p)=\community(p')
$$
(See also Corollary~\ref{corr.community.intersects.community}, which considers similar properties for $p$ and $p'$ that are not necessarily regular.)
\end{prop}
\begin{proof}
We prove two implications.
\begin{itemize}
\item
Suppose there exists $p''\in\community(p)\cap\community(p')$.
By Corollary~\ref{corr.p.p'.regular.community} ($p''$ is regular and) $\community(p)=\community(p'')=\community(p')$.
\item
Suppose $\community(p)=\community(p')$.
By assumption $p\in\community(p)$, so $p\in\community(p')$.
Thus $p\in\community(p)\cap\community(p')$.
\qedhere\end{itemize}
\end{proof}

Corollary~\ref{corr.topen.partition.char} is a simple characterisation of regular semitopological spaces (it is also a kind of continuation to Lemma~\ref{lemm.intertwined.space.regular}(\ref{item.intertwined.space.regular.2})):
\begin{corr}
\label{corr.topen.partition.char}
Suppose $(\ns P,\opens)$ is a semitopology.
Then the following are equivalent:
\begin{enumerate*}
\item\label{item.topen.partition.char.1}
$(\ns P,\opens)$ is regular.
\item\label{item.topen.partition.char.2}
$\ns P$ partitions into topen sets: there exists some set of topen sets $\mathcal T$ such that $\atopen\notbetween\atopen'$ for every $\atopen,\atopen'\in\mathcal T$ and $\ns P=\bigcup\mathcal T$.
\item\label{item.topen.partition.char.3}
Every $X\subseteq\ns P$ has a cover of topen sets: there exists some set of topen sets $\mathcal T$ such that $X\subseteq\bigcup\mathcal T$.
\end{enumerate*}
\end{corr}
\begin{proof}
The proof is routine from the machinery that we already have.
We prove equivalence of parts~\ref{item.topen.partition.char.1} and~\ref{item.topen.partition.char.2}:
\begin{enumerate}
\item
Suppose $(\ns P,\opens)$ is regular, meaning by Definition~\ref{defn.tn}(\ref{item.regular.S}\&\ref{item.regular.point}) that $p\in\community(p)\in\topens$ for every $p\in\ns P$.
We set $\mathcal T=\{\community(p) \mid p\in\ns P\}$.
By assumption this covers $\ns P$ in topens, and by Proposition~\ref{prop.community.partition} the cover is a partition. 
\item
Suppose $\mathcal T$ is a topen partition of $\ns P$.
By definition for every point $p$ there exists $T\in\mathcal T$ such that $p\in T$ and so $p$ has a topen neighbourhood.
By Theorem~\ref{thrm.max.cc.char}(\ref{char.some.topen}\&\ref{char.p.regular}) $p$ is regular.
\end{enumerate}
We prove equivalence of parts~\ref{item.topen.partition.char.2} and~\ref{item.topen.partition.char.3}:
\begin{enumerate}
\item
Suppose $\mathcal T$ is a topen partition of $\ns P$, and suppose $X\subseteq\mathcal P$.
Then trivially $X\subseteq\bigcup\mathcal T$.
\item
Suppose every $X\subseteq\ns P$ has a cover of topen sets.
Then $\ns P$ has a cover of topen sets; write it $\mathcal T$.
By Corollary~\ref{corr.max.cc} we may assume without loss of generality that $\mathcal T$ is a partition, and we are done.
\qedhere\end{enumerate} 
\end{proof}

\begin{rmrk}
\label{rmrk.the.moral}
The moral we take from the results and examples above (and those to follow) is that the world we are entering has rather different well-behavedness criteria than those familiar from the study of typical Hausdorff topologies like $\mathbb R$.
Put crudely: 
\begin{enumerate*}
\item
`Bad' spaces are spaces that are not regular.

$\mathbb R$ with its usual topology (which is also a semitopology) is an example of a `bad' semitopology; it is not even quasiregular.
\item
`Good' spaces are spaces that are regular.

The supermajority and all-but-one semitopologies from Example~\ref{xmpl.semitopologies}(\ref{item.supermajority}\&\ref{item.counterexample.X-x}) are typical examples of `good' semitopologies; both are intertwined spaces (Notation~\ref{nttn.intertwined.space}).
\item
Corollary~\ref{corr.topen.partition.char} shows that the `good' spaces are just the (disjoint, possibly infinite) unions of intertwined spaces.
\end{enumerate*}
\end{rmrk}

\jamiesubsection{Examples of communities and (ir)regular points}
\label{subsect.irregular}

By Definition~\ref{defn.tn} a point $p$ is regular when its community is a topen neighbourhood.
Then a point is \emph{not} regular when its community is \emph{not} a topen neighbourhood of $p$. 
We saw one example of this in Lemma~\ref{lemm.two.intertwined}.
In this subsection we take a moment to investigate the possible behaviour in more detail.

\begin{figure}
\vspace{-1em}
\centering
\includegraphics[width=0.35\columnwidth]{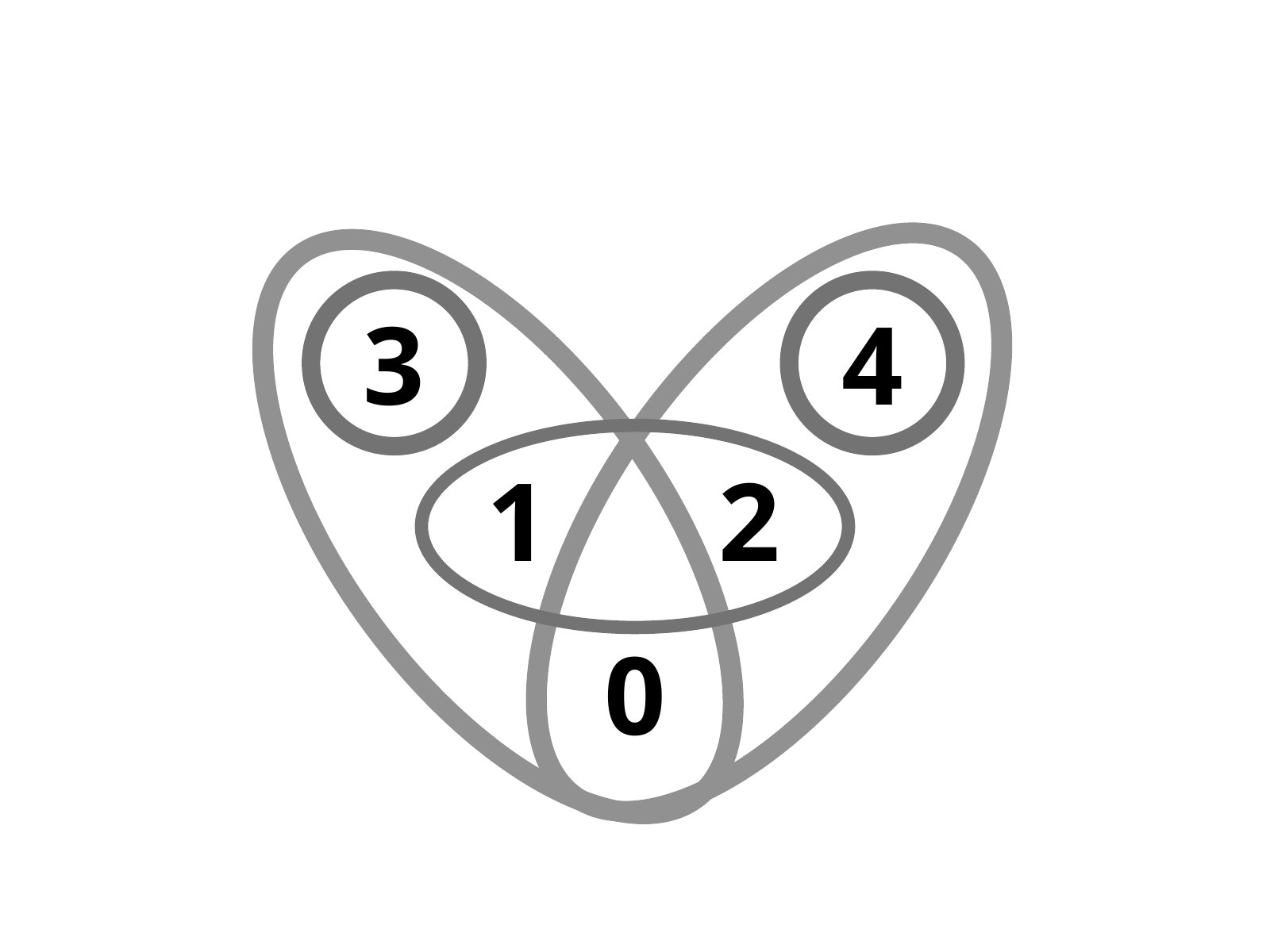}
\includegraphics[width=0.31\columnwidth]{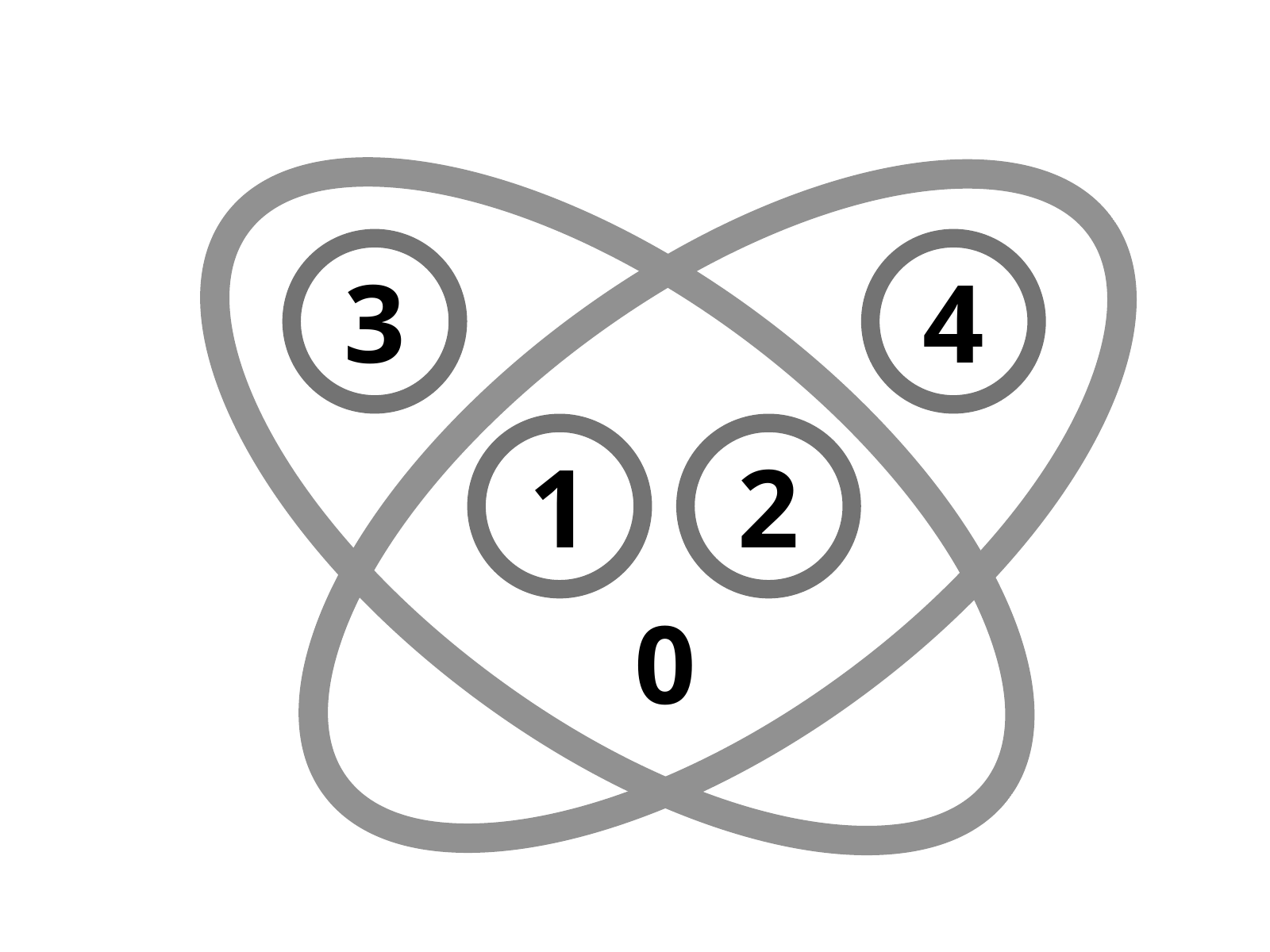}
\vspace{-0em}
\caption{Illustration of Example~\ref{xmpl.p.not.regular}(\ref{item.p.not.regular.01234}\&\ref{item.p.not.regular.01234b})}
\label{fig.irregular}
\end{figure}

\begin{xmpl}
\label{xmpl.p.not.regular}
\leavevmode
\begin{enumerate}
\item\label{item.p.not.regular.R}
We noted in Example~\ref{xmpl.p.not.regular}(\ref{item.wr.6}) and Lemma~\ref{lemm.wr.r.no}(\ref{item.wr.r.not.quasiregular}) that for $\mathbb R$ the real numbers with its usual topology, every $p\in\mathbb R$ is not regular. 
Then
$\intertwined{x}=\{x\}$ and $\community(x)=\varnothing$ for every $x\in\mathbb R$.
\item\label{item.p.not.regular.012}
We continue the semitopology from Example~\ref{xmpl.cc}(\ref{item.cc.two.regular}) (illustrated in Figure~\ref{fig.012}, top-left diagram), as used in Lemma~\ref{lemm.two.intertwined}:
\begin{itemize*}
\item
$\ns P=\{0,1,2\}$.
\item
$\opens$ is generated by $\{0\}$ and $\{2\}$. 
\end{itemize*}
Then:
\begin{itemize*}
\item
$\intertwined{0}=\{0,1\}$ and $\community(0)=\interior(\intertwined{0})=\{0\}$. 
\item
$\intertwined{2}=\{1,2\}$ and $\community(2)=\interior(\intertwined{2})=\{2\}$. 
\item
$\intertwined{1}=\{0,1,2\}$ and $\community(1)=\{0,1,2\}$. 
\end{itemize*}
\item\label{item.point.not.regular.but.community.is.topen}\label{item.p.not.regular.01234}
We take, as illustrated in Figure~\ref{fig.irregular} (left-hand diagram):
\begin{itemize*}
\item
$\ns P=\{0,1,2,3,4\}$.
\item
$\opens$ is generated by $\{1,2\}$, $\{0,1,3\}$, $\{0,2,4\}$, $\{3\}$, and $\{4\}$.
\end{itemize*}
Then:
\begin{itemize*}
\item
$\intertwined{x}=\{0,1,2\}$ and $\community(x)=\interior(\intertwined{x})=\{1,2\}$ for $x\in\{0,1,2\}$.
\item
$\intertwined{x}=\{x\}=\community(x)$ for $x\in\{3,4\}$.
\end{itemize*}
(We return to this example in Example~\ref{xmpl.p.not.regular.2}(\ref{item.p.not.regular.01234.2}), and we will also use it in the proof of Lemma~\ref{lemm.kernel.non-implications}.)
\item\label{item.p.not.regular.01234b}
We take, as illustrated in Figure~\ref{fig.irregular} (right-hand diagram):
\begin{itemize*}
\item
$\ns P=\{0,1,2,3,4\}$.
\item
$\opens$ is generated by $\{1\}$, $\{2\}$, $\{3\}$, $\{4\}$, $\{0, 1, 2, 3\}$, and $\{0, 1, 2, 4\}$. 
\end{itemize*}
Then:
\begin{itemize*}
\item
$\intertwined{0}=\{0,1,2\}$ and $\community(0)=\{1,2\}$.
\item
$\community(0)$ is not transitive and consists of two distinct topens $\{1\}$ and $\{2\}$.
\item
$0\notin\community(0)$. 
\end{itemize*}
See Remark~\ref{rmrk.indeed.two.closed.neighbourhoods} for further discussion of this example.
\item
The reader can also look ahead to Example~\ref{xmpl.two.topen.examples}.
In Example~\ref{xmpl.two.topen.examples}(\ref{item.two.topen.examples.1}), every point $p$ is regular and $\community(p)=\mathbb Q^2$.
In Example~\ref{xmpl.two.topen.examples}(\ref{item.two.topen.examples.2}), no point $p$ is regular and $\community(p)=\varnothing\subseteq\mathbb Q^2$.
\end{enumerate}
\end{xmpl}

\begin{lemm}
\label{lemm.p.not.regular}
Suppose $(\ns P,\opens)$ is a semitopology and $p\in\ns P$.
Then precisely one of the following possibilities must hold, and each one is possible: 
\begin{enumerate*}
\item
$p$ is regular: $p\in\community(p)$ and $\community(p)$ is topen (nonempty, open, and transitive). 
\item
$\community(p)$ is topen, but $p\notin\community(p)$. 
\item
$\community(p)=\varnothing$.
\item
$\community(p)$ is open but not transitive.
(Both $p\in\community(p)$ and $p\notin\community(p)$ are possible.)
\end{enumerate*}
\end{lemm}
\begin{proof} 
\leavevmode\begin{enumerate}
\item
To see that $p$ can be regular, consider $\ns P=\{0\}$ with the discrete topology.
Then $p\in\community(p)=\{0\}$.
\item
To see that it is possible for $\community(p)$ to be topen but $p$ is not in it, consider Example~\ref{xmpl.p.not.regular}(\ref{item.p.not.regular.01234}).
There, $\ns P=\{0,1,2,3,4\}$ and $\intertwined{0}=\{0,1,2\}$ and $\community(0)=\{1,2\}$.
Then $\community(0)$ is topen, but $0\notin\community(0)$.

(Another, slightly more compact but more distant, example is $p=\ast$ in the lower-right semitopology in Figure~\ref{fig.012}.)
\item
To see that $\community(p)=\varnothing$ is possible, consider Example~\ref{xmpl.p.not.regular}(\ref{item.p.not.regular.R}) (the real numbers $\mathbb R$ with its usual topology).
Then by Remark~\ref{rmrk.not.hausdorff} $\intertwined{r}=\{r\}$ and so $\community(x)=\interior(\{r\})=\varnothing$.
(See also Example~\ref{xmpl.two.topen.examples}(\ref{item.two.topen.examples.2}) for a more elaborate example.) 
\item
To see that it is possible for $\community(p)$ to be an open neighbourhood of $p$ but not transitive, see Example~\ref{xmpl.p.not.regular}(\ref{item.p.not.regular.012}).
There, $\ns P=\{0,1,2\}$ and $1\in \intertwined{1}=\{0,1,2\}=\community(1)$, but $\{0,1,2\}$ is not transitive (it contains two disjoint topens: $\{0\}$ and $\{2\}$).

To see that it is possible for $\community(p)$ to be open and nonempty yet not contain $p$ and not be transitive, see Example~\ref{xmpl.p.not.regular}(\ref{item.p.not.regular.01234b}) for $p=0$, and see also Remark~\ref{rmrk.indeed.two.closed.neighbourhoods} for a discussion of the connection with minimal closed neighbourhoods.
\end{enumerate}
The possibilities above are clearly mutually exclusive and exhaustive.
\end{proof}

\jamiesection{Closed sets}
\label{sect.closed.sets}

\jamiesubsection{Closed sets}
\label{subsect.closed.sets.basics}

\begin{rmrk}
\label{rmrk.computing.closures}
In Subsection~\ref{subsect.closed.sets.basics} we check that some familiar properties of closures carry over from topologies to semitopologies.
There are no technical surprises, but this is in itself a mathematical result that needs to be checked. 
From Subsection~\ref{subsect.trans.clos} and the following Subsections we will study the close relation between closures and sets of intertwined points. 

First, we spare a few words on why closures are particularly interesting in semitopologies:
\begin{enumerate*}
\item
A participant may wish to compute a quorum that it can be confident of remaining in agreement with, where algorithms succeed.
The notion of maximal topen from Definition~\ref{defn.transitive}(\ref{transitive.max.cc}) provides this, as discussed in Remark~\ref{rmrk.transitive.correlated} --- but computing maximal topens is hard since the definition involves quantifications over all open sets and there may be many of them.
\item
However, computing \emph{closures} is quite tractable in the right circumstances (see Section~\ref{sect.witness} and Remark~\ref{rmrk.computing.closed.sets}), so a characterisation of maximal topens using closed sets and closures is of practical interest.
\item
Closures are significant for other reasons too: see the discussions in Remarks~\ref{rmrk.fundamental.consensus} and~\ref{rmrk.why.top.closure}, and see the later material in Section~\ref{sect.dense} that considers \emph{dense sets}.
\end{enumerate*}
Thus, and just as is the case in topology, closures are an interesting object of study. 
\end{rmrk}

\begin{defn}
\label{defn.closure}
Suppose $(\ns P,\opens)$ is a semitopology and suppose $p\in\ns P$ and $P\subseteq\ns P$.
Then:
\begin{enumerate*}
\item\label{item.closure}
Define $\closure{P}\subseteq\ns P$ the \deffont{closure of $P$} to be the set of points $p$ such that every open neighbourhood of $p$ intersects $P$.
In symbols using Notation~\ref{nttn.between}: 
$$
\closure{P} = \{ p'\in\ns P \mid \Forall{O{\in}\opens} p'\in O \limp P\between O\} .
$$
\item\label{item.closure.p}
As is standard, we may write $\closure{p}$ for $\closure{\{p\}}$.
Unpacking definitions for reference:
$$
\closure{p} = \{ p'\in\ns P \mid \Forall{O{\in}\opens} p'\in O \limp p\in O\} .
$$
\end{enumerate*}
\end{defn}

\begin{lemm}
\label{lemm.closure.monotone}
Suppose $(\ns P,\opens)$ is a semitopology and suppose $P,P'\subseteq\ns P$.
Then taking the closure of a set is: 
\begin{enumerate*}
\item\label{closure.monotone}
\emph{Monotone:}\quad If $P\subseteq P'$ then $\closure{P}\subseteq\closure{P'}$.
\item\label{closure.increasing}
\emph{Increasing:}\quad $P\subseteq\closure{P}$.
\item\label{closure.idempotent}
\emph{Idempotent:}\quad $\closure{P}=\closure{\closure{P}}$.
\end{enumerate*}
\end{lemm}
\begin{proof}
By routine calculations from Definition~\ref{defn.closure}.
\end{proof}

\begin{lemm}
\label{lemm.closure.open.char}
Suppose $(\ns P,\opens)$ is a semitopology and $P\subseteq\ns P$ and $O\in\opens$.
Then 
$$
P\between O
\quad\text{if and only if}\quad 
\closure{P}\between O.
$$
\end{lemm}
\begin{proof}
Suppose $P\between O$.
Then $\closure{P}\between O$ using Lemma~\ref{lemm.closure.monotone}(\ref{closure.increasing}).

Suppose $\closure{P}\between O$.
Pick $p\in \closure{P}\cap O$.
By construction of $\closure{P}$ in Definition~\ref{defn.closure} $p\in O\limp P\between O$.
It follows that $P\between O$ as required.
\end{proof}

\begin{defn}
\label{defn.closed}
Suppose $(\ns P,\opens)$ is a semitopology and suppose $C\subseteq\ns P$.
\begin{enumerate*}
\item\label{item.closed.set}
Call $C$ a \deffont{closed set} when $C=\closure{C}$.
\item
Call $C$ a \deffont{clopen set} when $C$ is closed and open.
\item
Write $\closed$ for the set of \deffont[closed sets $\closed$]{closed sets} (as we wrote $\opens$ for the open sets; the ambient semitopology will always be clear or understood).
\end{enumerate*}
\end{defn}

\begin{lemm}
\label{lemm.closure.closed}
Suppose $(\ns P,\opens)$ is a semitopology and suppose $P\subseteq\ns P$.
Then $\closure{P}$ is closed and contains $P$.
In symbols:
$$
P\subseteq \closure{P}\in\closed .
$$ 
\end{lemm}
\begin{proof}
From Definition~\ref{defn.closed}(\ref{item.closed.set}) and Lemma~\ref{lemm.closure.monotone}(\ref{closure.increasing} \& \ref{closure.idempotent}).
\end{proof}

\begin{xmpl}\leavevmode
\begin{enumerate}
\item
Take $\ns P=\{0,1\}$ and $\opens=\{\varnothing, \{0\}, \{0,1\}\}$.
Then the reader can verify that:
\begin{itemize*}
\item
$\{0\}$ is open.
\item
The closure of $\{1\}$ is $\{1\}$ and $\{1\}$ is closed.
\item
The closure of $\{0\}$ is $\{0,1\}$.
\item
$\varnothing$ and $\{0,1\}$ are the only clopen sets.
\end{itemize*}
\item
Now take $\ns P=\{0,1\}$ and $\opens=\{\varnothing, \{0\}, \{1\}, \{0,1\}\}$.\footnote{Following Definition~\ref{defn.value.assignment} and Example~\ref{xmpl.semitopologies}(\ref{item.boolean.discrete}), this is just $\{0,1\}$ with the \emph{discrete semitopology}.}
Then the reader can verify that:
\begin{itemize*}
\item
Every set is clopen.
\item
The closure of every set is itself.
\end{itemize*}
\end{enumerate}
\end{xmpl}

\begin{rmrk}
There are two standard definitions for when a set is closed: when it is equal to its closure (as per Definition~\ref{defn.closed}(\ref{item.closed.set})), and when it is the complement of an open set.
In topology these are equivalent.
We do need to check that the same holds in semitopology, but as it turns out the proof is routine:
\end{rmrk}

\begin{lemm}
\label{lemm.closed.complement.open}
Suppose $(\ns P,\opens)$ is a semitopology.
Then:
\begin{enumerate*}
\item\label{item.closed.complement.open.1}
Suppose $C\in\closed$ is closed (by Definition~\ref{defn.closed}: $C=\closure{C}$).
Then $\ns P\setminus C$ is open.
\item\label{item.closed.complement.open.2}
Suppose $O\in\opens$ is open.
Then $\ns P\setminus O$ is closed (by Definition~\ref{defn.closed}: $\closure{\ns P\setminus O}=\ns P\setminus O$).
\end{enumerate*}
\end{lemm}
\begin{proof}
\leavevmode
\begin{enumerate}
\item
Suppose $p\in \ns P\setminus C$.
Since $C=\closure{C}$, we have $p\in\ns P\setminus\closure{C}$.
Unpacking Definition~\ref{defn.closure}, this means precisely that there exists $O_p\in\opens$ with $p\in O_p \notbetween C$.
We use Lemma~\ref{lemm.open.is.open}. 
\item
Suppose $O\in\opens$.
Combining Lemma~\ref{lemm.open.is.open} with Definition~\ref{defn.closure} 
it follows that $O\notbetween \closure{\ns P\setminus O}$ so that $\closure{\ns P\setminus O}\subseteq\ns P\setminus O$.
Furthermore, by Lemma~\ref{lemm.closure.monotone}(\ref{closure.increasing}) $\ns P\setminus O\subseteq\closure{\ns P\setminus O}$.
\qedhere\end{enumerate}
\end{proof}

\begin{corr}
\label{corr.closed.complement.union}
If $C\in\closed$ then $\ns P\setminus C=\bigcup_{O\in\opens} O\notbetween C$.
\end{corr}
\begin{proof}
By Lemma~\ref{lemm.closed.complement.open}(\ref{item.closed.complement.open.1}) $\ns P\setminus C\subseteq\bigcup_{O\in\opens} O\notbetween C$.
Conversely, if $O\notbetween C$ then $O\subseteq\ns P\setminus C$ by Definition~\ref{defn.closure}(\ref{item.closure}). 
\end{proof}

\begin{corr}
\label{corr.closure.closure}
Suppose $(\ns P,\opens)$ is a semitopology and $P\subseteq\ns P$ and $\mathcal C\subseteq\powerset(\ns P)$.
Then:
\begin{enumerate*}
\item
$\varnothing$ and $\ns P$ are closed.
\item\label{closure.closure.cap}
If every $C\in\mathcal C$ is closed, then $\bigcap\mathcal C$ is closed.
Or succinctly in symbols:
$$
\mathcal C\subseteq\closed \limp \bigcap\mathcal C\in\closed .
$$
\item\label{item.closure.as.intersection}
$\closure{P}$ is equal to the intersection of all the closed sets that contain it.
In symbols:
$$
\closure{P}=\bigcap\{C\in\closed \mid P\subseteq C\}. 
$$
\end{enumerate*}
\end{corr}
\begin{proof}
\leavevmode
\begin{enumerate}
\item
Immediate from Lemma~\ref{lemm.closed.complement.open}(\ref{item.closed.complement.open.2}).
\item
From Lemma~\ref{lemm.closed.complement.open} and Definition~\ref{defn.semitopology}(\ref{semitopology.empty.and.universe}\&\ref{semitopology.unions}).
\item
By Lemma~\ref{lemm.closure.closed} $\bigcap\{C\in\closed \mid P\subseteq C\}\subseteq\closure{P}$.
By construction $P\subseteq\bigcap\{C\in\closed \mid P\subseteq C\}$, and using Lemma~\ref{lemm.closure.monotone}(\ref{closure.monotone}) and part~\ref{item.closure.as.intersection} of this result we have
$$
\closure{P} 
\stackrel{L\ref{lemm.closure.monotone}(\ref{closure.monotone})}\subseteq 
\closure{\bigcap\{C\in\closed \mid P\subseteq C\}} 
\stackrel{pt.2}= 
\bigcap\{C\in\closed \mid P\subseteq C\} .
$$ 
\qedhere\end{enumerate}
\end{proof}

The usual characterisation of continuity in terms of inverse images of closed sets being closed, remains valid:
\begin{corr}
\label{corr.alternative.cont.closed}
Suppose $(\ns P,\opens)$ and $(\ns P',\opens')$ are semitopological spaces (Definition~\ref{defn.semitopology}) and suppose $\avaluation:\ns P\to\ns P'$ is a function.
Then the following are equivalent:
\begin{enumerate*}
\item
$\avaluation$ is continuous, meaning by Definition~\ref{defn.continuity}(\ref{item.continuous.function}) that $\avaluation^\mone(O')\in\opens$ for every $O'\in\opens'$.
\item
$\avaluation^\mone(C')\in\closed$ for every $C'\in\closed'$.
\end{enumerate*}
\end{corr}
\begin{proof}
By routine calculations as for topologies, using Lemma~\ref{lemm.closed.complement.open} and the fact that the inverse image of a complement is the complement of the inverse image; see~\cite[Theorem~7.2, page~44]{willard:gent} or~\cite[Proposition~1.4.1(iv), page~28]{engelking:gent}.
\end{proof}

\jamiesubsection{Duality between closure and interior}

The usual dualities between closures and interiors remain valid in semitopologies.
There are no surprises but this still needs to be checked, so we spell out the details:
\begin{lemm}
\label{lemm.closure.interior}
Suppose $(\ns P,\opens)$ is a semitopology and $O\in\opens$ and $C\in\closed$.
Then:
\begin{enumerate*}
\item\label{item.closure.interior.open}
$O\subseteq\interior(\closure{O})$.  The inclusion may be strict.
\item\label{item.closure.interior.closed}
$\closure{\interior(C)}\subseteq C$.  The inclusion may be strict.
\item\label{item.closure.interior.complement.closure}
$\interior(\ns P\setminus O)=\ns P\setminus\closure{O}$.
\item\label{item.closure.interior.complement.interior}
$\closure{\ns P\setminus C}=\ns P\setminus\interior(C)$. 
\end{enumerate*}
\end{lemm}
\begin{proof}
The reasoning is just as for topologies, but we spell out the details:
\begin{enumerate}
\item
By Lemma~\ref{lemm.closure.monotone}(\ref{closure.increasing}) $O\subseteq\closure{O}$.
By Corollary~\ref{corr.interior.monotone} $\interior(O)\subseteq\interior(\closure{O})$.
By Lemma~\ref{lemm.interior.open} $O=\interior(O)$, so we are done.

For an example of the strict inclusion, consider $\mathbb R$ with the usual topology (which is also a semitopology) and take $O=(0,1)\cup(1,2)$.
Then $O\subsetneq\interior(\closure{O})=(0,2)$.
\item
By Lemma~\ref{lemm.interior.open} $\interior(C)\subseteq C$.
By Lemma~\ref{lemm.closure.monotone}(\ref{closure.monotone}) $\closure{\interior(C)}\subseteq\closure{C}$.
By Definition~\ref{defn.closed}(\ref{item.closed.set}) (since we assumed $C\in\closed$) $\closure{C}=C$, so we are done.

For an example of the strict inclusion, consider $\mathbb R$ with the usual topology and take $C=\{0\}$.
Then $\closure{\interior(C)}=\varnothing\subsetneq C$.
\item
Consider some $p'\in\ns P$.
By Definition~\ref{defn.interior} $p'\in \interior(\ns P\setminus O)$ when there exists some $O'\in\opens$ such that $p'\in O'\notbetween O$.
By definition in Definition~\ref{defn.closure}(\ref{item.closure}) this happens precisely when $p'\notin\closure{O}$. 
\item
By Definition~\ref{defn.closure}(\ref{item.closure}), $p'\notin \closure{\ns P\setminus C}$ precisely when there exists some $O'\in\opens$ such that $p'\in O'\notbetween \ns P\setminus C$.
By facts of sets this means precisely that $p'\in O'\subseteq C$.
By Definition~\ref{defn.interior} this means precisely that $p'\in\interior(C)$.
\qedhere\end{enumerate}
\end{proof}

\begin{corr}
\label{corr.ic.ci}
Suppose $(\ns P,\opens)$ is a semitopology and 
$O\in\opens$ and $C\in\closed$.
Then:
\begin{enumerate*}
\item
$\closure{O} = \closure{\interior(\closure{O})}$. 
\item
$\interior(C)=\interior(\closure{\interior(C)})$.
\end{enumerate*}
\end{corr}
\begin{proof}
We use Lemma~\ref{lemm.closure.interior}(\ref{item.closure.interior.open}\&\ref{item.closure.interior.complement.closure}) along with Lemma~\ref{lemm.closure.monotone}(\ref{closure.monotone}) and Corollary~\ref{corr.interior.monotone}: 
$$
\begin{array}{r@{\ }c@{\ }c@{\ }c@{\ }ll}
\closure{O}
&\stackrel{L\ref{lemm.closure.interior}(\ref{item.closure.interior.open})\&L\ref{lemm.closure.monotone}(\ref{closure.monotone})}\subseteq&
\closure{\interior(\closure{O})}
&\stackrel{L\ref{lemm.closure.interior}(\ref{item.closure.interior.closed})}\subseteq&
\interior(\closure{O})
\\
\interior(C)
&\stackrel{L\ref{lemm.closure.interior}(\ref{item.closure.interior.open})}\subseteq&
\interior(\closure{\interior(C)})
&\stackrel{L\ref{lemm.closure.interior}(\ref{item.closure.interior.closed})\&C\ref{corr.interior.monotone}}\subseteq&
\interior(C)
\end{array}
$$
\end{proof}

\jamiesubsection{Transitivity and closure}
\label{subsect.trans.clos}

We explore how the topological closure operation interacts with taking transitive sets.
\begin{lemm}
\label{lemm.open.consensus}
Suppose $(\ns P,\opens)$ is a semitopology and $T\subseteq\ns P$ is transitive and $O\in\opens$.
Then 
$$
\atopen\between O
\quad\text{implies}\quad
\closure{T}\subseteq\closure{O}.
$$
\end{lemm}
\begin{proof}
Unpacking Definition~\ref{defn.closure}
we have:
$$
\begin{array}{r@{\ }l}
p'\in\closure{T}\liff&\Forall{O'{\in}\opens}p'\in O'\limp O'\between \atopen 
\qquad\text{and}
\\
p'\in\closure{O}\liff&\Forall{O'{\in}\opens}p'\in O'\limp O'\between O
.
\end{array}
$$
It would suffice to prove $O'\between \atopen\limp O'\between O$ for any $O'\in\opens$.

So suppose $O'\between \atopen$.
By assumption $\atopen\between O$ and by transitivity of $\atopen$ (Definition~\ref{defn.transitive}) $O'\between O$.
\end{proof}

\begin{prop}
\label{prop.open.consensus}
\label{prop.open.strong-consensus}
Suppose $(\ns P,\opens)$ is a semitopology and $\atopen\in\topens$ and $O\in\opens$.
Then the following are equivalent:
$$
\atopen\between O
\quad\text{if and only if}\quad
\atopen\subseteq\closure{\atopen}\subseteq \closure{O}
.
$$
\end{prop}
\begin{proof}
We prove two implications:
\begin{itemize}
\item
Suppose $\atopen\between O$.
By Lemma~\ref{lemm.open.consensus} $\closure{\atopen}\subseteq\closure{O}$.
By Lemma~\ref{lemm.closure.monotone}(\ref{closure.increasing}) (as standard) $\atopen\subseteq\closure{\atopen}$. 
\item
Suppose $\atopen\subseteq\closure{\atopen}\subseteq\closure{O}$.
Then $\atopen\between\closure{O}$ and by Lemma~\ref{lemm.closure.open.char} (since $\atopen$ is nonempty (and transitive) and open) also $\atopen\between O$.
\qedhere\end{itemize}
\end{proof}

\begin{rmrk}
\label{rmrk.gradecast}
In retrospect we can see the imprint of topens (Definition~\ref{defn.transitive}) in previous work, if we look at things in a certain way.
Many consensus algorithms have the property that once consensus is established in a quorum $O$, it propagates to $\closure{O}$.

This is apparent (for example) in the Grade-Cast algorithm~\cite{feldman_optimal_1988}, in which participants assign a confidence grade of 0, 1 or 2 to their output and must ensure that if any participant outputs $v$ with grade 2 then all must output $v$ with grade at least 1.
In this algorithm, if a participant finds that all its quorums intersect some set $S$ that unanimously supports value $v$, then the participant assigns grade at least 1 to $v$.
From our point of view here, this is just taking a closure in the style we discussed in Remark~\ref{rmrk.computing.closures}.
If $T$ unanimously supports $v$ and participants communicate enough, then eventually every member of $\closure{T}$ assigns grade at least 1 to $v$.
Thus, Proposition~\ref{prop.open.strong-consensus} suggests that, to convince a topen to agree on a value, we can first convince an open neighbourhood that intersects the topen, and then use Grade-Cast to convince the closure of that open set and thus in particular the topen which we know must be contained in that closure. 
\end{rmrk}

\begin{rmrk}
Later on we will revisit these ideas and fit them into a nice general framework having to do with dense subsets.
See Lemma~\ref{lemm.strongly.dense.for.closure} and Proposition~\ref{prop.most.general}. 
\end{rmrk}

We conclude with an easy observation which will be useful later.
Recall from Notation~\ref{nttn.intertwined.space} the notion of an intertwined space being one such that all nonempty open sets intersect.
Then we have:
\begin{lemm}
\label{lemm.intertwined.iff.closure}
Suppose $(\ns P,\opens)$ is a semitopology and suppose $\atopen\in\topens$.
Then the following are equivalent:
\begin{enumerate*}
\item
$\ns P$ is intertwined.
\item
$\closure{\atopen}=\ns P$.
\end{enumerate*}
\end{lemm}
\begin{proof}
Suppose $\closure{\atopen}=\ns P$ and consider any $O,O'\in\opens$.
Unpacking Definition~\ref{defn.closure}(\ref{item.closure}) it follows that $O\between\atopen\between O'$.
By transitivity of $\atopen$ (Definition~\ref{defn.transitive}(\ref{transitive.transitive})) $O\between O'$ as required.

Suppose $(\ns P,\opens)$ is intertwined.
By Lemma~\ref{lemm.intertwined.space} every nonempty open set is topen, thus $\ns P$ is topen, and $\ns P=\closure{\atopen}$ follows by Lemma~\ref{lemm.open.consensus}. 
\end{proof}

\jamiesubsection{Closed neighbourhoods and intertwined points}
\label{subsect.closed.neighbourhoods}

\jamiesubsubsection{Definition and basic properties}

\begin{defn}
\label{defn.cn}
Suppose $(\ns P,\opens)$ is a semitopology.
We generalise Definition~\ref{defn.open.neighbourhood} as follows:
\begin{enumerate*}
\item\label{item.neighbourhood.of.p}
Call $P\subseteq\ns P$ a \deffont{neighbourhood} when it contains an open set (i.e. when $\interior(P)\neq\varnothing$), and call $P$ a \deffont{neighbourhood of $p$} when $p\in\ns P$ and $P$ contains an open neighbourhood of $p$ (i.e. when $p\in\interior(P)$).
In particular:
\item\label{item.closed.neighbourhood.of.p}
$C\subseteq\ns P$ is a \deffont{closed neighbourhood of $p\in\ns P$} when $C$ is closed and $p\in\interior(C)$.
\item\label{item.closed.neighbourhood}
$C\subseteq\ns P$ is a \deffont{closed neighbourhood} when $C$ is closed and $\interior(C)\neq\varnothing$.
\end{enumerate*} 
\end{defn}

\begin{rmrk}
\leavevmode
\begin{enumerate}
\item
If $C$ is a closed neighbourhood of $p$ in the sense of Definition~\ref{defn.cn}(\ref{item.closed.neighbourhood.of.p}) then $C$ is a closed neighbourhood in the sense of Definition~\ref{defn.cn}(\ref{item.closed.neighbourhood}), just because if $p\in\interior(C)$ then $\interior(C)\neq\varnothing$. 
\item
$p\in C$ is not enough for $C$ to be a closed neighbourhood of $p$;
we require the stronger condition $p\in\interior(C)$.

For instance take $\ns P=\{0,1\}$ and $\opens=\{\varnothing,\{1\},\ns P\}$ (the Sierpi\'nski space; see Figure~\ref{fig.sierpinski}), and consider $p=0$ and $C=\{0\}$.
Then $p\in C$ but $p\not\oldin\interior(C)=\varnothing$, so that $C$ is not a closed neighbourhood of $p$. 
\end{enumerate}
\end{rmrk}

Recall from Definition~\ref{defn.intertwined.points} the notions of $p\intertwinedwith p'$ and $\intertwined{p}$.
Proposition~\ref{prop.intertwined.as.closure} packages up our material for convenient use in later results. 
\begin{prop}
\label{prop.intertwined.as.closure}
Suppose $(\ns P,\opens)$ is a semitopology and $p,p'\in\ns P$.
Then:
\begin{enumerate*}
\item\label{item.intertwined.as.closure.1}
We can characterise when $p'$ is intertwined with $p$ as follows: 
$$
p\intertwinedwith p' 
\quad\text{if and only if}\quad
\Forall{O{\in}\opens} p\in O\limp p'\in\closure{O} .
$$
\item\label{item.intertwined.as.intersection.of.closures}
As a corollary,
$$
\intertwined{p} = \bigcap\{\closure{O} \mid p\in O\in\opens\}.
$$
\item\label{intertwined.as.closure.closed}
Equivalently:
$$
\begin{array}{r@{\ }l@{\qquad}l}
\intertwined{p}
=& \bigcap\{C\in\closed \mid p\in \interior(C) \}
\\
=&
\bigcap\{C\in\tf{Closed} \mid C\text{ a closed neighbourhood of }p\}
&\text{Definition~\ref{defn.cn}}.
\end{array}
$$
Thus in particular, if $C$ is a closed neighbourhood of $p$ then $\intertwined{p}\subseteq C$.
\item\label{intertwined.p.closed}
$\intertwined{p}$ is closed and $\ns P\setminus\intertwined{p}$ is open.
\end{enumerate*}
\end{prop}
\begin{proof}
\leavevmode
\begin{enumerate}
\item
We just rearrange Definition~\ref{defn.intertwined.points}.
So
$$
\Forall{O,O'\in\opens}((p\in O\land p'\in O') \limp O\between O')
$$
rearranges to
$$
\Forall{O\in\opens}(p\in O\limp \Forall{O'\in\opens} (p'\in O' \limp O\between O')) . 
$$
We now observe from Definition~\ref{defn.closure} that this is precisely
$$
\Forall{O\in\opens}(p\in O\limp p'\in\closure{O}).
$$
\item
We just rephrase part~\ref{item.intertwined.as.closure.1} of this result.
\item
Using part~\ref{item.intertwined.as.intersection.of.closures} of this result it would suffice to prove
$$
\bigcap\{\closure{O}\mid p\in O\in\opens\} = \bigcap\{C\in\closed \mid p\in \interior(C) \} .
$$
We will do this by proving that for each $O$-component on the left there is a $C$ on the right with $C\subseteq\closure{O}$; and for each $C$-component on the right there is an $O$ on the left with $\closure{O}\subseteq C$:
\begin{itemize}
\item
Consider some $O\in\opens$ with $p\in O$.

We set $C=\closure{O}$, so that trivially $C\subseteq\closure{O}$.
By Lemma~\ref{lemm.closure.interior}(\ref{item.closure.interior.open}) $O\subseteq\interior(\closure{O})$, so $p\in\interior(C)$.
\item
Consider some $C\in\closed$ such that $p\in\interior(C)$.

We set $O=\interior(C)$.
Then $p\in O$, and by Lemma~\ref{lemm.closure.interior}(\ref{item.closure.interior.closed}) $\closure{O}\subseteq C$.
\end{itemize}
\item
Part~\ref{intertwined.as.closure.closed} of this result exhibits $\intertwined{p}$ as an intersection of closed sets, and by Corollary~\ref{corr.closure.closure}(\ref{closure.closure.cap}) this is closed.
By Lemma~\ref{lemm.closed.complement.open}(\ref{item.closed.complement.open.1}) its complement $\ns P\setminus\intertwined{p}$ is open.
\qedhere\end{enumerate}
\end{proof}

\begin{defn}
\label{defn.nbhd.system}
\label{defn.nbhd}
Suppose $(\ns P,\opens)$ is a semitopology and $p\in\ns P$.
\begin{enumerate*}
\item
Write $\nbhd(p)=\{O\in\opens\mid p\in\opens\}$ and call this the \deffont[open neighbourhood system $\nbhd(p)$]{open neighbourhood system} of $p\in\ns P$. 
\item
Write $\nbhd^c(p)=\{C\in\closed\mid p\in\closed\}$ and call this the \deffont[closed neighbourhood system $\nbhd^c(p)$]{closed neighbourhood system}\index{$\nbhd^c(p)$ (closed neighbourhood system of a point)} of $p\in\ns P$.
\end{enumerate*}
\end{defn}

\begin{rmrk}
\label{rmrk.nbhd.concise}
As standard, we can use Definition~\ref{defn.nbhd} to rewrite the definition of $\avaluation$ being continuous at $p$ (Definition~\ref{defn.continuity}(\ref{item.continuous.function.at.p})) as
$$
\Forall{O'{\in}\nbhd(f(p))}\Exists{O{\in}\nbhd(p)} O\subseteq f^\mone(O') .
$$
\end{rmrk}

\begin{figure}
\vspace{-1em}
\centering
\includegraphics[align=c,width=0.3\columnwidth,trim={50 0 50 0},clip]{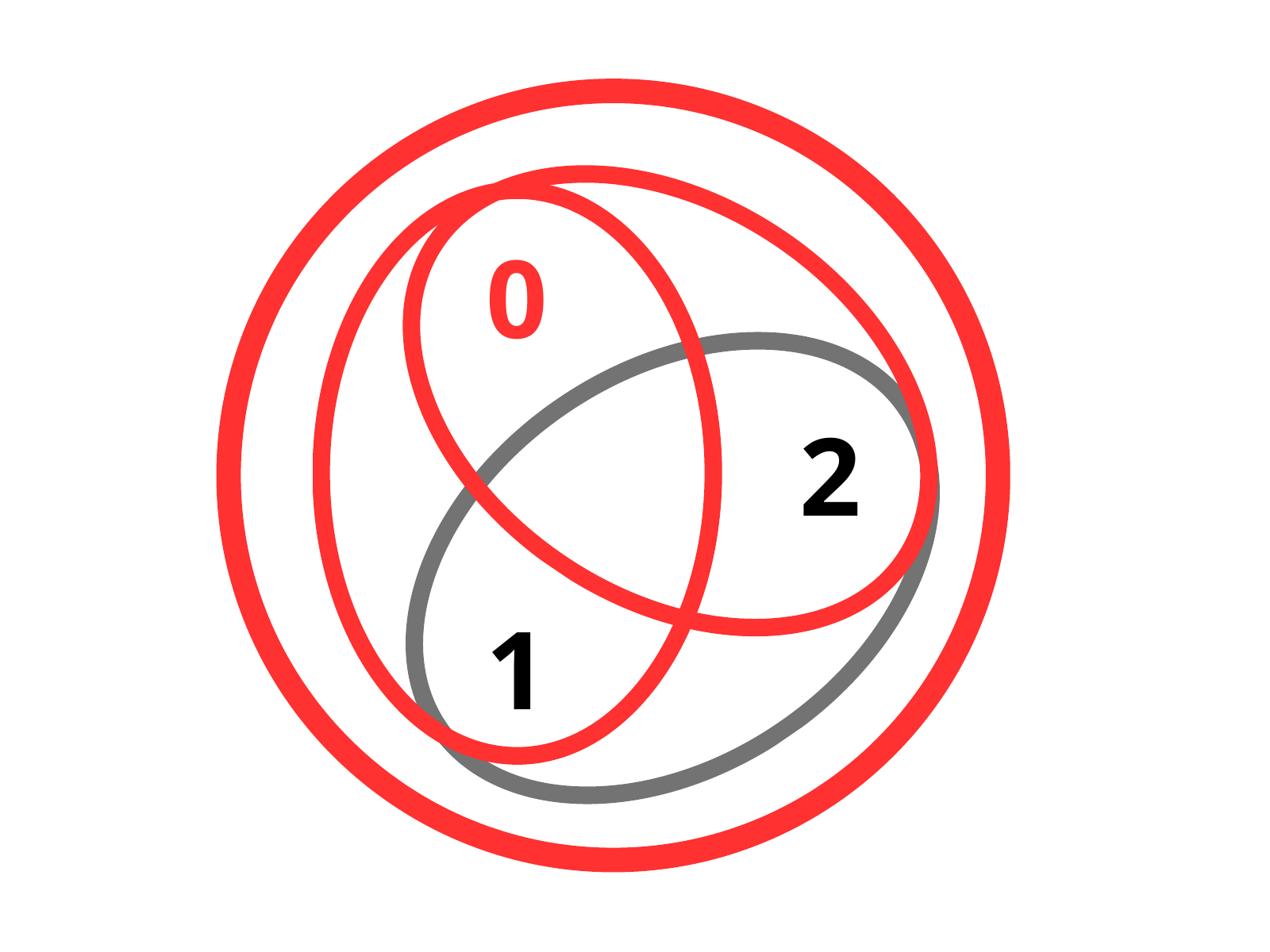}
\quad
\includegraphics[align=c,width=0.32\columnwidth,trim={50 0 50 0},clip]{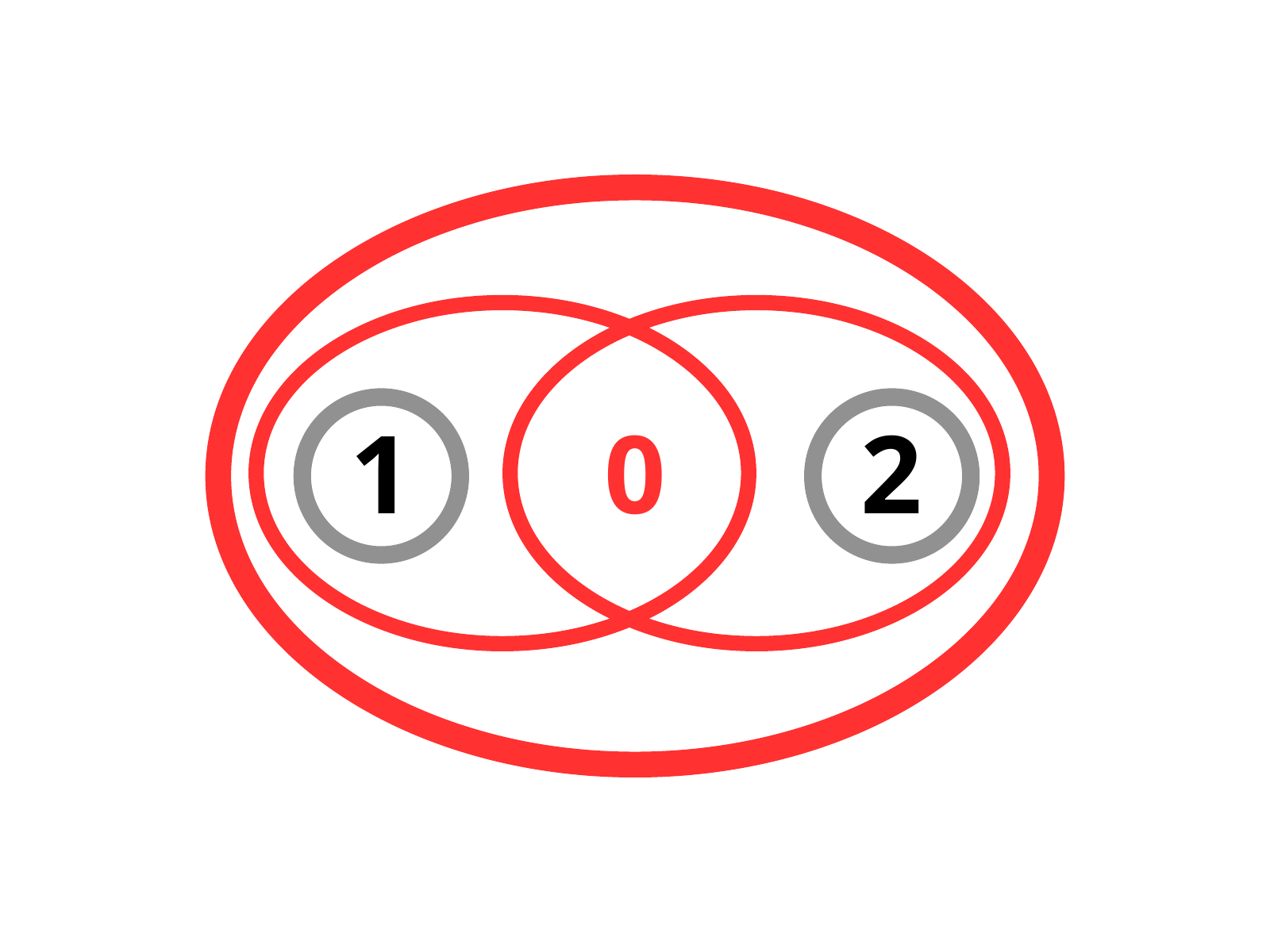}
\quad
\includegraphics[align=c,width=0.28\columnwidth,trim={50 0 50 0},clip]{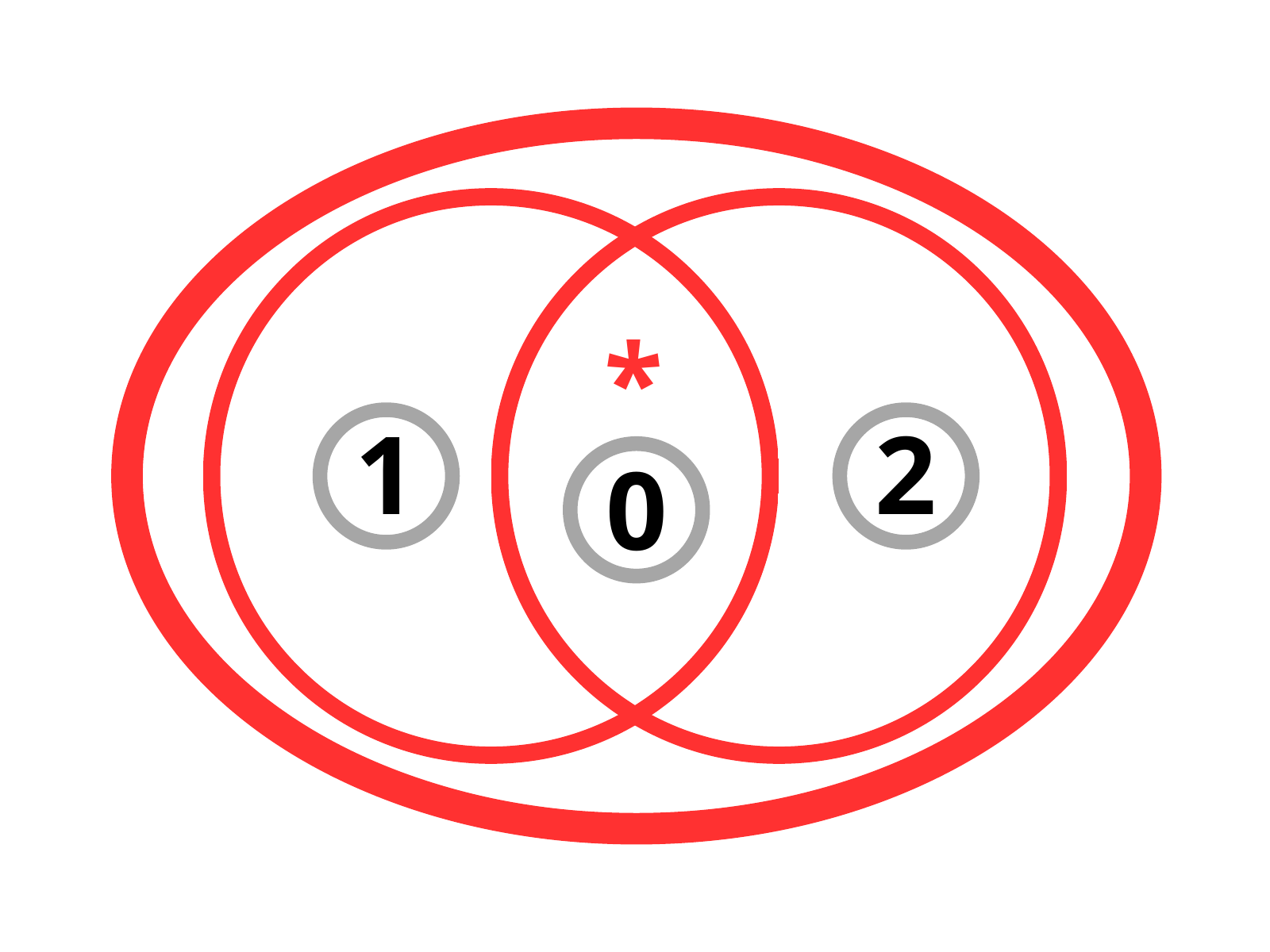}
\caption{Examples of open neighbourhoods (Remarks~\ref{rmrk.nbhd.filter} and~\ref{rmrk.no.meet})}
\label{fig.nbhd}
\end{figure}

\begin{figure}
\vspace{-1em}
\centering
\includegraphics[width=0.4\columnwidth]{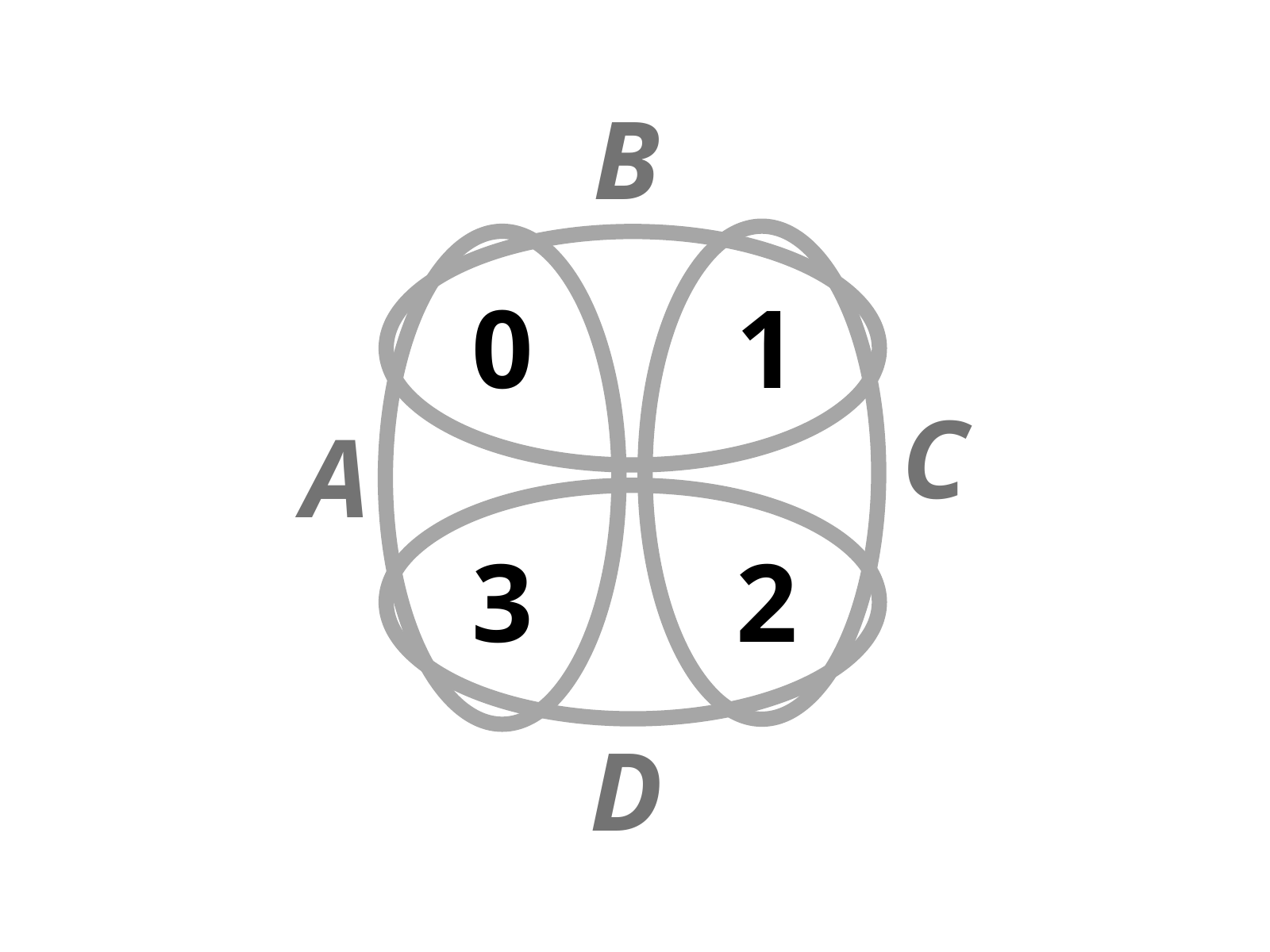}
\caption{An unconflicted, irregular space (Proposition~\ref{prop.unconflicted.irregular}) in which every minimal closed neighbourhood has a non-transitive open interior (Example~\ref{xmpl.not.intertwined})}
\label{fig.square.diagram}
\end{figure}

\begin{figure}
\centering
\includegraphics[width=0.4\columnwidth,trim={50 150 50 150},clip]{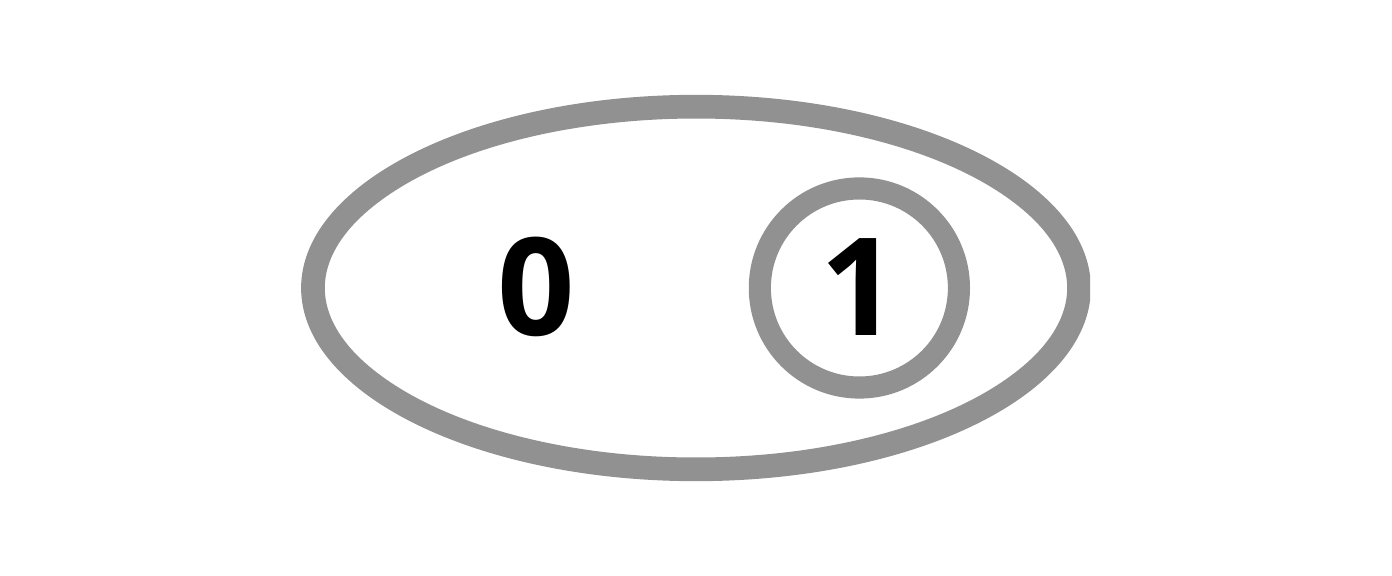}
\caption{The Sierpi\'nski space $\tf{Sk}$ (Example~\ref{xmpl.sk})}
\label{fig.sierpinski}
\end{figure}

\begin{rmrk}
\label{rmrk.nbhd.filter}
If $(\ns P,\opens)$ is a topology, then $\nbhd(p)$ is a filter (a nonempty up-closed down-directed set) and this is often called the \emph{neighbourhood filter} of $p$.

We are working with semitopologies, so $\opens$ is not necessarily closed under intersections, and $\nbhd(p)$ is not necessarily a filter.
Figure~\ref{fig.nbhd} illustrates examples of this: e.g. in the left-hand example $\{0,1\},\{0,2\}\in \nbhd(0)$ but $\{0\}\notin\nbhd(0)$, since $\{0\}$ is not an open set.
\end{rmrk}

\begin{rmrk}
\label{rmrk.cluster.convergence.2}
We can relate Proposition~\ref{prop.intertwined.as.closure} to a concept from topology. 
Following standard terminology (\cite[Definition~2, page~69]{bourbaki:gent1} or \cite[page~52]{engelking:gent}), a \deffont{cluster point} $p\in\ns P$ of $\mathcal O\subseteq\opens$ is one such that every open neighbourhood of $p$ intersects every $O\in\mathcal O$.
Then Proposition~\ref{prop.intertwined.as.closure}(\ref{item.intertwined.as.intersection.of.closures}) identifies $\intertwined{p}$ as the set of cluster points of $\nbhd(p)\subseteq\opens$.
\end{rmrk}

\jamiesubsubsection{Application to characterise (quasi/weak) regularity}

\begin{rmrk}
\label{rmrk.how.weakly.regular}
Recall that Theorem~\ref{thrm.max.cc.char} characterised regularity in multiple ways, including as the existence of a greatest topen neighbourhood. 
Proposition~\ref{prop.views.of.regularity} below does something similar, for quasiregularity and weak regularity and the existence of closed neighbourhoods (Definition~\ref{defn.cn}), and Theorem~\ref{thrm.up.down.char} is a result in the same style, for regularity.

Here, for the reader's convenience, is a summary of the relevant results:
\begin{enumerate*}
\item
Proposition~\ref{prop.views.of.quasiregularity}:\ 
$p$ is quasiregular when $\intertwined{p}$ is a closed neighbourhood.
\item
Proposition~\ref{prop.views.of.regularity}:\ 
$p$ is weakly regular when $\intertwined{p}$ is a closed neighbourhood of $p$.
\item
Theorem~\ref{thrm.up.down.char}:\ 
$p$ is regular when $\intertwined{p}$ is a closed neighbourhood of $p$ and is a minimal closed neighbourhood.
\end{enumerate*}
\end{rmrk}

\begin{prop}
\label{prop.views.of.quasiregularity}
Suppose $(\ns P,\opens)$ is a semitopology and $p\in\ns P$.
Then the following are equivalent:
\begin{enumerate*}
\item
$p$ is quasiregular, or in full: $\community(p)\neq\varnothing$ (Definition~\ref{defn.tn}(\ref{item.quasiregular.point})).
\item
$\intertwined{p}$ is a closed neighbourhood (Definition~\ref{defn.cn}(\ref{item.closed.neighbourhood})).
\end{enumerate*}
\end{prop}
\begin{proof}
By construction in Definition~\ref{defn.tn}(\ref{item.tn}), $\community(p)=\interior(\intertwined{p})$.
So $\community(p)\neq\varnothing$ means precisely that $\intertwined{p}$ is a closed neighbourhood.
\end{proof}

\begin{prop}
\label{prop.views.of.regularity}
Suppose $(\ns P,\opens)$ is a semitopology and $p\in\ns P$.
Then the following are equivalent:
\begin{enumerate*}
\item\label{item.views.of.regularity.wr}
$p$ is weakly regular, or in full: $p\in\community(p)$ (Definition~\ref{defn.tn}(\ref{item.weakly.regular.point})).
\item\label{item.intertwined.p.closed.neighbourhood.of.p}
$\intertwined{p}$ is a closed neighbourhood of $p$ (Definition~\ref{defn.cn}(\ref{item.closed.neighbourhood.of.p})).
\item\label{item.views.of.regularity.cn}
The poset of closed neighbourhoods of $p$ ordered by subset inclusion, has a least element.
\item\label{item.intertwined.p.least.in.poset.closed.neighbourhoods.of.p}
$\intertwined{p}$ is least in the poset of closed neighbourhoods of $p$ ordered by subset inclusion.
\end{enumerate*}
\end{prop}
\begin{proof}
We prove a cycle of implications:
\begin{itemize}
\item
Suppose 
$p\in\interior(\intertwined{p})$.
By Proposition~\ref{prop.intertwined.as.closure}(\ref{intertwined.p.closed}) $\intertwined{p}$ is closed, so this makes it a closed neighbourhood of $p$ as per Definition~\ref{defn.cn}.
\item
Suppose $\intertwined{p}$ is a closed neighbourhood of $p$.
By Proposition~\ref{prop.intertwined.as.closure}(\ref{intertwined.as.closure.closed}) 
$\intertwined{p}$ is the intersection of \emph{all} closed neighbourhoods of $p$, and it follows that this poset has $\intertwined{p}$ as a least element.
\item
Assume the poset of closed neighbourhoods of $p$ has a least element; write it $C$.
So $C=\bigcap\{C'\in\tf{Closed}\mid C'\text{ is a closed neighbourhood of }p\}$ and thus by Proposition~\ref{prop.intertwined.as.closure}(\ref{intertwined.as.closure.closed}) $C=\intertwined{p}$.
\item
If $\intertwined{p}$ is least in the poset of closed neighbourhoods of $p$ ordered by subset inclusion, then in particular $\intertwined{p}$ is a closed neighbourhood of $p$ and it follows from Definition~\ref{defn.cn} that $p\in\interior(\intertwined{p})$. 
\qedhere\end{itemize}
\end{proof}

Recall from Definition~\ref{defn.tn} that $\community(p)=\interior(\intertwined{p})$:
\begin{lemm}
\label{lemm.closure.community.subset}
Suppose $(\ns P,\opens)$ is a semitopology and $p\in\ns P$.
Then $\closure{\community(p)}\subseteq\intertwined{p}$.
\end{lemm}
\begin{proof}
By Proposition~\ref{prop.intertwined.as.closure}(\ref{intertwined.p.closed}) $\intertwined{p}$ is closed; we use Lemma~\ref{lemm.closure.interior}(\ref{item.closure.interior.closed}).
\end{proof}

\begin{thrm}
\label{thrm.pKp}
Suppose $(\ns P,\opens)$ is a semitopology and $p\in\ns P$.
Then:
\begin{enumerate*}
\item\label{item.pKp.1}
If $p$ is weakly regular then $\closure{\community(p)}=\intertwined{p}$.
In symbols:
$$
p\in\community(p)
\quad\text{implies}\quad \closure{\community(p)}=\intertwined{p}.
$$
\item\label{item.closure.community.p.intertwined}
As an immediate corollary, if $p$ is regular then $\closure{\community(p)}=\intertwined{p}$.
\end{enumerate*}
\end{thrm}
\begin{proof}
We consider each part in turn:
\begin{enumerate}
\item
If $p\in\community(p)=\interior(\intertwined{p})$ then $\closure{\community(p)}$ is a closed neighbourhood of $p$, so by Proposition~\ref{prop.intertwined.as.closure}(\ref{intertwined.as.closure.closed}) $\intertwined{p}\subseteq\closure{\community(p)}$.
By Lemma~\ref{lemm.closure.community.subset} $\closure{\community(p)}\subseteq\intertwined{p}$.
\item
By Lemma~\ref{lemm.wr.r}(\ref{item.r.implies.wr}) if $p$ is regular then it is weakly regular.
We use part~\ref{item.pKp.1} of this result. 
\qedhere\end{enumerate}
\end{proof}

We can combine Theorem~\ref{thrm.pKp} with Corollary~\ref{corr.regular.is.regular}: 
\begin{corr}
\label{corr.corr.pKp}
Suppose $(\ns P,\opens)$ is a semitopology and $p\in\ns P$. 
Then the following are equivalent:
\begin{enumerate*}
\item
$p$ is regular.
\item
$p$ is weakly regular and $\intertwined{p}=\intertwined{p'}$ \ for every $p'\in\community(p)$.
\end{enumerate*} 
\end{corr}
\begin{proof}
Suppose $p$ is regular and $p'\in\community(p)$.
Then $p$ is weakly regular by Lemma~\ref{lemm.wr.r}(\ref{item.r.implies.wr}), and $\community(p)=\community(p')$ by Corollary~\ref{corr.regular.is.regular}, and $\intertwined{p}=\intertwined{p'}$ by Theorem~\ref{thrm.pKp}.

Suppose $p$ is weakly regular and $\intertwined{p}=\intertwined{p'}$ for every $p'\in\community(p)$.
By Definition~\ref{defn.tn}(\ref{item.tn}) also $\community(p)=\interior(\intertwined{p})=\interior(\intertwined{p'})=\community(p')$ for every $p'\in\community(p)$, and by Corollary~\ref{corr.regular.is.regular} $p$ is regular.
\end{proof}

\begin{rmrk}
Note a subtlety to Corollary~\ref{corr.corr.pKp}: it is possible for $p$ to be regular, yet it is not the case that $\intertwined{p}=\intertwined{p'}$ for every $p'\in\intertwined{p}$ (rather than for every $p'\in\community(p)$).
For an example consider the top-left semitopology in Figure~\ref{fig.012}, taking $p=0$ and $p'=1$; then $1\in\intertwined{0}$ but $\intertwined{0}=\{0,1\}$ and $\intertwined{1}=\{0,1,2\}$.

To understand why this happens the interested reader can look ahead to Subsection~\ref{subsect.reg.tra.int}: in the terminology of that Subsection, $p'$ needs to be \emph{unconflicted} in Corollaries~\ref{corr.regular.is.regular} and~\ref{corr.corr.pKp}. 
\end{rmrk}

\jamiesubsection{Intersections of communities with open sets}

\begin{rmrk}[An observation about consensus]
\label{rmrk.fundamental.consensus}
Proposition~\ref{prop.regular.closure} and Lemma~\ref{lemm.regular.between} tell us some interesting and useful things: 
\begin{itemize*}
\item
Suppose a weakly regular $p$ wants to convince its community $\community(p)$ of some belief.
How might it proceed?

By Proposition~\ref{prop.regular.closure} it would suffice to seed one of the open neighbourhoods in its community with that belief, and then compute a \emph{topological closure} of that open set; in Remark~\ref{rmrk.why.top.closure} we discuss why topological closures are particularly interesting. 
\item
Suppose $p$ is regular, so it is a member of a transitive open neighbourhood, and $p$ wants to convince its community $\community(p)$ of some belief.

By Lemma~\ref{lemm.regular.between} $p$ need only convince \emph{some} open set that intersects its community (this open set need not even contain $p$), and then compute a topological closure as in the previous point.
\end{itemize*}
\end{rmrk}

\begin{lemm}
\label{lemm.regular.between}
Suppose $(\ns P,\opens)$ is a semitopology and $p\in\ns P$ is regular (so $p\in\community(p)\in\topens$).
Suppose $O\in\opens$.
Then
$$
p\in O\between \community(p)
\quad\text{implies}\quad 
\community(p)\subseteq\intertwined{p}\subseteq\closure{O}.
$$
In word:
\begin{quote}
If an open set intersects the community of a regular point, then that community is included in the closure of the open set.
\end{quote}
\end{lemm}
\begin{proof}
Suppose $p$ is regular, so $p\in\community(p)\in\topens$, and suppose $p\in O\between\community(p)$.
By Proposition~\ref{prop.open.strong-consensus} $\community(p)\subseteq\closure{\community(p)}\subseteq\closure{O}$.
By Theorem~\ref{thrm.pKp} $\closure{\community(p)}=\intertwined{p}$, and putting this together we get 
$$
\community(p)\subseteq\intertwined{p}\subseteq\closure{O}
$$ 
as required.
\end{proof}

Proposition~\ref{prop.regular.closure} generalises Theorem~\ref{thrm.pKp}, and is proved using it.
We regain Theorem~\ref{thrm.pKp} as the special case where $O=\community(p)$: 
\begin{prop}
\label{prop.regular.closure}
Suppose $(\ns P,\opens)$ is a semitopology and $p\in\ns P$ is weakly regular (so $p\in\community(p)\in\opens$).
Suppose $O\in\opens$.
Then:
\begin{enumerate*}
\item\label{item.regular.closure.1}
$p\in O\subseteq\community(p)$ implies
$\intertwined{p}=\closure{O}$.
\item\label{item.regular.closure.2}
As a corollary, $p\in O\subseteq\intertwined{p}$ implies
$\intertwined{p}=\closure{O}$.
\end{enumerate*}
\end{prop}
\begin{proof}
If $p\in O\subseteq\community(p)$ then $p\in\community(p)$ and using Theorem~\ref{thrm.pKp} $\closure{\community(p)}\subseteq\intertwined{p}$.
Since $O\subseteq\community(p)$ also $\closure{O}\subseteq\intertwined{p}$.
Also, by Proposition~\ref{prop.intertwined.as.closure}(\ref{item.intertwined.as.intersection.of.closures}) (since $p\in O\in\opens$) $\intertwined{p}\subseteq\closure{O}$.

For the corollary, we note that if $O$ is open then $O\subseteq\interior(\intertwined{p})=\community(p)$ if and only if $O\subseteq\intertwined{p}$.
\end{proof}

\begin{rmrk}
Note in Proposition~\ref{prop.regular.closure} that it really matters that $p\in O$ --- that is, that $O$ is an open neighbourhood \emph{of $p$} and not just an open set in $\intertwined{p}$.

To see why, consider the example in Lemma~\ref{lemm.two.intertwined} (illustrated in Figure~\ref{fig.012}, top-left diagram): so $\ns P=\{0,1,2\}$ and $\opens=\{\varnothing,\ns P,\{0\},\{2\}\}$.
Note that:
\begin{itemize*}
\item
$\intertwined{1}=\{0,1,2\}$.
\item
If we set $O=\{0\}\subseteq\{0,1,2\}$ then this is open, but $\closure{O}=\{0,1\}\neq\{0,1,2\}$.
\item
If we set $O=\{0,1,2\}\subseteq\{0,1,2\}$ then $\closure{O}=\{0,1,2\}$.
\end{itemize*}
\end{rmrk}

\begin{rmrk}
\label{rmrk.why.top.closure}
Topological closures will matter because we will develop a theory of computable semitopologies which will (amongst other things) deliver a distributed algorithm to compute closures (see Remark~\ref{rmrk.computing.closed.sets}).

Thus, putting together the results above with the witness semitopology machinery to come in Definition~\ref{defn.trust.topology} onwards, we can say that from the point of view of a regular participant $p$, Proposition~\ref{prop.regular.closure} and Lemma~\ref{lemm.regular.between} reduce the problem of 
\begin{quote}
$p$ wishes to progress with value $v$
\end{quote}
to the simpler problem of 
\begin{quote}
$p$ wishes to find an open set that intersects with the community of $p$, and work with this open set to agree on $v$ (which open set does not matter; $p$ can try several until one works).
\end{quote}
Once this is done, the distributed algorithm will safely propagate the belief across the network.

Note that no forking is possible above (this is when a distributed system that was in agreement, partitions into subsets that are committed to incompatible values); all the action is in finding and convincing the $O\between \community(p)$, and then the rest is automatic.

More discussion of this when we develop the notion of a \emph{kernel} in Section~\ref{sect.kernels}.
\end{rmrk}

\jamiesubsection{Regularity, maximal topens, \& minimal closed neighbourhoods}
\label{subsect.reg.max.min}

\begin{rmrk}
\label{rmrk.arc}
Recall we have seen an arc of results which 
\begin{itemize*}
\item
started with Theorem~\ref{thrm.max.cc.char} and Corollary~\ref{corr.regular.is.regular} --- characterisations of regularity %
in terms of maximal topens --- and 
\item
passed through Proposition~\ref{prop.views.of.regularity} --- characterisation of weak regularity $p\in\community(p)\in\opens$ in terms of minimal closed neighbourhoods.
\end{itemize*}
We are now ready to complete this arc by stating and proving Theorem~\ref{thrm.up.down.char}.
This establishes a pleasing --- and not-at-all-obvious --- duality between `has a maximal topen neighbourhood' and `has a minimal closed neighbourhood'.
\end{rmrk}

\begin{thrm}
\label{thrm.up.down.char}
Suppose $(\ns P,\opens)$ is a semitopology and $p\in\ns P$.
Then the following are equivalent:
\begin{enumerate*}
\item\label{item.up.down.char.regular}
$p$ is regular.
\item\label{item.up.down.char.max}
$\community(p)$ is a maximal/greatest topen neighbourhood of $p$.
\item\label{item.up.down.char.wr.mcn}
$p$ is weakly regular (meaning that $p\in\community(p)=\interior(\intertwined{p})$) and $\intertwined{p}$ is a minimal closed neighbourhood (Definition~\ref{defn.cn}).\footnote{We really do mean ``$\intertwined{p}$ is minimal amongst closed neighbourhoods'' and \emph{not} the weaker condition ``$\intertwined{p}$ is minimal amongst closed neighbourhoods of $p$''!  That weaker condition is treated in Proposition~\ref{prop.views.of.regularity}.  See Remark~\ref{rmrk.don't.misread}.}
\end{enumerate*}
\end{thrm}
\begin{proof}
Equivalence of parts~\ref{item.up.down.char.regular} and~\ref{item.up.down.char.max} is just Theorem~\ref{thrm.max.cc.char}(\ref{char.Kp.greatest.topen}).

For equivalence of parts~\ref{item.up.down.char.max} and~\ref{item.up.down.char.wr.mcn} we prove two implications:
\begin{itemize}
\item
Suppose $p$ is regular.
By Lemma~\ref{lemm.wr.r}(\ref{item.r.implies.wr}) $p$ is weakly regular.
Now consider a closed neighbourhood $C'\subseteq \intertwined{p}$.
Note that $C'$ has a nonempty interior by Definition~\ref{defn.cn}(\ref{item.closed.neighbourhood}), so pick any $p'$ such that
$$
p'\in\interior(C')\subseteq C'\subseteq\intertwined{p} .
$$
It follows that $p'\in\interior(\intertwined{p})=\community(p)$, and $p$ is regular, so by Corollary~\ref{corr.corr.pKp} $\intertwined{p'}=\intertwined{p}$, 
and then by Proposition~\ref{prop.views.of.regularity}(\ref{item.intertwined.p.closed.neighbourhood.of.p}\&\ref{item.intertwined.p.least.in.poset.closed.neighbourhoods.of.p}) (since $p'{\in}\interior(C')$) $\intertwined{p'}\subseteq C'$.
Putting this all together we have
$$
\intertwined{p}=\intertwined{p'} \subseteq C' \subseteq\intertwined{p},
$$
so that $C'=\intertwined{p}$ as required.
\item
Suppose $p$ is weakly regular and suppose $\intertwined{p}$ is minimal in the poset of closed neighbourhoods ordered by subset inclusion.

Consider some $p'\in\community(p)$.
By Proposition~\ref{prop.intertwined.as.closure}(\ref{intertwined.as.closure.closed}) $\intertwined{p'}\subseteq\intertwined{p}$, and by minimality it follows that $\intertwined{p'}=\intertwined{p}$.
Thus also $\community(p')=\community(p)$.

Now $p'\in\community(p)$ was arbitrary, so by Corollary~\ref{corr.regular.is.regular} $p$ is regular as required.  
\qedhere\end{itemize}
\end{proof}

\begin{rmrk}
\label{rmrk.indeed.two.closed.neighbourhoods}
Recall Example~\ref{xmpl.p.not.regular}(\ref{item.p.not.regular.01234b}), as illustrated in Figure~\ref{fig.irregular} (right-hand diagram).
This has a point $0$ whose community $\community(0)=\{1,2\}$ is not a single topen (it contains two topens: $\{1\}$ and $\{2\}$).

A corollary of Theorem~\ref{thrm.up.down.char} is that $\intertwined{0}=\{0,1,2\}$ cannot be a minimal closed neighbourhood, because if it were then $0$ would be regular and $\community(0)$ would be a maximal topen neighbourhood of $0$.

We check, and see that indeed, $\intertwined{0}$ contains \emph{two} distinct minimal closed neighbourhoods: $\{0,1\}$ and $\{0,2\}$.
\end{rmrk}

\begin{rmrk}
\label{rmrk.don't.misread}
Theorem~\ref{thrm.up.down.char}(\ref{item.up.down.char.wr.mcn}) looks like Proposition~\ref{prop.views.of.regularity}(\ref{item.intertwined.p.least.in.poset.closed.neighbourhoods.of.p}), but
\begin{itemize*}
\item
Proposition~\ref{prop.views.of.regularity}(\ref{item.intertwined.p.least.in.poset.closed.neighbourhoods.of.p}) regards the \emph{poset of closed neighbourhoods of $p$} (closed sets with a nonempty open interior that contains $p$),
\item
Theorem~\ref{thrm.up.down.char}(\ref{item.up.down.char.wr.mcn}) regards the \emph{poset of all closed neighbourhoods} (closed sets with a nonempty open interior, not necessarily including $p$).
\end{itemize*}
So the condition used in Theorem~\ref{thrm.up.down.char}(\ref{item.up.down.char.wr.mcn}) is strictly stronger than the condition used in Proposition~\ref{prop.views.of.regularity}(\ref{item.intertwined.p.least.in.poset.closed.neighbourhoods.of.p}).
Correspondingly, the regularity condition in Theorem~\ref{thrm.up.down.char}(\ref{item.up.down.char.regular}) can be written as $p\in\community(p)\in\topens$, and (as noted in Lemma~\ref{lemm.wr.r} and Example~\ref{xmpl.wr}(\ref{item.wr.2})) this is strictly stronger than the condition $p\in\community(p)$ used in Proposition~\ref{prop.views.of.regularity}(\ref{item.views.of.regularity.wr}). 
\end{rmrk}

Corollary~\ref{corr.anti-hausdorff} makes Remark~\ref{rmrk.not.hausdorff} (intertwined is the opposite of Hausdorff) a little more precise:
\begin{corr}
\label{corr.anti-hausdorff}
Suppose $(\ns P,\opens)$ is a Hausdorff semitopology (so every two points have a pair of disjoint neighbourhoods).
Then if $p\in\ns P$ is regular, then $\{p\}$ is clopen.
\end{corr}
\begin{proof}
Suppose $\ns P$ is Hausdorff and consider $p\in \ns P$.
By Remark~\ref{rmrk.not.hausdorff} $\intertwined{p}=\{p\}$. 
From Theorem~\ref{thrm.up.down.char}(\ref{item.up.down.char.wr.mcn}) $\{p\}$ is closed and has a nonempty open interior which must therefore also be equal to $\{p\}$.
By Corollary~\ref{corr.when.singleton.topen} (or from Theorem~\ref{thrm.up.down.char}(\ref{item.up.down.char.max})) this interior is transitive.
\end{proof}

\begin{prop}
\label{prop.max.topen.min.closed}
Suppose $(\ns P,\opens)$ is a semitopology.
Then:
\begin{enumerate*}
\item\label{item.max.topen.min.closed.1}
Every maximal topen is equal to the interior of a minimal closed neighbourhood.
\item\label{item.max.topen.min.closed.2}
The converse implication holds if $(\ns P,\opens)$ is a topology, but need not hold in the more general case that $(\ns P,\opens)$ is a semitopology: there may exist a minimal closed neighbourhood whose interior is not topen.
\end{enumerate*}
\end{prop}
\begin{proof}
\leavevmode
\begin{enumerate}
\item
Suppose $\atopen$ is a maximal topen.
By Definition~\ref{defn.transitive}(\ref{transitive.cc}) $\atopen$ is nonempty, so choose $p\in \atopen$.
By Proposition~\ref{prop.intertwined.as.closure}(\ref{intertwined.p.closed}) $\intertwined{p}$ is closed, and using Theorem~\ref{thrm.max.cc.char} 
$$
p\in \atopen=\community(p)=\interior(\intertwined{p})\subseteq\intertwined{p}.
$$
Thus $p$ is weakly regular and by Proposition~\ref{prop.views.of.regularity}(\ref{item.views.of.regularity.wr}\&\ref{item.intertwined.p.least.in.poset.closed.neighbourhoods.of.p}) $\intertwined{p}$ is a least closed neighbourhood of $p$.
\item
It suffices to provide a counterexample.
This is Example~\ref{xmpl.not.intertwined} below.
However, we also provide here a breaking `proof', which throws light on precisely what Example~\ref{xmpl.not.intertwined} is breaking, and illustrates what the difference between semitopology and topology can mean in practical proof.

Suppose $\atopen=\interior(C)$ is the nonempty open interior of some minimal closed neighbourhood $C$: we will try (and fail) to show that this is transitive.
By Proposition~\ref{prop.cc.char} it suffices to prove that $p\intertwinedwith p'$ for every $p,p'\in \atopen$.

So suppose $p\in O$ and $p'\in O'$ and $O\notbetween O'$.
By Definition~\ref{defn.closure}(\ref{item.closure}) $p'\notin\closure{O}$, so that $\closure{O}\cap C\subseteq C$ is a strictly smaller closed set.
Also, $O\cap C$ is nonempty because it contains $p$.

If $(\ns P,\opens)$ is a topology then we are done, because $O\cap\atopen=\interior(O\cap C)$ would necessarily be open, contradicting our assumption that $C$ is a minimal closed neighbourhood. 

However, if $(\ns P,\opens)$ is a semitopology then this does not necessarily follow: $O\cap\atopen$ need not be open, and we cannot proceed.
\qedhere\end{enumerate}
\end{proof}

\begin{lemm}
\label{lemm.square.diagram.not.qr}
Consider the semitopology illustrated in Figure~\ref{fig.square.diagram}.
So:
\begin{itemize}
\item
$\ns P = \{0, 1, 2, 3\}$.
\item
$\opens$ is generated by $\{A,B,C,D\}$ where: 
$$
A=\{3, 0\}, 
\quad
B=\{0, 1\},
\quad
C=\{1, 2\},
\quad\text{and}\quad
D=\{2, 3\}.
$$
\end{itemize}
Then for every $p\in\ns P$ we have:
\begin{enumerate*}
\item\label{item.square.diagram.not.qr.1}
$p$ is intertwined only with itself.
\item\label{item.square.diagram.not.qr.2}
$\community(p)=\varnothing$.
\end{enumerate*}
\end{lemm}
\begin{proof}
Part~\ref{item.square.diagram.not.qr.1} is by routine calculations from Definition~\ref{defn.intertwined.points}(\ref{intertwined.defn}).
Part~\ref{item.square.diagram.not.qr.2} follows, noting that $\interior(\{p\})=\varnothing$ for every $p\in\ns P$.
\end{proof}

\begin{xmpl}
\label{xmpl.not.intertwined}
The semitopology illustrated in Figure~\ref{fig.square.diagram}, and specified in Lemma~\ref{lemm.square.diagram.not.qr},
contains sets that are minimal amongst closed sets with a nonempty interior, yet that interior is not topen:
\begin{itemize*}
\item
$A$, $B$, $C$, and $D$ are clopen, because $C$ is the complement of $A$ and $D$ is the complement of $B$, so they are their own interior.
\item
$A$ is a minimal closed neighbourhood (which is also open, being $A$ itself), because 
\begin{itemize*}
\item
$A=\{3, 0\}$ is closed because it is the complement of $C$, and it is its own interior, and 
\item
its two nonempty subsets $\{3\}$ and $\{0\}$ are closed (being the complement of $B\cup C$ and $C\cup D$ respectively) but they have empty open interior because $\{3\}$ and $\{0\}$ are not open.
\end{itemize*} 
\item
$A$ is not transitive because $3$ and $0$ are not intertwined: $3\in D$ and $0\in B$ and $B\cap D=\varnothing$.
\item
Similarly $B$, $C$, and $D$ are minimal closed neighbourhoods, which are also open, and they are not transitive.
\end{itemize*}
We further note that:
\begin{enumerate*}
\item
$\closure{0}=\{0\}$, because its complement is equal to $C\cup D$ (Definition~\ref{defn.closure}; Lemma~\ref{lemm.closed.complement.open}).
Similarly for every other point in $\ns P$.
\item
$\intertwined{0}=\{0\}$, as noted in Lemma~\ref{lemm.square.diagram.not.qr}.
Similarly for every other point in $\ns P$.
\item\label{item.square.diagram.not.regular}
$\community(0)=\interior(\intertwined{0})=\varnothing$ as noted in Lemma~\ref{lemm.square.diagram.not.qr},
so that $0$ is not regular (Definition~\ref{defn.tn}(\ref{item.tn})), and $0$ is not even weakly regular or quasiregular.
Similarly for every other point in $\ns P$.
\item
$0$ has \emph{two} minimal closed neighbourhoods: $A$ and $B$.
Similarly for every other point in $\ns P$.
\end{enumerate*}
This illustrates that $\intertwined{p}\subsetneq C$ is possible, where $C$ is a minimal closed neighbourhood of $p$.
\end{xmpl}

\begin{rmrk}
The results and discussions above tell us something interesting above and beyond the specific mathematical facts which they express.

They demonstrate that points being intertwined (the $p\intertwinedwith p'$ from Definition~\ref{defn.intertwined.points}) is a distinct \emph{semitopological} notion. 
A reader familiar with topology might be tempted to identify maximal topens with interiors of minimal closed neighbourhood (so that in view of Proposition~\ref{prop.cc.char}, being intertwined would be topologically characterised just as two points being in the interior of the same minimal closed neighbourhood).

This works in topologies, but we see from Example~\ref{xmpl.not.intertwined} that in semitopologies being intertwined has its own distinct identity.
\end{rmrk}

We conclude with one more example, showing how an (apparently?) slight change to a semitopology can make a big difference to its intertwinedness:
\begin{xmpl}
\label{xmpl.two.topen.examples}
\leavevmode
\begin{enumerate*}
\item\label{item.two.topen.examples.1}
$\mathbb Q^2$ with open sets generated by any covering collection of pairwise non-parallel \deffont{rational lines} --- meaning a set of solutions to a linear equation $a.x\plus b.y=c$ for $a$, $b$, and $c$ integers --- is a semitopology.

This consists of a single (maximal) topen: lines are pairwise non-parallel, so any two lines intersect and (looking to Proposition~\ref{prop.cc.char}) all points are intertwined.
There is only one closed set with a nonempty open interior, which is the whole space.
\item\label{item.two.topen.examples.2}
$\mathbb Q^2$ with open sets generated by all (possibly parallel) rational lines, is a semitopology.
It has no topen sets and (looking to Proposition~\ref{prop.cc.char}) no two distinct points are intertwined.

For any line $l$, its complement $\mathbb Q^2\setminus l$ is a closed set, given by the union of all the lines parallel to $l$.
Thus every closed set is also an open set, and vice versa, and every line $l$ is an example of a minimal closed neighbourhood (itself), whose interior is not a topen. 
\end{enumerate*}
\end{xmpl}

\jamiesubsection{More on minimal closed neighbourhoods}

We make good use of closed neighbourhoods, and in particular minimal closed neighbourhoods, in Subsection~\ref{subsect.reg.max.min} and elsewhere.
We take a moment to give a pleasing alternative characterisation of this useful concept. 

\jamiesubsubsection{Regular open/closed sets}

\begin{rmrk}
The terminology `regular open/closed set' is from the topological literature.
It is not directly related to terminology `regular point' from Definition~\ref{defn.tn}(\ref{item.regular.point}), which comes from semitopologies.
However, it turns out that a mathematical connection does exist between these two notions. 
We outline some theory of regular open/closed sets, and then demonstrate the connections to what we have seen in our semitopological world. 
\end{rmrk}

\begin{defn}
\label{defn.regular.open.set}
Suppose $(\ns P,\opens)$ is a semitopology.
Recall some standard terminology from topology~\cite[Exercise~3D, page~29]{willard:gent}:
\begin{enumerate*}
\item
We call an open set $O\in\opens$ a \deffont{regular open set} when $O=\interior(\closure{O})$.
\item
We call a closed set $C\in\closed$ a \deffont{regular closed set} when $C=\closure{\interior(C)}$.
\item
Write $\regularOpens$ and $\regularClosed$ for the sets of regular open and regular closed sets respectively.
\end{enumerate*}
\end{defn}

\begin{lemm}
\label{lemm.ic.ci.regular}
Suppose $(\ns P,\opens)$ is a semitopology and $O\in\opens$ and $C\in\closed$.
Then:
\begin{enumerate*}
\item\label{item.ic.ci.regular.open}
$\interior(C)$ is a regular open set.
\item\label{item.ic.ci.regular.closed}
$\closure{O}$ is a regular closed set.
\end{enumerate*}
\end{lemm}
\begin{proof}
Direct from Definition~\ref{defn.regular.open.set} and Corollary~\ref{corr.ic.ci}.
\end{proof}

\begin{corr}
\label{corr.community.regular.open}
Suppose $(\ns P,\opens)$ is a semitopology and $p\in\ns P$.
Then $\community(p)\in\regularOpens$. 
\end{corr}
\begin{proof}
We just combine Lemma~\ref{lemm.ic.ci.regular}(\ref{item.ic.ci.regular.open}) with Proposition~\ref{prop.intertwined.as.closure}(\ref{intertwined.p.closed}).
\end{proof}

\begin{corr}
\label{corr.interior.closure.regular}
Suppose $(\ns P,\opens)$ is a semitopology and $O\in\opens$.
Then $\interior(\closure{O})$ is a regular open set.
\end{corr}
\begin{proof}
By Lemma~\ref{lemm.closure.closed} $\closure{O}$ is closed, and by Lemma~\ref{lemm.ic.ci.regular} $\interior(\closure{O})$ is regular open. 
\end{proof}

The regular open and the regular closed sets are the same thing, up to an easy and natural bijection: 
\begin{corr}
\label{corr.ro=rc}
Suppose $(\ns P,\opens)$ is a semitopology.
Then 
\begin{itemize*}
\item
the topological closure map $\closure{\text{-}}$ and 
\item
the topological interior map $\interior(\text{-})$ 
\end{itemize*}
define a bijection of posets between $\regularOpens$ and $\regularClosed$ ordered by subset inclusion. 
\end{corr}
\begin{proof}
By Lemma~\ref{lemm.ic.ci.regular}, $\closure{\text{-}}$ and $\interior(\text{-})$ map between $\regularOpens$ to $\regularClosed$.
Now we note that the regularity property from Definition~\ref{defn.regular.open.set}, which states that $\interior(\closure{O})=O$ when $O\in\regularOpens$ and $\closure{\interior(C)}=C$ when $C\in\regularClosed$, expresses precisely that these maps are inverse.

They are maps of posets by Corollary~\ref{corr.interior.monotone} and Lemma~\ref{lemm.closure.monotone}(\ref{closure.increasing}). 
\end{proof}

\begin{lemm}
\label{lemm.regular.open.closed}
Suppose $(\ns P,\opens)$ is a semitopology and $O\in\opens$ and $C\in\closed$.
Then:
\begin{enumerate*}
\item
$O$ is a regular open set if and only if $\ns P\setminus O$ is a regular closed set if and only if $\closure{O}$ is a regular closed set.
\item
$C$ is a regular closed set if and only if $\ns P\setminus C$ is a regular open set if and only if $\interior(C)$ is a regular open set.
\end{enumerate*}
\end{lemm} 
\begin{proof}
By routine calculations from the definitions using parts~\ref{item.closure.interior.complement.closure} and~\ref{item.closure.interior.complement.interior} of Lemma~\ref{lemm.closure.interior}.
\end{proof}

\jamiesubsubsection{Intersections of regular open sets}

An easy observation about open sets will be useful:
\begin{lemm}
\label{lemm.clint.between}
Suppose $(\ns P,\opens)$ is a semitopology and $O,O'\in\opens$.
Then the following are equivalent:
\begin{enumerate*}
\item\label{item.client.between.1} 
$O\between O'$.
\item\label{item.client.between.2} 
$O\between\interior(\closure{O'})$.
\item\label{item.client.between.3} 
$\interior(\closure{O})\between\interior(\closure{O'})$.
\end{enumerate*}
\end{lemm}
\begin{proof}
First we prove the equivalence of parts~\ref{item.client.between.1} and~\ref{item.client.between.2}:
\begin{enumerate}
\item
Suppose $O\between O'$.
By Lemma~\ref{lemm.closure.interior}(\ref{item.closure.interior.open}) $O\between \interior(\closure{O'})$.
\item
Suppose there is some $p\in O\cap\interior(\closure{O'})$.
Then $O$ is an open neighbourhood of $p$ and $p\in\closure{O'}$, so by Definition~\ref{defn.closure}(\ref{item.closure}) $O\between O'$ as required.\footnote{Lemma~\ref{lemm.closure.using.nbhd.intersections} packages this argument up nicely with some slick notation, which we have not yet set up.}
\end{enumerate}
Equivalence of parts~\ref{item.client.between.1} and~\ref{item.client.between.3} then follows easily by two applications of the equivalence of parts~\ref{item.client.between.1} and~\ref{item.client.between.2}.
\end{proof}

\begin{rmrk}
\label{rmrk.pi-base}
Lemma~\ref{lemm.clint.between} is true in topologies as well, but it is not prominent in the literature.
Two standard reference works~\cite{engelking:gent,willard:gent} do not seem to mention it.
It appears as equation~10 in Theorem~1.37 of~\cite{koppelberg:hanba1}, and as a lemma in $\pi$-base\footnoteref{https://topology.pi-base.org/theorems/T000420}{https://web.archive.org/web/20240108192930/https://topology.pi-base.org/theorems/T000420} (thanks to the mathematics StackExchange community for the pointers).  
We mention this to note an interesting contrast: this result is as true in topologies as it is in semitopologies, but somehow, it \emph{matters} more in the latter than the former.
\end{rmrk}

\begin{corr}
\label{corr.nonintersect.nonintersect.regular}
Suppose $(\ns P,\opens)$ is a semitopology and $p,p'\in\ns P$.
Then the following conditions are equivalent:
\begin{enumerate*}
\item\label{item.nonintersect.nonintersect.regular.1}
$p$ and $p'$ have a nonintersecting pair of open neighbourhoods.
\item\label{item.nonintersect.nonintersect.regular.2}
$p$ and $p'$ have a nonintersecting pair of regular open neighbourhoods.
\end{enumerate*}
\end{corr}
\begin{proof}
Part~\ref{item.nonintersect.nonintersect.regular.2} clearly implies part~\ref{item.nonintersect.nonintersect.regular.1}, since a regular open set is an open set.
Part~\ref{item.nonintersect.nonintersect.regular.1} implies part~\ref{item.nonintersect.nonintersect.regular.2} using Lemma~\ref{lemm.clint.between} and Corollary~\ref{corr.interior.closure.regular}.
\end{proof}

\begin{rmrk}
\label{rmrk.intertwined.with.regular.opens}
In Definition~\ref{defn.intertwined.points}(\ref{item.p.intertwinedwith.p'}) we defined $p\intertwinedwith p'$ in terms of open neighbourhoods of $p$ and $p'$ as follows:
$$
\Forall{O,O'{\in}\opens} (p\in O\land p'\in O') \limp O\between O' .
$$ 
In the light of Corollary~\ref{corr.nonintersect.nonintersect.regular}, we could just as well have defined it just in terms of regular open neighbourhoods: 
$$
\Forall{O,O'{\in}\regularOpens} (p\in O\land p'\in O') \limp O\between O' .
$$ 
Mathematically, for what we have needed so far, this latter characterisation is not needed.
However, it is easy to think of scenarios in which it might be useful.
In particular, \emph{computationally} it could make sense to restrict to the regular open sets, simply because there are fewer of them. 
\end{rmrk}

\jamiesubsubsection{Minimal nonempty regular closed sets are precisely the minimal closed neighbourhoods}

\begin{lemm}
\label{lemm.lcn.nrc}
Suppose $(\ns P,\opens)$ is a semitopology and $C\in\closed$.
Then:
\begin{enumerate*}
\item\label{item.lcn.nrc.1}
If $C$ is a minimal closed neighbourhood (a closed set with a nonempty open interior), then $C$ is a nonempty regular closed set (Definition~\ref{defn.regular.open.set}).
\item\label{item.lcn.nrc.2}
If $C$ is a nonempty regular closed set then $C$ is a closed neighbourhood (Definition~\ref{defn.cn}).
\end{enumerate*}
\end{lemm}
\begin{proof}
We consider each part in turn:
\begin{enumerate}
\item
\emph{Suppose $C$ is a minimal closed neighbourhood.}

Write $O'=\interior(C)$ and $C'=\closure{O'}=\closure{\interior(C)}$.
Because $C$ is a closed neighbourhood, by Definition~\ref{defn.cn} $O'\neq\varnothing$.
By Lemma~\ref{lemm.closure.closed} $C'\in\closed$.
Using Corollary~\ref{corr.ic.ci} $\interior(C')=\interior(\closure{\interior(C)})=\interior(C)=O'\neq\varnothing$, so that $C'$ is a closed neighbourhood, and by minimality $C'=C$.
But then $C=\closure{\interior(C)}$ so $C$ is regular, as required.
\item
\emph{Suppose $C$ is a nonempty regular closed set,} so that $\varnothing\neq C=\closure{\interior(C)}$.

It follows that $\interior(C)\neq\varnothing$ and this means precisely that $C$ is a closed neighbourhood. 
\qedhere\end{enumerate}
\end{proof}

In Theorem~\ref{thrm.up.down.char} we characterised the point $p$ being regular in terms of minimal closed neighbourhoods.
We can now characterise the minimal closed neighbourhoods in terms of something topologically familiar:
\begin{prop}
\label{prop.lnrc.lcn}
Suppose $(\ns P,\opens)$ is a semitopology and $C\in\closed$.
Then the following are equivalent:
\begin{enumerate*}
\item
$C$ is a minimal nonempty regular closed set. 
\item
$C$ is a minimal closed neighbourhood. 
\end{enumerate*}
\end{prop}
\begin{proof}
We prove two implications:
\begin{itemize}
\item
\emph{Suppose $C$ is a minimal closed neighbourhood.}

By Lemma~\ref{lemm.lcn.nrc}(\ref{item.lcn.nrc.1}) $C$ is a nonempty regular closed set.
Furthermore by Lemma~\ref{lemm.lcn.nrc}(\ref{item.lcn.nrc.2}) if $C'\subseteq C$ is any other nonempty regular closed set contained in $C$, then it is a closed neighbourhood, and by minimality it is equal to $C$.
Thus, $C$ is minimal.
\item
\emph{Suppose $C$ is a minimal nonempty regular closed set.}

By Lemma~\ref{lemm.lcn.nrc}(\ref{item.lcn.nrc.2}) $C$ is a closed neighbourhood.
Furthermore by Lemma~\ref{lemm.lcn.nrc}(\ref{item.lcn.nrc.1}) if $C'\subseteq C$ is any other closed neighbourhood then it is a nonempty regular closed set, and by minimality it is equal to $C$.
\qedhere\end{itemize}
\end{proof}

\jamiesubsection{How are $\intertwined{p}$ and $\closure{p}$ related?}

\begin{rmrk}
\label{rmrk.re-read.closure}
Recall the definitions of $\intertwined{p}$ and $\closure{p}$:
\begin{itemize*}
\item
The set $\closure{p}$ is the \emph{closure} of $p$.

By Definition~\ref{defn.closure} this is the set of $p'$ such that every open neighbourhood $O'\ni p'$ intersects with $\{p\}$.
By Definition~\ref{defn.closed} $\closure{p}$ is closed.
\item
The set $\intertwined{p}$ is the set of points \emph{intertwined} with $p$.

By Definition~\ref{defn.intertwined.points}(\ref{intertwined.defn}) this is the set of $p'$ such that every open neighbourhood $O'\ni p'$ intersects with every open neighbourhood $O \ni p$. 
By Proposition~\ref{prop.intertwined.as.closure}(\ref{intertwined.p.closed}) $\intertwined{p}$ is closed.
\end{itemize*}
So we see that $\closure{p}$ and $\intertwined{p}$ give us two canonical ways of generating a closed set from a point $p\in \ns P$. 
This invites a question: 
\begin{quote}
\emph{How are $\intertwined{p}$ and $\closure{p}$ related?}
\end{quote}
\end{rmrk}

Lemma~\ref{lemm.char.not.intertwined} rephrases Remark~\ref{rmrk.re-read.closure} more precisely by looking at it through sets complements.
We will use it in Lemma~\ref{lemm.cast.comp}(\ref{item.cast.comp.nbhd}):
\begin{lemm}
\label{lemm.char.not.intertwined}
Suppose $(\ns P,\opens)$ is a semitopology and $p\in\ns P$.
Then:
\begin{enumerate*}
\item
$\ns P\setminus\closure{p} = \bigcup \{O\in\opens \mid p\notin O\}\oldin\opens$.
\item\label{item.intertwined.open.avoid}
$\ns P\setminus\intertwined{p} = \bigcup\{O'\in\opens \mid \Exists{O{\in}\opens} p\in O\land O'\notbetween O\}\oldin\opens$.
\item
$\ns P\setminus\intertwined{p} = \bigcup\{O\in\opens \mid p\notin \closure{O}\}\oldin\opens$.
\end{enumerate*}
In words, we can say: $\ns P\setminus\closure{p}$ is the union of the open sets such that $p$ avoids them, and $\ns P\setminus\intertwined{p}$ is the union of the open sets such that $p$ avoids their closures.
\end{lemm} 
\begin{proof}
\leavevmode
\begin{enumerate*}
\item
Immediate from Definitions~\ref{defn.intertwined.points} and~\ref{defn.closure}.\footnote{A longer proof via Corollary~\ref{corr.closure.closure}(\ref{item.closure.as.intersection}) and Lemma~\ref{lemm.closed.complement.open} is also possible.}
Openness is from Definition~\ref{defn.semitopology}(\ref{semitopology.unions}).
\item
By a routine argument direct from Definition~\ref{defn.intertwined.points}. 
Openness is from Definition~\ref{defn.semitopology}(\ref{semitopology.unions}).
\item
Rephrasing part~\ref{item.intertwined.open.avoid} of this result using Definition~\ref{defn.closure}(\ref{item.closure}).
\qedhere\end{enumerate*}
\end{proof}

\begin{prop}
\label{prop.closure.intertwined}
Suppose $(\ns P,\opens)$ is a semitopology and $p\in\ns P$.
Then:
\begin{enumerate*}
\item\label{item.closure.intertwined.1}
$\closure{p}\subseteq \intertwined{p}$.
\item\label{item.closure.intertwined.2}
The subset inclusion may be strict; that is, $\closure{p}\subsetneq\intertwined{p}$ is possible --- even if $p$ is regular (Definition~\ref{defn.tn}(\ref{item.regular.point})).
\item\label{item.closure.intertwined.3}
If $\interior(\closure{p})\neq\varnothing$ (so $\closure{p}$ has a nonempty interior)
then 
$\closure{p}=\intertwined{p}$.
\end{enumerate*}
\end{prop}
\begin{proof}
\leavevmode
\begin{enumerate}
\item
We reason as follows:
$$
\begin{array}{r@{\ }l@{\quad}l}
\closure{p}=&
\closure{\{p\}}
&\text{Definition~\ref{defn.closure}(\ref{item.closure.p})}
\\
=&
\bigcap\{C\in\closed \mid p\in C\}
&\text{Corollary~\ref{corr.closure.closure}(\ref{item.closure.as.intersection})}
\\
\subseteq&
\bigcap\{C\in\closed \mid p\in\interior(C)\}
&\text{Fact of intersections}
\\
=&
\intertwined{p} 
&\text{Proposition~\ref{prop.intertwined.as.closure}(\ref{intertwined.as.closure.closed})}
\end{array}
$$
\item
Example~\ref{xmpl.closure.101} below shows that $\closure{p}\subsetneq\intertwined{p}$ is possible for $p$ regular. 
\item
Write $O=\interior(\closure{p})$.
By standard topological reasoning, $\closure{p}$ is the complement of the union of the open sets that do not contain $p$, and $O=\interior(\closure{p})$ is the greatest open set such that $\Forall{O'{\in}\opens}O\between O'\limp p\in O'$.  
We assumed that $O$ is nonempty, so $O\between O$, thus $p\in O$.

Then by part~\ref{item.closure.intertwined.1} of this result $p\in O\subseteq\closure{p}\subseteq\intertwined{p}$, and by Proposition~\ref{prop.regular.closure}(\ref{item.regular.closure.2}) $\intertwined{p}=\closure{O}$.
Using more standard topological reasoning (since $O\neq\varnothing$) $\closure{O}=\closure{p}$, and the result follows.
\qedhere\end{enumerate}
\end{proof}

\begin{xmpl}
\label{xmpl.closure.101}
\label{xmpl.sk}
Define $\tf{Sk}$ the \deffont{Sierpi\'nski space}~\cite[Example~3.2(e)]{willard:gent} by $\ns P=\{0,1\}$ and $\opens=\{\varnothing,\{1\},\{0,1\}\}$, as illustrated in Figure~\ref{fig.sierpinski}. 
Then:
\begin{itemize*}
\item
$\closure{0}=\{0\}$ (because $\{1\}$ is open), but
\item
$\intertwined{0}=\{0,1\}$ (because every open neighbourhood of $0$ intersects with every open neighbourhood of $1$). 
\end{itemize*}
Thus we see that $\closure{0}=\{0\}\subsetneq\{0,1\}=\intertwined{0}$, and $0$ is regular since $0\in\interior(\intertwined{0})=\{0,1\}\in\topens$.
(The Sierpi\'nski space is also a topology and is a known space.  We come back to it in Section~\ref{sect.three}.)
\end{xmpl}

\begin{rmrk}
We have one loose end left.
We know from Theorem~\ref{thrm.up.down.char}(\ref{item.up.down.char.wr.mcn}) that $\intertwined{p}$ is a minimal closed neighbourhood (closed set with nonempty open interior) when $p$ is regular. 
We also know from Proposition~\ref{prop.closure.intertwined} that $\closure{p}\subsetneq\intertwined{p}$ is possible, even if $p$ is regular.

So a closed \emph{neighbourhood} in between $\closure{p}$ and $\intertwined{p}$ is impossible by minimality, but can there be any closed \emph{sets} (not necessarily having a nonempty open interior) in between $\closure{p}$ and $\intertwined{p}$?

Somewhat counterintuitively perhaps, this is possible: 
\end{rmrk}

\begin{lemm}
Suppose $(\ns P,\opens)$ is a semitopology and $p\in\ns P$. 
Then it is possible for there to exist a closed set $C\subseteq\ns P$ with $\closure{p}\subsetneq C\subsetneq\intertwined{p}$, even if $p$ is regular.
\end{lemm}
\begin{proof}
It suffices to provide an example.
Consider $\mathbb N$ with the semitopology whose open sets are generated by 
\begin{itemize*}
\item
final segments $n_\geq=\{n'\in\mathbb N\mid n'\geq n\}$ for $n\in\mathbb N$ (cf. Example~\ref{xmpl.meet-irreducible}(\ref{item.final.N})), and 
\item
$\{0,1,2,3,4,5,6,7,8,9\}$.
\end{itemize*} 
The reader can check that $\closure{0}=\{0\}$ and $\intertwined{0}=\{0,1,2,3,4,5,6,7,8,9\}$.
However, there are also eight closed sets $\{0,1\}$, $\{0,1,2\}$, \dots, $\{0,1,2,3,\dots,8\}$ in between $\closure{0}$ and $\intertwined{0}$. 
\end{proof}

We will study $\intertwined{p}$ further but to make more progress we need the notion of a(n un)conflicted point.
This is an important idea in its own right:

\jamiesection{(Un)conflicted points: transitivity of $\intertwinedwith$}
\label{sect.unconflicted.point}

\jamiesubsection{The basic definition} 
\label{subsect.reg.tra.int}

In Lemma~\ref{lemm.intertwined.not.transitive} we asked whether the `is intertwined with' relation $\intertwinedwith$ from Definition~\ref{defn.intertwined.points}(\ref{item.p.intertwinedwith.p'}) is transitive --- answer: not necessarily.

Transitivity of $\intertwinedwith$ is a natural condition.
We now have enough machinery to study it in more detail, and this will help us gain a deeper understanding of the properties of not-necessarily-regular points.

\begin{defn}
\label{defn.conflicted}
Suppose $(\ns P,\opens)$ is a semitopology.
\begin{enumerate*}
\item\label{item.conflicted.point}
Call $p$ a \deffont{conflicted point} when there exist $p'$ and $p''$ such that $p'\intertwinedwith p$ and $p\intertwinedwith p''$ yet $\neg(p'\intertwinedwith p'')$.
\item\label{item.unconflicted}
If $p'\intertwinedwith p\intertwinedwith p''$ implies $p'\intertwinedwith p''$ always, then call $p$ an \deffont{unconflicted point}.
\item
Continuing Definition~\ref{defn.tn}(\ref{item.regular.S}), if $P\subseteq\ns P$ and every $p\in P$ is conflicted/unconflicted, then we may call $P$ a \deffont{conflicted/unconflicted set} respectively. 
\end{enumerate*}
\end{defn}

\begin{xmpl}
\label{xmpl.conflicted.points}
We consider some examples:
\begin{enumerate*}
\item\label{item.example.of.conflicted.point}
In Figure~\ref{fig.012} top-left diagram, $0$ and $2$ are unconflicted and intertwined with themselves, and $1$ is conflicted (being intertwined with $0$, $1$, and $2$).

If the reader wants to know what a conflicted point looks like: it looks like $1$. 
\item 
In Figure~\ref{fig.012} top-right diagram, $0$ and $2$ are unconflicted and intertwined with themselves, and $1$ is conflicted (being intertwined with $0$, $1$, and $2$).
\item
In Figure~\ref{fig.012} lower-left diagram, $0$ and $1$ are unconflicted and intertwined with themselves, and $3$ and $4$ are unconflicted and intertwined with themselves, and $2$ is conflicted (being intertwined with $0$, $1$, $2$, $3$, and $4$).
\item
In Figure~\ref{fig.012} lower-right diagram, all points are unconflicted, and $0$ and $2$ are intertwined just with themselves, and $1$ and $\ast$ are intertwined with one another.
\item
In Figure~\ref{fig.square.diagram}, all points are unconflicted and intertwined only with themselves.
\end{enumerate*}
\end{xmpl}

So $p$ is conflicted when it witnesses a counterexample to $\intertwinedwith$ being transitive.
We start with an easy lemma (we will use this later, but we mention it now for Remark~\ref{rmrk.intertwined.unconflicted.in.context}):
\begin{lemm}
\label{lemm.unconflicted.char}
Suppose $(\ns P,\opens)$ is a semitopology and $p\in\ns P$.
Then the following are equivalent:
\begin{enumerate*}
\item\label{item.unconflicted.char.1}
$p$ is unconflicted.
\item\label{item.unconflicted.p.in.q}
If $q\in\ns P$ and $p\in\intertwined{q}$ then $\intertwined{p}\subseteq\intertwined{q}$. 
\item\label{item.p'.in.unconflicted.p}
$\intertwined{p}\subseteq\intertwined{p'}$ for every $p'\in\intertwined{p}$.
\item\label{item.unconflicted.as.least}
$\intertwined{p}$ is least in the set $\{\intertwined{p'}\mid p\intertwinedwith p'\}$ ordered by subset inclusion.
\end{enumerate*}
\end{lemm}
\begin{proof}
The proof is just by pushing definitions around in a cycle of implications.
\begin{itemize}
\item
\emph{Part~\ref{item.unconflicted.char.1} implies part~\ref{item.unconflicted.p.in.q}.}

Suppose $p$ is unconflicted.
Consider $q\in\ns P$ such that $p\in\intertwined{q}$, and consider any $p'\in\intertwined{p}$.
Unpacking definitions we have that $p'\intertwinedwith p\intertwinedwith q$ and so $p'\intertwinedwith q$, thus $p'\in\intertwined{q}$ as required.
\item
\emph{Part~\ref{item.unconflicted.p.in.q} implies part~\ref{item.p'.in.unconflicted.p}.}

From the fact that $p'\in\intertwined{p}$ if and only if $p'\intertwinedwith p$ if and only if $p\in\intertwined{p'}$.
\item
\emph{Part~\ref{item.p'.in.unconflicted.p} implies part~\ref{item.unconflicted.as.least}.}

Part~\ref{item.unconflicted.as.least} just rephrases part~\ref{item.p'.in.unconflicted.p}.
\item
\emph{Part~\ref{item.unconflicted.as.least} implies part~\ref{item.unconflicted.char.1}.}

Suppose $\intertwined{p}$ is $\subseteq$-least in $\{\intertwined{p'}\mid p\intertwinedwith p'\}$ and suppose $p''\intertwinedwith p\intertwinedwith p'$.
Then $p''\in\intertwined{p}\subseteq\intertwined{p'}$, so $p''\intertwinedwith p'$ as required.
\qedhere\end{itemize}
\end{proof}

\begin{rmrk}
\label{rmrk.intertwined.unconflicted.in.context}
Lemma~\ref{lemm.unconflicted.char} is just an exercise in reformulating definitions, but part~\ref{item.unconflicted.as.least} of the result helps us to contrast the property of being unconflicted, with structurally similar 
characterisations of \emph{weak regularity} and of \emph{regularity} in Proposition~\ref{prop.views.of.regularity} and Theorem~\ref{thrm.up.down.char} respectively.
For the reader's convenience we collect them here --- all sets below are ordered by subset inclusion:
\begin{enumerate}
\item
$p$ is unconflicted when \emph{$\intertwined{p}$ is least in $\{\intertwined{p'}\mid p\intertwinedwith p'\}$}. 
\item
$p$ is weakly regular when \emph{$\intertwined{p}$ is least amongst closed neighbourhoods of $p$}.

See Proposition~\ref{prop.views.of.regularity} and recall from Definition~\ref{defn.cn} that a closed neighbourhood of $p$ is a closed set $C$ such that $p\in\interior(C)$.
\item 
$p$ is regular when \emph{$\intertwined{p}$ is a closed neighbourhood of $p$ and minimal amongst all closed neighbourhoods}.

See Theorem~\ref{thrm.up.down.char} and recall that a closed neighbourhood is any closed set with a nonempty interior (not necessarily containing $p$).
\end{enumerate}
We know from Lemma~\ref{lemm.wr.r}(\ref{item.r.implies.wr}) that regular implies weakly regular. 
We now consider how these properties relate to being unconflicted.
\end{rmrk}

\jamiesubsection{Regular = weakly regular + unconflicted}
\label{subsect.r=wr+uc}

\begin{prop}
\label{prop.unconflicted.irregular}
Suppose $(\ns P,\opens)$ is a semitopology and $p\in\ns P$.
Then:
\begin{enumerate*}
\item\label{item.reg.implies.unconflicted}
If $p$ is regular then it is unconflicted.

Equivalently by the contrapositive: if $p$ is conflicted then it is not regular.
\item\label{item.unconflicted.irregular.2}
$p$ may be unconflicted and neither quasiregular, weakly regular, nor regular.
\item\label{item.unconflicted.irregular.3}
There exists a semitopological space such that 
\begin{itemize*}
\item
every point is unconflicted (so $\intertwinedwith$ is a transitive relation) yet 
\item
every point has empty community, so that the space is irregular, not weakly regular, and not quasiregular.%
\footnote{See also Proposition~\ref{prop.conflicted.weakly.regular}.}
\end{itemize*}
\end{enumerate*}
\end{prop}
\begin{proof}
We consider each part in turn:
\begin{enumerate}
\item
So consider $q\intertwinedwith p \intertwinedwith q'$.
We must show that $q\intertwinedwith q'$, so consider open neighbourhoods $Q\ni q$ and $Q'\ni q'$.
By assumption $p$ is regular, so unpacking Definition~\ref{defn.tn}(\ref{item.regular.point}) $p\in\community(p)\in\topens$.
From
$$
q\intertwinedwith p\intertwinedwith q'
\quad\text{if follows that}\quad
Q\between \community(p)\between Q',
$$
and by transitivity of $\community(p)$ (Definition~\ref{defn.transitive}(\ref{transitive.transitive})) we have $Q\between Q'$ as required.
\item
Consider the semitopology illustrated in Figure~\ref{fig.square.diagram}.
By Lemma~\ref{lemm.square.diagram.not.qr} the point $0$ is trivially unconflicted (because it is intertwined only with itself), but it is also neither quasiregular, weakly regular, nor regular, because its community is the empty set. 
See also Example~\ref{xmpl.boundary.examples}. 
\item
As for the previous part, noting that the same holds of points $1$, $2$, and $3$ in Figure~\ref{fig.square.diagram}.
\qedhere\end{enumerate}
\end{proof}

We can combine Proposition~\ref{prop.unconflicted.irregular} with a previous result Lemma~\ref{lemm.wr.r} to get a precise and attractive relation between being 
\begin{itemize*}
\item
regular (Definition~\ref{defn.tn}(\ref{item.regular.point})), 
\item
weakly regular (Definition~\ref{defn.tn}(\ref{item.weakly.regular.point})), and 
\item
unconflicted (Definition~\ref{defn.conflicted}), 
\end{itemize*}
as follows:
\begin{thrm}
\label{thrm.r=wr+uc}
Suppose $(\ns P,\opens)$ is a semitopology and $p\in\ns P$.
Then the following are equivalent:
\begin{itemize*}
\item
$p$ is regular.
\item
$p$ is weakly regular and unconflicted.
\end{itemize*}
More succinctly we can write: \emph{regular = weakly regular + unconflicted}.\footnote{See also a similar result Theorem~\ref{thrm.regular=qr+sc}, and a discussion in Remark~\ref{rmrk.two.char.r}.}
\end{thrm}
\begin{proof}
We prove two implications:
\begin{itemize}
\item
If $p$ is regular then it is weakly regular by Lemma~\ref{lemm.wr.r} and unconflicted by Proposition~\ref{prop.unconflicted.irregular}(\ref{item.reg.implies.unconflicted}). 
\item
Suppose $p$ is weakly regular and unconflicted.
By Definition~\ref{defn.tn}(\ref{item.weakly.regular.point}) $p\in\community(p)$ and by Lemma~\ref{lemm.three.transitive} it would suffice to show that $q\intertwinedwith q'$ for any $q,q'\in\community(p)$.

So consider $q,q'\in\community(p)$.
Now by Definition~\ref{defn.tn}(\ref{item.tn}) $\community(p)=\interior(\intertwined{p})$ so in particular $q,q'\in\intertwined{p}$.
Thus $q\intertwinedwith p\intertwinedwith q'$, and since $p$ is unconflicted $q\intertwinedwith q'$ as required.
\qedhere\end{itemize}
\end{proof}

We can use Theorem~\ref{thrm.r=wr+uc} to derive simple global well-behavedness conditions on spaces, as follows: 
\begin{corr}
Suppose $(\ns P,\opens)$ is a semitopology.
Then:
\begin{enumerate*}
\item
If the $\intertwinedwith$ relation is transitive (i.e. if every point is unconflicted) then a point is regular if and only if it is weakly regular.
\item
If every point is weakly regular (i.e. if $p\in\community(p)$ always) then a point is regular if and only if it is unconflicted.
\end{enumerate*} 
\end{corr}
\begin{proof}
Immediate from Theorem~\ref{thrm.r=wr+uc}. 
\end{proof}

\jamiesubsection{The boundary of $\intertwined{p}$}
\label{subsect.boundary.intertwined}

In this short Subsection we ask what points on the topological boundary of $\intertwined{p}$ can look like:
\begin{nttn}
\label{nttn.boundary}
Suppose $(\ns P,\opens)$ is a semitopology and $P\subseteq\ns P$.
\begin{enumerate*}
\item
As standard, we define 
$$
\f{boundary}(P) = P\setminus\interior(P)
$$ 
and we call this the \deffont{boundary of $P$}.
\item
In the case that $P=\intertwined{p}$ for $p\in\ns P$ then 
$$
\f{boundary}(\intertwined{p})=\intertwined{p}\setminus\interior(\intertwined{p})=\intertwined{p}\setminus\community(p).
$$
\end{enumerate*}
\end{nttn}

Points in the boundary of $\intertwined{p}$ are \emph{not} regular points:
\begin{prop}
\label{prop.boundary.points.not.regular}
\label{prop.char.boundary}
Suppose $(\ns P,\opens)$ is a semitopology and $p,q\in\ns P$ and $q\in\intertwined{p}$.
Then:
\begin{enumerate*}
\item\label{item.char.boundary.1}
If $q$ is regular then $q\in\community(p)=\interior(\intertwined{p})$.
\item\label{item.char.boundary.2}
If $q$ is regular then $q\notin\boundary(\intertwined{p})$.
\item\label{item.char.boundary.3}
If $q\in\boundary(\intertwined{p})$ then $q$ is either conflicted or not weakly regular (or both).
\end{enumerate*}
\end{prop}
\begin{proof}
We consider each part in turn:
\begin{enumerate}
\item
Suppose $q$ is regular.
By Theorem~\ref{thrm.r=wr+uc} $q$ is unconflicted, so that by Lemma~\ref{lemm.unconflicted.char}(\ref{item.p'.in.unconflicted.p}) $\intertwined{q}\subseteq\intertwined{p}$; and also $q$ is weakly regular, so that $q\in\community(q)\in\opens$ and $\community(q)\subseteq\intertwined{q}\subseteq\intertwined{p}$.
Thus $\community(q)$ is an open neighbourhood of $q$ that is contained in $\intertwined{p}$ and thus $q\in\interior(\intertwined{p})$ as required.
\item
This just repeats part~\ref{item.char.boundary.2} of this result, recalling from Notation~\ref{nttn.boundary} that $q\in\boundary(\intertwined{p})$ if and only if $q\notin\interior(\intertwined{p})$.
\item
This is just the contrapositive of part~\ref{item.char.boundary.2}, combined with Theorem~\ref{thrm.r=wr+uc}.
\qedhere\end{enumerate}
\end{proof}

\begin{figure}
\vspace{-1em}
\centering
\includegraphics[width=0.32\columnwidth,trim={50 20 50 20},clip]{diagrams/counterexample-1.pdf}
\includegraphics[width=0.32\columnwidth,trim={50 20 50 20},clip]{diagrams/012a.pdf}
\includegraphics[width=0.30\columnwidth,trim={50 20 50 20},clip]{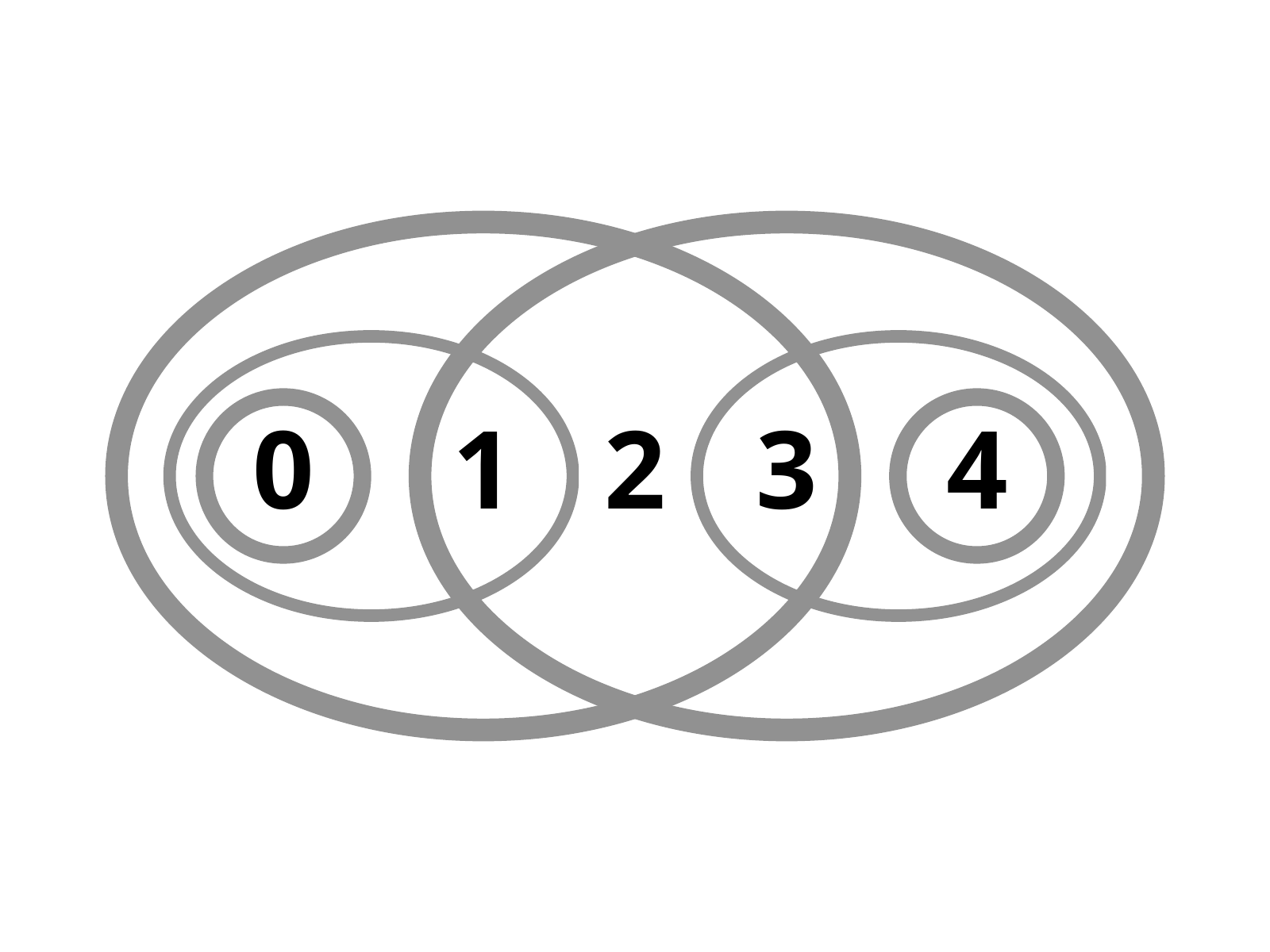}
\caption{Examples of boundary points (Example~\ref{xmpl.boundary.examples}).}
\label{fig.boundaries}
\end{figure}

\begin{xmpl}
\label{xmpl.boundary.examples}
Proposition~\ref{prop.char.boundary}(\ref{item.char.boundary.3}) tells us that points on the topological boundary of $\intertwined{p}$ are either conflicted, or not weakly regular, or perhaps both.
It remains to show that all options are possible.
It suffices to provide examples: 
\begin{enumerate*}
\item\label{item.boundary.examples.1}
In Figure~\ref{fig.boundaries} (left-hand diagram) the point $\ast$ is on the boundary of $\intertwined{1}=\{\ast,1\}$.
It is unconflicted (being intertwined just with itself and $1$), and not weakly regular (since $\ast\notin\community(\ast)=\{1\}$). 
\item\label{item.boundary.examples.2}
In Figure~\ref{fig.boundaries} (middle diagram) the point $1$ is on the boundary of $\intertwined{0}=\{0,1\}$.
It is conflicted (since $0\intertwinedwith 1\intertwinedwith 2$ yet $0\notintertwinedwith 2$) and it is weakly regular (since $1\in\community(1)=\{0,1,2\}$).\footnote{This semitopology is also in Figure~\ref{fig.012}.  We reproduce it here for the reader's convenience so that the examples are side-by-side.  

This particular semitopology is very important for other reasons: it is isomorphic to $\THREE$ (Definition~\ref{defn.3.top}); and, it is a counterexample for the plausible-seeming-yet-false non-result that $\Exists{p{\in}\ns P}\ns P=\intertwined{p}$ implies that $\ns P$ is intertwined: $\ns P=\intertwined{1}$ yet $0\notintertwinedwith 2$.
The version of this non-result that \emph{does} hold is Lemma~\ref{lemm.intertwined.iff.closure}, which is also used in Remark~\ref{rmrk.practical.intertwined} in a discussion of fast intertwinedness checking.}
\item\label{item.boundary.examples.3}
In Figure~\ref{fig.boundaries} (right-hand diagram) the point $2$ is conflicted (since $1\intertwinedwith 2\intertwinedwith 3$ yet $1\notintertwinedwith 3$) and it is not weakly regular, or even quasiregular (since $\community(2)=\interior(\{1,2,3\})=\varnothing$).
\end{enumerate*} 
Example~\ref{item.boundary.examples.3} above illustrates a boundary point that does two things --- be conflicted \emph{and} be non-weakly-regular --- even though examples~\ref{item.boundary.examples.1} and~\ref{item.boundary.examples.2} already provide examples of boundary points that each do one of these (but not the other).
It would also be nice to be able to build an example that does two (bad) things by composing two smaller examples that do one (bad) thing each --- e.g. by suitably composing examples~\ref{item.boundary.examples.1} and~\ref{item.boundary.examples.2} above.
In fact this is easy to do using products of semitopologies, but we need a little more machinery for that; see Corollary~\ref{corr.conflicted.and.not.wr}.
\end{xmpl}

We consider the special case of \emph{regular} spaces (we will pick this thread up again in Subsection~\ref{subsect.boundaries.of.closed.sets} after we have built more machinery):
\begin{corr}
\label{corr.bgp}
Suppose $(\ns P,\opens)$ is a semitopology and $p\in\ns P$. 
Then:
\begin{enumerate*}
\item\label{item.bgp.1}
If the set $\intertwined{p}$ is regular, then $\boundary(\intertwined{p})=\varnothing$ and $\intertwined{p}$ is clopen (closed and open) and transitive.
\item\label{item.bgp.2}
If $\ns P$ is a regular space (so every point in it is regular) then $\ns P$ partitions into clopen transitive components given by $\{\intertwined{p} \mid p\in\ns P\}$.
\end{enumerate*}
\end{corr}
\begin{proof}
\leavevmode
\begin{enumerate}
\item
By Proposition~\ref{prop.char.boundary} $\intertwined{p}=\interior(\intertwined{p})$, so by Lemma~\ref{lemm.interior.open} $\intertwined{p}$ is open.
By Proposition~\ref{prop.intertwined.as.closure}(\ref{intertwined.p.closed}) $\intertwined{p}$ is closed.
By Definition~\ref{defn.tn}(\ref{item.regular.point}) $p\in\community(p)=\interior(\intertwined{p})\in\topens$.
It follows that $\intertwined{p}$ is (topen and therefore) transitive.
\item
By part~\ref{item.bgp.1} of this result each $\intertwined{p}$ is a clopen transitive set.
Using Theorem~\ref{thrm.r=wr+uc} every point is unconflicted and it follows that if $\intertwined{p}\between\intertwined{p'}$ then $\intertwined{p}=\intertwined{p'}$. 
\qedhere\end{enumerate}
\end{proof}

\jamiesubsection{The intertwined preorder}

\jamiesubsubsection{Definition and properties}

\begin{rmrk}
Recall the \emph{specialisation preorder} on points from topology, defined by 
$$
p\leq p'
\quad\text{when}\quad
\closure{p}\subseteq\closure{p'}.
$$
In words: we order points $p$ by subset inclusion on their closure $\closure{p}$.

This can also be defined on semitopologies of course, but we will also find a similar preorder interesting, which is defined using $\intertwined{p}$ instead of $\closure{p}$ (Definition~\ref{defn.intertwined.preorder}).
Recall that:
\begin{itemize*}
\item
$\closure{p}$ is a closed set and is equal to the intersection of all the closed sets containing $p$, and 
\item
$\intertwined{p}$ is also a closed set (Proposition~\ref{prop.intertwined.as.closure}(\ref{intertwined.p.closed}))
and it is the intersection of all the closed neighbourhoods of $p$ (closed sets with an interior that contains $p$; see Definition~\ref{defn.cn} and Proposition~\ref{prop.intertwined.as.closure}(\ref{intertwined.as.closure.closed})).
\end{itemize*}
\end{rmrk}

\begin{defn}
\label{defn.intertwined.preorder}
Suppose $(\ns P,\opens)$ is a semitopology.
\begin{enumerate}
\item
Define the \deffont[intertwined preorder $p\leqk p'$]{intertwined preorder}\index{$p\leqk p'$ (intertwined preorder on points)} on points $p,p'\in\ns P$ by:
$$
p\leqk p'
\quad\text{when}\quad
\intertwined{p}\subseteq\intertwined{p'}.
$$
As standard, we may write $p'\geqk p$ when $p\leqk p'$ (pronounced `$p'$ is intertwined-less / intertwined-greater than $p$').

Calling this the `intertwined preorder' does not refer to the ordering being intertwined in any sense; it just means that we order on $\intertwined{p}$ (which is read `intertwined-$p$').
\item\label{item.intertwined-bounded}
Call $(\ns P,\opens)$ an \deffont{$\intertwinedwith$-complete semitopology}\index{intertwined-complete semitopology} (read `\deffont{intertwined-complete}') when 
for every subset $P\subseteq\ns P$ that is totally ordered by $\leqk$, 
there exists some $p\in\ns P$ such that $\intertwined{p}\subseteq \bigcap_i\{\intertwined{p}\mid p\in P\}$.
\end{enumerate}
\end{defn}

\begin{rmrk}
\label{rmrk.intertwinedwith-bounded.natural}
Being $\intertwinedwith$-complete (Definition~\ref{defn.intertwined.preorder}(\ref{item.intertwined-bounded})) is a plausible well-behavedness condition, because important classes of semitopologies are $\intertwinedwith$-complete: 
\begin{enumerate*}
\item
Finite semitopologies, since a descending chain of subsets of a finite set is terminating.
Note that real systems are finite, so assuming that a semitopology is $\intertwinedwith$-complete is justifiable just on these practical grounds.
\item\label{item.intertwinedwith-bounded.natural.chain-bounded}
The strongly chain-complete semitopologies which we consider later in Definition~\ref{defn.chain-complete} are $\intertwinedwith$-complete; see Lemma~\ref{lemm.chain-bounded.implies.intertwinedwith.bounded}. 
\end{enumerate*}
For now, it suffices to just work with what we need for this subsection, which is being $\intertwinedwith$-complete. 
\end{rmrk}

\begin{rmrk}
There is also the \deffont[community preorder $p\leq_K p'$]{community preorder}\index{$p\leq_K p'$ (community preorder on points)} defined such that $p\leq_K p'$ when $\community(p)\subseteq\community(p')$, which is related to $p\leq p'$ via the fact that by definition $\community(p)=\interior(\intertwined{p})$, so that $\leq_K$ is a coarser relation (meaning: it relates more points).
There is an argument that this would sit more nicely with the condition $q\in\community(p)$ in Lemma~\ref{lemm.weakly.regular.community}, but ordering on $\community(p)$ would relate all points with empty community, e.g. all of the points in Figure~\ref{fig.square.diagram}, and would slightly obfuscate the parallel with the specialisation preorder. 
This strikes us as unintuitive, so we prefer to preorder on $\intertwined{p}$. 
\end{rmrk}

\begin{lemm}
\label{lemm.weakly.regular.community}
Suppose $(\ns P,\opens)$ is a semitopology and $p,q\in\ns P$. 
Then:
\begin{enumerate*}
\item\label{item.weakly.regular.community.1}
If $q\in\community(p)$ then $q\leqk p$ (meaning that $\intertwined{q}\subseteq\intertwined{p}$).
\item\label{item.weakly.regular.community.2}
If $q\in\community(p)$ then $\community(q)\subseteq \community(p)$.
\end{enumerate*}
\end{lemm}
\begin{proof}
We consider each part in turn:
\begin{enumerate}
\item
Suppose $q\in\community(p)$ and recall from Lemma~\ref{lemm.two.intertwined}(\ref{item.two.intertwined.1})
that $\community(p)\in\opens$, which means that $\closure{\community(p)}$ is a closed neighbourhood of $q$.
We use Proposition~\ref{prop.intertwined.as.closure}(\ref{item.intertwined.as.intersection.of.closures}) and Lemma~\ref{lemm.closure.community.subset}:\footnote{If the reader prefers a proof by concrete calculations, it runs as follows:
Suppose $p'\in\community(p)$, so that in particular $p'\intertwinedwith p$.
We wish to prove that $\intertwined{p'}\subseteq\intertwined{p}$.
So consider $p''\intertwinedwith p'$; we will show that $p''\intertwinedwith p$, i.e. that every pair of open neighbourhoods of $p''$ and $p$ must intersect.
Consider a pair of open neighbourhoods $p''\in O''\in\opens$ and $p\in O\in\opens$.
We note that $O''\between \community(p)$, because $p'\in\community(p)\in\opens$ and $p''\intertwinedwith p'$.
Choose $q\in\community(p)\cap O''$. 
Now $q\intertwinedwith p$ and $q\in O''$ and $p\in O$, and we conclude that $O''\between O$ as required.
}
$$
\intertwined{q} 
\stackrel{P\ref{prop.intertwined.as.closure}(\ref{item.intertwined.as.intersection.of.closures})}{\subseteq} 
\closure{\community(p)} 
\stackrel{L\ref{lemm.closure.community.subset}}{\subseteq} 
\intertwined{p}.
$$
\item
Suppose $q\in\community(p)$.
By part~\ref{item.weakly.regular.community.1} of this result and Definition~\ref{defn.intertwined.preorder} $\intertwined{q}\subseteq\intertwined{p}$.
It is a fact that then $\interior(\intertwined{q})\subseteq\interior(\intertwined{p})$.
By Definition~\ref{defn.tn}(\ref{item.tn}) therefore $\community(q)\subseteq\community(p)$ as required.
\qedhere\end{enumerate}
\end{proof}

In the rest of this Subsection we develop corollaries of Lemma~\ref{lemm.weakly.regular.community} (and compare this with Proposition~\ref{prop.community.partition}):
\begin{corr}
\label{corr.community.intersects.community}
Suppose $(\ns P,\opens)$ is a semitopology and $q,q'\in\ns P$.
Then:
\begin{enumerate*}
\item\label{item.community.intersects.community.1}
If $\community(q)\between\community(q')$ then $q\intertwinedwith q'$.
\item\label{item.community.intersects.community.2}
If $q$ and $q'$ are weakly regular (so that $q\in\community(q)$ and $q'\in\community(q')$) then
$$
q\intertwinedwith q'
\quad\text{if and only if}\quad
\community(q)\between\community(q').
$$
\end{enumerate*}
\end{corr}
\begin{proof} 
We consider each part in turn:
\begin{enumerate}
\item
Suppose $r\in\community(q)\cap\community(q')$.
Then $\intertwined{r}\subseteq\intertwined{q}\cap\intertwined{q'}$ using Lemma~\ref{lemm.weakly.regular.community}(\ref{item.weakly.regular.community.1}).
But $q\in\intertwined{r}$, so $q\in\intertwined{q'}$, and thus $q\intertwinedwith q'$.
\item
If $q$ and $q'$ are weakly regular and $q\intertwinedwith q'$ then $\community(q)\between\community(q')$ follows from Definition~\ref{defn.intertwined.points}(\ref{item.p.intertwinedwith.p'}).
The result follows from this and from part~\ref{item.community.intersects.community.1} of this result.
\qedhere\end{enumerate}
\end{proof}

Theorem~\ref{thrm.K-regular} is somewhat reminiscent of the \emph{hairy ball theorem}:\footnote{This famous result states that every tangent vector field on a sphere of even dimension --- this being the surface of a ball of odd dimension --- must vanish at at least one point.  Intuitively, if we consider a `hairy ball' in three-dimensional space and we try to comb its hairs so they all lie smoothly flat (with no discontinuities in direction), then at least one of the hairs is pointing straight up (i.e. its projection onto the ball is zero).  A nice combinatorial proof is in \cite{doi:10.1080/00029890.2004.11920120}.} 
\begin{thrm}
\label{thrm.K-regular}
Suppose $(\ns P,\opens)$ is an $\intertwinedwith$-complete quasiregular semitopology.\footnote{Definition~\ref{defn.tn}(\ref{item.quasiregular.point}): a semitopology that is $\intertwinedwith$-complete and whose every point has a nonempty community.}
Then:
\begin{enumerate*}
\item\label{item.K-regular.1}
For every $p\in\ns P$ there exists some regular $q\in\community(p)$.
\item\label{item.K-regular.2}
$\ns P$ contains a regular point.
\end{enumerate*}
(See also Proposition~\ref{prop.up.down.wr}, which gives similar result for weak regularity.)
\end{thrm}
\begin{proof}
We consider each part in turn:
\begin{enumerate}
\item
Consider the subset $\{p'\in\ns P \mid p'\leqk p\}\subseteq\ns P$ ordered by $\leqk $.
Using Zorn's lemma (on $\geqk$), this contains a $\leqk$-minimal element $q'$.
By assumption of quasiregularity $\community(q')\neq\varnothing$, so choose $q\in\community(q')$.
By Lemma~\ref{lemm.weakly.regular.community}(\ref{item.weakly.regular.community.1}) $\intertwined{q}\subseteq\intertwined{q'}$ and by $\leqk$-minimality $\intertwined{q}=\intertwined{q'}$ and it follows that $q\in\community(q)$.
Thus $q$ is weakly regular.
Applying similar reasoning to $p'\in\community(q)$ we deduce that $\intertwined{p'}=\intertwined{q}$, and thus $\community(p')=\community(q)$, for every $p'\in\community(q)$, and so by Corollary~\ref{corr.corr.pKp} $q$ is regular.
\item
Choose any $p\in\ns P$, and use part~\ref{item.K-regular.1} of this result.
\qedhere\end{enumerate}
\end{proof}

\begin{rmrk}
We care about the existence of regular points as these are the ones that are well-behaved with respect to our semitopological model. 
A semitopology with a regular point is one that --- in some idealised mathematical sense --- is capable of some collaboration somewhere to take some action.

So Theorem~\ref{thrm.K-regular} can be read as a guarantee that, provided the semitopology is $\intertwinedwith$-complete and quasiregular, there exists somebody, somewhere, who can make sense of their local network and progress to act.
This a mathematical guarantee and not an engineering one, much as is the hairy ball theorem of which the result reminds us. 
\end{rmrk} 

\jamiesubsubsection{Application to quasiregular conflicted spaces}

In Proposition~\ref{prop.unconflicted.irregular}(\ref{item.unconflicted.irregular.3}) we saw an example of an unconflicted irregular space (illustrated in Figure~\ref{fig.square.diagram}): this is a space in which every point is unconflicted but not weakly regular.
In this subsection we consider a dual case, of a conflicted quasiregular space: a space in which every point is conflicted yet quasiregular.

One question is: does such a creature even exist?
The answer is: 
\begin{itemize*}
\item
no, in the finite case (Corollary~\ref{corr.no.finite.wr.c}); and 
\item
yes, in the infinite case (Proposition~\ref{prop.conflicted.weakly.regular}).
\end{itemize*}

\begin{prop}
\label{prop.weakly.regular.to.regular}
Suppose $(\ns P,\opens)$ is a finite quasiregular semitopology (so $\ns P$ is finite and every $p\in\ns P$ is quasiregular) --- in particular this holds if the semitopology is weakly regular.
Then:
\begin{enumerate*}
\item
For every $p\in\ns P$ there exist some regular $q\in\community(p)$. 
\item
$\ns P$ contains a regular point.
\end{enumerate*}
In words we can say: every finite quasiregular semitopology contains a regular point.
\end{prop}
\begin{proof}
From Theorem~\ref{thrm.K-regular}, since `is finite' implies `is $\intertwinedwith$-complete'.%
\footnote{The proof of Theorem~\ref{thrm.K-regular} uses Zorn's lemma.  A longer, direct proof of Proposition~\ref{prop.weakly.regular.to.regular} is also possible, by explicit induction on size of sets.}
\end{proof}

\begin{corr}
\label{corr.no.finite.wr.c}
There exists no finite quasiregular conflicted semitopology (i.e. a semitopology with finitely many points, each of which is quasiregular but conflicted).
\end{corr}
\begin{proof}
Suppose $(\ns P,\opens)$ is finite and quasiregular.
By Proposition~\ref{prop.weakly.regular.to.regular} it contains a regular $q\in\ns P$ and by Proposition~\ref{prop.unconflicted.irregular}(\ref{item.reg.implies.unconflicted}) $q$ is unconflicted. 
\end{proof}

\begin{figure}
\centering
\includegraphics[width=0.6\columnwidth]{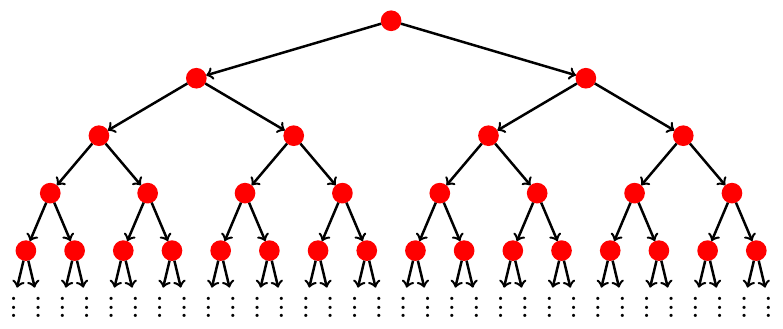}
\caption{A weakly regular, conflicted space (Proposition~\ref{prop.conflicted.weakly.regular}); the opens are the down-closed sets}
\label{fig.weakly-regular.conflicted}
\end{figure}

Corollary~\ref{corr.no.finite.wr.c} applies to finite semitopologies because these are necessarily $\intertwinedwith$-complete.
The infinite case is different, as we shall now observe:
\begin{prop}
\label{prop.conflicted.weakly.regular}
There exists an infinite quasiregular --- indeed it is also weakly regular --- conflicted semitopology $(\ns P,\opens)$.

In more detail:
\begin{itemize*}
\item
every $p\in\ns P$ is weakly regular (so $p\in\community(p)\in\opens$; see Definition~\ref{defn.tn}(\ref{item.weakly.regular.point})) yet 
\item
every $p\in\ns P$ is conflicted (so $\intertwinedwith$ is not transitive at $p$; Definition~\ref{defn.conflicted}(\ref{item.conflicted.point})).
\end{itemize*}
Furthermore: $\ns P$ is a topology\footnote{Forward reference: it is also a witness semitopology.  See Lemma~\ref{lemm.w.cwr}.} and contains no topen sets.
\end{prop}
\begin{proof}
Take $\ns P=[01]^*$ to be the set of words (possibly empty finite lists) from $0$ and $1$.
For $w,w'\in\ns P$ write $w\leq w'$ when $w$ is an initial segment of $w'$ and define 
$$
w_\geq = \{w' \mid w\leq w'\}
\quad\text{and}\quad
w_\leq = \{w' \mid w'\leq w\}.
$$
Let open sets be generated as (possibly empty) unions of the $w_\geq$.
This space is illustrated in Figure~\ref{fig.weakly-regular.conflicted}; open sets are down-closed subsets. 

The reader can check that $\neg(w0\intertwinedwith w1)$, because $w0_\geq\cap w1_\geq=\varnothing$, and that $w\intertwinedwith w'$ when $w\leq w'$ or $w'\leq w$.
It follows from the above that 
$$
\intertwined{w}=w_\geq\cup w_\leq
\quad\text{and}\quad 
\community(w)=\interior(\intertwined{w})=w_\geq,
$$
and since $w\in w_\geq$ every $w$ is weakly regular. 
Yet every $w$ is also conflicted, because $w0\intertwinedwith w \intertwinedwith w1$ yet $\neg(w0\intertwinedwith w1)$. 

This example is a topology, because an intersection of down-closed sets is still down-closed.
It escapes the constraints of Theorem~\ref{thrm.K-regular} by not being $\intertwinedwith$-complete.
It contains no topen sets because if it did contain some topen $\atopen$ then by Theorem~\ref{thrm.max.cc.char}(\ref{char.p.regular}\&\ref{char.some.topen}) there would exist a regular $p\in\atopen$ in $\ns P$.
\end{proof}

\jamiesubsubsection{(Un)conflicted points and boundaries of closed sets}
\label{subsect.boundaries.of.closed.sets}

Recall from Definition~\ref{defn.cn} that a closed neighbourhood is a closed set with a nonempty interior, and recall that $\intertwined{p}$ --- the set of points intertwined with $p$ from Definition~\ref{defn.intertwined.points} --- is characterised using closed neighbourhoods in Proposition~\ref{prop.closure.intertwined}, as the intersection of all closed neighbourhoods that have $p$ in their interior.

This leads to the question of whether the theory of $\intertwined{p}$ might \emph{be} a theory of closed neighbourhoods.
The answer seems to be no: $\intertwined{p}$ has its own distinct character, as the results and counterexamples below will briefly illustrate. 

\begin{figure}
\vspace{-2em}
\centering
\subcaptionbox{Regular boundary point of closed neighbourhood that is not intertwined with its interior (Lemma~\ref{lemm.ab12}(\ref{item.ab12.2}))}{\includegraphics[width=0.4\columnwidth,trim={50 60 50 50},clip]{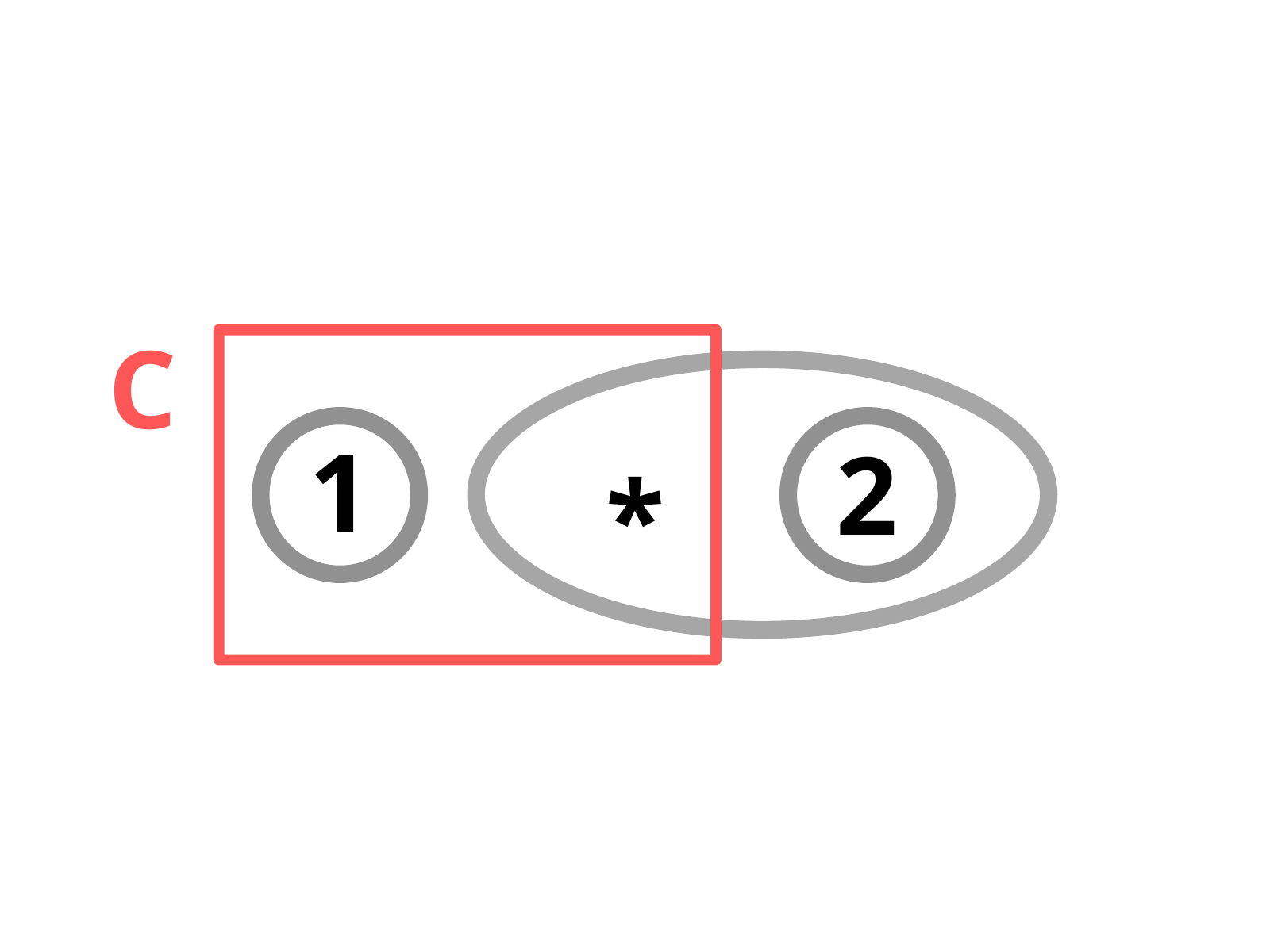}}
\qquad
\subcaptionbox{Regular point in kissing set of closed neighbourhoods, not intertwined with interiors (Corollary~\ref{corr.ab123}(\ref{item.ab123.2}))}{\includegraphics[width=0.4\columnwidth,trim={50 20 50 50},clip]{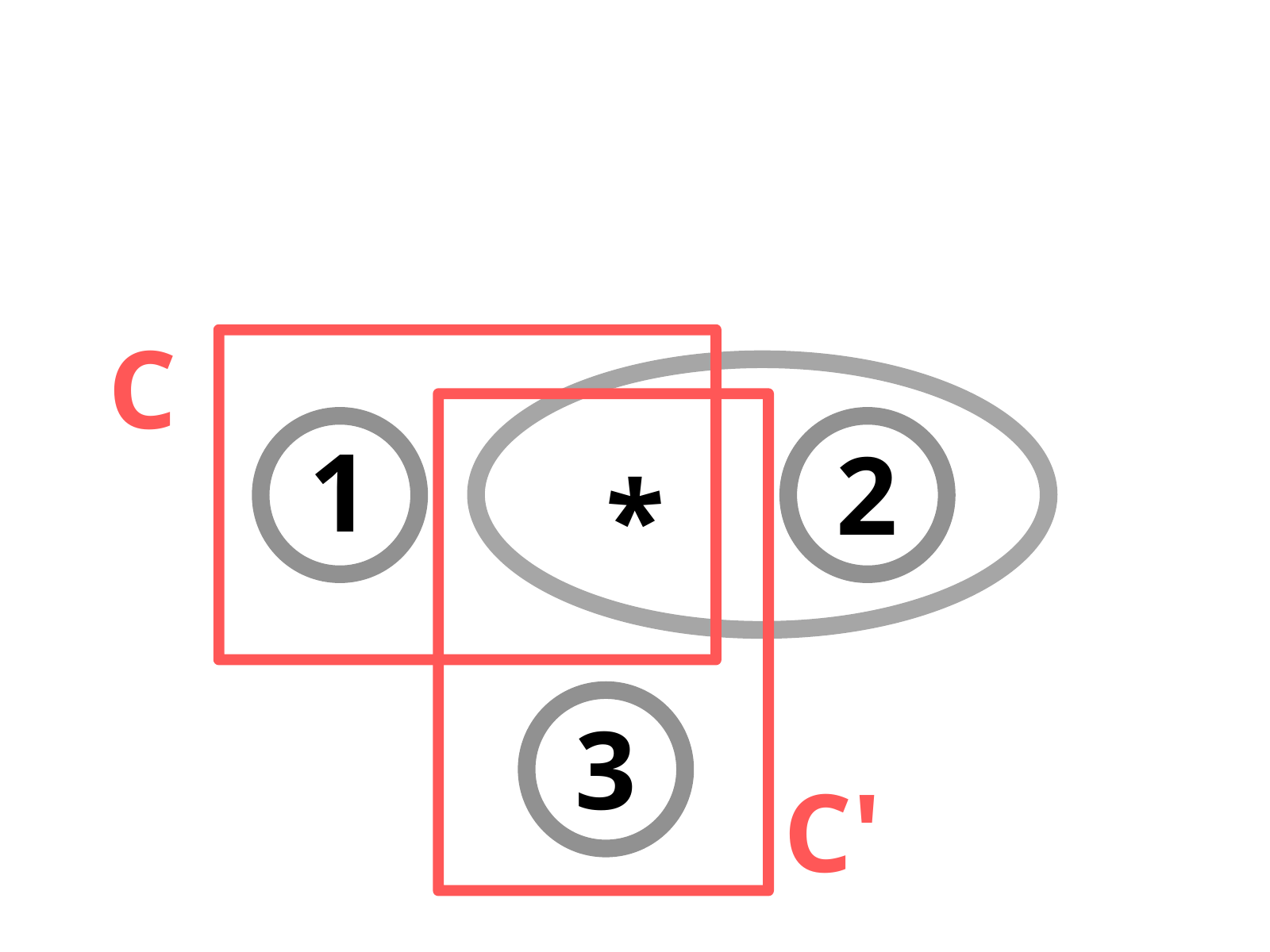}}
\caption{Two counterexamples}
\label{fig.Ast12}
\end{figure}

For instance: in view of Proposition~\ref{prop.closure.intertwined} characterising $\intertwined{p}$ as an intersection of closed neighbourhoods of $p$, might it be the case that for $C$ a closed neighbourhood, $C=\bigcup\{\intertwined{p} \mid p\in\interior(C)\}$.
In words: is a closed neighbourhood $C$ the union of the points intertwined with its interior? 
This turns out to be only half true:
\begin{lemm}
\label{lemm.ab12}
Suppose $(\ns P,\opens)$ is a semitopology and $C\in\closed$ is a closed neighbourhood.
Then: 
\begin{enumerate*}
\item\label{item.ab12.1}
$\bigcup\{\intertwined{p} \mid p\in\interior(C)\}\subseteq C$.
\item\label{item.ab12.2}
This subset inclusion may be strict: it is possible for $p\in\ns P$ to be on the boundary of a closed neighbourhood $C$, but not intertwined with any point in that neighbourhood's interior.
This is true even if $\ns P$ is a regular space (meaning that every $p\in\ns P$ is regular).
\end{enumerate*}
\end{lemm}
\begin{proof}
We consider each part in turn:
\begin{enumerate}
\item
If $p\in\interior(C)$ then $\intertwined{p}\subseteq C$ by Proposition~\ref{prop.intertwined.as.closure}(\ref{intertwined.as.closure.closed}).
\item
We provide a counterexample, as illustrated in Figure~\ref{fig.Ast12} (left-hand diagram): 
\begin{itemize*}
\item
$\ns P=\{\ast, 1, 2\}$.
\item
Open sets are generated by $\{1\}$, $\{2\}$, and $\{\ast,2\}$.
\item
We set $p=\ast$ and $C=\{1,\ast\}$.
\end{itemize*}
Then the reader can check that $\interior(C)=\{1\}$ $\intertwined{\ast}=\{\ast,2\}$ and $\ast\notintertwinedwith 2$ and every point in $\ns P$ is regular.
\qedhere\end{enumerate}
\end{proof}

\begin{defn}
Suppose $(\ns P,\opens)$ is a semitopology and $P,P'\subseteq\ns P$.
Then
define 
$$
\f{kiss}(P,P')=\boundary(P)\cap \boundary(P')
$$ 
and call this the \deffont{kissing set of $P$ and $P'$}.
\end{defn}

\begin{lemm}
\label{lemm.kissing.conflict}
Suppose $(\ns P,\opens)$ is a semitopology.
Then the following are equivalent:
\begin{itemize*}
\item
$p$ is conflicted.
\item
There exist $q,q'\in\ns P$ such that $q\notintertwinedwith q'$ and $p\in\kiss(\intertwined{q},\intertwined{q'})$.
\item
There exist $q,q'\in\ns P$ such that $q\notintertwinedwith q'$ and $p\in\intertwined{q}\cap\intertwined{q'}$.
\end{itemize*}
\end{lemm}
\begin{proof}
We prove a cycle of implications:
\begin{itemize}
\item
\emph{Suppose $p$ is conflicted.}\quad

Then there exist $q,q'\in\ns P$ such that $q\intertwinedwith p\intertwinedwith q'$ yet $q\notintertwinedwith q'$.
Rephrasing this, we obtain that $p\in\intertwined{q}\cap\intertwined{q'}$.

We need to check that $p\notin\community(q)$ and $p\notin\community(q')$.
We prove $p\notin\community(q)$ by contradiction ($p\notin\community(q')$ follows by identical reasoning).
Suppose $p\in\community(q)$.
Then by Lemma~\ref{lemm.weakly.regular.community}(\ref{item.weakly.regular.community.1}) $\intertwined{p}\subseteq\intertwined{q}$.
But $q'\in\intertwined{p}$, so $q'\in\intertwined{q}$, so $q'\intertwinedwith q$, contradicting our assumption.
\item
\emph{Suppose $q\notintertwinedwith q'$ and $p\in\boundary(\intertwined{q})\cap\boundary(\intertwined{q'})$.}

Then certainly $p\in\intertwined{q}\cap\intertwined{q'}$.
\item
\emph{Suppose $q\notintertwinedwith q'$ and $p\in\intertwined{q}\cap\intertwined{q'}$.}

Then $q\intertwinedwith p\intertwinedwith q'$ and $q\notintertwinedwith q'$, which is precisely what it means to be conflicted.
\qedhere\end{itemize}
\end{proof}

We can look at Definition~\ref{defn.conflicted} and Lemma~\ref{lemm.kissing.conflict} and conjecture that a point $p$ is conflicted if and only if it is in the kissing set of a pair of distinct closed sets.
Again, this is half true:
\begin{corr}
\label{corr.ab123}
Suppose $(\ns P,\opens)$ is a semitopology and $p\in\ns P$.
Then:
\begin{enumerate*}
\item\label{item.ab123.1}
If $p$ is conflicted then there exist a pair of closed sets such that $p\in\kiss(C,C')$.
\item\label{item.ab123.2}
The reverse implication need not hold: it is possible for $p$ to be in the kissing set of a pair of closed sets $C$ and $C'$, yet $p$ is unconflicted.
This is even possible if the space is regular (meaning that every point in the space is regular, including $p$) and $C$ and $C'$ are closed neighbourhoods.
\end{enumerate*}
\end{corr}
\begin{proof}
We consider each part in turn:
\begin{enumerate}
\item
If $p$ is conflicted then we use Lemma~\ref{lemm.kissing.conflict} and Proposition~\ref{prop.intertwined.as.closure}(\ref{intertwined.p.closed}).
\item
We provide a counterexample, as illustrated in Figure~\ref{fig.Ast12} (right-hand diagram): 
\begin{itemize*}
\item
$\ns P=\{\ast, 1, 2, 3\}$.
\item
Open sets are generated by $\{1\}$, $\{2\}$, $\{3\}$, and $\{\ast, 2\}$. 
\item
We set $p=\ast$ and $C=\{\ast,1\}$ and $C'=\{\ast, 3\}$.
\end{itemize*}
Note that $\ast$ is regular (being intertwined with itself and $2$), and $C$ and $C'$ are closed neighbourhoods that kiss at $\ast$, and $1$, $2$, and $3$ are also regular. 
\qedhere\end{enumerate}
\end{proof}

\jamiesubsection{Regular = quasiregular + hypertransitive}

\begin{rmrk}
In Theorem~\ref{thrm.r=wr+uc} we characterised regularity in terms of weak regularity and being unconflicted.
Regularity and weak regularity are two of the regularity properties considered in Definition~\ref{defn.tn}, but there is also a third: \emph{quasiregularity}.
This raises the question whether there might be some other property $X$ such that regular = quasiregular + $X$?\footnote{By Lemma~\ref{lemm.wr.r}(\ref{item.wr.implies.qr}) being weakly regular is a stronger condition than being quasiregular, thus we would expect $X$ to be stronger than being unconflicted.  And indeed this will be so: see Lemma~\ref{lemm.regular.sc}(\ref{item.sc.implies.uc}).}

Yes there is, and we develop it in this Subsection, culminating with Theorem~\ref{thrm.regular=qr+sc}.
\end{rmrk}

\jamiesubsubsection{Hypertransitivity}

\begin{nttn}
\label{nttn.between.nbhd}
Suppose $(\ns P,\opens)$ is a semitopology and $O'\in\opens$ and $\mathcal O\subseteq\opens$.
\begin{enumerate*}
\item\label{item.between.nbhd.1}
Write $O'\between\mathcal O$, or equivalently $\mathcal O\between O'$, when $O'\between O$ for every $O\in\mathcal O$.
In symbols:
$$
O'\between\mathcal O
\quad\text{when}\quad
\Forall{O{\in}\mathcal O}O'\between O .
$$
\item\label{item.between.nbhd}
As a special case of part~\ref{item.between.nbhd.1} above taking $\mathcal O=\nbhd(p)$ (Definition~\ref{defn.nbhd.system}), if $p\in\ns P$ then write $O'\between\nbhd(p)$, or equivalently $\nbhd(p)\between O'$, when $O'\between O$ for every $O\in\opens$ such that $p\in O$. 
\end{enumerate*}
\end{nttn}

\begin{lemm}
\label{lemm.closure.using.nbhd.intersections}
Suppose $(\ns P,\opens)$ is a semitopology and $p\in\ns P$ and $O'\in\opens$.
Then 
$$
p\in\closure{O'}
\quad\text{if and only if}\quad 
O'\between\nbhd(p) .
$$
See also Corollary~\ref{corr.closure.using.covers}.
\end{lemm}
\begin{proof}
This just rephrases Definition~\ref{defn.closure}(\ref{item.closure}). 
\end{proof}

\begin{defn}
\label{defn.sc}
Suppose $(\ns P,\opens)$ is a semitopology.
Call $p\in\ns P$ a \deffont{hypertransitive point} when for every $O',O''\in\opens$, 
$$
O'\between\nbhd(p)\between O''
\quad\text{implies}\quad O'\between O''.
$$
Call $(\ns P,\opens)$ a \deffont{hypertransitive semitopology} when every $p\in\ns P$ is hypertransitive.
\end{defn}

\begin{lemm}
\label{lemm.sc.op.reg.op}
Suppose $(\ns P,\opens)$ is a semitopology and $p\in\ns P$.
Then the following are equivalent:
\begin{enumerate*}
\item\label{item.sc.op.reg.op.1}
$p$ is hypertransitive.
\item\label{item.sc.op.reg.op.2}
For every pair of open sets $O',O''\in\opens$, $p\in \closure{O'}\cap \closure{O''}$ implies $O'\between O''$.
\item\label{item.sc.op.reg.op.3}
For every pair of \emph{regular} open sets $O',O''\in\regularOpens$, $p\in \closure{O'}\cap \closure{O''}$ implies $O'\between O''$ (cf. Remark~\ref{rmrk.intertwined.with.regular.opens}).
\end{enumerate*}
\end{lemm}
\begin{proof}
For the equivalence of parts~\ref{item.sc.op.reg.op.1} and~\ref{item.sc.op.reg.op.2} we reason as follows:
\begin{itemize*}
\item
Suppose $p$ is hypertransitive and suppose $p\in\closure{O'}$ and $p\in\closure{O''}$.
By Lemma~\ref{lemm.closure.using.nbhd.intersections} it follows that $O'\between\nbhd(p)\between O''$.
By hypertransitivity, $O'\between O''$ as required.
\item
Suppose for every $O,O'\in\opens$, $p\in\closure{O}\cap\closure{O'}$ implies $O'\between O''$, and suppose $O'\between\nbhd(p)\between O''$.
By Lemma~\ref{lemm.closure.using.nbhd.intersections} $p\in\closure{O}\cap\closure{O'}$ and therefore $O'\between O''$.
\end{itemize*}
For the equivalence of parts~\ref{item.sc.op.reg.op.2} and~\ref{item.sc.op.reg.op.3} we reason as follows: 
\begin{itemize*}
\item
Part~\ref{item.sc.op.reg.op.2} implies part~\ref{item.sc.op.reg.op.3} follows since every open regular set is also an open set.
\item
To show part~\ref{item.sc.op.reg.op.3} implies part~\ref{item.sc.op.reg.op.2}, suppose for every pair of regular opens $O',O''\in\regularOpens$, $p\in \closure{O'}\cap \closure{O''}$ implies $O'\between O''$, and suppose $O',O''\in\opens$ are two open sets that are not necessarily regular, and suppose $p\in\closure{O'}\cap\closure{O''}$.
We must show that $O'\between O''$.

Write $P'=\interior(\closure{O'})$ and $P''=\interior(\closure{O''})$ and note by Lemmas~\ref{lemm.ic.ci.regular} and~\ref{lemm.closure.closed} that $P'$ and $P''$ are regular open sets and $\closure{P'}=\closure{O'}$ and $\closure{P''}=\closure{O''}$.
Then $\closure{P'}\between\closure{P''}$, so $P'\between P''$, and $O'\between O''$ follows from Lemma~\ref{lemm.clint.between}
\qedhere\end{itemize*}
\end{proof}

\jamiesubsubsection{The equivalence}

\begin{lemm}
\label{lemm.regular.sc}
Suppose $(\ns P,\opens)$ is a semitopology and $p\in\ns P$.
Then:
\begin{enumerate*}
\item\label{item.r.implies.sc}
If $p$ is regular then it is hypertransitive.
\item\label{item.sc.implies.uc}
If $p$ is hypertransitive then it is unconflicted.
\item
The reverse implication need not hold: it is possible for $p$ to be unconflicted but not hypertransitive.
\item
It is possible for $p$ to be hypertransitive (and unconflicted), but not quasiregular (and thus not weakly regular or regular).
\end{enumerate*}
\end{lemm}
\begin{proof}
We consider each part:
\begin{enumerate}
\item
Suppose $p$ is regular and $O,O'\in\opens$ and $O\between\nbhd(p)\between O'$.
By Definition~\ref{defn.tn}(\ref{item.regular.point}) (since $p$ is regular) $\community(p)$ is a topen (= open and transitive) neighbourhood of $p$.
Therefore by transitivity $O\between O'$ as required. 
\item
Suppose $p$ is hypertransitive and suppose $p',p''\in\ns P$ and $p'\intertwinedwith p\intertwinedwith p''$.
Now consider $p'\in O'\in\opens$ and $p''\in O''\in\opens$.
By our intertwinedness assumptions we have that $O'\between\nbhd(p)\between O''$.
But $p$ is hypertransitive, so $O'\between O''$ as required.
\item
It suffices to provide a counterexample.
Consider the bottom right semitopology in Figure~\ref{fig.012}, and take $p=\ast$ and $O'=\{1\}$ and $O''=\{0,2\}$.
Note that:
\begin{itemize*}
\item
$\ast$ is unconflicted, since it is intertwined only with itself and $1$.
\item
$O'$ and $O'$ intersect every open neighbourhood of $\ast$, but $O'\notbetween O''$, so $\ast$ is not hypertransitive.
\end{itemize*} 
\item
It suffices to provide an example.
Consider the semitopology illustrated in Figure~\ref{fig.012}, top-right diagram; so $\ns P=\{0,1,2\}$ and $\opens=\{\varnothing,\{0\},\{2\},\{1,2\},\{0,1\},\{0,1,2\}\}$.
The reader can check that $p=1$ is hypertransitive, but $\intertwined{1}=\{1\}$ and $\community(1)=\varnothing$ so $p$ is not quasiregular.
\qedhere\end{enumerate}
\end{proof}

(Yet) another characterisation of being quasiregular will be helpful:
\begin{lemm}
\label{lemm.quasiregular.iff.between}
Suppose $(\ns P,\opens)$ is a semitopology and $p\in\ns P$.
Then the following conditions are equivalent:
\begin{enumerate*}
\item\label{item.quasiregular.iff.between.1}
$p$ is quasiregular (meaning by Definition~\ref{defn.tn}(\ref{item.quasiregular.point}) that $\community(p)\neq\varnothing$).
\item\label{item.quasiregular.iff.between.2}
$\community(p)\between\nbhd(p)$ (meaning by Notation~\ref{nttn.between.nbhd}(\ref{item.between.nbhd}) that $\community(p)\between O$ for every $O\in\nbhd(p)$).
\item\label{item.quasiregular.iff.between.3}
$p\in\closure{\community(p)}$.
\end{enumerate*}
\end{lemm}
\begin{proof}
Equivalence of parts~\ref{item.quasiregular.iff.between.2} and~\ref{item.quasiregular.iff.between.3} is immediate from Lemma~\ref{lemm.closure.using.nbhd.intersections}.

For equivalence of parts~\ref{item.quasiregular.iff.between.1} and~\ref{item.quasiregular.iff.between.2}, we prove two implications:
\begin{itemize}
\item
Suppose $p$ is quasiregular, meaning by Definition~\ref{defn.tn}(\ref{item.quasiregular.point}) that $\community(p)\neq\varnothing$.
Pick some $p'\in\community(p)$ (it does not matter which).
It follows by construction in Definitions~\ref{defn.intertwined.points}(\ref{intertwined.defn}) and~\ref{defn.tn}(\ref{item.tn}) and Lemma~\ref{lemm.interior.open} that $p'\intertwinedwith p$, so that $p'\in\community(p)$. 
Using Definition~\ref{defn.intertwined.points}(\ref{item.p.intertwinedwith.p'}) it follows that $\community(p)\between O$ for every $O\in\nbhd(p)$, as required.
\item
Suppose $\community(p)\between\nbhd(p)$.
Then in particular $\community(p)\between\ns P$ (because $p\in\ns P\in\opens$), and by Notation~\ref{nttn.between}(\ref{item.between}) it follows that $\community(p)\neq\varnothing$.
\qedhere\end{itemize}
\end{proof}

Compare and contrast Theorem~\ref{thrm.regular=qr+sc} with Theorem~\ref{thrm.r=wr+uc}:
\begin{thrm}
\label{thrm.regular=qr+sc}
Suppose $(\ns P,\opens)$ is a semitopology and $p\in\ns P$.
Then the following are equivalent:
\begin{enumerate*}
\item
$p$ is regular.
\item
$p$ is quasiregular and hypertransitive.
\end{enumerate*}
\end{thrm}
\begin{proof}
We consider two implications:
\begin{itemize}
\item
\emph{Suppose $p$ is regular.}\quad

Then $p$ is quasiregular by Lemma~\ref{lemm.wr.r}(\ref{item.r.implies.wr}\&\ref{item.wr.implies.qr}), and hypertransitive by Lemma~\ref{lemm.regular.sc}(\ref{item.r.implies.sc}). 
\item
\emph{Suppose $p$ is quasiregular and hypertransitive.}\quad

By Lemma~\ref{lemm.regular.sc}(\ref{item.sc.implies.uc}) $p$ is unconflicted.
If we can prove that $p$ is weakly regular (meaning by Definition~\ref{defn.tn}(\ref{item.weakly.regular.point}) that $p\in\community(p)$), then by Theorem~\ref{thrm.r=wr+uc} it would follow that $p$ is regular as required.
Thus, it would suffice to show that $p\in\community(p)$, thus that there is an open neighbourhood of points with which $p$ is intertwined.

Write $O''=\interior(\ns P\setminus\community(p))$.
We have two subcases to consider:
\begin{itemize*}
\item
\emph{Suppose $\nbhd(p)\between O''$.}\quad

By Lemma~\ref{lemm.quasiregular.iff.between} (since $p$ is quasiregular) we have that $\community(p)\between\nbhd(p)$.
Thus $\community(p)\between\nbhd(p)\between O''$, and by hypertransitivity of $p$ it follows that $\community(p)\between O''$.
But this contradicts the construction of $O''$ as being a subset of $\ns P\setminus\community(p)$, so this case is impossible.
\item
\emph{Suppose $\nbhd(p)\notbetween O''$.}\quad
Then there exists some $O\in\nbhd(p)$ such that $O\notbetween O''$, and it follows that $O\subseteq\community(p)$ so that $p\in\community(p)$ as required.
\end{itemize*}
Thus $p$ is weakly regular, as required.
\qedhere\end{itemize}
\end{proof}

\begin{rmrk}
\label{rmrk.two.char.r}
So we have obtained two nice characterisations of regularity of points from Definition~\ref{defn.tn}(\ref{item.regular.point}):
\begin{enumerate*}
\item
Regular = weakly regular + unconflicted, by Theorem~\ref{thrm.r=wr+uc}. 
\item
Regular = quasiregular + hypertransitive, by Theorem~\ref{thrm.regular=qr+sc}. 
\end{enumerate*}
It remains an open problem to check whether there is some natural property $X'$ such that regular = indirectly regular + $X'$ (see Definition~\ref{defn.indirectly.regular}).
Regularity properties are important to us and characterising them is a theme for us: 
see in particular Theorems~\ref{thrm.r=wr+sc} and~\ref{thrm.r=wr+sc.st}, and the discussions in Remarks~\ref{rmrk.subtly.different}, \ref{rmrk.summary.of.sc}(\ref{item.r=wr+sc.natural}) and~\ref{rmrk.mcn}.
\end{rmrk}

\jamiesection{The product semitopology} 
\label{sect.product}

Products of semitopologies can be defined just as for topologies.
We do this in Definition~\ref{defn.product.semitopology}, then study how semitopological properties --- like being a (maximal) topen or being a regular point --- interact with products.

\jamiesubsection{Basic definitions and results (shared with topologies)} 

\begin{defn}
\label{defn.squares}
Suppose $\ns P_1$ and $\ns P_2$ are sets and suppose $P_1\subseteq\ns P_1$ and $P_2\subseteq\ns P_2$.
Then:
\begin{enumerate}
\item
Call the set 
$$
P_1{\times}P_2=\{(p_1,p_2)\mid p_1\in P_1,\ p_2\in P_2\}
$$ 
a \deffont[square $P_1{\times}P_2$]{square},\index{$P_1{\times}P_2$ (square)} and 
\item
call $P_1$ and $P_2$ the \deffont{sides of the square}. 
\end{enumerate}
\end{defn}

\begin{defn}[Product semitopology]
\label{defn.product.semitopology}
Suppose $(\ns P_1,\opens_1)$ and $(\ns P_2,\opens_2)$ are semitopologies.
\begin{enumerate}
\item
As for topologies, define the \deffont{product semitopology $\ns P_1\times\ns P_2$} such that:
\begin{itemize*}
\item
The set of points is the Cartesian product $\ns P_1\times\ns P_2$.
\item
Open sets are (possibly empty, possibly infinite) unions of squares $O_1{\times} O_2$ for $O_1\in\opens_1$ and $O_2\in\opens_2$.
By abuse of notation we may write this set $\opens_1\times\opens_2$.
\end{itemize*}
\item
Define the \deffont[first projection $\pi_1:\ns P_1{\times}\ns P_2{\to} \ns P_1$]{first projection}\index{$\pi_1$ (first projection)} $\pi_1:\ns P_1\times\ns P_2\to \ns P_1$ 
and the \deffont[second projection $\pi_2:\ns P_1{\times}\ns P_2{\to} \ns P_1$]{second projection}\index{$\pi_2$ (second projection)} $\pi_2:\ns P_1\times\ns P_2\to\ns P_2$ as usual such that $\pi_1(p_1,p_2)=p_1$ and $\pi_2(p_1,p_2)=p_2$.
\item
For this Subsection, if $X$ is a set and $f$ is a function on $X$ then we define the \deffont{pointwise application $f(X)$} by
$$
f(X) =\{f(x)\mid x\in X\}.
$$
In particular we will use this notation for pointwise application of $\pi_1$ and $\pi_2$ to subsets $P\subseteq\ns P_1\times\ns P_2$.
\end{enumerate}
\end{defn}

\begin{lemm}
\label{lemm.proj.cont}
Suppose $(\ns P_1,\opens_1)$ and $(\ns P_2,\opens_2)$ are semitopologies.
Then the first and second projections $\pi_1$ and $\pi_2$ from Definition~\ref{defn.product.semitopology} are
\begin{enumerate*}
\item\label{item.proj.cont.1}
continuous (inverse image of open set is open / inverse image of closed set is closed), and
\item\label{item.proj.cont.2}
open (pointwise image of open set is open).
\end{enumerate*}
\end{lemm}
\begin{proof}
By routine calculations, as for topologies; see for example~\cite{engelking:gent}, page~79, just before Example~2.3.10.
\end{proof}

Lemma~\ref{lemm.point.closure.square} below is a special case of a general result from topology~\cite[Lemma~2.3.3, page~78]{engelking:gent} that (in our terminology from Definition~\ref{defn.squares}) the closure of a square is the square of the closure of its sides.
We do need to check that this still works for semitopologies, and it does:
\begin{lemm}
\label{lemm.point.closure.square}
Suppose $(\ns P_1,\opens_1)$ and $(\ns P_2,\opens_2)$ are semitopologies and $p_1\in\ns P_1$ and $p_2\in\ns P_2$.
Then
$$
\closure{(p_1,p_2)}=\closure{p_1}{\times}\closure{p_2}.
$$
\end{lemm}
\begin{proof}
The closure of a set is the complement of the largest open set disjoint from it.\footnote{That is: the complement of the interior of the complement.} 
By construction in Definition~\ref{defn.product.semitopology}, open sets in the product topology are unions of squares of opens, and the result now just follows noting that for $O_1\in\opens_1$ and $O_2\in\opens_2$, $(p_1,p_2)\in O_1{\times}O_2$ if and only if $p_1\in O_1$ and $p_2\in O_2$. 
\end{proof}

\jamiesubsection{Componentwise composition of semitopological properties}
\label{subsect.componentwise.composition}

We prove a sequence of results checking how properties such as being intertwined, regular, weakly, regular, and conflicted relate between a product space and the component spaces.
Most notably perhaps, we show that `being intertwined', `being regular', `being weakly regular', and `being conflicted' hold componentwise --- i.e. the results have the form 
\begin{quote}
``$(\ns P_1,\ns P_2)$ has property $\phi$ if and only if $\ns P_1$ and $\ns P_2$ have $\phi$''.
\end{quote}
We will then use this to generate examples with complex behaviour that is obtained by composing the behaviour of their (simpler) components: see in particular Corollary~\ref{corr.conflicted.and.not.wr} and Theorem~\ref{thrm.nitpicked}.

\begin{lemm}[Intersecting squares is componentwise]
\label{lemm.intersecting.squares}
Suppose $(\ns P_1,\opens_1)$ and $(\ns P_2,\opens_2)$ are semitopologies and suppose $O,O'\in\opens_1{\times}\opens_2$ are squares. 
Then
$$
O\between O'
\quad\text{if and only if}\quad
\pi_1(O)\between\pi_1(O') \ \land\ \pi_2(O)\between\pi_2(O') .
$$
\end{lemm}
\begin{proof}
By routine sets calculations, noting that since $O$ and $O'$ are squares by definition $O=\pi_1(O)\times\pi_2(O)$ and $O'=\pi_1(O')\times\pi_2(O')$.
\end{proof}

\begin{prop}[Intertwined is componentwise]
\label{prop.product.intertwined}
Suppose $(\ns P_1,\opens_1)$ and $(\ns P_2,\opens_2)$ are semitopologies.
Then:
\begin{enumerate*}
\item\label{item.product.intertwined.1} 
$(p_1,p_2)\between (p_1',p_2')$ if and only if $p_1\between p_1' \ \land\  p_2\between p_2'$.
\item\label{item.product.intertwined.2} 
As an immediate corollary, $\intertwined{(p_1,p_2)} = \intertwined{{p_1}}\times\intertwined{{p_2}}$.
\end{enumerate*}
\end{prop}
\begin{proof}
For part~\ref{item.product.intertwined.1} of this result we prove two implications:
\begin{itemize}
\item
\emph{Suppose $p_1\between p_1'$ and $p_2\between p_2'$.}\quad

Consider two open neighbourhoods $O\ni (p_1,p_2)$ and $O'\ni (p_1',p_2')$.
We wish to show that $O\between O'$.

Without loss of generality we may assume that $O$ and $O'$ are squares, since: opens are unions of squares so we just choose squares in $O$ and $O'$ that contain $(p_1,p_2)$ and $(p_1',p_2')$ respectively.
Thus, $O=O_1{\times} O_2$ and $O'=O_1'{\times} O_2'$.

Now $p_1\in O_1$ and $p_1'\in O_1'$ and $p_1\between p_1'$, thus $O_1\between O_1'$.
Similarly for $p_2$ and $p_2'$.
We use Lemma~\ref{lemm.intersecting.squares}.
\item
\emph{Suppose $(p_1,p_2)\between (p_1',p_2')$.}\quad

Then in particular all square open neighbourhoods intersect, and by Lemma~\ref{lemm.intersecting.squares} so must their sides.
\end{itemize}
Part~\ref{item.product.intertwined.2}  just rephrases part~\ref{item.product.intertwined.1}  of this result using Definition~\ref{defn.intertwined.points}(\ref{intertwined.defn}).
\end{proof}

\begin{corr}[(Maximal) topen is componentwise]
\label{corr.topens.are.squares}
Suppose $(\ns P_1,\opens_1)$ and $(\ns P_2,\opens_2)$ are semitopologies and $\atopen\in\opens_1{\times}\opens_2$ is a square.
Then for each of `a topen' / `a maximal topen' below, the following are equivalent:
\begin{itemize*}
\item
$\atopen\in\opens_1{\times}\opens_2$ is a topen / a maximal topen in $\ns P_1\times\ns P_2$.
\item
The sides $\pi_1(\atopen)$ and $\pi_2(\atopen)$ of $\atopen$ are topens / maximal topens in $\ns P_1\times\ns P_2$.
\end{itemize*}
\end{corr}
\begin{proof}
First, we consider the versions without `maximal':
\begin{enumerate}
\item
\emph{Suppose $\atopen\in\opens_1{\times}\opens_2$ is a topen in $\ns P_1\times\ns P_2$.}

By Lemma~\ref{lemm.proj.cont}(\ref{item.proj.cont.2}) its sides $\pi_1(\atopen)$ and $\pi_2(\atopen)$ are open.
Now consider $p_1,p_1'\in\pi_1(\atopen)$ and choose any $p_2\in\pi_2(\atopen)$.
We know $(p_1,p_2)\between(p_1',p_2)$ must hold, because both points are in $\atopen$ and by Proposition~\ref{prop.cc.char} all points in $\atopen$ are intertwined.
By Proposition~\ref{prop.product.intertwined}(\ref{item.product.intertwined.1}) it follows that $p_1\between p_1'$.
Since $p_1$ and $p_1'$ were arbitrary in $\pi_1(\atopen)$ it follows using Proposition~\ref{prop.cc.char} again that $\pi_1(\atopen)$ is topen. 

The reasoning for $\pi_2(\atopen)$ is precisely similar.
\item
\emph{Suppose $T_1\in\opens_1$ and $T_2\in\opens_2$ are topen in $\ns P_1\times\ns P_2$.}

By construction in Definition~\ref{defn.product.semitopology} the square $T_1{\times}T_2$ is open, and it follows using Proposition~\ref{prop.product.intertwined}(\ref{item.product.intertwined.1}) and Proposition~\ref{prop.cc.char} that $T_1{\times}T_2$ is topen.
\end{enumerate}
We now consider maximality:
\begin{enumerate}
\item
\emph{Suppose $\atopen\in\opens_1{\times}\opens_2$ is a maximal topen in $\ns P_1\times\ns P_2$.}

By our reasoning above its sides are topens, but if those sides were not maximal topens --- so at least one of them is included in a strictly larger topen --- then, again using our reasoning above, we could use obtain a larger topen square in $\opens_1{\times}\opens_2$, contradicting maximality of $\atopen$.
\item 
\emph{Suppose $T_1\in\opens_1$ and $T_2\in\opens_2$ are maximal topens in $\ns P_1$ and $\ns P_2$.}

By our reasoning above the square $T_1{\times}T_2$ is a topen.
If it were not a maximal topen --- so it is included in some strictly larger topen $\atopen$ --- then by our reasoning above $\pi_1(\atopen)$ and $\pi_2(\atopen)$ are also topens and one of them would have to be larger than $T_1$ or $T_2$, contradicting their maximality.
\qedhere\end{enumerate}
\end{proof}

\begin{corr}[Regular is componentwise]
\label{corr.regular.pair}
Suppose $(\ns P_1,\opens_1)$ and $(\ns P_2,\opens_2)$ are semitopologies and $p_1\in\ns P_1$ and $p_2\in\ns P_2$.
Then the following are equivalent:
\begin{itemize*}
\item
$(p_1,p_2)$ is regular in $\ns P_1\times\ns P_2$. 
\item
$p_1$ is regular in $\ns P_1$ and $p_2$ is regular in $\ns P_2$.
\end{itemize*}
\end{corr}
\begin{proof}
Suppose $(p_1,p_2)$ is regular.
By Theorem~\ref{thrm.max.cc.char}(\ref{char.p.regular}\&\ref{char.some.topen}) it has a topen neighbourhood $\atopen$.
Using Corollary~\ref{corr.topens.are.squares} $\pi_1(\atopen)$ and $\pi_2(\atopen)$ are topen neighbourhoods of $p_1$ and $p_2$ respectively.
By Theorem~\ref{thrm.max.cc.char}(\ref{char.p.regular}\&\ref{char.some.topen}) $p_1$ and $p_2$ are regular.

If conversely $p_1$ and $p_2$ are regular then we just reverse the reasoning of the previous paragraph. 
\end{proof}

Proposition~\ref{prop.wr.pair} does for `is conflicted' and `is weakly regular' what Corollary~\ref{corr.regular.pair} does for `is regular'.
With the machinery we now have, the argument is straightforward:
\begin{prop}[Unconflicted \& weakly regular is componentwise]
\label{prop.wr.pair}
Suppose $(\ns P_1,\opens_1)$ and $(\ns P_2,\opens_2)$ are semitopologies and suppose $p_1\in\ns P_1$ and $p_2\in\ns P_2$.
Then:
\begin{enumerate*}
\item\label{item.wr.pair.1}
$(p_1,p_2)$ is unconflicted in $\ns P_1\times\ns P_2$ if and only if $p_1$ is unconflicted in $\ns P_1$ and $p_2$ is unconflicted in $\ns P_2$.
\item\label{item.wr.pair.2}
$(p_1,p_2)$ is weakly regular in $\ns P_1\times\ns P_2$ if and only if $p_1$ is weakly regular in $\ns P_1$ and $p_2$ is weakly regular in $\ns P_2$.
\end{enumerate*}
\end{prop}
\begin{proof}
For part~\ref{item.wr.pair.1} we prove two implications:
\begin{itemize}
\item
\emph{Suppose $(p_1,p_2)$ is unconflicted.}
We will show that $p_1$ is unconflicted (the case of $p_2$ is precisely similar).

Consider $p',p''\in \ns P_1$ and suppose $p'\intertwinedwith p_1\intertwinedwith p''$.
Using Proposition~\ref{prop.product.intertwined}(\ref{item.product.intertwined.1}) $(p',p_2)\intertwinedwith (p_1,p_2)\intertwinedwith (p'',p_2)$, by transitivity (since we assumed $(p_1,p_2)$ is unconflicted) $(p',p_2)\intertwinedwith (p'',p_2)$, and using Proposition~\ref{prop.product.intertwined}(\ref{item.product.intertwined.1}) we conclude that $p'\intertwinedwith p''$ as required.

\emph{Suppose $p_1$ and $p_2$ are unconflicted.}
We will assume $(p_1',p_2')\intertwinedwith (p_1,p_2)\intertwinedwith (p_1'',p_2'')$ and prove $(p_1',p_2')\intertwinedwith(p_1'',p_2'')$.

Using Proposition~\ref{prop.product.intertwined}(\ref{item.product.intertwined.1}) $p_1'\intertwinedwith p_1\intertwinedwith p_1''$ and by transitivity (since we assumed $p_1$ is unconflicted) we have $p_1'\intertwinedwith p_1''$.
Similarly $p_2'\intertwinedwith p_2''$, and using Proposition~\ref{prop.product.intertwined}(\ref{item.product.intertwined.1}) $(p_1',p_2')\intertwinedwith (p_1'',p_2'')$ as required. 
\end{itemize}
Part~\ref{item.wr.pair.2} follows by routine reasoning just combining part~\ref{item.wr.pair.1} of this result and Corollary~\ref{corr.regular.pair} with Theorem~\ref{thrm.r=wr+uc}.  
\end{proof}

We now have the machinery that we need to make good on a promise made at the end of Example~\ref{xmpl.boundary.examples}:
\begin{corr}
\label{corr.conflicted.and.not.wr}
There exists a semitopology $(\ns P,\opens)$ and points $p,q\in\ns P$ such that 
\begin{itemize*}
\item
$q$ is on the boundary of $\intertwined{p}$ and 
\item
$q$ is conflicted and not weakly regular.
\end{itemize*}
\end{corr}
\begin{proof}
We already know this from Example~\ref{xmpl.boundary.examples}(\ref{item.boundary.examples.3}), as illustrated in the right-hand diagram in Figure~\ref{fig.boundaries},
but now we can give a more principled construction:
we let $(\ns P_1,\opens_1)$ and $(\ns P_2,\opens_2)$ be examples~\ref{item.boundary.examples.1} and~\ref{item.boundary.examples.2} from Example~\ref{xmpl.boundary.examples}, as illustrated in Figure~\ref{fig.boundaries} (left-hand and middle diagram).

The point $\ast\in\ns P_1$ is on the boundary of $\intertwined{1}$ and it is 
unconflicted and not weakly regular.
The point $1\in\ns P_2$ is on the boundary of $\intertwined{0}$ and it is 
conflicted and weakly regular.
It follows from Proposition~\ref{prop.wr.pair} that $(\ast,1)$ is conflicted and not weakly regular.

By Proposition~\ref{prop.product.intertwined}(\ref{item.product.intertwined.2}) $\intertwined{(1,0)}=\intertwined{1}\times\intertwined{0}$, and by some routine topological calculation we see that $(\ast,1)$ is on the boundary of this set.
\end{proof}

\jamiesubsection{Minimal closed neighbourhoods, and a counterexample}

We continue the development of Subsection~\ref{subsect.componentwise.composition} and the example in Corollary~\ref{corr.conflicted.and.not.wr} with some slightly more technical results, leading up to another example. 

\begin{lemm}
\label{lemm.minimal.squares}
Suppose that:
\begin{itemize*}
\item
$(\ns P_1,\opens_1)$ and $(\ns P_2,\opens_2)$ are semitopologies. 
\item
$C$ is a square (Definition~\ref{defn.squares}) in $\ns P_1\times\ns P_2$. 
\end{itemize*}
Then 
\begin{itemize*}
\item
if $C$ is a minimal closed neighbourhood in $\ns P_1\times\ns P_2$, 
\item
then the sides of $C$, $C_1=\pi_1(C)$ and $C_2=\pi_2(C)$, are minimal closed neighbourhoods in $\ns P_1$ and $\ns P_2$ respectively.
\end{itemize*}
\end{lemm}
\begin{proof}
Suppose $C$ is a square minimal closed neighbourhood, and consider $C_1'\subseteq C_1$ a closed neighbourhood in $\ns P_1$.
We will show that $C_1'=C_1$ (the argument for $\ns P_2$ is no different).
Using Lemma~\ref{lemm.proj.cont}, $C_1'{\times}C_2$ is a closed neighbourhood in $\ns P_1\times\ns P_2$.
By routine sets calculations and minimality we have that $C_1'{\times}C_2 = C_1{\times}C_2$, and it follows that $C_1'=C_1$.
\end{proof}

\begin{corr}
\label{corr.neighbourhood.up}
Suppose that:
\begin{itemize*}
\item
$(\ns P_1,\opens_1)$ and $(\ns P_2,\opens_2)$ are semitopologies.
\item
$p_2\in\ns P_2$.
\item
$\intertwined{{p_2}}$ is not a minimal closed neighbourhood of $p_2$.
\end{itemize*}
Then for every $p_1\in\ns P_1$ and for every $C$ a minimal closed neighbourhood of $(p_1,p_2)$, we have that $\intertwined{(p_1,p_2)}\subsetneq C$. 
\end{corr}
\begin{proof}
By Proposition~\ref{prop.product.intertwined}(\ref{item.product.intertwined.2}) $\intertwined{(p_1,p_2)}=\intertwined{{p_1}}{\times}\intertwined{{p_2}}$ and by Proposition~\ref{prop.intertwined.as.closure}(\ref{intertwined.as.closure.closed}) $\intertwined{(p_1,p_2)}\subseteq C$.

If $C=\intertwined{(p_1,p_2)}=\intertwined{{p_1}}{\times}\intertwined{{p_2}}$ then by 
Lemma~\ref{lemm.minimal.squares} its side $\intertwined{{p_2}}$ is a minimal closed neighbourhood of $p_2$, but we assumed this is not the case.
Thus, $\intertwined{(p_1,p_2)}\subsetneq C$ as required.
\end{proof}

\begin{rmrk}
Recall that Proposition~\ref{prop.closure.intertwined}(\ref{item.closure.intertwined.1}) shows that $\closure{p}\subseteq\intertwined{p}$, and Example~\ref{xmpl.closure.101} shows that this inclusion may be strict by giving a semitopology in which $\closure{p}\subsetneq\intertwined{p}$.
Recall also that it follows from Proposition~\ref{prop.intertwined.as.closure}(\ref{intertwined.as.closure.closed}) that $\intertwined{p}\subseteq C$ for any $C$ a (minimal) closed neighbourhood of $p$, and Example~\ref{xmpl.not.intertwined} shows that this inclusion may be strict by giving a semitopology in which $\intertwined{p}\subsetneq C$ for $C$ a minimal closed neighbourhood of $p$.
What we have not done so far is show that both inclusions may be strict \emph{for a single $p$}: we can now apply what we have shown about the product semitopology in this Subsection, to `glue' our examples together: 
\end{rmrk}

\begin{thrm}
\label{thrm.nitpicked}
There exists a semitopology $(\ns P,\opens)$ and a $p\in\ns P$ and a minimal closed neighbourhood $C$ of $p$ such that the inclusions below are strict: 
$$
\closure{p} \subsetneq \intertwined{p} \subsetneq C .
$$ 
\end{thrm}
\begin{proof}
Let $(\ns P_1,\opens_1)$ be the semitopology from Example~\ref{xmpl.closure.101}, and $(\ns P_2,\opens_2)$ be that from Example~\ref{xmpl.not.intertwined}.
We set:
\begin{itemize*}
\item
$(\ns P,\opens)=(\ns P_1,\opens_1)\times(\ns P_2,\opens_2)$, the product semitopology.
\item
$p_1=1\in\ns P_1$, for which $\closure{1}\subsetneq\intertwined{1}=\{0,1\}$, and 
\item
$p_2=(0,0)$, which has a minimal closed neighbourhood $A=\{(0,0),(1,0)\}$ which is not equal to $\intertwined{{p_2}}=\intertwined{(0,0)}=\{(0,0)\}$, and
\item
$C=\{0,1\}{\times}\{(0,0), (1,0)\}$.
\end{itemize*}
We show that $\closure{(p_1,p_2)}\subsetneq\intertwined{(p_1,p_2)}$, as follows:
$$
\begin{array}{r@{\ }l@{\qquad}l}
\closure{(p_1,p_2)}
=&
\closure{p_1}{\times}\closure{p_2}
&\text{Lemma~\ref{lemm.point.closure.square}}
\\
\subsetneq&
\intertwined{{p_1}}{\times}\closure{p_2}
&\closure{1}\subsetneq \intertwined{1}
\\
\subseteq&
\intertwined{{p_1}}{\times}\intertwined{{p_2}}
&\text{Proposition~\ref{prop.closure.intertwined}(\ref{item.closure.intertwined.1})} .
\end{array}
$$
Furthermore, by Corollary~\ref{corr.neighbourhood.up} $\intertwined{(p_1,p_2)}\subsetneq C$, because $\intertwined{{p_2}}\subsetneq A$.
\qedhere
\end{proof}

\begin{figure}
\vspace{-1em}
\centering
\includegraphics[width=0.4\columnwidth]{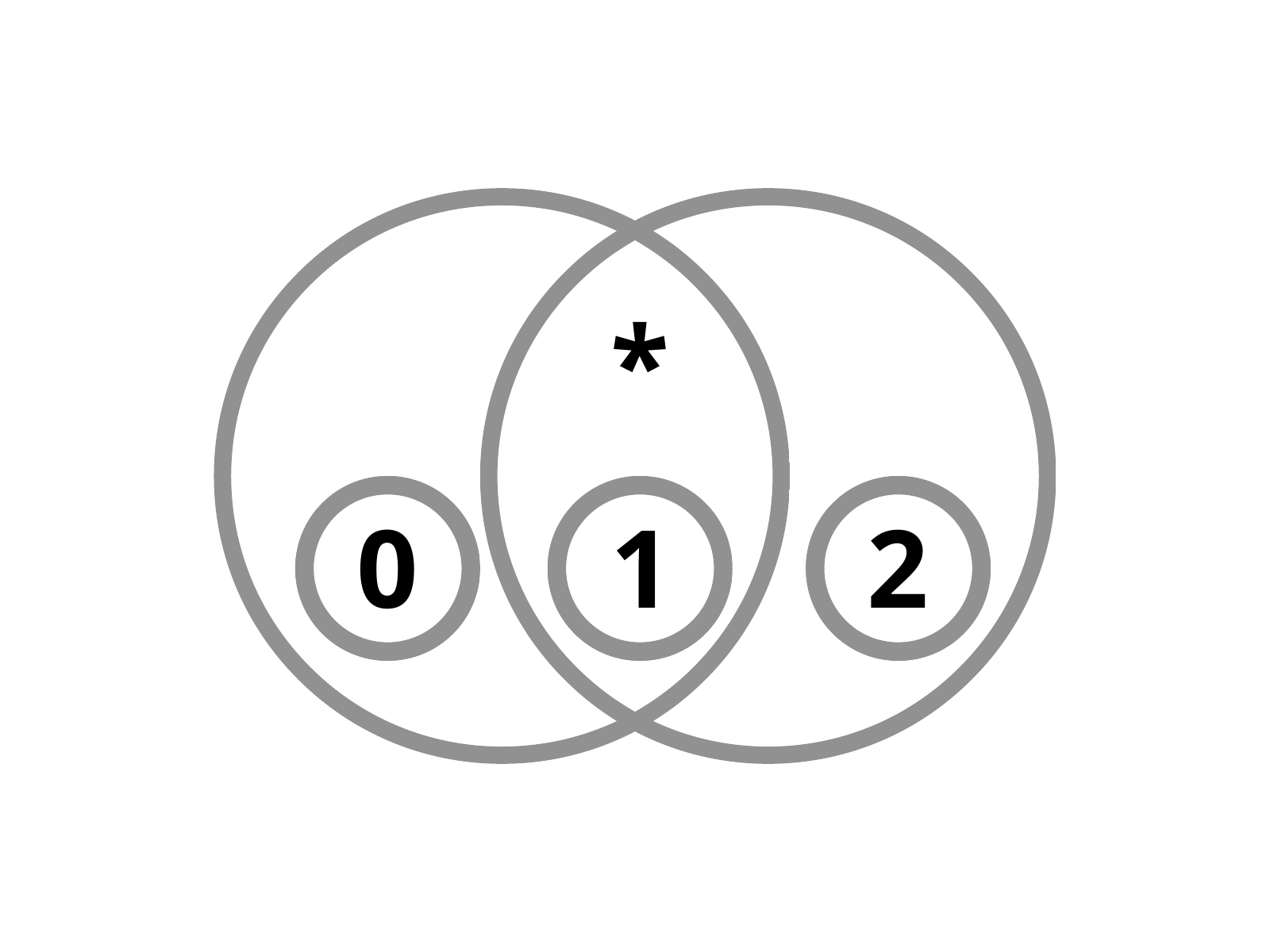}
\vspace{-1em}
\caption{Example~\ref{xmpl.nitpick}: $\closure{\ast}\subsetneq\intertwined{\ast}\subsetneq \{0,1,\ast\}$}
\label{fig.nitpick}
\end{figure}

\begin{xmpl}
\label{xmpl.nitpick}
We now give a smaller, but less compositional, example for Theorem~\ref{thrm.nitpicked}.
Set 
\begin{itemize*}
\item
$\ns P=\{0,1,2,\ast\}$ and 
\item
let $\opens$ be generated by $\{0\}$, $\{1\}$, $\{2\}$ (so $\{0,1,2\}$ has the discrete semitopology) and by $\{0,1,\ast\}$, and $\{1,2,\ast\}$,
\end{itemize*}
as illustrated in Figure~\ref{fig.nitpick} (we used this same example in Figure~\ref{fig.boundaries}, left-hand diagram, and it was also one of our first examples of topens in Figure~\ref{fig.012}).
Then: 
\begin{itemize*}
\item
$\closure{\ast}=\{\ast\}$, because $\{0,1,2\}$ is open.
\item
$\intertwined{\ast}=\{1,\ast\}$, since $\{1,2,\ast\}$ is disjoint from $\{0\}$ and $\{0,1,\ast\}$ is disjoint from $\{2\}$.
\item
$\{0,1,\ast\}$ and $\{1,2,\ast\}$ are distinct minimal closed neighbourhoods of $\ast$, with open interiors $\{0,1\}$ and $\{1,2\}$ respectively.
\end{itemize*}
\end{xmpl}

\jamiesection{The witnesses semitopology}
\label{sect.witness}

\jamiesubsection{Discussion}

\begin{rmrk}
\label{rmrk.local}
In this Section, we turn to the problem of computing with semitopologies.
We want two things from our maths: 
\begin{itemize*}
\item
that it will deliver algorithms; and also 
\item
that these algorithms should be \textbf{local}, by which we mean \emph{executable by points knowing only information near (local) to them, by communicating with local peers}.
\end{itemize*}
In particular, a local algorithm should not assume that points can globally synchronise or agree.\footnote{Indeed, to do this would be to assume a solution to the problem that semitopologies were created to study.}

We now note that our notion of `open neighbourhood of a point' from semitopologies is not \emph{a priori} particularly local.
The simplest illustration is perhaps to note that $(\ns P,\opens)=(\mathbb N,\{\varnothing,\mathbb N\})$ expresses that points coordinate on whether they all agree, but the lack of locality shows up in the mathematics in other, perhaps unexpected ways, because we can encode nontrivial information in the structure of open sets.
Consider the following example of a semitopology with (by design) poor algorithmic behaviour:
\end{rmrk}

\begin{xmpl}
\label{xmpl.uncomputable.semitopology}
Let the \deffont{uncomputable semitopology} have 
\begin{itemize*}
\item
$\ns P=\mathbb N$ and 
\item
open sets generated as unions of \emph{uncomputable subsets} of $\mathbb N$.
\end{itemize*}
(Call a subset $U\subseteq\mathbb N$ \emph{uncomputable} when there is no algorithm that inputs $n\in\mathbb N$ and returns `true' if $n\in U$ and `false' if $n\notin U$.)
This is a semitopology.
It is not a topology, since the intersection of two uncomputable subsets need not be uncomputable. 
By construction, no algorithm can compute its open sets.
\end{xmpl}

\begin{rmrk}
\label{rmrk.computational.content}
Example~\ref{xmpl.uncomputable.semitopology} just comes from the fact that the definition of semitopologies involves a subset of the (powerset of) $\mathbb N$.
This is not unusual, and the existence of such uncomputable subsets is well-known~\cite[Theorem~XVIII, page~360]{church:unspen}.

What we should do now is determine and study algorithmically tractable semitopologies.
So: what is an appropriate and useful definition?
 
In this Section will identify a class of algorithmically tractable semitopologies, and furthermore this in the strong sense that the definition is clean, makes a novel connection to declarative programming, and from it we extract distributed and local algorithms in the sense discussed above.
To do this, we will define witnessed sets (Definition~\ref{defn.witnessed.set}) and show that they determine computationally tractable semitopologies in a sense made formal by results including 
\begin{itemize*}
\item
Propositions~\ref{prop.open.limit} and~\ref{prop.lim.is.closure} (which show that algorithms exist to compute open and closed sets) and 
\item
the remarkable Theorem~\ref{thrm.lim.O.open} (which shows intuitively that witness semitopologies behave locally like finite sets, even if they are globally infinite). 
\end{itemize*}
\end{rmrk}

The impatient reader can jump to Remarks~\ref{rmrk.computing.open.sets} and~\ref{rmrk.computing.closed.sets}, where we describe these algorithms.
They are described at a high level, but what matters is that they exist, and
what is nice about them is that they correspond to natural (semi)topological operations.

\jamiesubsection{The witness function and semitopology}

\begin{nttn}
\label{nttn.finpow}
We extend Notation~\ref{nttn.powerset}.
Suppose $\ns P$ is a set.
\begin{enumerate*}
\item
Call a nonempty subset of $\ns P$ a \deffont{witness-set}, and write $\powerset_{\neq\varnothing}(\ns P)$ for the set of witness-sets (nonempty subsets) of $\ns P$.
\item
Write $\finpow(\ns P)$ for the finite powerset of $\ns P$ (the set of finite subsets of $\ns P$).
\item
Write $\finpow_{\neq\varnothing}(\ns P)$ for the finite powerset of $\ns P$ (the set of finite subsets of $\ns P$).
\item
Write 
$$
\mathcal W(\ns P)=\finpow_{\neq\varnothing}(\powerset_{\neq\varnothing}(\ns P)) 
$$ 
(finite sets of witness-sets of $\ns P$), and call $\mathcal W(\ns P)$ the \deffont{witnessing universe of $\ns P$}. 
\end{enumerate*}
\end{nttn}

\begin{defn}
\label{defn.witnessed.set}
Suppose $\ns P$ is a set. 
Then:
\begin{enumerate*}
\item\label{witness.function}
A \deffont[witness function $\witness:\ns P\to\mathcal W(\ns P)$]{witness function} on $\ns P$ is a function
$$
\witness:\ns P\to\mathcal W(\ns P)=\finpow_{\neq\varnothing}(\powerset_{\neq\varnothing}(\ns P)) . 
$$
Intuitively, a witness function assigns to each $p\in\ns P$ finitely many witness-sets. 
We call each $w\in\witness(p)$ a \deffont[witness-set $w\in\witness(p)$]{witness-set} for $p$.
\item\label{witness.witness}
A \deffont[witnessed set $(\ns P,\witness)$]{witnessed set} is a pair $(\ns P,\witness)$ of a set and a witness function on that set.
\end{enumerate*}
\end{defn}

\begin{rmrk}\leavevmode
\begin{enumerate}
\item
A witness function $\witness$ gives rise to a relation $\witness\subseteq\ns P\times\powerset_{\neq\varnothing}(\ns P)$ by taking 
$$
p\mathrel{\witness}w \quad\text{when}\quad
w\in\witness(p).
$$
\item
If we read the relation from right to left then for each $w\in\witness(p)$ we can read $w$ as an abstract notion of `potential set of witness for the beliefs of $p$'. 
\item
The nonemptiness conditions implies that every $p$ is witnessed by some nonempty $\varnothing\neq w\in\witness(p)$ --- even if $w$ is just equal to $\{p\}$. 
\end{enumerate} 
\end{rmrk}

\begin{defn}
\label{defn.blocking.set}
Suppose $\witness:\ns P\to\mathcal W(\ns P)$ is a witness function on a finite set $\ns P$ (Definition~\ref{defn.witnessed.set}), and suppose $p\in\ns P$ and $P\subseteq\ns P$. 
\begin{enumerate*}
\item\label{item.p.blocks.P}
Define $p\blocks{\witness} P$, or synonymously $P\blocks{\witness} p$, by: 
$$
p\blocks{\witness} P \quad\text{when}\quad \Forall{w{\in}\witness(p)} w\between P
$$
and say that $P$ \deffont[blocks ($P$ blocks $p$)]{blocks} $p$, and call $P$ a \deffont{blocking set for $p$}.
 
In words: $P$ blocks $p$ when it intersects with all of $p$'s witness-sets.
\item\label{item.blocking.enables}
Define $p\enabledby{\witness} P$, or synonymously $P\enables{\witness} p$, by 
$$
p\enabledby{\witness} P \quad\text{when}\quad \Exists{w{\in}\witness(p)} w\subseteq P 
$$
and say that $P$ \deffont[enables ($P$ enables $p$)]{enables} $p$, and call $P$ an \deffont{enabling set for $p$}.
 
In words: $P$ enables $p$ when it contains at least one of $p$'s witness-sets.
\end{enumerate*}
\end{defn}

\begin{defn}
\label{defn.trust.topology}
Suppose $\witness:\ns P\to\mathcal W(\ns P)$ is a witness function on $\ns P$ (Definition~\ref{defn.witnessed.set}).
\begin{enumerate*}
\item\label{item.w.witnessed}
Call $O\subseteq\ns P$ a \deffont{($\witness$-)open set} when
$$
\Forall{p\in\ns P} p\in O \limp p\enabledby{\witness} O
$$
In words, $O$ is open when it enables its own elements.\footnote{Note that if $p\in O$ then $O$ need not contain \emph{every} enabling witness-set of $p$.  In Definition~\ref{defn.blocking.set}(\ref{item.blocking.enables}) $p\enabledby{\witness} O$ is existential, that $O$ contains \emph{some} witness-set of $p$.}
\item\label{item.w.blocking}
Call $C\subseteq\ns P$ a \deffont{($\witness$-)closed set} when
$$
\Forall{p\in\ns P} p\blocks{\witness} C \limp p\in C .
$$
In words, $C$ is closed when it contains every element that it blocks.
\item\label{item.witness.semitopology}
Let the \deffont{witness semitopology $\opens(\witness)$}\index{$\opens(\witness)$} on $\ns P$ be the set of $\witness$-open sets.
In symbols:
$$
\begin{array}{r@{\ }l}
\opens(\witness) 
=& \{O\subseteq\ns P \mid \text{$O$ is $\witness$-open} \} 
\\
=& \{ O\subseteq \ns P \mid \Forall{p\in \ns P} p\in O \limp p\enabledby{\witness} O\}
\\
=& \{ O\subseteq \ns P \mid \Forall{p\in O} \Exists{w{\in}\witness(p)}w\subseteq O \}.
\end{array}
$$ 
We also define \deffont{$\closed(\witness)$} by:
$$
\begin{array}{r@{\ }l}
\closed(\witness) 
=& \{C\subseteq\ns P \mid \text{$C$ is $\witness$-closed} \} 
\\
=& \{ C\subseteq \ns P \mid \Forall{p\in \ns P} p\blocks{\witness} P \limp p\in C\} .
\end{array}
$$ 
\end{enumerate*}
\end{defn}

By Lemma~\ref{lemm.closed.complement.open}, being open and being closed are dual.
We make the elementary observation that $\enabledby{\witness}$ and $\blocks{\witness}$, and $\opens(\witness)$ and $\closed(\witness)$, 
do indeed match up as they should:
\begin{lemm}
\label{lemm.enabled.iff.complement.blocked}
Suppose $\witness:\ns P\to\mathcal W(\ns P)$ is a witness function on $\ns P$ and suppose $p\in\ns P$ and $P\subseteq\ns P$. 
Then 
\begin{enumerate*}
\item
$p\blocks{\witness} P$ if and only if $p\notenable{\witness} \ns P{\setminus} P$.
\item
$p\enabledby{\witness} P$ if and only if $p\nblocks{\witness} \ns P{\setminus} P$.
\item
$P\in\opens(\witness)$ if and only if $\ns P\setminus P\in\closed(\witness)$, 
and 
\\
$P\in\closed(\witness)$ if and only if $\ns P\setminus P\in\opens(\witness)$. 
\end{enumerate*}
\end{lemm}
\begin{proof}
By routine calculations from Definition~\ref{defn.trust.topology}.
\end{proof}

\begin{lemm}
\label{lemm.union.of.witness.opens.is.open}
Suppose $\witness:\ns P\to\mathcal W(\ns P)$ is a witness function on a finite set $\ns P$, and suppose $\mathcal O\subseteq\powerset(\ns P)$.
Then if every $O\in\mathcal O$ is open in the sense of Definition~\ref{defn.blocking.set}(\ref{item.blocking.enables}), then $\bigcup\mathcal O$ is also open.
\end{lemm}
\begin{proof}
Suppose $p\in\bigcup\mathcal O$.
Then $p\in O$ for some $O\in\mathcal O$.
By openness of $O$, $p$ is enabled by some $w\in\witness(p)$ such that $w\subseteq O$.
But then also $w\subseteq\bigcup\mathcal O$, so we are done.
\end{proof}

\begin{corr}
\label{corr.witness.semitopology.is.semitopology}
Suppose $\witness:\ns P\to\mathcal W(\ns P)$ is a witness function on a finite set $\ns P$.
Then $\opens(\witness)$ from Definition~\ref{defn.trust.topology} makes $\ns P$ into a semitopology in the sense of Definition~\ref{defn.semitopology}.
\end{corr}
\begin{proof}
Unpacking conditions~\ref{semitopology.empty.and.universe} and~\ref{semitopology.unions} of Definition~\ref{defn.semitopology}, we must check that $\varnothing$ and $\ns P$ are open --- which is routine --- and that an arbitrary union of open sets is open --- which is Lemma~\ref{lemm.union.of.witness.opens.is.open}.
So we are done.
\end{proof}

\begin{rmrk}
There is design freedom, whether we want to include (or exclude) $p\in w\in\witness(p)$: Definition~\ref{defn.witnessed.set} makes no commitment either way.
Lemma~\ref{lemm.might.as.well} is an easy observation that expresses a precise mathematical sense in which this choice does not really matter; so we can choose whatever is most convenient for a particular case.
We will use Lemma~\ref{lemm.might.as.well} later, to prove Lemma~\ref{lemm.more-than-one}.
\end{rmrk}

\begin{lemm}
\label{lemm.might.as.well}
Suppose $(\ns P,\witness)$ is a witnessed set. 
Let $\witness'$ and $\witness''$ be defined by\footnote{The case-split in $\witness''$ is required just because witness function in Definition~\ref{defn.witnessed.set}(\ref{witness.function}) must return a finite set of \emph{nonempty} sets.}
$$
\begin{array}{r@{\ }l}
\witness'(p)=&\{w\cup\{p\} \mid w\in\witness(p)\} 
\\
\witness''(p)=&\{w\setminus\{p\} \mid w\in\witness(p)\land w\neq\{p\}\}\cup\{w \mid w\in\witness(p)\land w=\{p\}\} 
\end{array}
$$
Then $(\ns P,\witness')$ and $(\ns P,\witness'')$ are also witnessed sets, and they generate the same witness semitopology as does $(\ns P,\witness)$.
\end{lemm}
\begin{proof}
By a routine calculation.
\end{proof}

\jamiesubsection{Examples} 

\begin{rmrk}
\leavevmode
\begin{itemize*}
\item
Sometimes, proving the existence of a witness function $\witness$ to generate a given semitopology $(\ns P,\opens)$ as a witness semitopology (Definition~\ref{defn.trust.topology}) is fairly straightforward.
Lemma~\ref{lemm.finite.witness} gives a natural example of this.
\item
Sometimes, the existence of a witness function is less evident.
Lemma~\ref{lemm.all-but-one} illustrates one example of a non-obvious witness function for a semitopology, and Lemma~\ref{lemm.more-than-one} conversely illustrates an apparently not dissimilar semitopology, but for which no witness function exists.
\end{itemize*}
\end{rmrk}

\begin{lemm}
\label{lemm.finite.witness}
Suppose $(\ns P,\opens)$ is a \emph{finite} semitopology (so $\ns P$ is finite, and so is $\opens$). 
Then $(\ns P,\opens)$ can be generated as witness semitopology.
Thus: every finite semitopology is also a witness semitopology for a witnessed set.\footnote{The reader might consider Lemma~\ref{lemm.finite.witness} to be a satisfactory answer to the open problem we describe later in Remark~\ref{rmrk.two.open.problems}, since all semitopologies realisable in the real world are finite.
We are not so sure --- even if all you care about is physically realisable semitopologies --- for reasons outlined in Remark~\ref{rmrk.why.infinite}.}  
\end{lemm}
\begin{proof}
Set $\witness(p)=\{O\in\opens \mid p\in O\}$.
The reader can check that this satisfies the finiteness conditions on a witness function in Definition~\ref{defn.witnessed.set}; it remains to show that $\opens(\witness)=\opens$.
If $X\in\opens(\witness)$ then by Definition~\ref{defn.trust.topology}(\ref{item.w.witnessed}) $X$ is a union of open sets, and thus $X\in\opens$. 
Conversely, if $O\in\opens$ then $O\in\opens(\witness)$ because each $p\in O$ is witnessed by $O$.
\end{proof}

\begin{lemm}
\label{lemm.all-but-one}
Consider the \emph{all-but-one semitopology} on $\mathbb Z$ from Example~\ref{xmpl.semitopologies}(\ref{item.counterexample.X-x}): 
\begin{itemize*}
\item
$\ns P=\mathbb Z$ and 
\item
$\opens = \{ \varnothing,\ \mathbb Z\} \cup \{\mathbb Z{\setminus}\{i\}\mid i\in\mathbb Z\}$.
\end{itemize*}
Then a witness function for this semitopology is: 
$$
\witness(i)=\{ \{i\minus 1, i\plus 1\},\ \mathbb Z\setminus\{i\plus 1\},\ \mathbb Z\setminus\{i\minus 1\} \} 
$$
\end{lemm}
\begin{proof}
We prove that $\opens=\opens(\witness)$ by checking two subset inclusions.
\begin{itemize}
\item
\emph{We check that if $O\in\opens$ then $O\in\opens(\witness)$:}

If $O=\varnothing$ or $O=\mathbb Z$ then there is nothing to prove.
So suppose $O = \mathbb Z\setminus \{i\}$.

We must show that every $j\in O$ is witnessed by some element $w_j\in\witness(j)$.
This is routine:
\begin{itemize*}
\item
For $j\not\oldin\{i\minus 1,i\plus 1\}$ we use witness-set $\{j\minus 1,j\plus 1\}$; 
\item
for $j=i\minus 1$ we use witness-set $\mathbb Z\setminus \{j\plus 1\}$; and 
\item
for $j=i\plus 1$ use witness-set $\mathbb Z\setminus \{j\minus 1\}$.
\end{itemize*}
\item
\emph{We check that if $O\in\opens(\witness)$ then $O\in\opens$.}

If $O=\varnothing$ or $O = \mathbb Z$ then there is nothing to prove.
So suppose $O\not\oldin\{\varnothing,\mathbb Z\}$.

Then there exists an $i\in\mathbb Z$ such that $i\in O$ and $\{i\minus 1,i\plus 1\}\not\subseteq O$.
We assumed $O\in\opens(\witness)$, so one of the following must hold:
\begin{itemize*}
\item
$\{i\minus 1,i\plus 1\}\subseteq O$, which we assumed is not the case, or
\item
$i\plus 1\not\oldin O$ and $\mathbb Z\setminus\{i\plus 1\}\subseteq O$, so we are done because, with $O\neq \mathbb Z$, it must be that $O=\mathbb Z\setminus\{i\plus 1\}$, or
\item
$i\minus 1\not\oldin O$ and $\mathbb Z\setminus\{i\minus 1\}\subseteq O$, and again we are done.
\qedhere\end{itemize*}
\end{itemize}
\end{proof}

\begin{lemm}
\label{lemm.w.cwr}
A witness function for the semitopology used in Proposition~\ref{prop.conflicted.weakly.regular}, as illustrated in Figure~\ref{fig.weakly-regular.conflicted}, is
$$
\witness(w)=\{\{w0,w1\}\}.
$$
\end{lemm}
\begin{proof}
Setting $\witness(w)=\{\{w0,w1\}\}$ just expresses that if $w\in O$ then $w0,w1\in O$, i.e. that $O$ is down-closed --- for `down' as illustrated in Figure~\ref{fig.weakly-regular.conflicted}.
\end{proof}

Lemma~\ref{lemm.more-than-one} will provide a key counterexample later in Lemma~\ref{lemm.elaborating}:
\begin{lemm}
\label{lemm.more-than-one}
Consider the \emph{more-than-one semitopology} on $\mathbb N$ from Example~\ref{xmpl.semitopologies}(\ref{item.counterexample.more-than-one}): so $X=\mathbb N$ and opens have the form $\varnothing$ or any set of cardinality more than one (i.e. containing at least two elements).
There is no witness function for this semitopology.
\end{lemm}
\begin{proof}
Suppose some such witness function $\witness$ exists.
Using Lemma~\ref{lemm.might.as.well} we may assume without loss of generality that $n\in w$ for every $w\in\witness(n)$, for every $n\in\mathbb N$ (that is, $n\in\bigcap\witness(n)$ always).
Furthermore because no singletons are open, we know that $\{n\}\notin\witness(n)$ for every $n\in\mathbb N$.

Now consider two distinct $n\neq n'\oldin\mathbb N$.
We know that $\{n,n'\}$ is open, so it follows that one of the following must hold: 
\begin{enumerate*}
\item
\emph{Suppose $\{n,n'\}\in\witness(n)$ and $\{n,n'\}\notin\witness(n')$.}\quad

This is impossible because $\{n'\}\notin\witness(n')$ and $\witness(n')$ is not empty, so $\{n,n'\}$ could not be open. 
\item
\emph{Suppose $\{n,n'\}\in\witness(n')$ and $\{n,n'\}\notin\witness(n)$.}\quad

This is also impossible because $\{n\}\notin\witness(n)$ and $\witness(n)$ is not empty, so $\{n,n'\}$ could not be open. 
\item
It follows that $\{n,n'\}\in\witness(n)$ and $\{n,n'\}\in\witness(n')$.
\end{enumerate*}
It follows that $\{n,n'\}\in\witness(n)$ for \emph{every} $n'$ other than $n$.
But this contradicts finiteness of $\witness(n)$.
\end{proof}

\jamiesubsection{Computing open and closed sets in witness semitopologies}

\jamiesubsubsection{Computing open sets: $X$ is open when $X\witno X$}

\begin{defn}
\label{defn.prec}
Suppose that $(\ns P,\witness)$ is a witnessed set (Definition~\ref{defn.witnessed.set}) and $X,X'\subseteq\ns P$.
Define the \deffont{witness ordering $X\witno X'$} by 
$$
X\witno X'
\quad\text{when}\quad
X\subseteq X' 
\ \land\  
\Forall{p{\in}X}\Exists{w{\in}\witness(p)}w\subseteq X' .
$$ 
If $X\witno X$ then call $X$ a \deffont{$\witno$-fixedpoint}. 
\end{defn}

\begin{rmrk}
Intuitively, $X\witno X'$ when $X'$ extends $X$ with (at least) one witness-set for every element $p\in X$.
\end{rmrk} 

\begin{lemm}
Suppose $(\ns P,\witness)$ is a witnessed set, and recall the witness ordering $\witno$ from Definition~\ref{defn.prec}.
Then:
\begin{enumerate*}
\item
If $X\witno X'$ then $X\subseteq X'$, or in symbols: ${\witno}\subseteq{\subseteq}$.
\item
$\witno$ is a transitive ($X\witno X'\witno X''$ implies $X\witno X''$) and antisymmetric ($X\witno X'$ and $X'\witno X$ implies $X=X'$) relation on $\powerset(\ns P)$. 
\end{enumerate*}
\end{lemm}
\begin{proof}
By routine calculations from Definition~\ref{defn.prec}.
\end{proof}

\begin{lemm}
\label{lemm.char.prec.open}
\label{lemm.obvious.open}
Suppose $(\ns P,\witness)$ is a witnessed set. 
Then the following are equivalent:
\begin{itemize*}
\item
$O$ is open in the witness semitopology (Definition~\ref{defn.trust.topology}).
\item
$O$ is a $\witno$-fixedpoint (Definition~\ref{defn.prec}).
\end{itemize*}
In symbols:
$$
\opens(\witness) = \{ X\subseteq\ns P \mid X\witno X\} .
$$
\end{lemm}
\begin{proof}
Being a $\witno$-fixedpoint in Definition~\ref{defn.prec} --- every point in $O$ is witnessed by a subset of $O$ --- simply reformulates the openness condition from Definition~\ref{defn.trust.topology}.
\end{proof}

\begin{prop}
\label{prop.open.limit}
\label{prop.prec.upchain}
Suppose $(\ns P,\witness)$ is a witnessed set and suppose $\mathcal X=(X_0\witno X_1\witno\dots)$ is a countably ascending $\witno$-chain.
Write $\bigcup \mathcal X$ for the union $\bigcup_i X_i$ of the elements in $\mathcal X$.
Then:
\begin{enumerate*}
\item\label{item.open.limit.1}
$\bigcup\mathcal X$ is a $\witno$-limit for $\mathcal X$:\ \ $\Forall{i}X_i\witno\bigcup\mathcal X$.
\item\label{item.open.limit.2}
$\bigcup\mathcal X$ is a $\witno$-fixedpoint and so (by Lemma~\ref{lemm.char.prec.open}) is open: $\bigcup\mathcal X\witno\bigcup\mathcal X\in\opens(\witness)$.
\end{enumerate*} 
\end{prop}
\begin{proof}
\leavevmode
\begin{enumerate}
\item
We must show that if $p\in X_i$ then $w\subseteq\bigcup\mathcal X$ for some $w\in\witness(p)$.
But this is automatic from the fact that $X_i\witno X_{i\plus 1}\subseteq\bigcup\mathcal X$.
\item
From part~\ref{item.open.limit.1} noting that if $p\in\bigcup\mathcal X$ then $p\in X_i$ for some $i$.
\qedhere\end{enumerate}
\end{proof}

\begin{rmrk}[Computing open sets]
\label{rmrk.computing.open.sets}
Proposition~\ref{prop.open.limit} and Lemma~\ref{lemm.obvious.open} above are not complicated\footnote{This is a feature and did not happen by accident: it required design effort.} 
and they say something important: in the \emph{witness} semitopology, open sets can be computed with a simple loop that accumulates a set of points; and for each point in the set so far, add some choice of witness-set of that point to the set (if one is not already present); repeat until we reach a fixed point; then return the result.

In more detail, to compute an open set in the witness semitopology: 
\begin{enumerate*}
\item
Nondeterministically choose an initial $R_0$ --- in particular, to compute an open neighbourhood of $p\in\ns P$ we can set $R_0=\{p\}$.
\item
Given $R_i$, for each $p\in R_i$ nondeterministically pick some witness-set $w(p)\in\witness(p)$ and set $R_{i\plus 1}=R_i\cup\bigcup_{p\in R_i}w(p)$.
\item
If $R_{i\plus 1}=R_i$ then terminate with result $R_i$; otherwise loop back to~2.
\end{enumerate*}
This algorithm is nondeterministic and could run forever if $\ns P$ is infinite, but it is an algorithm and it is local in the sense of Remark~\ref{rmrk.local}.
We continue this thread in Remarks~\ref{rmrk.computing.closed.sets} and~\ref{rmrk.algorithms}. 
\end{rmrk}

\jamiesubsubsection{Computing closed sets using limit points: $\closure{P}=\f{lim}(P)$}

\begin{defn}
\label{defn.O.between.R}
Suppose $P$ is a set and $\mathcal W$ is a set (or a sequence) of sets.
Define $P\between \mathcal W$ by
$$
P\between\mathcal W
\quad\text{when}\quad
\Forall{W{\in}\mathcal W}P\between W .
$$
In words: $P\between \mathcal W$ when $P$ intersects with every $W\in\mathcal W$.
\end{defn}

\begin{defn}
\label{defn.admires}
Suppose $(\ns P,\witness)$ is a witnessed set and $P\subseteq\ns P$.
Define $\f{lim}_w(P)$ by
$$
\f{lim}_w(P) = P \cup \{p\in\ns P \mid P\between\witness(p) \} .
$$ 
In words: $\f{lim}_w(P)$ is the set of points $p$ whose every witness-set contains a $P$-element.

We iterate this:
$$
\begin{array}{r@{\ }l@{\quad}l}
\f{lim}_0(P) =& P
\\
\f{lim}_{i\plus 1}(P)=&\f{lim}_w(\f{lim}_i(P))
\\
\f{lim}(P)=&\bigcup_{n\geq 0}\f{lim}_n(P)
\end{array}
$$
We call $\f{lim}(P)$ the set of \deffont[limit points of $P$ ($\f{lim}(P)$)]{limit points of $P$}.
\end{defn}

\begin{rmrk}
In Definition~\ref{defn.witnessed.set}(\ref{witness.function}) we insisted that $\witness(p)$ is nonempty for every point $p$.
This avoids a degenerate situation in the definition of $\f{lim}_w(P)$ in Definition~\ref{defn.admires} above in which the condition $P\between \witness(p)$ is vacuously satisfied by a $p$ with empty $\witness(p)$ (i.e. by a $p$ with no witness sets).
Definition~\ref{defn.witnessed.set}(\ref{witness.function}) excludes this by insisting that $p$ has to have at least one witness, even if it is just $\witness(p)=\{\{p\}\}$.
\end{rmrk}

\begin{lemm}
\label{lemm.R.subset.lim.R}
Suppose $(\ns P,\witness)$ is a witnessed set and $P\subseteq\ns P$.
Then 
$$
P\subseteq\f{lim}(P).
$$
\end{lemm}
\begin{proof}
It is a fact of Definition~\ref{defn.admires} that $P=\f{lim}_0(P)\subseteq\f{lim}_1(P)\subseteq\f{lim}(P)$. 
\end{proof}

\begin{lemm}
\label{lemm.limit.is.open}
Suppose $(\ns P,\witness)$ is a witnessed set and $p\in\ns P$ and $P\subseteq\ns P$.
Then:
\begin{enumerate*}
\item\label{item.limit.is.open.1}
If $\f{lim}(P)\between\witness(p)$ (Definition~\ref{defn.O.between.R}) then $p\in \f{lim}(P)$.
\item\label{item.limit.is.open.2}
By the contrapositive and expanding Definition~\ref{defn.O.between.R},
$$
p\in \ns P\setminus\f{lim}(P)
\quad\text{implies}\quad
\Exists{w{\in}\witness(p)}w\cap \f{lim}(P)=\varnothing. 
$$
\end{enumerate*}
\end{lemm}
\begin{proof}
Suppose $\f{lim}(P)\between\witness(p)$.
Unpacking Definitions~\ref{defn.O.between.R} and~\ref{defn.admires} it follows that for every $w{\in}\witness(p)$ there exists $n_w\geq 0$ such that $\f{lim}_{n_w}(P)\between w$.
Now by Definition~\ref{defn.witnessed.set}(\ref{witness.function}) $\witness(p)$ --- the set of witness-sets to $p$ --- is finite, and 
it follows that for some/any $n$ greater than the maximum of all the $n_w$, we have $\f{lim}_n(P)\between\witness(p)$.
Thus $p\in\f{lim}_w(\f{lim}_n(P))\subseteq\f{lim}(P)$ as required.
\end{proof}

\begin{lemm}
\label{lemm.O.between.lim.R}
Suppose $(\ns P,\witness)$ is a witnessed set and $p\in\ns P$ and $P\subseteq\ns P$ and $O\in\opens(\witness)$.
Then:
\begin{enumerate*}
\item\label{item.O.between.lim.R.1}
If $O\between \f{lim}_w(P)$ then $O\between P$.
\item\label{item.O.between.lim.R.2}
If $O\between \f{lim}(P)$ then $O\between P$.
\item\label{item.O.between.lim.R.3}
As a corollary, if $O\cap P=\varnothing$ then $O\cap\f{lim}(P)=\varnothing$.
\end{enumerate*}
\end{lemm}
\begin{proof}
\leavevmode
\begin{enumerate}
\item
Consider $p\in\ns P$ such that $p\in O$ and $p\in\f{lim}_w(P)$.
By assumption there exists $w\in\witness(p)$ such that $w\subseteq O$.
Also by assumption $w\between P$.
It follows that $O\between P$ as required.
\item
If $O\between\f{lim}(P)$ then $O\between\f{lim}_n(P)$ for some finite $n\geq 0$.
By a routine induction using part~\ref{item.O.between.lim.R.1} of this result, it follows that $O\between P$ as required.
\item
This is just the contrapositive of part~\ref{item.O.between.lim.R.2} of this result, noting that $O\between P$ when $O\cap P=\varnothing$ by Notation~\ref{nttn.between}, and similarly for $O\between\f{lim}(P)$.
\qedhere\end{enumerate}
\end{proof}

\begin{prop}
\label{prop.lim.is.closure}
Suppose $(\ns P,\witness)$ is a witnessed set and suppose $P\subseteq\ns P$.
Then:
$$
\f{lim}(P) = \closure{P} .
$$
In words: the set of limit points of $P$ from Definition~\ref{defn.admires} is equal to the topological closure of $P$ from Definition~\ref{defn.closure}.
\end{prop}
\begin{proof}
We prove two implications:
\begin{itemize}
\item
\emph{Suppose $p\notin\closure{P}$.}\quad

Then there exists some $p\in O\in\opens(\witness)$ such that $O\cap P=\varnothing$.
Thus by Lemma~\ref{lemm.O.between.lim.R}(\ref{item.O.between.lim.R.3}) also $O\cap\f{lim}(P)=\varnothing$. 
\item
\emph{Suppose $p\notin \f{lim}(P)$.}\quad

By Definition~\ref{defn.closure} we need to exhibit an $p\in O\in\opens(\witness)$ that is disjoint from $P$, and since $P\subseteq\f{lim}(P)$ by Lemma~\ref{lemm.R.subset.lim.R}, 
it would suffice to exhibit $p\in O\in\opens(\witness)$ that is disjoint from $\f{lim}(P)$.
We set 
$$
O=\ns P\setminus \f{lim}(P).
$$
Lemma~\ref{lemm.limit.is.open}(\ref{item.limit.is.open.2}) expresses precisely that this is an open set in the witness semitopology, and by construction it is disjoint from $\f{lim}(P)$.
\qedhere\end{itemize}
\end{proof}

\begin{rmrk}[Computing closed sets]
\label{rmrk.computing.closed.sets}
As in Remark~\ref{rmrk.computing.open.sets} we see that in the \emph{witness} semitopology, closed sets can be computed with a simple loop that accumulates a set of points so far: and for each point in the space, if all of its witness-sets intersect with the set of points so far, add that point to the set so far; repeat until we reach a fixed point; return the result. 

In more detail, to compute a closed set in the witness semitopology: 
\begin{enumerate*}
\item
Nondeterministically choose an initial $P_0$ --- in particular, to compute a closed set containing $p\in\ns P$ we can set $P_0=\{p\}$.
\item
Given $P_i$, for every $p\in \ns P$ check if $w\between P_i$ for every witness-set $w\in\witness(p)$ and collect these $p$ into a set $B_i$.
Set $P_{i\plus 1}=P_i\cup B_i$.
\item
If $P_{i\plus 1}=P_i$ then terminate with result $P_i$; otherwise loop back to~2.
\end{enumerate*}

This algorithm could run forever if $\ns P$ is infinite, but it is an algorithm and it is local in the sense of Remark~\ref{rmrk.local}.
Note that quantification over every point is local in the sense of Remark~\ref{rmrk.local}, in spite of the quantification over all $p\in\ns P$ in step~2 above: participants would listen for queries from peers on the channel ``I am trying to compute an open set; here is my $R_i$; do you want to join it?''.
\end{rmrk}

\begin{rmrk}[Summing up]
\label{rmrk.distributed.witnesses}
It might at first appear that working with semitopologies would require some form of prior coordination: e.g.\ for participants to at least have common knowledge of their shared, minimal open neighbourhoods.
For, consider a new participant $p$ joining a system based on semitopology: how is $p$ supposed to know which are the open sets?

Surprisingly, we have seen that \emph{witness semitopologies can be built without any coordination}.
Each participant just unilaterally chooses a set of witness-sets.
As discussed in Remarks~\ref{rmrk.computing.open.sets} and \ref{rmrk.computing.closed.sets}, and even in an infinite semitopology, 
a participant can compute open and closed sets --- they do not have to, but they can if they wish to spend the bandwidth --- by exploring witness-sets using nondeterministic algorithms. 

We make no claims to efficiency (we have not even set up machinery to measure what that would mean) but what matters is that for witness semitopologies such procedures exist, in contrast e.g.\,to the uncomputable semitopology from Example~\ref{xmpl.uncomputable.semitopology}.

In the next subsection we offer an interpretation of witness functions that in some sense explains why this should be so,
and gives a new intuition of why witness semitopologies are amenable to decentralised computation in the style that we require.
\end{rmrk}

\jamiesubsection{Declarative content of witness semitopologies}
\label{subsect.declarative.witness}

\jamiesubsubsection{Witnessed sets and Horn clause theories}

\begin{rmrk} 
\label{rmrk.algorithms}
Recall that a \emph{sequential space} is one in which the sets closed under convergent sequences, are precisely the closed sets.
Proposition~\ref{prop.lim.is.closure} ($\f{lim}(P) = \closure{P}$) looks, just a bit, like a sequential space closure result.
Looking more closely, we see that the similarity comes from the fact that the definition uses an $\omega$-iteration that is, just a little, reminiscent of a converging $\omega$-sequence of points. 
Perhaps surprisingly, we can make this resemblance into something much more precise, as follows: 
\end{rmrk}

\begin{defn}
\label{defn.logic}
Suppose $(\ns P,\witness)$ is a finite witnessed set (so $\ns P$ is a finite set).
\begin{enumerate}
\item
Let the \deffont{derived logic $\tf{Prop}(\ns P,\witness)$} be a propositional syntax with connectives $\tbot$, $\ttop$, $\tor$, $\tand$, and $\timp$ over a set of \deffont[atomic proposition symbols $\bar{\ns P}$]{atomic proposition symbols} $\bar{\ns P}=\{\bar p\mid p\in\ns P\}$.

Note that $\bar p$ is just a symbol in our formal syntax; there is one such for each point $p\in\ns P$.
\item\label{item.logic.axiom}
For each $p\in\ns P$ define an \deffont{axiom $\barwitness(p)$} by\footnote{Below, $\tand$ and $\tor$ denote a finite list of $\tand$ and $\tor$ connectives.  We use this instead of $\bigwedge$ and $\bigvee$ to emphasise that this is formal syntax in $\tf{Prop}(\ns P,\witness)$.}
$$
\barwitness(p) = \bigl( \tand_{w\in\witness(p)} \tor_{q\in w} \bar q\bigr) \timp \bar p 
$$
and collect these axioms into a set
$$
\tf{Ax}(\ns P,\witness) = \{\barwitness(p)\mid p\in\ns P\}.
$$
\item
A \deffont{sequent $\Phi\barcent\Psi$}\index{$\Phi\barcent\Psi$ (sequent)} is a pair of finite sets of propositions in the syntax of $\tf{Prop}(\ns P,\witness)$.
\item
Call $\Phi\barcent\Psi$ a \deffont{derivable sequent} when $\Phi,\tf{Ax}(\ns P,\witness)\cent\Psi$ is derivable in propositional logic.
\item\label{item.S.deductively.closed}
If $S\subseteq\ns P$ write $\bar S=\{\bar p\mid p\in S\}$.
Call $\bar S$ a \deffont[model (for $\tf{Ax}(\ns P,\witness)$)]{model} or \deffont[answer set (for $\tf{Ax}(\ns P,\witness)$)]{answer set} for $\tf{Ax}(\ns P,\witness)$ when 
$$
\Forall{p{\in}\ns P}(\bar S\barcent \bar p \limp \bar p\in\bar S).
$$
\end{enumerate}
\end{defn}

\begin{prop}[Declarative interpretation]
\label{prop.declarative.interpretation}
Suppose $(\ns P,\witness)$ is a finite witnessed set and $C\subseteq\ns P$.
Then the following are equivalent: 
\begin{itemize*}
\item
$C$ is closed in the witness semitopology (Definition~\ref{defn.trust.topology}).
\item
$\bar C$ is a model (Definition~\ref{defn.logic}(\ref{item.S.deductively.closed})).
\end{itemize*}
\end{prop}
\begin{proof}
By Definitions~\ref{defn.trust.topology}(\ref{item.w.blocking}) and~\ref{defn.logic}(\ref{item.logic.axiom}),
the condition in Definition~\ref{defn.logic}(\ref{item.S.deductively.closed}) for $\bar C$ to be a model precisely expresses the property that $C$ is closed.
\end{proof}

\begin{corr}
\label{corr.finite.witness.horn}
Every finite semitopology can be exhibited as the set of (set complements of) models of a propositional Horn clause theory.
\end{corr}
\begin{proof}
Lemma~\ref{lemm.finite.witness} shows how to exhibit a finite semitopology as the witness semitopology of a witnessed set, and Proposition~\ref{prop.declarative.interpretation} shows how to interpret that witnessed set as a Horn clause theory in a propositional logic.
\end{proof}

\begin{rmrk}
\label{rmrk.horn}
An axiom $\barwitness(p)$ consists of a propositional goal implied by a conjunction of disjunctions of (unnegated) propositional goals.  
This fits the Horn clause syntax from Section~3 of~\cite{miller:unipfl}, and it can be translated into a more restricted Prolog-like syntax if required, just by expanding the disjuncts into multiple clauses using the \rulefont{\tor L} rule.\footnote{An example makes the point: $((\bar q\tor\bar q')\tand\bar q'') \timp\bar p$ is equivalent to two simpler clauses $(\bar q\tand \bar q'')\timp\bar p$ and $(\bar q'\tand\bar q'')\timp \bar p$; for more details see~\cite{miller:unipfl}.} 

Thus closed sets --- and so also open sets, which are their complements --- can be computed from the axioms $\tf{Ax}(\ns P,\witness)$ by asking a suitable propositional solver to compute \emph{models}. 
Answer Set Programming (ASP) tool is one such tool~\cite{lifshitz:whaasp,lifshitz:anssp}.
Thus:
\begin{itemize*}
\item
We can view the algorithm for computing closed sets described in Remark~\ref{rmrk.computing.closed.sets} as `just' (see next Remark) an ASP solver for the Horn clause theory $\tf{Ax}(\ns P,\witness)$ in the logic $\tf{Prop}(\ns P,\witness)$.
\item
Conversely, we can view this Subsection as observing that the set of all solutions to a finite Horn clause theory has a semitopological structure, via witnessed sets.
\end{itemize*}
\end{rmrk}

\begin{rmrk}
Proposition~\ref{prop.declarative.interpretation} 
is not a `proof' that we should, or even could, actually use an ASP solver to do this.

Proposition~\ref{prop.declarative.interpretation} assumes complete and up-to-date information on the witness function.
Mathematically this is fine, just as writing `consider an uncomputable subset of $\mathbb N$' is mathematically fine --- we can prove that this exists.
As a \emph{computational} statement about possible implementations, this is more problematic, because a point of working with distributed systems is that we do not suppose that a participant could ever collect a global snapshot of the network state; and if they somehow did, it could become out-of-date; and in any case, in the presence of failing or adversarial participants it could be inaccurate.
So just because there \emph{is} a network state at some point in time, does not mean we have access to it.

Even mathematically, Proposition~\ref{prop.declarative.interpretation} is not the full story of (witness) semitopologies: 
\begin{itemize*}
\item
it concerns \emph{finite} semitopologies, whereas we are also interested in \emph{infinite} ones (see Remark~\ref{rmrk.why.infinite}); and
\item
the questions we ask in the mathematics --- especially the second-order ones such as ``Are these two points intertwined?'' or ``Find a maximal topen neighbourhood of this point, or confirm that none such exists.'' --- have not been considered in declarative programming, so far as we know. 
\end{itemize*} 
So it is important to appreciate that while Proposition~\ref{prop.declarative.interpretation} characterises closed and open sets in a witness semitopology in terms of solutions to Horn clause theories, and so helps us to understand what these sets really are at a mathematical level, this is not in and of itself automatically useful to actually turning such a semitopology into working network --- for that, we need algorithms like that described in Remark~\ref{rmrk.computing.closed.sets} --- nor is it a full mathematical account of all the facts of interest about semitopologies.

One practical use case where the correspondence with declarative programming might be immediately useful would be a monitoring tool, especially one testing mathematical properties to detect leading indicators of network malfunction.
Thus, for a network that is operating well and not changing too quickly, it would be feasible to traverse the network collecting information, and then use something like an ASP solver as part of a monitoring tool to compute the closed and open sets and so monitor properties such as the current intertwinedness of the network.
This is fine, so long as the reader is clear that a (centralised) network monitor that works in good conditions is not the same thing as the robust decentralised network itself.
\end{rmrk}

\jamiesubsubsection{Witnessed sets and topologies}

If finite semitopologies can be thought of as sets of solutions to Horn clause theories via witnessed sets, as outlined in Corollary~\ref{corr.finite.witness.horn}, what do finite topologies correspond to? 
We will find answers just by unrolling definitions and doing some simple reasoning, but the results are perhaps illuminating and a little bit surprising:

\begin{defn}
\label{defn.deterministic}
\leavevmode
\begin{enumerate*}
\item
Call a semitopology $(\ns P,\opens)$ a \deffont[deterministic semitopology]{deterministic} when each point $p$ has a unique least open neighbourhood $p\in M_p\in\opens$.
\item
Call a Horn clause theory (in the sense used in Definition~\ref{defn.logic}(\ref{item.logic.axiom})) \deffont[deterministic Horn clause theory]{deterministic} when for each propositional atom $\bar p\in\bar{\ns P}$ there exists at most one axiom in which $\bar p$ appears in its head.\footnote{The \emph{head} of the axiom is its final propositional atom, written to the right-hand side of the $\timp$ in Definition~\ref{defn.logic}(\ref{item.logic.axiom}).} 
\item
Call a witnessed set $(\ns P,\witness)$ (Definition~\ref{defn.witnessed.set}(\ref{witness.function})) \deffont[deterministic witnessed set]{deterministic} when for each point $p$, $\witness(p)$ is a singleton set; thus $\witness(p)=\{W_p\}$.\footnote{$W_p$ (the witness-set to $p$) is not necessarily equal to $M_p$ (the least open set containing $p$).  The witness function \emph{generates} a witness semitopology, but is not necessarily \emph{equal} to it.}
In words: $\witness$ is deterministic when \emph{every point has precisely one (possibly empty) witness-set}. 
\end{enumerate*}
\end{defn}

\begin{rmrk}
Recall the algorithms for computing open and closed sets from witness functions from Remarks~\ref{rmrk.computing.open.sets} and~\ref{rmrk.computing.closed.sets}.
When $\witness$ is deterministic, the algorithms simplify: there is precisely one witness-set to each point, and this removes the \emph{nondeterminism} from the algorithms and they become deterministic --- as our choice of name in Definition~\ref{defn.deterministic} suggests. 
\end{rmrk}

\begin{lemm}
\label{lemm.top.det}
Suppose $(\ns P,\opens)$ is a finite semitopology.
Then the following are equivalent:
\begin{enumerate*}
\item
$(\ns P,\opens)$ is a topology (intersections of open sets are open).
\item
$(\ns P,\opens)$ is a deterministic semitopology (every $p\in\ns P$ has a unique least open neighbourhood $M_p\in\opens$).
\end{enumerate*}
\end{lemm}
\begin{proof}
Suppose $(\ns P,\opens)$ is a topology and consider some $p\in\ns P$.
We must find a least open neighbourhood $p\in M_p$.
We just set $M_p=\bigcap\{O\in\opens \mid p\in O\}$.
Open sets in topologies are closed under finite unions, so $M_p$ is an open neighbourhood of $p$, and by construction it is least.

Suppose $(\ns P,\opens)$ is deterministic and consider $O,O'\in\opens$.
We must show that $O\cap O'$ is open.
We just note that $O\cap O'=\bigcup\{M_p \mid p\in O\cap O'\}$.
This is a union of open sets and so an open set, and by construction it contains $O\cap O'$.
But also it is contained in $O\cap O'$, since if $p\in O$ then $M_p\subseteq O$, and similarly for $O'$.
\end{proof}

\begin{rmrk}
Returning to the terminology \emph{deterministic} in Definition~\ref{defn.deterministic} above: when we are doing resolution in the Horn clause theory, and when we are building an open set using the algorithm in Remark~\ref{rmrk.computing.open.sets}, there is only ever one witness/clause for each point.
Thus resolution never has to backtrack; and building the open set never has to make any choices.
\end{rmrk}

\begin{lemm}
\label{lemm.top.horn}
Suppose $(\ns P,\opens)$ is a finite semitopology.
Then the following are equivalent:
\begin{itemize*}
\item
$(\ns P,\opens)$ is a topology.
\item
$(\ns P,\opens)=(\ns P,\opens(\witness))$ for some deterministic witness function $\witness$ on $\ns P$.
\end{itemize*}
\end{lemm}
\begin{proof}
Suppose $(\ns P,\opens)$ is a topology.
We just modify the construction from Lemma~\ref{lemm.finite.witness} and set $\witness(p)=\{M_p\}$.
The reader can check that $\opens=\opens(\witness)$. 

Conversely, suppose $\opens=\opens(\witness)$ for deterministic $\witness$, and write $\witness(p)=\{W_p\}$.
Now consider $O,O'\in\opens$; we need to show that $O\cap O'\in\opens$.
By the construction of the witness semitopology in Definition~\ref{defn.trust.topology}(\ref{item.witness.semitopology}) it would suffice to show that if $p\in O\cap O'$ then $W_p\subseteq O\cap O'$.
But this is immediate, since $O,O'\in\opens(\witness)$ so that if $p\in O$ then $W_p\subseteq O$, and similarly for $O'$.
\end{proof}

\begin{prop}
\label{prop.det}
Suppose $(\ns P,\opens)$ is a finite semitopology.
Then the following are equivalent:
\begin{itemize*}
\item
$(\ns P,\opens)$ is a topology.
\item
$(\ns P,\opens)$ is a deterministic semitopology. 
\item
$(\ns P,\opens)$ is the witness semitopology of a deterministic witness function.
\end{itemize*}
\end{prop}
\begin{proof}
We combine Lemmas~\ref{lemm.top.det} and~\ref{lemm.top.horn}. 
\end{proof}

\begin{rmrk}
The definitions and proofs in this Subsection are quite easy, but they capture a nice intuition which is not immediately obvious from just looking at the definitions: 
\begin{itemize*}
\item
Finite semitopologies correspond to computation with nondeterminism and backtracking.
\item
Finite topologies correspond to computation that does not require backtracking.
\end{itemize*}
Proposition~\ref{prop.det} makes this intuition formal up to a point, but it is not the full story.
What is missing is that a semitopology may have more than one presentation as the witness semitopology of a witnessed set.\footnote{Let's spell that out: it is possible for $\opens=\opens(\witness)=\opens(\witness')$ for distinct $\witness$ and $\witness'$.}
In particular, it is possible to create a non-deterministic witness function that generates a topology; intuitively, just because there might be a choice of witness-set, does not mean that the choice makes any difference to the final result.
Put another way: \emph{determinism} ensures that backtracking is impossible, but nondeterminism not necessarily imply that it is \emph{required}.\footnote{Think of reducing a simply-typed $\lambda$-calculus term; there are many reduction paths, but they all lead to the same normal form.}

We speculate that Proposition~\ref{prop.det} could be strengthened to show that topologies correspond to Horn clause theories that (may not be deterministic in the sense of Definition~\ref{defn.deterministic}, but that) do not require backtracking.
We leave this for future work. 
\end{rmrk}

\jamiesection{(Strongly) chain-complete semitopologies}
\label{sect.cc.cb}

\jamiesubsection{Definition and discussion}

\begin{rmrk}
\label{rmrk.not.chain-complete}
Just as for topologies, in semitopologies it is not true in general that the intersection of a descending chain of open sets is open.

Consider $\mathbb N$ with the semitopology generated by $O\subseteq\mathbb N$ such that $\{0\}\subsetneq O$.
Then $(\{0\}\cup i_\geq\mid i\geq 1)$ where $i_\geq=\{i'\mid i'\geq i\}$ is a descending chain of open sets, but its intersection $\{0\}$ is not open.

For the special case of \emph{witness} semitopologies, we can say something considerably stronger, as we shall see in Definition~\ref{defn.chain-complete} and Theorem~\ref{thrm.lim.O.open}.
\end{rmrk}

Recall from Definition~\ref{defn.ascending.chains} the notion of an ascending/descending \emph{chain of sets}:
\begin{defn}
\label{defn.chain-complete}
\leavevmode
\begin{enumerate*}
\item\label{item.chain-complete}
Call a semitopology \deffont[chain-complete semitopology]{chain-complete} when for every descending chain of open sets $\mathcal O\subseteq\opens$ (Definition~\ref{defn.ascending.chains}), its intersection $\bigcap\mathcal O$ is open.
\item\label{item.chain-bounded}
Call a semitopology \deffont[strongly chain-complete semitopology]{strongly chain-complete} when for every nonempty descending chain of nonempty open sets $\mathcal O\subseteq\opens_{\neq\varnothing}$, its intersection $\bigcap\mathcal O$ is open and nonempty.\footnote{We insist the chain is nonempty to exclude the pathological case of an empty chain over the semitopology $(\varnothing,\{\varnothing\})$.} 
\end{enumerate*}
\end{defn}

\begin{rmrk}[Chain-completeness in context]
\label{rmrk.chain-completeness.in.context}
We make a few general observations about Definition~\ref{defn.chain-complete} in the context of topology: 
\begin{enumerate}
\item
The strong chain-completeness condition (every descending chain of nonempty open sets is nonempty and open) is reminiscent of, though different from, a standard \emph{compactness} condition on metric spaces, that every descending chain of nonempty closed sets should be nonempty and closed.
\item
Call a topological space \deffont[Alexandrov topology]{Alexandrov} when its open sets are closed under arbitrary (and not just finite) intersections.

In the case that a semitopology $(\ns P,\opens)$ is a topology (so open sets are closed under finite intersections), and assuming that open sets can be well-ordered, being chain-complete is equivalent to being Alexandrov.
Clearly, an Alexandrov space is chain-complete; and conversely if we have an infinite collection of open sets in a chain-complete topology then (assuming that this collection can be well-ordered) we obtain their intersection by a transfinite induction taking limits of infinite descending chains of intersections.

The Alexandrov condition is unnatural in semitopologies in the sense that we do not assume even that finite intersections exist, so there is no finite-intersections condition to strengthen to the infinite case.
However, the chain-completeness condition \emph{is} natural in semitopologies, so in the light of the previous paragraph we could argue that chain-completeness is to semitopologies as being Alexandrov is to topologies. 

This is an intuitive observation, not a mathematical one, but it may help to guide the reader's intuitions.
\item
Strong chain-completeness has a stronger flavour of finiteness than chain-completeness.

For example: a strongly chain-complete space can contain only finitely many disjoint open sets --- since otherwise it would be easy to form an infinite descending chain of open sets with an empty intersection --- so, in the light of the topen partitioning result in Theorem~\ref{thrm.topen.partition}, we see that the topen partition of a strongly chain-complete semitopology is \emph{actually finite}.
\end{enumerate}
\end{rmrk}

\begin{rmrk}
\label{rmrk.cc.natural}
Definition~\ref{defn.chain-complete} abstracts two useful properties of two important classes of semitopologies:
\begin{enumerate*}
\item
Every finite semitopology is strongly chain-complete, because a strictly descending chain of finite sets is finite.\footnote{For the record, it is easy to come up with other conditions.  For instance, an even stronger condition is that a descending chain of open sets strictly above some $O\in\opens$ has an open intersection that is also strictly above $O$ (we recover the strong chain-completeness condition just by restricting $O$ to be equal to $\varnothing$).  This is a very reasonable thing to say: it is in a footnote and not the main text just because we have not (yet) found a direct use for it.  In contrast, strong chain-completeness turns out to be natural and useful, so we focus on that.} 
\item
Every witness semitopology is chain-complete; we will prove this shortly, in Theorem~\ref{thrm.lim.O.open}.
\end{enumerate*}
More discussion of these points is in Remark~\ref{rmrk.plausible.abstraction}.
The main mathematical/technical properties that come out of a semitopology being chain-complete and strongly chain-complete are respectively:
\begin{itemize*}
\item
Lemma~\ref{lemm.zorn.for.open.covers} and Corollary~\ref{corr.cover.exists} (existence of open covers), and 
\item
Lemma~\ref{lemm.zorn.for.atoms} and Corollary~\ref{corr.atom.exists} (existence of open atoms) respectively.
\end{itemize*}
However, before we come to that, we will set up some machinery and check some useful properties.
\end{rmrk}

\jamiesubsection{Elementary properties of the definition}

\begin{lemm}
\label{lemm.strongly.chain-complete.implies.chain-complete}
Suppose $(\ns P,\opens)$ is a semitopology.
Then:
\begin{enumerate*}
\item
If $(\ns P,\opens)$ is strongly chain-complete, then it is chain-complete.
\item
The reverse implication need not hold: it is possible for a semitopology to be chain-complete but not strongly chain-complete.
\item
Not every semitopology is chain-complete.
\end{enumerate*} 
\end{lemm}
\begin{proof}
We consider each part in turn:
\begin{enumerate}
\item
Consider a descending chain of open sets $\mathcal O$.
If one of the elements in $\mathcal O$ is empty then $\bigcap\mathcal O=\varnothing$ and $\varnothing\in\opens$ so we are done.
If all of the elements in $\mathcal O$ are nonempty then by chain-completeness $\bigcap\mathcal O$ is nonempty and open, and thus in particular it is open.
\item
A counterexample is $(\mathbb N,\powerset(\mathbb N))$ (the discrete semitopology on the infinite set of natural numbers).
Then $i_\geq=\{i'\mid i'\geq i\}$ for $i\geq 0$ is a descending chain of nonempty open sets whose intersection $\varnothing$ is open but not nonempty.
\item
This just repeats Remark~\ref{rmrk.not.chain-complete}, which gives an easy counterexample.
\qedhere\end{enumerate} 
\end{proof}

\begin{xmpl}
\leavevmode
\begin{enumerate}
\item
The \emph{all-but-one} and \emph{more-than-one} semitopologies (see Examples~\ref{xmpl.semitopologies}(\ref{item.counterexample.X-x}\&\ref{item.counterexample.more-than-one})) are (strongly) chain-complete.
\item
The closed interval $[\minus 1,1]$ with its usual topology is not chain-complete (and not strongly chain-complete): e.g. $\{(\minus 1/i, 1/i) \mid i\geq 1\}$ is a descending chain of open sets but its intersection $\{0\}$ is not open.
Similarly for the two semitopologies on $\mathbb Q^2$ in Example~\ref{xmpl.two.topen.examples}.

(Looking ahead just for a moment to Theorem~\ref{thrm.lim.O.open}, this tells us that these semitopologies cannot be generated by witness functions.)
\end{enumerate}
\end{xmpl}

\begin{lemm}
\leavevmode
Suppose $(\ns P,\opens)$ is a semitopology.
Then:
\begin{enumerate*}
\item
$\ns P$ is chain-complete if and only if the union of any ascending chain of closed sets, is closed.
\item
$\ns P$ is strongly chain-complete if and only if the union of any ascending chain of closed sets that are not equal to $\ns P$, is closed and not equal to $\ns P$. 
\end{enumerate*}
\end{lemm}
\begin{proof}
Direct from Definition~\ref{defn.chain-complete} using Lemma~\ref{lemm.closed.complement.open}, which notes that closed sets are the complements of open sets (just as for topologies).
\end{proof}

\jamiesubsection{Consequences of being strongly chain-complete}

Being strongly chain-complete is a useful well-behavedness condition.
We consider some of its consequences. 

\jamiesubsubsection{Strongly chain-complete implies $\intertwinedwith$-complete}

We saw a chain-completeness condition before: $\intertwinedwith$-completeness from Definition~\ref{defn.intertwined.preorder}(\ref{item.intertwined-bounded}).
As promised in Remark~\ref{rmrk.intertwinedwith-bounded.natural}(\ref{item.intertwinedwith-bounded.natural.chain-bounded}), we now note that strongly chain-complete semitopologies are also $\intertwinedwith$-complete:
\begin{lemm}
\label{lemm.chain-bounded.implies.intertwinedwith.bounded}
Suppose $(\ns P,\opens)$ is a quasiregular semitopology.
Then if $(\ns P,\opens)$ is strongly chain-complete then it is $\intertwinedwith$-complete (Definition~\ref{defn.intertwined.preorder}(\ref{item.intertwined-bounded})). 
\end{lemm}
\begin{proof}
Suppose we have a $\geqk$-descending chain of points $p_1\geqk p_2\geqk \dots$.

Since $\ns P$ is quasiregular, $\community(p_i)\in\opens_{\neq\varnothing}$ for every $i$.
Write $I=\bigcap_i \community(p_i)$.

Since $\ns P$ is strongly chain-complete (Definition~\ref{defn.chain-complete}(\ref{item.chain-bounded})), $I\in\opens_{\neq\varnothing}$ (we need \emph{strong} chain-completeness here to know that $I$ is not just open but also nonempty).
Choose some $p\in I$.

It follows from Lemma~\ref{lemm.weakly.regular.community}(\ref{item.weakly.regular.community.1}) that $p\leqk p_i$ for every $i$, thus $p$ is a $\leqk$-lower bound for the chain.
\end{proof}

\jamiesubsubsection{Indirectly regular points: inherent properties}

In Definition~\ref{defn.tn} we saw three regularity conditions on points: quasiregular, weakly regular, and regular.
We now add a fourth condition to this mix: \emph{indirect regularity}.
A point is indirectly regular when it is intertwined with a regular point; intuitively, if regular points are `nice' then an indirectly regular point is a point that is not necessarily nice itself, but it is intertwined with a point that \emph{is} nice.\footnote{Like in the movies: where a gangster falls in love with a nice person; the gangster may not stop being a gangster, but they now have a moral compass, if only indirectly.}
It is not at all obvious that this should have anything to do with strong chain-completeness, but it does: a punchline of this Subsection will come in Remark~\ref{rmrk.linear.regularity}, where we note that if a semitopology in strongly chain-complete then indirect regularity slots in particularly nicely with the three regularity conditions from Definition~\ref{defn.tn}.
We now set about building the machinery we need to tell this story:
\begin{defn}
\label{defn.indirectly.regular}
Suppose $(\ns P,\opens)$ is a semitopology.
Call $p$ an \deffont{indirectly regular point} when $p\intertwinedwith q$ for some regular $q$.
\end{defn}

\begin{lemm}
Suppose $(\ns P,\opens)$ is a semitopology and $p\in\ns P$.
Then $p$ is indirectly regular if and only if $p$ is in the closure of a topen set.
\end{lemm}
\begin{proof}
We prove two implications:
\begin{itemize}
\item
\emph{Suppose $p$ is indirectly regular; so $p\intertwinedwith q$ for some regular $q\in\ns P$.}
 
By Definition~\ref{defn.tn}(\ref{item.regular.point}) $q\in\community(q)\in\topens$.
By Definition~\ref{defn.intertwined.points}(\ref{intertwined.defn}) (since $p\intertwinedwith q$) $p\in\intertwined{q}$, and by Theorem~\ref{thrm.pKp}(\ref{item.closure.community.p.intertwined}) $\intertwined{q}=\closure{\community(q)}$.
Thus $p$ is in the closure of the topen set $\community(q)$.
\item
\emph{Suppose $p\in\closure{\atopen}$ for some $\atopen\in\topens$.}

Choose any $q\in\atopen$.
Note by Theorem~\ref{thrm.max.cc.char}(\ref{char.some.topen}) that $q$ is regular; we will now show that $p\intertwinedwith q$.

Consider a pair of open neighbourhoods $p\in O\in\opens$ and $q\in O'\in\opens$.
Then $O\between\atopen$ (because $p\in O$ and $p\in\closure{\atopen}$), and $\atopen\between O'$ (because $q\in\atopen\cap O'$).
By transitivity of $\atopen$, $O\between O'$.
It follows that $p\intertwinedwith q$ as required.
\qedhere\end{itemize}
\end{proof}

We will need Lemma~\ref{lemm.intertwined.community.subset} below.
We can think of this as a version of Lemma~\ref{lemm.weakly.regular.community} where we know that one of the points is regular:
\begin{lemm}
\label{lemm.intertwined.community.subset}
Suppose $(\ns P,\opens)$ is a semitopology and $p,q\in\ns P$ and $q$ is regular.
Then:
\begin{enumerate*}
\item
If $q\intertwinedwith p$ then $\community(q)\subseteq\intertwined{p}$. 
\item
As a corollary, if $p\in\intertwined{q}$ or $q\in\intertwined{p}$, then $\community(q)\subseteq\intertwined{p}$. 
\end{enumerate*}
\end{lemm}
\begin{proof}
The corollary follows because by Definition~\ref{defn.intertwined.points}(\ref{intertwined.defn}), $p\in\intertwined{q}$ and $q\in\intertwined{p}$ are both equivalent to $p\intertwinedwith q$. 

So suppose $q$ is regular and $q\in\intertwined{p}$.
By Definition~\ref{defn.tn}(\ref{item.regular.point}) $q\in\community(q)\in\topens$ and by Theorem~\ref{thrm.r=wr+uc} $q$ is unconflicted.
Consider any other $q'\in\community(q)$; unpacking Definition~\ref{defn.tn}(\ref{item.tn}) and~\ref{defn.intertwined.points}(\ref{intertwined.defn}) $p\intertwinedwith q\intertwinedwith q'$ and so by Definition~\ref{defn.conflicted}(\ref{item.unconflicted}) (since $q$ is unconflicted) $p\intertwinedwith q'$.
Thus $\community(q)\subseteq\intertwined{p}$ as required.
\end{proof}

\begin{corr}
Suppose $(\ns P,\opens)$ is a semitopology and $p\in\ns P$.
Then the following are equivalent:
\begin{enumerate*}
\item
$p$ is indirectly regular.
\item
$\intertwined{p}$ contains a topen set.
\end{enumerate*}
\end{corr}
\begin{proof}
We prove two implications:
\begin{itemize}
\item
\emph{Suppose there exists $\atopen\in\topens$ such that $\atopen\subseteq\intertwined{p}$.}

Take any $q\in\atopen$.
By Theorem~\ref{thrm.max.cc.char}(\ref{char.some.topen}) $q$ is regular, and by Definition~\ref{defn.intertwined.points}(\ref{intertwined.defn}) $q\intertwinedwith p$.
\item
\emph{Suppose $p$ is indirectly regular.}

By Definition~\ref{defn.indirectly.regular} $p\intertwinedwith q$ for some regular $q\in\ns P$.
By Lemma~\ref{lemm.intertwined.community.subset} $\community(q)\subseteq\intertwined{p}$.
By Definition~\ref{defn.tn}(\ref{item.regular.point}) $\community(q)\in\topens$.
\qedhere\end{itemize}
\end{proof}

\jamiesubsubsection{Indirectly regular points in the context of other regularity properties}

We can now continue the observations made in Remark~\ref{rmrk.r.wr.qr}:
\begin{rmrk}
\label{rmrk.linear.regularity}
This Subsection develops a sequence of results that are interesting in themselves, but also taken together with Lemma~\ref{lemm.wr.r} they indicate that in a strongly chain-complete semitopology, our regularity conditions organise into a nice list ordered by increasing strength as follows:
\begin{itemize*}
\item
Being quasiregular (having a nonempty community).
\item
Being indirectly regular (intertwined with a regular point / being on the boundary of a topen set).
\item
Being weakly regular (being an element of your community).
\item
Being regular (being an element of your \emph{topen} community).
\end{itemize*}
As a diagram, in strongly chain-complete semitopologies we get the following \emph{chain of implications}: 
$$
\text{quasiregular}
\quad\limp\quad
\text{indirectly regular}
\quad\limp\quad
\text{weakly regular}
\quad\limp\quad
\text{regular} .
$$
If the semitopology is not strongly chain-complete, then (by Lemma~\ref{lemm.wr.r} we still have the other implications, but) being indirectly regular does not fall so neatly in line.

We read this as evidence that the \emph{strongly chain-complete semitopologies} are a particularly natural class of semitopologies for us to study, and they are a useful abstraction of the finite semitopologies (much as e.g. Alexandrov topologies, or compact topologies, capture aspects of finiteness for topologies).
\end{rmrk}
 
\begin{prop}
\label{prop.up.down.wr}
Suppose $(\ns P,\opens)$ is a strongly chain-complete semitopology (by Remark~\ref{rmrk.cc.natural} this holds in particular if $\ns P$ is finite).
Then:
\begin{enumerate*}
\item\label{item.up.down.wr.ir}
If $p\in\ns P$ is weakly regular then $p$ is indirectly regular.
\item
The converse implication need not hold: it is possible for $(\ns P,\opens)$ to be strongly chain-complete, and even actually \emph{finite}, and $p\in\ns P$ is indirectly regular yet not weakly regular.  
\item
If $(\ns P,\opens)$ is not strongly chain-complete then the implication in part~\ref{item.up.down.wr.ir} might fail: it is possible for $p\in\ns P$ to be weakly regular but not indirectly regular, even if $(\ns P,\opens)$ is chain-complete (but not strongly chain-complete).
\end{enumerate*}
\end{prop}
\begin{proof}
We consider each part in turn:
\begin{enumerate}
\item
Suppose $p\in\ns P$ is weakly regular.
From Proposition~\ref{prop.views.of.regularity} $\intertwined{p}$ is a closed neighbourhood (a closed set with a nonempty open interior).
Using strong chain-completeness and Zorn's lemma on $\supseteq$, the set of closed neighbourhoods that are subsets of $\intertwined{p}$ contains a minimal closed neighbourhood $C\subseteq\intertwined{p}$ (we need \emph{strong} chain-completeness to ensure that $C$ has a \emph{nonempty} open interior).
Take $q\in \interior(C)$; By Theorem~\ref{thrm.up.down.char} $q$ is regular.
\item
For a counterexample consider point $\ast$ in Figure~\ref{fig.nitpick}.  Then $\community(\ast)=\{1\}$ and $1$ is regular, but $\ast\notin\community(\ast)$ so $\ast$ is not weakly regular. 

(It \emph{does} follow from existence of a regular $q\in\community(p)$ that $p$ is quasiregular, but only because existence of \emph{any} (not necessarily regular) $q\in\community(p)$ means precisely that $\community(p)\neq\varnothing$.)
\item
A counterexample is in Figure~\ref{fig.weakly-regular.conflicted}.
\qedhere\end{enumerate}
\end{proof}

\begin{rmrk}
Proposition~\ref{prop.up.down.wr} above is just an easy corollary of Theorem~\ref{thrm.up.down.char}.
We can think of this as another version of the `hairy ball' result that we saw in Theorem~\ref{thrm.K-regular}, but for the case of a weakly regular point, instead of for a quasiregular space.

Recall that we care about regular points because these are (for our purposes) well-behaved: they have a topen neighbourhood (Theorem~\ref{thrm.max.cc.char}), by which fact local consensus is guaranteed where algorithms succeed (Remark~\ref{rmrk.transitive.correlated}).
Thus the interest of Proposition~\ref{prop.up.down.wr} is that it provides certain guarantees of progress; a weakly regular point may not be able to progress (even if algorithms succeed), but it guaranteed to be intertwined with some well-behaved regular point.
\end{rmrk}

\begin{figure}
\centering
\includegraphics[width=0.4\columnwidth,trim={50 50 50 50},clip]{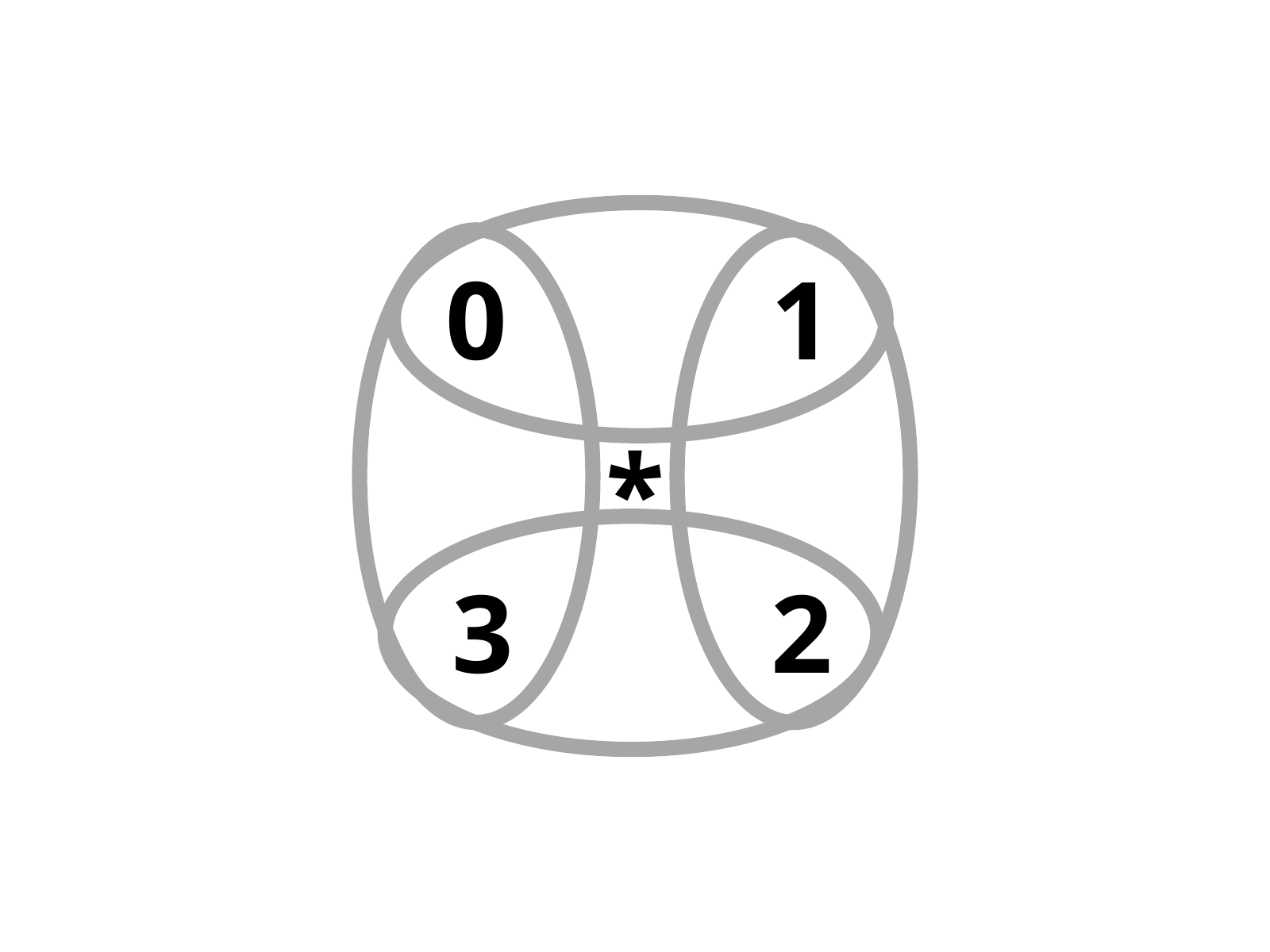}
\caption{Lemma~\ref{lemm.indirectly.regular.to.quasiregular}(\ref{item.indirectly.regular.to.quasiregular.2}): a point $\ast$ that is quasiregular but not indirectly regular}
\label{fig.ir.not.qr}
\end{figure}

\begin{lemm}
\label{lemm.indirectly.regular.to.quasiregular}
Suppose $(\ns P,\opens)$ is a semitopology and suppose $p\in\ns P$.
Then:
\begin{enumerate*}
\item\label{item.indirectly.regular.to.quasiregular.1}
If $p$ is indirectly regular then $p$ is quasiregular.
\item\label{item.indirectly.regular.to.quasiregular.2}
The converse implication need not hold: it is possible for $p$ to be quasiregular but not indirectly regular.
\end{enumerate*}
\end{lemm}
\begin{proof}
We consider each part in turn:
\begin{enumerate}
\item
Suppose $p$ is indirectly regular.
By Definition~\ref{defn.indirectly.regular} $p\in\intertwined{q}$ for some regular $q$.
By Lemma~\ref{lemm.wr.r}(\ref{item.r.implies.wr}\&\ref{item.wr.implies.qr}) $q$ is quasiregular, meaning that $\varnothing\neq\community(q)$.
By Lemma~\ref{lemm.intertwined.community.subset} $\community(q)\subseteq\community(p)$,
so that $\community(p)$ is nonempty and $p$ is quasiregular.
\item
It suffices to provide a counterexample.
Consider the point $\ast$ in the semitopology illustrated in Figure~\ref{fig.ir.not.qr}:\footnote{This is an elaboration of the semitopology we have already seen in Figure~\ref{fig.square.diagram}).}
\begin{itemize}
\item
$\ns P = \{\ast, 0, 1, 2, 3\}$.
\item
$\opens$ is generated by $\{\{3,0\},\ \{0,1\},\ \{1,2\},\ \{2,3\}\}$;
note that the only open neighbourhood of $\ast$ is all of $\ns P$.
\end{itemize}
The reader can check that $\intertwined{\ast}=\ns P$, so $\community(\ast)=\ns P\neq\varnothing$ so $\ast$ is quasiregular.
However, the reader can also check that no point in this space is regular, so $\ast$ is not intertwined with any regular point.
\qedhere\end{enumerate}
\end{proof}

This completes the chain of implications from Remark~\ref{rmrk.linear.regularity}.

\jamiesubsection{Witness semitopologies are chain-complete}

\begin{thrm}
\label{thrm.lim.O.open}
Suppose that
$(\ns P,\witness)$ is a witnessed set.
Then the witness semitopology $\opens(\witness)$ from Definition~\ref{defn.trust.topology} is chain-complete.

Unpacking this we can say:
\begin{quote}
In a witness semitopology, intersections of descending chains of open sets are open, and unions of ascending chains of closed sets are closed.
\end{quote}
\end{thrm}
\begin{proof}
Consider a chain of open sets $\mathcal O\subseteq\opens$. 
There are three cases:
\begin{itemize}
\item
\emph{Suppose $\bigcap\mathcal O=\varnothing$.}\quad

We note that $\varnothing\in\opens$ (Definition~\ref{defn.semitopology}(\ref{semitopology.empty.and.universe})) and we are done.
\item
\emph{Suppose $\mathcal O$ has a least element $O$.}\quad

Then $O=\bigcap\mathcal O$ and $O\in\opens$ and we are done.
\item
\emph{Suppose $\mathcal O\neq\varnothing$ and $\mathcal O$ has no least element.}\quad

Then note that $\mathcal O$ is infinite.
Consider some $p\in\bigcap\mathcal O$.
By construction of the witness semitopology (Definition~\ref{defn.trust.topology}) for each $O\in\mathcal O$ there exists a witness-set $w_O\in\witness(p)$ such that $w_O\subseteq O$.
Now by Definition~\ref{defn.witnessed.set}(\ref{witness.function}) $\witness(p)$ is finite, so by the pigeonhole principle, there exists some $w\in\witness(p)$ such that $w\subseteq O$ for for every $O\in\mathcal O$,
and thus $w\subseteq\bigcap\mathcal O$. 

Now $p$ in the previous paragraph was arbitrary, so we have shown that if $p\in\bigcap\mathcal O$ then also there exists $w\in\witness(p)$ such that $w\subseteq\bigcap\mathcal O$. 
It follows by construction of the witness semitopology in Definition~\ref{defn.trust.topology} that $\bigcap\mathcal O$ is open as required.
\qedhere\end{itemize}
\end{proof}

\begin{lemm}
\label{lemm.elaborating}
The reverse implication in Theorem~\ref{thrm.lim.O.open} does not hold: there exists a chain-complete semitopology (indeed, it is also strongly chain-complete) that is not generated as the witness semitopology of a witnessed set.
\end{lemm}
\begin{proof}
It is a fact that the more-than-one semitopology on $\mathbb N$ (having open sets generated by distinct pairs $\{i,i'\}\subseteq\ns P$; see Example~\ref{xmpl.semitopologies}(\ref{item.counterexample.more-than-one})) is strongly chain-complete, but by Lemma~\ref{lemm.more-than-one} is is not generated by a witness function.
\end{proof}

\begin{rmrk}
Elaborating further on Lemma~\ref{lemm.elaborating}, suppose $(\ns P,\opens)$ is a chain-complete semitopology.
Then to every $p$ we can assign a nonempty set $\mathcal O_p$ of \emph{covers} (minimal open sets containing $p$; see Definition~\ref{defn.open.covers}).

Can we obtain a witness function just by setting $\witness(p)=\mathcal O_p$?
No: because $p$ need not have finitely many covers, and Definition~\ref{defn.witnessed.set} insists on a \emph{finite} set of (possibly infinite) nonempty witness-sets.\footnote{See Example~\ref{xmpl.semitopologies}(\ref{item.counterexample.X-x}) for an example of a semitopology containing points with infinitely many covers, though interestingly, this \emph{can} be generated by a witness function, as noted in Lemma~\ref{lemm.all-but-one}.}  

We could allow an infinite set of witness-sets in Definition~\ref{defn.witnessed.set}, but at a price:
\begin{itemize*}
\item
The proof of Theorem~\ref{thrm.lim.O.open} depends on the pigeonhole principle, which uses finiteness of the set of witness-sets.
\item
The proof of Lemma~\ref{lemm.limit.is.open} depends on the set of witness-sets being finite, and this is required for Proposition~\ref{prop.lim.is.closure}.
\end{itemize*}
\end{rmrk}

\begin{rmrk}
\label{rmrk.characterise.witness.semitopologies}
Theorem~\ref{thrm.lim.O.open} shows that witness semitopologies are chain-complete, but Lemma~\ref{lemm.elaborating} demonstrates that this cannot precisely characterise witness semitopologies.
Might there be another way?

We might look at Corollary~\ref{corr.cover.exists} (open covers exist), cross-reference with Definition~\ref{defn.witnessed.set}(\ref{witness.function}) (every $p$ has only finitely many witness-sets), and ask if we might characterise witness semitopologies as those topologies that are chain-complete \emph{and} every $p$ has finitely many open covers (Definition~\ref{defn.open.covers}(\ref{item.open.cover})).

No: by Lemma~\ref{lemm.all-but-one}, the all-but-one semitopology from Example~\ref{xmpl.semitopologies}(\ref{item.counterexample.X-x}) is a witness semitopology, and if the underlying set of points is infinite then every point has infinitely many covers.
See also Remark~\ref{rmrk.two.open.problems}(\ref{item.two.open.problems.1}).
\end{rmrk}

\begin{prop}
\label{prop.finite.chain-bounded}
\leavevmode
\begin{enumerate*}
\item\label{item.finite.chain-bounded.1}
Not every witness semitopology (Definition~\ref{defn.trust.topology}) is strongly chain-complete (Definition~\ref{defn.chain-complete}(\ref{item.chain-bounded})).
\item
Part~\ref{item.finite.chain-bounded.1} holds even if we restrict the witness function 
$\witness:\ns P\to\finpow_{\neq\varnothing}(\powerset_{\neq\varnothing}(\ns P))$ in Definition~\ref{defn.witnessed.set}(\ref{witness.function}) to return a finite set of finite witness-sets, so that 
$\witness:\ns P\to\finpow_{\neq\varnothing}(\finpow_{\neq\varnothing}(\ns P))$.
\item
Every \emph{finite} semitopology (this includes every finite witness semitopology) is strongly chain-complete. 
\end{enumerate*}
\end{prop}
\begin{proof}
\leavevmode
\begin{enumerate}
\item
It suffices to provide a counterexample.
Consider $\mathbb N$ with witness function $\witness(i)=\{\{i\plus 1\}\}$.
This generates a semitopology with open sets generated by $i_\geq=\{i'\in\mathbb N\mid i'\geq i\}$.
Then $(i_\geq\mid i\in\mathbb N)$ is a descending chain of open sets with an open, but empty, intersection.
\item
We just use the counterexample in part~\ref{item.finite.chain-bounded.1}.
\item
We noted already in Remark~\ref{rmrk.cc.natural} that if the semitopology is finite then every descending chain of open sets is eventually stationary; so we just take the final element in the chain. 
\qedhere\end{enumerate}
\end{proof}

\begin{rmrk}
\label{rmrk.plausible.abstraction}
We are particularly interested in the concrete example of finite witnessed semitopologies, since these are the ones that we can actually implement.
But we can ask what it is about this class of examples that makes them mathematically well-behaved; what essential algebraic features might we identify here?
Proposition~\ref{prop.finite.chain-bounded} suggests that being strongly chain-complete may be a suitable mathematical abstraction: 
\begin{itemize*}
\item
by Proposition~\ref{prop.finite.chain-bounded} the abstraction is both non-trivial and sound (not every witness semitopology is strongly chain-complete, but every \emph{finite} witness semitopology is), and 
\item
Theorem~\ref{thrm.lim.O.open} asserts that for a (possibly infinite) $(\ns P,\opens)$, any convergence using a descending sequence of open sets has a flavour of being `locally finite' in the sense of being guaranteed to have a nonempty open intersection.\footnote{There is also a computational interpretation to (strong) chain-completeness: think of a descending chain of open sets as a computation that computes to narrow down possibilities to smaller and smaller nonempty open sets, then this possibly infinite computation does deliver a final answer that is a (nonempty) open set.} 
\end{itemize*}
So strongly chain-complete semitopologies are a plausible abstraction of finite witness semitopologies.\footnote{There may be more than one such abstraction; identifying one candidate does not mean there may not be others.  For example, both \emph{rings} and \emph{models of first-order arithmetic} are valid abstractions of the notion of `number'.  Which of these mathematical structures we work with, depends on which aspects of the concrete thing we are interested in studying.}
The test is now to explore the theory of strongly chain-complete semitopologies.
Key results are Corollaries~\ref{corr.cover.exists} and~\ref{corr.atom.exists}, which ensure that in a strongly chain-complete semitopology, open covers and atoms always exist, and we will build from there.
\end{rmrk}

\begin{rmrk}[Why infinity?]
\label{rmrk.why.infinite}
\label{rmrk.infinite}
Following on from Remark~\ref{rmrk.plausible.abstraction},
we sometimes get asked, especially by engineers, why we care about infinite models when all practical computer networks are finite.

A simple answer is that we do this for the same reason that Python (and many other programming languages) have a datatype for infinite precision integers.
Any given execution will only compute numbers in a finite subset this infinity, but since we may not be able to predict how large this subset is, it is natural to support the notion of an infinite datatype.
Note this this holds for data, not just datatypes: e.g. Python accommodates values for $\pi$, $e$, and $j$ even though these are not rational numbers, and for infinite streams and may other `infinite' objects.\footnote{I once struggled to convince a Computer Science undergraduate student that $1/3$ is finite.  The blockage was that the student only believed in the \texttt{float} datatype, and the decimal expansion of $1/3$ as $0.333\dots$ is infinite.  This deadlock was broken by inviting the student to implement a base-3 float type.  The deeper point here is that what we consider `infinite' may depend on what representation we assume as primitive.  We see something similar in model theory, where we may distinguish between \emph{internal} and \emph{external} notions of size in a model.  The bottom line is: obsessing about size can become a dead end; we also need to pay attention to what seems elegant and natural, i.e. to what our brains want --- and then model \emph{that}.} 

However there is another reason: participants cannot depend on an exhaustive search of the full network ever terminating (nor that even an attempt at this would be cost-effective), so this requires a theory and algorithms that make sense on at least countably infinitely many points.

In fact, arguably the natural cardinality for semitopology is at least \emph{uncountable}, since for a participant on a system with network latency, the system is not just unbounded, but also unenumerable.
This is another reason that Theorem~\ref{thrm.lim.O.open} is remarkable.
See also Remark~\ref{rmrk.finiteness.and.compactness}.
\end{rmrk}

\jamiesubsection{Minimal sets: open covers and atoms}

\jamiesubsubsection{Open covers (minimal open neighbourhoods)}

First, some useful notation:
\begin{nttn}
\label{nttn.gtrdot}
Suppose $(\ns P,\opens)$ is a semitopology and $P\subseteq\ns P$.
Write 
$$
O\gtrdot P \quad\text{and synonymously}\quad P\lessdot O
$$
when $O$ is a minimal nonempty open set containing $P$.
In symbols:
$$
O\gtrdot P
\quad\text{when}\quad
O\neq\varnothing\land P\subseteq O\land \Forall{O'{\in}\opens_{\neq\varnothing}}(P\subseteq O'\subseteq O \limp O'=O).
$$
We may combine $\gtrdot$ with other relations for compactness.
For example:
\begin{itemize*}
\item
$p\in O\gtrdot P$ is shorthand for $p\in O \land O\gtrdot P$; and 
\item
$P\supseteq O\gtrdot Q$ is shorthand for $O\subseteq P\land O\gtrdot Q$.
\end{itemize*}
\end{nttn}

Definition~\ref{defn.open.covers} collects some (standard) terminology.
\begin{defn}
\label{defn.open.covers}
\leavevmode
Suppose $(\ns P,\opens)$ is a semitopology and $p\in\ns P$.
\begin{enumerate}
\item\label{item.open.cover.1}
Call $O\in\opens$ an \deffont{(open) neighbourhood of $p$} when $p\in O$.
\item\label{item.open.cover}
Call $O\in\opens$ an \deffont{(open) cover of $p$}, write 
$$
O\gtrdot p
\quad\text{and/or}\quad 
p\lessdot O, 
$$
and say that $O$ \deffont[covers ($O$ covers $p$;\ $O\gtrdot\{p\}$)]{covers} $p$, when $O\gtrdot\{p\}$ (Notation~\ref{nttn.gtrdot}).

In words using the terminology of part~\ref{item.open.cover.1}: $O\gtrdot p$ when $O$ is a minimal open neighbourhood of $p$.
\item\label{item.covers.p}
Write $\f{Covers}(p)$ for the set of \deffont[open covers of $p$ (set of;\ $\f{Covers}(p)$)]{open covers of $p$}, and if $O'\in\opens$ then write $\f{Covers}_{O'}(p)$ for the open covers of $p$ that are subsets of $O'$.
In symbols:
$$
\f{Covers}(p)=\{O\in\opens \mid p\lessdot O\} 
\quad\text{and}\quad
\f{Covers}_{O'}(p)=\{O\in\opens \mid p\lessdot O\subseteq O'\} 
.
$$
Note of course that $\f{Covers}(p)=\f{Covers}_{\ns P}(p)$.
\end{enumerate}
\end{defn}

\begin{lemm}
\label{lemm.zorn.for.open.covers}
Suppose $(\ns P,\opens)$ is a strongly chain-complete semitopology and suppose $\varnothing\neq\mathcal O\subseteq\opens_{\neq\varnothing}$ is a nonempty set of nonempty open sets that is $\subseteq$-down-closed (meaning that if $\varnothing\neq O'\subseteq O\in\mathcal O$ then $O'\in\mathcal O$).

Then $\mathcal O$ contains a $\subseteq$-minimal element. 
\end{lemm}
\begin{proof}
We use Zorn's Lemma~\cite{jech:axic,campbell:orizl}:
By strong chain-completeness, $\mathcal O$ ordered by the \emph{superset} relation (the reverse of the subset inclusion relation), contains limits, and so upper bounds, of ascending chains.
By Zorn's Lemma, $\mathcal O$ contains a $\supseteq$-maximal element.
This is the required $\subseteq$-minimal element.
\end{proof}

\begin{corr}[Existence of open covers]
\label{corr.cover.exists}
Suppose $(\ns P,\opens)$ is a chain-complete semitopology and $p\in\ns P$.\footnote{Note that we only require \emph{chain-completeness} here (Definition~\ref{defn.chain-complete}(\ref{item.chain-complete})), not strong chain-completeness (Definition~\ref{defn.chain-complete}(\ref{item.chain-bounded})).}
Then:
\begin{enumerate*}
\item\label{item.cover.exists.1}
If $p\in O'\in\opens$ then $O'$ contains an open cover of $p$.
In symbols: 
$$
\f{Covers}_{O'}(p) \neq\varnothing,
\quad\text{equivalently}\quad
\Exists{O\in\opens} p\lessdot O\subseteq O' .
$$
\item\label{item.cover.exists.2}
$p$ has at least one open cover.
In symbols:
$$
\f{Covers}(p) \neq\varnothing, 
\quad\text{equivalently}\quad
\Exists{O\in\opens} p\lessdot O .
$$
\end{enumerate*}
\end{corr}
\begin{proof}
Direct from Lemma~\ref{lemm.zorn.for.open.covers}, considering $\{O\in\opens \mid p\in O\subseteq O'\}$ (nonempty because it contains $O'$), and then setting $O'=\ns P$.
\end{proof}

We can apply Corollary~\ref{corr.cover.exists} in an elementary way to extend Lemma~\ref{lemm.closure.using.nbhd.intersections}:
\begin{corr}
\label{corr.closure.using.covers}
Suppose $(\ns P,\opens)$ is a chain-complete semitopology (in particular this holds if $\ns P$ is finite) and suppose $p\in\ns P$ and $O'\in\opens$.
Then the following are equivalent:
\begin{enumerate*}
\item\label{item.closure.using.covers.1} 
$p\in\closure{O'}$.
\item\label{item.closure.using.covers.2} 
$O'\between\nbhd(p)$.
\item\label{item.closure.using.covers.3} 
$O'\between\f{Covers}(p)$.
\end{enumerate*}
\end{corr}
\begin{proof}
Parts~\ref{item.closure.using.covers.1} and~\ref{item.closure.using.covers.2} just repeat Lemma~\ref{lemm.closure.using.nbhd.intersections}.
Equivalence of parts~\ref{item.closure.using.covers.2} and~\ref{item.closure.using.covers.3} is routine as follows:
\begin{itemize*}
\item
If $O'\between\nbhd(p)$ then $O'\between\f{Covers}(p)$, since by construction $\f{Covers}(p)\subseteq\nbhd(p)$.
\item
Suppose $O'\between\f{Covers}(p)$ and consider some $O\in\nbhd(p)$.
By Corollary~\ref{corr.cover.exists}(\ref{item.cover.exists.1}) $O$ contains some cover $p\lessdot O''\subseteq O$, and since $O'\between O''$ also $O'\between O$.
Thus $O'\between\nbhd(p)$.
\qedhere
\end{itemize*}
\end{proof}

\begin{rmrk}
\label{rmrk.uncomputable.semitopology}
Recall that our semitopological analysis of consensus is all about continuity and value assignments being locally constant --- as per Definitions~\ref{defn.continuity}(\ref{item.continuous.function.at.p}) and~\ref{defn.value.assignment} and results like Lemma~\ref{lemm.open.lc} --- and these discussions are about the open neighbourhoods of $p$.
Thus, to understand consensus at $p$ we need to understand its open neighbourhoods.

Corollary~\ref{corr.cover.exists} tells us that in a witness semitopology, 
we can simplify and just consider the open covers of $p$.
This is because if a continuous function $f:\ns P\to\ns P'$ such that $f(p)=p'\in O'$ is continuous at $p\in\ns P$, then using continuity and Corollary~\ref{corr.cover.exists} there exists some open cover $p\lessdot P\subseteq f^\mone(O')$.

Turning this around, if we want to \emph{create} consensus around $p$ --- perhaps as part of a consensus algorithm --- it suffices to find some open cover of $p$, and convince that cover.
This fact is all the more powerful because Corollary~\ref{corr.cover.exists} does not assume that $\ns P$ is finite: it is a fact of witness semitopologies of any cardinality.
\end{rmrk}

\jamiesubsubsection{Atoms (minimal nonempty open sets)}

\begin{defn}
\label{defn.atomic.open.set}
Suppose $(\ns P,\opens)$ is a semitopology.
\begin{enumerate}
\item\label{item.atom}
Call $A\in\opens$ an \deffont{(open) atom} when $A$ is a minimal nonempty open set.\footnote{An open atom covers every point that it contains, but an open cover for a point $p$ need not be an atom, since it may contain a smaller open set --- just not one that contains $p$.  See Example~\ref{xmpl.p.not.in.O}(\ref{item.atom.need.not.contain.p}).} 
In symbols using Notation~\ref{nttn.gtrdot} this is: 
$$
A\gtrdot\varnothing \quad\text{and synonymously}\quad \varnothing\lessdot A.
$$
\item\label{item.atoms.of.P}
If $P\subseteq\ns P$ then write $\f{Atoms}(P)$ for the atoms that are subsets of $P$. 
In symbols:
$$
\f{Atoms}(P) = \{A\in\opens \mid \varnothing\lessdot A \subseteq P\} .
$$
\end{enumerate}
\end{defn}

\begin{lemm}
\label{lemm.zorn.for.atoms}
Suppose $(\ns P,\opens)$ is a strongly chain-complete semitopology and suppose $\varnothing\neq\mathcal O\subseteq\opens_{\neq\varnothing}$ is a nonempty set of nonempty open sets that is $\subseteq$-down-closed (meaning that if $\varnothing\neq O'\subseteq O\in\mathcal O$ then $O'\in\mathcal O$).

Then $\mathcal O$ contains an atom.
\end{lemm}
\begin{proof}
Just from Lemma~\ref{lemm.zorn.for.open.covers}, noting that an atom is precisely a $\subseteq$-minimal nonempty open set.
\end{proof}

\begin{corr}[Existence of atoms]
\label{corr.atom.exists}
Suppose $(\ns P,\opens)$ is a strongly chain-complete semitopology and $O\in\opens_{\neq\varnothing}$ is a nonempty open set.
Then $O$ contains an atom.
In symbols:
$$
\f{Atoms}(O)\neq\varnothing .
$$
\end{corr}
\begin{proof}
From Lemma~\ref{lemm.zorn.for.atoms}, considering $\{O'\in\opens \mid \varnothing\neq O'\subseteq O\}$ (which is nonempty because it contains $O$).
\end{proof}

\begin{rmrk}
\label{rmrk.topology.crush}
A simple observation is that if $(\ns P,\opens)$ is a strongly chain-complete topology --- thus, a strongly chain-complete semitopology whose opens are closed under finite intersections --- then the atom that exists by Corollary~\ref{corr.atom.exists} is unique, simply because if we have atoms $A$ and $A'$ then $A\cap A'$ is less than both and so by minimality must be equal to both.
See also Lemma~\ref{lemm.kernel.crush}.
\end{rmrk}

\jamiesubsubsection{Discussion}

\begin{rmrk}[Origin of terminology]
\label{rmrk.why.cover}
\leavevmode
\begin{enumerate*}
\item
The terminology ``$O$ covers $p$'' in Definition~\ref{defn.open.covers}(\ref{item.open.cover})
is adapted from order theory (see e.g.~\cite[\S 1.14]{priestley:intlo}), where we say that $y$ covers $x$ when $y>x$ and there exists no $z$ such that $y>z>x$.
\item
The terminology ``$A$ is an atom'' in Definition~\ref{defn.atomic.open.set}(\ref{item.atom})
is also adapted from order theory (see e.g.~\cite[\S 5.2]{priestley:intlo}), where we call $x$ an atom when it is a least element not equal to $\bot$ (i.e. when $x$ covers $\bot$).
\end{enumerate*}
\end{rmrk}

\begin{xmpl}[(Counter)examples of atoms and open covers]
\label{xmpl.p.not.in.O}
\leavevmode
\begin{enumerate}
\item
\emph{$p$ can be in multiple distinct atoms (minimal nonempty open sets), and/or open covers (minimal open sets that contain $p$).}

For instance, consider $\mathbb N$ with the semitopology generated by $1_\leq=\{0,1\}$ and $1_\geq = \{1,2,3,\dots\}$.
Then $1\in 1_\leq$ and $1\in 1_\geq$, and $1_\leq$ and $1_\geq$ are distinct minimal open sets (and also open covers of $1$).

A topology would compress this down to nothing: if $\{0,1\}$ is open and $\{1,2,3,\dots\}$ is open then their intersection $\{1\}$ would be open, and this would be the unique least open set containing $1$.
Because open sets in semitopologies are not necessarily closed under intersection, semitopologies permit richer structure.
\item
\emph{An open cover $O$ of $p$ is a minimal open set that contains $p$ --- but $O$ need not be an atom (a minimal nonempty open set).} 

Consider $\mathbb N$ with the semitopology generated by $i_\geq = \{i'\in\mathbb N \mid i'\geq i\}$.
Then $\f{Covers}(i)=\{i_\geq\}$ but (with this semitopology) $\f{Atoms}(\ns P)=\varnothing$; there are no least nonempty open sets.
\item\label{item.atom.need.not.contain.p}
\emph{An atom $A\in\f{atoms}(p)$ is a minimal nonempty open set that is a subset of a minimal open set that contains $p$ --- but $A$ need not contain $p$.}

For instance, consider $\mathbb N$ with the semitopology generated by $i_\leq = \{i'\in\mathbb N \mid i'\leq i\}$.
Then $\f{atoms}(i)=\{\{0\}\}$ for every $i$, because with this semitopology $\f{Atoms}(\mathbb N)=\{\{0\}\}$ and each $i$ is covered by $i_\leq$, and $\{0\}\subseteq i_\leq$.
However, we only have $i\in\{0\}$ when $i=0$.
\end{enumerate}
\end{xmpl}

\begin{rmrk}[Two open problems]
\label{rmrk.two.open.problems}
\leavevmode
\begin{enumerate}
\item\label{item.two.open.problems.1}
\emph{Topological characterisation of witness semitopologies.}

Following on from Remark~\ref{rmrk.characterise.witness.semitopologies}, we have seen that witness semitopologies are chain-complete, but that this does not precisely characterise witness semitopologies.
A topological characterisation of witness semitopologies, or a proof that such a characterisation is impossible, remains an open problem. 
To this end, the material in Subsection~\ref{subsect.declarative.witness} may be relevant, which relates witness semitopologies to a Turing-complete model of computation.
\item
\emph{Conditions on witness functions to guarantee (quasi)regularity.}

It remains an open problem to investigate conditions on witness functions to guarantee that every point is quasiregular.
In view of Proposition~\ref{prop.weakly.regular.to.regular} and Corollary~\ref{corr.no.finite.wr.c}, such conditions would suffice to guarantee the existence of a regular point in the finite case.
Regular points are well-behaved, so a system with at least one regular point is a system that is in some sense `somewhere sensible'. 
\end{enumerate}
\end{rmrk}

\jamiesection{Kernels: the atoms in a community}
\label{sect.kernels}

\jamiesubsection{Definition and examples}

\begin{rmrk}
\label{rmrk.arrow}
We have studied $\community(p)$ the community of a point and have seen that is has a rich mathematics. 
We also know from results (like Theorem~\ref{thrm.correlated}) and discussions (like Remark~\ref{rmrk.fundamental.consensus}) that to understand consensus in a semitopology, we have to understand its communities. 

It is now interesting to look at the atoms in a community (Definition~\ref{defn.atomic.open.set}; minimal nonempty open sets).
As we shall see later, the atoms in a community dictate its ability to act --- see e.g. Corollary~\ref{corr.boundary.kernel}, which is reminiscent of \emph{Arrow's theorem} from social choice theory --- so that understanding $\community(p)$ is, in a sense we will make formal, much the same thing as understanding the atoms in $\community(p)$ (see e.g. Proposition~\ref{prop.kernel.weakly.strongly.dense}).

Kernels are also interesting in and of themselves, so we start in this Section by studying them (culminating, out of several results, with Propositions~\ref{prop.kernel.atoms.subset.intersect} and~\ref{prop.kernel.O}).
\end{rmrk}

\begin{defn}
\label{defn.kernel}
Suppose $(\ns P,\opens)$ is a semitopology and $p\in\ns P$.
\begin{enumerate*}
\item\label{item.kernel}
Define $\kernel(p)$ the \deffont[kernel of $p$ ($\kernel(p)$)]{kernel of $p$} to be the union of the atoms in its community.
We give equivalent formulations which we may use as convenient:
$$
\begin{array}{r@{\ }l}
\kernel(p) 
=& \bigcup\{A\in\f{Atoms}(\ns P) \mid A\subseteq \intertwined{p} \}  
\\
=& \bigcup\{A\in\f{Atoms}(\ns P) \mid A\subseteq \community(p)\}  
\\
=& \bigcup\{A\in\opens \mid \varnothing\lessdot A\subseteq \community(p)\}  .
\end{array}
$$
Above, $\varnothing\lessdot A$ is just another way of saying that $A$ is an atom (minimal nonempty open set; see Definition~\ref{defn.atomic.open.set}), and $A\subseteq\intertwined{p}$ if and only if $A\subseteq\community(p)$ because $A$ is open and $\community(p)$ is just the open interior of $\intertwined{p}$ (Definition~\ref{defn.tn}(\ref{item.tn})).
\item\label{item.kernel.atom}
If $A$ is an atom that is a subset of $\kernel(p)$ (in symbols: $\varnothing\lessdot A\subseteq\kernel(p)$) then we might call $A$ a \deffont{kernel atom of $p$}.
\item\label{item.kernel.P}
Extend $\kernel$ to subsets $P\subseteq\ns P$ by taking a sets union:
$$
\kernel(P) = \bigcup\{\kernel(p) \mid p\in P\} .
$$
\end{enumerate*}
\end{defn}

\begin{figure}
\vspace{-1em}
\centering
\includegraphics[width=0.35\columnwidth]{diagrams/universal-counterexample.pdf}
\includegraphics[width=0.35\columnwidth]{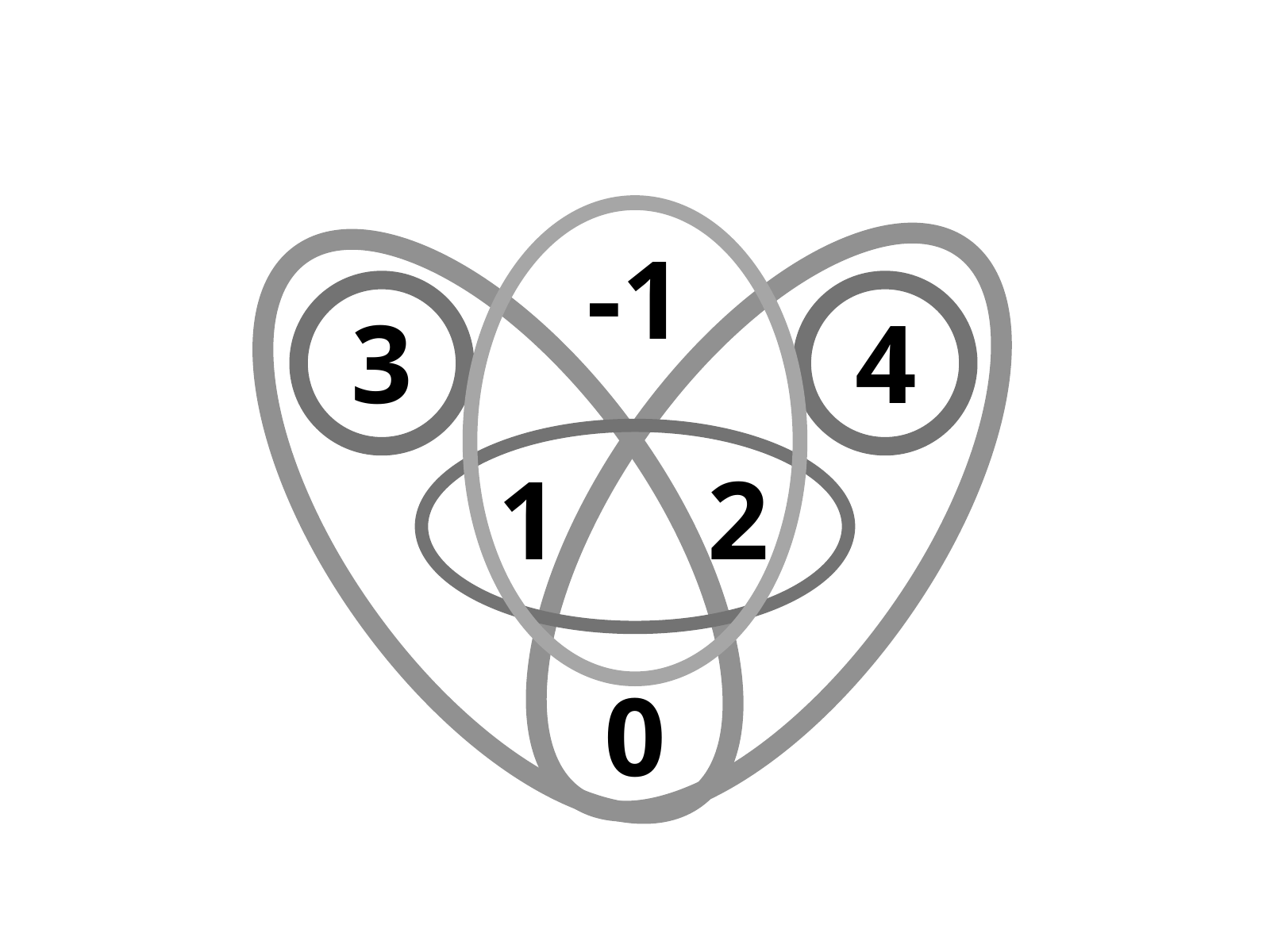}
\vspace{-0em}
\caption{Illustration of Example~\ref{xmpl.p.not.regular.2}(3\&4)}
\label{fig.irregular.2}
\end{figure}

We return to and extend Example~\ref{xmpl.p.not.regular}, and we include details of the kernels:
\begin{xmpl}
\label{xmpl.p.not.regular.2}
\leavevmode
\begin{enumerate}
\item\label{item.p.not.regular.R.2}
Take $\ns P$ to be $\mathbb R$ the real numbers, with its usual topology (which is also a semitopology), as per Example~\ref{xmpl.p.not.regular}(\ref{item.p.not.regular.R}).
Then:
\begin{itemize*}
\item
$\intertwined{x}=\{x\}$ and $\community(x)=\varnothing$ for every $x\in\mathbb R$.
\item
$\kernel(x)=\varnothing$ for every $x\in\mathbb R$.
\item
$\kernel(\mathbb R)=\varnothing$.
\end{itemize*}
\item\label{item.p.not.regular.012.2}
We take, as per Example~\ref{xmpl.p.not.regular}(\ref{item.p.not.regular.012}) and as illustrated in Figure~\ref{fig.012}, top-left diagram:
\begin{itemize*}
\item
$\ns P=\{0,1,2\}$.
\item
$\opens$ is generated by $\{0\}$ and $\{2\}$. 
\end{itemize*}
Then:
\begin{itemize*}
\item
$\intertwined{0}=\{0,1\}$ and $\community(0)=\interior(\intertwined{0})=\{0\}=\kernel(0)$. 
\item
$\intertwined{2}=\{1,2\}$ and $\community(2)=\interior(\intertwined{2})=\{2\}=\kernel(2)$. 
\item
$\intertwined{1}=\{0,1,2\}$ and $\community(1)=\{0,1,2\}$ and $\kernel(1)=\{0,2\}$. 
\item
$\kernel(\ns P)=\{0,2\}$.
\end{itemize*}
\item\label{item.p.not.regular.01234.2}
We take, as per Example~\ref{xmpl.p.not.regular}(\ref{item.p.not.regular.01234}), as illustrated in Figure~\ref{fig.irregular}, and as reproduced for convenience here in Figure~\ref{fig.irregular.2} (left-hand diagram):
\begin{itemize*}
\item
$\ns P=\{0,1,2,3,4\}$.
\item
$\opens$ is generated by $\{1,2\}$, $\{0,1,3\}$, $\{0,2,4\}$, $\{3\}$, and $\{4\}$.
\end{itemize*}
Then:
\begin{itemize*}
\item
$\intertwined{x}=\{0,1,2\}$ and $\community(x)=\interior(\intertwined{x})=\{1,2\}$ for $x\in\{0,1,2\}$.
\item
$\intertwined{x}=\{x\}=\community(x)$ for $x\in\{3,4\}$.
\item
$\kernel(x)=\{1,2\}$ for $x\in\{0,1,2\}$. 
\item
$\kernel(x)=\{x\}$ for $x\in\{3,4\}$. 
\item
$\kernel(\ns P)=\{1,2,3,4\}$.
\item
By construction $\kernel(\ns P)\subseteq\bigcup\f{Atoms}(\ns P)$, but but we see here that the inclusion may be strict, since e.g. $\{0,1,3\}$ is an atom in this example but $0\notin\kernel(\ns P)$.
\end{itemize*}
\item\label{item.p.not.regular.01234.2b}
We add one point to part~\ref{item.p.not.regular.01234.2} of this example, $\minus 1$, which is intertwined with $0$, $1$, and $2$ but is not in a minimal nonempty open set, as illustrated in Figure~\ref{fig.irregular.2} (right-hand diagram):
\begin{itemize*}
\item
$\ns P=\{\minus 1,0,1,2,3,4\}$.
\item
$\opens$ is generated by $\{\minus 1,1,2\}$, $\{1,2\}$, $\{0,1,3\}$, $\{0,2,4\}$, $\{3\}$, and $\{4\}$.
\end{itemize*}
Then:
\begin{itemize*}
\item
$\intertwined{x}=\{\minus 1,0,1,2\}$ and $\community(x)=\interior(\intertwined{x})=\{\minus 1,1,2\}$ for $x\in\{\minus 1,0,1,2\}$.
\item
$\intertwined{x}=\{x\}=\community(x)$ for $x\in\{3,4\}$.
\item
$\kernel(x)=\{1,2\}$ for $x\in\{\minus 1,0,1,2\}$. 
\item
$\kernel(x)=\{x\}$ for $x\in\{3,4\}$. 
\item
$\kernel(\ns P)=\{1,2,3,4\}$.
\item
By construction $\kernel(p)\subseteq\community(p)\subseteq\intertwined{p}$, but the inclusions may be strict.
For instance:
$$
\begin{array}{r@{\ }l}
\kernel(0)=\kernel(\minus 1)=\{1,2\}
\subsetneq& 
\community(0)=\community(\minus 1)=\{\minus 1,1,2\}
\\
\subsetneq&
\intertwined{0}=\intertwined{\minus 1}=\{\minus 1,0,1,2\}.
\end{array}
$$
\end{itemize*}
\item\label{item.nonempty.community.empty.kernel}
We take $\ns P=\mathbb N$, with the semitopology (also a topology) generated by final subsets $n_\geq = \{n'\in\mathbb N\mid n'\geq n\}$ for $n\in\mathbb N$.
Then $\intertwined{n}=\mathbb N=\community(n)$ for every $n\in\mathbb N$, and $\kernel(n)=\varnothing$ (because there is no minimal nonempty open set). 
\end{enumerate}
\end{xmpl}

We warm up with a couple of simple lemmas:
\begin{lemm}
\label{lemm.kernel.atoms.intersect}
Suppose $(\ns P,\opens)$ is a semitopology and $p\in\ns P$ is a regular point.

Then all kernel atoms of $p$ intersect, or in symbols:
$$
\varnothing\lessdot A,A'\subseteq\kernel(p) \quad\text{implies}\quad A\between A'.
$$
\end{lemm}
\begin{proof}
By construction in Definition~\ref{defn.kernel}(\ref{item.kernel}) $A,A'\subseteq\kernel(p)\subseteq\community(p)$.
By regularity (Definition~\ref{defn.tn}(\ref{item.regular.point})) $\community(p)$ is transitive.
Then $A\between \community(p)\between A'$ and by transitivity (Definition~\ref{defn.transitive}) it follows that $A\between A'$. 
\end{proof}

\begin{lemm}
\label{lemm.kernel.crush}
Suppose $(\ns P,\opens)$ is a \emph{topology} --- thus: a semitopology whose open sets are closed under intersections --- and $p\in\ns P$ is regular.
Then one of the following holds:
\begin{itemize*}
\item
$\kernel(p)=\varnothing$.
\item
$\kernel(p)=A$ for some atom $\varnothing\lessdot A$.
\end{itemize*}
\end{lemm}
\begin{proof}
Suppose $\kernel(p)\neq\varnothing$, and suppose there exist two atoms $A,A'\subseteq\kernel(p)$.
Then (just as already noted in Remark~\ref{rmrk.topology.crush}) $A\cap A'$ is an open set.
It is not empty because $A\between A'$ by Lemma~\ref{lemm.kernel.atoms.intersect}.
By minimality, $A=A\cap A'=A'$.
Thus, being a topology crushes Definition~\ref{defn.kernel} down to be at most a single atom.
\end{proof}

\jamiesubsection{Characterisations of the kernel}

We open with a non-implication:
\begin{lemm}
\label{lemm.reg.ker.0}
Suppose $(\ns P,\opens)$ is a semitopology and $p\in\ns P$ is regular (so $p\in\community(p)\in\topens$).
Then it is not necessarily the case that $\kernel(p)\neq\varnothing$.
\end{lemm}
\begin{proof}
A counterexample is Example~\ref{xmpl.p.not.regular.2}(\ref{item.nonempty.community.empty.kernel}).
In full: we consider $\mathbb N$ with the semitopology generated by $n_\geq = \{n'\in\mathbb N\mid n'\geq n\}$ for $n\in\mathbb N$.
Then $\intertwined{n}=\mathbb N=\community(n)$ for every $n\in\mathbb N$, but $\kernel(n)=\varnothing$ because there is no minimal nonempty open set.
\end{proof}

We can exclude the case noted in the proof of Lemma~\ref{lemm.reg.ker.0} by restricting to strongly chain-complete semitopologies. 
\begin{lemm}
\label{lemm.nonempty.community}
Suppose $(\ns P,\opens)$ is a strongly chain-complete semitopology --- in particular, this holds if $\ns P$ is finite by Proposition~\ref{prop.finite.chain-bounded} --- and $p\in\ns P$.
Then:
\begin{enumerate*}
\item\label{item.nonempty.community.1}
$\community(p)=\varnothing$ if and only if $\kernel(p)=\varnothing$, and equivalently $\community(p)\neq\varnothing$ if and only if $\kernel(p)\neq\varnothing$.

In words: $p$ has a nonempty community if and only if it has a nonempty kernel.
\item\label{nonempty.community.topen}
If $p$ is regular then $\kernel(p)$ is a topen subset of $\community(p)$ (nonempty transitive and open, see Definition~\ref{defn.transitive}(\ref{transitive.cc})).

(See also Lemma~\ref{lemm.topen.max.min}, which proves a stronger version of this property for the kernel atoms of a regular $p$.)
\end{enumerate*}
\end{lemm}
\begin{proof}
\leavevmode
\begin{enumerate}
\item
Suppose $\varnothing\neq\community(p)=\interior(\intertwined{p})$.
Then by Corollary~\ref{corr.atom.exists} (since $\ns P$ is strongly chain-complete) $\community(p)$ contains at least one atom $A$, which is a subset of $\community(p)$ by construction, and so $A\in\kernel(p)$.

Conversely, if there exists an atom $A\in\kernel(p)$ then (since an atom is by assumption a nonempty set) we have $\varnothing\neq A\subseteq\community(p)$. 
\item
Suppose $p$ is regular.
Unpacking Definition~\ref{defn.tn}(\ref{item.regular.point}) this means that $p\in\community(p)$.
Thus in particular $\community(p)\neq\varnothing$, and by part~\ref{item.nonempty.community.1} of this result $\kernel(p)\neq\varnothing$. 

So $\kernel(p)$ is a nonempty subset of $\community(p)$.
By Theorem~\ref{thrm.max.cc.char} $\community(p)$ is a (maximal) topen, and by Lemma~\ref{lemm.transitive.subset}(\ref{item.transitive.subset.2}) $\varnothing\neq\kernel(p)\subseteq\community(p)$ is topen as required.
\qedhere\end{enumerate}
\end{proof}

\begin{rmrk}
\label{rmrk.kernel.not.neighbourhood.of.p}
Note in Lemma~\ref{lemm.nonempty.community}(\ref{nonempty.community.topen}) that $\kernel(p)$ need not be a topen neighbourhood of $p$, simply because $p$ (even if it is regular) might generate a topen community $\community(p)$ but need not necessarily be in an atom in that community.
See Example~\ref{xmpl.p.not.regular.2}(\ref{item.p.not.regular.01234.2b}) taking $p=0$ or $p=\minus 1$, or Lemma~\ref{lemm.kernel.non-implications}(\ref{item.kernel.non-implications.2}).
\end{rmrk}

We complement Lemma~\ref{lemm.nonempty.community} with some non-implications:
\begin{lemm}
\label{lemm.kernel.non-implications}
Suppose $(\ns P,\opens)$ is a semitopology.
Then:
\begin{enumerate*}
\item\label{item.kernel.non-implications.1}
$\kernel(p)\neq\varnothing$ does not imply that $p$ is regular.
\item\label{item.kernel.non-implications.2}
$\kernel(p)$ topen does not imply that $p$ is regular. 
\end{enumerate*}
\end{lemm}
\begin{proof}
\leavevmode
\begin{enumerate}
\item
See Example~\ref{xmpl.p.not.regular}(\ref{item.p.not.regular.012}):
then $\kernel(1)=\{0,2\}\neq\varnothing$ but (as noted in Lemma~\ref{lemm.p.not.regular}) $1$ is not regular. 
\item
See Example~\ref{xmpl.p.not.regular}(\ref{item.point.not.regular.but.community.is.topen}): so $\ns P=\{0,1,2,3,4\}$ and $\opens$ is generated by $\{1,2\}$, $\{0,1,3\}$, $\{0,2,4\}$, $\{3\}$, and $\{4\}$ and $\kernel(0)=\community(0)=\{1,2\}$, and this is (nonempty and) topen, but $0$ is not regular since $0\not\oldin\community(0)=\{1,2\}$.
\item
Take $\ns P=\{0,1\}$ and set $\opens=\{\varnothing,\{0\},\{0,1\}\}$.
Then $1$ is regular but $1\not\oldin\kernel(p)=\{0\}$.
\qedhere\end{enumerate}
\end{proof}

\begin{lemm}
\label{lemm.topen.max.min}
Suppose $(\ns P,\opens)$ is a semitopology and $p\in\ns P$ is a regular point and $A\subseteq\community(p)$.
Then the following are equivalent:
\begin{enumerate*}
\item
$A$ is a kernel atom of $p$ ($\varnothing\lessdot A\subseteq\community(p)$).
\item
$A$ is a minimal topen in $\community(p)$.
\end{enumerate*}
\end{lemm}
\begin{proof}
We prove two implications:
\begin{itemize}
\item
Suppose $A$ is a kernel atom of $p$.
By assumption in Definition~\ref{defn.kernel}(\ref{item.kernel.atom}) it is an atom (a minimal nonempty open set) in $\community(p)$, and by Lemma~\ref{lemm.transitive.subset}(\ref{item.transitive.subset.2}) it is topen; so it is necessarily a minimal topen.
\item
Suppose $A$ is a minimal topen in $\community(p)$ and suppose $A'\subseteq A$ is any nonempty open set.
By Lemma~\ref{lemm.transitive.subset}(\ref{item.transitive.subset.2}) $A'$ is topen, so by minimality $A=A'$.
Thus, $A$ is an atom in $\community(p)$, and so is a kernel atom of $p$.
\qedhere\end{itemize}
\end{proof}

\begin{rmrk}
\label{rmrk.topen.minimax}
The proof of Lemma~\ref{lemm.topen.max.min} above is elementary given our results so far, but it makes a useful observation. 
Recall from Theorem~\ref{thrm.max.cc.char} that if $p$ is regular (so $p\in\community(p)\in\topens$) then $\community(p)$ is a maximal topen, and recall from Definition~\ref{defn.kernel} that a kernel atom is an atom (i.e. a minimal nonempty open set) in $\community(p)$.
So we can read Lemma~\ref{lemm.topen.max.min} as follows: 
\begin{quote}
\emph{A kernel atom is a minimal topen inside a maximal topen.}
\end{quote}
Thus for regular $p$, $\kernel(p)$ and $\community(p)$ are in some sense dual: the community of $p$ is the maximal topen containing $p$, and the kernel of $p$ is the union of the minimal topens inside that maximal topen.
\end{rmrk}

So Lemma~\ref{lemm.topen.max.min} tells us that for regular $p$, the kernel atoms of $p$ are the minimal topens that are subsets of the community of $p$.
Proposition~\ref{prop.kernel.atoms.subset.intersect} strengthens this to show that in fact, the kernel atoms of regular $p$ are also the minimal topens that even \emph{intersect} with the community of $p$ (the significance of this to consensus is discussed in Remark~\ref{rmrk.fundamental.consensus}):
\begin{prop}
\label{prop.kernel.atoms.subset.intersect}
Suppose that:
\begin{itemize*}
\item
$(\ns P,\opens)$ is a semitopology.\footnote{We do not seem to need $\ns P$ to be (strongly) chain-complete here.  This is simply because we normally use strong chain-completeness to ensure that atoms and open covers exist, but in this result this is assumed.}
\item
$A\in\f{Atoms}(\ns P)$ is an atom.
\item
$p\in\ns P$ is a regular point (so by Definition~\ref{defn.tn}(\ref{item.tn}) $p\in\community(p)\in\topens$).
\item
$O\in\topens$ and $O\between\community(p)$, so $O$ is any topen set that intersects the community of $p$ (at least one such exists, by regularity, namely $\community(p)$ itself).\footnote{By Proposition~\ref{prop.topen.intersect.subset}, this is equivalent to $O\subseteq\community(p)$.  We use $O\between\community(p)$ because it yields a stronger form of the result.}
\end{itemize*}
Then the following are all equivalent:
\begin{enumerate*}
\item
$A\subseteq\kernel(p)$.

In words: $A$ is a kernel atom of $p$.
\item
$A\subseteq\community(p)$.

In words: $A$ is an atom in the community of $p$.
\item
$A$ is topen and $A\between O$.

In words: $A$ is topen and intersects $O$.
\end{enumerate*}
In particular, if $A$ is a topen atom\footnote{An atom is a minimal nonempty open set, and a topen is a nonempty open transitive set; so saying `topen atom' is just the same as saying `transitive atom'.} then we have:
$$
A\subseteq \kernel(p) 
\ \liff\ 
A\subseteq \community(p) 
\ \liff\ 
A\between O
\ \liff\ 
A\between\kernel(p) 
\ \liff\ 
A\between\community(p) . 
$$
\end{prop}
\begin{proof}
We consider a cycle of implications:
\begin{itemize}
\item
Suppose $A\subseteq\kernel(p)$.
By construction in Definition~\ref{defn.kernel}(\ref{item.kernel}) $\kernel(p)\subseteq\community(p)$, so $A\subseteq\community(p)$.
\item
Suppose $A\subseteq\community(p)$.
By Definition~\ref{defn.atomic.open.set}(\ref{item.atom}) (since $A$ is an atom) $A$ is nonempty.
Then $A\between\community(p)\between O$.
By regularity $\community(p)$ is topen, so by transitivity (Definition~\ref{defn.transitive}) $A\between O$ as required.
\item
Suppose $A$ is topen and $A\between O$.
By assumption $A\between O\between\community(p)$ so by transitivity of $O$, $A\between\community(p)$.
By Proposition~\ref{prop.topen.intersect.subset} $A\subseteq\community(p)$ and it follows from Definition~\ref{defn.kernel}(\ref{item.kernel.atom}) that $A\subseteq\kernel(p)$ as required.
\end{itemize}
The equivalence 
$$
A\between O\liff A\between\kernel(p) \liff A\between\community(p)
$$
then follows routinely from the above, noting the equivalence $A\between O\liff A\subseteq\community(p)$ and choosing $O=\kernel(p)$ or $O=\community(p)$.
\end{proof}

\begin{corr}
\label{corr.kernel.as.topen.atoms.intersecting.O}
Suppose $(\ns P,\opens)$ is a semitopology and $p\in\ns P$ and $O\in\topens$ and $p\in O$ (so $p$ is regular and $O$ is some topen neighbourhood of $p$).
Then 
$$
\kernel(p) = \bigcup\{A{\in}\topens(\ns P) \mid \varnothing\lessdot A\between O\}.
$$
In words: for any topen neighbourhood $O$ of $p$, $\kernel(p)$ is equal to the union of the topen atoms that intersect that neighbourhood. 
\end{corr}
\begin{proof}
Unpacking Definition~\ref{defn.kernel}(\ref{item.kernel}), $\kernel(p)$ is the union of atoms $A\subseteq\community(p)$.
We use Proposition~\ref{prop.kernel.atoms.subset.intersect}. 
\end{proof}

Lemma~\ref{lemm.atom.is.kernel.atom} explicitly characterises the union of all kernels as the union of all transitive atoms, which (given the results above) is what one might expect:
\begin{lemm}
\label{lemm.atom.is.kernel.atom}
Suppose $(\ns P,\opens)$ is a semitopology.
Then:
\begin{enumerate*}
\item\label{item.atom.is.kernel.atom.1}
If $A\subseteq\ns P$ is a transitive atom then $A\subseteq\kernel(p)$ for every $p\in A$.

In words we can say: a transitive atom is a kernel atom for any points that it contains.
\item
$\kernel(\ns P)$ is the union of the transitive atoms in $\ns P$.
\end{enumerate*}
\end{lemm}
\begin{proof}
\leavevmode
\begin{enumerate}
\item
If $p\in A\in\topens$ then $A$ is a topen neighbourhood for $p$. 
By Theorem~\ref{thrm.max.cc.char} $p\in A\subseteq \community(p)$.
But then by construction $A$ is an atom in $\community(p)$ so by Definition~\ref{defn.kernel}(\ref{item.kernel}) $A\subseteq\kernel(p)$.
\item
It follows from Lemma~\ref{lemm.topen.max.min} and Definition~\ref{defn.kernel}(\ref{item.kernel.P}) that every atom in $\kernel(\ns P)$ is (topen and so) transitive.
Conversely by part~\ref{item.atom.is.kernel.atom.1} of this result every transitive atom is in the kernel of the community of its points. 
\qedhere\end{enumerate}
\end{proof}

\jamiesubsection{Further properties of kernels}

\jamiesubsubsection{Intersections between the kernel of $p$ and its open neighbourhoods}

Lemma~\ref{lemm.ker.intersect} is quite easy to prove by following definitions and applying transitivity properties, but it makes a useful point:
\begin{lemm}
\label{lemm.ker.intersect}
Suppose that:
\begin{itemize*}
\item
$(\ns P,\opens)$ is a semitopology.
\item
$p\in\ns P$ is a regular point.
\item
$A$ is a kernel atom for $p$.  In symbols: $\varnothing\lessdot A\subseteq\kernel(p)$.
\end{itemize*} 
Then 
$$
\Forall{O{\in}\opens} p\in O \limp A\between O .
$$
In words: 
\begin{quote}
Every open neighbourhood of a regular $p$ intersects every kernel atom of $p$.\footnote{This property is a bit subtle, because it is not necessarily the case that $p\in\kernel(p)$ (cf. Remark~\ref{rmrk.kernel.not.neighbourhood.of.p}).  So a kernel atom $\varnothing\lessdot A\subseteq\kernel(p)$ is is not itself necessarily a neighbourhood of $p$, but it still has a property of `oversight' over $p$ in the sense that it intersects with every quorum (open neighbourhood) that $p$ has.}
\end{quote} 
\end{lemm}
\begin{proof}
By our assumption that $p$ is regular we have $p\in\community(p)\in\topens$ (Definition~\ref{defn.tn}(\ref{item.regular.point})), and we assumed $p\in O$, so $O\between\community(p)$.

Also by assumption $A\between\community(p)$, since $A\subseteq\kernel(p)\subseteq\community(p)$ by Definition~\ref{defn.kernel}(\ref{item.kernel.atom}\&\ref{item.kernel}).
Thus $O\between\community(p)\between A$.
Now $\community(p)$ is topen, thus it is transitive (Definition~\ref{defn.transitive}(\ref{transitive.transitive})) and so $O\between A$ as required. 
\end{proof}

\begin{prop}
\label{prop.kernel.O}
$(\ns P,\opens)$ is a semitopology and $p$ is regular and $p\in O\in\opens$.
Then:
\begin{enumerate*}
\item 
The kernel of $p$ is a subset of the union of the atoms intersecting $O$.
In symbols:
$$
\kernel(p) \subseteq \bigcup\{A{\in}\opens \mid \varnothing\lessdot A\between O\} = \bigcup\{A{\in}\f{Atoms}(\ns P) \mid A\between O\}.
$$
\item
The inclusion may be strict, even if $O$ is an open cover of $p$ (in symbols: $O\gtrdot p$).
\item
The inclusion may be strict, even if $O\gtrdot p$ is a topen (transitive open) cover of $p$.
\item\label{item.kernel.O.4}
If $O$ is a topen cover of $p$, then the kernel of $p$ is precisely equal to the union of the transitive atoms intersecting $O$.
In symbols: 
$$
\kernel(p) = \bigcup\{A{\in}\topens(\ns P) \mid \varnothing\lessdot A\between O\}.
$$
\end{enumerate*}
\end{prop}
\begin{proof}
\leavevmode
\begin{enumerate}
\item
For the inclusion we just combine Lemma~\ref{lemm.ker.intersect} with Definition~\ref{defn.kernel}(\ref{item.kernel}).
\item
To see how the inclusion may be strict, see Example~\ref{xmpl.strict.kernel.O}(\ref{item.strict.kernel.O.1}).
\item
To see how the inclusion may be strict, even for transitive $O$, see Example~\ref{xmpl.strict.kernel.O}(\ref{item.strict.kernel.O.2}).
\item
This just repeats Corollary~\ref{corr.kernel.as.topen.atoms.intersecting.O}.
\qedhere
\end{enumerate}
\end{proof}

\begin{figure}
\vspace{-1em}
\centering
\includegraphics[width=0.35\columnwidth]{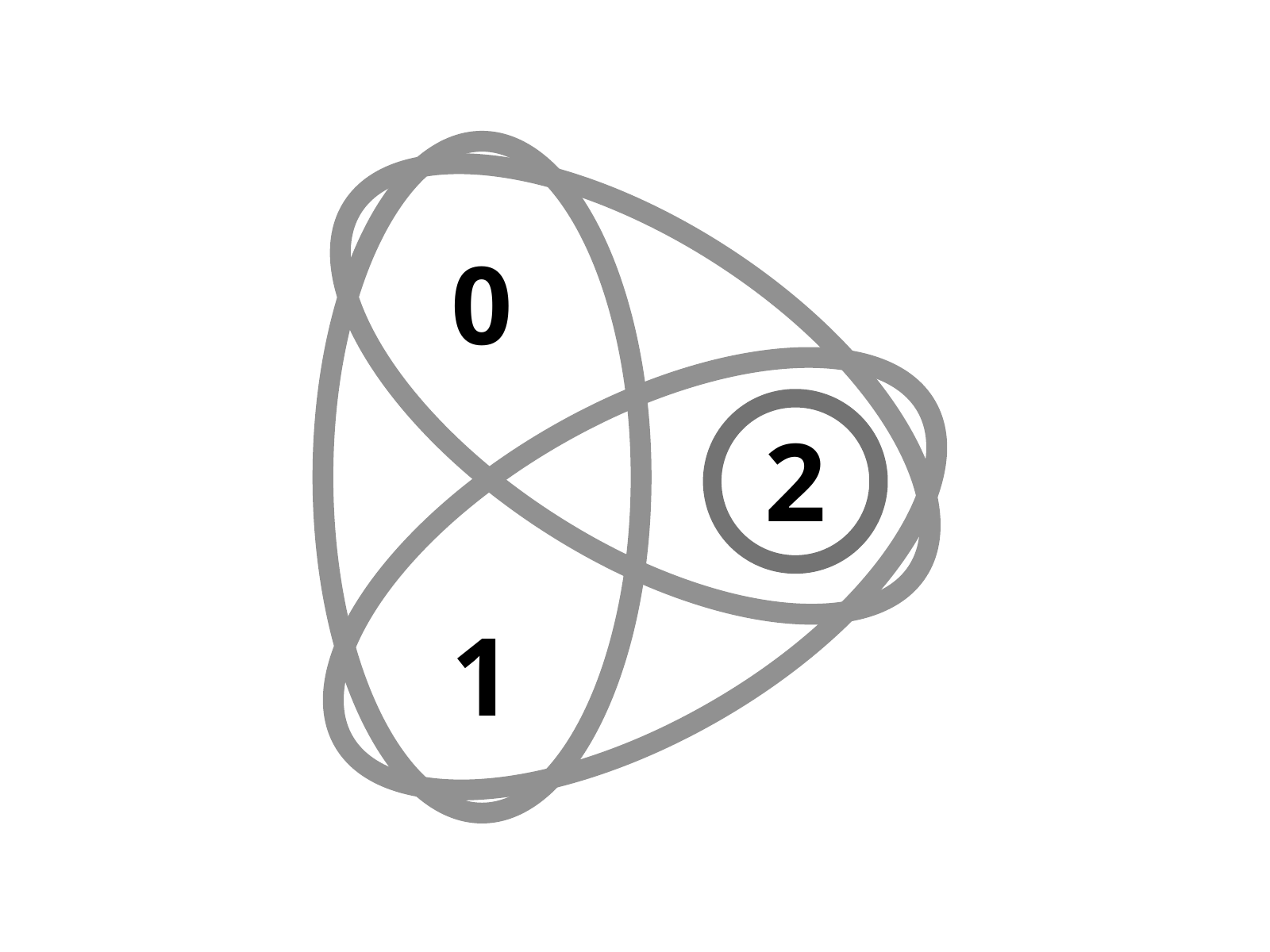}
\hspace{-2em}
\includegraphics[width=0.35\columnwidth]{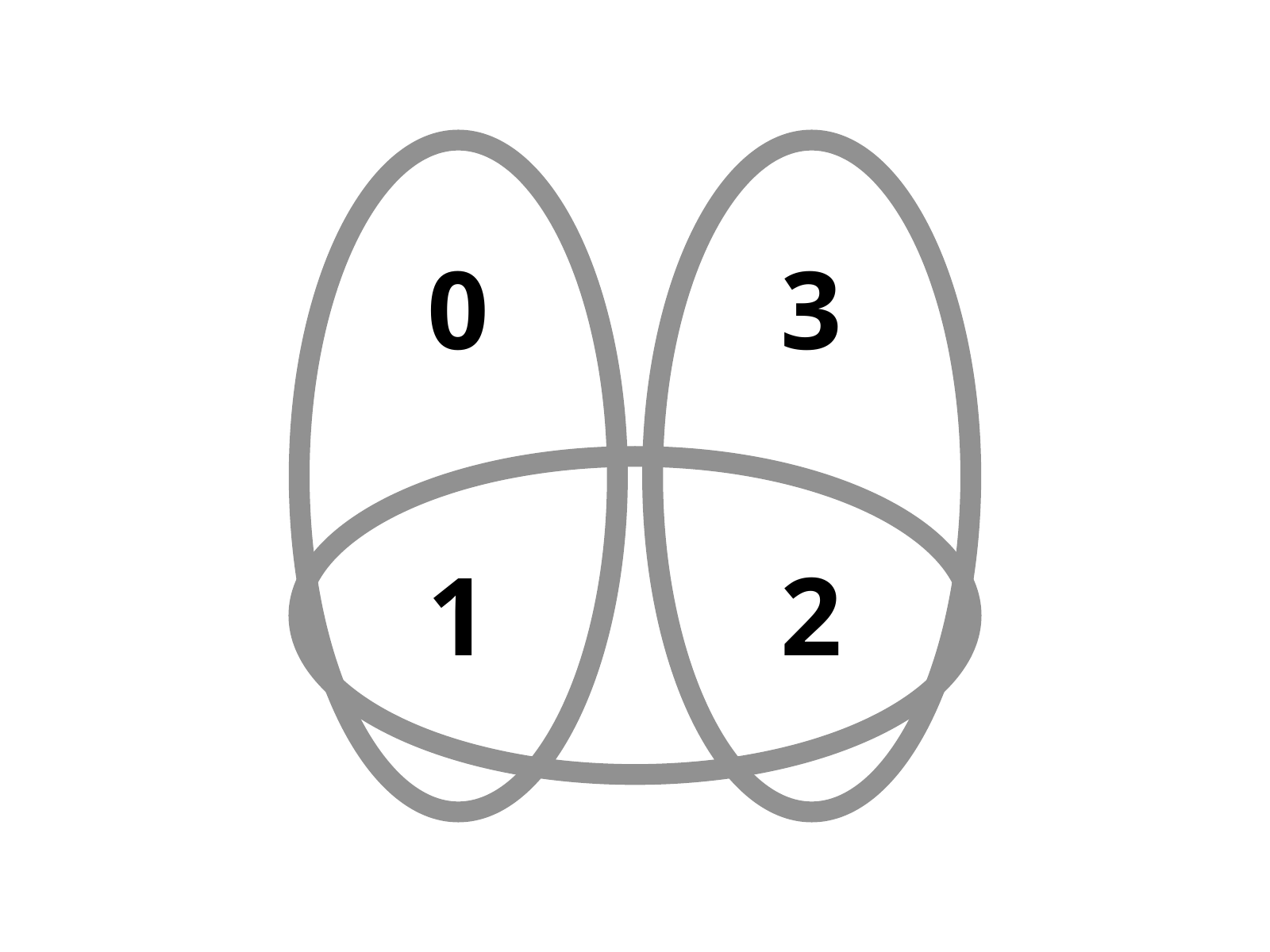}
\caption{The semitopologies in Example~\ref{xmpl.strict.kernel.O}}
\label{fig.ovals}
\end{figure}

\begin{xmpl}
\label{xmpl.strict.kernel.O}
\leavevmode
\begin{enumerate}
\item\label{item.strict.kernel.O.1}
Take $\ns P=\{0,1,2\}$ and let opens be generated by $\{0,1\}$ and $\{0,2\}$ and $\{1,2\}$ and $\{2\}$, as illustrated in Figure~\ref{fig.ovals} (left-hand diagram).

Set $p=1$ and $O=\{1,2\}$.
Then we can calculate that:
\begin{itemize*}
\item
$0$, $1$, and $2$ are all regular.
\item
The community and kernel of $1$ and $0$ are equal to $\{0,1\}$ --- $2$ is not intertwined with $0$ or $1$ because $\{2\}\cap\{0,1\}=\varnothing$.
\item
The community and kernel of $2$ are equal to $\{2\}$. 
\item
$\{1,2\}$ is an open cover of $1$.
\item
The union of the atoms that intersect with $\{1,2\}$ is the whole space $\{0,1,2\}$.
\end{itemize*}
Thus $\kernel(1) =\{0,1\}\subsetneq \bigcup\{A{\in}\f{Atoms}(\ns P) \mid A\between \{1,2\}\}=\{0,1\}\cup\{2\}=\{0,1,2\}$.
\item\label{item.strict.kernel.O.2}
Take $\ns P=\{0,1,2,3\}$ and let opens be generated by $\{0,1\}$ and $\{1,2\}$ and $\{2,3\}$, as illustrated in Figure~\ref{fig.ovals} (right-hand diagram). 

Set $p=1$ and $O=\{0,1\}$.
Then we can calculate that:
\begin{itemize*}
\item
$\ns P$ splits into two disjoint topen sets: $\{0,1\}$ and $\{2,3\}$.
So $O$ is topen.
\item
The community and kernel of $0$ and $1$ are equal to $\{0,1\}$ --- $2$ is not intertwined with $0$ or $1$ because $\{2,3\}\cap\{0,1\}=\varnothing$.
So $\{1,2\}$ is an atom, but it is not transitive.
\item
The community and kernel of $2$ and $3$ are equal to $\{2,3\}$. 
\item
$\{0,1\}$ is an open cover of $1$.
\end{itemize*}
Thus $\kernel(1) =\{0,1\}\subsetneq \bigcup\{A{\in}\f{Atoms}(\ns P) \mid A\between \{0,1\}\}=\{0,1\}\cup\{1,2\}=\{0,1,2\}$.
\end{enumerate}
\end{xmpl}

\begin{rmrk}[Algorithmic content of Proposition~\ref{prop.kernel.O}]
Proposition~\ref{prop.kernel.O} 
reduces the problem of computing kernels to the problem of identifying transitive sets.\footnote{We considered \emph{that} question in results including Proposition~\ref{prop.views.of.regularity} and Theorem~\ref{thrm.up.down.char}.}
Once we have this, an algorithm for computing $\kernel(p)$ for regular $p$ follows:
\begin{itemize*}
\item
Compute a transitive open neighbourhood $O$ of $p$ --- for example using the algorithm outlined in 
Remark~\ref{rmrk.computing.open.sets} to compute open neighbourhoods of $p$, and testing until we find one that is transitive.
At least one transitive cover of $p$ exists, by our assumption that $p$ is regular.
\item
For each $p'\in O$, compute all the atoms that contain $p'$ --- for example by computing the open neighbourhoods of $p'$ and checking which are atoms, and are transitive.
\end{itemize*}
By Proposition~\ref{prop.kernel.O}(\ref{item.kernel.O.4}), this collection of transitive atoms that intersect with $O$, will return the kernel atoms of $p$. 
\end{rmrk}

We conclude by noting a non-result:
\begin{lemm}
Suppose $(\ns P,\opens)$ is a semitopology and $p\in\ns P$ is regular.
Recall from Theorem~\ref{thrm.up.down.char} that $\community(p)$ the community of $p$ is the greatest transitive open neighbourhood of $p$, so that any \emph{transitive} open neighbourhood of $p$ is contained in the community of $p$.

However, there may still exist a non-transitive open cover of $p$ that is not contained in the community of $p$.
\end{lemm}
\begin{proof}
It suffices to provide a counterexample, and as it happens we have just considered one.
Consider Example~\ref{xmpl.strict.kernel.O}(\ref{item.strict.kernel.O.2}), as illustrated in Figure~\ref{fig.ovals} (right-hand diagram).
Then $\intertwined{1}=\community(1)=\{0,1\}$ and $\{1,2\}$ is an open cover of $1$ and $\{1,2\}\not\subseteq\{0,1\}$. 
\end{proof}

\jamiesubsubsection{Idempotence properties of the kernel and community}

\begin{rmrk}
In Definitions~\ref{defn.tn}(\ref{item.community.P}) and~\ref{defn.kernel}(\ref{item.kernel.P}) we extend the notions of community and kernel of a set of points, using sets union.
This allows us to take the community of a community $\community(\community(p))$, then kernel of a kernel $\kernel(\kernel(p))$, and so forth.
Does doing this add any information?
We would hope not --- but we need to check.

In this Subsection we take check this for regular points, and see that they display good behaviour (e.g.: the community of the community is just the community, and so forth).
The proofs also illuminate how regularity condition ensures good behaviour.
\end{rmrk}

\begin{lemm}
\label{lemm.double.community}
Suppose $(\ns P,\opens)$ is a semitopology and suppose $p\in\ns P$ is regular.
Then
$$
\community(\community(p))=\community(p).
$$
\end{lemm}
\begin{proof}
We prove two subset inclusions:
\begin{itemize}
\item
Suppose $q\in\community(\community(p))$, so unpacking Definition~\ref{defn.tn}(\ref{item.community.P}) there exists $p'\in\community(p)$ such that $q\in\community(p')$.
By Corollary~\ref{corr.regular.is.regular} (since $p$ is regular) $\community(p')=\community(p)$, so $q\in\community(p)$.

$q$ was arbitrary, and it follows that $\community(\community(p))\subseteq\community(p)$.
\item
Suppose $q\in\community(p)$.
Then by Corollary~\ref{corr.regular.is.regular} (since $p$ is regular) $\community(q)=\community(p)\in\topens$, so in particular $q\in\community(q)$.

$q$ was arbitrary, and it follows that $\community(p)\subseteq\community(\community(p))$. 
\qedhere\end{itemize}
\end{proof}

\begin{corr}
\label{corr.community.kernel.p}
Suppose $(\ns P,\opens)$ is a semitopology and $p\in\ns P$ is regular.
Suppose further that $\kernel(p)\neq\varnothing$ (if $\ns P$ is strongly chain-complete or finite then by Lemma~\ref{lemm.nonempty.community}(\ref{nonempty.community.topen}) and Proposition~\ref{prop.finite.chain-bounded} $\kernel(p)\neq\varnothing$ is guaranteed).
Then
$$
\community(p)=\community(\kernel(p)).
$$
\end{corr}
\begin{proof}
Suppose $q\in\community(p)$ and pick any $k\in\kernel(p)\subseteq\community(p)$.
Then $k\in\community(p)$ so by Corollary~\ref{corr.regular.is.regular} $\community(p)=\community(k)$ so $q\in\community(k)$.
Thus $\community(p)\subseteq\community(\kernel(p))$.

Furthermore $\community(\kernel(p))\subseteq\community(\community(p))$ is a structural fact of Definition~\ref{defn.tn}(\ref{item.community.P}) and the fact, noted above, that $\kernel(p)\subseteq\community(p)$.

We finish with Lemma~\ref{lemm.double.community}:
$$
\community(p)\subseteq\community(\kernel(p))\subseteq\community(\community(p)) \stackrel{L\ref{lemm.double.community}}= \community(p) .
\qedhere$$ 
\end{proof}

\begin{lemm}
\label{lemm.kernel.community}
Suppose $(\ns P,\opens)$ is a semitopology and suppose $p\in\ns P$ is regular.
Then 
$$
\kernel(\community(p))=\kernel(p).
$$
\end{lemm}
\begin{proof}
Unpacking Definition~\ref{defn.kernel}, $\kernel(\community(p))=\bigcup\{\kernel(p')\mid p'\in\community(p)\}$ and for each $p'\in\community(p)$ we have $\kernel(p')=\bigcup\{A\subseteq\ns P \mid \varnothing\lessdot A\subseteq \community(p')\}$. 
By Corollary~\ref{corr.regular.is.regular}, $\community(p)=\community(p')$ for every $p'\in\community(p)$, and threading this equality through the definitions above, we obtain the result.
\end{proof}

\begin{lemm}
Suppose $(\ns P,\opens)$ is a semitopology and suppose $p\in\ns P$ is regular.
Then
$$
\kernel(p)=\kernel(\kernel(p)).
$$
\end{lemm}
\begin{proof}
If $\kernel(p)=\varnothing$ then the result is immediate.
So suppose $\kernel(p)\neq\varnothing$.
We show two subset inclusions.
\begin{itemize}
\item
To prove $\kernel(p)\subseteq\kernel(\kernel(p))$ we can reason as follows:
$$
\begin{array}{r@{\ }l@{\quad}l}
\kernel(\kernel(p))\subseteq&\kernel(\community(p))
&\kernel(p){\subseteq}\community(p), \text{Def.~\ref{defn.kernel}(\ref{item.kernel.P})}
\\
=&\kernel(p) &\text{Lemma~\ref{lemm.kernel.community}}
\end{array}
$$
\item 
To prove $\kernel(\kernel(p))\subseteq\kernel(p)$ we note that a kernel is a union of atoms in Definition~\ref{defn.kernel}(\ref{item.kernel}), and we reason as follows, for an atom $\varnothing\lessdot A$ (which exists because $\kernel(p)$ is a union of atoms and we assumed $\kernel(p)\neq\varnothing$):
$$
\begin{array}{r@{\ }l@{\quad}l}
A\subseteq\kernel(p) 
\liff& A\subseteq\community(p) 
&\text{Definition~\ref{defn.kernel}(\ref{item.kernel})}
\\
\liff& A\subseteq\community(\kernel(p)) 
&\community(p)=\community(\kernel(p))=\community(\community(p))
\\
\liff& A\subseteq\kernel(\kernel(p))
&\text{Definition~\ref{defn.kernel}(\ref{item.kernel})} .
\end{array}
$$
Above, $\community(p)=\community(\kernel(p))=\community(\community(p))$ follows from Lemma~\ref{lemm.double.community} and Corollary~\ref{corr.community.kernel.p} (since we assumed $\kernel(p)\neq\varnothing$).
\qedhere\end{itemize}
\end{proof}

\jamiesection{Dense subsets \& continuous extensions}
\label{sect.dense}

\jamiesubsection{Definition and basic properties}

\begin{rmrk}
Suppose $(\ns P,\opens)$ is a semitopology and suppose $\varnothing\neq D\subseteq P\oldin\opens$ ($D$ need not be open).
The following four standard definitions of what it means for $D$ to be \emph{dense} in $P$ are equivalent in topology: 
\begin{enumerate*}
\item
Every nonempty open subset of $P$ intersects $D$.
\item
The interior of $P\setminus D$ is empty.
\item
Every open subset that intersects $P$, intersects $D$.
\item
$\closure{D}=\closure{P}$.
\end{enumerate*}
We shall see that in semitopologies, these definitions split into two groups.
\end{rmrk}

\begin{defn}
\label{defn.dense}
Suppose $(\ns P,\opens)$ is a semitopology and suppose $\varnothing\neq D\subseteq P\oldin\opens$ ($D$ need not be open).
Then:\index{dense set}\index{dense set!strongly dense}\index{dense set!weakly dense}
\begin{enumerate*}
\item\label{item.dense}
Call $D$ \deffont{weakly dense in $P$} when 
$$
\Forall{O\in\opens} \varnothing\neq O\subseteq P \limp D\between O . 
$$
In words: 
\begin{quote}
$D$ is weakly dense in $P$ when every nonempty open subset of $P$ intersects $D$.
\end{quote}
\item\label{item.strongly.dense}
Call $D$ \deffont{strongly dense in $P$} when 
$$
\Forall{O\in\opens} P\between O \limp D\between O . 
$$
In words: 
\begin{quote}
$D$ is strongly dense in $P$ when every open set that intersects $P$, intersects $D$.\footnote{We do not need to explicitly state that $O$ is nonempty because if $O$ is empty then $O\between P$ is false.}
\end{quote}
\item\label{item.dense.neighbourhood}
If $D$ is strongly dense in $P$ and $\interior(D)\neq\varnothing$ then, following Definition~\ref{defn.cn}(\ref{item.neighbourhood.of.p}), we may call $D$ a \deffont{strongly dense neighbourhood in $P$}. 
\end{enumerate*}
\end{defn}

In a topology, the two notions of being \emph{dense} described in Definition~\ref{defn.dense} above are equivalent.
A semitopology permits richer structure, because we do not insist that intersections of open sets be open, and thus it discriminates more finely between the definitions:
\begin{lemm}
\label{lemm.sd.implies.wd}
Suppose $(\ns P,\opens)$ is a semitopology and $\varnothing\neq D\subseteq P\in\opens$.
Then:
\begin{enumerate*}
\item\label{item.sd.implies.wd.1}
If $D$ is strongly dense in $P$ then $D$ is weakly dense in $P$.
\item\label{item.sd.implies.wd.2}
In a topology, the reverse implication holds; but 
\item\label{item.sd.implies.wd.3}
in a semitopology the reverse implication need not hold: it may be that $D$ is weakly dense but not strongly dense in $P$.
\end{enumerate*}
\end{lemm}
\begin{proof}
We consider each part in turn:
\begin{enumerate}
\item
If a nonempty open set is a subset of $P$ then it intersects with $P$.
It follows that if $D$ intersects every nonempty open set that intersects $P$, then it certainly intersects every nonempty open set that is a subset of $P$. 
\item
Suppose $(\ns P,\opens)$ is a topology and suppose $D$ is weakly dense in $P$ and $O\between P$.
Then $\varnothing\neq O\cap P\between P$, and because (this being a topology) $O\cap P\in\opens$ we have that $O\cap P\between D$ and so $O\between D$ as required.
\item
It suffices to provide a counterexample.
Consider the top-right semitopology in Figure~\ref{fig.012} and take $D=\{0\}$ and $P=\{0,1\}$.
Then $D$ is weakly dense in $P$ (because $D$ intersects $\{0\}$ and $\{0,1\}$) but $D$ is not strongly dense in $P$ (because $D$ does not intersect $\{1,2\}$).
\qedhere\end{enumerate}
\end{proof}

We can rearrange the definitions to obtain more abstract characterisations of weakly and strongly dense:
\begin{prop}
\label{prop.wd.sd.iff}
Suppose $(\ns P,\opens)$ is a semitopology and suppose $\varnothing\neq D\subseteq P\oldin\opens$ ($D$ need not be open).
Then:
\begin{enumerate*}
\item\label{item.wd.sd.iff.1}
$D$ is weakly dense in $P$ if and only if $\interior(P\setminus D)=\varnothing$.
\item\label{item.wd.sd.iff.2}
$D$ is strongly dense in $P$ if and only if $\closure{D}=\closure{P}$.
\end{enumerate*}
\end{prop}
\begin{proof}
For each part we prove two implications:
\begin{enumerate}
\item
$\interior(P\setminus D)=\varnothing$ means precisely that there is no nonempty open subset of $P\setminus D$, i.e. that every nonempty subset of $P$ intersects $D$.
But this is just the definition of $D$ being weakly dense in $P$ from Definition~\ref{defn.dense}(\ref{item.dense}).
\item
Since $D\subseteq P$, also $\closure{D}\subseteq\closure{P}$.

To prove $\closure{P}\subseteq\closure{D}$ it suffices to prove $\ns P\setminus\closure{D}\subseteq\ns P\setminus\closure{P}$.
By Corollary~\ref{corr.closed.complement.union} $\ns P\setminus \closure{D}$ is the union of the open sets that do not intersect $D$, and $\ns P\setminus\closure{P}$ is the union of the open sets that do not intersect $P$.
So $\ns P\setminus\closure{D}\subseteq\ns P\setminus\closure{P}$ when for every open set $O\in\opens$, if $O$ does not intersect $D$ then $O$ does not intersect $P$.
This is just the contrapositive of the property of $D$ being strongly dense in $P$ from Definition~\ref{defn.dense}(\ref{item.strongly.dense}).
\qedhere\end{enumerate}
\end{proof}

\begin{corr}
\label{corr.wd.meets.atoms}
Suppose $(\ns P,\opens)$ is a strongly chain-complete semitopology and $\varnothing\neq D\subseteq P\oldin\opens$ ($D$ need not be open).
Then the following are equivalent:
\begin{enumerate*}
\item
$D$ is weakly dense in $P$.
\item
$D$ intersects every atom $\varnothing\lessdot A\subseteq P$ in $P$.
\end{enumerate*}
In symbols using Definitions~\ref{defn.atomic.open.set} and~\ref{defn.O.between.R} we can write: 
$$
\text{$D$ weakly dense in $P$} 
\quad\liff\quad
D\between\f{Atoms}(P).
$$
\end{corr}
\begin{proof}
We prove two implications:
\begin{itemize}
\item
Suppose $D$ is weakly dense in $P$.
By Definition~\ref{defn.dense}(\ref{item.dense}) this means that $D$ intersects every open $O\subseteq P$.
In particular, $D$ intersects every atom $\varnothing\lessdot A\subseteq P$.
\item
Conversely, suppose $D$ intersects every atom $\varnothing\lessdot A\subseteq P$ and suppose $O\subseteq P$ is open.
By Corollary~\ref{corr.atom.exists} (since $\ns P$ is strongly chain-complete) there exists an atom $\varnothing\lessdot A\subseteq O$ and by assumption $D\between A$, thus also $D\between O$ as required.
\qedhere\end{itemize}
\end{proof}

\jamiesubsection{Dense subsets of topen sets}

\begin{lemm}
\label{lemm.open.O.dense.in.topen} 
Suppose $(\ns P,\opens)$ is a semitopology and suppose $\atopen\in\topens$ and $O\in\opens_{\neq\varnothing}$ and $O\subseteq \atopen$. 
Then $O$ is strongly dense in $\atopen$.

In words: any nonempty open subset of a topen set is strongly dense.
\end{lemm}
\begin{proof}
Suppose $\atopen\between O'\in\opens$.
Thus $O\between \atopen\between O'$ and by transitivity of $\atopen$ (since $\atopen$ is topen; see Definition~\ref{defn.transitive}) we have $O\between O'$ as required.
\end{proof}

\begin{corr}
Suppose $(\ns P,\opens)$ is a semitopology and suppose $\varnothing\neq D\subseteq \atopen\in\topens$.
Then \emph{precisely one} of the following holds:
\begin{itemize*}
\item
$D$ is weakly dense in $\atopen$.
\item
$\atopen\setminus D$ is a strongly dense neighbourhood in $\atopen$ (Definition~\ref{defn.dense}(\ref{item.dense.neighbourhood})).
\end{itemize*}
As a corollary, \emph{precisely one} of the following holds:
\begin{itemize*}
\item
$D$ is a strongly dense neighbourhood in $\atopen$.
\item
$\atopen\setminus D$ is weakly dense in $\atopen$.
\end{itemize*}
\end{corr}
\begin{proof}
For the first part we reason as follows:
\begin{itemize}
\item
If $D$ is weakly dense in $\atopen$ then by Proposition~\ref{prop.wd.sd.iff}(\ref{item.wd.sd.iff.1}) $\interior(\atopen\setminus D)=\varnothing$, so following Definition~\ref{defn.dense}(\ref{item.dense.neighbourhood}) $\atopen\setminus D$ is not a strongly dense neighbourhood.
\item
If $D$ is not weakly dense in $\atopen$ then by Proposition~\ref{prop.wd.sd.iff}(\ref{item.wd.sd.iff.1}) $\interior(\atopen\setminus D)\neq\varnothing$.
By Lemma~\ref{lemm.open.O.dense.in.topen} (since $\atopen$ is topen)
$\interior(\atopen\setminus D)$ is strongly dense in $\atopen$, thus so is $\atopen\setminus D$.
It follows from Definition~\ref{defn.dense}(\ref{item.dense.neighbourhood}) that $\atopen\setminus D$ is a strongly dense neighbourhood in $\atopen$, as required.
\end{itemize}
For the corollary, write $D=\atopen\setminus D'$, so that $D'=\atopen\setminus D$.
We use the first part of this result, for $D'$.
\end{proof}

\begin{corr}
\label{corr.sd.char}
Suppose $(\ns P,\opens)$ is a strongly chain-complete semitopology and suppose $\varnothing\neq D\subseteq \atopen\in\topens$. 
Then the following are equivalent:
\begin{itemize*}
\item
$D$ is a strongly dense neighbourhood in $\atopen$.
\item
$\interior(D)\neq\varnothing$.
\item
$D$ contains an atom, or in symbols: $\varnothing\lessdot A\subseteq D$.
\end{itemize*}
\end{corr} 
\begin{proof}
We prove a cycle of implications:
\begin{itemize*}
\item
If $D$ is a strongly dense neighbourhood in $\atopen$ then $\interior(D)\neq\varnothing$ direct from Definition~\ref{defn.dense}(\ref{item.dense.neighbourhood}).
\item
If $\interior(D)\neq\varnothing$ then there exists an atom $\varnothing\lessdot A\subseteq \interior(D)\subseteq D$ by Corollary~\ref{corr.atom.exists} (since $\atopen$ is strongly chain-complete).
\item
If $\varnothing\lessdot A\subseteq D$ then using Lemma~\ref{lemm.open.O.dense.in.topen} (since $\atopen$ is topen) $D$ is strongly dense in $\atopen$.
\qedhere\end{itemize*}
\end{proof}

\jamiesubsection{Explaining kernels}
\label{subsect.explaining.kernels}

\begin{nttn}
Suppose $(\ns P,\opens)$ is a semitopology and $\varnothing\neq D\subseteq P\subseteq\ns P$.
Then:
\begin{enumerate}
\item
Call $D$ \deffont{minimally weakly dense in $P$} when:
\begin{itemize*}
\item
$D$ is weakly dense in $P$, and 
\item
if $\varnothing\neq D'\subseteq D$ and $D'$ is weakly dense in $P$, then $D'=D$.
\end{itemize*}
\item
Call $D$ a \deffont{minimally strongly dense open subset of $P$} when:
\begin{itemize*}
\item
$D\in\opens$,
\item
$D$ is a strongly dense subset of $P$, and 
\item
if $\varnothing\neq D'\subseteq D$ and $D'$ is a strongly dense open subset of $P$, then $D'=D$.
\end{itemize*}
\end{enumerate}
\end{nttn}

Recall from Definition~\ref{defn.atomic.open.set}(\ref{item.atoms.of.P}) that $\f{Atoms}(P)=\{A\in\opens \mid \varnothing\lessdot A\subseteq P\}$.
\begin{prop}
\label{prop.kernel.weakly.strongly.dense}
Suppose $(\ns P,\opens)$ is a strongly chain-complete semitopology and suppose $P\in\opens$.
Then:
\begin{enumerate*}
\item\label{item.kernel.weakly.strongly.dense.1}
$\bigcup\f{Atoms}(P)$ is equal to the sets union of the minimal weakly dense subsets of $P$.\footnote{This sentence is potentially confusing, because $\bigcup\f{Atoms}(P)$ is itself a sets union of the atoms in $P$.  So what is being asserted is that the sets union of the atoms in $P$, is equal to the sets union of the minimal weakly dense subsets of $P$.  However, these minimal weakly dense subsets are not necessarily atoms, and the atoms are not necessarily minimal weakly dense subsets.} 
\item\label{item.kernel.weakly.strongly.dense.2}
If furthermore $P$ is transitive (so that $P\in\topens$) then $\bigcup\f{Atoms}(P)$ is equal to the sets union of the minimal strongly dense neighbourhoods in $P$. 
\item\label{item.kernel.weakly.strongly.dense}
If $p\in\ns P$ is regular then $\kernel(p)$ is equal to the union of the minimal weakly dense subsets of $\community(p)$ and also to the union of the minimal strongly dense neighbourhoods in $\community(p)$.
\end{enumerate*}
\end{prop}
\begin{proof}
We consider each part in turn:
\begin{enumerate}
\item
For brevity, write $\mathcal A=\f{Atoms}(P)$.
By Corollary~\ref{corr.wd.meets.atoms} (since $\ns P$ is strongly chain-complete), if $D$ is weakly dense in $P$ then so is $D\cap \bigcup\mathcal A$.
Thus, the \emph{minimal} weakly dense subsets of $P$ are all contained in $\bigcup\mathcal A$.

It remains to show that every $p\in S$ is contained in \emph{some} minimal weakly dense subset of $P$.
Fix one such $p$.
We need some notation: if $X\subseteq P$ then write $\atmclosure{X}$\index{$\atmclosure{X}$ (atoms in $P$ intersecting $X$)} for the union of the set of atoms in $P$ that intersect $X$. 
We now argue by transfinite induction (a clear and accessible presentation is in~\cite{emerson:trai}, or see~\cite[Section~6]{johnstone:notlst}) to generate a minimal weakly dense $D\subseteq P$ that contains $p$:
\begin{itemize*}
\item
Set $p_0=p$ and set $A_0$ to be some atom in $P$ that contains $p$; one such exists, since we chose $p\in\bigcup\mathcal A$.
\item
Suppose $p_{\alpha'}$ and $A_{\alpha'}$  are defined for all $\alpha'<\alpha$.
If $\bigcup\mathcal A\subseteq\atmclosure{\{p_{\alpha'}\mid \alpha'<\alpha\}}$ --- so that $\bigcup\mathcal A=\atmclosure{\{p_{\alpha'}\mid \alpha'<\alpha\}}$ --- then stop.
Otherwise, pick an atom $A_\alpha$ of $P$ that is not a subset, and choose some $p_\alpha\in A_\alpha\setminus \atmclosure{\{p_{\alpha'}\mid \alpha'<\alpha\}}$.

Note that $A_\alpha\subseteq\atmclosure{\{p_{\alpha'}\mid \alpha'\leq\alpha\}}$.
\item
Continue transfinitely until we stop, and set $D=\{p_{\alpha'}\mid \alpha'<\alpha\}$.
By construction, $\atmclosure{D}=\bigcup\mathcal A$.
\end{itemize*}
Then $D$ is weakly dense in $P$ by Corollary~\ref{corr.wd.meets.atoms}, and it is minimal because if we remove any $p_{\alpha'}$ from $D$ to obtain a smaller $D'=D\setminus\{p_{\alpha'}\}$ then by construction $A_{\alpha'}\not\subseteq\atmclosure{D'}$. 
\item 
It follows from Corollary~\ref{corr.sd.char} that $D$ is a minimal strongly dense neighbourhood in $P$ precisely when $D$ is equal to an atom in $P$.
The result follows.
\item
$\community(p)$ is transitive by Definition~\ref{defn.tn}(\ref{item.regular.point}), and $\kernel(p)=\bigcup\f{Atoms}(\community(p))$ by Definition~\ref{defn.kernel}(\ref{item.kernel}).
We use parts~\ref{item.kernel.weakly.strongly.dense.1} and~\ref{item.kernel.weakly.strongly.dense.2} of this result.
\qedhere\end{enumerate}
\end{proof}

\begin{rmrk}
Proposition~\ref{prop.kernel.weakly.strongly.dense}(\ref{item.kernel.weakly.strongly.dense}) gives some independent explanation for why $\kernel(p)$ --- the atoms in the community of $p$, as studied in Section~\ref{sect.kernels} --- is interesting.
$\kernel(p)$ identifies where the minimal weakly dense and strongly dense subsets of $\community(p)$ are located.
\end{rmrk}

\jamiesubsection{Unifying is-transitive and is-strongly-dense-in}
\label{subsect.unifying.trans.dense}

It turns out that transitivity and denseness are closely related:
in this Subsection we explore their relationship.

\begin{rmrk}
Consider the following three notions:
\begin{enumerate*}
\item
\emph{$D$ is strongly dense in $P$} from Definition~\ref{defn.dense}(\ref{item.strongly.dense}).
\item
\emph{$P$ is transitive} from Definition~\ref{defn.transitive}(\ref{transitive.transitive}).
\item
\emph{$P$ is strongly transitive} from Definition~\ref{defn.strongly.transitive}(\ref{item.strongly.transitive}).
\end{enumerate*}
Notice that while the definitions are different, they share a `family resemblance'. 
Can we identify a common ancestor for them; some definition that naturally subsumes them into a most general principle?

Yes: it is easy to see that items~1 and~3 above are very closely related --- see Lemma~\ref{lemm.strongly.dense.strongly.transitive} --- and then we will prove that all three definitions listed above are special instances of a general definition --- see Definition~\ref{defn.transitive.wrt} and Proposition~\ref{prop.most.general}.
\end{rmrk}

\begin{lemm}
\label{lemm.strongly.dense.strongly.transitive}
Suppose $(\ns P,\opens)$ is a semitopology and $P\subseteq\ns P$.
Then the following are equivalent:\footnote{Cf. also Lemma~\ref{lemm.meet-irreducible}.}
\begin{itemize*}
\item
$P$ is strongly transitive.
\item
$O\cap P$ is strongly dense in $P$, for every $O{\in}\opens$ such that $O\between P$ (meaning that $O\cap P\neq\varnothing$).
\end{itemize*}
In words we can say:
\begin{quote}
$P$ is strongly transitive when every nontrivial open intersection with $P$ is strongly dense. 
\end{quote}
\end{lemm}
\begin{proof}
Unpacking Definition~\ref{defn.strongly.transitive}(\ref{item.strongly.transitive}), $P$ is strongly transitive when $O\between P\between O'$ implies $O\cap P\between O'\cap P$.
Unpacking Definition~\ref{defn.dense}(\ref{item.strongly.dense}), $O\cap P$ is strongly dense in $P$ when $P\between O'$ implies $O\cap P\between O'$ --- and this is clearly equivalent to $O\cap P\between O'\cap P$.
The result now follows by routine reasoning.
\end{proof}

We can generalise the notion of strongly dense from Definition~\ref{defn.dense}(\ref{item.strongly.dense}) from $D\subseteq P$ to \emph{any} $D$.
\begin{defn}
\label{defn.transitive.wrt}
Suppose $(\ns P,\opens)$ is a semitopology and $D,P\subseteq\ns P$.

Call $D$ \deffont{strongly dense for $P$} when
$$
\Forall{O{\in}\opens}P\between O\limp D\between O.
$$
\end{defn}

We state the obvious:
\begin{lemm}
$\varnothing\neq D\subseteq P$ is strongly dense \emph{in} $P$ in the sense of Definition~\ref{defn.dense}(\ref{item.strongly.dense}) if and only if it is strongly dense \emph{for} $P$ in the sense of Definition~\ref{defn.transitive.wrt}.
\end{lemm}
\begin{proof}
The definitions are identical where they overlap.
The only difference is that Definition~\ref{defn.dense}(\ref{item.strongly.dense}) assumes a nonempty subset of $P$, whereas Definition~\ref{defn.transitive.wrt} assumes a nonempty set that intersects (but is not necessarily a subset of) $P$. 
\end{proof}

\begin{lemm}
\label{lemm.strongly.dense.for.closure}
Suppose $(\ns P,\opens)$ is a semitopology and $D,P\subseteq\ns P$.
Then the following are equivalent:
\begin{itemize*}
\item
$D$ is strongly dense for $P$.
\item
$\closure{P}\subseteq\closure{D}$.\footnote{Compare with Proposition~\ref{prop.open.strong-consensus}.}
\end{itemize*}
\end{lemm}
\begin{proof}
Suppose $D$ is strongly dense for $P$.
Then $O\in\opens$ does not intersect with $P$ if and only if $O$ does not intersect with $D$, and it follows (just as in the proof of Proposition~\ref{prop.wd.sd.iff}(\ref{item.wd.sd.iff.2})) that $\ns P\setminus\closure{D}\subseteq\ns P\setminus\closure{P}$, and so that $\closure{P}\subseteq\closure{D}$.

Conversely, if $\closure{P}\subseteq\closure{D}$ then $\ns P\setminus\closure{D}\subseteq\ns P\setminus\closure{P}$ and it follows that if $O\in\opens$ does not intersect with $P$ then $O$ does not intersect with $D$, and thus that $D$ is strongly dense for $P$.
\end{proof}

\begin{prop}
\label{prop.most.general}
Suppose $(\ns P,\opens)$ is a semitopology and $\atopen\subseteq\ns P$.
Then the following are equivalent:
\begin{itemize*}
\item
$\atopen$ is transitive.
\item
$O$ is strongly dense for $\atopen$, for every $\atopen\between O\in\opens$.\footnote{Compare with Lemma~\ref{lemm.strongly.dense.strongly.transitive}.}
\end{itemize*}
\end{prop}
\begin{proof}
We unpack Definition~\ref{defn.transitive.wrt} and see that a condition that $\atopen$ is transitive with respect to every $\atopen\between O\in\opens$ is precisely what Definition~\ref{defn.transitive}(\ref{transitive.transitive}) asserts, namely: for every $O\in\opens$ such that $O\between \atopen$,
$$
\Forall{O'{\in}\opens} \atopen\between O' \limp O\between O'.
\qedhere
$$
\end{proof}

\begin{rmrk}
\label{rmrk.strongly.dense.for}
In topology it makes less sense to talk about $D$ being dense in $P$ for $D\not\subseteq P$, since we can just consider $D\cap P$ --- and if $D$ and $P$ are open then so is $D\cap P$.
In semitopology the following happens: 
\begin{itemize*}
\item
The notion of \emph{dense in} splits into two distinct concepts (\emph{weakly dense in} and \emph{strongly dense in}), as we saw in Definition~\ref{defn.dense} and the subsequent discussion.
\item
The notion of \emph{strongly dense in} generalises to a notion that we call \emph{strongly dense for}, which has the same definition but just weakens a precondition that $D\subseteq P$.
\end{itemize*}
Given the above, we then see from Lemma~\ref{lemm.strongly.dense.strongly.transitive} and Proposition~\ref{prop.most.general}
that the notions of \emph{transitive} and \emph{strongly transitive} from Definitions~\ref{defn.transitive}(\ref{transitive.transitive}) and~\ref{defn.strongly.transitive}(\ref{item.strongly.transitive}) lend themselves to being naturally expressed in terms of strongly-dense-for.
\end{rmrk}

\jamiesubsection{Towards a continuous extension result} 
\label{subsect.towards.ce}

\begin{rmrk}
\label{rmrk.top.ce}
Topology has a family of results on \emph{continuous extensions}\index{continuous extension (in topology)} of functions: a nice historical survey is in~\cite{gutev:simecu}. 
Here is an example, adapted from~\cite[Theorem~24.1.15]{erdman:protac}:
\begin{quote}
Suppose $f:B\to\mathbb R$ is uniformly continuous and suppose $B$ is a dense subset of $A$.
Then $f$ has a unique extension to a continuous function $g:A\to\mathbb R$.
\end{quote}
This is true in the world of topologies: but what might correspond to this in the semitopological world?

A direct translation to semitopologies seems unlikely,\footnote{Except trivially that we can restrict to those semitopologies that are also topologies (i.e. for which intersections of open sets are open).} 
because we have seen from Definition~\ref{defn.dense} and Lemma~\ref{lemm.sd.implies.wd} and the subsequent discussion and results how the notion of `is dense in' behaves differently for semitopologies in general, so that the very premise of the topological result above is now up for interpretation.\footnote{There are other differences.  For instance we care a lot about value assignments --- maps to discrete semitopologies --- rather than maps to $\mathbb R$.
Of course we could try to generalise from value assignments to more general examples, but as we shall see, even this `simple' case of value assignments is more than rich enough to raise some canonical questions.} 

But, the impossibility of a direct translation only opens up an even more interesting question: whether we can find definitions and well-behavedness conditions on semitopological spaces that reflect the spirit of the corresponding topological results, but without assuming that intersections of open sets are open.

We shall see that this is possible and we propose a suitable result below in Definition~\ref{defn.continuous.extension}.
However, before we come to that, we will sketch a design space of \emph{failing} definitions and counterexamples --- and so put our working definition in its proper design context.

We map to semitopologies of values, so (the spirit of) uniform continuity is automatic, and we concentrate (to begin with) on being strongly dense in rather than weakly dense in, since by Lemma~\ref{lemm.sd.implies.wd}(\ref{item.sd.implies.wd.1}) the former implies the latter:
\begin{enumerate}
\item
\textbf{Candidate definition 1}.
\begin{quote}
Suppose $(\ns P,\opens)$ is a semitopology and suppose $f:\ns P\to\tf{Val}$ is a value assignment that is continuous on $D\subseteq\ns P$, and suppose $D$ is a strongly dense subset of $\ns P$.
Then $f$ has a unique extension to a continuous function $g:\ns P\to\tf{Val}$.
\end{quote}
This does not work: 
\begin{itemize*}
\item
Take $(\ns P,\opens)$ to be the top-left example in Figure~\ref{fig.012} and 
\item
$\tf{Val}=\{0,1\}$ with the discrete semitopology.
\item
Define $f:\ns P\to\tf{Val}$ such that $f(0)=0$ and $f(1)=1$ and $f(2)=1$ and 
\item
set $D=\{0,2\}$ and $P=\ns P$.
\end{itemize*}
Note that $D$ is a strongly dense open subset of $\ns P$, and $f$ is continuous on $D$.

However, $f$ cannot be continuously extended to a $g$ that is continuous at $1$.
We note that $1$ is conflicted and intertwined with two distinct topens: $\{0\}$ and $\{2\}$.
Looking at this example we see that Candidate definition~1 is unreasonable: \emph{of course} we cannot extend $f$ continuously to $1$, because $1$ is intertwined with two distinct topen sets on which $f$ takes distinct values.
The natural solution is just to exclude conflicted points since they may be, as the terminology suggests, conflicted:
\item
\textbf{Candidate definition 2.}\quad
\begin{quote}
Suppose $(\ns P,\opens)$ is a semitopology and suppose $f:\ns P\to\tf{Val}$ is a value assignment that is continuous on $D\subseteq\ns P$, and suppose $D$ is a strongly dense subset of $\ns P$.
Then $f$ has a unique extension to a function $g:\ns P\to\tf{Val}$ that is continuous at all \emph{unconflicted} points.
\end{quote}
This does not work:
\begin{itemize*}
\item
Take $(\ns P,\opens)$ to be the semitopology in Figure~\ref{fig.nitpick} and 
\item
$\tf{Val}=\{0,1,2\}$ with the discrete semitopology. 
\item
Define $f:\ns P\to\tf{Val}$ such that $f(0)=0$ and $f(1)=1$ and $f(2)=2$ and $f(\ast)=0$ and 
\item
set $D=\{0,1,2\}$ and $P=\ns P$.
\end{itemize*}
Note that $D$ is a strongly dense open subset of $\ns P$ and $f$ is continuous on $D$.

Note that $\ast$ is unconflicted (because $\intertwined{\ast}=\{\ast, 1\}$). 
However, $f$ cannot be continuously extended to a $g$ that is continuous at $\ast$.
\item\label{item.semi-regular}
Candidate definition~2 is even more telling than it might appear.
Note that $\ast$ is unconflicted and quasiregular (Definition~\ref{defn.tn}(\ref{item.quasiregular.point})), because $\community(\ast)=\{1\}\neq\varnothing$.

Thus we could not even rescue Candidate definition~2 above by insisting that points be not only unconflicted but also quasiregular (i.e. unconflicted and having a nonempty community).
The next natural step up from this is to be unconflicted and weakly regular, which by Corollary~\ref{corr.corr.pKp} leads us to regular points.
\end{enumerate}
We can now state a definition and result that work:
\end{rmrk}

\begin{defn}
\label{defn.continuous.extension}
Suppose $f,g:(\ns P,\opens)\to\tf{Val}$ are value assignments (Definition~\ref{defn.value.assignment}(\ref{item.value.assignment})) and suppose $P\subseteq\ns P$.
\begin{enumerate}
\item\label{item.continuous.extension.1}
Say that $g$ \deffont{continuously extends}\index{continuous extension (in semitopology)} $f$ to regular points in $P$ when:
\begin{itemize*}
\item
If $f$ is continuous at $p\in P$ then $f(p)=g(p)$.
\item
$g$ is continuous on every regular $p\in P$ (Definition~\ref{defn.tn}(\ref{item.regular.point})).
\end{itemize*}
\item\label{item.continuous.extension.2}
Say that $g$ is a \deffont{unique continuous extension} of $f$ to regular points in $P$ when for any other continuous extension $g'$ of $f$ to $P$, we have $g(p)=g'(p)$ for every regular $p\in P$.
\end{enumerate}
\end{defn}

\begin{rmrk}[Justification for regular points]
\label{rmrk.cont.uniq}
Note that `continuously extends' and `uniquely' in Definition~\ref{defn.continuous.extension} both apply to to \emph{regular} points in $P$ only.
By the examples in Remark~\ref{rmrk.top.ce} it would not be reasonable to expect unique continuous extensions on non-regular points.
This gives a retrospective justification for the theories of topens and regular points that we develop (see Definitions~\ref{defn.transitive} and~\ref{defn.tn}): regularity arises as a natural condition for a semitopological continuous extension result.\footnote{This is not an exclusive claim. Other reasonable conditions or generalisations might also exist, for example the \emph{hypertwined} points discussed starting from Definition~\ref{defn.ht.ce}(\ref{item.ht}).  But being hypertwined is a more complex notion, and note by Remark~\ref{rmrk.simple.intertwinedX} that intertwined and hypertwined amount to the same thing on regular points.  So it seems likely that Definition~\ref{defn.continuous.extension} is a canonical point in the design space.  Investigating this further is future work.} 
\end{rmrk}

\begin{prop}
\label{prop.cet}
Suppose $(\ns P,\opens)$ is a semitopology and $f:(\ns P,\opens)\to\tf{Val}$ is a value assignment and suppose $D, P\subseteq\ns P$.
Then:
\begin{enumerate*}
\item\label{item.cet.1}
If $f$ is continuous on $D$ then $f$ can be continuously extended to all regular points in $\ns P$.
\item\label{item.cet.2}
If $D$ is strongly dense for $P$ (Definition~\ref{defn.transitive.wrt}) then this extension is unique on $P$ in the sense of Definition~\ref{defn.continuous.extension}(\ref{item.continuous.extension.2}).
\end{enumerate*}
\end{prop}
\begin{proof}
Choose some fixed but arbitrary \emph{default value} $v\in\tf{Val}$ and for this choice of $v$ define $g$ by cases as follows:
\begin{itemize}
\item
\emph{Suppose $f$ is continuous at $p$.}

We set $g(p)=f(p)$.
\item
\emph{Suppose $f$ is not continuous at $p$ and $\community(p)\between D$.}

Choose some $d\in\community(p)\cap D$ and set $g(p)=f(d)$.

If $p$ is intertwined with two points $d$ and $d'$ then (because $p$ is regular and so unconflicted by Theorem~\ref{thrm.r=wr+uc}) $d\intertwinedwith d'$ and their open neighbourhoods of continuity intersect, so that $f(d)=f(d')$.
\item
\emph{Suppose $f$ is not continuous at $p$ and $\community(p)\notbetween D$.}

If $D$ is strongly dense for $P$ then this case cannot happen because $P\between\community(p)$ so by the strong dense property $D\between\community(p)$ (see Definition~\ref{defn.dense}(\ref{item.strongly.dense})).

Otherwise, we set $g(p)=v$ (so $g(p)$ is the fixed but arbitrary default value). 
\end{itemize}
We now show that if $p\in \ns P$ is regular, then $g$ is continuous at $p$.
The proof is again by cases:
\begin{itemize}
\item
If $f$ is continuous at $p$ then $g(p)=f(p)$ and so $g$ is continuous at $p$.
\item
If $f$ is not continuous at $p$ and $d\in \community(p)\cap D$ then $g(p)=f(d)$.
Thus (since we assumed that $p$ is regular) $p\in \community(p)\subseteq g^\mone(g(p))$, so that $g$ is continuous at $p$.
\item
If $f$ is not continuous at $p$ and $\community(p)\cap D=\varnothing$ then $g(p)=v$.
Thus (since we assumed that $p$ is regular) $p\in \community(p)\subseteq g^\mone(v)$, so that $g$ is continuous at $p$.
\end{itemize}
If $D$ is strongly dense for $P$ then uniqueness follows by routine reasoning from the above, using Theorem~\ref{thrm.correlated}.
\end{proof}

In view of Lemma~\ref{lemm.strongly.dense.for.closure} we can more succinctly rephrase Proposition~\ref{prop.cet} as follows:
\begin{corr}
\label{corr.cet.closed}
Suppose $(\ns P,\opens)$ is a semitopology and $f:(\ns P,\opens)\to\tf{Val}$ is a value assignment. 
Then if $f$ is continuous on $D\subseteq \ns P$, then $f$ can be continuously extended to all regular points in $\closure{D}$.
\end{corr}
\begin{proof}
Direct from Proposition~\ref{prop.cet} taking $P=\closure{D}$ and using Lemma~\ref{lemm.strongly.dense.for.closure}. 
\end{proof}

\begin{figure}
\vspace{-1em}
\centering
\includegraphics[width=0.35\columnwidth,trim={50 50 50 50},clip]{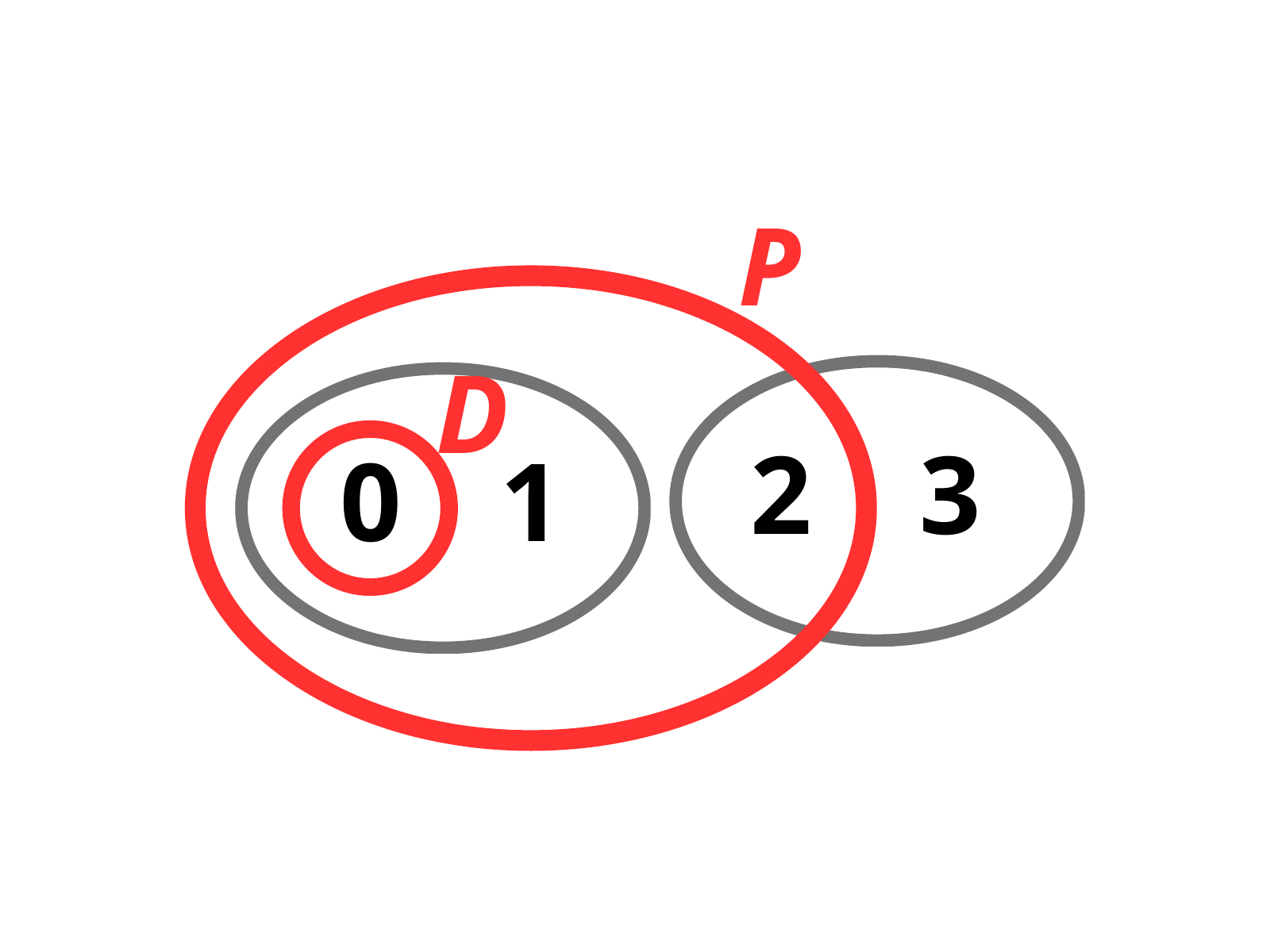}
\caption{A weakly dense subset is not enough for uniqueness (Remark~\ref{rmrk.wd.not.enough}(\ref{item.not.enough.for.uniqueness}))}
\label{fig.wd-not-enough}
\end{figure}

\begin{rmrk}
\label{rmrk.wd.not.enough}
\leavevmode
\begin{enumerate}
\item
There are a few subtleties to Corollary~\ref{corr.cet.closed}.
The result actually tells us that there exists an open set $O\in\opens$ such that $D\subseteq\closure{D}\subseteq O$, and $f$ continuously extends to some $g:\ns P\to\tf{Val}$ that is continuous at $O$.
This is because if $g$ is continuous at $p\in\closure{D}$, then it is by definition continuous on some open neighbourhood of $p$.
\item
Similarly, the condition that $f$ be continuous on $D$ is equivalent to insisting that $f$ be continuous on an \emph{open} $D$.
\item\label{item.not.enough.for.uniqueness}
The condition of $D$ being strongly dense in $P$ is required for uniqueness in Proposition~\ref{prop.cet}(\ref{item.cet.2}).
Being weakly dense is not enough.
For, consider the semitopology illustrated in Figure~\ref{fig.wd-not-enough}, such that:
\begin{itemize*}
\item
$\ns P=\{0,1,2,3\}$.
\item
$\opens$ is generated by $D=\{0\}$, $\{0,1\}$, $P=\{0,1,2\}$, and $\{2,3\}$.
\end{itemize*} 
Then we have $D\subseteq P\subseteq\ns P$, and we even have that $D,P\in\opens$ and every point in the space is regular, making this is a particularly well-behaved example.
This semitopology is \emph{not} a topology, because $P\cap\{2,3\}\not\oldin\opens$; we will exploit this fact in a moment. 
The reader can check that 
\begin{itemize*}
\item
$D$ is weakly dense in $P$ (because $D$ intersects every open subset of $P$), but 
\item
$D$ is not strongly dense in $P$ (because $D$ does not intersect $\{2,3\}$), and 
\end{itemize*}
the value assignment $f:\ns P\to\mathbb B$ mapping $0$ to $\bot$ and every other point to $\top$ has two continuous extensions to all of $P$: $g$ mapping all points to $\bot$, and $g'$ mapping $0$ and $1$ to $\bot$ and $2$ and $3$ to $\top$.
\end{enumerate}
\end{rmrk}

\jamiesubsection{Kernels determine values of continuous extensions} 
\label{subsect.kerels.determine}

\begin{rmrk}
In Subsection~\ref{subsect.towards.ce} we considered continuous extensions in a semitopological context.
We concluded with Corollary~\ref{corr.cet.closed}, which showed how to extend a value assignment $f:\ns P\to\tf{Val}$ that is continuous on some $D$, to a $g$ that is continuous on $D$ and on the regular points in $\closure{D}$.
We also discussed why this result is designed as it is and why it seems likely to be optimal within a certain design space as outlined in Remark~\ref{rmrk.top.ce}.

However, our study of semitopologies is motivated by systems that may be distributed over several participants.
This means that we also care about intermediate continuous extensions of $f$; i.e. about $g$ that continuously extend $f$ but not necessarily on all of $\closure{D}$. 

The mathematics in this subsection is in some sense a `pointwise' dual to the `setwise' mathematics in Subsection~\ref{subsect.towards.ce}.
Perhaps surprisingly, we shall see that when developed pointwise, the details are different: contrast Theorem~\ref{thrm.p.kernel.p} and Corollary~\ref{corr.boundary.kernel} with Proposition~\ref{prop.cet} and Corollary~\ref{corr.cet.closed}; they are similar, but they are not the same.
\end{rmrk} 

Recall that:
\begin{itemize*}
\item
A point $p$ is \emph{regular} when its community (which is the interior of its intertwined points) is a topen neighbourhood of $p$; see Definition~\ref{defn.tn}(\ref{item.regular.point}).
\item
$\kernel(p)$ is the union of the kernel atoms of $p$ (minimal nonempty open sets in the community of $p$); see Definition~\ref{defn.kernel}(\ref{item.kernel}).
\item
A value assignment $f:\ns P\to\tf{Val}$ is a mapping from $\ns P$ to some set of values $\tf{Val}$ having the discrete semitopology; see Definition~\ref{defn.value.assignment}.
\end{itemize*}
We now consider how the value of $f$ on kernel atoms influences the value of $f$ at regular points.

\begin{defn}
\label{defn.regular.f}
Suppose $(\ns P,\opens)$ is a semitopology and $f,g:\ns P\to\tf{Val}$ are value assignments (Definition~\ref{defn.value.assignment}).\footnote{This definition makes sense for $f$ mapping $\ns P$ to any semitopology $\ns Q$, but (for now) we will only care about the case when $\ns Q$ is a discrete semitopology so that $f$ is a value assignment.} 
\begin{enumerate*}
\item\label{item.regular.f.1}
Call $f$ \deffont{confident at $p\in\ns P$} when $f$ is continuous on some atom $\varnothing\lessdot A\subseteq\kernel(p)$.
\item\label{item.regular.f.2}
Call $f$ \deffont{unanimous at $p\in\ns P$} when $f$ is continuous on all of $\community(p)$.
\item
We generalise Definition~\ref{defn.continuous.extension} and write $f\leq g$, and call $g$ a \deffont{(partial) continuous extension of $f$}, when for every $p\in\ns P$, if $f$ is continuous at $p$ then $g$ is continuous at $p$ and $f(p)=g(p)$.\footnote{Definition~\ref{defn.continuous.extension} is interested in a $g$ that continuously extends $f$ all at once from $D$ to $\closure{D}$, which is fine mathematically but less helpful computationally.  The definition here refines this concept and is interested in the space of all possible $g$ such that $f\leq g$, which more accurately reflects how $g$ might be computed, in stages, on a network.}

It is routine to check that $\leq$ is a preorder (reflexive, transitive).
\end{enumerate*}
\end{defn}

\begin{rmrk}
Intuitively, $f$ is `confident' at $p$ when the value we obtain if we continuously extend $f$ to $p$, is already determined.
It may be that a result has been determined by some part of the system, but not yet fully propagated to the whole system.\footnote{For instance: $f$ may know the results of an election, but not yet have told point $p$; whereas some $g\geq f$ may represent a state in which this result has been correctly propagated to $p$.  Similarly, technology allows us to determine the weather tomorrow based on weather data that was collected this morning, but that is not the same thing as knowing what the weather will be: a supercomputer needs to run calculations, and the data needs to be broadcast, and put on a webpage and sent down a cable and rendered to a computer screen, and so on.}

We make this formal in Theorem~\ref{thrm.p.kernel.p}(\ref{item.fg.leq.cc}).
We start with an easy lemma:
\end{rmrk}

\begin{lemm}
\label{lemm.unanimous.implies.confident}
Suppose $(\ns P,\opens)$ is a strongly chain-complete semitopology and $f,g:\ns P\to\tf{Val}$ are value assignments.
Then:
\begin{enumerate*}
\item\label{item.unanimous.implies.confident.1}
If $f\leq g$ and $f$ is confident/unanimous at $p$ then $g$ is also confident/unanimous at $p$.
\item\label{item.unanimous.implies.confident.2}
If $f$ is unanimous at $p\in\ns P$ then it is confident at $p$.
\end{enumerate*}
\end{lemm}
\begin{proof}
By simple arguments from the definitions:
\begin{enumerate}
\item
Just unpacking definitions: if $f$ is continuous on some $\varnothing\lessdot A\subseteq\kernel(p)$ then so is $g$; and if $f$ is continuous on $\community(p)$ then so is $g$.
\item
Suppose $f$ is unanimous at $p$, meaning that $f$ is continuous on $\community(p)$.
Then $f$ is also continuous on some kernel atom in $\community(p)$ --- some such kernel atom exists by Corollary~\ref{corr.atom.exists}, since $\ns P$ is strongly chain-complete.
Thus $f$ is confident at $p$, as required.
\qedhere\end{enumerate}
\end{proof}

For brevity in what follows, a notation will be useful:
\begin{nttn}
\label{nttn.f.const}
Suppose $\ns P$ is a set and $f$ is some function on $\ns P$ and $\varnothing\neq A\subseteq\ns P$.
Suppose further that it is known that $f$ is constant on $A$.
In symbols:
$$
\Exists{c}\Forall{p{\in}A}f(p)=c.
$$
Then we may write $f(A)$ for the unique constant value that $f(p)$ takes as $p$ ranges over $A$. %
\end{nttn}

\begin{rmrk}
\label{rmrk.machinery.together}
Suppose $(\ns P,\opens)$ is a semitopology and $f:\ns P\to\tf{Val}$ is a value assignment.
Then:
\begin{enumerate}
\item
Suppose $f$ is confident at $p\in\ns P$.

By Definition~\ref{defn.regular.f} $f$ is continuous on some kernel atom $\varnothing\lessdot A\subseteq\kernel(p)$.
By Lemma~\ref{lemm.topen.max.min} $A$ is transitive, so by Theorem~\ref{thrm.correlated} (since $f$ is continuous on $A$) $f$ is constant on $A$, and thus it makes sense to use Notation~\ref{nttn.f.const} and write $f(A)$ to denote the (unique) value of $f$ on $A$. 
\item
Likewise if $f$ is unanimous at $p$ then we can sensibly write 
$f(\community(p))$.
\item\label{item.machinery.together.3}
Just for this paragraph call $f$ \emph{doubly confident} at $p$ when $f$ is continuous on two distinct kernel atoms $\varnothing\lessdot A\neq A'\subseteq\kernel(p)$ of $p$.
Suppose $f$ is doubly confident at $p$ and suppose $p$ is regular; so by the previous paragraph $f(A)$ and $f(A')$ are both well-defined.
Now $A\between A'$ by Lemma~\ref{lemm.kernel.atoms.intersect}, so $f(A)=f(A')$.

Thus being doubly confident at $p$ is the same as just being confident, provided that $p$ is regular so that all of its kernel atoms intersect.
\end{enumerate}
\end{rmrk}

Remark~\ref{rmrk.machinery.together} brings us to a notation:
\begin{defn}
\label{defn.fp.limit}
Suppose $(\ns P,\opens)$ is a semitopology and $f:\ns P\to\tf{Val}$ is a value assignment and $p\in\ns P$ is a regular point.
Then define $\limitat{p} f$ the \deffont[limit of $f$ at $p$ ($\limitat{p} f$)]{limit of $f$ at $p$} by 
$$
\textstyle\limitat{p} f = f(A) 
$$ 
where $\varnothing\lessdot A\subseteq\kernel(p)$ is some/any (by Remark~\ref{rmrk.machinery.together}(\ref{item.machinery.together.3}) writing `some/any' makes sense) kernel atom of $p$ on which $f$ is continuous.
The justification for calling this value the \emph{limit} of $f$ at $p$ is below in Theorem~\ref{thrm.p.kernel.p}, culminating with part~\ref{item.fg.leq.cc} of that result. 
\end{defn}

Recall that Theorem~\ref{thrm.correlated} asserted that continuous value assignments are constant on transitive sets.
We can now prove a more general result along the same lines: 
\begin{thrm}
\label{thrm.p.kernel.p}
Suppose that:
\begin{itemize*}
\item
$(\ns P,\opens)$ is a semitopology.
\item
$p\in\ns P$ is regular. 
\item
$f,g:\ns P\to\tf{Val}$ are value assignments to some set of values $\tf{Val}$. 
\end{itemize*}
Then:
\begin{enumerate}
\item\label{item.p.kernel.p.1}
If $f$ is confident at $p$ (Definition~\ref{defn.regular.f}(\ref{item.regular.f.1})) then
$$
f\leq g
\quad\text{implies}\quad 
\limitat{p} f = \limitat{p} g .
$$
\item\label{item.ff.cc}
If $f$ is confident at $p$ and also $f$ is continuous at $p$ (Definition~\ref{defn.continuity}) then
$$
f(p)=\limitat{p} f.
$$
\item\label{item.fg.leq.cc}
Combining parts~\ref{item.p.kernel.p.1} and~\ref{item.ff.cc} of this result, if $f$ is confident at $p$ and $g$ is continuous at $p$ then
$$
f\leq g 
\quad\text{implies}\quad 
g(p)=\limitat{p} f.
$$
In words: the limit value of an $f$ confident at $p$, is the value of any sufficiently continuous extension of $f$ --- where `sufficiently continuous' means `continuous at $p$'.
\item\label{item.gpf}
As a corollary using Lemma~\ref{lemm.unanimous.implies.confident}, if $f$ is unanimous at $p$ (Definition~\ref{defn.regular.f}(\ref{item.regular.f.2})) and $g$ is continuous at $p$, then 
$$
f\leq g 
\quad\text{implies}\quad 
g(p) = \limitat{p} f = f(\community(p)).
$$
\end{enumerate}
\end{thrm}
\begin{proof}
We reason as follows:
\begin{enumerate}
\item
Since $f$ is confident at $p$, there exists some kernel atom $\varnothing\lessdot A\subseteq\kernel(p)$ 
on which $f$ is continuous (by Remark~\ref{rmrk.machinery.together}(\ref{item.machinery.together.3}) it does not matter which one).
Since $f\leq g$, $g$ is also continuous at $A$.
It follows from Definition~\ref{defn.fp.limit} that $\limitat{p} f = \limitat{p} g$.
\item
Since $f$ is confident at $p$, $f$ is continuous on some kernel atom $\varnothing\lessdot A\subseteq\kernel(p)$. 
Since $f$ is continuous at $p$, $f$ is continuous on some open neighbourhood $p\in O\in\opens$.
By Lemma~\ref{lemm.ker.intersect} $O\between A$, and using Corollary~\ref{corr.correlated.intersect} we have that $f(p) = f(O) = f(A) = \limitat{p} f$ as required. 
\item
Direct from parts~\ref{item.p.kernel.p.1} and~\ref{item.ff.cc} of this result, using Lemma~\ref{lemm.unanimous.implies.confident}(\ref{item.unanimous.implies.confident.1}) to note that $g$ is confident at $p$ because $f$ is and $f\leq g$.
\item
Suppose $f$ is unanimous at $p$.
Then by Lemma~\ref{lemm.unanimous.implies.confident} $f$ is confident at $p$, and we use part~\ref{item.fg.leq.cc} of this result.
\qedhere\end{enumerate}
\end{proof}

\begin{nttn}
Suppose $f$ is a function on sets, and $X$ is a set.
Recall that we write $f|_X$ for the function obtained by \deffont[restriction (of a function to a set)]{restricting} $f$ to $X$ (so that $\f{domain}(f|_X)=\f{domain}(f)\cap X$).
\end{nttn}

\begin{rmrk}
We can use Theorem~\ref{thrm.p.kernel.p} to obtain a result that seems to us similar in spirit to Arrow's theorem~\cite{fey:strpat} from social choice theory, in the sense that $\kernel(p)$ is identified as a `dictator set' for $\community(p)$ (the technical details seem to be different):
\end{rmrk}

\begin{corr}
\label{corr.boundary.kernel}
Suppose that:
\begin{itemize*}
\item
$(\ns P,\opens)$ is a 
semitopology.
\item
$p\oldin\ns P$ is regular.
\item
$f,f':\ns P\to\tf{Val}$ are value assignments to some set of values $\tf{Val}$. 
\item
$f$ and $f'$ are continuous and confident at $p$. 
\end{itemize*}
Then
$$
f|_{\kernel(p)}=f'|_{\kernel(p)}
\quad\text{implies}\quad
f(p)=f'(p).
$$
In words we can say:
\begin{quote}
Confident continuous values at regular points are determined by their kernel.
\end{quote}
Note that we assume that $f$ and $f'$ are equal on $\kernel(p)$, but they do not need to be continuous on all of $\kernel(p)$; they only need to be continuous (and confident) at $p$.
\end{corr}
\begin{proof}
By confidence of $f$ and $f'$ 
there exist
\begin{itemize*}
\item
a kernel atom $\varnothing\lessdot A\subseteq\kernel(p)$ on which $f$ is continuous and so (as discussed in Remark~\ref{rmrk.machinery.together}) on which $f$ is constant with value $f(A)=\limitat{p} f$, and 
\item
a kernel atom $\varnothing\lessdot A'\subseteq\kernel(p)$ on which $f'$ is continuous and so constant with value $f'(A')=\limitat{p} f'$.
\end{itemize*}
It may be that $A\neq A'$, but by Lemma~\ref{lemm.kernel.atoms.intersect} they intersect --- in symbols: $A\between A'$ --- so that $\limitat{p} f= \limitat{p} f'$.\footnote{If we only had $f|_O=f'|_O$ for some open set $O$ that intersects the kernel (so $O\between\kernel(p)$), then the reasoning would break down at this point. 
We would still know that $A\between A'$ but we would not necessarily know that $O\between (A\cap A')$ so that $f(A)=f(A')$ and $\limitat{p} f=\limitat{p} f'$.  (Remember that we have not assumed continuity on all of $\kernel(p)$, so $f$ and $f'$ might not be constant on $\kernel(p)$.)}

We use Theorem~\ref{thrm.p.kernel.p} and the above to reason as follows: 
$$
\begin{array}[b]{r@{\ }l@{\qquad}l}
f(p) =& \limitat{p} f 
&\text{Theorem~\ref{thrm.p.kernel.p}(\ref{item.ff.cc})} 
\\
=& \limitat{p} f' 
&\limitat{p} f {=} f(A),\ \limitat{p} f' {=} f'(A'),\ A\between A'
\\
=& f'(p) 
&\text{Theorem~\ref{thrm.p.kernel.p}(\ref{item.ff.cc})}
\end{array}
\qedhere$$
\end{proof}

\jamiepart{Semiframes: algebra and duality}
\label{part.2}
\jamiesection{Semiframes: compatible complete semilattices} 
\label{sect.semiframes}

\jamiesubsection{Complete join-semilattices, and morphisms between them} 

\begin{rmrk}[Setting the scene]
We have studied point-set semitopologies; now the challenge is to give an algebraic account of them.

A straightforward reading of the definition of a semitopology in Definition~\ref{defn.semitopology} is that a semitopology is a complete semilattice (under sets unions) in a powerset, so the algebraic version of this should be just a complete semilattice.
This turns out to be wrong: what we need is a \emph{compatible} complete semilattice, which we call a \emph{semiframe}. 
Precise details are in Definition~\ref{defn.semiframe}.

We proceed in three steps:
\begin{enumerate*}
\item
Between here and Section~\ref{sect.duality} we develop a duality between semitopologies and semiframes, which deliberately echoes the classic duality between topologies and frames.
\item
Then in Section~\ref{sect.closer.look.at.semifilters} we give algebraic versions of the antiseparation properties.
\item
Finally, in Section~\ref{sect.graphs} we (briefly) consider alternative representations, using graphs.
\end{enumerate*}
Taken together, this gives a fairly comprehensive algebraic treatment of the material we have seen thus far, complementing the point-set approach taken until now. 
\end{rmrk}

\begin{rmrk}
\label{rmrk.amazing}
Something amazing will happen below.
We have the compatibility relation $\ast$ in Subsection~\ref{subsect.compatibility.relation}.
This arises naturally in two ways: is is a natural algebraic abstraction of $\between$ in semitopologies (sets intersection) which we have used to express well-behavedness properties such as intertwinedness and regularity, but it will also be key to our categorical duality result. 

These motivations for $\ast$ are independent: the categorical duality does not require regularity, and the regularity properties do not require a duality --- and yet when we study both well-behavedness and duality, the same structures emerge.
\end{rmrk}

We recall some (mostly standard) definitions and facts:
\begin{defn}
\label{defn.complete.semilattice}
\leavevmode
\begin{enumerate*}
\item
A \deffont{poset} $(\ns X,\leq)$ is a set $\ns X$ of \deffont[elements (of a poset)]{elements} and a relation ${\leq}\subseteq\ns X\times\ns X$ that is transitive, reflexive, and antisymmetric.
\item
A poset $(\ns X,\leq)$ is a \deffont{complete join-semilattice} when every $X\subseteq\ns X$ ($X$ may be empty or equal to all of $\ns X$) has a least upper bound --- or \deffont[join (of elements in a poset)]{join} --- $\bigvee X\in\ns X$.

All the semilattices we consider will be join (rather than meet) semilattices, so we may omit the word `join' and just call this a \emph{complete semilattice} henceforth. 
\item\label{item.bot.X}
If $(\ns X,\leq)$ is a complete semilattice then we may write 
$$
\tbot_{\ns X}=\bigvee\varnothing.
$$
By the least upper bound property, $\tbot_{\ns X}\leq x$ for every $x\in\ns X$.
\item\label{item.top.X}
If $(\ns X,\leq)$ is a complete semilattice then we may write 
$$
\ttop_{\ns X}=\bigvee\ns X.
$$
By the least upper bound property, $x\leq \ttop_{\ns X}$ for every $x\in\ns X$.
\end{enumerate*}
\end{defn}

\begin{defn}
\label{defn.complete.semilattice.morphism}
Suppose $(\ns X',\leq')$ and $(\ns X,\leq)$ are complete join-semilattices.
Define a \deffont{morphism} $g:(\ns X',\leq')\to(\ns X,\leq)$ to be a function $\ns X'\to\ns X$ that commutes with joins, and sends $\ttop_{\ns X'}$ to $\ttop_{\ns X}$.
That is:
\begin{enumerate*}
\item\label{item.semilattice.morphism.join}
If $X'\subseteq\ns X'$ then $g(\bigvee X')=\bigvee_{x'{\in}\ns X'}g(x')$.
\item\label{item.semilattice.morphism.top}
$g(\ttop_{\ns X'})=\ttop_{\ns X}$.
\end{enumerate*} 
\end{defn}

\begin{rmrk}
In Definition~\ref{defn.complete.semilattice}(\ref{item.semilattice.morphism.top}) we insist that $g(\ttop_{\ns X'})=\ttop_{\ns X}$; i.e. we want our notion of morphism to preserve the top element.

This does not follow from Definition~\ref{defn.complete.semilattice}(\ref{item.semilattice.morphism.join}), because $g$ need not be surjective onto $\ns X$, so we need to add this as a separate condition.
Contrast with $g(\tbot_{\ns X})=\tbot_{\ns X'}$, which does follow from Definition~\ref{defn.complete.semilattice}(\ref{item.semilattice.morphism.join}), because $\tbot_{\ns X}$ is the least upper bound of $\varnothing$.

We want $g(\ttop_{\ns X'})=\ttop_{\ns X}$ because our intended model is that $(\ns X,\leq)=(\opens,\subseteq)$ is the semilattice of open sets of a semitopology $(\ns P,\opens)$, and similarly for $(\ns X',\leq')$, and $g$ is equal to $f^\mone$ where $f:(\ns P,\opens)\to(\ns P',\opens')$ is a continuous function. 
\end{rmrk}

We recall a standard result:
\begin{lemm}
\label{lemm.semi.hom.mon}
Suppose $(\ns X,\leq)$ is a complete join-semilattice. 
Then:
\begin{enumerate*}
\item\label{item.semi.hom.mon.1}
If $x_1,x_2\in\ns X$ then $x_1\leq x_2$ if and only if $x_1\tor x_2 = x_2$.
\item\label{item.semi.hom.mon.2}
If $f:(\ns X,\leq)\to(\ns X',\leq')$ is a semilattice morphism 
(Definition~\ref{defn.complete.semilattice.morphism})
then $f$ is a \deffont{monotone morphism}: if $x_1\leq x_2$ then $f(x_1)\leq f(x_2)$, for every $x_1,x_2\in\ns X$.
\end{enumerate*}
\end{lemm}
\begin{proof}
We consider each part in turn:
\begin{enumerate}
\item
Suppose $x_1\leq x_2$.
By the definition of a least upper bound, this means precisely that $x_2$ is a least upper bound for $\{x_1,x_2\}$.
It follows that $x_1\tor x_2=x_2$.
The converse implication follows just by reversing this reasoning.
\item
Suppose $x_1\leq x_2$.
By part~\ref{item.semi.hom.mon.1} of this result $x_1\tor x_2=x_2$, so $f(x_1\tor x_2)=f(x_2)$.
By Definition~\ref{defn.complete.semilattice.morphism} $f(x_1)\tor f(x_2)=f(x_2)$.
By part~\ref{item.semi.hom.mon.1} of this result $f(x_1)\leq f(x_2)$.
\qedhere\end{enumerate}
\end{proof}

\begin{rmrk}
As the reader may know, \emph{frames} and \emph{locales} are the same thing: the category of locales is just the categorical opposite of the category of frames.
So every time we write `semiframe', the reader can safely read `semilocale'; these are two names for essentially the same structure up to reversing arrows.
The literature on frames and locales is huge: the interested reader can consult two classic texts~\cite{johnstone:stos,maclane:sheglf}; more recent (and very readable) presentations include~\cite{picado:fraltw,picado:seppft}.
\end{rmrk}

\jamiesubsection{The compatibility relation} 
\label{subsect.compatibility.relation}

Definition~\ref{defn.compatibility.relation} is a simple idea, but so far as we are aware it is novel:
\begin{defn}
\label{defn.compatibility.relation}
Suppose $(\ns X,\leq)$ is a complete semilattice.
A \deffont{compatibility relation ${\ast}\subseteq\ns X\times\ns X$}\index{$x\ast x'$ (compatibility relation on elements)} is a relation on $\ns X$ such that: 
\begin{enumerate*}
\item\label{item.compatible.symmetric}
$\ast$ is \emph{symmetric}, so if $x,x'\in\ns X$ then 
$$
x\ast x' 
\quad\text{if and only if}\quad
x'\ast x. 
$$
\item\label{item.compatible.reflexive}
$\ast$ is a \deffont{properly reflexive relation},\footnote{`Properly reflexive' is a loose riff on terminologies like `proper subset of' or `proper ideal of a ring'.  We might also call this `non-$\tbot$ reflexive', which is descriptive, but perhaps a bit of a mouthful.}
 by which we mean 
$$
\Forall{x\in\ns X{\setminus}\{\tbot_{\ns X}\}} x\ast x .
$$
Note that it will follow from the axioms of a compatibility relation that $x\ast x\liff x\neq\tbot_{\ns X}$; see Lemma~\ref{lemm.tbot.incompatible}(\ref{item.properly.reflexive.iff}).
\item\label{item.compatible.distributive}
$\ast$ satisfies a \deffont[distributive law (for compatibility relation)]{distributive law}, that 
if $x\in \ns X$ and $X'\subseteq\ns X$ then
$$
x\ast\bigvee X' \liff \Exists{x'{\in}X'} x\ast x' .
$$
\end{enumerate*}
Thus we can say:
\begin{quote}
a compatibility relation ${\ast}\subseteq\ns X\times\ns X$ is a symmetric properly reflexive completely distributive relation on $\ns X$. 
\end{quote}
When $x\ast x'$ holds, we may call $x$ and $x'$ \deffont{compatible elements}.
\end{defn}

\begin{rmrk}
\label{rmrk.compatibility.intuition}
The compatibility relation $\ast$ is what it is, but we take a moment to discuss some intuitions, and to put it in the context of some natural generalisations:
\begin{enumerate}
\item
We can think of $\ast$ as an \emph{abstract intersection}.

It lets us observe whether $x$ and $x'$ intersect --- but without having to explicitly represent this intersection as a meet $x\tand x'$ in the semilattice itself.

We call $\ast$ a \emph{compatibility relation} following an intuition of $x,x'\in\ns X$ as observations, and $x\ast x'$ holds when there is some possible world at which it is possible to observe $x$ and $x'$ together.
More on this in Example~\ref{xmpl.simple.concrete.model}.
\item
We can think of $\ast$ as a \emph{generalised intersection}; so our notion of semiframe in Definition~\ref{defn.semiframe}
is an instance of a frame with a \emph{generalised} meet.

We will concentrate on the case where $x \ast x'$ measures whether $x$ and $x'$ intersect, but there are other possibilities.
Here are some natural ways to proceed:
\begin{enumerate}
\item
$(\ns X,\cti)$ is a complete join-semilattice and ${\ast} : (\ns X\times\ns X)\to \ns X$ is any commutative distributive map.
For concreteness, we can set $x\ast x' \in\{\tbot_{\ns X},\ttop_{\ns X}\}\subseteq\ns X$.
\item 
$(\ns X,\cti)$ is a complete join-semilattice and ${\ast} : (\ns X\times\ns X)\to \mathbb N$ is any commutative distributive map.
We think of $x\ast x'$ as returning the \emph{size} of the intersection of $x$ and $x'$.
\item
Any complete join-semilattice $(\ns X,\cti)$ is of course a (generalised) semiframe by taking $x\ast x' = \bigvee\{x'' \mid x''\cti x,\ x''\cti x'\}$.
\item
We can generalise further, in more than one direction.
We would take $(\ns X,\cti)$ and $(\ns X',\cti')$ to be complete join-semilattices and ${\ast} : (\ns X\times\ns X)\to \ns X'$ to be any commutative distributive map (which generalises the above).
We could also take $\ns X$ to be a cocomplete symmetric monoidal category~\cite[Section~VII]{maclane:catwm}: a category with all colimits and with a (symmetric) monoid action $\ast$ that distributes over (commutes with) colimits. 
\end{enumerate}
\end{enumerate}
See also Remark~\ref{rmrk.generalising.ast}.
\end{rmrk}

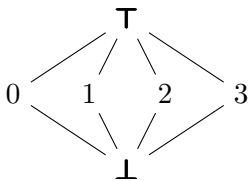
\begin{figure}
\begin{center}
\begin{tikzpicture}[scale=.55]
  \node (one) at (0,2) {$\ttop$};
  \node (a) at (-3,0) {$0$};
  \node (b) at (-1,0) {$1$};
  \node (c) at (1,0) {$2$};
  \node (d) at (3,0) {$3$};
  \node (zero) at (0,-2) {$\tbot$};
  \draw (zero) -- (a) -- (one) -- (b) -- (zero) -- (c) -- (one) -- (d) -- (zero);
\end{tikzpicture}
\end{center}
\caption{A semiframe with no abstract points (Lemma~\ref{lemm.no.converge})}
\label{fig.ast.no.tand}
\end{figure}

\begin{lemm}
\label{lemm.compatibility.monotone}
Suppose $(\ns X,\leq)$ is a complete semilattice and suppose ${\ast}\subseteq\ns X\times\ns X$ is a compatibility relation on $\ns X$.
Then:
\begin{enumerate*}
\item\label{item.ast.monotone}
$\ast$ is monotone on both arguments.

That is: if $x_1\ast x_2$ and $x_1\cti x_1'$ and $x_2\cti x_2'$, then $x_1'\ast x_2'$. 
\item\label{item.ast.lower.bound}
If $x_1,x_2\in\ns X$ have a non-$\tbot$ lower bound $\tbot_{\ns X}\lneq x\leq x_1,x_2$, then $x_1\ast x_2$.

In words we can write: $\ast$ reflects non-$\tbot$ lower bounds.
\item
The converse implication to part~\ref{item.ast.lower.bound} need not hold: it may be that $x_1\ast x_2$ ($x_1$ and $x_2$ are compatible) but the greatest lower bound of $\{x_1,x_2\}$ is $\tbot$.
\end{enumerate*}
\end{lemm}
\begin{proof}
We consider each part in turn:
\begin{enumerate}
\item
We argue much as for Lemma~\ref{lemm.semi.hom.mon}(\ref{item.semi.hom.mon.1}).
Suppose $x_1\ast x_2$ and $x_1\cti x_1'$ and $x_2\cti x_2'$.
By Lemma~\ref{lemm.semi.hom.mon} $x_1\tor x_1'=x_1'$ and $x_2\tor x_2'=x_2'$.
It follows using distributivity and commutativity (Definition~\ref{defn.compatibility.relation}(\ref{item.compatible.distributive}\&\ref{item.compatible.symmetric})) that $x_1\ast x_2$ implies that $(x_1\tor x_1')\ast (x_2\ast x_2')$, and thus that $x_1'\ast x_2'$ as required.
\item
Suppose $\tbot_{\ns X}\neq x \leq x_1,x_2$, so $x$ is a non-$\tbot_{\ns X}$ lower bound.
By assumption $\ast$ is properly reflexive (Definition~\ref{defn.compatibility.relation}(\ref{item.compatible.reflexive})) so (since $x\neq\tbot_{\ns X}$) $x\ast x$.
By part~\ref{item.ast.monotone} of this result it follows that $x_1\ast x_2$ as required.
\item
It suffices to provide a counterexample.
Define $(\ns X,\cti,\ast)$ by: 
\begin{itemize*}
\item
$\ns X = \{\tbot,0,1,\ttop\}$.
\item
$\tbot\cti 0,1 \cti \ttop$ and $\neg(0\cti 1)$ and $\neg(1\cti 0)$.
\item
$x\ast x'$ for every $\tbot\neq x,x'\in\ns X$.
\end{itemize*}
We note that $0\ast 1$ but the greatest lower bound of $\{0,1\}$ is $\tbot$.
We will revisit a slightly more elaborate version of this counterexample in Figure~\ref{fig.ast.no.tand}.
\qedhere\end{enumerate}
\end{proof}

\jamiesubsection{The definition of a semiframe}

\begin{defn}
\label{defn.semiframe}
A \deffont{semiframe} is a tuple $(\ns X,\cti,\ast)$ such that 
\begin{enumerate*}
\item
$(\ns X,\cti)$ is a complete semilattice (Definition~\ref{defn.complete.semilattice}), and 
\item
$\ast$ is a compatibility relation on it (Definition~\ref{defn.compatibility.relation}).
\end{enumerate*}
Slightly abusing terminology, we can say that 
\begin{quote}
semiframe = \emph{compatible complete semilattice}.
\end{quote}
\end{defn}

Semiframes are new, so far as we know, but they are a natural idea.
We consider some elementary ways to generate examples, starting with arguably the simplest possible instance:
\begin{xmpl}[The empty semiframe]
\label{xmpl.empty.semiframe}
Suppose $(\ns X,\cti,\ast)$ is a semiframe.
\begin{enumerate*}
\item
If $\ns X$ is a singleton set, so that $\ns X=\{\bullet\}$ for some element $\bullet$, then we call $(\ns X,\cti,\ast)$ the \deffont{empty semiframe} or \deffont{singleton semiframe}.

Then necessarily $\bullet=\tbot_{\ns X}=\ttop_{\ns X}$ and $\bullet\cti\bullet$ and $\neg(\bullet\ast\bullet)$.
\item\label{item.nonempty.semiframe}
If $\ns X$ has more than one element then we call $(\ns X,\cti,\ast)$ a \deffont{nonempty semiframe}.
Then necessarily $\tbot_{\ns X}\neq\ttop_{\ns X}$.
\end{enumerate*}
Thus, $(\ns X,\cti,\ast)$ is nonempty if and only if $\tbot_{\ns X}\neq\ttop_{\ns X}$.
We call a singleton semiframe \emph{empty}, because this corresponds to the semiframe of open sets of the empty topology, which has no points and one open set, $\varnothing$.
\end{xmpl}

Example~\ref{xmpl.simple.concrete.model} continues Remark~\ref{rmrk.compatibility.intuition}:
\begin{xmpl}
\label{xmpl.simple.concrete.model}
\leavevmode
\begin{enumerate*}
\item\label{item.open.semiframe}
Suppose $(\ns P,\opens)$ is a semitopology.
Then the reader can check that the \emph{semiframe of open sets} $(\ns P,\subseteq,\between)$ is a semiframe.
We will study this example in detail; see Definition~\ref{defn.semi.to.dg} and Lemma~\ref{lemm.Fr.semiframe}.
\item
Suppose $(\ns X,\leq,\tbot,\ttop)$ is a frame (a complete lattice such that meets distribute over arbitrary joins).
Then $(\ns X,\leq,\ast)$ is a semiframe, where $x\ast x'$ when $x\tand x'\neq\tbot$.\footnote{Just being a complete lattice is not enough; it has to be distributive as well.  Consider $\omega\plus 1=\mathbb N\cup\{\omega\}$ with its usual ordering, augmented with an element $d$ such that $0\leq d\leq\omega$.  Then $\omega=\bigvee\mathbb N$ and $d\ast \omega$, but $\neg(d\ast n)$ for every $n\in\mathbb N$.} 
\item
Take $\ns X=\{\tbot,0,1,\ttop\}$ with $\tbot\cti 0\cti\ttop$ and $\tbot\cti 1\cti \ttop$ (so $0$ and $1$ are incomparable).
There are two possible semiframe structures on this, characterised by choosing $0\ast 1$ or $\neg(0\ast 1)$.
\item
See also the semiframes used in Lemmas~\ref{lemm.no.converge}.
\end{enumerate*}
\end{xmpl}

Definition~\ref{defn.semi.to.dg} is just an example of semiframes for now, though we will see much more of it later: 
\begin{defn}[{\bf Semitopology $\to$ semiframe}]
\label{defn.semi.to.dg}
Suppose $(\ns P,\opens)$ is a semitopology.
Define the \deffont{semiframe of open sets $\tf{Fr}(\ns P,\opens)$} (cf. Example~\ref{xmpl.simple.concrete.model}(\ref{item.open.semiframe}))
by:
\begin{enumerate*}
\item
$\tf{Fr}(\ns P,\opens)$ has elements open sets $O\in\opens$.
\item
$\cti$ is subset inclusion.
\item\label{item.semiframe.ast}
$\ast$ is $\between$ (sets intersection).
\end{enumerate*}
We may write 
$$
(\opens,\subseteq,\between)
\quad\text{as a synonym for}\quad
\tf{Fr}(\ns P,\opens) .
$$
\end{defn}

\begin{lemm}
\label{lemm.Fr.semiframe}
Suppose $(\ns P,\opens)$ is a semitopology.
Then $(\opens,\subseteq,\between)$ is indeed a semiframe.
\end{lemm}
\begin{proof}
As per Definition~\ref{defn.semiframe} we must show that $\opens$ is a complete semilattice (Definition~\ref{defn.complete.semilattice})
and $\between$ is a compatibility relation (Definition~\ref{defn.compatibility.relation})
--- symmetric, properly reflexive, and distributive and satisfies a distributive law that if $O\between\bigcup\mathcal O'$ then $O\between O'$ for some $O'\in\mathcal O'$. 
These are all facts of sets.
\end{proof}

\begin{rmrk}
\label{rmrk.setting.the.scene.semiframes}
Definition~\ref{defn.semi.to.dg} and Lemma~\ref{lemm.Fr.semiframe} are the start of our development.
Once we have built more machinery, we will have a pair of translations:
\begin{itemize*}
\item
Definition~\ref{defn.semi.to.dg} and Lemma~\ref{lemm.Fr.semiframe} go from semitopologies to semiframes.
\item
Definition~\ref{defn.st.g} and Lemma~\ref{lemm.St.semitop} go from semiframes to semitopologies.
\end{itemize*}
These translations are part of a dual pair of functors between categories of semitopologies and semiframes, as described in Definitions~\ref{defn.morphism.semitopologies} and~\ref{defn.category.of.spatial.graphs} and Proposition~\ref{prop.semitop.adjunction}.

Semitopologies are (relatively) concrete: we have concrete points and open sets that are sets of points.
Semiframes are more abstract: we have a join-complete semilattice, and a compatibility relation.
The duality we are about to build will show how these two worlds interact and reflect each other.
\end{rmrk}

We conclude with a simple technical lemma:
\begin{lemm}
\label{lemm.tbot.incompatible}
Suppose $(\ns X,\leq,\ast)$ is a semiframe (a complete semilattice with a compatibility relation) and $x\in\ns X$.
Then:
\begin{enumerate*}
\item\label{item.tbot.incompatible.tbot}
$\neg(x\ast\tbot_{\ns X})$ and in particular $\neg(\tbot_{\ns X}\ast\tbot_{\ns X})$.
\item\label{item.properly.reflexive.iff}
$x\ast x$ if and only if $x\neq\tbot_{\ns X}$.
\item
$x\ast\ttop_{\ns X}$ if and only if $x\neq\tbot_{\ns X}$.
\item
$\ttop_{\ns X}\ast\ttop_{\ns X}$ holds precisely if $\ns X$ is nonempty (Example~\ref{xmpl.empty.semiframe}).
\end{enumerate*}
\end{lemm}
\begin{proof}
We consider each part in turn:
\begin{enumerate}
\item
Recall from Definition~\ref{defn.complete.semilattice}(\ref{item.bot.X}) that $\tbot_{\ns X}=\bigvee\varnothing$.
By distributivity (Definition~\ref{defn.compatibility.relation}(\ref{item.compatible.distributive}))
$$
x\ast\tbot_{\ns X}\liff \Exists{x'\in \varnothing}x\ast x' \liff \bot.
$$
\item
We just combine part~\ref{item.tbot.incompatible.tbot} of this result with 
Definition~\ref{defn.compatibility.relation}(\ref{item.compatible.reflexive}).
\item
Suppose $x\neq\tbot_{\ns X}$.
Then $\tbot_{\ns X}\lneq x\leq x\leq\ttop_{\ns X}$, and by Lemma~\ref{lemm.compatibility.monotone}(\ref{item.ast.lower.bound}) $x\ast\ttop_{\ns X}$.

Suppose $x=\tbot_{\ns X}$.
Then $\neg(x\ast\ttop_{\ns X})$ by combining commutativity of $\ast$ (Definition~\ref{defn.compatibility.relation}(\ref{item.compatible.symmetric})) with part~\ref{item.tbot.incompatible.tbot} of this result.
\item
If $\ns X$ is nonempty then by Example~\ref{xmpl.empty.semiframe} $\tbot_{\ns X}\neq\ttop_{\ns X}$ and so $\ttop_{\ns X}\ast\ttop_{\ns X}$ holds by part~\ref{item.properly.reflexive.iff} of this result.
However, in the degenerate case that $\ns X$ has one element then $\ttop_{\ns X}=\tbot_{\ns X}$ and $\ttop_{\ns X}\ast\ttop_{\ns X}$ does not hold. 
\qedhere\end{enumerate}
\end{proof}

\begin{rmrk}
Recall from \cite[Definition~5.22, page~128]{priestley:intlo} that if $\ns X$ is a lattice, then the \deffont{pseudocomplement} to $x\in\ns X$ is $x^*=\bigvee\{x'\in\ns X\mid x'\wedge x=\tbot\}$.
A semiframe $(\ns X,\cti,\ast)$ naturally supports a notion of pseudocomplement for $x\in\ns X$, given by 
$$
x^c=\bigvee\{x'\in\ns X\mid \neg(x'\ast x)\}.
$$
It is easy to prove that $\neg(x^c\ast x)$, arguing by contradiction: if $x^c\ast x$ then $\bigvee\{x'\mid \neg(x'\ast x)\}\ast x$, and by distributivity (Definition~\ref{defn.compatibility.relation}(\ref{item.compatible.distributive})) there exists $x'\in\ns X$ such that $x'\ast x$ and $\neg(x'\ast x)$, a contradiction.

Note that it may be that $(x^c)^c\lneq x$.
For example, in the semiframe illustrated in Figure~\ref{fig.ast.no.tand}, $0^c=\bigvee\{1,2,3\}=\ttop$ and $(0^c)^c=\tbot\lneq 0$ (this behaviour will be familiar to the reader who has seen, for example, double negation in intuitionistic logic).

$x^c$ and related constructions will be useful later, in Definition~\ref{defn.cast} and Lemma~\ref{lemm.cast.comp}. 
\end{rmrk}

\jamiesection{Semifilters \& abstract points}
\label{sect.semifilters.and.points}

\jamiesubsection{The basic definition, and discussion}

\begin{defn}
\label{defn.point}
Suppose $(\ns X,\cti,\ast)$ is a semiframe and suppose $\afilter\subseteq\ns X$.
Then: 
\begin{enumerate*}
\item\label{item.prime}
Call $\afilter$ \deffont[prime subset of a semiframe]{prime} when for every $x,x'\in\ns X$,
$$
x\tor x'\in \afilter 
\quad\text{implies}\quad
x\in \afilter\lor x'\in \afilter .
$$
\item\label{item.completely.prime}
Call $\afilter$ \deffont[completely prime subset of a semiframe]{completely prime} when for every (possibly empty) $X\subseteq\ns X$,
$$
\bigvee X\in \afilter
\quad\text{implies}\quad
\Exists{x{\in}X}x\in \afilter.
$$
(This condition is used in Lemma~\ref{lemm.op.commutes.with.joins}, which is needed for Lemma~\ref{lemm.Op.unions}.) 
\item\label{item.up-closed}
Call $\afilter$ \deffont[up-closed subset of a semiframe]{up-closed} when $x\in \afilter$ and $x\cti x'$ implies $x'\in \afilter$.
\item\label{item.weak.clique}
Call $\afilter$ \deffont[compatible subset of a semiframe]{compatible} when its elements are \deffont[pairwise compatible subset of a semiframe]{pairwise compatible}, by which we mean that $x\ast x'$ for every $x, x'\in \afilter$.
\item\label{item.semifilter}
A \deffont{semifilter} is a nonempty, up-closed, compatible subset $\afilter\subseteq\ns X$.
\item\label{item.maximal.semifilter}
Call $\afilter\subseteq\ns X$ a \deffont{maximal semifilter} when it is a semifilter and is contained in no strictly greater semifilter.
\item\label{item.abstract.point}
An \deffont[abstract point (completely prime semifilter)]{abstract point} is a completely prime semifilter.
\item
Write 
$$
\tf{Points}(\ns X,\cti,\ast)
$$
for the set of abstract points of $(\ns X,\cti,\ast)$.
\end{enumerate*}
\end{defn}

\begin{nttn}
We will generally write $\afilter\subseteq\ns X$ for a subset of $\ns X$ that is intended to be a semifilter, or for which in most cases of interest $\afilter$ is a semifilter.
We will generally write $\apoint\subseteq\ns X$ when the subset is intended to be an abstract point, or when in most cases of interest $\apoint$ is an abstract point.
\end{nttn}

\begin{rmrk}
\label{rmrk.no.meet}
\emph{Note on design:}
The notion of semifilter from Definition~\ref{defn.point} is, obviously, based on the standard notion of filter~\cite[I.2.2, page~12]{johnstone:stos}.
We just replace the \emph{closure under binary meets} condition 
\begin{quote}
`if $x,x'\in \afilter$ then $x\tand x'\in\afilter$' 
\end{quote}
with a weaker \emph{compatibility condition}
\begin{quote}
`if $x,x'\in\afilter$ then $x\ast x'$'.
\end{quote}
This is in keeping with our move from frames to semiframes, which weakens from $\tand$ to the compatibility relation $\ast$.

Note that a semifilter or abstract point need not be directed:
\begin{enumerate*}
\item
Consider $\nbhd(0)$ in the (semiframes of open sets of the) semitopologies in the left-hand and middle examples in Figure~\ref{fig.nbhd}.
In both cases, $\{0,1\},\{0,2\}\in\nbhd(0)$ but $\{0\}\notin\nbhd(0)$ because $\{0\}$ is not an open set.
\item
Consider $\{0,1,2\}$ with the discrete semitopology (so every set is open).
Then the set of all two- or three-element subsets $\{\{0,1\},\{1,2\},\{2,0\},\{0,1,2\}\}$ is a semifilter, but it is not closed under sets intersections because it does not contain $\{0\}$, $\{1\}$, or $\{2\}$.
\end{enumerate*}
This second example is particularly interesting.
As the reader may know, the intuition of a filter in topology is a set of \emph{approximations}.
But this example is clearly not approximating anything --- after all, we are in the discrete semitopology and there is no need to approximate anything since we can just take a singleton set!
This suggests that a better intuition for semiframe is a set of \emph{collaborations}; in this case, of $0$ with $1$, $1$ with $2$, and $2$ with $0$.

Thus in particular, the standard result in frames that a proper finite filter\footnote{Recall that a proper filter is a filter that does not contain $\tbot$.} has a non-$\tbot$ least element (obtained as the finite meet of all the elements in the filter), does not hold for semifilters in semiframes.
See also Remark~\ref{rmrk.other.properties} and Proposition~\ref{prop.non.tbot.lower.bound}(\ref{item.tbot.lower.bound.semifilter}).
\end{rmrk}

\begin{xmpl}
\label{xmpl.abstract.point}
Suppose $(\ns X,\cti,\ast)$ is a semiframe.
We recall some (standard) facts about abstract points, which carry over from topologies and frames:
\begin{enumerate*}
\item\label{item.xmpl.nbhd.abstract.point}
Suppose $(\ns P,\opens)$ is a semitopology and $(\ns X,\cti,\ast)=(\opens,\subseteq,\between)$. 
By Lemma~\ref{lemm.Fr.semiframe}, $(\ns X,\cti,\ast)$ is a semiframe. 

If $p\in\ns P$ then the \emph{neighbourhood system} 
$$
\nbhd(p)=\{O\in\opens \mid p\in O\}
$$ 
from Definition~\ref{defn.nbhd} is an abstract point: see Proposition~\ref{prop.nbhd.iff}.
Intuitively, $\nbhd(p)$ abstractly represents $p$ as the set of all of its open approximations in $\opens$.
\item
\label{item.xmpl.abstract.point.2}
Suppose $(\ns P,\opens)$ is a semitopology.
Then $(\opens,\subseteq,\between)$ could contain an abstract point that is not the neighbourhood semifilter $\nbhd(p)$ of a point $p\in\ns P$.
 
Set $\ns X=\{\mathbb Q\}\cup\{(\pi\minus q,\pi\plus q)\subseteq\mathbb Q \mid q\in\mathbb Q_{\geq 0}\}$ (the set of all symmetric open intervals around $\pi$ in the rational numbers $\mathbb Q$), and set ${\cti}={\subseteq}$ and ${\ast}={\between}$. 

Set $P=\ns X\setminus\{\varnothing\}$ to be the set of all \emph{nonempty} symmetric open intervals around $\pi$.
Note that $\pi\notin\mathbb Q$, but $P$ is a set of open sets `approximating' $\pi$. 
\item
\label{item.xmpl.abstract.point.3}
We mention one more (standard) example.
Consider $\mathbb N$ with the \deffont{final segment semitopology} such that opens are either $\varnothing$ or sets $n_\geq = \{n'\in\mathbb N \mid n'\geq n\}$.
Then $\{n_\geq \mid n\in\mathbb N\}$ is an abstract point.
Intuitively, this approximates a point at infinity, which we can understand as $\omega$. 
\end{enumerate*}
\end{xmpl}

\begin{lemm}
\label{lemm.compatible.not.tbot}
Suppose $(\ns X,\cti,\ast)$ is a semiframe and suppose $\afilter\subseteq\ns X$ is compatible.
Then $\tbot_{\ns X}\notin\afilter$.
\end{lemm}
\begin{proof}
By compatibility, $x\ast x$ for every $x\in\afilter$.
We use Lemma~\ref{lemm.tbot.incompatible}(\ref{item.tbot.incompatible.tbot}).
\end{proof}

\begin{rmrk}
\label{rmrk.other.properties}
We continue Remark~\ref{rmrk.no.meet}.

As the reader may know, a semiframe still has greatest lower bounds, because we can build them as $x\tand x' = \bigvee \{x'' \mid x''\cti x,\ x''\cti x'\}$.
It is just that this greatest lower bound may be unhelpful.
To see why, consider again the examples in Figure~\ref{fig.nbhd}.
In the left-hand and middle examples in Figure~\ref{fig.nbhd}, the greatest lower bound of $\{0,1\}$ and $\{0,2\}$ exists in the semiframe of open sets: but it is $\varnothing$ the empty set in the left-hand and middle example, not $\{0\}$.
In the right-hand example, the greatest lower bound of $\{0,\ast,1\}$ and $\{0,\ast,2\}$ is $\{0\}$, not $\{0,\ast\}$.

So the reader could ask whether perhaps we should add the following weakened meet-closure condition to the definition of semifilters (and thus to abstract points):
\begin{quote}
\emph{If $x, x'\in \afilter$ and $x\tand x'\neq\tbot$ then $x\tand x'\in \afilter$.}
\end{quote}
Intuitively, this insists that semifilters are closed under \emph{non-$\tbot$} greatest lower bounds. 
However, there are two problems with this:
\begin{itemize*}
\item
It would break our categorical duality proof in the construction of $g^\circ$ in Lemma~\ref{lemm.gcirc.well-defined}; see the discussion in Remark~\ref{rmrk.further.restrictions.on.points}.
This technical difficulty may be superable, but \dots
\item
\dots the condition is probably not what we want anyway.
It would mean that the set of open neighbourhoods of $\ast$ in the right-hand example of Figure~\ref{fig.nbhd}, would not be a semifilter, because it contains $\{0,\ast,1\}$ and $\{0,\ast,2\}$ but not its (non-$\varnothing$) greatest lower bound $\{0\}$.
\end{itemize*}
\end{rmrk}

\jamiesubsection{Properties of semifilters}

\jamiesubsubsection{Things that are familiar from filters}
\label{subsect.familiar}

\begin{lemm}
\label{lemm.P.top}
Suppose $(\ns X,\cti,\ast)$ is a semiframe and $\afilter\subseteq\ns X$ is a semifilter.
Then:
\begin{enumerate*}
\item\label{item.P.yes.top}
$\ttop_{\ns X}\in \afilter$.
\item\label{item.P.no.bot}
$\tbot_{\ns X}\notin \afilter$.
\end{enumerate*}
\end{lemm}
\begin{proof}
We consider each part in turn:
\begin{enumerate}
\item
By nonemptiness (Definition~\ref{defn.point}(\ref{item.abstract.point})) $\afilter$ is nonempty, so there exists some $x\in \afilter$.
By definition $x\cti \ttop_{\ns X}$.
By up-closure (Definition~\ref{defn.point}(\ref{item.up-closed})) $\ttop_{\ns X}\in \afilter$ follows. 
\item
By assumption in Definition~\ref{defn.point}(\ref{item.weak.clique}) elements in $\afilter$ are pairwise compatible (so $x\ast x$ for every $x\in\afilter$).
We use Lemma~\ref{lemm.compatible.not.tbot}.
\qedhere\end{enumerate} 
\end{proof}

\begin{lemm}
\label{lemm.maxfilter.vs.primefilter}
Suppose $(\ns X,\cti,\ast)$ is a semiframe.
It is possible for a semifilter $\afilter\subseteq\ns X$ to be completely prime but not maximal.
\end{lemm}
\begin{proof}
We give a standard example (which also works for frames and filters).
Take $\ns P=\{0,1\}$ and $\opens=\{\varnothing, \{0\}, \{0,1\}\}$.
Then $\apoint'=\{\{0,1\}\}$ is an abstract point, but it is not a maximal semifilter (it is not even a maximal abstract point) since $\apoint'$ is contained in the strictly larger semifilter $\{\{0\},\{0,1\}\}$ (which is itself also a strictly larger abstract point).
\end{proof}

\begin{lemm}
\label{lemm.prime.completely.prime.finite}
If $(\ns X,\cti,\ast)$ is a finite semiframe (meaning that $\ns X$ is finite) then the properties of 
\begin{itemize*}
\item
being a prime semifilter (Definition~\ref{defn.point}(\ref{item.prime})) and 
\item
being a completely prime semifilter (Definition~\ref{defn.point}(\ref{item.completely.prime})), 
\end{itemize*}
coincide.
\end{lemm}
\begin{proof}
This is almost trivial, except that if $X=\varnothing$ in the condition for being completely prime then we get that $\tbot_{\ns X}\notin P$ --- but we know that anyway from Lemma~\ref{lemm.P.top}(\ref{item.P.no.bot}), from the compatibility condition on semifilters.
\end{proof}

\begin{lemm}
\label{lemm.zorn.maximal.semifilter}
Suppose $(\ns X,\cti,\ast)$ is a semiframe.
Then:
\begin{enumerate*}
\item
The union of an ascending chain of semifilters in $\ns X$, is a semifilter in $\ns X$.
\item
As a corollary, every semifilter $F\subseteq\ns X$ is contained in some maximal semifilter $F'\subseteq\ns X$ (assuming Zorn's lemma).
\end{enumerate*}
\end{lemm} 
\begin{proof}
We consider each part in turn:
\begin{enumerate}
\item
By a straightforward verification of the conditions of being a semifilter from Definition~\ref{defn.point}(\ref{item.semifilter}).
\item
Direct application of Zorn's lemma.
\qedhere\end{enumerate}
\end{proof}

\begin{rmrk}
\label{rmrk.tbot.not.in}
\leavevmode
\begin{enumerate*}
\item\label{item.tbot.not.in.1}
Lemma~\ref{lemm.P.top}(\ref{item.P.no.bot}) has a small twist to it.
In the theory of \emph{filters}, it does not follow from the property of being nonempty, up-closed, and closed under finite meets, that $\tbot_{\ns X}\notin\afilter$; this must be added as a distinct condition if required.

In contrast, we see in the proof of Lemma~\ref{lemm.P.top}(\ref{item.P.no.bot}) that for semifilters, $\tbot_{\ns X}\notin\afilter$ follows from the compatibility condition.
\item\label{item.tbot.not.in.2}
Lemma~\ref{lemm.prime.completely.prime.finite} matters in particular to us here, because we are particularly interested in abstracting the behaviours of finite semitopologies, because our original motivation for looking at both of these structures comes from looking at real networks, which are finite.\footnote{This is carefully worded.  We care about \emph{abstracting} properties of finite semitopologies, but we should not restrict to considering \emph{only} semitopologies and semiframes that are actually finite!  See Remark~\ref{rmrk.why.infinite}.}
\end{enumerate*}
\end{rmrk}

\jamiesubsubsection{Things that are different from filters}
\label{subsect.things.that.are.different}

\begin{rmrk}
Obviously, by definition semifilters are necessarily compatible but not necessarily closed under meets.
But aside from this fact, we have so far seen semiframes and semifilters behave more-or-less like frames and filters, modulo small details like that mentioned in Remark~\ref{rmrk.tbot.not.in}(\ref{item.tbot.not.in.1}).

But there are also differences, as we will now briefly explore.
In the theory of (finite) frames, the following facts hold:
\begin{enumerate*}
\item
\emph{Every proper filter $\afilter$ has a greatest lower bound $x$, and $\afilter=x^\cti=\{x'\mid x\cti x'\}$.}

Just take $x=\bigwedge\afilter$ the meet of all of its (finitely many) elements.
This is not $\tbot$, by the filter's finite intersection property.
\item
\emph{Every proper filter can be extended to a maximal filter.}\footnote{A proper filter is a filter that does not contain $\tbot$.  A maximal filter is a filter that is maximal amongst proper filters.}

Just extend using Zorn's lemma (as in Lemma~\ref{lemm.zorn.maximal.semifilter}).
\item
\emph{Every maximal filter is completely prime.}

It is a fact of finite frames that a maximal filter is prime,\footnote{A succinct proof is in Wikipedia~\cite{wiki:Ideal_(order_theory)}.} and since we assume the frame is finite, it is also completely prime.
\item
\emph{Every non-$\tbot$ element $x\neq\tbot_{\ns X}$ in a finite frame is contained in some abstract point.}

Just form $\{x'\mid x\cti x'\}$, observe it is a filter, form a maximal filter above it, and we get an abstract point. 
\item
\emph{As a corollary, if the frame is nonempty (so $\tbot\neq\ttop$; see Example~\ref{xmpl.empty.semiframe}) then it has at least one abstract point.} 
\end{enumerate*}
In Lemma~\ref{lemm.no.converge} and Proposition~\ref{prop.non.tbot.lower.bound} we consider some corresponding \emph{non-properties} of (finite) semiframes.
\end{rmrk}

\begin{lemm}
\label{lemm.no.converge}
Suppose $(\ns X,\cti,\ast)$ is a semiframe.
It is possible for $\tf{Points}(\ns X,\cti,\ast)$ to be empty, even if $(\ns X,\cti,\ast)$ is nonempty (Example~\ref{xmpl.empty.semiframe}(\ref{item.nonempty.semiframe})).
This is possible even if $\ns X$ is finite, and even if $\ns X$ is infinite. 
\end{lemm}
\begin{proof}
It suffices to provide an example.
We define a semiframe as below, and as illustrated in Figure~\ref{fig.ast.no.tand}:
\begin{itemize*}
\item
$\ns X=\{\tbot,0,1,2,3,\ttop\}$.
\item
Let $x\cti x'$ when $x=x'$ or $x=\tbot$ or $x'=\ttop$.
\item
Let $x\ast x'$ when $x\tand x'\neq\tbot$.\footnote{Unpacking what that means, we obtain this: $x\neq\tbot \land x=x'$ or $x\neq\tbot \land x'=\ttop$ or $x'\neq\tbot \land x=\ttop$.

This definition for $\ast$ is what we need for our counterexample, but other choices for $\ast$ also yield valid semiframes.  For example, we can set $x\ast x'$ when $x,x'\neq\tbot$.
} 
\end{itemize*}
Then $(\ns X,\cti,\ast)$ has no abstract points.

For suppose $P$ is one such. 
By Lemma~\ref{lemm.P.top} $\ttop\in P$.
Note that $\ttop=0\tor 1=2\tor 3$.
By assumption $P$ is completely prime, we know that $0\in P \lor 1\in P$, and also $2\in P\lor 3\in P$.
But this is impossible because $0$, $1$, $2$, and $3$ are not compatible.

For the infinite case, we just increase the width of the semiframe by taking $\ns X=\{\tbot\}\cup\mathbb N\cup \{\ttop\}$.
\end{proof}

\begin{prop}
\label{prop.non.tbot.lower.bound}
Suppose $(\ns X,\cti,\ast)$ is a semiframe and $\afilter\subseteq\ns X$ is a semifilter.
Then:
\begin{enumerate*}
\item\label{item.tbot.lower.bound.semifilter}
It is not necessarily the case that $\afilter$ has a non-$\tbot$ greatest lower bound (even if $\ns X$ is finite).
\item
Every semifilter can be extended to a maximal semifilter, but \dots
\item\label{item.max.not.prime}
\dots this maximal semifilter is not necessarily prime (even if $\ns X$ is finite).
\item
There may exist a non-$\tbot$ element $x\neq\tbot_{\ns X}$ that is contained in no abstract point.
\end{enumerate*}
\end{prop}
\begin{proof}
We consider each part in turn:
\begin{enumerate}
\item
Consider $(\powerset(\{0,1,2\}),\subseteq,\between)$ and take 
$$
\afilter = \{\{0,1\},\ \{1,2\},\ \{0,2\},\ \{0,1,2\}\} .
$$
The greatest lower bound of $\afilter$ is $\varnothing$.
\item
This is Lemma~\ref{lemm.zorn.maximal.semifilter}.
\item
$\afilter$ from part~\ref{item.tbot.lower.bound.semifilter} of this result is maximal, and it cannot be extended to a point $P\supseteq\afilter$.
Figure~\ref{fig.ast.no.tand} gives another counterexample, and in a rather interesting way: the semitopology has four maximal semifilters $\{i,\ttop\}$ for $i\in\{0,1,2,3\}$, but by Lemma~\ref{lemm.no.converge} it has no prime semifilters at all.\footnote{See also a discussion of the design of the notion of semifilter in Remarks~\ref{rmrk.other.properties} and~\ref{rmrk.further.restrictions.on.points}.} 
\item
We just take $x=0\in\ns X$ from the example in 
Lemma~\ref{lemm.no.converge}
(see Figure~\ref{fig.ast.no.tand}).
Since this semiframe has no abstract points at all, there is no abstract point that contains $x$.
\qedhere\end{enumerate}
\end{proof}

\begin{rmrk}
For now, we will just read Proposition~\ref{prop.non.tbot.lower.bound} as a caution not to assume that semiframes and semifilters behave like frames and filters.
Sometimes they do, and sometimes they don't; we have to check.

We now proceed to build our categorical duality, culminating with Theorem~\ref{thrm.categorical.duality.semiframes}.
Once that machinery is constructed, 
we will continue our study of the fine structure of semifilters in Section~\ref{sect.closer.look.at.semifilters}.
\end{rmrk}

\jamiesubsection{Sets of abstract points}

\begin{defn}
\label{defn.Op}
Suppose $(\ns X,\cti,\ast)$ is a semiframe and recall $\tf{Points}(\ns X,\cti,\ast)$ from Definition~\ref{defn.point}(\ref{item.abstract.point}).
Define a map $\f{Op}:\ns X\to\powerset(\tf{Points}(\ns X,\cti,\ast))$ by
$$
\f{Op}(x) = \{P\in\tf{Points}(\ns X,\cti,\ast) \mid x\in P\} . 
$$
\end{defn}

\begin{lemm}
\label{lemm.op.commutes.with.joins}
Suppose $(\ns X,\cti,\ast)$ is a semiframe and $X\subseteq\ns X$.
Then
$$
\f{Op}(\bigvee X)=\bigcup_{x\in X}\f{Op}(x) .
$$
In words: we can say that $\f{Op}$ commutes with joins, and that $\f{Op}$ commutes with taking least upper bounds. 
\end{lemm}
\begin{proof}
Suppose $P\in\tf{Points}(\ns X,\cti,\ast)$.
We reason as follows:
$$
\begin{array}[b]{r@{\ }l@{\qquad}l}
P\in\f{Op}(\bigvee X)
\liff& 
\bigvee X\in P
&\text{Definition~\ref{defn.Op}}
\\
\liff&
\Exists{x{\in}X}x\in P
&\text{Definition~\ref{defn.point}(\ref{item.completely.prime})}
\\
\liff&
P\in\bigcup_{x\in X}\f{Op}(x)
&\text{Definition~\ref{defn.Op}}
\end{array}
\qedhere
$$
\end{proof}

\begin{prop}
\label{prop.semiframe.to.Op}
Suppose $(\ns X,\cti,\ast)$ is a semiframe and $x,x'\in\ns X$. 
Then:
\begin{enumerate*}
\item\label{item.semiframe.to.Op.subset}
If $x\cti x'$ then $\f{Op}(x)\subseteq \f{Op}(x')$. 
\item\label{item.semiframe.to.Op.between}
If $\f{Op}(x)\between \f{Op}(x')$ then $x\ast x'$. 
\item\label{item.semiframe.to.Op.top}
$\f{Op}(\ttop_{\ns X})=\tf{Points}(\ns X,\cti,\ast)$ 
and
$\f{Op}(\tbot_{\ns X})=\varnothing$.
\item\label{item.semiframe.to.Op.bigvee}
$\f{Op}(\bigvee X)=\bigcup_{x\in X}\f{Op}(x)$ for $X\subseteq\ns X$.
\end{enumerate*}
\end{prop}
\begin{proof}
We consider each part in turn:
\begin{enumerate}
\item
\emph{We prove that $x\cti x'$ implies $\f{Op}(x)\subseteq \f{Op}(x')$.}
 
Suppose $x\cti x'$, and consider some abstract point $P\in\f{Op}(x)$.
By Definition~\ref{defn.Op} $x\in P$, and by up-closure of $P$ (Definition~\ref{defn.point}(\ref{item.up-closed})) $x'\in P$, so by Definition~\ref{defn.Op} $P\in\f{Op}(x')$.
$P$ was arbitrary, and it follows that $\f{Op}(x)\subseteq\f{Op}(x')$. 
\item
\emph{We prove that $\f{Op}(x)\between \f{Op}(x')$ implies $x\ast x'$.}

Suppose there exists an abstract point $P\in\f{Op}(x)\cap\f{Op}(x')$.
By Definition~\ref{defn.Op} $x,x'\in P$, and by compatibility of $P$ (Definition~\ref{defn.point}(\ref{item.weak.clique})) $x\ast x'$. 
\item
Unpacking Definition~\ref{defn.Op}, it suffices to show that $\ttop_{\ns X}\in P$ and $\tbot_{\ns X}\notin P$, for every abstract point $P\in\tf{Points}(\ns X,\cti,\ast)$.
This is from Lemma~\ref{lemm.P.top}(\ref{item.P.yes.top}).
\item
This is just Lemma~\ref{lemm.op.commutes.with.joins}.
\qedhere\end{enumerate}
\end{proof}

\begin{rmrk}
\label{rmrk.not.enough.points}
Proposition~\ref{prop.semiframe.to.Op} carries a clear suggestion that 
$(\{\f{Op}(x) \mid x\in\ns X\},\subseteq,\between)$ is trying, in some sense, to be an isomorphic copy of $(\ns X,\cti,\ast)$.
Lemma~\ref{lemm.Op.may.fail} notes that it may not quite manage this, because there may not be enough points (indeed, there may not be any abstract points at all).
This will (just as for topologies and frames) lead us to the notion of a \emph{spatial} semiframe in Definition~\ref{defn.spatial.graph} and Proposition~\ref{prop.Op.subseteq}.
\end{rmrk}

\begin{lemm}
\label{lemm.Op.may.fail}
The converse implications in Proposition~\ref{prop.semiframe.to.Op}(\ref{item.semiframe.to.Op.subset}\&\ref{item.semiframe.to.Op.between}) need not hold.
That is:
\begin{enumerate*}
\item
There exists a semiframe $(\ns X,\cti,\ast)$ and $x,x'\in\ns X$ such that $\f{Op}(x)\subseteq\f{Op}(x')$ yet $x\not\cti x'$.
\item 
There exists a semiframe $(\ns X,\cti,\ast)$ and $x,x'\in\ns X$ such that $x\ast x'$ yet $\f{Op}(x)\notbetween\f{Op}(x')$. 
\end{enumerate*}
\end{lemm}
\begin{proof}
The example from Lemma~\ref{lemm.no.converge} (as illustrated in Figure~\ref{fig.ast.no.tand}) is a counterexample for both cases: 
\begin{itemize*}
\item
$\f{Op}(0)\subseteq\f{Op}(1)$ because both are equal to the empty set, yet $0\not\cti 1$; and 
\item
$\ttop\ast\ttop$ yet $\f{Op}(\ttop)\notbetween\f{Op}(\ttop)$.
\qedhere\end{itemize*}
\end{proof}

\jamiesubsection{The semitopology of abstract points}

Recall from Definition~\ref{defn.point}(\ref{item.abstract.point}) that an abstract point in a semiframe $(\ns X,\cti,\ast)$ is a nonempty up-closed compatible completely prime subset of $\ns X$, and recall from Definition~\ref{defn.Op} that 
$$
\f{Op}(x) = \{P\in \tf{Points}(\ns X,\cti,\ast) \mid x\in P\},
$$
or in words: $\f{Op}(x)$ is the set of abstract points that contain $x$.
\begin{defn}
\label{defn.OpX}
Suppose $(\ns X,\cti,\ast)$ is a semiframe.
Then define $\tf{Op}(\ns X,\cti,\ast)$ by
$$
\tf{Op}(\ns X,\cti,\ast) = \{\f{Op}(x) \mid x\in\ns X\} .
$$ 
\end{defn}

\begin{lemm}
\label{lemm.Op.unions}
Suppose $(\ns X,\cti,\ast)$ is a semiframe.
Then:
\begin{enumerate*}
\item\label{item.Op.unions.1}
$\tf{Op}(\ns X,\cti,\ast)$ from Definition~\ref{defn.OpX} is closed under arbitrary sets union.
\item\label{item.Op.unions.2}
As a corollary, $(\tf{Op}(\ns X,\cti,\ast),\subseteq)$ (in words: $\tf{Op}(\ns X,\cti,\ast)$ ordered by subset inclusion) is a complete join-semilattice.
\end{enumerate*}
\end{lemm}
\begin{proof}
Part~\ref{item.Op.unions.1} is just Lemma~\ref{lemm.op.commutes.with.joins}.
The corollary part~\ref{item.Op.unions.2} is just a fact, since $\tf{Op}(\ns X,\cti,\ast)\subseteq\powerset(\tf{Points}(\ns X,\cti,\ast))$, and sets union is the join (least upper bound) in the powerset lattice. 
\end{proof}

Recall from Definition~\ref{defn.semi.to.dg} and Lemma~\ref{lemm.Fr.semiframe} that we showed how to go from a semitopology $(\ns P,\opens)$ to a semiframe $(\opens,\subseteq,\between)$.
We now show how to go in the other direction: 
\begin{defn}[{\bf Semiframe $\to$ semitopology}]
\label{defn.st.g}
Suppose $(\ns X,\cti,\ast)$ is a semiframe. 
Define the \deffont{semitopology of abstract points $\tf{St}(\ns X,\cti,\ast)$}\index{$\tf{St}(\ns X,\cti,\ast)$ (semitopology of abstract points)} by
$$
\tf{St}(\ns X,\cti,\ast) = \bigl(\tf{Points}(\ns X,\cti,\ast), \tf{Op}(\ns X,\cti,\ast)\bigr) .
$$
Unpacking this a little:
\begin{enumerate*}
\item\label{item.st.pt}
The set of points of $\tf{St}(\ns X,\cti,\ast)$ is the set of abstract points $\tf{Points}(\ns X,\cti,\ast)$ from Definition~\ref{defn.point}(\ref{item.abstract.point}) --- namely, the completely prime nonempty up-closed compatible subsets of $\ns X$.\footnote{There are no guarantees in general about \emph{how many} abstract points exist; e.g. Lemma~\ref{lemm.no.converge} gives an example of a semiframe that has no abstract points at all and so maps to the empty semitopology.  Later on in Definition~\ref{defn.spatial.graph} we consider conditions to ensure the existence of abstract points.}
\item\label{item.st.op}
Open sets $\tf{Opens}(\ns X,\cti,\ast)$ are the $\f{Op}(x)$ from Definition~\ref{defn.Op}:
$$
\f{Op}(x)=\{P\in \tf{Points}(\ns X,\cti,\ast) \mid x\in P\} . 
$$  
\end{enumerate*}
\end{defn}

\begin{lemm}
\label{lemm.St.semitop}
Suppose $(\ns X,\cti,\ast)$ is a semiframe. 
Then $\tf{St}(\ns X,\cti,\ast)$ from Definition~\ref{defn.st.g} is indeed a semitopology.
\end{lemm}
\begin{proof}
From conditions~\ref{semitopology.empty.and.universe} and~\ref{semitopology.unions} of Definition~\ref{defn.semitopology}, we need to check that 
$\tf{Op}(\ns X,\cti,\ast)$ contains $\varnothing$ and $\tf{Points}(\ns X,\cti,\ast)$ and is closed under arbitrary unions.
This is from Proposition~\ref{prop.semiframe.to.Op}(\ref{item.semiframe.to.Op.top}\&\ref{item.semiframe.to.Op.bigvee}).
\end{proof}

Recall from Definitions~\ref{defn.st.g} and~\ref{defn.semi.to.dg} that $\tf{St}(\ns X,\cti,\ast)$ is a semitopology, and $\tf{Fr}\,\tf{St}(\ns X,\cti,\ast)$ is a semiframe each of whose elements is the set of abstract points of $(\ns X,\cti,\ast)$ that contain some $x\in\ns X$: 
\begin{lemm}
\label{lemm.st.opens.generator}
Suppose $(\ns X,\cti,\ast)$ is a semiframe. 
Then
$\f{Op}:(\ns X,\cti,\ast) \to \tf{Fr}\,\tf{St}(\ns X,\cti,\ast)$ is surjective.
\end{lemm}
\begin{proof}
Direct from Definition~\ref{defn.st.g}(\ref{item.st.op}).
\end{proof}

We conclude with Definition~\ref{defn.top.ind} and Proposition~\ref{prop.top.ind.eq}, which are standard properties of the construction in Definition~\ref{defn.st.g}.

\begin{defn}
\label{defn.topind}
\label{defn.top.ind}
Suppose $(\ns P,\opens)$ is a semitopology and $p,p'\in\ns P$.
Define $p\topind p'$\index{$p\topind p'$ (topologically indistinguishable points)} by
$$
p\topind p'
\quad\text{when}\quad
\Forall{O\in\opens} p\in O\liff p'\in O .
$$
We recall some standard terminology from topology: 
\begin{enumerate*}
\item\label{item.top.ind.1}
Call $p$ and $p'$ \deffont[topologically indistinguishable points]{topologically indistinguishable} when $p\topind p'$. 
\item\label{item.top.ind.2}
Call $p$ and $p'$ \deffont[topologically distinguishable]{topologically distinguishable} when $\neg(p\topind p')$ (so there exists some $O\in\opens$ such that $p\in O\land p'\notin O$ or $p\notin O\land p'\in O$).
\item\label{item.T0.space}
Call $(\ns P,\opens)$ a \deffont{$T_0$ space} when points are topologically indistinguishable precisely when they are equal, or in symbols: ${\topind} = {=}$.
\end{enumerate*}
\end{defn}

\begin{prop}
\label{prop.top.ind.eq}
Suppose $(\ns X,\cti,\ast)$ is a semiframe.
Then $\tf{St}(\ns X,\cti,\ast)$ (Definition~\ref{defn.st.g}) is a $T_0$ space.
\end{prop}
\begin{proof}
Suppose $P,P'\in\tf{Points}(\ns X,\cti,\ast)$.
Unpacking Definition~\ref{defn.point}(\ref{item.abstract.point}), this means that $P$ and $P'$ are completely prime nonempty up-closed compatible subsets of $\ns X$.

It is immediate that $P=P'$ implies $P\topind P'$. 

Now suppose $P\topind P'$ in $\tf{St}(\ns X,\cti,\ast)$; to prove $P=P'$ it would suffice to show that $x\in P \liff x\in P'$, for arbitrary $x\in\ns X$.
By Definition~\ref{defn.st.g}(\ref{item.st.op}), every open set in $\tf{St}(\ns X,\cti,\ast)$ has the form $\f{Op}(x)$ for some $x\in\ns X$.
We reason as follows: 
$$
\begin{array}{r@{\ }l@{\qquad}l}
x\in P \liff&
P\in\f{Op}(x) 
&\text{Definition~\ref{defn.Op}}
\\
\liff&
P'\in\f{Op}(x)
&\text{$P$, $P'$ top. indisting.}
\\
\liff&
x\in P' 
&\text{Definition~\ref{defn.Op}}
\end{array}
$$
Since $x$ was arbitrary and $P,P'\subseteq\opens$, it follows that $P=P'$ as required.
\end{proof}

\jamiesection{Spatial semiframes \& sober semitopologies}
\label{sect.spatial.and.sober}

\jamiesubsection{Definition of spatial semiframes}

\begin{rmrk}
We continue Remark~\ref{rmrk.not.enough.points}.
We saw in Example~\ref{xmpl.abstract.point}(\ref{item.xmpl.abstract.point.2}\&\ref{item.xmpl.abstract.point.3}) that there may be \emph{more} abstract points than there are concrete points, and in Remark~\ref{rmrk.not.enough.points} that there may also be \emph{fewer}.

In the theory of frames, the condition of being \emph{spatial} means that the abstract points and concrete points correspond.
We imitate this terminology for a corresponding definition on semiframes: 
\end{rmrk}

\begin{defn}[{\bf Spatial semiframe}]
\label{defn.spatial.graph}
Call a semiframe $(\ns X,\cti,\ast)$ \deffont[spatial semiframe]{spatial} when:
\begin{enumerate*}
\item\label{item.spatial.iff}
$\f{Op}(x)\subseteq \f{Op}(x')$ implies $x\cti x'$, for every $x,x'\in\ns X$. 
\item\label{item.spatial.ast.iff}
$x\ast x'$ implies $\f{Op}(x)\between \f{Op}(x')$, for every $x,x'\in\ns X$. 
\end{enumerate*}
\end{defn}

\begin{rmrk}
Not every semiframe is spatial, just as not every frame is spatial.
Lemma~\ref{lemm.no.converge} gives an example of a semiframe that is not spatial because it has no points at all, as illustrated in Figure~\ref{fig.ast.no.tand}.
\end{rmrk}

We check that the conditions in Definition~\ref{defn.spatial.graph} correctly strengthen the implications in Proposition~\ref{prop.semiframe.to.Op} to become logical equivalences:
\begin{prop}
\label{prop.Op.subseteq}
Suppose $(\ns X,\cti,\ast)$ is a spatial semiframe and $x,x'\in\ns X$.
Then:
\begin{enumerate*}
\item\label{item.Op.spatial.cti}
$x\cti x'$ if and only if $\f{Op}(x)\subseteq \f{Op}(x')$. 
\item\label{item.Op.spatial.ast}
$x\ast x'$ if and only if $\f{Op}(x)\between \f{Op}(x')$.
\item\label{item.Op.spatial.inj}
$x= x'$ if and only if $\f{Op}(x)=\f{Op}(x')$. 
\item\label{item.Op.top.all.points}
$\f{Op}(\ttop_{\ns X})=\tf{Points}(\ns X,\cti,\ast)$ 
and
$\f{Op}(\tbot_{\ns X})=\varnothing$.
\item\label{item.Op.vee}
$\f{Op}(\bigvee X)=\bigcup_{x\in X}\f{Op}(x)$ for $X\subseteq\ns X$.
\end{enumerate*}
\end{prop}
\begin{proof}
We consider each part in turn:
\begin{enumerate}
\item
\emph{We prove that $x\cti x'$ if and only if $\f{Op}(x)\subseteq \f{Op}(x')$.}
 
The right-to-left implication is direct from Definition~\ref{defn.spatial.graph}(\ref{item.spatial.iff}).
The left-to-right implication is Proposition~\ref{prop.semiframe.to.Op}(\ref{item.semiframe.to.Op.subset}).
\item
\emph{We prove that $x\ast x'$ if and only if $\f{Op}(x)\between \f{Op}(x')$.}

The left-to-right implication is direct from Definition~\ref{defn.spatial.graph}(\ref{item.spatial.ast.iff}).
The right-to-left implication is Proposition~\ref{prop.semiframe.to.Op}(\ref{item.semiframe.to.Op.between}).
\item
\emph{We prove that $x= x'$ if and only if $\f{Op}(x)= \f{Op}(x')$.}

If $x=x'$ then $\f{Op}(x)=\f{Op}(x')$ is immediate.
If $\f{Op}(x)=\f{Op}(x')$ then $\f{Op}(x)\subseteq\f{Op}(x')$ and $\f{Op}(x')\subseteq\f{Op}(x)$.
By part~\ref{item.Op.spatial.cti} of this result (or direct from Definition~\ref{defn.spatial.graph}(\ref{item.spatial.iff})) $x\cti x'$ and $x'\cti x$.
By antisymmetry of $\cti$ it follows that $x=x'$.
\item
This is just Proposition~\ref{prop.semiframe.to.Op}(\ref{item.semiframe.to.Op.top})
\item
This is just Lemma~\ref{lemm.op.commutes.with.joins}.
\qedhere\end{enumerate}
\end{proof}

Definition~\ref{defn.semiframe.iso} will be useful in a moment:\footnote{More on this topic later on in Definition~\ref{defn.category.of.spatial.graphs}, when we build the category of semiframes.}
\begin{defn}
\label{defn.semiframe.iso}
Suppose $(\ns X,\cti,\ast)$ and $(\ns X',\cti',\ast')$ are semiframes.
Then an \deffont[isomorphism between semiframes]{isomorphism} between them is a function $g:\ns X\to\ns X'$ such that:
\begin{enumerate*}
\item\label{item.semiframe.iso.1}
$g$ is a bijection between $\ns X$ and $\ns X'$.
\item\label{item.semiframe.iso.2}
$x_1\cti x_2$ if and only if $g(x_1)\cti g(x_2)$.
\item\label{item.semiframe.iso.3}
$x_1\ast x_2$ if and only if $g(x_1)\ast g(x_2)$.
\end{enumerate*}
\end{defn}

\begin{lemm}
\label{lemm.iso.semiframe.top}
Suppose $(\ns X,\cti,\ast)$ and $(\ns X',\cti',\ast')$ are semiframes and $g:\ns X\to\ns X'$ is an isomorphism between them.
Then $g(\tbot_{\ns X})=g(\tbot_{\ns X'})$ and $g(\ttop_{\ns X})=\ttop_{\ns X'}$.
\end{lemm}
\begin{proof}
By construction $\tbot_{\ns X}\leq x$ for every $x\in\ns X$.
It follows from Definition~\ref{defn.semiframe.iso}(\ref{item.semiframe.iso.2}) that $g(\tbot_{\ns X})\leq g(x)$ for every $x\in\ns X$; but $g$ is a bijection, so $g(\tbot_{\ns X})\leq x'$ for every $x'\in\ns X'$.
It follows that $g(\tbot_{\ns X})=\tbot_{\ns X'}$.

By similar reasoning we conclude that $g(\ttop_{\ns X})=\ttop_{\ns X'}$.
\end{proof}

\begin{rmrk}
Suppose $(\ns X,\cti,\ast)$ is a semiframe and recall from Definition~\ref{defn.OpX} that
$\tf{Op}(\ns X,\cti,\ast)=\{\f{Op}(x) \mid x\in\ns X\}$.
Then the intuitive content of Proposition~\ref{prop.Op.subseteq} is that a semiframe $(\ns X,\cti,\ast)$ is spatial when  
$(\ns X,\cti,\ast)$ is isomorphic (in the sense made formal by Definition~\ref{defn.semiframe.iso}) to
$(\tf{Op}(\ns X,\cti,\ast),\subseteq,\between)$.

And, because $\f{Op}(\ttop_{\ns X})=\tf{Points}(\ns X,\cti,\ast)$ we can write a slogan:
\begin{quote}
\emph{A semiframe is spatial when it is (up to isomorphism) generated by its abstract points.}
\end{quote}
We will go on to prove in Proposition~\ref{prop.Gr.P.spatial} that every semitopology generates a spatial semiframe --- and in Theorem~\ref{thrm.categorical.duality.semiframes} we will tighten and extend the slogan above to a full categorical duality. 
\end{rmrk}

\jamiesubsection{The neighbourhood semifilter $\nbhd(p)$} 

\jamiesubsubsection{The definition and basic lemma}

Recall from Definition~\ref{defn.nbhd} that $\nbhd(p)=\{O\in\opens \mid p\in O\}$.
\begin{prop}
\label{prop.nbhd.iff}
Suppose $(\ns P,\opens)$ is a semitopology and $p\in\ns P$ and $O\in\opens$.
Then:
\begin{enumerate*}
\item\label{item.nbhd.point}
$\nbhd(p)$ (Definition~\ref{defn.nbhd}) is an abstract point (a completely prime semifilter) in the semiframe $\tf{Fr}(\ns P,\opens)$ (Definition~\ref{defn.semi.to.dg}).
In symbols:
$$
\nbhd:\ns P\to \tf{Points}(\tf{Fr}(\ns P,\opens)) .
$$
\item\label{item.nbhd.iff}
The following are equivalent:
$$
\nbhd(p)\in\f{Op}(O)
\quad\liff\quad
O\in \nbhd(p)
\quad\liff\quad
p\in O.
$$
\item\label{item.nbhd.mone}
We have an equality:
$$
\nbhd^\mone(\f{Op}(O))=O.
$$
\end{enumerate*}
\end{prop}
\begin{proof}
We consider each part in turn:
\begin{enumerate}
\item
From Definition~\ref{defn.point}(\ref{item.abstract.point}), we must check that $\nbhd(p)$ is a nonempty, completely prime, up-closed, and compatible subset of $\opens$ when considered as a semiframe as per Definition~\ref{defn.semi.to.dg}.
All properties are by facts of sets; we give brief details:
\begin{itemize*}
\item
$\nbhd(p)$ is nonempty because $p\in \ns P\in\opens$.
\item
$\nbhd(p)$ is completely prime because it is a fact of sets that if $P\subseteq\opens$ and $p\in \bigcup P$ then $p\in O$ for some $O\in P$.
\item
$\nbhd(p)$ is up-closed because it is a fact of sets that if $p\in O$ and $O\subseteq O'$ then $p\in O'$.
\item
$\nbhd(p)$ is compatible because if $p\in O$ and $p\in O'$ then $O\between O'$.
\end{itemize*}
\item
By Definition~\ref{defn.Op}, $\f{Op}(O)$ is precisely the set of abstract points $P$ that contain $O$, and by part~\ref{item.nbhd.point} of this result $\nbhd(p)$ is one of those points.
By Definition~\ref{defn.nbhd}, $\nbhd(p)$ is precisely the set of open sets that contain $p$.
The equivalence follows.
\item
We reason as follows:
$$
\begin{array}[b]{r@{\ }l@{\quad}l}
p\in\nbhd^\mone(\f{Op}(O))
\liff&
\nbhd(p)\in\f{Op}(O)
&\text{Fact of function inverse}
\\
\liff&
p\in O
&\text{Part~2 of this result}
\end{array}
\qedhere$$
\end{enumerate}
\end{proof}

\begin{corr}
\label{corr.op.sub.between}
Suppose $(\ns P,\opens)$ is a semitopology and $O,O'\in\opens$.
Then:
\begin{enumerate*}
\item
$\f{Op}(O)\subseteq\f{Op}(O')$ if and only if $O\subseteq O'$.
\item
$\f{Op}(O)\between \f{Op}(O')$ if and only if $O\between O'$.
\item
As a corollary, $\nbhd^\mone(\varnothing)=\varnothing$ and $\nbhd^\mone(\tf{Points}(\opens,\subseteq,\between))=\ns P$; i.e. $\nbhd^\mone$ maps the bottom/top element to the bottom/top element.
\end{enumerate*}
\end{corr}
\begin{proof}
We consider each part in turn:
\begin{enumerate}
\item
If $\f{Op}(O)\subseteq\f{Op}(O')$ then $\nbhd^\mone(\f{Op}(O))\subseteq\nbhd^\mone(\f{Op}(O'))$ by facts of inverse images, and $O\subseteq O'$ follows by Proposition~\ref{prop.nbhd.iff}(\ref{item.nbhd.mone}).

If $O\subseteq O'$ then $\f{Op}(O)\subseteq\f{Op}(O')$ by Proposition~\ref{prop.semiframe.to.Op}(\ref{item.semiframe.to.Op.subset}).
\item 
If $O\between O'$ then there exists some point $p\in\ns P$ with $p\in O\cap O'$.
By Proposition~\ref{prop.nbhd.iff}(\ref{item.nbhd.point}) $\nbhd(p)$ is an abstract point, and by Proposition~\ref{prop.nbhd.iff}(\ref{item.nbhd.iff}) $\nbhd(p)\in\f{Op}(O)\cap\f{Op}(O')$; thus $\f{Op}(O)\between\f{Op}(O')$.

If $\f{Op}(O)\between\f{Op}(O')$ then $O\between O'$ by Proposition~\ref{prop.semiframe.to.Op}(\ref{item.semiframe.to.Op.between}).
\item
Routine from Proposition~\ref{prop.semiframe.to.Op}(\ref{item.semiframe.to.Op.top}) (or from Lemma~\ref{lemm.iso.semiframe.top}).
\qedhere\end{enumerate}
\end{proof}

\jamiesubsubsection{Application to semiframes of open sets}

\begin{prop}
\label{prop.nbhd.mone.bijects}
Suppose $(\ns P,\opens)$ is a semitopology.
Then:
\begin{enumerate*}
\item\label{item.nbhd.mone.bijects.1}
$\nbhd^\mone$ bijects open sets of $\tf{St}(\opens,\subseteq,\between)$ (as defined in Definition~\ref{defn.st.g}(\ref{item.st.op})), with open sets of $(\ns P,\opens)$, taking $\f{Op}(O)$ to $O$.
\item\label{item.nbhd.iso}
$\nbhd^\mone$ is an isomorphism between the semiframe of open sets of $\tf{St}(\opens,\subseteq,\between)$, and the semiframe of open sets of $(\ns P,\opens)$ (Definition~\ref{defn.semiframe.iso}).
\end{enumerate*}
\end{prop}
\begin{proof}
We consider each part in turn:
\begin{enumerate}
\item
Unpacking Definition~\ref{defn.st.g}(\ref{item.st.op}), an open set in $\tf{St}\,\tf{Fr}(\ns P,\opens)$ has the form $\f{Op}(O)$ for some $O\in\opens$.
By Proposition~\ref{prop.nbhd.iff}(\ref{item.nbhd.mone}) $\nbhd^\mone(\f{Op}(O))=O$, and so $\nbhd^\mone$ is surjective and injective.
\item
Unpacking Definition~\ref{defn.semiframe.iso} it suffices to check that:
\begin{itemize*}
\item
$\nbhd^\mone$ is a bijection, and maps $\f{Op}(O)$ to $O$. 
\item
$\f{Op}(O)\subseteq\f{Op}(O')$ if and only if $O\subseteq O'$.
\item
$\f{Op}(O)\between \f{Op}(O')$ if and only if $O\between O'$.
\end{itemize*}
The first condition is part~\ref{item.nbhd.mone.bijects.1} of this result; the second and third are from Corollary~\ref{corr.op.sub.between}.
\qedhere\end{enumerate}
\end{proof}

\begin{prop}
\label{prop.Gr.P.spatial}
Suppose $(\ns P,\opens)$ is a semitopology.
Then the semiframe $\tf{Fr}(\ns P,\opens)=(\opens,\subseteq,\between)$ from Definition~\ref{defn.semi.to.dg} is spatial.
\end{prop}
\begin{proof}
The properties required by Definition~\ref{defn.spatial.graph} are that $\f{Op}(O)\subseteq\f{Op}(O')$ implies $O\subseteq O'$, and $O\between O'$ implies $\f{Op}(O)\between\f{Op}(O')$.
Both of these are immediate from Proposition~\ref{prop.nbhd.mone.bijects}(\ref{item.nbhd.iso}).
\end{proof}

\jamiesubsubsection{Application to characterise $T_0$ spaces}

\begin{lemm}
\label{lemm.nbhd.top.ind}
Suppose $(\ns P,\opens)$ is a semitopology and $p,p'\in\ns P$.
Then the following are equivalent:
\begin{enumerate*}
\item\label{item.nbhd.top.ind.1}
$\nbhd(p)=\nbhd(p')$\ (cf. also Lemma~\ref{lemm.intertwined.sober})
\item\label{item.nbhd.top.ind.2}
$\Forall{O{\in}\opens}p\in O\liff p\in O'$
\item\label{item.nbhd.top.ind.3}
$p\topind p'$\ (Definition~\ref{defn.topind}: $p$ and $p'$ are topologically indistinguishable in $(\ns P,\opens)$).
\item\label{item.nbhd.top.ind.4}
$\nbhd(p)\topind \nbhd(p')$\ ($\nbhd(p)$ and $\nbhd(p')$ are topologically indistinguishable as --- by Proposition~\ref{prop.nbhd.iff}(\ref{item.nbhd.point}) --- abstract points in $\tf{St}\,\tf{Fr}(\ns P,\opens)$). 
\end{enumerate*}
\end{lemm}
\begin{proof}
Equivalence of parts~\ref{item.nbhd.top.ind.1} and~\ref{item.nbhd.top.ind.2} is direct from Definition~\ref{defn.nbhd}.
Equivalence of parts~\ref{item.nbhd.top.ind.2} and~\ref{item.nbhd.top.ind.3} is just Definition~\ref{defn.top.ind}(\ref{item.top.ind.1}).
Equivalence of parts~\ref{item.nbhd.top.ind.4} and~\ref{item.nbhd.top.ind.1} is from Proposition~\ref{prop.top.ind.eq}.
\end{proof}

\begin{corr}
\label{corr.T0.nbhd.inj}
Suppose $(\ns P,\opens)$ is a semitopology.
Then the following are equivalent:
\begin{enumerate*}
\item
$(\ns P,\opens)$ is $T_0$ (Definition~\ref{defn.top.ind}(\ref{item.T0.space})).
\item
$\nbhd:\ns P\to \tf{Points}(\opens,\subseteq,\between)$ is injective.
\end{enumerate*}
\end{corr}
\begin{proof}
Suppose $(\ns P,\opens)$ is $T_0$, and suppose $\nbhd(p)=\nbhd(p')$.
By Lemma~\ref{lemm.nbhd.top.ind}(\ref{item.nbhd.top.ind.1}\&\ref{item.nbhd.top.ind.3}) $p\topind p'$. 
By Definition~\ref{defn.top.ind}(\ref{item.top.ind.2}) $p=p'$.
Since $p$ and $p'$ were arbitrary, $\nbhd$ is injective.

Suppose $\nbhd$ is injective.
Reversing the reasoning of the previous paragraph, we deduce that $(\ns P,\opens)$ is $T_0$.
\end{proof}

\jamiesubsection{Sober semitopologies}
\label{subsect.sober.semitopologies}

Recall from Proposition~\ref{prop.Gr.P.spatial} that if we go from a semitopology $(\ns P,\opens)$ to a semiframe $(\opens,\subseteq,\between)$, then the result is not just any old semiframe --- it is a \emph{spatial} one. 

We now investigate what happens when we go from a semiframe to a semitopology using Definition~\ref{defn.st.g}.

\jamiesubsubsection{The definition and a key result}
 
\begin{defn}
\label{defn.sober.semitopology}
Call a semitopology $(\ns P,\opens)$ \deffont[sober semitopology]{sober} when every abstract point $P$ of $\tf{Fr}(\ns P,\opens)$ --- i.e. every completely prime nonempty up-closed compatible set of open sets --- is equal to the neighbourhood semifilter $\nbhd(p)$ of some unique $p\in\ns P$. 
Equivalently, $(\ns P,\opens)$ is sober when $\nbhd:\ns P\to\tf{Points}(\tf{Fr}(\ns P,\opens))$ (Definition~\ref{defn.point}(\ref{item.abstract.point})) is a bijection. 
\end{defn}

\begin{rmrk}
\label{rmrk.enough.points}
A bijection is a map that is injective and a surjective.
We noted in Corollary~\ref{corr.T0.nbhd.inj} that a space is $T_0$ when $\nbhd$ is injective.
So the sobriety condition can be thought of as having two parts: 
\begin{itemize*}
\item
$\nbhd$ is injective and the space is $T_0$, so it intuitively contains no `unnecessary' duplicates of points;
\item
$\nbhd$ is surjective, so the space contains `enough' points that there is (precisely) one concrete point for every abstract point.\footnote{`Unnecessary' and `enough' are in scare quotes here because these are subjective terms.  For example, if points represent computer servers on a network then we might consider it a \emph{feature} to not be $T_0$ by having multiple points that are topologically indistinguishable --- e.g. for backup, or to reduce latency --- and likewise, we might consider it a feature to not have one concrete point for every abstract point, if this avoids redundancies.  There is no contradiction here: a computer network based on a small non-sober space with multiple backups of what it has, may be a more efficient and reliable system than one based on a larger non-sober space that does not back up its servers but is full of redundant points.
And, this smaller non-sober space may present itself to the user abstractly as the larger, sober space. 

Users may even forget about the computation that goes on under the hood of this abstraction, as illustrated by the following \emph{true story:} I had a paper presenting an efficient algorithm rejected because it `lacked motivation'.  Why?  Because the algorithm was unnecessary: the reviewer claimed, apparently with a straight face, that guessing the answer until you got it right was computationally equivalent. 
}
\end{itemize*} 
\end{rmrk}

We start with a very simple example of sober semitopologies:
\begin{lemm}
\label{lemm.discrete.sober}
Suppose $\ns P$ is any set.
Then the discrete semitopology $(\ns P,\powerset(\opens))$ is sober.
\end{lemm}
\begin{proof}
Consider an abstract point $P\subseteq\opens$ (completely prime nonempty up-closed and compatible, as per Definition~\ref{defn.point}(\ref{item.abstract.point})).
Then $\ns P\in P$ and $\ns P=\bigvee\{\{p\}\mid p\in\ns P\}$.
Since $P$ is completely prime, $\{p\}\in P$ for some $p\in\ns P$.
It follows easily that $P=\nbhd(p)$.
\end{proof}

\begin{xmpl}
\label{xmpl.sober.non-sober}
We give some more examples of sober and non-sober semitopologies.
\begin{enumerate}
\item
Take $\ns P=\{0,1\}$ and $\opens=\{\varnothing,\{0,1\}\}$.
This has one abstract point $P=\{\{0,1\}\}$ but two concrete points $0$ and $1$.
It is therefore not sober.
\item
Take $\ns P=\{0,1\}$ and $\opens=\{\varnothing,\{1\},\{0,1\}\}$.
This has two abstract points 
$$
\{\{1\},\{0,1\}\}
\quad\text{and}\quad \{\{0,1\}\}
$$ 
corresponding to two concrete points $0$ and $1$.
It is sober.
\item
Take $\ns P=\mathbb N$ with the final topology; so $O\in\opens$ when $O=\varnothing$ or $O=n_\geq$ for some $n\in\mathbb N$, where $n_\geq = \{n'\in\mathbb N \mid n'\geq n\}$. 
Take $P=\{n_\geq \mid n\in\mathbb N\}$.
The reader can check that this is an abstract point (up-closed, completely prime, compatible); however $P$ is not the neighbourhood semifilter of any $n\in\mathbb N$.
Thus this space is not sober.
\item\label{item.sober.R}
$\mathbb R$ with its usual topology (which is also a semitopology) is sober.

This is a known result for topologies, but Remark~\ref{rmrk.no.meet} (and also the later Remark~\ref{rmrk.sober.not.hausdorff}) caution us that we cannot take this for granted, so we sketch the proof. 
 
Suppose $P$ is an abstract point; we wish to exhibit a unique $p\in\mathbb R$ such that $P=\nbhd(p)$.

We cover $\mathbb R$ with open intervals of radius $1$ by writing $\mathbb R=\bigcup\{\openinterval{r\minus 0.5,r\plus 0.5} \mid r\in\mathbb R\}$, and we use the completely prime property to find (at least one) such open interval that is in $P$; write it $O_1\in P$.
We then cover $O_1$ with open intervals of radius at most $1/2$ by writing $O_1=\bigcup\{O_1\cap\openinterval{r\minus 0.25,r\plus 0.25}\mid r\in O_1\}$, and we iterate to obtain a sequence $(O_i\mid i\in\mathbb N)\subseteq P$.
This converges to some unique $p\in\mathbb R$.
We check that $P=\nbhd(p)$:
\begin{itemize*}
\item
Suppose $O\in\opens$ is such that $p\in O$.
Because $p\in O$, there exists some $\epsilon$ such that $\openinterval{p\minus\epsilon,p\plus\epsilon}\subseteq O$.
It follows that for $i>1/\epsilon$ we have $O_i\subseteq O$ and thus $O\in P$ by up-closedness. 
\item
Suppose $O'\in\opens$ is such that $p\notin O'$.
For a sufficiently large $i$ we have $O_i\notintersectswith O'$, so by compatibility it follows that $O'\notin P$.
\end{itemize*}
\item
$\mathbb Q$ is sober.
The argument is much as for $\mathbb R$ above.
We have to work just a little harder because the $p$ we obtain need not be rational, but the arguments on open intervals remain valid.
See a brief discussion in point~\ref{item.rounded.sober} in Subsection~\ref{subsect.open.problems}.
\end{enumerate}
\end{xmpl}

\begin{prop}
\label{prop.spatial.gives.sober}
Suppose $(\ns X,\cti,\ast)$ is a semiframe. 
Then $\tf{St}(\ns X,\cti,\ast)$ from Definition~\ref{defn.st.g} is a sober semitopology.
\end{prop}
\begin{proof}
We know from Lemma~\ref{lemm.St.semitop} that $\tf{St}(\ns X,\cti,\ast)$ is a semitopology.
The issue is whether it is sober; thus by Definition~\ref{defn.sober.semitopology} we wish to exhibit every abstract point $P$ of $\tf{Fr}\,\tf{St}(\ns X,\cti,\ast)$ as a neighbourhood semifilter $\nbhd(p)$ for some unique abstract point $p$ of $(\ns X,\cti,\ast)$.
The calculations to do so are routine, but we give details.

Fix some abstract point $P$ of $\tf{Fr}\,\tf{St}(\ns X,\cti,\ast)$.
By Definition~\ref{defn.point}(\ref{item.abstract.point}), 
$P$ 
is a completely prime nonempty up-closed set of intersecting open sets in the semitopology $\tf{St}(\ns X,\cti,\ast)$, and by Definition~\ref{defn.st.g}(\ref{item.st.op}) each open set in $\tf{St}(\ns X,\cti,\ast)$ has the form $\f{Op}(x)=\{p\in\tf{Points}(\ns X,\cti,\ast) \mid x\in p\}$ for some $x\in\ns X$. 

We define $p\subseteq\ns X$ as follows:
$$
p=\{x\in\ns X \mid \f{Op}(x)\in P\} \subseteq\ns X .
$$
By construction we have that $x\in p$ if and only if $\f{Op}(x)\in P$, and so
$$
\begin{array}{r@{\ }l@{\qquad}l}
\nbhd(p) 
=&
\{\f{Op}(x) \mid p\in \f{Op}(x)\}
&\text{Definition~\ref{defn.nbhd}}
\\
=&
\{\f{Op}(x) \mid x\in p\}
&\text{Definition~\ref{defn.Op}}
\\
=&
\{\f{Op}(x) \mid \f{Op}(x)\in P\}
&\text{Construction of $p$}
\\
=&
P 
&\text{Fact}
.
\end{array}
$$
Now $P$ is completely prime, nonempty, up-closed, and compatible and it follows by elementary calculations using Proposition~\ref{prop.Op.subseteq} that $p$ is also completely prime, nonempty, up-closed, and compatible --- so $p$ is an abstract point of $(\ns X,\cti,\ast)$.

So we have that
$$
p\in\tf{Point}(\ns X,\cti,\ast)
\quad\text{and}\quad
P = \f{nbhd}(p) .
$$
To prove uniqueness of $p$, suppose $p'$ is any other abstract point such that $P=\nbhd(p')$.
We follow the definitions: $\f{Op}(x)\in \nbhd(p') \liff \f{Op}(x)\in \nbhd(p)$, and thus by Definition~\ref{defn.nbhd} $p'\in\f{Op}(x) \liff p\in\f{Op}(x)$, and thus by Definition~\ref{defn.Op} $x\in p'\liff x\in p$, and thus $p'=p$. 
\end{proof}

\jamiesubsubsection{Sober topologies contrasted with sober semitopologies}
\label{subsect.sober.top.contrast}

\begin{figure}
\vspace{-1em}
\centering
\subcaptionbox{Finite $T_0$ (and also $T_1$) semitopology that is not sober (Lemma~\ref{lemm.T0.not.sober})}{\includegraphics[width=0.35\columnwidth,trim={50 20 50 20},clip]{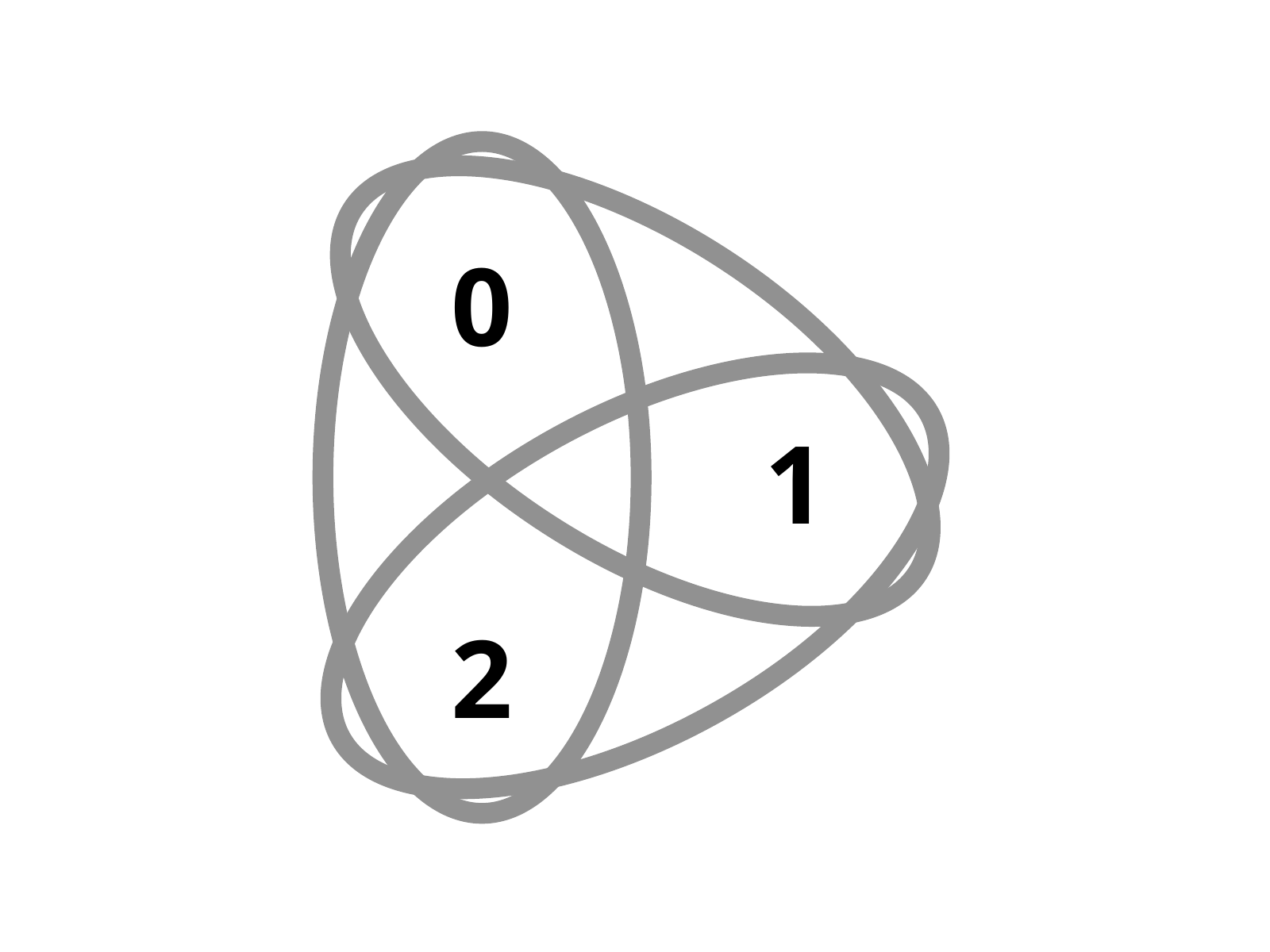}}
\qquad
\subcaptionbox{Hausdorff semitopology that is not sober (Lemma~\ref{lemm.hausdorff.not.sober})}{\includegraphics[width=0.4\columnwidth,trim={50 20 0 20},clip]{diagrams/square-diagram.pdf}}
\caption{Two counterexamples for sobriety}
\label{fig.012-triangle}
\end{figure}

\begin{figure}
\vspace{-1em}
\centering
\subcaptionbox{Soberification of left-hand example in Figure~\ref{fig.012-triangle}}{\includegraphics[width=0.35\columnwidth,trim={50 20 50 20},clip]{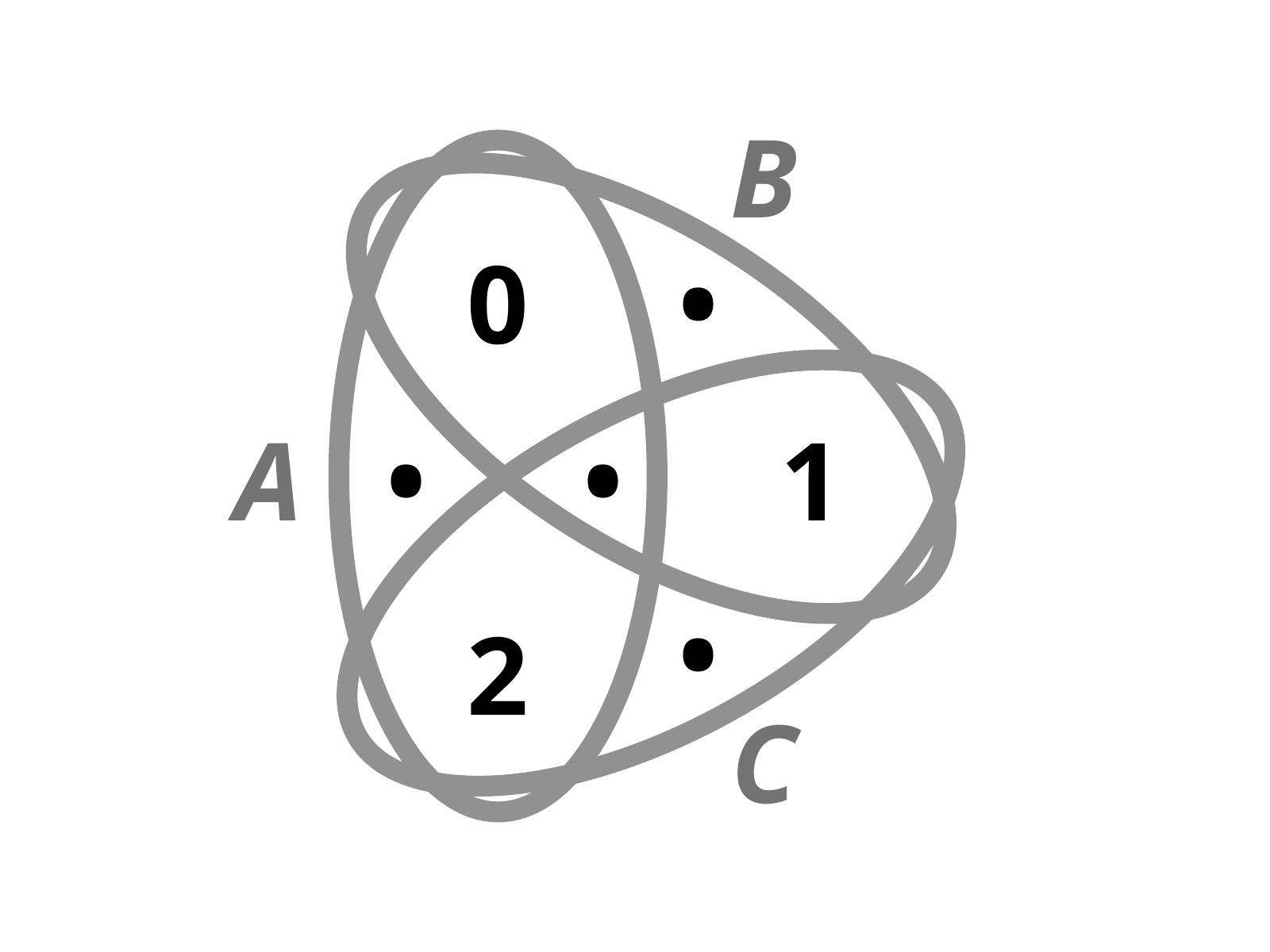}}
\qquad
\subcaptionbox{Soberification of right-hand example in Figure~\ref{fig.012-triangle}}{\includegraphics[width=0.4\columnwidth,trim={50 20 0 20},clip]{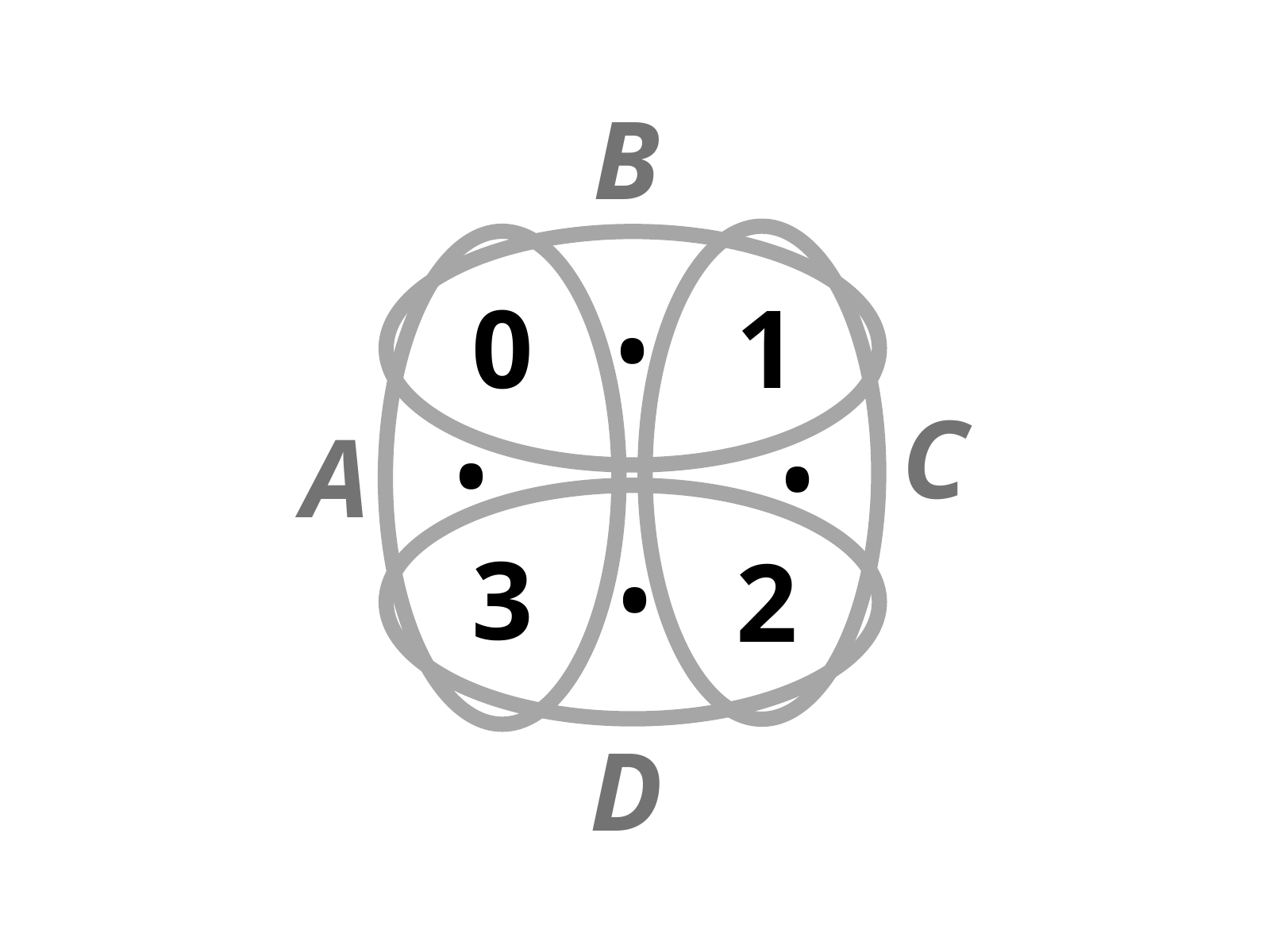}}
\caption{Example soberifications (Remark~\ref{rmrk.soberified.examples})}
\label{fig.012-triangle-sober}
\end{figure}

We will need Notation~\ref{nttn.irreducible.closed.set} for Remark~\ref{rmrk.topology.separation.axioms}:
\begin{nttn}
\label{nttn.irreducible.closed.set}
Call a closed set \deffont[irreducible closed set]{irreducible} when it cannot be written as the union of two proper closed subsets.
\end{nttn}

\begin{rmrk}
\label{rmrk.topology.separation.axioms}
Topology has a wealth of separation actions.
Three of them are: $T_0$ (distinct points have distinct neighbourhood (semi)filters); $T_1$ (distinct points have distinct open neighbourhoods); and $T_2$, also known as the Hausdorff condition (distinct points have disjoint open neighbourhoods) --- see Remark~\ref{rmrk.T0-T2} for formal statements.
In the case of topologies, the following is known about sobriety:
\begin{enumerate*}
\item
Every finite $T_0$ (and thus every finite $T_1$) topological space is sober. 
\item
Every $T_2$/Hausdorff space (including infinite ones) is sober~\cite[page~475, Theorem~3]{maclane:sheglf}. 
\item
A topological space is sober if and only if every nonempty irreducible closed set is the closure of a unique point~\cite[page~475]{maclane:sheglf}. 
\end{enumerate*}
The situation for semitopologies is different, as we explore in the rest of this Subsection.
\end{rmrk}

\begin{lemm}
\label{lemm.T0.not.sober}
\leavevmode
\begin{enumerate*}
\item
It is not necessarily the case that a finite $T_0$ semitopology (or even a finite $T_1$ semitopology) is sober (Definition~\ref{defn.sober.semitopology}).
\item
It is not necessarily the case that if every nonempty irreducible closed set is the closure of a unique point, then a semitopology is sober.
\end{enumerate*}
These non-implications hold even if the semitopology is regular (so $p\in\community(p)\in\topens$ for every $p$; see Definition~\ref{defn.tn}(\ref{item.regular.point})).
\end{lemm}
\begin{proof}
We provide a semitopology that is a counterexample for both parts.

Consider the left-hand semitopology illustrated in Figure~\ref{fig.012-triangle}, so that:\footnote{Recall that we used this example, for different purposes, in Lemma~\ref{lemm.cc.subspaces} and Figure~\ref{fig.not-strong-topen}.}
\begin{itemize*}
\item
$\ns P=\{0,1,2\}$, and 
\item
$\opens=\{\varnothing,\{0,1\},\{1,2\},\{0,2\},\{0,1,2\}\}$.
\end{itemize*}
We note that:
\begin{itemize*}
\item
$(\ns P,\opens)$ is $T_0$ and $T_1$.
\item
$(\ns P,\opens)$ is regular because all points are intertwined, so that $\community(p)=\ns P$ for every $p\in\ns P$.
\item
The nonempty irreducible closed sets are $\{0\}$ (which is the complement of $\{1,2\}$), $\{1\}$, and $\{2\}$.
Since these are singleton sets, they are certainly the closures of unique points.
\end{itemize*}
So $(\ns P,\opens)$ is $T_0$, regular, and irreducible closed sets are the closures of unique points.

We take as our semifilter $P=\opens\setminus\{\varnothing\}$.
The reader can check that $P$ is completely prime (Definition~\ref{defn.point}(\ref{item.completely.prime})), nonempty, up-closed, and compatible ($P$ is also the greatest semifilter); but, $P$ is not the neighbourhood semifilter of $0$, $1$, or $2$ in $\ns P$.
Thus, $(\ns P,\opens)$ is not sober. 
\end{proof}

\begin{rmrk}
The counterexample used in Lemma~\ref{lemm.T0.not.sober} generalises, as follows: the reader can check that the \emph{all-but-one} semitopology from Example~\ref{xmpl.semitopologies}(\ref{item.counterexample.X-x}) on three or more points (so open sets are generated by $\ns P\setminus\{p\}$ for every $p\in\ns P$) has similar behaviour.
\end{rmrk}

In topology, every Hausdorff space is sober.
In semitopologies, this implication does not hold, and in a rather strong sense:
\begin{lemm}
\label{lemm.hausdorff.not.sober}
\leavevmode
\begin{enumerate*}
\item\label{item.hausdorff.not.sober}
It is not necessarily the case that if a semitopology is Hausdorff, then it is sober. 
\item
Every quasiregular Hausdorff semitopology is (discrete and therefore) sober.
\end{enumerate*}
\end{lemm}
\begin{proof}
We consider each part in turn:
\begin{enumerate}
\item
It suffices to give a counterexample. 
Consider the right-hand semitopology illustrated in Figure~\ref{fig.012-triangle} (which we also used, for different purposes, in Figure~\ref{fig.square.diagram}), so that:
\begin{itemize*}
\item
$\ns P=\{0,1,2,3\}$, and 
\item
$\opens$ is generated by $X=\{\{3,0\},\{0,1\},\{1,2\},\{2,3\}\}$.
\end{itemize*}
This is Hausdorff, but it is not sober: the reader can check that the up-closure $\{3,0\}^\leq\subseteq\opens$ 
is nonempty, up-closed, compatible, and completely prime, but it is not the neighbourhood filter of any $p\in\ns P$.
\item
From Lemmas~\ref{lemm.quasiregular.hausdorff} (quasiregular Hausdorff is discrete) and~\ref{lemm.discrete.sober} (discrete is sober).
\qedhere\end{enumerate}
\end{proof}

\begin{rmrk}
\label{rmrk.sober.not.hausdorff}
A bit more discussion of Lemma~\ref{lemm.hausdorff.not.sober}.
\begin{enumerate}
\item
The space used in the counterexample for part~\ref{item.hausdorff.not.sober} is Hausdorff, $T_1$, and unconflicted (Definition~\ref{defn.conflicted}(\ref{item.unconflicted})).
It is not quasiregular (Definition~\ref{defn.tn}(\ref{item.quasiregular.point})) because the community of every point is empty; see Proposition~\ref{prop.unconflicted.irregular}.
\item
The implication holds if we add quasiregularity as a condition: every quasiregular Hausdorff space is sober.
But, this holds for very bad reasons, because by Lemma~\ref{lemm.quasiregular.hausdorff} every quasiregular Hausdorff space is discrete.
\item
Thus, the non-implication discussed in Lemma~\ref{lemm.hausdorff.not.sober} is informative and tells us something interesting about semitopological sobriety.
Semitopological sobriety is not just a weak form of topological sobriety, and it has its own distinct personality.
\end{enumerate}
\end{rmrk}

\begin{rmrk}
\label{rmrk.soberified.examples}
We can inject the examples illustrated in Figure~\ref{fig.012-triangle} (used in Lemmas~\ref{lemm.T0.not.sober} and~\ref{lemm.hausdorff.not.sober}) into \emph{soberified} versions of the spaces that are sober and have an isomorphic lattice of open sets, as illustrated in Figure~\ref{fig.012-triangle-sober}:
\begin{enumerate*}
\item
The left-hand semitopology has abstract points (completely prime semifilters; see Definition~\ref{defn.point}(\ref{item.completely.prime})) generated as the $\subseteq$-up-closures of the following sets: $\{A\}$, $\{B\}$, $\{C\}$, $\{A,B\}$, $\{B,C\}$, $\{C,A\}$, and $\{A,B,C\}$.
Of these, $\{A,B\}^\subseteq=\nbhd(0)$, $\{B,C\}^\subseteq=\nbhd(1)$, and $\{C,A\}^\subseteq=\nbhd(2)$.
The other completely prime semifilters are not generated as the neighbourhood semifilters of any point in the original space, so we add points as illustrated using $\bullet$ in 
the left-hand diagram in Figure~\ref{fig.012-triangle-sober}.
This semitopology is sober, and has the same semiframe of open sets.
\item
For the right-hand example, we again add a $\bullet$ point for every abstract point in the original space that is not already the neighbourhood semifilter of a point in the original space.
These abstract points are generated as the $\subseteq$-up-closures of $\{A\}$, $\{B\}$, $\{C\}$, and $\{D\}$.

There is no need to add a $\bullet$ for the abstract point generated as the $\subseteq$-up-closure of $\{A,B\}$, because $\{A,B\}^\subseteq=\nbhd(0)$.
Similarly $\{B,C\}^\subseteq=\nbhd(1)$, $\{C,D\}^\subseteq=\nbhd(2)$, and $\{D,A\}^\subseteq=\nbhd(3)$.
Note that $\{A,B,C\}$ does not generate an abstract point because it is not compatible: $A\notbetween C$.
Similarly for $\{B,C,D\}$, $\{C,D,A\}$, $\{D,A,C\}$, and $\{A,B,C,D\}$.
\end{enumerate*}
These soberified spaces are instances of a general construction described in Theorem~\ref{thrm.nbhd.morphism}.
And, continuing the observation made in Remark~\ref{rmrk.sober.not.hausdorff}, note that neither of these spaces, with their extra points, are Hausdorff.
\end{rmrk}

\jamiesection{Four categories \& functors between them}
\label{sect.duality}

\jamiesubsection{The categories $\tf{Sober}/\tf{SemiTop}$ of (sober) semitopologies}

\begin{defn}
\label{defn.morphism.semitopologies}
\leavevmode
\begin{enumerate*}
\item\label{item.morphism.st}
Suppose $(\ns P,\opens)$ and $(\ns P',\opens')$ are semitopologies and $f:\ns P\to\ns P'$ is any function.
Then call $f$ a \deffont{morphism of semitopologies} when
$f$ is continuous, by which we mean (as standard) that 
$$
O'\in\opens'
\quad\text{implies}\quad
f^\mone(O')\in\opens .
$$ 
\item
Define $\tf{SemiTop}$ the \deffont[category of semitopologies $\tf{SemiTop}$]{category of semitopologies} such that: 
\begin{itemize*}
\item
objects are semitopologies, and 
\item
arrows are morphisms of semitopologies (continuous maps on points).\footnote{A discussion of possible alternatives, for future work, is in Remark~\ref{rmrk.more.conditions}.  See also Remarks~\ref{rmrk.no.meet} and~\ref{rmrk.other.properties}.}
\end{itemize*}
\item\label{item.cat.sober}
Write $\tf{Sober}$ for the \deffont[category of sober semitopologies $\tf{Sober}$]{category of sober semitopologies} and continuous functions between them.
By construction, $\tf{Sober}$ is the full subcategory of $\tf{SemiTop}$, on its sober semitopologies. 
\end{enumerate*}
\end{defn}

\begin{rmrk}
\label{rmrk.reading.nbhd.morphism}
For convenience reading Theorem~\ref{thrm.nbhd.morphism} we recall some facts:
\begin{enumerate*}
\item
The \emph{semiframe} 
$$
\tf{Fr}(\ns P,\opens)=(\opens,\subseteq,\between)
$$ 
from Definition~\ref{defn.semi.to.dg} has elements open sets $O\in\opens$, preordered by subset inclusion and with a compatibility relation given by sets intersection.

It is spatial, by Proposition~\ref{prop.Gr.P.spatial}.
\item
An abstract point $P$ in $\tf{Points}(\tf{Fr}(\ns P,\opens))$ is a completely prime nonempty up-closed compatible subset of $\opens$.
\item\label{item.describe.StFr}
$\tf{St}\,\tf{Fr}(\ns P,\opens)$ is by Definition~\ref{defn.st.g} a semitopology such that:
\begin{enumerate*}
\item
Its set of points is $\tf{Points}(\tf{Fr}(\ns P,\opens))$; the set of abstract points in $\tf{Fr}(\ns P,\opens)=(\opens,\subseteq,\between)$, the semilattice of open sets of $(\ns P,\opens)$, and 
\item
Its open sets are given by the $\f{Op}(O)$, for $O\in\opens$. 
\end{enumerate*}
It is sober, by Proposition~\ref{prop.spatial.gives.sober}. 
\end{enumerate*}
\end{rmrk}

Continuing Remark~\ref{rmrk.reading.nbhd.morphism}, a notation will be useful:
\begin{nttn}
\label{nttn.soberify}
Suppose $(\ns P,\opens)$ is a semitopology.
Then define
$$
\tf{Soberify}(\ns P,\opens) = \tf{St}\,\tf{Fr}(\ns P,\opens).
$$
We may use $\tf{Soberify}(\ns P,\opens)$ and $\tf{St}\,\tf{Fr}(\ns P,\opens)$ interchangeably, depending on whether we want to emphasise ``this is a sober semitopology obtained from $(\ns P,\opens)$'' or ``this is $\tf{St}$ acting on $\tf{Fr}(\ns P,\opens)=(\opens,\subseteq,\between)$''.
\end{nttn}

\begin{thrm}
\label{thrm.nbhd.morphism}
Suppose $(\ns P,\opens)$ is a semitopology.
Then 
\begin{enumerate*}
\item
$\nbhd:\ns P\to \tf{Points}(\tf{Fr}(\ns P,\opens))$ is a morphism of semitopologies from 
$(\ns P,\opens)$ to $\tf{St}\,\tf{Fr}(\ns P,\opens)=\tf{Soberify}(\ns P,\opens)$
\item
taking $(\ns P,\opens)$ to a sober semitopology $\tf{Soberify}(\ns P,\opens)$, such that
\item\label{item.nbhd.morphism.is.iso}
$\nbhd^\mone$ induces a bijection on open sets by mapping $\f{Op}(O)$ to $O$, and furthermore this is an isomorphism of the semiframes of open sets, in the sense of Definition~\ref{defn.semiframe.iso}.
\end{enumerate*}
\end{thrm}
\begin{proof}
We consider each part in turn:
\begin{enumerate}
\item
Following Definition~\ref{defn.morphism.semitopologies} 
we must show that $\nbhd$ is continuous (inverse images of open sets are open) from $(\ns P,\opens)$ to $\tf{Soberify}(\ns P,\opens)$.
So following Definition~\ref{defn.st.g}(\ref{item.st.op}), consider $\f{Op}(O)\in\opens(\tf{Soberify}(\ns P,\opens))$.
By Proposition~\ref{prop.nbhd.iff}(\ref{item.nbhd.mone}) 
$$
\nbhd^\mone(\f{Op}(O))=O\in\opens.
$$
Continuity follows.
\item
$\tf{Soberify}(\ns P,\opens)$ is sober by Proposition~\ref{prop.spatial.gives.sober}.
\item
This is Proposition~\ref{prop.nbhd.mone.bijects}.
\qedhere\end{enumerate}
\end{proof}

\begin{rmrk}
\label{rmrk.nbhd.summary}
We can summarise Theorem~\ref{thrm.nbhd.morphism} as follows: 
\begin{enumerate*}
\item
By construction, the kernel of the $\nbhd$ function (the relation determined by which points it maps to equal elements) is topological indistinguishability $\topind$. 
\item
We can think of $\tf{St}\,\tf{Fr}(\ns P,\opens)$ as being obtained from $(\ns P,\opens)$ by 
\begin{enumerate*}
\item
quotienting topologically equivalent points to obtain a $T_0$ space, and then
\item
adding extra points to make it sober.
\end{enumerate*}
See also the discussion in Remark~\ref{rmrk.enough.points} about what it means to have `enough' points.
\item
This is done without affecting the semiframe of open sets (up to isomorphism), with the semiframe bijection given by $\nbhd^\mone$.
\end{enumerate*}
In this sense, we can view $\tf{St}\,\tf{Fr}(\ns P,\opens)$ as a \deffont[soberification $\tf{St}\,\tf{Fr}(\ns P,\opens)$]{soberification}\index{$\tf{St}\,\tf{Fr}(\ns P,\opens)$ (soberification)} of $(\ns P,\opens)$.
\end{rmrk}

\jamiesubsection{The categories $\tf{Spatial}/\tf{SemiFrame}$ of (spatial) semiframes}

\begin{defn}
\label{defn.category.of.spatial.graphs}
\leavevmode
\begin{enumerate*}
\item\label{item.category.spatial.morphism}
Suppose $(\ns X,\cti,\ast)$ and $(\ns X',\cti',\ast')$ are semiframes (Definition~\ref{defn.semiframe}) and $g:\ns X\to\ns X'$ is any function. 
Then call $g$ a \deffont{morphism of semiframes} when:
\begin{enumerate*}
\item\label{item.g.semilattice.morphism}
$g$ is a morphism of complete semilattices (Definition~\ref{defn.complete.semilattice.morphism}).
\item\label{item.g.compatible}
$g$ is \deffont[compatibility (of morphism of semiframes)]{compatible}, 
by which we mean that $g(x')\ast g(x'')$ implies $x'\ast x''$ for every $x',x''\in\ns X'$. 
\end{enumerate*}
\item\label{item.semilat}
We define $\tf{SemiFrame}$ the \deffont[category of semiframes $\tf{SemiFrame}$]{category of semiframes} such that:
\begin{itemize*}
\item
objects are semiframes, and 
\item
arrows are morphisms of semiframes.
\end{itemize*}
\item\label{item.spatial}
Write $\tf{Spatial}$ for the \deffont[category of spatial semiframes $\tf{Spatial}$]{category of spatial semiframes} and semiframe morphisms between them.
By construction, $\tf{Spatial}$ is the full subcategory of $\tf{SemiFrame}$, on its spatial semiframes (Definition~\ref{defn.spatial.graph}). 
\end{enumerate*}
\end{defn}

\begin{lemm}
\label{lemm.Op.morphism}
Suppose $(\ns X,\cti,\ast)$ is a semiframe. 
Then $\f{Op}:(\ns X,\cti,\ast)\to\tf{Fr}\,\tf{St}(\ns X,\cti,\ast)$ is a morphism of semiframes and is surjective on underlying sets.
\end{lemm}
\begin{proof}
Following Definition~\ref{defn.category.of.spatial.graphs}(\ref{item.category.spatial.morphism})
we must show that 
\begin{itemize*}
\item
$\f{Op}$ is a semilattice morphism (Definition~\ref{defn.complete.semilattice.morphism}) --- commutes with joins and maps $\ttop_{\ns X}$ to $\tf{Points}(\ns X,\cti,\ast)$) --- and 
\item
is compatible with the compatibility relation $\ast$, and
\item
we must show that $\f{Op}$ is surjective.
\end{itemize*}
We consider each property in turn:
\begin{itemize*}
\item
\emph{$\f{Op}$ is a semilattice morphism.}

$\f{Op}(\bigvee X)=\bigvee_{x\in X} \f{Op}(x)$ by Lemma~\ref{lemm.op.commutes.with.joins}, and $\f{Op}(\ttop_{\ns X})=\tf{Points}(\ns X,\cti,\ast)$ by Proposition~\ref{prop.semiframe.to.Op}(\ref{item.semiframe.to.Op.top}).
\item
\emph{$\f{Op}$ is compatible with $\ast$.}

Unpacking Definition~\ref{defn.category.of.spatial.graphs}(\ref{item.g.compatible}), we must show that $\f{Op}(x)\between \f{Op}(x')$ implies $x\ast x'$.
We use Proposition~\ref{prop.semiframe.to.Op}(\ref{item.semiframe.to.Op.between}).
\item
\emph{$\f{Op}$ is surjective}
\ \dots \ by Lemma~\ref{lemm.st.opens.generator}.
\qedhere
\end{itemize*}
\end{proof}

\jamiesubsection{Functoriality of the maps}

\begin{defn}
\label{defn.g.circ}
Suppose $g:(\ns X',\cti',\ast')\to(\ns X,\cti,\ast)\oldin \tf{SemiFrame}$.
Define a mapping $g^\circ:\tf{St}(\ns X,\cti,\ast)\to\tf{St}(\ns X',\cti',\ast')$ by 
$$
\begin{array}{r@{\ }l@{\quad}c@{\quad}l}
g^\circ :&\tf{Points}(\ns X,\cti,\ast) &\longrightarrow& \tf{Points}(\ns X',\cti',\ast')
\\
& P &\longmapsto& P'=\{x'\in\ns X' \mid g(x')\in P\}.
\end{array}
$$
\end{defn}

\begin{rmrk}
We will show that $g^\circ$ from Definition~\ref{defn.g.circ} is an arrow in $\tf{SemiTop}$.
We will need to prove the following:
\begin{itemize*}
\item
If $P\in\tf{Points}(\ns X,\cti,\ast)$ then $g^\circ(P)\in\tf{Points}(\ns X',\cti',\ast')$.
\item
$g^\circ$ is a morphism of semitopologies.
\end{itemize*}
We do this in Lemmas~\ref{lemm.gcirc.well-defined} and~\ref{lemm.g.circ.continuous} respectively.
\end{rmrk}

\begin{lemm}[$g^\circ$ well-defined]
\label{lemm.gcirc.well-defined}
Suppose $g:(\ns X',\cti',\ast')\to(\ns X,\cti,\ast) \oldin \tf{SemiFrame}$ and suppose $P\in\tf{Points}(\ns X,\cti,\ast)$.
Then $g^\circ(P)$ from Definition~\ref{defn.g.circ} is indeed in $\tf{Points}(\ns X',\cti',\ast')$ --- and thus $g^\circ$ is well-defined function from $\tf{Points}(\ns X,\cti,\ast)$ to $\tf{Points}(\ns X',\cti',\ast')$.
\end{lemm}
\begin{proof}
For brevity write
$$
P'=\{x'\in\ns X'\mid g(x')\in P\} .
$$ 
We must check that $P'$ is a completely prime nonempty up-closed compatible subset of $\ns X'$.
We consider each property in turn:
\begin{enumerate}
\item
\emph{$P'$ is completely prime.}\quad

Consider some $X'\subseteq P'$ and suppose $g(\bigvee X')\in P$.
By Definition~\ref{defn.category.of.spatial.graphs}(\ref{item.g.semilattice.morphism}) $g$ is a semilattice homomorphism, so by Definition~\ref{defn.complete.semilattice}(\ref{item.semilattice.morphism.top}) $g(\bigvee X')=\bigvee_{x'\in X'}g(x')$.
Thus $\bigvee_{x'\in X'} g(x')\in P$.
By assumption $P$ is completely prime, so $g(x')\in P$ for some $x'\in X'$.
Thus $x'\in P'$ for that $x'$.
Since $X'$ was arbitrary, it follows that $P'$ is completely prime.
\item
\emph{$P'$ is nonempty.}\quad

By assumption $g$ is an arrow in $\tf{SemiFrame}$ (i.e. a semiframe morphism) and unpacking Definition~\ref{defn.category.of.spatial.graphs}(\ref{item.g.semilattice.morphism}) it follows that it is a semilattice homomorphism.
In particular by Definition~\ref{defn.complete.semilattice}(\ref{item.semilattice.morphism.top}) $g(\ttop_{\ns X'})=\ttop_{\ns X}$, and 
by Lemma~\ref{lemm.P.top}(\ref{item.P.yes.top}) $\ttop_{\ns X}\in P$.
Thus $\ttop_{\ns X'}\in P'$, so $P'$ is nonempty.
\item
\emph{$P'$ is up-closed.}\quad

Suppose $x'\in P'$ and $x'\leq x''$.
By construction $g(x')\in P$.
By Lemma~\ref{lemm.semi.hom.mon} (because $g$ is a semilattice morphism by Definition~\ref{defn.category.of.spatial.graphs}(\ref{item.g.semilattice.morphism})) $g$ is monotone, so $g(x')\leq g(x'')$.
By assumption in Definition~\ref{defn.point}(\ref{item.up-closed}) $P$ is up-closed, so that $g(x'')\in P$ and thus $x''\in P'$ as required.
\item
\emph{$P'$ is compatible.}\quad

Suppose $x',x''\in P'$.
Thus $g(x'),g(x'')\in P$.
By assumption in Definition~\ref{defn.point}(\ref{item.weak.clique}) $P$ is compatible, so $g(x')\ast g(x'')$.
By compatibility of $g$ (Definition~\ref{defn.category.of.spatial.graphs}(\ref{item.g.compatible})) it follows that $x'\ast x''$.
Thus $P'$ is compatible.
\qedhere\end{enumerate}
\end{proof}

\begin{rmrk}
\label{rmrk.further.restrictions.on.points}
\emph{Note on design:}
If we want to impose further conditions on being an abstract point (such as those discussed in Remark~\ref{rmrk.other.properties}) then 
Lemma~\ref{lemm.gcirc.well-defined} would need to be extended to show that these further conditions are preserved by the $g^\circ$ operation, so that for $P\in\tf{Points}(\ns X,\cti,\ast)$ an abstract point in $(\ns X,\cti,\ast)$, $g^\circ(P)=\{x'\in\ns X'\mid g(x')\in P\}$ is an abstract point in $(\ns X',\cti',\ast')$. 

For example: consider what would happen if we add the extra condition on semifilters from Remark~\ref{rmrk.other.properties}.
Then the $P'$ defined in the proof of Lemma~\ref{lemm.gcirc.well-defined} above might not be closed under this additional condition (it will be if $g$ is surjective).
This could be mended by closing $P'$ under greatest lower bounds that are not $\tbot$, but that in turn might compromise the property of being completely prime.
These comments are not a proof that the problems would be insuperable; but they suggest that complexity would be added.
For now, we prefer to keep things simple!
\end{rmrk}

\begin{lemm}
\label{lemm.g.circ.inv}
Suppose $g:(\ns X',\cti',\ast')\to(\ns X,\cti,\ast) \oldin \tf{SemiFrame}$, and suppose $x'\in\ns X'$.
Then 
$$
(g^\circ)^\mone(\f{Op}(x'))=\f{Op}(g(x')).
$$
\end{lemm}
\begin{proof}
Consider an abstract point $P\in\tf{Point}(\tf{Gr}(\ns X',\cti',\ast'))$.
We just chase definitions:
$$
\begin{array}{r@{\ }l@{\quad}l}
P\in (g^\circ)^\mone(\f{Op}(x'))
\liff& 
g^\circ(P)\in \f{Op}(x')
&\text{Fact of inverse image}
\\
\liff&
x'\in g^\circ(P)
&\text{Definition~\ref{defn.Op}}
\\
\liff&
g(x')\in P 
&\text{Definition~\ref{defn.g.circ}}
\\
\liff&
P\in \f{Op}(g(x')).
&\text{Definition~\ref{defn.Op}}
\end{array}
$$
The choice of $P$ was arbitrary, so $(g^\circ)^\mone(\f{Op}(x'))=\f{Op}(g(x'))$ as required.
\end{proof}

\begin{lemm}[$g^\circ$ continuous]
\label{lemm.g.circ.continuous}
Suppose $g:(\ns X',\cti',\ast')\to(\ns X,\cti,\ast) \oldin \tf{SemiFrame}$. 
Then $g^\circ:\tf{St}(\ns X,\cti,\ast)\to\tf{St}(\ns X',\cti',\ast')$ is continuous:
$$
(g^\circ)^\mone(\mathcal O')\in\opens(\tf{St}(\ns X,\cti,\ast))
$$
for every $\mathcal O'\in\opens(\tf{St}(\ns X',\cti',\ast'))$. 
\end{lemm}
\begin{proof}
By Definition~\ref{defn.st.g}, $\mathcal O'=\f{Op}(x')$ for some $x'\in\ns X'$.
By Lemma~\ref{lemm.g.circ.inv} $(g^\circ)^\mone(\f{Op}(x'))=\f{Op}(g(x'))$.
By Definition~\ref{defn.st.g}(\ref{item.st.op}) $\f{Op}(g(x'))\in\opens(\tf{St}(\ns X,\cti,\ast))$.
\end{proof}

\begin{prop}[Functoriality]
\label{prop.semitop.adjunction}
\leavevmode
\begin{enumerate*}
\item
Suppose $f:(\ns P,\opens)\to(\ns P',\opens') \oldin \tf{SemiTop}$ (so $f$ is a continuous map on underlying points).

Then $f^\mone$ is an arrow $\tf{Fr}(\ns P',\opens')\to\tf{Fr}(\ns P,\opens)$ in $\tf{SemiFrame}$.
\item
Suppose $g:(\ns X',\cti',\ast')\to(\ns X,\cti,\ast) \oldin \tf{SemiFrame}$.

Then $g^\circ$ from Definition~\ref{defn.g.circ} is an arrow $\tf{St}(\ns X,\cti,\ast)\to\tf{St}(\ns X',\cti',\ast')$ in $\tf{SemiTop}$. 
\item
The assignments $f\mapsto f^\mone$ and $g\mapsto g^\circ$ are \deffont[functorial map]{functorial} --- they map identity maps to identity maps, and commute with function composition.
\end{enumerate*}
\end{prop}
\begin{proof}
We consider each part in turn:
\begin{enumerate}
\item
Following Definition~\ref{defn.category.of.spatial.graphs}, we must check that $f^\mone$ is a morphism of semiframes.
We just unpack what this means and see that the required properties are just facts of taking inverse images:
\begin{itemize*}
\item
\emph{$f^\mone$ commutes with joins, i.e. with $\bigcup$.}\quad

This is a fact of inverse images.
\item
\emph{$f^\mone$ maps $\ttop_{\tf{Fr}(\ns P',\opens')}=\ns P'$ to $\ttop_{\tf{Fr}(\ns P,\opens)}=\ns P$.}\quad

This is a fact of inverse images.
\item
\emph{$f^\mone$ is compatible, meaning that $f^\mone(O')\between f^\mone(O'')$ implies $O'\between O''$.}\quad 

This is a fact of inverse images.
\end{itemize*}
\item
We must check that $g^\circ$ is continuous.
This is Lemma~\ref{lemm.g.circ.continuous}.
\item
Checking functoriality is routine, but we sketch the reasoning anyway:
\begin{itemize*}
\item
Consider the identity function $\id$ on some semitopology $(\ns P,\opens)$.
Then $\id^\mone$ should be the identity function on $(\opens,\subseteq,\between)$. 
It is.
\item
Consider $f:(\ns P,\opens)\to(\ns P',\opens')$ and $f':(\ns P',\opens')\to(\ns P'',\opens'')$.
Then $(f'\circ f)^\mone$ should be equal to $f^\mone\circ (f')^\mone$.
It is.
\item
Consider the identity function $\id$ on $(\ns X,\cti,\ast)$.
Then $\id^\circ$ should be the identity function on $\tf{Points}(\ns X,\cti,\ast)$.
We look at Definition~\ref{defn.g.circ} and see that this amounts to checking that $P=\{x\in\ns X \mid \id(x)\in P\}$.
It is.
\item
Consider $g:(\ns X,\cti,\ast)\to(\ns X',\cti',\ast')$ and $g':(\ns X',\cti',\ast')\to(\ns X'',\cti'',\ast'')$ and consider some $P''\in\tf{Points}(\ns X'',\cti'',\ast'')$.
Then $(g'\circ g)^\circ(P'')$ should be equal to $(g^\circ\circ (g')^\circ)(P'')$.
We look at Definition~\ref{defn.g.circ} and see that this amounts to checking that 
$\{x\in\ns X \mid g'(g(x))\in P''\}=\{x\in\ns X \mid g(x)\in P'\}$ 
where
$P'=\{x'\in\ns X' \mid g'(x')\in P''\}$.
Unpacking these definitions, we see that the equality does indeed hold.
\qedhere\end{itemize*}
\end{enumerate}
\end{proof}

\jamiesubsection{Sober semitopologies are dual to spatial semiframes}

We can now state the duality result between $\tf{Sober}$ and $\tf{Spatial}$: 
\begin{thrm}
\label{thrm.categorical.duality.semiframes}
The maps $\tf{St}$ (Definition~\ref{defn.st.g}) and $\tf{Fr}$ (Definition~\ref{defn.semi.to.dg}), with actions on arrows as described in Proposition~\ref{prop.semitop.adjunction}, form a categorical duality between the categories
\begin{itemize*}
\item
$\tf{Sober}$ of sober semitopologies (Definition~\ref{defn.sober.semitopology}) and continuous compatible morphisms between them; and 
\item
$\tf{Spatial}$ of spatial semiframes and morphisms between them (Definition~\ref{defn.category.of.spatial.graphs}(\ref{item.spatial})).
\end{itemize*}
\end{thrm}
\begin{proof}
There are various things to check:
\begin{itemize}
\item
Proposition~\ref{prop.spatial.gives.sober} shows that $\tf{St}$ maps spatial semiframes to sober semitopologies.
\item
Proposition~\ref{prop.Gr.P.spatial} shows that $\tf{Fr}$ maps sober semitopologies to spatial semiframes.
\item
By Proposition~\ref{prop.semitop.adjunction} the maps $f\mapsto f^\mone$ (inverse image) and $g\mapsto g^\circ$ (Definition~\ref{defn.g.circ}) are functorial.
\item
The equivalence morphisms are given by the bijections $\f{Op}$ and $\nbhd$:
\begin{itemize*}
\item
$\f{Op}$ is from Definition~\ref{defn.Op}.
By Lemma~\ref{lemm.Op.morphism} $\f{Op}$ is a morphism $(\ns X,\cti,\ast)\to\tf{Fr}\,\tf{St}(\ns X,\cti,\ast)$ in $\tf{Spatial}$ that is surjective on underlying sets.
Injectivity is from Proposition~\ref{prop.Op.subseteq}(\ref{item.Op.spatial.inj}).
\item
$\nbhd$ is from Definition~\ref{defn.nbhd}.
By Theorem~\ref{thrm.nbhd.morphism} $\nbhd$ is a morphism $(\ns P,\opens)\to \tf{St}\,\tf{Fr}(\ns P,\opens)$ in $\tf{Sober}$.
It is a bijection on underlying sets by the sobriety condition in Definition~\ref{defn.sober.semitopology}.
\end{itemize*}
\end{itemize}
Finally, we must check naturality of $\f{Op}$ and $\nbhd$, which means (as standard) checking commutativity of the following diagrams:
$$
\begin{tikzcd}
(\ns P,\opens) \arrow{r}{\nbhd} \arrow{d}{f} 
& 
\tf{St}\,\tf{Fr}(\ns P,\opens) \arrow{d}{(f^\mone)^\circ}
\\
(\ns P',\opens') \arrow{r}{\nbhd} 
&
\tf{St}\,\tf{Fr}(\ns P',\opens') 
\end{tikzcd}
\qquad
\begin{tikzcd}
(\ns X,\cti,\ast) \arrow{r}{\f{Op}} \arrow{d}{g} 
& 
\tf{Fr}\,\tf{St}(\ns X,\cti,\ast) \arrow{d}{(g^\circ)^\mone}
\\
(\ns X',\cti',\ast') \arrow{r}{\f{Op}} 
&
\tf{Fr}\,\tf{St}(\ns X',\cti',\ast') 
\end{tikzcd}
$$
We proceed as follows:
\begin{itemize}
\item
Suppose $g:(\ns X',\cti',\ast')\to(\ns X,\cti,\ast)$ in $\tf{Spatial}$, so that 
$g^\circ:\tf{St}(\ns X,\cti,\ast)\to\tf{St}(\ns X',\cti',\ast')$ in $\tf{Sober}$ and
$(g^\circ)^\mone:\tf{Fr}\,\tf{St}(\ns X',\cti',\ast')\to \tf{Fr}\,\tf{St}(\ns X,\cti,\ast)$ in $\tf{Spatial}$.
To prove naturality we must check that
$$
(g^\circ)^\mone(\f{Op}(x)) = \f{Op}(g(x)) 
$$
for every $x\in\ns X$.
This is just Lemma~\ref{lemm.g.circ.inv}.
\item
Suppose $f:(\ns P,\opens)\to(\ns P',\opens')$ in $\tf{SemiTop}$, so that 
$f^\mone:\tf{Fr}(\ns P',\opens')\to\tf{Fr}(\ns P,\opens)$ in $\tf{Spatial}$ and
$(f^\mone)^\circ:\tf{St}\,\tf{Fr}(\ns P,\opens)\to \tf{St}\,\tf{Fr}(\ns P',\opens')$ in $\tf{SemiTop}$.
To prove naturality we must check that
$$
(f^\mone)^\circ(\nbhd(p)) = \nbhd(f(p)) .
$$
We just chase definitions, for an open set $O'\in\opens'$:
$$
\begin{array}{r@{\ }l@{\qquad}l}
O'\in (f^\mone)^\circ(\nbhd(p))
\liff&
f^\mone(O') \in \nbhd(p)
&\text{Definition~\ref{defn.g.circ}}
\\
\liff&
p\in f^\mone(O')
&\text{Definition~\ref{defn.nbhd}}
\\
\liff&
f(p)\in O'
&\text{Inverse image}
\\
\liff&
O'\in\nbhd(f(p)) 
&\text{Definition~\ref{defn.nbhd}}
.
\end{array}
$$
\qedhere\end{itemize}
\end{proof}

\begin{rmrk}
We review the background to Theorem~\ref{thrm.categorical.duality.semiframes}:
\begin{enumerate*}
\item
A semitopology $(\ns P,\opens)$ is a set of points $\ns P$ and a set of open sets $\opens\subseteq\powerset(\ns P)$ that contains $\ns P$ and is closed under arbitrary (possibly empty) unions (Definition~\ref{defn.semitopology}).
\item
A morphism between semitopologies is a continuous function, just as for topologies (Definition~\ref{defn.morphism.semitopologies}(\ref{item.morphism.st})).
\item
A semiframe $(\ns X,\cti,\ast)$ is a complete join-semilattice $(\ns X,\cti)$ with a properly reflexive distributive \emph{compatibility relation} $\ast$ (Definition~\ref{defn.semiframe}).
\item
A morphism between semiframes is a morphism of complete join-semilattices with $\ttop$ that is compatible with the compatibility relation (Definition~\ref{defn.category.of.spatial.graphs}(\ref{item.category.spatial.morphism})).
\item
An \emph{abstract point} of a semitopology $(\ns P,\opens)$ is a completely prime nonempty up-closed compatible subset $P\subseteq\opens$ (Definition~\ref{defn.point}(\ref{item.abstract.point})).
\item
A semitopology is \emph{sober} when the neighbourhood semifilter map $p\in\ns P\mapsto \nbhd(p)=\{O\in\opens \mid p\in O\}$ is injective and surjective between the points of $\ns P$ and the abstract points of $\ns P$ (Definition~\ref{defn.sober.semitopology}).
\item
By Theorem~\ref{thrm.nbhd.morphism}, and as discussed in Remark~\ref{rmrk.nbhd.summary}, every (possibly non-sober) semitopology $(\ns P,\opens)$ maps into its \emph{soberification} $\tf{St}\,\tf{Fr}(\ns P,\opens)$, which has an isomorphic semiframe of open sets.
So even if our semitopology $(\ns P,\opens)$ is not sober, there is a standard recipe to make it so. 
\item
A semiframe is \emph{spatial} when $x\in\ns X \mapsto \f{Op}(x)=\{P\in\tf{Point} \mid x\in P\}$ respects $\cti$ and $\ast$ in senses make formal in
Definition~\ref{defn.spatial.graph} and Proposition~\ref{prop.Op.subseteq}.
\item 
Sober semitopologies and continuous functions between them, and spatial semiframes and semiframe morphisms between them, are categorically dual (Theorem~\ref{thrm.categorical.duality.semiframes}).
\end{enumerate*}
\end{rmrk}

\begin{rmrk}
\label{rmrk.categorical.duality}
A \emph{categorical duality} between two categories $\mathbb C$ and $\mathbb D$ is an equivalence between $\mathbb C$ and $\mathbb D^{op}$; this is an adjoint pair of functors whose unit and counit are natural isomorphisms.
See~\cite[IV.4]{maclane:catwm}.\footnote{The Wikipedia page is also exceptionally clear~\cite{wiki:Equivalence_of_categories}.}

There are many duality results in the literature.
The duality between topologies and frames is described (for example) in \cite[page~479, Corollary~4]{maclane:sheglf}.
A duality between distributive lattices and coherent spaces is in \cite[page~66]{johnstone:stos}.
There is the classic duality by Stone between Boolean algebras and compact Hausdorff spaces with a basis of clopen sets~\cite{stone:therba,johnstone:stos}. 
An encyclopedic treatment is in \cite{caramello:toptas}, with a rather good overview in Example~2.9 on page~17.

Theorem~\ref{thrm.categorical.duality.semiframes} appends another item to this extensive canon.
It also constructively moves us forward in studying semitopologies, because it gives us an algebraic treatment of semitopologies, and a formal framework for studying morphisms between semitopologies.
For instance: taking morphisms to be continuous functions is sensible not just because this is also how things work for topologies, but also because this is what is categorically dual to the ${\cti}/{\ast}$-homomorphisms between semiframes (Definition~\ref{defn.category.of.spatial.graphs}). 
And of course, if we become interested in different notions of semitopology morphism (a flavour of these is in Remark~\ref{rmrk.more.conditions}) then the algebraic framework gives us a distinct mathematical light with which to inspect and evaluate them. 

Note what Theorem~\ref{thrm.categorical.duality.semiframes} does \emph{not} do: it does not give a duality between all semitopologies and all semiframes; it gives a duality between sober semitopologies and spatial semiframes.
This in itself is nothing new --- the topological duality is just the same ---
but what is interesting is that our motivation for studying semitopologies comes from practical network systems.
These tend to be (finite) non-sober semitopologies --- non-sober, because a guarantee of sobriety cannot be enforced, and anyway it is precisely the point of the exercise to achieve coordination, \emph{without} explicitly requiring every possible constellation of cooperating agents to be explicitly represented by a point.

It is true that by Theorem~\ref{thrm.nbhd.morphism} every non-sober $T_0$ semitopology can be embedded into a sober one without affecting the semiframe of open sets, but this makes the system to which it corresponds larger, by adding points.
Thus the duality in Theorem~\ref{thrm.categorical.duality.semiframes} is a mathematical statement, but not necessarily a practical one --- and this is as expected, because we knew that this is an abstract result.
$\nbhd$ maps a point to a set of (open) sets; and $\f{Op}$ maps an (open) set of points to a set of sets of (open) sets.
Of course these might not be computationally optimal.
\end{rmrk}

\begin{rmrk}
We have constructed an algebraic representation of semitopologies --- but this is not the last word on representing semitopologies.
Other methodologies are also illuminating, and because our motivation comes from computer systems that are networks of machines, we are particularly interested in representations based on ideas from graphs. 
We will investigate these in Section~\ref{sect.graphs}. 
\end{rmrk}

\jamiesection{Well-behavedness conditions, dually}
\label{sect.closer.look.at.semifilters}

We want to understand semifilters better, and in particular we want to understand how properties of semifilters and abstract points correspond to the well-behavedness properties which we found useful in studying semitopologies --- for example \emph{topens}, \emph{regularity}, and being \emph{unconflicted} (Definitions~\ref{defn.transitive},\ \ref{defn.tn} and~\ref{defn.conflicted}). 

\jamiesubsection{(Maximal) semifilters and transitive elements}

\begin{rmrk}[Semifilters are not filters]
We know that semifilters do not necessarily behave like filters.
For instance:
\begin{enumerate*}
\item
It is possible for a finite semifilter to have more than one minimal element, because the closure under binary meets condition of filters is replaced by a weaker compatibility condition (see also Remarks~\ref{rmrk.no.meet} and~\ref{rmrk.other.properties}).
\item
There are more semifilters than proper filters --- even if the underlying space is a topology.
For example, the discrete semitopology on $\{0,1,2\}$ (whose open sets are all subsets of the space) is a topology.
Every proper filter in this space is a semifilter, but it also has a semifilter $\bigl\{\{0,1\},\{1,2\},\{2,0\},\{0,1,2\}\bigr\}$ (see the top-left diagram in Figure~\ref{fig.012-triangle}) and this is not a filter.
\end{enumerate*}
More on this in Subsection~\ref{subsect.things.that.are.different}.

In summary: semifilters are different and we cannot necessarily take their behaviour for granted without checking it.
We now examine them in more detail.
\end{rmrk}

We start with some easy definitions and results:
\begin{nttn}
\label{nttn.X.ast.Y}
Suppose $(\ns X,\cti,\ast)$ is a semiframe and $X,Y\subseteq\ns X$ and $x\in\ns X$.
Then we generalise $x\ast y$ to $x\ast Y$, $X\ast y$, and $X\ast Y$ as follows:
\begin{enumerate*}
\item\label{item.x.ast.Y}
Write $x\ast Y$ for $\Forall{y{\in}Y}x\ast y$. 
\item\label{item.X.ast.y}
Write $X\ast y$ for $\Forall{x{\in}X}x\ast y$. 
\item\label{item.X.ast.Y}
Write $X\ast Y$ for $\Forall{x{\in}X}\Forall{y{\in}Y}x\ast y$. 
\end{enumerate*}
We read $x\ast Y$ as `$x$ is \deffont[compatibility (of $x$ with $Y$:\ $x\ast Y$)]{compatible} with $Y$', and similarly for $X\ast y$ and $X\ast Y$.
\end{nttn}

\begin{rmrk}
\label{rmrk.promise.ast.int.char}
We will see later on in Lemma~\ref{lemm.intertwined.sober} that
$X\ast X'$ generalises $p\intertwinedwith p'$, in the sense that if $X=\nbhd(p)$ and $X'=\nbhd(p')$, then $p\intertwinedwith p'$ if and only if $\nbhd(p)\ast\nbhd(p')$.
\end{rmrk}

\begin{lemm}[Characterisation of maximal semifilters]
\label{lemm.char.maxfilter}
Suppose $(\ns X,\cti,\ast)$ is a semiframe and $\afilter\subseteq\ns X$ is a semifilter.
Then the following conditions are equivalent:
\begin{enumerate*}
\item
$\afilter$ is maximal.
\item
For every $x\in\ns X$, $x\ast\afilter$ if and only if $x\in\afilter$. 
\end{enumerate*}
\end{lemm}
\begin{proof}
We prove two implications:
\begin{itemize}
\item
\emph{Suppose $\afilter$ is a maximal semifilter.}

Suppose $x\in\afilter$.
Then $x\ast\afilter$ is immediate from 
Notation~\ref{nttn.X.ast.Y}(\ref{item.x.ast.Y}) and semifilter compatibility (Definition~\ref{defn.point}(\ref{item.weak.clique})).

Suppose $x\ast\afilter$; thus by Notation~\ref{nttn.X.ast.Y}(\ref{item.x.ast.Y}) $x$ is compatible with (every element of) $\afilter$.
We note that the $\cti$-up-closure of $\{x\}\cup\afilter$ is a semifilter (nonempty, up-closed, compatible).
By maximality, $x\in\afilter$.
\item
\emph{Suppose $x\ast\afilter$ if and only if $x\in\afilter$, for every $x\in\ns X$.}

Suppose $\afilter'$ is a semifilter and $\afilter\subseteq\afilter'$.
Consider $x'\in\afilter'$.
Then $x\ast\afilter$ by compatibility of $\afilter'$, and so $x\in\afilter$.
Thus, $\afilter'\subseteq\afilter$. 
\qedhere\end{itemize}
\end{proof}

\begin{defn}
\label{defn.x.transitive}
Suppose $(\ns X,\cti,\ast)$ is a semiframe and $x\in\ns X$.
Call $x$ \deffont[transitive element (in a semiframe)]{transitive} when:
\begin{enumerate*}
\item
$x\neq\tbot_{\ns X}$.
\item
$x'\ast x\ast x''$ implies $x'\ast x''$, for every $x',x''\in\ns X$.
\end{enumerate*}
\end{defn}

`Being topen' in semitopologies (Definition~\ref{defn.transitive}(\ref{transitive.cc})) corresponds to `being transitive' in semiframes (Definition~\ref{defn.x.transitive}): 
\begin{lemm}[Characterisation of topen sets]
\label{lemm.topen.transitive}
Suppose $(\ns P,\opens)$ is a semitopology and $O\in\opens$.
Then the following are equivalent:
\begin{enumerate*}
\item
$O$ is topen in $(\ns P,\opens)$ in the sense of Definition~\ref{defn.transitive}(\ref{transitive.cc}).
\item
$O$ is transitive in $(\opens,\subseteq,\between)$ in the sense of Definition~\ref{defn.x.transitive}.\footnote{\emph{Confusing terminology alert:}  Definition~\ref{defn.transitive}(\ref{transitive.transitive}) also has a notion of \emph{transitive set}.  The notion of transitive set is well-defined for a set that may not be open.  In the world of semiframes, we just have elements of the semiframe (which correspond, intuitively, to open sets).  Thus \emph{transitive} semiframe elements correspond to (nonempty) transitive open sets of a semitopology, which are called \emph{topens}.}
\end{enumerate*}
\end{lemm}
\begin{proof}
We unpack the definitions and note that the condition for being topen --- being a nonempty open set that is transitive for $\between$ --- is identical to the condition for being transitive in $(\opens,\subseteq,\between)$ --- being a non-$\tbot_{\opens}$ element that is transitive for ${\ast}={\between}$.
\end{proof}

\jamiesubsection{The compatibility system $x^\ast$}

\begin{defn}
\label{defn.x.ast}
Suppose $(\ns X,\cti,\ast)$ is a semiframe and $x\in\ns X$.
Then define $x^\ast$ the \deffont[compatibility system (of an element:\ $x^\ast$)]{compatibility system}\index{$x^\ast$ (compatibility system of a semiframe element)} of $x$ by
$$
x^\ast=\{x'\in\ns X\mid x'\ast x\} .
$$
\end{defn}

\begin{lemm}
\label{lemm.bigvee.ast.union}
Suppose $(\ns X,\cti,\ast)$ is a semiframe and $X\subseteq\ns X$.
Then $(\bigvee X)^\ast = \bigcup_{x{\in}X} x^\ast$.
\end{lemm}
\begin{proof}
We just follow the definitions:
$$
\begin{array}[b]{r@{\ }l@{\qquad}l}
y\in(\bigvee X)^\ast
\liff&
y\ast \bigvee X
&\text{Definition~\ref{defn.x.ast}}
\\
\liff&
\Exists{x{\in}X}y\ast x
&\text{Definition~\ref{defn.compatibility.relation}(\ref{item.compatible.distributive})}
\\
\liff&
\Exists{x{\in}X}y\in x^\ast
&\text{Definition~\ref{defn.x.ast}}
\\
\liff&
y\in\bigcup_{x{\in}X} x^\ast 
&\text{Fact of sets}
\end{array}
\qedhere$$
\end{proof}

\begin{lemm}
\label{lemm.x.ast.cycle}
Suppose $(\ns X,\cti,\ast)$ is a semiframe and $x\in\ns X$ is transitive.
Then the following are equivalent for every $y\in\ns X$:
$$
y\ast x
\quad\liff\quad
y\in x^\ast
\quad\liff\quad
y\ast x^\ast .
$$
\end{lemm}
\begin{proof}
We prove a cycle of implications:
\begin{itemize*}
\item
Suppose $y\ast x$.
Then $y\in x^\ast$ is direct from Definition~\ref{defn.x.ast}.
\item
Suppose $y\in x^\ast$.
Then $y\ast x^\ast$ --- meaning by Notation~\ref{nttn.X.ast.Y}(\ref{item.x.ast.Y}) that $y\ast x'$ for every $x'\in x^\ast$ --- follows by transitivity of $x$.
\item
Suppose $y\ast x^\ast$.
By proper reflexivity of $\ast$ (Definition~\ref{defn.compatibility.relation}(\ref{item.compatible.reflexive}); since $x\neq\tbot_{\ns X}$) $x\in x^\ast$, and $y\ast x$ follows.
\qedhere\end{itemize*} 
\end{proof}

\begin{prop}
\label{prop.trans.cps}
Suppose $(\ns X,\cti,\ast)$ is a semiframe and suppose $\tbot_{\ns X}\neq x\in\ns X$.
Then the following are equivalent:
\begin{enumerate*}
\item\label{item.cps.transitive}
$x$ is transitive. 
\item\label{item.cps.point}
$x^\ast$ is a completely prime semifilter (i.e. an abstract point).
\item\label{item.cps.semifilter}
$x^\ast$ is a semifilter.
\item\label{item.cps.compatible}
$x^\ast$ is compatible.
\item\label{item.cps.maximal}
$x^\ast$ is a maximal semifilter.
\end{enumerate*}
\end{prop}
\begin{proof}
We first prove a cycle of implications between parts~\ref{item.cps.transitive}, \ref{item.cps.point}, \ref{item.cps.semifilter}, and~\ref{item.cps.compatible}:
\begin{enumerate}
\item
Suppose $x$ is transitive.
We need to check that $x^\ast$ is nonempty, up-closed, compatible, and completely prime.
We consider each property in turn:
\begin{itemize*}
\item
$x\ast x$ by proper reflexivity of $\ast$ (Definition~\ref{defn.compatibility.relation}(\ref{item.compatible.reflexive}); since $x\neq\tbot_{\ns X}$), so $x\in x^\ast$.
\item
It follows from monotonicity of $\ast$ (Lemma~\ref{lemm.compatibility.monotone}(\ref{item.ast.monotone})) that if $x'\cti x''$ and $x\ast x'$ then $x\ast x''$.
\item
Suppose $x'\ast x\ast x''$.
By transitivity of $x$ (Definition~\ref{defn.x.transitive}), $x'\ast x''$.
\item
Suppose $x\ast\bigvee X'$; then by distributivity of $\ast$ (Definition~\ref{defn.compatibility.relation}(\ref{item.compatible.distributive}))
$x\ast x'$ for some $x'\in X'$.
\end{itemize*}
\item
If $x^\ast$ is a completely prime semifilter, then it is certainly a semifilter.
\item
If $x^\ast$ is a semifilter, then it is compatible (Definition~\ref{defn.point}(\ref{item.semifilter})\&\ref{item.weak.clique}).
\item
Suppose $x^\ast$ is compatible (Definition~\ref{defn.point}(\ref{item.weak.clique})) and suppose $x'\ast x\ast x''$.
By Lemma~\ref{lemm.x.ast.cycle} $x',x''\in x^\ast$, and by compatibility of $x^\ast$ we have $x'\ast x''$.
Thus, $x$ is transitive.
\end{enumerate}
To conclude, we prove two implications between parts~\ref{item.cps.compatible} and~\ref{item.cps.maximal}:
\begin{itemize}
\item
Suppose $x^\ast$ is a semifilter.
By equivalence of parts~\ref{item.cps.semifilter} and~\ref{item.cps.transitive} of this result, $x$ is transitive, and so 
using Lemma~\ref{lemm.x.ast.cycle} 
$x'\ast x^\ast$ if and only if $x'\in x^\ast$.
By Lemma~\ref{lemm.char.maxfilter}, $x^\ast$ is maximal.
\item
Clearly, if $x^\ast$ is a maximal semifilter then it is a semifilter.
\qedhere\end{itemize}
\end{proof}

\jamiesubsection{The compatibility system $\afilter^\ast$}

\jamiesubsubsection{Basic definitions and results}

\begin{defn}
\label{defn.X.ast}
Suppose $(\ns X,\cti,\ast)$ is a semiframe and suppose $\afilter\subseteq\ns X$ ($\afilter$ may be a semifilter, but the definition does not depend on this).
Define $\afilter^\ast$ the \deffont[compatibility system (of a set;\ $\afilter^\ast$)]{compatibility system}\index{$\afilter^\ast$ (compatibility system of a set)} of $\afilter$ by
$$
\afilter^\ast  = \{x'\in\ns X \mid x'\ast\afilter\}
$$
Unpacking Notation~\ref{nttn.X.ast.Y}(\ref{item.x.ast.Y}), and combining with Definition~\ref{defn.x.ast}, we can write:
$$
\afilter^\ast  
=
\{x'\in\ns X \mid x'\ast\afilter\}
=
\{x'\in\ns X \mid \Forall{x{\in}\afilter}x'\ast x\}
=
\bigcap\{x^\ast \mid x\in \afilter \} . 
$$
\end{defn}

Lemma~\ref{lemm.nbhd.ast.char} presents one easy and useful example of Definition~\ref{defn.X.ast}:
\begin{lemm}
\label{lemm.nbhd.ast.char}
Suppose $(\ns P,\opens)$ is a semitopology and suppose $p\in\ns P$ and $O'\in\opens$.
Then: 
$$
\begin{array}{l@{\ \liff\ }l}
O'\in\nbhd(p)^\ast
&
\Forall{O{\in}\opens}p\in O\limp O'\between O
\\
O'\notin\nbhd(p)^\ast
&
\Exists{O{\in}\opens}p\in O\land O'\notbetween O .
\end{array}
$$
\end{lemm}
\begin{proof}
We just unpack Definitions~\ref{defn.nbhd} and~\ref{defn.X.ast}.
\end{proof}

\begin{lemm}
\label{lemm.X.ast.up-closed}
Suppose $(\ns X,\cti,\ast)$ is a semiframe and $\afilter\subseteq\ns X$.
Then
$\afilter^\ast$ is up-closed.
\end{lemm}
\begin{proof}
This is just from Definition~\ref{defn.X.ast} and monotonicity of $\ast$ (Lemma~\ref{lemm.compatibility.monotone}(\ref{item.ast.monotone})).
\end{proof}

\begin{lemm}
\label{lemm.afilter.subset.afilter.ast}
Suppose $(\ns X,\cti,\ast)$ is a semiframe and $\afilter\subseteq\ns X$ is a semifilter.
Then:
\begin{enumerate*}
\item\label{item.afilter.subset.afilter.ast.1}
If $x\in \afilter$ then $\afilter \subseteq x^\ast$.
\item\label{item.afilter.subset.afilter.ast.2}
As a corollary, $\afilter\subseteq\afilter^\ast$.
\end{enumerate*}
\end{lemm}
\begin{proof}
Suppose $x\in\afilter$.
By compatibility of $\afilter$ (Definition~\ref{defn.point}(\ref{item.weak.clique})), $x'\ast x$ for every $x'\in\afilter$.
It follows from Definition~\ref{defn.x.ast} that $\afilter\subseteq x^\ast$.
The corollary is immediate from Definition~\ref{defn.X.ast}.
\end{proof}

We can use Lemma~\ref{lemm.afilter.subset.afilter.ast} and Definition~\ref{defn.X.ast} to give a more succinct rendering of Lemma~\ref{lemm.char.maxfilter}:
\begin{corr}
\label{corr.new.char.maxfilter}
Suppose $(\ns X,\cti,\ast)$ is a semiframe and $\afilter\subseteq\ns X$ is a semifilter.
Then the following are equivalent:
\begin{enumerate*}
\item\label{item.new.char.maxfilter.1}
$\afilter$ is maximal.
\item\label{item.new.char.maxfilter.2}
$\afilter^\ast=\afilter$.
\item\label{item.new.char.maxfilter.3}
$\afilter^\ast\subseteq\afilter$.
\end{enumerate*}
\end{corr}
\begin{proof}
Equivalence of parts~\ref{item.new.char.maxfilter.1} and~\ref{item.new.char.maxfilter.2} just repeats Lemma~\ref{lemm.char.maxfilter} using Definition~\ref{defn.X.ast}.
To prove equivalence of parts~\ref{item.new.char.maxfilter.2} and~\ref{item.new.char.maxfilter.3} we use use Lemma~\ref{lemm.afilter.subset.afilter.ast}(\ref{item.afilter.subset.afilter.ast.2}).
\end{proof}

\jamiesubsubsection{Strong compatibility: when $\afilter^\ast$ is a semifilter}

Proposition~\ref{prop.trans.cps} relates good properties of $x$ (transitivity) to good properties of its compatibility system $x^\ast$ (e.g. being compatible).
It will be helpful to ask similar questions of $\afilter^\ast$.
What good properties are of interest for $\afilter^\ast$, and what conditions can we impose on $\afilter$ to guarantee them?

\begin{defn}
\label{defn.F.strongly.compatible}
Suppose $(\ns X,\cti,\ast)$ is a semiframe.
Then:
\begin{enumerate*}
\item\label{item.strongly.compatible.filter}
Call $\afilter\subseteq\ns X$ \deffont[strong compatibility (of a set)]{strongly compatible} when $\afilter^\ast$ is nonempty and compatible.
\item\label{item.strongly.compatible.filter.space}
Call $(\ns X,\cti,\ast)$ \deffont[strongly compatible semiframe]{strongly compatible} when every abstract point (completely prime semifilter) $\apoint\subseteq\ns X$ is strongly compatible.
\end{enumerate*}
\end{defn}

\begin{rmrk}
\label{rmrk.what.does.strongly.compatible.mean}
For the reader's convenience we unpack Definition~\ref{defn.F.strongly.compatible}.
\begin{enumerate}
\item
By Definition~\ref{defn.point}(\ref{item.weak.clique}), $\afilter^\ast$ is compatible when $x\ast x'$ for every $x,x'\in\afilter^\ast$.
Combining this with Definition~\ref{defn.X.ast} and Notation~\ref{nttn.X.ast.Y}, $\afilter^\ast$ is compatible when $x\ast \afilter\ast x'$ implies $x\ast x'$, for every $x,x'\in\ns X$.
Thus, $\afilter$ is strongly compatible when 
$$
\Forall{x,x'{\in}\ns X}\ x\ast\afilter\ast x' \limp x\ast x'.
$$
\item
$(\ns X,\cti,\ast)$ is strongly compatible when every abstract point $\apoint\in\tf{Point}(\ns X,\cti,\ast)$ is strongly compatible in the sense just given above.
\end{enumerate}
\end{rmrk}

\begin{lemm}
\label{lemm.ht.sc.eq}
Suppose $(\ns P,\opens)$ is a semitopology and $p\in\ns P$.
Recall from Definition~\ref{defn.semi.to.dg}(\ref{item.semiframe.ast}) and Lemma~\ref{lemm.Fr.semiframe} that $(\opens,\subseteq,\between)$ is a semiframe.
Then the following are equivalent:
\begin{enumerate*}
\item
The point $p\in\ns P$ is hypertransitive in the sense of Definition~\ref{defn.sc}.
\item
The semifilter $\nbhd(p)\subseteq\opens$ is strongly compatible in the sense of Definition~\ref{defn.F.strongly.compatible}.
\end{enumerate*} 
\end{lemm}
\begin{proof}
Remark~\ref{rmrk.what.does.strongly.compatible.mean} notes that the condition in Definition~\ref{defn.sc} is precisely the condition for $\nbhd(p)$ to be strongly compatible. 
\end{proof}

\begin{rmrk}
\label{rmrk.semiframes.caution}
Given Lemma~\ref{lemm.ht.sc.eq}, the reader might ask why we do not just call a strongly compatible semifilter `hypertransitive'.

There is a case for doing so, but caution is required:
strong compatibility of semiframes is not \emph{quite} the same thing as hypertransitivity of points.
Every point $p$ generates a semifilter $\nbhd(p)$, but there may be more semifilters than there are points, and this makes the strong compatibility condition subtly different from the hypertransitivity condition. 
We shall see the effects of this in Lemma~\ref{lemm.r=wr+sc}(\ref{item.wr.kt.1b}), and in Theorem~\ref{thrm.r=wr+sc} (see Remark~\ref{rmrk.subtly.different} for a brief discussion), and then again in Definition~\ref{defn.strongly.compatible.semitopology} where we define a notion of \emph{strongly compatible semitopology} (essentially: all of its semifilters are strongly compatible), which is not the same thing as the space being hypertransitive (essentially: all of its points are hypertransitive). 

Therefore, we maintain a terminological distinction: \emph{points} are hypertransitive, \emph{semiframes} are strongly compatible.
The notions are related, but not quite the same thing.
\end{rmrk}

\begin{lemm}
\label{lemm.ast.semifilter.compatible}
Suppose $(\ns X,\cti,\ast)$ is a semiframe and suppose $\afilter\subseteq\ns X$ is nonempty.
Then the following are equivalent:
\begin{enumerate*}
\item\label{item.ast.semifilter.compatible.1}
$\afilter^\ast$ is a semifilter.
\item\label{item.ast.semifilter.compatible.2}
$\afilter^\ast$ is compatible.
\item\label{item.ast.semifilter.compatible.3}
$\afilter$ is strongly compatible.
\end{enumerate*}
\end{lemm}
\begin{proof}
Equivalence of parts~\ref{item.ast.semifilter.compatible.2} and~\ref{item.ast.semifilter.compatible.3} is just Definition~\ref{defn.F.strongly.compatible}.
For equivalence of parts~\ref{item.ast.semifilter.compatible.1} and~\ref{item.ast.semifilter.compatible.2} we prove two implications:
\begin{itemize}
\item
Suppose $\afilter^\ast$ is a semifilter.
Then $\afilter^\ast$ is compatible by assumption in Definition~\ref{defn.point}(\ref{item.semifilter}).
\item
Suppose $\afilter^\ast$ is compatible.
Then $\afilter^\ast$ is up-closed by Lemma~\ref{lemm.X.ast.up-closed}, and nonempty by Lemma~\ref{lemm.afilter.subset.afilter.ast}(\ref{item.afilter.subset.afilter.ast.2}) (since $\afilter$ is nonempty).
Thus, by Definition~\ref{defn.point}(\ref{item.semifilter}) $\afilter^\ast$ is a semifilter.
\qedhere\end{itemize}
\end{proof}

\begin{lemm}
\label{lemm.not.necessarily.strongly.compatible}
Suppose $(\ns X,\cti,\ast)$ is a semiframe and suppose $\afilter\subseteq\ns X$.
Then it is not necessarily the case that $\afilter^\ast$ is a semifilter.

This non-implication holds even in strong well-behavedness conditions:  that $(\ns X,\cti,\ast)$ is spatial and $\afilter$ is an abstract point (a completely prime semifilter).
\end{lemm}
\begin{proof}
It suffices to provide a counterexample.
Let $(\ns P,\opens)=(\{0,1,2\},\{\varnothing,\{0\},\{2\},\ns P\})$, as illustrated in the top-left semitopology in Figure~\ref{fig.012}.
Take $(\ns X,\cti,\ast)=(\opens,\subseteq,\between)$ (which is spatial by Proposition~\ref{prop.Gr.P.spatial}) and set $\afilter=\nbhd(1)=\{0,1,2\}$.
Then $\nbhd(1)^\ast=\{\{0\},\{2\},\{0,1,2\}\}$, and this is not compatible because $\{0\}\notbetween\{2\}$.\footnote{$1$ is also a \emph{conflicted} point; see Example~\ref{xmpl.conflicted.points}(\ref{item.example.of.conflicted.point}).  This is no accident: by Lemma~\ref{lemm.regular.sc}(\ref{item.sc.implies.uc}) if $p$ is conflicted then it is not hypertransitive, and by Lemma~\ref{lemm.ht.sc.eq} it follows that $\nbhd(p)^\ast$ is not compatible.}
\end{proof}

\begin{rmrk}
Lemma~\ref{lemm.not.necessarily.strongly.compatible} gives an example of a semifilter $\afilter$ that is not strongly compatible (i.e. such that $\afilter^\ast$ is not a semifilter).
Note that in this example both the space and $\afilter$ are well-behaved.
This raises the question of finding sufficient (though perhaps not necessary) criteria for strong compatibility.
We conclude with Proposition~\ref{prop.x.transitive.afilter.ast} which provides one such criterion; it will be useful later in Lemma~\ref{lemm.r=wr+sc} and Theorem~\ref{thrm.r=wr+sc.st}.
\end{rmrk}

\begin{figure}
\vspace{-1em}
\centering
\includegraphics[width=0.4\columnwidth,trim={0 100 0 100},clip]{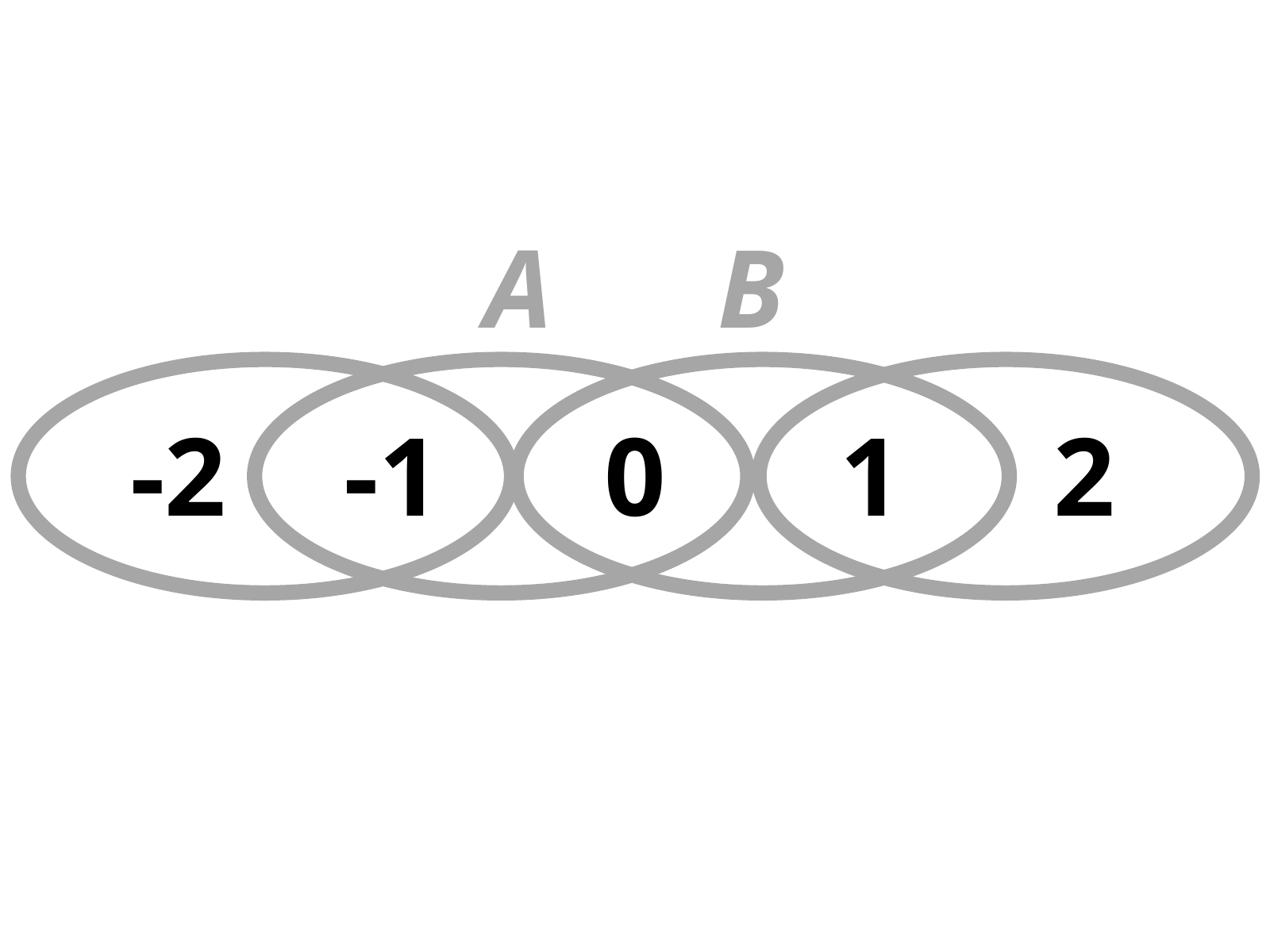}
\vspace{-1em}
\caption{Strongly compatible filter that contains no transitive element}
\label{fig.strong.compat.no.transitive}
\end{figure}

Proposition~\ref{prop.x.transitive.afilter.ast} bears a family resemblance to Theorem~\ref{thrm.max.cc.char} (if a point has a topen neighbourhood then it is regular):
\begin{prop}
\label{prop.x.transitive.afilter.ast}
Suppose $(\ns X,\cti,\ast)$ is a semiframe and $\afilter\subseteq\ns X$ is a semifilter.
Then:
\begin{enumerate*}
\item
If $\afilter$ contains a transitive element then $\afilter$ is strongly compatible.
\item
The converse implication need not hold: it may be that $\afilter$ is strongly compatible yet $\afilter$ contains no transitive element.
\end{enumerate*}
\end{prop}
\begin{proof}
We consider each part in turn:
\begin{enumerate}
\item
Suppose $x\in\afilter$ is transitive.
By Lemma~\ref{lemm.ast.semifilter.compatible} it would suffice to show that $\afilter^\ast$ is compatible (Definition~\ref{defn.point}(\ref{item.weak.clique})).
So consider $y\ast\afilter\ast y'$.
Then $y\ast x\ast y'$ and by transitivity $y\ast y'$.
Thus $\afilter^\ast$ is compatible.
\item
It suffices to provide a counterexample.
We take, as illustrated in Figure~\ref{fig.strong.compat.no.transitive},
\begin{itemize*}
\item
$\ns P=\{\minus 2,\minus 1,0,1,2\}$ and 
\item
we let $\opens$ be generated by $\{i,i\plus 1\}$ for $\minus 2\leq i\leq 1$ (unordered pairs of adjacent numbers).
\end{itemize*}
Write $A=\{\minus 1,0\}$ and $B=\{0,1\}$ and let $\afilter$ be the up-closure of $\{A,B\}$. %
Note that $A$ and $B$ are not transitive (i.e. not topen).
The reader can check that $\afilter^\ast=\afilter$ (e.g. $\{1,2\}\notin\afilter^\ast$ because $\{1,2\}\notbetween \{\minus 1,0\}\in\afilter$), but $\afilter$ contains no transitive element.
\qedhere\end{enumerate}
\end{proof}

\jamiesubsection{Semiframe characterisation of community}

\begin{rmrk}
We saw the notion of $\community(p)$ the \emph{community} of a point in Definition~\ref{defn.tn}(\ref{item.tn}).
In this Subsection we construct an analogue to it in semiframes.
We will give two characterisations: one in Definition~\ref{defn.abstract.community}, and another in Proposition~\ref{prop.framecommunity.universal}.
\end{rmrk}

We will mostly be interested in Definition~\ref{defn.cast} when $\afilter$ is a semifilter, but the definition does not require this: 
\begin{defn}
\label{defn.cast}
Suppose $(\ns X,\cti,\ast)$ is a semiframe and $\afilter\subseteq\ns X$ and $x\in\ns X$.
Then define $\cclo{\afilter}\in\ns X$, $\cast{\afilter}\in\ns X$, and $\cast{x}\in\ns X$ by
$$
\begin{array}{r@{\ }l}
\cclo{\afilter} =& \bigvee \{y\in\ns X \mid y\notin \afilter\} 
\\
\cast{\afilter} =& \cclo{(\afilter^\ast)} 
\\
\cast{x} =& \cclo{(x^\ast)} .
\end{array}
$$ 
\end{defn}

\begin{rmrk}
\label{rmrk.cast.simpler}
We unpack the definitions of $\cast{\afilter}$ and $\cast{x}$: 
$$
\begin{array}{r@{\ }l@{\qquad}l}
\cast{\afilter} 
=& \cclo{(\afilter^\ast)} 
&\text{Definition~\ref{defn.cast}}
\\
=& \bigvee \{y\in\ns X \mid y\notin \afilter^\ast\} 
&\text{Definition~\ref{defn.cast}}
\\
=& \bigvee \{y\in\ns X \mid \neg(y\ast \afilter)\} 
&\text{Definition~\ref{defn.X.ast}}
\\[2ex]
\cast{x} 
=& \cclo{(x^\ast)}
&\text{Definition~\ref{defn.cast}}
\\
=& \bigvee \{y\in\ns X \mid y\notin x^\ast \} 
&\text{Definition~\ref{defn.cast}}
\\
=& \bigvee \{y\in\ns X \mid \neg(y\ast x) \} .
&\text{Definition~\ref{defn.x.ast}}
\end{array}
$$ 
\end{rmrk}

Lemma~\ref{lemm.cast.comp} will be useful, and gives some intuition for $\cclo{(\text{-})}$ and $\cast{(\text{-})}$ by unpacking their concrete meaning in the special case of a semiframe of open sets of a semitopology:
\begin{lemm}
\label{lemm.cast.comp}
Suppose $(\ns P,\opens)$ is a semitopology and $p\in\ns P$ and $O\in\opens$.
Then:
\begin{enumerate*}
\item\label{item.cast.comp.cclo}
$\cclo{\nbhd(p)}=\ns P\setminus\closure{p}$.
\item\label{item.cast.comp.nbhd}
$\cast{\nbhd(p)}=\ns P\setminus\intertwined{p}$.
\item\label{item.cast.comp.O}
$\cast{O}=\ns P\setminus\closure{O}=\interior(\ns P\setminus O)$.
\end{enumerate*}
\end{lemm}
\begin{proof}
We consider each part in turn:
\begin{enumerate}
\item
It is a fact of Definition~\ref{defn.closure} that $\ns P\setminus\closure{p}=\bigcup\{O'\in\opens \mid p\notin O'\}$.
By Proposition~\ref{prop.nbhd.iff}(\ref{item.nbhd.iff}) $p\notin O'$ if and only if $O'\notin\nbhd(p)$.
\item
It is a fact of Definition~\ref{defn.intertwined.points}, which is spelled out in Lemma~\ref{lemm.char.not.intertwined}(\ref{item.intertwined.open.avoid}), that $\ns P\setminus\intertwined{p}=\bigcup\{O'\in\opens \mid \Exists{O{\in}\opens} p\in O \land O'\notbetween O\}$.
By Lemma~\ref{lemm.nbhd.ast.char} $\Exists{O{\in}\opens} p\in O\land O'\notbetween O$ precisely when $O'\notin\nbhd(p)^\ast$. 
\item
By Definitions~\ref{defn.cast} and~\ref{defn.X.ast} we have
$$
\cclo{O}=\cast{(O^\ast)}=\bigcup\{O'{\in}\opens \mid O'\notin O^\ast\} =\bigcup\{O'{\in}\opens \mid O'\notbetween O\} .
$$
The result then follows by routine reasoning on closures (Definition~\ref{defn.closure}).
\qedhere\end{enumerate}
\end{proof}

\begin{defn}
\label{defn.abstract.community}
Suppose $(\ns X,\cti,\ast)$ is a semiframe and $\afilter\subseteq\ns X$.
Then define $\framecommunity(\afilter)\in\ns X$ the \deffont[abstract community (of a set:\ $\framecommunity(\afilter)$)]{abstract community}\index{$\framecommunity(\afilter)$ (abstract community of a set)} of $\afilter$ by
$$
\framecommunity(\afilter)=\cast{(\cast{\afilter})} \in \ns X.
$$
(For a more direct characterisation, see Proposition~\ref{prop.framecommunity.universal}.)
\end{defn}

\begin{prop}
\label{prop.framecommunity.nbhd.community}
Suppose $(\ns P,\opens)$ is a semitopology and $p\in\ns P$.
Then 
$$
\framecommunity(\nbhd(p))=\community(p) .
$$
In words: the abstract community of the abstract point $\nbhd(p)$ in $(\opens,\subseteq,\between)$, is identical to the community of $p$.
\end{prop}
\begin{proof}
We reason as follows:
$$
\begin{array}[b]{r@{\ }l@{\qquad}l}
\framecommunity(\nbhd(p))
=&
\cast{(\cast{\nbhd(p)})}
&\text{Definition~\ref{defn.abstract.community}}
\\
=&
\cast{(\ns P\setminus\intertwined{p})}
&\text{Lemma~\ref{lemm.cast.comp}(\ref{item.cast.comp.nbhd})}
\\
=&
\interior(\ns P\setminus (\ns P\setminus\intertwined{p}))
&\text{Lemma~\ref{lemm.cast.comp}(\ref{item.cast.comp.O})}
\\
=&
\interior(\intertwined{p})
&\text{Fact of sets}
\\
=&
\community(p)
&\text{Definition~\ref{defn.tn}(\ref{item.tn})}
\end{array}
\qedhere$$
\end{proof}

We can also give a more direct characterisation of the abstract community from Definition~\ref{defn.abstract.community}:
\begin{prop}
\label{prop.framecommunity.universal}
Suppose $(\ns X,\cti,\ast)$ is a semiframe and $\afilter\subseteq\ns X$.
Then 
$$
\framecommunity(\afilter)=\bigvee\{x\in\ns X \mid x^\ast\subseteq\afilter^\ast\} ,
$$
and $\framecommunity(\afilter)$ is the greatest element in $\ns X$ such that $\framecommunity(\afilter)^\ast\subseteq\afilter^\ast$.
\end{prop}
\begin{proof}
We follow the definitions: 
$$
\begin{array}{r@{\ }l@{\qquad}l}
\cast{(\cast{\afilter})}
=&
\bigvee\{x\in\ns X\mid \neg (x\ast \cast{\afilter})\} 
&\text{Remark~\ref{rmrk.cast.simpler}}
\\
=&
\bigvee\{x\in\ns X\mid \neg (x\ast \bigvee\{y \mid \neg (y\ast\afilter)\})\}
&\text{Remark~\ref{rmrk.cast.simpler}}
\\
=&
\bigvee\{x\in\ns X\mid \neg \Exists{y{\in}\ns X} (x\ast y \land \neg (y\ast\afilter))\}
&\text{Definition~\ref{defn.compatibility.relation}(\ref{item.compatible.distributive})}
\\
=&
\bigvee\{x\in\ns X \mid \Forall{y{\in}\ns X} y\ast x \limp y\ast\afilter\}
&\text{Fact of logic}
\\
=&
\bigvee\{x\in\ns X \mid x^\ast\subseteq \afilter^\ast \}
&\text{Definitions~\ref{defn.x.ast} \&~\ref{defn.X.ast}}
\end{array}
$$
To see that $\framecommunity(\afilter)$ is the greatest element such that $\framecommunity(\afilter)^\ast\subseteq\afilter^\ast$, we note from Lemma~\ref{lemm.bigvee.ast.union} that
$$
\framecommunity(\afilter)^\ast = \bigcup \{x^\ast \mid x{\in}\ns X,\ x^\ast\subseteq\afilter^\ast\} .
\qedhere$$
\end{proof}

\jamiesubsection{Semiframe characterisation of regularity}

We now have enough to generalise the notions of quasiregularity, weak regularity, and regularity from semitopologies (Definition~\ref{defn.tn} parts~\ref{item.quasiregular.point}, \ref{item.weakly.regular.point}, and~\ref{item.regular.point}) to semiframes:
\begin{defn}
\label{defn.afilter.regular}
Suppose $(\ns X,\cti,\ast)$ is a semiframe and $\afilter\subseteq\ns X$ is a semifilter.
\begin{enumerate}
\item\label{item.afilter.quasiregular}
Call $\afilter$ \deffont[quasiregular semifilter]{quasiregular} when $\framecommunity(\afilter)\neq\tbot_{\ns X}$.

Thus, there exists some $x\in\ns X$ such that $x^\ast\subseteq \afilter^\ast$.
\item\label{item.afilter.weakly.regular}
Call $\afilter$ \deffont[weakly regular semifilter]{weakly regular} when $\framecommunity(\afilter)\in\afilter$.
\item\label{item.afilter.regular}
Call $\afilter$ \deffont[regular semifilter]{regular} when $\framecommunity(\afilter)\in\afilter$ and $\framecommunity(\afilter)$ is transitive.
\end{enumerate}
\end{defn}

Lemma~\ref{lemm.afilter.regular.imp} does for semiframes what Lemma~\ref{lemm.wr.r} does for semitopologies:
\begin{lemm}
\label{lemm.afilter.regular.imp}
Suppose $(\ns X,\cti,\ast)$ is a semiframe and $\afilter\subseteq\ns X$ is a semifilter.
Then:
\begin{enumerate*}
\item
If $\afilter$ is regular then it is weakly regular.
\item
If $\afilter$ is weakly regular then it is quasiregular.
\end{enumerate*}
(The converse implications need not hold, and it is possible for $\afilter$ to not be quasiregular:
it is convenient to defer the proofs to Corollary~\ref{corr.quasiregular.no.converse}.)
\end{lemm}
\begin{proof}
The proofs are easy:
If $\framecommunity(\afilter)\in\afilter$ and $\framecommunity(\afilter)$ is transitive, then certainly $\framecommunity(\afilter)\in\afilter$.
If $\framecommunity(\afilter)\in\afilter$ then by Lemma~\ref{lemm.P.top}(\ref{item.P.no.bot}) $\framecommunity(\afilter)\neq\tbot_{\ns X}$.
\end{proof}

\begin{lemm}
\label{lemm.r=wr+sc}
Suppose $(\ns X,\cti,\ast)$ is a semiframe and $\afilter\subseteq\ns X$ is a semifilter.
Then:
\begin{enumerate*}
\item\label{item.wr.kt.1}
If $\afilter$ is quasiregular and strongly compatible then $\framecommunity(\afilter)$ is transitive.
\item\label{item.wr.kt.1b}
The converse implication need not hold: it is possible for $\afilter$ to be quasiregular and $\framecommunity(\afilter)$ to be transitive, yet $\afilter$ is not strongly compatible.
\item\label{item.wr.kt.2}
If $\afilter$ is weakly regular and $\framecommunity(\afilter)$ is transitive then $\afilter$ is strongly compatible.
\item\label{item.wr.kt.iff}
If $\afilter$ is weakly regular, then $\framecommunity(\afilter)$ is transitive if and only if $\afilter$ is strongly compatible.
\end{enumerate*}
\end{lemm}
\begin{proof}
We consider each part in turn:
\begin{enumerate}
\item
Suppose $\afilter$ is quasiregular and strongly compatible.
 
By quasiregularity $\tbot_{\ns X}\neq\framecommunity(\afilter)$.
By Proposition~\ref{prop.framecommunity.universal} $\framecommunity(\afilter)^\ast\subseteq\afilter^\ast$.
By strong compatibility $\afilter^\ast$ is a semifilter and so in particular $\afilter^\ast$ is compatible.
It follows from Proposition~\ref{prop.trans.cps}(\ref{item.cps.transitive}\&\ref{item.cps.compatible}) that $\framecommunity(\afilter)$ is transitive, as required.
\item
It suffices to provide a counterexample.
Let $(\mathbb R,\opens)$ be the real numbers with their usual topology, and let $(\mathbb R,\opens')$ be the topology generated by $\opens\cup\{\{0\}\}$ --- in words: we add $\{0\}$ as an open set.
 
Let $\afilter$ be the semifilter of all $\opens$-open neighbourhoods of $0$.
$\afilter^\ast$ is the set of $\opens'$-open sets that intersect every $\opens$-open neighbourhood of $0$.
This is not compatible, because it contains $\openinterval{0,}$ (the set of numbers strictly less than $0$) and $\openinterval{,0}$ (the set of numbers strictly greater than $0$), and these do not intersect.
Using Proposition~\ref{prop.framecommunity.universal}, we calculate that $\framecommunity(\afilter)=\{0\}$; this is transitive because it is a singleton set.

So $\afilter$ is quasiregular, $\framecommunity(\afilter)$ is transitive, yet $\afilter$ is not strongly compatible.
\item
Suppose $\framecommunity(\afilter)$ is transitive and suppose $\afilter$ is weakly regular, so $\framecommunity(\afilter)\in\afilter$.
By Proposition~\ref{prop.x.transitive.afilter.ast} $\afilter$ is strongly compatible.
\item
From parts~\ref{item.wr.kt.1} and~\ref{item.wr.kt.2} of this result, noting from Lemma~\ref{lemm.afilter.regular.imp} that if $\afilter$ is weakly regular then it is quasiregular.
\qedhere\end{enumerate}
\end{proof}

\begin{thrm}
\label{thrm.r=wr+sc}
Suppose $(\ns X,\cti,\ast)$ is a semiframe and $\afilter\subseteq\ns X$ is a semifilter.
Then $\afilter$ is regular if and only if $\afilter$ is weakly regular and strongly compatible.
We can write this succinctly as follows:
\begin{quote}
Regular = weakly regular + strongly compatible.\footnote{Compare this slogan with the version for semitopologies in Theorem~\ref{thrm.r=wr+uc}.}
\end{quote}
\end{thrm}
\begin{proof}
Suppose $\afilter$ is weakly regular and strongly compatible.
By Lemma~\ref{lemm.r=wr+sc}(\ref{item.wr.kt.iff}) $\framecommunity(\afilter)$ is transitive, and by Definition~\ref{defn.afilter.regular}(\ref{item.afilter.regular}) $\afilter$ is regular.

For the converse implication we just reverse the reasoning above.
\end{proof}

\begin{rmrk}
\label{rmrk.subtly.different}
In Theorem~\ref{thrm.regular=qr+sc} we characterised regularity of points in terms of quasiregularity and being hypertransitive.
In view of Lemma~\ref{lemm.ht.sc.eq} we might expect Theorem~\ref{thrm.r=wr+sc} to read `regular = quasiregular + strongly compatible'.
But this is false, as per the discussion in Remark~\ref{rmrk.semiframes.caution} and the counterexample in Lemma~\ref{lemm.r=wr+sc}(\ref{item.wr.kt.1b}).
Thus, the semiframes results are subtly different from those governing point-set semitopologies. 
\end{rmrk}

\jamiesubsection{Semiframe characterisation of (quasi/weak)regularity}
	
The direct translation in Definition~\ref{defn.afilter.regular} of parts~\ref{item.quasiregular.point}, \ref{item.weakly.regular.point}, and~\ref{item.regular.point} of Definition~\ref{defn.tn}, along with the machinery we have now built, makes Lemma~\ref{lemm.match.up} easy to prove:
\begin{lemm}
\label{lemm.match.up}
Suppose $(\ns P,\opens)$ is a semitopology and $p\in\ns P$.
Recall from Definition~\ref{defn.nbhd} and Proposition~\ref{prop.nbhd.iff}(\ref{item.nbhd.point})
that $\nbhd(p)=\{O\in\opens \mid p\in O\}$ is a (completely prime) semifilter.
Then:
\begin{enumerate*}
\item
$p$ is quasiregular in the sense of Definition~\ref{defn.tn}(\ref{item.quasiregular.point})
if and only if 
$\nbhd(p)$ is quasiregular in the sense of Definition~\ref{defn.afilter.regular}(\ref{item.afilter.quasiregular}).
\item
$p$ is weakly regular in the sense of Definition~\ref{defn.tn}(\ref{item.weakly.regular.point}) if and only if $\nbhd(p)$ is weakly regular in the sense of Definition~\ref{defn.afilter.regular}(\ref{item.afilter.weakly.regular}).
\item
$p$ is regular in the sense of Definition~\ref{defn.tn}(\ref{item.regular.point}) if and only if $\nbhd(p)$ is regular in the sense of Definition~\ref{defn.afilter.regular}(\ref{item.afilter.regular}).
\end{enumerate*}
\end{lemm}
\begin{proof}
We consider each part in turn:
\begin{enumerate}
\item
Suppose $p$ is quasiregular.
By Definition~\ref{defn.tn}(\ref{item.quasiregular.point}) $\community(p)\neq\varnothing$.
By Proposition~\ref{prop.framecommunity.nbhd.community} $\framecommunity(\nbhd(p))\neq\varnothing=\tbot_{\opens}$.
By Definition~\ref{defn.afilter.regular}(\ref{item.afilter.quasiregular}) $\nbhd(p)$ is quasiregular.

The reverse implication follows just reversing the reasoning above.
\item
Suppose $p$ is weakly regular.
By Definition~\ref{defn.tn}(\ref{item.weakly.regular.point}) $p\in\community(p)$.
By Definition~\ref{defn.nbhd} $\community(p)\in\nbhd(p)$.
By Proposition~\ref{prop.framecommunity.nbhd.community} $\framecommunity(\nbhd(p))\in\nbhd(p)$ as required.

The reverse implication follows just reversing the reasoning above.
\item
Suppose $p$ is regular.
By Definition~\ref{defn.tn}(\ref{item.regular.point}) $p\in\community(p)\in\topens$.
By Definition~\ref{defn.nbhd} and Proposition~\ref{prop.framecommunity.nbhd.community} $\framecommunity(\nbhd(p))\in\nbhd(p)$.
By Proposition~\ref{prop.framecommunity.nbhd.community} and Lemma~\ref{lemm.topen.transitive} $\framecommunity(\nbhd(p))$ is transitive. 

The reverse implication follows just reversing the reasoning above.
\qedhere\end{enumerate}
\end{proof}

\begin{prop}
\label{prop.regular.match.up}
Suppose $(\ns P,\opens)$ is a semitopology and $p\in\ns P$.
Then 
\begin{itemize*}
\item
$p$ is quasiregular / weakly regular / regular in $(\ns P,\opens)$ in the sense of Definition~\ref{defn.tn} 
\\
if and only if 
\item
$\nbhd(p)$ is quasiregular / weakly regular / regular in $\tf{Soberify}(\ns P,\opens)$ in the sense of Definition~\ref{defn.afilter.regular}.
\end{itemize*}
\end{prop} 
\begin{proof}
We consider just the case of regularity; quasiregularity and weak regularity are no different.

Suppose $p$ is regular.
By Definition~\ref{defn.tn}(\ref{item.regular.point}) $p\in \community(p)\in\topens$.
It follows from Lemma~\ref{lemm.topen.transitive} that $\community(p)$ is transitive in $(\opens,\subseteq,\between)$, and from Proposition~\ref{prop.nbhd.iff}(\ref{item.nbhd.iff}) that $\community(p)\in\nbhd(p)$.
It follows from Proposition~\ref{prop.framecommunity.nbhd.community} that $\nbhd(p)$ is regular in the sense of Definition~\ref{defn.afilter.regular}(\ref{item.afilter.regular}).
\end{proof}

\begin{corr}
\label{corr.quasiregular.no.converse}
Suppose $(\ns X,\cti,\ast)$ is a semiframe and $\afilter\subseteq\ns X$ is a semifilter.
Then the converse implications in Lemma~\ref{lemm.afilter.regular.imp} need not hold: $\afilter$ may be quasiregular but not regular, and it may be weakly regular but not regular, and it may not even be quasiregular.
\end{corr}
\begin{proof}
It suffices to provide counterexamples.
We easily obtain these by using Proposition~\ref{prop.regular.match.up} to consider $\nbhd(p)$ for $p\in\ns P$ as used in Lemma~\ref{lemm.wr.r.no}.
\end{proof}

\jamiesubsection{Characterisation of being intertwined}

This Subsection continues Remark~\ref{rmrk.promise.ast.int.char}.

The notion of points being intertwined from Definition~\ref{defn.intertwined.points}(\ref{item.p.intertwinedwith.p'}) generalises in semiframes to the notion of semifilters being compatible:
\begin{lemm}
\label{lemm.intertwined.sober}
Suppose $(\ns P,\opens)$ is a semitopology and $p,p'\in\ns P$.
Then 
$$
p\intertwinedwith p'
\quad\liff\quad
\nbhd(p)\ast\nbhd(p')
\quad\liff\quad
\nbhd(p)\intertwinedwith\nbhd(p') .
$$
For clarity and precision we unpack this.
The following are equivalent: 
\begin{enumerate*}
\item\label{item.intertwined.sober.pp'.intertwinedwith}
$p\intertwinedwith p'$ in the semitopology $(\ns P,\opens)$ (Definition~\ref{defn.intertwined.points}(\ref{item.p.intertwinedwith.p'})).

In words: the point $p$ is intertwined with the point $p'$.
\item\label{item.intertwined.sober.nbhd.ast}
$\nbhd(p)\ast\nbhd(p')$ in the semiframe $(\opens,\subseteq,\between)$ (Notation~\ref{nttn.X.ast.Y}(\ref{item.X.ast.Y})).

In words: the abstract point $\nbhd(p)$ is compatible with the abstract point $\nbhd(p')$.
\item\label{item.intertwined.sober.3}
$\nbhd(p)\intertwinedwith\nbhd(p')$ in the semitopology $\tf{St}(\opens,\subseteq,\between)$ (Definition~\ref{defn.intertwined.points}(\ref{item.p.intertwinedwith.p'})). 

In words: the point $\nbhd(p)$ is intertwined with the point $\nbhd(p')$.
\end{enumerate*}
\end{lemm}
\begin{proof}
We unpack definitions:
\begin{itemize*}
\item
By Definition~\ref{defn.intertwined.points}(\ref{item.p.intertwinedwith.p'}) $p\intertwinedwith p'$ when for every pair of open neighbourhoods $p\in O$ and $p'\in O'$ we have $O\between O'$.
\item
By Notation~\ref{nttn.X.ast.Y}(\ref{item.X.ast.Y}) $\nbhd(p)\ast\nbhd(p')$ when for every $O\in\nbhd(p)$ and $O'\in\nbhd(p')$ we have $O\ast O'$.

By Proposition~\ref{prop.nbhd.iff}(\ref{item.nbhd.iff}) we can simplify this to: $p\in O$ and $p'\in O'$ implies $O\ast O'$.
\item
By Definition~\ref{defn.intertwined.points}(\ref{item.p.intertwinedwith.p'}) and Theorem~\ref{thrm.nbhd.morphism}, $\nbhd(p)\intertwinedwith \nbhd(p')$ when:
for every pair of open neighbourhoods $\nbhd(p)\in \f{Op}(O)$ and $\nbhd(p')\in \f{Op}(O')$ we have $\f{Op}(O)\between \f{Op}(O')$.

By Proposition~\ref{prop.nbhd.iff}(\ref{item.nbhd.iff}) we can simplify this to: $p\in O$ and $p'\in O'$ implies $\f{Op}(O)\between \f{Op}(O')$.

By Proposition~\ref{prop.semiframe.to.Op}(\ref{item.semiframe.to.Op.between}) we can simplify this further to: $p\in O$ and $p'\in O'$ implies $O\ast O'$.
\end{itemize*}
But by definition, the compatibility relation $\ast$ of $(\opens,\subseteq,\between)$ is $\between$, so $O\ast O'$ and $O\between O'$ are the same assertion.
The equivalences follow. 
\end{proof}

The property of being intertwined is preserved and reflected when we use $\nbhd$ to map to the soberified space:
\begin{corr}
\label{corr.intertwined.sober}
Suppose $(\ns P,\opens)$ is a semitopology and $p,p'\in\ns P$.
Then $p\intertwinedwith p'$ in $(\ns P,\opens)$ if and only if $\nbhd(p)\intertwinedwith\nbhd(p')$ in $\tf{Soberify}(\ns P,\opens)$. 
\end{corr}
\begin{proof}
This just reiterates the equivalence of parts~\ref{item.intertwined.sober.pp'.intertwinedwith} and~\ref{item.intertwined.sober.3} in Lemma~\ref{lemm.intertwined.sober}.
\end{proof}

\begin{prop}
\label{prop.conflicted.sober}
Suppose $(\ns P,\opens)$ is a semitopology.
Then:
\begin{enumerate*}
\item\label{item.conflicted.sober.1}
It may be that $(\ns P,\opens)$ is unconflicted (meaning that it contains no conflicted points), but the semitopology $\tf{Soberify}(\ns P,\opens)$ contains a conflicted point.
\item\label{item.conflicted.sober.2}
It may further be that $(\ns P,\opens)$ is unconflicted and $p\in\ns P$ is such that $\nbhd(p)$ is conflicted in the semitopology $\tf{Soberify}(\ns P,\opens)$.
\end{enumerate*}
We can summarise the two assertions above as follows:
\begin{enumerate*}
\item
Soberifying a space might introduce a conflicted point, even if none was originally present.
\item
Soberifying a space can make a point that was unconflicted, into a point that is conflicted.\footnote{If we stretch the English language, we might say that soberifying a space can conflictify one of its points.}
\end{enumerate*}
\end{prop}
\begin{proof}
It suffices to provide counterexamples.
\begin{enumerate}
\item
Consider the right-hand semitopology in Figure~\ref{fig.012-triangle}; this is unconflicted because every point is intertwined only with itself.
The soberification of this space is illustrated in the right-hand semitopology in Figure~\ref{fig.012-triangle-sober}.
Each of the extra points is intertwined with the two numbered points next to it; e.g. the extra point in the open set $A$ --- write it $\bullet_A$ (in-between $3$ and $0$) --- is intertwined with $0$ and $3$; so $3\intertwinedwith \bullet_A\intertwinedwith 0$.
However, the reader can check that $3\notintertwinedwith 0$.
Thus, $\bullet_A$ is conflicted.
\item
We define $(\ns P,\opens)$ by:
\begin{itemize}
\item
$\ns P = \openinterval{\minus 1,1}$ (real numbers between $\minus 1$ and $1$ exclusive).
\item
$\opens$ is generated by:
\begin{itemize*}
\item 
All open intervals that do not contain $0$; so this is open intervals $\openinterval{r_1,r_2}$ where $\minus 1\leq r_1<r_2\leq 0$ or $0\leq r_1<r_2\leq 1$.
\item
All of the open intervals $\openinterval{\minus 1/n,1/n}$, for $n\geq 2$. 
\end{itemize*}
\end{itemize}
The reader can check that:
\begin{itemize*}
\item
Points in this semitopology are intertwined only with themselves.
\item
The soberification includes four additional points, corresponding to completely prime semifilters $\minus 1/0$ generated by $\{\openinterval{\minus 1/n,0} \mid n\geq 2\}$ and $\plus 1/0$ generated by $\{\openinterval{0,1/n} \mid n\geq 2\}$, and to the endpoints $\minus 1$ and $i\plus 1$.
\item
$\minus 1/0$ and $\plus 1/0$ are intertwined with $0$, but are not intertwined with one another.
\end{itemize*}
Thus, $0$ is conflicted in $\tf{Soberify}(\ns P,\opens)$ but not in $(\ns P,\opens)$.
\qedhere\end{enumerate}
\end{proof}

\begin{rmrk}
Proposition~\ref{prop.conflicted.sober} may seem surprising in view of Corollary~\ref{corr.intertwined.sober}, but the key observation is that the soberified space may add points to the original space.
These points can add conflicting behaviour that is `hidden' in the completely prime semifilters of the original space.

Thus, Proposition~\ref{prop.conflicted.sober} shows that the property of `being unconflicted' \emph{cannot} be characterised purely in terms of the semiframe of open sets --- if it could be, then soberification would make no difference, by Theorem~\ref{thrm.nbhd.morphism}(\ref{item.nbhd.morphism.is.iso}).

There is nothing wrong with that, except that we are interested in well-behavedness conditions on semiframes.
We can now look for some other condition --- but one having to do purely with open sets --- that might play a similar role in the theory of (weak/quasi)regularity of semiframes, as being unconflicted does in theory of (weak/quasi)regularity of semitopologies.

We already saw a candidate for this in Theorem~\ref{thrm.r=wr+sc}: \emph{strong compatibility}.
We examine this next.
\end{rmrk}

\jamiesubsection{Strong compatibility in semitopologies}

\begin{rmrk}
Note that:
\begin{enumerate*}
\item
Theorem~\ref{thrm.r=wr+sc} characterises `regular' for semiframes as `weakly regular + strongly compatible'. 
\item
Theorem~\ref{thrm.r=wr+uc} characterises `regular' for semitopologies as `weakly regular + unconflicted'.
\end{enumerate*}
We know from results like Lemma~\ref{lemm.match.up} and Corollary~\ref{corr.intertwined.sober} that there are accurate correspondences between notions of regularity in semiframes and semitopologies.
This is by design, e.g. in Definition~\ref{defn.afilter.regular}; we designed the semiframe definitions so that semiframe regularity and semitopological regularity would match up closely.

Yet there are differences too, since Theorem~\ref{thrm.r=wr+sc} uses strong compatibility, and Theorem~\ref{thrm.r=wr+uc} uses being unconflicted.
What is the difference here, and why does it arise?

One answer is given by Proposition~\ref{prop.conflicted.sober}, which illustrates that the condition of `unconflicted' (which comes from semitopologies) does not sit comfortably with the `pointless' semiframe definitions. 
This raises the question of how strong compatibility (which comes from semiframes) translates into the context of semitopologies; and how this relates to being (un)conflicted?

We look into this now; see Remark~\ref{rmrk.summary.of.sc} for a summary.
\end{rmrk}

\begin{lemm}
\label{lemm.what.sc.point.means}
Suppose $(\ns P,\opens)$ is a semitopology and $p\in\ns P$.
Then the following are equivalent:
\begin{enumerate*}
\item\label{item.what.sc.point.means.1}
$p$ is hypertransitive (Definition~\ref{defn.sc}).
\item\label{item.what.sc.point.means.2}
$\nbhd(p)$ is strongly compatible.
\item\label{item.what.sc.point.means.3}
$\nbhd(p)^\ast$ is compatible.
\item\label{item.what.sc.point.means.4}
For every $O',O''\in\opens$, if $O'\ast \nbhd(p)\ast O''$ then $O'\between O''$.
\end{enumerate*}
(Above, $O'\ast\nbhd(p)$ follows Notation~\ref{nttn.X.ast.Y}(\ref{item.x.ast.Y}) and means that $O'\between O$ for every $p\in O\in\opens$, and similarly for $\nbhd(p)\ast O''$.)
\end{lemm}
\begin{proof}
Equivalence of parts~\ref{item.what.sc.point.means.1} and~\ref{item.what.sc.point.means.2} is just Lemma~\ref{lemm.ht.sc.eq}.
Equivalence of parts~\ref{item.what.sc.point.means.2} and~\ref{item.what.sc.point.means.3} is Definition~\ref{defn.F.strongly.compatible}(\ref{item.strongly.compatible.filter}).
For the equivalence of parts~\ref{item.what.sc.point.means.3} and~\ref{item.what.sc.point.means.4}, we just unpack what it means for $\nbhd(p)^\ast$ to be compatible (see Remark~\ref{rmrk.what.does.strongly.compatible.mean}).
\end{proof}

Lemma~\ref{lemm.sc.point.iff.soberified} shows that the situation outlined in Proposition~\ref{prop.conflicted.sober}(\ref{item.conflicted.sober.2}) cannot arise if we work with a strongly compatible point instead of an unconflicted one \dots
\begin{lemm}
\label{lemm.sc.point.iff.soberified}
Suppose $(\ns P,\opens)$ is a semitopology and $p,p'\in\ns P$.
Then the following are equivalent:
\begin{enumerate*}
\item
$p$ is hypertransitive in $(\ns P,\opens)$.
\item
$\nbhd(p)$ is hypertransitive in $\tf{Soberify}(\ns P,\opens)$ (Notation~\ref{nttn.soberify}).
\end{enumerate*}
\end{lemm}
\begin{proof}
Note that from Lemma~\ref{lemm.what.sc.point.means}, $p$ is hypertransitive in $(\ns P,\opens)$ when 
$$
(\Forall{O{\in}\opens} p\in O \limp O'\between O\between O'') \quad\text{implies}\quad  O'\between O''
$$
for every $O',O''\in\opens$. 
Also, from Definition~\ref{defn.st.g}(\ref{item.st.op}) and Lemma~\ref{lemm.what.sc.point.means}, $\nbhd(p)$ is hypertransitive in $\tf{Soberify}(\ns P,\opens)$ when 
\begin{multline*}
(\Forall{O{\in}\opens} \nbhd(p)\in\f{Op}(O) \limp \f{Op}(O')\between \f{Op}(O)\between \f{Op}(O''))
\\
\text{implies}\quad
\f{Op}(O')\between \f{Op}(O'') 
\end{multline*}
for every $\f{Op}(O'),\f{Op}(O'')\oldin\tf{Opens}(\tf{Soberify}(\ns P,\opens))$.

Now by Proposition~\ref{prop.nbhd.iff}(\ref{item.nbhd.iff}), $\nbhd(p)\in\f{Op}(O)$ if and only if $p\in O$, and 
by Corollary~\ref{corr.op.sub.between} $\f{Op}(O')\between\f{Op}(O)$ if and only if $O'\between O$, and $\f{Op}(O)\between\f{Op}(O'')$ if and only if $O\between O''$.
The result follows.
\end{proof}

\noindent\dots but, the situation outlined in Proposition~\ref{prop.conflicted.sober}(\ref{item.conflicted.sober.1}) \emph{can} arise, indeed we use the same counterexample:
\begin{lemm}
\label{lemm.sc.point.still.not.quite.right}
It may be that every point in $(\ns P,\opens)$ is hypertransitive, yet $\tf{Soberify}(\ns P,\opens)$ contains a point that is not hypertransitive.
\end{lemm}
\begin{proof}
The same counterexample as used in Proposition~\ref{prop.conflicted.sober}(\ref{item.conflicted.sober.1}) illustrates a space $(\ns P,\opens)$ such that every point in $(\ns P,\opens)$ is hypertransitive, but $\tf{Soberify}(\ns P,\opens)$ contains a point that is not hypertransitive.
We note that $\bullet_A$ (the extra point in-between $3$ and $0$) is not hypertransitive, because both $B$ and $D$ intersect with every open neighbourhood of $\bullet_A$, but $B$ does not intersect with $D$.
\end{proof}

The development above suggests that we define:
\begin{defn}
\label{defn.strongly.compatible.semitopology}
Call a semitopology $(\ns P,\opens)$ \deffont[strongly compatible semitopology]{strongly compatible} when $(\opens,\subseteq,\between)$ is strongly compatible in the sense of Definition~\ref{defn.F.strongly.compatible}(\ref{item.strongly.compatible.filter.space}).
\end{defn}

The proof of Proposition~\ref{prop.sc.semitop.robust} is then very easy:
\begin{prop}
\label{prop.sc.semitop.robust}
Suppose $(\ns P,\opens)$ is a semitopology.
Then the following are equivalent:
\begin{enumerate*}
\item\label{item.sc.semitop.robust.1}
$(\ns P,\opens)$ is strongly compatible in the sense of Definition~\ref{defn.strongly.compatible.semitopology}.
\item\label{item.sc.semitop.robust.2}
$\tf{Soberify}(\ns P,\opens)$ is strongly compatible in the sense of Definition~\ref{defn.strongly.compatible.semitopology}.
\item\label{item.sc.semitop.robust.3}
$\tf{Soberify}(\ns P,\opens)$ is strongly compatible in the sense of Definition~\ref{defn.F.strongly.compatible}(\ref{item.strongly.compatible.filter.space}).
\item\label{item.sc.semitop.robust.4}
$\tf{Soberify}(\ns P,\opens)$ is hypertransitive in the sense of Definition~\ref{defn.sc}.
\end{enumerate*}
\end{prop}
\begin{proof}
We unpack Definition~\ref{defn.strongly.compatible.semitopology} and note that strong compatibility of $(\ns P,\opens)$ is expressed purely as a property of its semiframe of open sets $(\opens,\subseteq,\between)$.
By Theorem~\ref{thrm.nbhd.morphism}(\ref{item.nbhd.morphism.is.iso}), the semiframe of open sets of $\tf{Soberify}(\ns P,\opens)$ is isomorphic to $(\opens,\subseteq,\between)$, via $\nbhd^\mone$.
Equivalence of parts~\ref{item.sc.semitop.robust.1} and~\ref{item.sc.semitop.robust.2} follows.

By Notation~\ref{nttn.soberify} and Remark~\ref{rmrk.reading.nbhd.morphism}(\ref{item.describe.StFr}), the points of $\tf{Soberify}(\ns P,\opens)$ are just abstract points of $(\opens,\subseteq,\between)$.
Equivalence of parts~\ref{item.sc.semitop.robust.2} and~\ref{item.sc.semitop.robust.3} follows. 

Equivalence of part~\ref{item.sc.semitop.robust.4} with the other parts follows using Lemmas~\ref{lemm.what.sc.point.means} and~\ref{lemm.sc.point.iff.soberified}.
\end{proof}

Recall from Definition~\ref{defn.tn}(\ref{item.regular.S}) that $(\ns P,\opens)$ being (weakly) regular means that every point in $(\ns P,\opens)$ is (weakly) regular.
Recall from Definition~\ref{defn.strongly.compatible.semitopology} that $(\ns P,\opens)$ being strongly compatible means that $(\opens,\subseteq,\between)=\tf{Fr}(\ns P,\opens)$ is strongly compatible in the sense of Definition~\ref{defn.F.strongly.compatible}(\ref{item.strongly.compatible.filter.space}).
We can now prove an analogue of Theorems~\ref{thrm.r=wr+uc} and~\ref{thrm.r=wr+sc}:
\begin{thrm}
\label{thrm.r=wr+sc.st}
Suppose $(\ns P,\opens)$ is a semitopology and $p\in\ns P$.
Then the following are equivalent:
\begin{enumerate*}
\item
$(\ns P,\opens)$ is regular.
\item
$(\ns P,\opens)$ is weakly regular and strongly compatible.
\end{enumerate*}
\end{thrm}
\begin{proof}
Suppose $(\ns P,\opens)$ is regular, meaning that every $p\in\ns P$ is regular.

By Theorems~\ref{thrm.r=wr+uc} and~\ref{thrm.regular=qr+sc} every $p\in\ns P$ is weakly regular and hypertransitive.
(So by Lemma~\ref{lemm.what.sc.point.means} every $\nbhd(p)$ is strongly compatible, and by Lemma~\ref{lemm.sc.point.iff.soberified} also hypertransitive.)
The definition of weak regularity for a space in Definition~\ref{defn.tn}(\ref{item.regular.S}) is pointwise, so it follows immediately that $(\ns P,\opens)$ is weakly regular.  

But, the definition of strong compatibility for a space in Definition~\ref{defn.strongly.compatible.semitopology} is on its semiframe of open sets, which may include abstract points not only of the form $\nbhd(p)$.  
It therefore does not follow immediately that $(\ns P,\opens)$ is strongly compatible; Lemma~\ref{lemm.sc.point.still.not.quite.right} contains a counterexample.

We can still prove that $(\ns P,\opens)$ is strongly compatible --- but we need to do a bit more work.

Unpacking Definition~\ref{defn.strongly.compatible.semitopology}, we must show that $(\opens,\subseteq,\between)$ is strongly compatible.
Unpacking Definition~\ref{defn.F.strongly.compatible}(\ref{item.strongly.compatible.filter.space}), we must show that every abstract point in $(\opens,\subseteq,\between)$ is strongly compatible.

So consider an abstract point $\apoint\subseteq\opens$.
By Corollary~\ref{corr.topen.partition.char} $\ns P$ has a topen partition $\mathcal T$, which means that: every $\atopen\in\mathcal T$ is topen; the elements of $\mathcal T$ are disjoint; and $\bigcup\mathcal T=\ns P$.

Now $\bigcup\mathcal T=\ns P\in\atopen$ by Definition~\ref{defn.point}(\ref{item.abstract.point}) and Lemma~\ref{lemm.P.top}(\ref{item.P.yes.top}), so by Definition~\ref{defn.point}(\ref{item.completely.prime}) there exists at least one (and in fact precisely one) $\atopen\in\mathcal T$ such that $\atopen\in\apoint$.
Now $\atopen$ is a transitive element in $\opens$, so by Proposition~\ref{prop.x.transitive.afilter.ast} $\apoint\subseteq\opens$ is strongly compatible as required.
\end{proof}

\begin{rmrk}
\label{rmrk.summary.of.sc}
We summarise what we have seen:
\begin{enumerate*}
\item
The notions of (quasi/weak)regularity match up particularly nicely between a semitopology and its soberification as a semiframe (Proposition~\ref{prop.regular.match.up}).
\item
We saw in Proposition~\ref{prop.conflicted.sober} that the notions of (un)conflicted point and unconflicted space from Definition~\ref{defn.conflicted}(\ref{item.unconflicted}) are not robust under forming soberification (Notation~\ref{nttn.soberify}).
From the point of view of a pointless methodology in semitopologies --- in which we seek to understand a semitopology $(\ns P,\opens)$ starting from its semiframe structure $(\opens,\subseteq,\between)$ --- this is a defect. 
\item
A pointwise notion of strong compatibility exists; by Lemma~\ref{lemm.what.sc.point.means} it is actually hypertransitivity from Definition~\ref{defn.sc}.  
This is preserved pointwise by soberification (Lemma~\ref{lemm.sc.point.iff.soberified}),
but soberification can still introduce \emph{extra} points, and it turns out that the property of a space being pointwise hypertransitive is still not robust under soberification because the extra points need not necessarily be hypertransitive; see Lemma~\ref{lemm.sc.point.still.not.quite.right}. 
\item\label{item.r=wr+sc.natural}
This motivates the notion of a \emph{strongly compatible} semitopology from Definition~\ref{defn.strongly.compatible.semitopology}; and then Proposition~\ref{prop.sc.semitop.robust} becomes easy.

Our larger point (no pun intended) is that the Definition and its corresponding Proposition are natural, \emph{and also} that the other design decisions are \emph{less} natural, as noted above.

Perhaps somewhat unexpectedly, `regular = weakly regular + strongly compatible' then works pointwise \emph{and} for the entire space; see Theorem~\ref{thrm.r=wr+sc.st}.
Thus Definition~\ref{defn.strongly.compatible.semitopology} has good properties and is natural from a pointless/semiframe/open sets perspective.
\end{enumerate*}
\end{rmrk}

\jamiesection{Graph representation of semitopologies}
\label{sect.graphs}

A substantial body of literature exists studying social networks as connected graphs.
A semitopology has the flavour of a social network, in the sense that it models voting and consensus on a distributed system.
It is therefore interesting to consider representations of semitopologies as graphs.
We will consider two ways to do this:
\begin{enumerate*}
\item
We can map a semitopology to the intersection graph of its open sets.
We discuss it in Subsection~\ref{subsect.semitopology.to.graph}.
This works well but loses information (Remark~\ref{rmrk.intersecting.topens}).
\item
We can use a \emph{subintersection} relation between sets.
We discuss this in Subsection~\ref{subsect.semiframe.to.graph}.
\end{enumerate*}

\jamiesubsection{From a semitopology to its intersection graph}
\label{subsect.semitopology.to.graph}

We start with a very simple representation of $(\ns P,\opens)$ obtained just as the \emph{intersection graph} of $\opens$ (Definition~\ref{defn.G}).
This is not necessarily the most detailed representation (we spell out why, with examples, in Remark~\ref{rmrk.intersecting.topens}), but it is still simple, direct, and nontrivial: 

\jamiesubsubsection{The basic definition}
\label{subsect.intersection.graph}

\begin{defn}
\label{defn.G}
Suppose $(\ns P,\opens)$ is a semitopology.
Define its \deffont{intersection graph $\tf{IntGr}(\ns P,\opens)$} by:
\begin{itemize*}
\item
The nodes of $\tf{IntGr}(\ns P,\opens)$ are nonempty open sets $O\in\opens_{\neq\varnothing}$.
\item
There is an edge $O\cw O'$ between $O$ and $O'$ when $O\between O'$.
\end{itemize*}
\end{defn}

\begin{rmrk}
\leavevmode
\begin{enumerate*}
\item
The notion of the \emph{intersection graph} of a set of sets is standard~\cite{wiki:Intersection_graph}.
The notion used in Definition~\ref{defn.G} is slightly different, in that we exclude the empty set.
This technical tweak is mildly useful, to give us Lemma~\ref{lemm.cw.reflexive}.
\item
If $(\ns P,\opens)$ is a semitopology and $\tf{IntrGr}(\ns P,\opens)$ is its intersection graph in the sense of Definition~\ref{defn.G}, then $O\cw O'$ is a synonym for $O\between O'$.
But, writing $O\cw O'$ suggests that we view $O$ and $O'$ as nodes.
\end{enumerate*}
\end{rmrk}

\begin{nttn}
For the rest of this Section we assume a fixed but arbitrary 
$$
G=\tf{IntGr}(\ns P,\opens)
$$ 
that is the open intersection graph of a semitopology $(\ns P,\opens)$. 
\end{nttn}

We start with an easy lemma:
\begin{lemm}
\label{lemm.cw.reflexive}
$O\cw O$ always (the graph $G$ is reflexive).
\end{lemm}
\begin{proof}
From Definition~\ref{defn.G}, noting that nodes are \emph{nonempty} open sets $O\in\opens$, and it is a fact of sets intersection that $O\between O$ when $O$ is nonempty.
\end{proof}

\jamiesubsubsection{The node preorder $\leq$}

\begin{defn}
\label{defn.Oleq}
Write $O\leq O'$\index{$O\leq O'$ (node preorder on a graph)}\index{node preorder $\leq$ (on a graph)} when for every $O''$, 
if $O\cw O''$ then $O'\cw O''$.
In symbols:
$$
O\leq O'
\quad\text{when}\quad
\Forall{O''}O\cw O'' \limp O'\cw O'' .
$$
\end{defn}

We note that $\leq$ from Definition~\ref{defn.Oleq} is a preorder (reflexive and transitive relation):
\begin{lemm}
\label{lemm.oleq.preorder}
\leavevmode
\begin{enumerate*}
\item\label{item.oleq.cw.refl}
$\leq$ is reflexive: $O\leq O$.
\item\label{item.oleq.cw.trans}
$\leq$ is transitive: if $O\leq O'\leq O''$ then $O\leq O''$.
\end{enumerate*}
\end{lemm}
\begin{proof}
By routine calculations.
\end{proof}

\begin{figure}
\vspace{-1em}
\centering
\includegraphics[width=0.4\columnwidth,trim={50 100 50 100},clip]{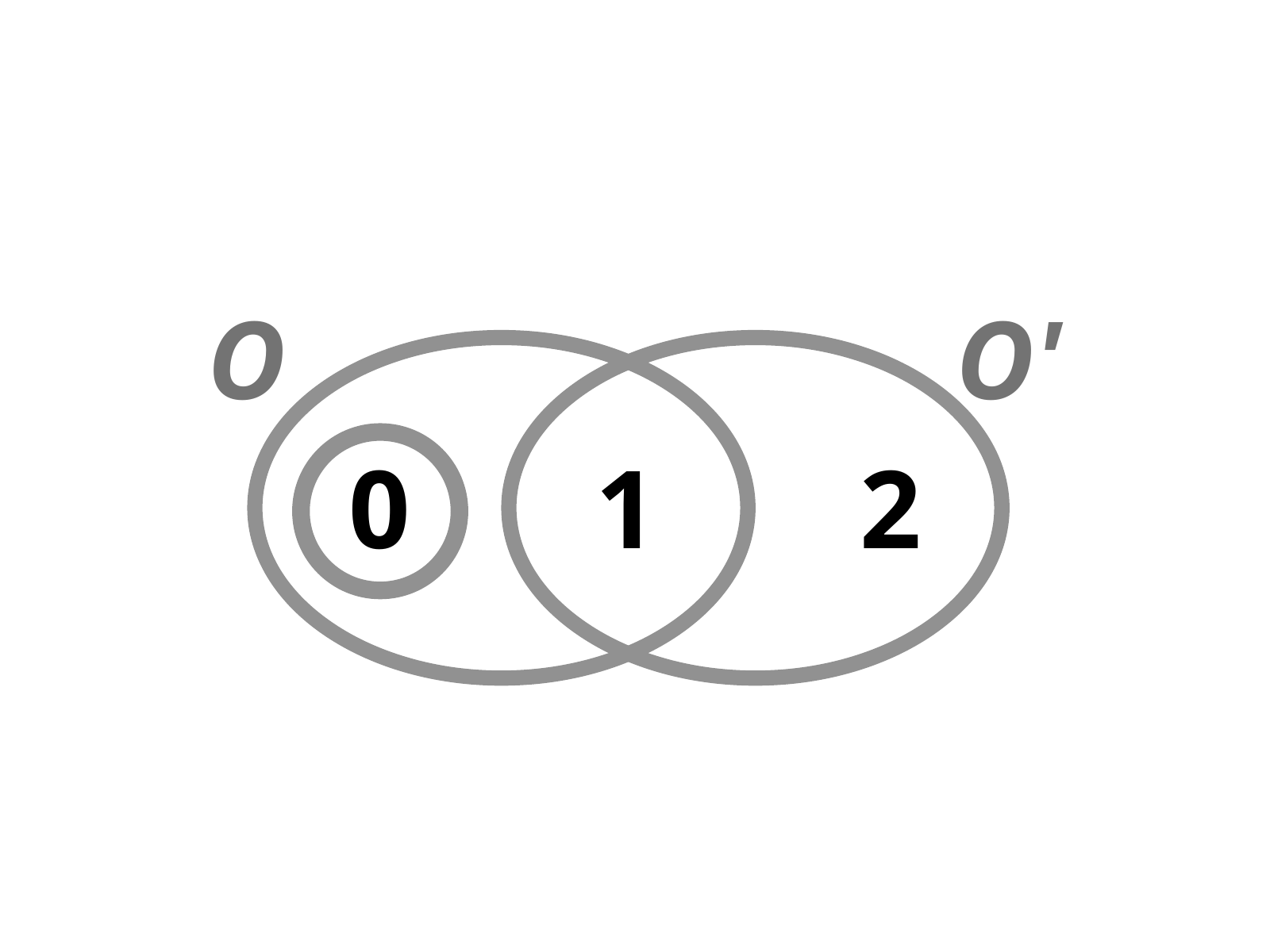}
\vspace{-1em}
\caption{$O\cw O'$ but $O\not\leq O'$ (Lemma~\ref{lemm.oleq.to.cw}(\ref{item.oleq.to.cw.2}))}
\label{fig.leq.counterexample}
\end{figure}

\begin{lemm}
\label{lemm.oleq.to.cw}
\leavevmode
\begin{enumerate*}
\item\label{item.oleq.to.cw.1}
If $O\leq O'$ then $O\cw O'$. 
\item\label{item.oleq.to.cw.2}
It is not in general the case that $O\cw O'$ implies $O\leq O'$ (cf. Proposition~\ref{prop.transitive.leq}(\ref{item.transitive.leq.1}\&\ref{item.transitive.leq.2})).
\end{enumerate*}
In symbols we can write: $\leq \subseteq \cw$ and $\cw\not\subseteq\leq$ in general.\footnote{It gets better: see Lemma~\ref{lemm.subseteq.leq}.}
\end{lemm}
\begin{proof}
We consider each part in turn:
\begin{enumerate}
\item
Suppose $O\leq O'$.
By Lemma~\ref{lemm.cw.reflexive} $O\cw O$, and it follows (since $O\leq O'$) that $O'\cw O$ as required.
\item
It suffices to give a counterexample.
Consider the semitopology $(\ns P,\opens)$ where $\ns P=\{0,1,2\}$ and $\opens$ is generated by $O=\{0,1\}$, $O'=\{1,2\}$, and $O''=\{0\}$, as illustrated in Figure~\ref{fig.leq.counterexample}.
Then $O\cw O'$ but $O\not\leq O'$ since $O\cw \{0\}$ but $O'\not\cw\{0\}$.
\qedhere\end{enumerate}
\end{proof}

\begin{rmrk}
Suppose $O\leq O'$, so that by Lemma~\ref{lemm.oleq.to.cw} also $O\cw O'$.
We can illustrate Definition~\ref{defn.Oleq} in the style of a categorical diagram --- 
$$
\begin{tikzcd}[column sep=30pt,row sep=20pt]
O 
\\
O'\arrow[r,dhdashedarrow,"\exists",swap]\arrow[u, dharrow,"\leq"] & O'' \arrow[ul, dharrow] 
\end{tikzcd}
$$
--- expressing that $O\leq O'$ holds when every arrow out of $O$ factorises through $O'$. 
\end{rmrk} 

\begin{lemm}[$\leq$ generalises $\subseteq$]
\label{lemm.subseteq.leq}
We have:
\begin{enumerate*}
\item\label{item.subseteq.leq.1}
If $O\subseteq O'$ then $O\leq O'$.
\item\label{item.subseteq.leq.2}
The converse implication need not hold: $O\leq O'$ does not necessarily imply $O\subseteq O'$.\footnote{So to sum up this and Lemma~\ref{lemm.oleq.to.cw}: $\subseteq\subseteq\leq\subseteq\cw$, and the inclusion may be strict: $\subseteq\subsetneq\leq$ in general.}
\end{enumerate*}
\end{lemm}
\begin{proof}
\leavevmode
\begin{enumerate}
\item
A fact of sets.
\item
It suffices to give a counterexample.
Set $\ns P=\{0,1\}$ and $\opens=\{\varnothing, \{0\}, \{0,1\}\}$.
This generates a very simple graph $G$ as follows:
$$
\begin{tikzcd}[column sep=20pt,row sep=20pt]
\{0\}\arrow[r, dharrow] & \{0,1\} 
\end{tikzcd}
$$
The reader can check that $\{0,1\}\leq \{0\}$, but $\{0,1\}\not\subseteq\{0\}$. 
\qedhere\end{enumerate}
\end{proof}

\jamiesubsubsection{Transitive elements}

\begin{defn}
\label{defn.transitive.node}
Call $T\in G$ \deffont[transitive node in a graph]{transitive} when for every $O,O'\in G$ we have that
$$
O\cw T\cw O'
\quad\text{implies}\quad
O\cw O' .
$$ 
In pictures:
$$
\begin{tikzcd}[column sep=28pt,row sep=30pt]
O \arrow[r, dharrow]\arrow[rr, dhdashedarrow, bend left = 45,"\exists"] 
&
T \arrow[r, dharrow]
&
O'
\end{tikzcd}
$$
\end{defn}

\begin{lemm}
Suppose $T\in G$ is a node.
Then the following are equivalent:
\begin{itemize*}
\item
$\atopen$ is transitive in the sense of Definition~\ref{defn.transitive}(\ref{transitive.transitive}).\footnote{Equivalently, $\atopen$ is topen (Definition~\ref{defn.transitive}(\ref{transitive.cc})), since every node in $G$ is a nonempty open.}
\item
$\atopen$ is transitive in the sense of Definition~\ref{defn.transitive.node}.
\end{itemize*} 
\end{lemm}
\begin{proof}
Just unfolding Definitions~\ref{defn.G} and~\ref{defn.transitive}.
\end{proof}

Lemmas~\ref{lemm.oleq.to.cw} and~\ref{lemm.subseteq.leq} suggest that $\leq$ generalises subset inclusion $\leq$. 
In this light, Proposition~\ref{prop.transitive.leq} makes transitive elements look like singleton sets:
\begin{prop}
\label{prop.transitive.leq}
The following conditions are equivalent:
\begin{enumerate*}
\item\label{item.transitive.leq.1}
$\atopen$ is transitive.
\item\label{item.transitive.leq.2}
$\Forall{O}(T\cw O \limp T\leq O)$.
\item\label{item.transitive.leq.3}
$\Forall{O}(T\cw O \liff T\leq O)$.
\end{enumerate*}
\end{prop}
\begin{proof}
For the top equivalence, we prove two implications:
\begin{itemize}
\item
\emph{The top-down implication.}

Suppose $\atopen$ is transitive and suppose $T\cw O$.
To prove $T\leq O'$ it suffices to consider any $O'$ and show that $T\cw O'$ implies $O\cw O'$.

But this is just from transitivity and the fact that $\cw$ is symmetric: $O\cw T\cw O'$ implies $O\cw O'$.
\item
\emph{The bottom-up implication.}

Suppose for every $O$, if $T\cw O$ then $T\leq O$, and suppose $O\cw T\cw O'$.
Because $T\leq O$ and $T\cw O'$, we have $O\cw O'$ as required. 
\end{itemize}
The lower equivalence then follows from Lemma~\ref{lemm.oleq.to.cw}.
\end{proof}

\begin{corr}
\label{corr.trans.leq}
\leavevmode
\begin{enumerate*}
\item
If $\atopen$ is transitive then $\atopen$ is $\leq$-minimal.
That is: 
$$
O\leq T
\quad\text{implies}\quad
T\leq O.
$$
\item
It is possible for $\atopen$ to be $\leq$-least (and thus $\leq$-minimal) and not transitive.
\end{enumerate*}
\end{corr}
\begin{proof}
\leavevmode
\begin{enumerate}
\item
Suppose $\atopen$ is transitive and $O\leq T$.
By Lemma~\ref{lemm.oleq.to.cw}(\ref{item.oleq.to.cw.1}) $O\cw T$ and by Proposition~\ref{prop.transitive.leq}(\ref{item.transitive.leq.2}) $T\leq O$.
\item
It suffices to provide a counterexample.
Consider the semitopology illustrated in Figure~\ref{fig.square.diagram}.
It is a fact that $A$ is not transitive, 
yet $A$ is $\leq$-least: $A\not\leq B$ (because $B\cw D$ yet $A\not\cw D$) and similarly $A\not\leq C$, and $A\not\leq D$ (because $A\not\cw D$).
\qedhere\end{enumerate}
\end{proof}

\begin{defn}
\label{defn.extensionally.equivalent}
Suppose $O,O'\in G$.
Define $O\approx O'$ when $O\leq O'\land O'\leq O$, and in this case call $O$ and $O'$ \deffont[extensionally equivalent (nodes in a graph:\ $O\approx O'$)]{extensionally equivalent}. 
It is easy to see from Definition~\ref{defn.Oleq} that
$$
O\approx O' \liff \Forall{O''}(O\cw O'' \liff O'\cw O'' ) .
$$
\end{defn}

\begin{corr}
\label{corr.ext.equal}
\leavevmode
\begin{enumerate*}
\item\label{item.ext.equal.1}
If $\atopen$ and $\atopen'$ are transitive (Definition~\ref{defn.transitive.node}) then the following are equivalent:
$$
T\leq T'
\quad\liff\quad
T'\leq T
\quad\liff\quad
T\cw T' 
\quad\liff\quad
T\approx T' .
$$
\item\label{item.ext.equal.2}
As a corollary, if $\atopen$ and $\atopen'$ are transitive then $T\cw T'$ if and only if $T\approx T'$.
\end{enumerate*}
\end{corr}
\begin{proof}
The left-hand equivalence is from Corollary~\ref{corr.trans.leq} (since $\atopen$ and $\atopen'$ are transitive). 
The middle equivalence is from Proposition~\ref{prop.transitive.leq}.
The right-hand equivalent follows from the left-hand and middle equivalences using Definitions~\ref{defn.Oleq} and~\ref{defn.extensionally.equivalent}.

The corollary just repeats the right-hand equivalence.
\end{proof}

\begin{figure}
\vspace{-1em}
\centering
\includegraphics[align=c,width=0.32\columnwidth,trim={100 100 100 100},clip]{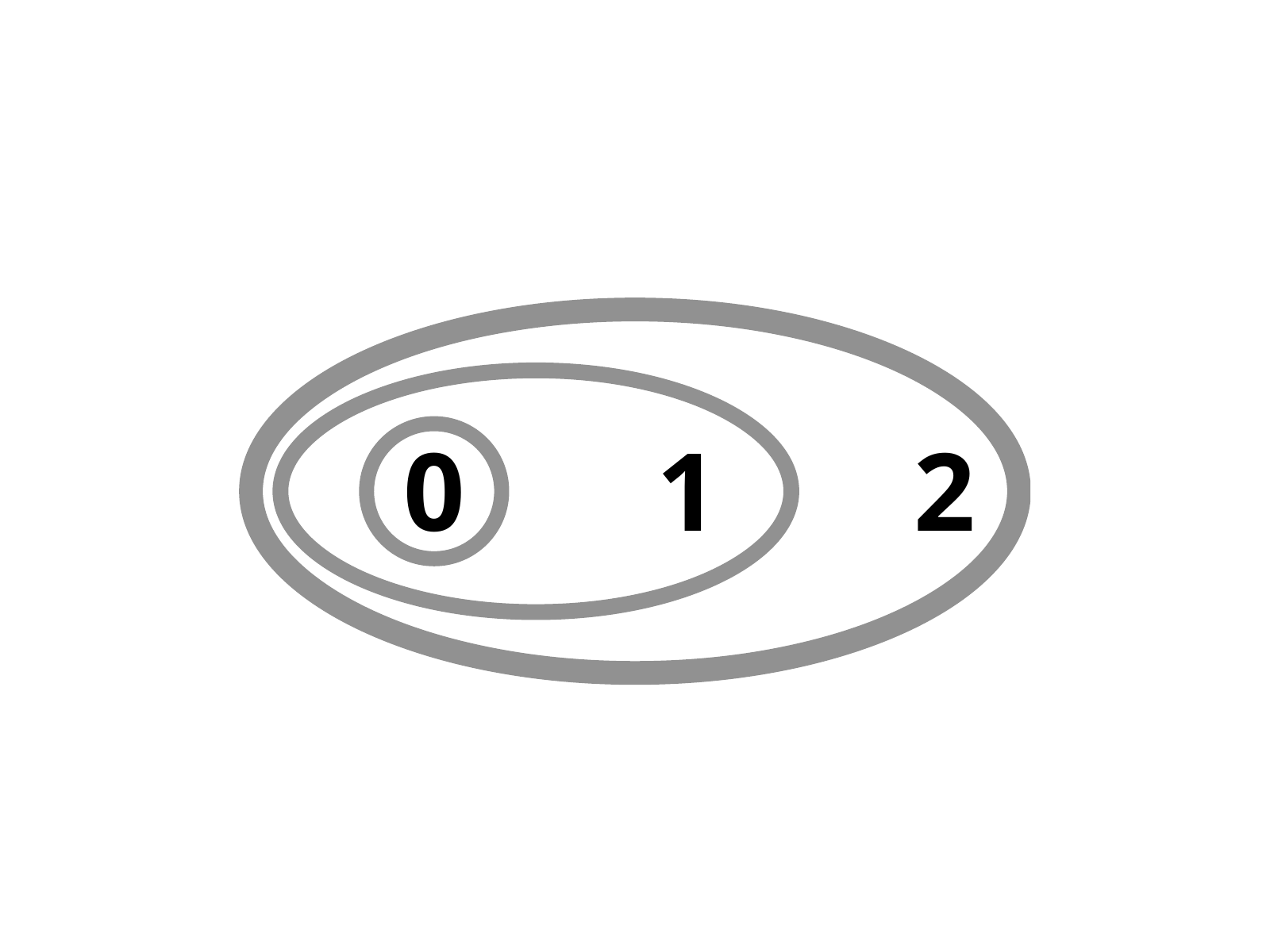}
\includegraphics[align=c,width=0.32\columnwidth,trim={100 100 100 100},clip]{diagrams/two-min.pdf}
\includegraphics[align=c,width=0.33\columnwidth,trim={100 100 100 100},clip]{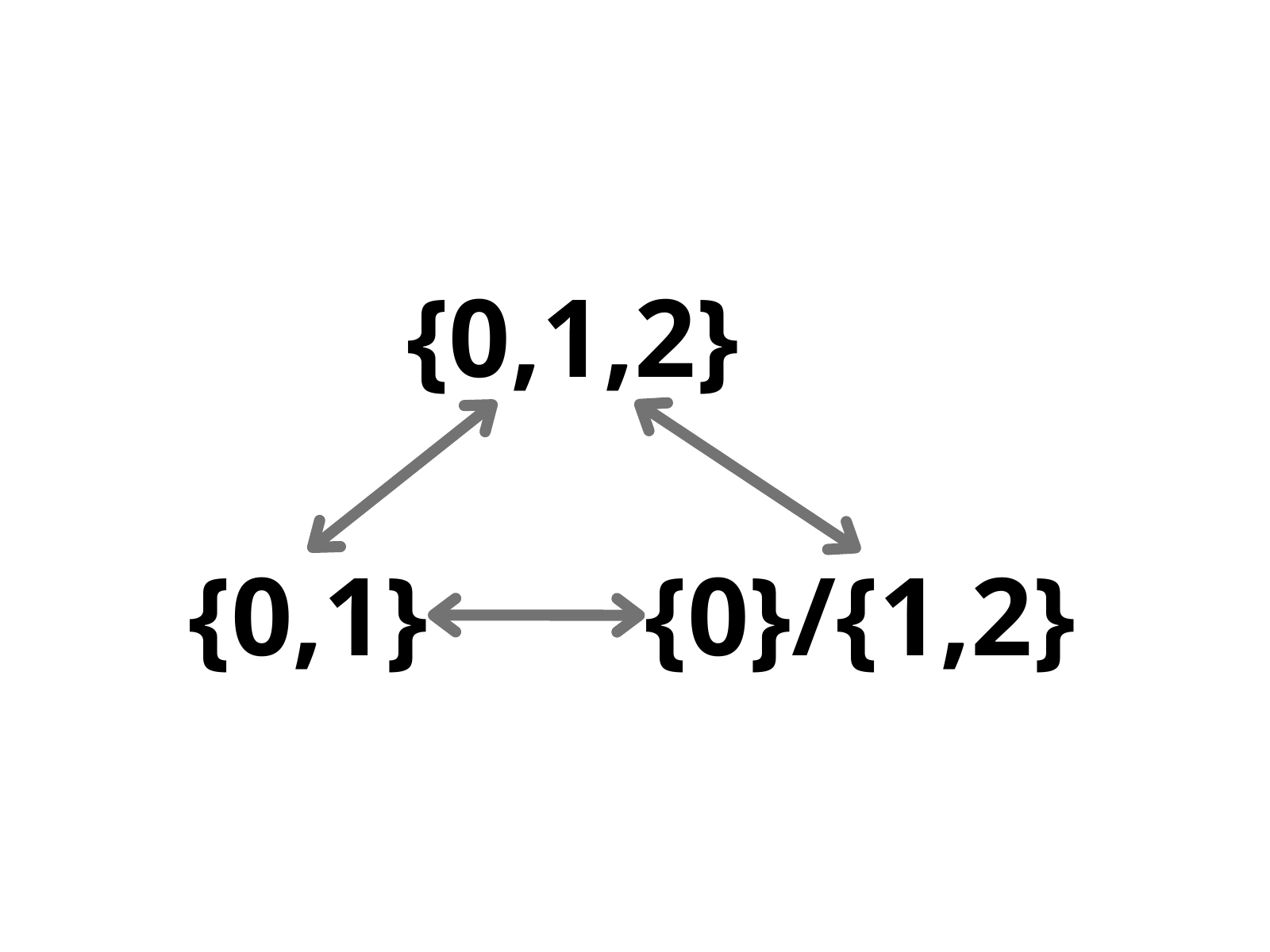}
\vspace{-1em}
\caption{Semitopologies with isomorphic intersection graphs (Remark~\ref{rmrk.intersecting.topens})}
\label{fig.three-growing}
\end{figure}

\begin{rmrk}[Intersection graph loses information]
\label{rmrk.intersecting.topens}
The proof of Corollary~\ref{corr.ext.equal}(\ref{item.ext.equal.2}) is not hard but it tells us something useful: the intersection graph identifies intersecting topens, and thus identifies a topen with the (by Corollary~\ref{corr.max.cc}) unique maximal topen that contains it.

Consider a semitopology $(\ns P,\opens)$ and its intersection graph $\tf{IntGr}(\ns P,\opens)$, and consider some regular point $p\in\community(p)$.
Recall from Theorem~\ref{thrm.max.cc.char} that $\community(p)$ is the greatest topen (transitive open) neighbourhood of $p$. 
Putting Corollary~\ref{corr.ext.equal}(\ref{item.ext.equal.2}), Theorem~\ref{thrm.max.cc.char}, and Lemma~\ref{lemm.transitive.transitive} together, we have that $\community(p)$ --- when considered as a node in the intersection graph of open sets --- is extensionally equivalent to each of its topen subsets, and also (equivalently) to any topen set that it intersects with.

So, if we were to build a functor from intersection graphs back to semitopologies, by forming a notion of abstract point and mapping a node to the set of abstract points that contain it, then Corollary~\ref{corr.ext.equal} tells us that this will map all connected transitive nodes down to a single point.
Thus, our intersection graph representation from Definition~\ref{defn.G} \emph{loses information}.

It is easy to generate examples of this kind of information loss.
The following clearly different semitopologies give rise to isomorphic intersection graphs, namely: the full graph on three points, representing three pairwise intersecting nonempty open sets.
\begin{enumerate*}
\item
$\ns P=\{0,1,2\}$ and $\opens=\bigl\{\varnothing,\ \{0\},\ \{0,1\},\ \{0,1,2\} \bigr\}$.
\item
$\ns P'=\{0,1,2\}$ and $\opens'=\bigl\{\varnothing,\ \{0,1\},\ \{1,2\},\ \{0,1,2\} \bigr\}$.
\end{enumerate*}
See Figure~\ref{fig.three-growing}; the intersection graph isomorphism is illustrated on the right (where we equate $\{0\}$ with $\{1,2\}$).
The left-hand and middle examples in the figure are of intersecting topens, consistent with Corollary~\ref{corr.ext.equal}(\ref{item.ext.equal.2}).

Whether this behaviour is a feature or a bug depends on what we want to accomplish --- but for our purposes of modelling networks, we may care to also consider a representation that retains more information. 
In the next Subsection we consider a slightly more elaborate graph representation, which is more discriminating.
\end{rmrk}

\jamiesubsection{From a semiframe to its subintersection graph}
\label{subsect.semiframe.to.graph}

\begin{rmrk}
In Remark~\ref{rmrk.intersecting.topens} we gave a natural representation of a semitopology as its intersection graph.
We noted in Corollary~\ref{corr.ext.equal} that this identifies open sets up to a notion of extensional equivalence $\approx$ given in Definition~\ref{defn.extensionally.equivalent}, and because topen sets are extensionally equivalent if and only if they intersect by Corollary~\ref{corr.ext.equal}, the intersection graph representation of semitopologies identifies two topens precisely when they intersect.

This is not wrong --- intersection topen sets \emph{are} extensionally equivalent, after all --- but suppose we want to retain a bit more detail.
How can we proceed?
\end{rmrk}

\jamiesubsubsection{The subintersection relation $\flanks$}

\begin{rmrk}
Notice that the notion of semiframe $(\ns X,\cti,\ast)$ from Definition~\ref{defn.semiframe} is based on \emph{two} structures on $\ns X$: 
\begin{itemize*}
\item
a semilattice relation $\cti$, and 
\item
a compatibility relation $\ast$.
\end{itemize*}
Correspondingly, our notion of semitopology observes \emph{two} properties of open sets: whether $O$ is a subset of $O'$, and whether $O$ intersects $O'$.

We can ask whether these two structures can be obtained from a single relation. 
The answer is yes (if we are also allowed to observe equality): we can combine $\cti$ and $\ast$ into a single relation and so obtain a graph structure, without the loss of information we noted of intersection graphs.
The definition is as follows:
\end{rmrk}

\begin{defn}
\label{defn.ct}
\leavevmode
\begin{enumerate}
\item
Suppose $\ns P$ is a set and $X\subseteq\ns P$.\footnote{We will be most interested in the case that $P$ is the set of points of a semitopology, but the definition does not depend on this.}
Define $X^c$ the \deffont[complement of $X$ ($X^c$)]{complement of $X$} by $X^c=\ns P\setminus X$.
\item\label{item.ct.flanks.set}
Suppose $\ns P$ is a set and $X,Y\subseteq\ns P$.
Define a relation $X\flanks Y$, read `$X$ \deffont[proper subintersection (in sets:\ $X{\flanks} Y$)]{properly subintersects} $Y$', as follows: 
$$
X\flanks Y
\quad\text{when}\quad
X\between Y\land X^c\between Y .
$$
When $X\flanks Y\lor X=Y$, we say that $X$ \deffont[subintersection (in sets:\ $X{\flanks} Y\lor X{=}Y$)]{subintersects}\index{$X\flanks Y$ (subintersection of sets)} $Y$.
\item\label{item.ct.flanks.semiframe}
Suppose $(\ns X,\cti,\ast)$ is a semiframe and $x,y\in\ns X$.
Define a relation $x\flanks y$, and say that $x$ \deffont[proper subintersection (in semiframes:\ $x{\flanks} y$)]{properly subintersects} $y$, \index{$x\flanks y$ (subintersection of semiframe elements)} when 
$$
x\flanks y 
\quad\text{when}\quad
x\ast y \land y\not\cti x .
$$
\end{enumerate}
\end{defn}

\begin{xmpl}
Set $P=\{0,1,2\}$.
Then:
\begin{enumerate*}
\item
$\{0\}$ properly subintersects $\{0,1\}$, because $\{0\}\between\{0,1\}$ and $\{0,1,2\}\setminus\{0\}=\{1,2\}\between \{0,1\}$.
Similarly, $\{1\}$ properly subintersects $\{0,1\}$.
\item
$\{0,2\}$ properly subintersects $\{0,1\}$, because $\{0,2\}\between\{0,1\}$ and $\{0,1,2\}\setminus\{0,2\}=\{1\}\between \{0,1\}$.
\item
$\{2\}$ does not subintersect $\{0,1\}$, because $\{2\}\notbetween\{0,1\}$.
\item
$\{0,1\}$ does not subintersect $\{0\}$ or $\{1\}$, because $\{2\}\notbetween\{0\}$ and $\{2\}\notbetween\{1\}$ and $\{2\}\notbetween\{0,1\}$.
\item
$\{0,1\}$ subintersects itself, but not properly because $\{0,1,2\}\setminus\{0,1\}\notbetween\{0,1\}$.
More generally $X$ subintersects itself by definition, but it does not properly subintersect itself (think: $X\subsetneq Y$ vs. $X\subseteq Y$).
\end{enumerate*}
\end{xmpl}

\begin{rmrk}[One property, and three non-properties]
It is easy to show that $X\flanks Y$ is positive (covariant) in its second argument: if $X\flanks Y$ and $Y\subseteq Y'$ then $X\flanks Y'$.
However, $X\flanks Y$ is neither positive nor negative in its first argument, and it does not commute with intersection in its right argument.

Take $X,Y\subseteq\{0,1,2,3\}$.
Then:
\begin{enumerate*}
\item
It is not the case that $X\flanks Y \land X\subseteq X'$ implies $X'\flanks Y$.

Take $X=\{0\}$ and $Y=\{0,1\}$ and $X'=\{0,1\}$.
\item 
It is not the case that $X\flanks Y \land X'\subseteq X$ implies $X'\flanks Y$.

Take $X=\{0,1\}$ and $Y=\{1,2\}$ and $X'=\{0\}$.
\item
It is not the case that $X\flanks Y \land X\flanks Y'$ implies $X\flanks (Y\cap Y')$.

Take $X=\{0,1\}$ and $Y=\{1,2\}$ and $Y'=\{1,3\}$.
\end{enumerate*}
\end{rmrk}

\begin{lemm}
\label{lemm.two.flanks}
Suppose $\ns P$ is a set and $X,Y\subseteq\ns P$.
Then the following are equivalent:
\begin{enumerate*}
\item
$X\flanks Y$ in the sense of Definition~\ref{defn.ct}(\ref{item.ct.flanks.set}).
\item
$X\between Y\land Y\not\subseteq X$.
\end{enumerate*}
In other words: $X\flanks Y$ for $X$ and $Y$ considered as sets as per Definition~\ref{defn.ct}(\ref{item.ct.flanks.set}), precisely when $X\flanks Y$ for $X$ and $Y$ considered as elements in the semiframe $(\powerset(\ns P),\subseteq,\between)$ as per Definition~\ref{defn.ct}(\ref{item.ct.flanks.semiframe}).
\end{lemm}
\begin{proof}
Routine, using the fact of sets that $Y\between X^c$ if and only if $Y\not\subseteq X$.
\end{proof}

\begin{corr}
Suppose $(\ns X,\cti,\ast)$ is a spatial semiframe and $x,y\in\ns X$.
Then 
$$
x\flanks y
\quad\text{if and only if}\quad
\f{Op}(x)\flanks \f{Op}(y).
$$
\end{corr}
\begin{proof}
Suppose $(\ns X,\cti,\ast)$ is a spatial semiframe.
By Proposition~\ref{prop.Op.subseteq}(\ref{item.Op.spatial.ast}\&\ref{item.Op.spatial.cti}) 
$x\ast y$ if and only if $\f{Op}(x)\between\f{Op}(y)$, and $x\cti y$ if and only if $\f{Op}(x)\subseteq\f{Op}(y)$.
We use Lemma~\ref{lemm.two.flanks}.
\end{proof}

\jamiesubsubsection{Recovering $\cti$ and $\ast$ from $\flanks$}

\begin{rmrk}
We can recover $\subseteq$ and $\between$ from $\flanks$.
We can also recover $\cti$ and $\ast$. 
We consider the construction for semiframes and $\cti$ and $\ast$, because it is the more general setting; the proofs for the concrete instance of $\subseteq$ and $\between$ are identical:
\end{rmrk}

\begin{prop}
\label{prop.recover.data}
Suppose $(\ns X,\cti,\ast)$ is a semiframe and suppose $x,y\neq\tbot_{\ns X}$.
Then:
\begin{enumerate*}
\item
$x\ast y$ if and only if $x=y\lor x\flanks y\lor y\flanks x$.
\item
$x\cti y$ if and only if $x=y\lor (x\flanks y \land \neg(y\flanks x))$.
\end{enumerate*}
\end{prop}
\begin{proof}
We consider each implication in turn:
\begin{itemize}
\item
\emph{We show that $x\ast y$ implies $x=y\lor x\flanks y\lor y\flanks x$.}

Suppose $x\ast y$.
By antisymmetry of $\cti$, either $x=y$ or $y\not\cti x$ or $x\not\cti x$.
The result follows.
\item
\emph{We show that $x=y\lor x\flanks y\lor y\flanks x$ implies $x\ast y$.}

By reversing the reasoning of the previous case.
\item
\emph{We show that $x=y\lor (x\flanks y \land \neg(y\flanks x))$ implies $x\cti y$.}\quad

Suppose $x=y\lor (x\flanks y \land \neg(y\flanks x))$.

If $x=y$ then $x\cti y$ and we are done.
If $x\neq y$ then we unpack Definition~\ref{defn.ct}(\ref{item.ct.flanks.semiframe}) and simplify as follows:
\begin{multline*}
x\flanks y\land \neg(y\flanks x)
\liff
x\ast y \land y\not\cti x \land (\neg(x\ast y) \lor x\cti y)
\liff
\\
x\ast y \land y\not\cti x \land x\cti y 
\limp 
x\cti y
\end{multline*}
\item
\emph{We show that $x\cti y$ implies $x=y\lor (x\flanks y \land \neg(y\flanks x))$.}\quad

Suppose $x\cti y$.
By assumption $x\neq\tbot$, so by Lemma~\ref{lemm.compatibility.monotone}(\ref{item.ast.lower.bound}) $x\ast y$.
If $x=y$ then we are done; otherwise by antisymmetry $y\not\cti x$ and again we are done. 
\qedhere\end{itemize}
\end{proof}

\begin{rmrk}
\label{rmrk.graphs.1-1}
It follows from the above that a semiframe $(\ns X,\cti,\ast)$ can be presented as a graph $\tf{Gr}(\ns X,\cti,\ast)$ such that:
\begin{itemize*}
\item
Nodes of the graph are elements $x\in\ns X$ such that $x\neq\tbot_{\ns X}$.
\item
There is an edge $x\to x'$ when $x\flanks x'$ in the sense of Definition~\ref{defn.ct}(\ref{item.ct.flanks.semiframe}) --- that is, when $x\ast x'\land x'\not\cti x$. 
\end{itemize*}
Similarly we can present a semitopology $(\ns P,\opens)$ as a graph $\tf{Gr}(\ns P,\opens)$ such that:
\begin{itemize*}
\item
Nodes of the graph are nonempty open sets $\varnothing\neq O\in\opens$.
\item
There is an edge $O\to O'$ when $O\flanks O'$ in the sense of Definition~\ref{defn.ct}(\ref{item.ct.flanks.set}) --- that is, when $O\between O'\land O^c\between O'$. 
\end{itemize*}
These presentations are equivalent in the following sense: if we start from $(\ns P,\opens)$ and consider it as a semiframe $(\opens,\subseteq,\between)$ (which is spatial by Proposition~\ref{prop.Gr.P.spatial}) and then map to a graph, then we get the same graph as if we just map direct from $(\ns P,\opens)$ to the graph.
In symbols we can write:
$$
\tf{Gr}(\ns P,\opens) = \tf{Gr}(\opens,\subseteq,\between) .
$$
By Proposition~\ref{prop.recover.data}, the mapping from $(\ns X,\cti,\ast)$ to $\tf{Gr}(\ns X,\cti,\ast)$ loses no information; we can view the graph as just a different way of presenting the same structure.
\end{rmrk}

\begin{rmrk}
\label{rmrk.different.notions.of.morphism}
Although $(\ns X,\cti,\ast)$ and $\tf{Gr}(\ns X,\cti,\ast)$ are in one-to-one correspondence as discussed in Remark~\ref{rmrk.graphs.1-1}, the representations suggest different notions of morphism.
\begin{itemize*}
\item
For a semiframe $(\ns X,\cti,\ast)$, the natural notion of morphism is a ${\cti}/{\ast}$-preserving map, in a suitable sense as defined in Definition~\ref{defn.category.of.spatial.graphs}(\ref{item.category.spatial.morphism}): $x\cti x'$ implies $g(x)\cti g(x')$, and $g(x)\ast g(x')$ implies $x\ast x'$.
\item
For a graph $(\ns G,\flanks)$, the natural notion of morphism is some notion of $\flanks$-preserving map, and this is \emph{not} necessarily the same as a ${\cti}/{\ast}$-preserving map, because $\flanks$ uses $\not\cti$.

If we still want to preserve notions of lattice structure and semifilter, then we look at how $\cti$ is defined from $\flanks$ in Proposition~\ref{prop.recover.data}, and see that it uses both $\flanks$ and $\neg\flanks$, and so in this case we may want a notion of morphism such that $x\flanks x'$ \emph{if and only if} $g(x)\flanks g(x')$.
Looked at from the point of view of $\cti$ and $\ast$, this suggests that $x\cti x'$ if and only if $g(x)\cti g(x')$, and $x\ast x'$ if and only if $g(x)\ast g(x')$.

Investigating the design space here is future work.
See also Remark~\ref{rmrk.more.conditions}.
\end{itemize*}
\end{rmrk}

\jamiepart{Logic and computation}
\label{part.3}
\jamiesection{Three-valued logic}
\label{sect.three}

\jamiesubsection{Three-valued logic, valuations, and continuity}

\begin{figure}
\centering
\includegraphics[width=0.4\columnwidth,trim={50 150 50 150},clip]{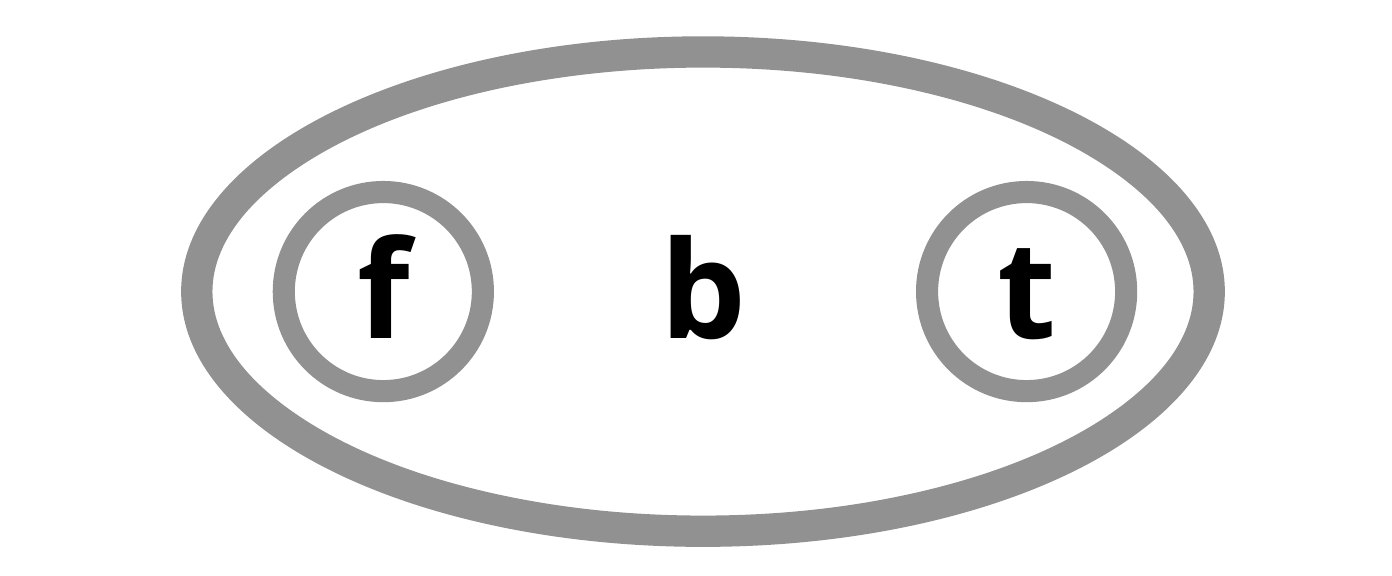}
\caption{The semitopology $\THREE$ / the set of truth values of three-valued logic}
\label{fig.THREE}
\end{figure}

\begin{rmrk}[Setting the scene]
\label{rmrk.three.scene}
Suppose $(\ns P,\opens)$ is a semitopology and suppose $p,p'\in\ns P$.
Then two foundational questions of semitopologies, as we apply them, are:
\begin{enumerate*}
\item
Determine whether $p\intertwinedwith p'$.
That is, determine whether every open neighbourhood of $p$ intersects with every open neighbourhood of $p'$.\footnote{This matters because being intertwined corresponds to the \emph{quorum intersection property} familiar from the traditional theory of consensus.}
\item
Determine whether a function $\avaluation$ on $\ns P$, whose values are known at $p$ and $p'$, can be continuously extended to a continuous function on the whole space.\footnote{Continuous extensions to topological closures matter because (within our semitopological framework) taking closures corresponds to propagating consensus. 
This has been a running theme in our maths; see (for example) Remarks~\ref{rmrk.fundamental.consensus} and~\ref{rmrk.why.top.closure} and results including those in Subsections~\ref{subsect.towards.ce} and~\ref{subsect.kerels.determine}.}
\end{enumerate*} 
Both questions are hard to answer if we just look at continuous functions from $\ns P$ to the discrete space $\{\tvF,\tvT\}$. 
For example: the only continuous maps to $\{\tvF,\tvT\}$ from the top-left semitopology in 
Figure~\ref{fig.012} ($\ns P=\{0,1,2\}$ and $\opens$ is generated by $\{0\}$ and $\{2\}$) are $\lambda p.\tvT$ and $\lambda p.\tvF$.
This is so uninformative that it does not even distinguish between that semitopology and a space with just one point.

An answer from topology is to work with the Sierpi\'nski space that we mentioned in Example~\ref{xmpl.sk}.
As is known, continuous functions in $\ns P\to\tf{Sk}$ biject with $\opens$.\footnote{The Wikipedia page~\cite{wiki:Sierpinski_space} has a brief but clear description.}
However, and perhaps surprisingly, this is still not enough!
As we noted in Remarks~\ref{rmrk.PtoP} and~\ref{rmrk.setting.the.scene.semiframes} a semitopology is not \emph{just} its collection of open sets. 
We also need to know how they intersect.

It turns out that what we need is the semitopology $\THREE$, as illustrated in Figure~\ref{fig.THREE}.
We can think of this as $\{\tvF,\tvT\}$ augmented with a third truth-value $\tvB$ (for `both'), or as two Sierpi\'nski spaces glued end-to-end.
Intuitively, a map $\avaluation:\ns P\to\THREE$ can encode whether a point is in some $O$, in some $O'$, or in both --- and this gives us the expressivity to express semitopological properties of interest.

As a domain of truth-values, $\THREE$ will let us use logic to describe properties of semitopologies, including e.g. being intertwined (see Subsection~\ref{subsect.logical.intertwined}).
This matters in particular because logic is a portal to computation: if we can state something in logic, then we can compute it --- using e.g. proof-search or a SAT solver. 

Thus the logical material that follows below, while a prototype, is still a substantive step in the direction of turning semitopologies into a practical and applicable specification and computational framework. 
\end{rmrk}

\jamiesubsubsection{The semitopology $\THREE$}

\begin{defn}
\label{defn.3.top}
\leavevmode
\begin{enumerate*}
\item
\label{item.3.top.open}
\label{item.3.top.closed}
Define a semitopology $(\THREE,\opens_\THREE)$,\index{THREE $\THREE$ (semitopology)}\index{$\THREE=\{\tvT,\tvB,\tvF\}$ (semitopology)} as illustrated in Figure~\ref{fig.THREE}, by 
$$
\THREE = \{\tvT,\tvB,\tvF\}
\quad\text{and}\quad
\opens_\THREE =\{\varnothing, \{\tvT\}, \{\tvF\}, \{\tvT,\tvF\}, \THREE\}.
$$
Closed sets of $(\THREE,\opens_\THREE)$ are 
$$
\closed_\THREE =\{\THREE, \{\tvT,\tvB\}, \{\tvF,\tvB\}, \{\tvB\}, \varnothing\}. 
$$
We may write $(\THREE,\opens_\THREE)$ just as $\THREE$.
\item\label{item.3.valuation}
Suppose $\ns P$ is a set.
Call a function $\avaluation:\ns P\to\THREE$ a \deffont[valuation $\avaluation:\ns P{\to}\THREE$]{valuation}. 
\item
Call a valuation $\avaluation:\ns P\to\THREE$ \deffont[continuous valuation $\avaluation:\ns P{\to}\THREE$]{continuous} when $\avaluation$ is continuous as a map of semitopologies from $(\ns P,\opens)$ to $(\THREE,\opens_\THREE)$, as per Definition~\ref{defn.continuity}(\ref{item.continuous.function}) (inverse image of an open set is an open set).
\end{enumerate*}
\end{defn}

\begin{rmrk}
$\THREE$ is essentially just the top-left example in Figure~\ref{fig.012}, where in that Figure we identify $0$ with $\tvF$, $1$ with $\tvB$, and $2$ with $\tvT$.
\end{rmrk}

\begin{lemm}
\label{lemm.when.cont.3}
Suppose $(\ns P,\opens)$ is a semitopology and $\avaluation:\ns P\to\THREE$ is a valuation.
Then the following are equivalent:
\begin{enumerate*}
\item
$\avaluation$ is continuous.
\item
$\avaluation^\mone\{\tvT\}$ and $\avaluation^\mone\{\tvF\}$ are open sets.
\item 
$\avaluation^\mone\{\tvT,\tvB\}$ and $\avaluation^\mone\{\tvF,\tvB\}$ are closed sets (see also Definition~\ref{defn.f^ment}).
\end{enumerate*}
\end{lemm}
\begin{proof}
By Corollary~\ref{corr.alternative.cont.closed}, $\avaluation$ is continuous when the inverse image of every open set is open, and also when the inverse image of every closed set is closed.

It follows from Definition~\ref{defn.3.top}(\ref{item.3.top.open}) that $\{\{\tvT\},\{\tvF\}\}$ is a subbasis for the open sets,\footnote{By which we mean a set that, together with $\ns P$, generates all open sets by forming arbitrary, possibly empty, unions.} and that $\{\{\tvT,\tvB\},\{\tvF,\tvB\}\}$ is a subbasis for the closed sets.\footnote{By which we mean a set that, together with $\varnothing$, generates all closed sets by forming arbitrary, possibly empty, intersections.}
The result follows by routine calculations.
\end{proof}

\jamiesubsubsection{Indicator functions and characteristic sets}

\begin{defn}
\label{defn.indicator.functions}
Suppose that:
\begin{itemize*}
\item
$(\ns P,\opens)$ is a semitopology.
\item
$O_\tvT,O_\tvF\in\opens$ and $C_{\tvT\tvB},C_{\tvF\tvB}\in\closed$.
\item
$O_\tvT\notbetween O_\tvF$ and $C_{\tvT\tvB}\cup C_{\tvF\tvB}=\ns P$ and $O_\tvT\subseteq C_{\tvT\tvB}$.
\end{itemize*}
Then define \deffont[indicator function $\delta$]{indicator functions} 
$$
\indicator{O_\tvT,O_\tvF},\ \indicator{O_\tvT,C_{\tvT\tvB}},\ \indicator{C_{\tvT\tvB},C_{\tvF\tvB}}:\ns P\to\THREE
$$ 
as follows:
$$
\begin{array}{r@{\ }l@{\qquad}r@{\ }l}
\indicator{O_\tvT,O_\tvF}(p) =& 
\begin{cases}
\tvT & p\in O_\tvT 
\\
\tvB & p\in \ns P\setminus (O_\tvT\cup O_\tvF)
\\
\tvF & p\in O_\tvF
\end{cases}
&
\indicator{O_\tvT,C_{\tvT\tvB}}(p) =& 
\begin{cases}
\tvT & p\in O_\tvT 
\\
\tvB & p\in C_{\tvT\tvB}\setminus O_\tvT
\\
\tvF & p\in \ns P\setminus C_{\tvT\tvB}
\end{cases}
\\[5ex]
\indicator{C_{\tvT\tvB},C_{\tvF\tvB}}(p) =& 
\begin{cases}
\tvT & p\in \ns P\setminus C_{\tvF\tvB} 
\\
\tvB & p\in C_{\tvT\tvB}\cap C_{\tvF\tvB}
\\
\tvF & p\in \ns P\setminus C_{\tvF\tvB}
\end{cases}
\end{array}
$$ 
\end{defn}

\begin{lemm}
\label{lemm.indicator.continuous}
Suppose $(\ns P,\opens)$ is a semitopology and suppose 
$$
O_\tvT,O_\tvF\in\opens, \quad
C_{\tvT\tvB},C_{\tvF\tvB}\in\closed, \quad
O_\tvT\notbetween O_\tvF, \quad
C_{\tvT\tvB}\cup C_{\tvF\tvB}=\ns P,
\quad\text{and}\quad
O_\tvT\subseteq C_{\tvT\tvB}.
$$
Then the indicator functions 
$$
\indicator{O_\tvT,O_\tvF},\ \indicator{O_\tvT,C_{\tvT\tvB}},\ \indicator{C_{\tvT\tvB},C_{\tvF\tvB}}\ \in\  \ns P\to\THREE
$$ 
from Definition~\ref{defn.indicator.functions} are all continuous from $(\ns P,\opens)$ to $\THREE$.
\end{lemm}
\begin{proof}
By Lemma~\ref{lemm.when.cont.3} it suffices to check that the inverse images of $\{\tvT\}$ and $\{\tvF\}$ are open, or that the inverse images of $\{\tvT,\tvB\}$ and $\{\tvF,\tvB\}$ are closed.
This follows by construction in Definition~\ref{defn.indicator.functions}, noting from Lemma~\ref{lemm.closed.complement.open}
that a set is open/closed if and only if its complement is closed/open. 
\end{proof}

\begin{defn}
\label{defn.char.function}
Suppose $(\ns P,\opens)$ is a semitopology and suppose $\avaluation:\ns P\to\THREE$. 
Then define its \deffont[characteristic sets of a valuation $\f{char}(\avaluation)$]{characteristic sets}  
$$
\f{char}_{OO}(\avaluation),\f{char}_{OC}(\avaluation),\f{char}_{CC}(\avaluation)\in\powerset(\ns P)\times\powerset(\ns P) . 
$$
as follows:
$$
\begin{array}{r@{\ }l}
\f{char}_{OO}(\avaluation)=&(\avaluation^\mone\{\tvT\},\avaluation^\mone\{\tvF\})
\\
\f{char}_{OC}(\avaluation)=&(\avaluation^\mone\{\tvT\},\avaluation^\mone\{\tvF\tvB\})
\\
\f{char}_{CC}(\avaluation)=&(\avaluation^\mone\{\tvT,\tvB\},\avaluation^\mone\{\tvF,\tvB\})
\end{array}
$$
\end{defn}

\begin{rmrk}
\label{rmrk.looking.forward.to.f.ment}
The pair of sets $\avaluation^\mone\{\tvT,\tvB\}$ and $\avaluation^\mone\{\tvF,\tvB\}$ will be particularly useful.
Later on in Definition~\ref{defn.f^ment} we will give them names --- $\avaluation^\ment$ and $\avaluation^{\ment\tneg}$ --- and study their properties.
\end{rmrk}

\begin{lemm}
\label{lemm.char.open}
Suppose $(\ns P,\opens)$ is a semitopology and suppose $\avaluation:\ns P\to\THREE$ is continuous.
Then we have:
\begin{enumerate*}
\item
$\f{char}_{OO}(\avaluation)=(O_{\tvT},O_{\tvF})\oldin\opens\times\opens$ and $O_\tvT\notbetween O_\tvF$.
\item
$\f{char}_{OC}(\avaluation)=(O_{\tvT},C_{\tvF\tvB})\oldin\opens\times\closed$ and $O_\tvT\subseteq C_{\tvF\tvB}$.
\item
$\f{char}_{CC}(\avaluation)=(C_{\tvT\tvB},C_{\tvF\tvB})\in\closed\times\closed$ and $C_{\tvT\tvB}\cup C_{\tvF\tvB}=\ns P$.
\end{enumerate*}
\end{lemm}
\begin{proof}
By Corollary~\ref{corr.alternative.cont.closed} the inverse image of an open set under a continuous function is open and that of a closed set is closed.
The conditions $O_\tvT\notbetween O_\tvF$, $O_\tvT\subseteq C_{\tvF\tvB}$, and $C_{\tvT\tvB}\cup C_{\tvF\tvB}=\ns P$ come from the fact that $\{\tvT\}\notbetween\{\tvF\}$, $\{\tvT\}\subseteq\{\tvT,\tvB\}$, and $\{\tvT,\tvB\}\cup\{\tvB,\tvF\}=\THREE$.
\end{proof}

\begin{prop}
\label{prop.what.is.an.indicator.function}
Suppose $(\ns P,\opens)$ is a semitopology.
Then $\indicator{\text{-}}$ and $\f{char}(\text{-})$ from Definitions~\ref{defn.indicator.functions} and~\ref{defn.char.function}
determine bijections between: 
\begin{enumerate*}
\item
Continuous valuations $\avaluation:(\ns P,\opens)\to \THREE$. 
\item
Ordered pairs of disjoint (possibly empty) open sets: 
\\
\quad $(O_\tvT,O_\tvF)\oldin\opens\times\opens$ such that $O_\tvT\notbetween O_\tvF$.
\item
Ordered pairs of an open set contained in a closed set: 
\\
\quad $(O_\tvT,C_{\tvT\tvB})\oldin\opens\times\closed$ such that $O_\tvT\subseteq C_{\tvT\tvB}$.

Note that $\closure{O_\tvT}\subseteq C_{\tvT\tvB}$ by construction, but we do not require an equality.
\item\label{item.indicated.by.two.closed.sets}
Ordered pairs of closed sets whose union is $\ns P$:
\\
\quad $(C_{\tvT\tvB},C_{\tvF\tvB})\in\closed\times\closed$ such that $C_{\tvT\tvB}\cup C_{\tvF\tvB}=\ns P$.
\end{enumerate*}
\end{prop}
\begin{proof}
From Lemmas~\ref{lemm.indicator.continuous} and~\ref{lemm.char.open} and by routine computations on sets.
\end{proof}

\begin{corr}
\label{corr.valuation.open.char}
Suppose $(\ns P,\opens)$ is a semitopology.
Then:
\begin{enumerate*}
\item\label{item.valuation.open.char}
The following are equivalent:
\begin{enumerate*}
\item
$O\in\opens$. 
\item
There exists a continuous valuation $\avaluation:\ns P\to\THREE$ such that $\avaluation^\mone\{\tvT\}=O$.
\item
There exists a continuous valuation $\avaluation:\ns P\to\THREE$ such that $\avaluation^\mone\{\tvF\}=O$.
\end{enumerate*}
\item
The following are equivalent (and see also Definition~\ref{defn.f^ment}):
\begin{enumerate*}
\item
$C\in\closed$.
\item
There exists a continuous $\avaluation:\ns P\to\THREE$ such that $\avaluation^\mone\{\tvT, \tvB\}=C$. 
\item
There exists a continuous $\avaluation:\ns P\to\THREE$ such that $\avaluation^\mone\{\tvF, \tvB\}=C$. 
\end{enumerate*}
\end{enumerate*}
\end{corr}
\begin{proof}
Routine using Proposition~\ref{prop.what.is.an.indicator.function}.
For example, we can map $O$ to $\indicator{O_\tvT O_\tvF}(O,\varnothing)$.
\end{proof}

\jamiesubsection{Three-valued truth-tables}

\begin{figure}
$$
\begin{array}{c@{\ \ }c@{\ \ }c@{\ \ }c@{\ \ }c@{\ \ }c}
\begin{array}{c c c c}
p\tnotor q & \tvT & \tvB & \tvF
\\
\tvT &  \tvT & \tvB & \tvF
\\
\tvB &  \tvT & \tvB & \underline{\tvB}
\\
\tvF &  \tvT & \tvT & \tvT
\end{array}
&
\begin{array}{cc}
\tneg p
\\
\tvT & \tvF
\\
\tvB & \tvB
\\
\tvF & \tvT
\\
\end{array}
&
\begin{array}{c c c c}
p\tlatticeiff q & \tvT & \tvB & \tvF
\\
\tvT &  \tvT & \tvB & \tvF
\\
\tvB &  \tvB & \tvB & \tvB
\\
\tvF &  \tvF & \tvB & \tvT
\end{array}
&
\begin{array}{c c c c}
p\tand q & \tvT & \tvB & \tvF
\\
\tvT &  \tvT & \tvB & \tvF
\\
\tvB &  \tvB & \tvB & \tvF 
\\
\tvF &  \tvF & \tvF & \tvF 
\end{array}
&
\begin{array}{c c c c}
p\tor q & \tvT & \tvB & \tvF
\\
\tvT &  \tvT & \tvT & \tvT
\\
\tvB &  \tvT & \tvB & \tvB
\\
\tvF &  \tvT & \tvB & \tvF
\end{array}
\\[7ex]
\begin{array}{c c c c}
p\hspace{1pt}\timp q & \tvT & \tvB & \tvF
\\
\tvT &  \tvT & \tvB & \tvF
\\
\tvB &  \tvT & \tvB & \underline{\tvF}
\\
\tvF &  \tvT & \tvT & \tvT
\end{array}
&
\begin{array}{cc}
p\hspace{1pt}\timp \tvF
\\
\tvT & \tvF
\\
\tvB & \tvF
\\
\tvF & \tvT
\end{array}
&
\begin{array}{c c c c}
p\hspace{0.5pt}\tiff q & \tvT & \tvB & \tvF
\\
\tvT &  \tvT & \tvB & \tvF
\\
\tvB &  \tvB & \tvB & \tvF
\\
\tvF &  \tvF & \tvF & \tvT
\end{array}
\\[7ex]
\begin{array}{cc}
\modT p 
\\
\tvT & \tvT
\\
\tvB & \tvF
\\
\tvF & \tvF
\end{array}
&
\begin{array}{cc}
\modTB p  
\\
\tvT & \tvT
\\
\tvB & \tvT
\\
\tvF & \tvF
\end{array}
&
\begin{array}{cc}
\modB p
\\
\tvT & \tvF
\\
\tvB & \tvT
\\
\tvF & \tvF
\end{array}
\end{array}
$$
\emph{Above, the vertical axis of a table indicates values for $p$; the horizontal axis (if nontrivial) denotes values for $q$.}\index{$\tnotor$, $\tlatticeiff$, $\tand$, $\tor$, $\timp$, $\tiff$, $\modT$, $\modTB$, $\modB$ (connectives and modalities in $\THREE$)}
\caption{Truth-tables for three-valued logic (Definition~\ref{defn.truth.table.connectives})}
\label{fig.3}
\end{figure}

\begin{figure}
$$
\begin{array}{r@{\ }l@{\qquad}r@{\ }l}
\tneg\tneg p =& p
\\
p\tor q =& \tneg(\tneg p\tand \tneg q)
\\
p\tand q =& \tneg(\tneg p\tor \tneg q)
\\
p\tnotor q =& (\tneg p)\tor q  
\\
p\tnotor \tvF =& \tneg p
\\
p\tlatticeiff q =& (p\tnotor q) \tand (q\tnotor p) = (\tneg p\tor q)\tand (p\tor \tneg q)
\\
p\timp q =& (\modTB p) \tnotor q = (\tneg\modTB p)\tor q = (\modT \tneg p)\tor q
\\
p\tiff q =& (p\timp q)\tand (q\timp p)
\\
\modT p =& \tneg\modTB\tneg p =(\tneg p)\timp \tvF 
\hspace{-6em}
\\
\modTB p =& \tneg\modT\tneg p = \tneg(p\timp\tvF) = (p\timp \tvF)\timp\tvF 
\hspace{-6em}
\\
\modB p =& \modTB (p\tand\tneg p) = \modTB p \tand \modTB\tneg p = \modTB p \tand \tneg\modT p
\hspace{-6em}
\\
\modT \modTB p =&\modTB p
\\
\modTB(p\timp q) =& (\modTB p)\timp \modTB q
\\
\tf M(p\circ q) =& \tf M p\circ \tf M q \qquad\quad (\tf M\in\{\modT,\modTB\},\ {\circ}\in\{\tand,\tor\})
\hspace{-6em}
\hspace{-6em}
\end{array}
$$
\caption{Some truth-table equivalences (Lemma~\ref{lemm.checking.truth.tables})}
\label{fig.3.equivalences}
\end{figure}

Having three truth-values gives us a great deal of extra structure over the two-valued case.
In this Subsection we survey the connectives that will be useful to us later. 

\jamiesubsubsection{Truth-tables of connectives}

\begin{defn}
\label{defn.truth.table.connectives}
We define 
\begin{enumerate*}
\item
unary functions $Q\oldin\unaryconnectives$ where $Q:\THREE\to \THREE$, and
\item
binary functions $Q\oldin\binaryconnectives$ where $Q:(\THREE\times\THREE) \to \THREE$ 
\end{enumerate*}
by the truth-tables in Figure~\ref{fig.3}. 
\end{defn}

\begin{lemm}
\label{lemm.checking.truth.tables}
The truth-tables in Figure~\ref{fig.3} are related as per the equalities in Figure~\ref{fig.3.equivalences}, where for the purposes of these equations, $p$ and $q$ are considered to range over elements of $\THREE$.
\end{lemm}
\begin{proof}
By checking truth-tables.
\end{proof}

\begin{rmrk}
\label{rmrk.comments.on.tt}
Definition~\ref{defn.truth.table.connectives} and Figure~\ref{fig.3} and Lemma~\ref{lemm.checking.truth.tables} and Figure~\ref{fig.3.equivalences} are elementary, but they express some useful observations:
\begin{enumerate*}
\item
Figure~\ref{fig.3} presents truth-tables for propositional connectives in a three-valued paraconsistent logic~\cite{wiki:Paraconsistent_logic}.
There is nothing particularly unusual here within the genre of three-valued logic: we have three truth-values, and the connectives do what they do.
\item\label{item.comments.on.tt.tneg}
$\tand$ and $\tor$ in Figure~\ref{fig.3} are least upper bound and greatest lower bound operators on $\THREE$ considered as a simple lattice with $\tvF\leq \tvB \leq \tvT$, and $\tneg$ inverts the lattice order.
More on this in Subsection~\ref{subsect.conj.and.disj}.
\item\label{item.material.implication}
The equivalence $p\tnotor q = (\tneg p)\tor q$ from Figure~\ref{fig.3.equivalences} characterises $\tnotor$ as a \deffont[material implication $p\tnotor q$]{material implication}~\cite{edgington:indc}.
Similarly, $p\tlatticeiff q = (p\tnotor q)\tand(q\tnotor q)$ is a \deffont[material equivalence $p\tlatticeiff q$]{material equivalence}.

The symbols $\tnotor$ and $\tlatticeiff$ for material implication and equivalence follow the (now standard) notation used in~\cite[page~7]{whitehead:prim1}.\footnote{The phrase \emph{material implication} itself goes back to Bertrand Russell in 1903~\cite[Section~37]{russell:prim}.  It is traditionally used to refer to $(\neg A) \lor B$ in classical logic (having two truth-values).  I am not aware of a standard terminology for this in multi-valued logics, but the generalisation of `material implication' to refer to `not-A or B' in a multi-valued logic (for whatever `not' and `or' are understood to mean in that logic), seems reasonable.}
\item
$\modT p$ and $\modTB p$ are de Morgan duals, as per the equivalences $\modT p = \tneg\modTB\tneg p$ and $\modTB p=\tneg\modT\tneg p$ in Figure~\ref{fig.3.equivalences}.
Looking at the truth-tables, $\modT$ acts as a modality that identifies $\tvT$, $\modTB$ is a modality that identifies being $\tvT$ or $\tvB$.
\item
The equivalence
$$
\tf M(p\circ q) = \tf M p\circ \tf M q \quad (\tf M\in\{\modT,\modTB\},\ {\circ}\in\{\tand,\tor\})
$$
is a little unexpected; it means in particular that we have these two properties:
$$
\modT(p\tor q)=\modT p\tor \modT q
\quad\text{and}\quad 
\modTB (p\tand q)=\modTB p\tand\modTB q.
$$
If we think of $\modT$ as a \emph{necessitation} (or `box') modality and $\modTB$ as a \emph{possibility} (or `diamond') modality, and look at these two properties in terms of Kripke structures~\cite{blackburn:modl}, then it is not hard to see that they are characteristic of Kripke frames in which each world sees precisely one world.\footnote{Our denotation for $\modT$ and $\modTB$ uses truth-values in $\THREE$, not Kripke structures.  We are just pointing out where such axioms can and do arise in Kripke semantics for modalities.}
\item
We take a moment to compute the equivalence $\modTB(p\timp q)=(\modTB p)\timp\modTB q$ from the other equivalences:
$$
\begin{array}{r@{\ }l}
\modTB(p\timp q)=&\modTB(\tneg p\tor q) = (\modTB\tneg \modTB p)\tor\modTB q = (\tneg\modT\modTB p)\tor\modTB q = (\tneg\modTB p)\tor\modTB q
\\
(\modTB p)\timp\modTB q=&(\tneg\modTB \modTB p)\tor\modTB q = (\tneg\modTB p)\tor\modTB q
\end{array}
$$
\end{enumerate*}
\end{rmrk}

\jamiesubsubsection{Conjunction and disjunction}
\label{subsect.conj.and.disj}

\begin{defn}
\label{defn.bigwedge.V}
Suppose $V\subseteq\THREE$ is a set of truth-values.
Define $\bigwedge V$ and $\bigvee V$ to be the least upper bound and greatest lower bound of $V$ in $\THREE$ considered as a lattice with $\tvF<\tvB<\tvT$.

As noted in Remark~\ref{rmrk.comments.on.tt}(\ref{item.comments.on.tt.tneg}), the truth-tables for $\tand$ and $\tor$ in Figure~\ref{fig.3} also compute least upper bounds and greater lower bounds, so that $v\tand v' = \bigwedge\{v,v'\}$ and $v\tor v' = \bigvee\{v,v'\}$.
We may elide the difference between $\tand$-the-binary-operator and its generalisation to $\bigwedge$ henceforth, and similarly for $\tor$ and $\bigvee$. 
\end{defn}

\begin{lemm}
\label{lemm.tand.tt.iff}
Suppose $V\subseteq\THREE$ is a set of truth-values.
Then:
\begin{enumerate*}
\item
$\bigwedge V \in\{\tvT,\tvB\}$ if and only if $V\subseteq \{\tvT,\tvB\}$. 
\item
$\bigwedge V = \tvF$ if and only if $V\between\{\tvF\}$ (if and only if $\tvF\in V$).
\item
$\bigvee V \in\{\tvT,\tvB\}$ if and only if $V\between \{\tvT,\tvB\}$. 
\item
$\bigvee V = \tvF$ if and only if $V\subseteq\{\tvF\}$.
\end{enumerate*}
\end{lemm}
\begin{proof}
A fact of Definition~\ref{defn.bigwedge.V}.
\end{proof}

\jamiesubsubsection{Implication(s)}

\begin{rmrk}[Two implications]
\label{rmrk.two.implications}
\leavevmode
Figure~\ref{fig.3} has two implication operators; $\tnotor$ (material implication) and also $\timp$ (we underline the differences in the truth-tables in Figure~\ref{fig.3}).

Having multiple implication connectives in paraconsistent logic is typical: e.g. the authors of~\cite{arieli:idepl} note the existence of \emph{sixteen} possible implications just in a three-valued logic~\cite[note~5, page~22]{arieli:idepl}. 

We chose to include $\tnotor$ and $\timp$ as primitive, from the sixteen implications available, because they will be especially useful.
See --- just for example --- Proposition~\ref{prop.timp.MP}(\ref{item.timp.is.timp}), Corollary~\ref{corr.pull.results.together}, Definitions~\ref{defn.Ax} and~\ref{defn.AxEx}, and Figure~\ref{fig.short.table}.

We make some simple observations:
\end{rmrk}
 
\begin{lemm}
\label{lemm.no.contrapositive}
For $v,v'\in\THREE$ we have:
\begin{enumerate*}
\item
$v\tnotor v' = \tneg v' \tnotor \tneg v$
\item
It is not necessarily the case that $v\timp v' = \tneg v'\timp \tneg v$.
\end{enumerate*}
In words: the contrapositive rule is valid for $\tnotor$, but not for $\timp$.
\end{lemm}
\begin{proof}
We compute and compare truth-tables; differences are underlined:
$$
\begin{array}{c@{\qquad}c} 
\begin{array}{c c c c}
p\tnotor q & \tvT & \tvB & \tvF
\\
\tvT &  \tvT & \tvB & \tvF
\\
\tvB &  \tvT & \tvB & \tvB
\\
\tvF &  \tvT & \tvT & \tvT
\end{array}
&
\begin{array}{c c c c}
\tneg q\tnotor\tneg p
     &  \tvT & \tvB & \tvF
\\
\tvT &  \tvT & \tvB & \tvF
\\
\tvB &  \tvT & \tvB & \tvB
\\
\tvF &  \tvT & \tvT & \tvT
\end{array}
\\[5ex]
\begin{array}{c c c c}
p\timp q & \tvT & \tvB & \tvF
\\
\tvT &  \tvT & \underline{\tvB} & \tvF
\\
\tvB &  \tvT & \tvB & \underline{\tvF}
\\
\tvF &  \tvT & \tvT & \tvT
\end{array}
&
\begin{array}{c c c c}
\tneg q\timp \tneg p
     &  \tvT & \tvB & \tvF
\\
\tvT &  \tvT & \underline{\tvF} & \tvF
\\
\tvB &  \tvT & \tvB & \underline{\tvB}
\\
\tvF &  \tvT & \tvT & \tvT
\end{array}
\end{array}
$$
\end{proof}

\begin{rmrk}
\label{rmrk.imp.modalities}
We noted in Figure~\ref{fig.3.equivalences} and Lemma~\ref{lemm.checking.truth.tables} that 
$$
p\timp q = (\modTB p)\tnotor q = (\tneg\modTB p)\tor q.
$$
A minimal set of propositional connectives for our logic could leave out $\timp$ --- and then perhaps leave out $\tnotor$ in favour of $\tor$ and $\tneg$, and arrive at (say) $\{\tvF, \tor,\tneg,\modTB\}$.
More discussion of $\timp$ and $\tnotor$ is in Subsection~\ref{subsect.implications}.
\end{rmrk}

\jamiesubsection{Predicates}

\begin{nttn}
Fix a countably infinite set of \deffont{variable symbols $\tf{Var}$}.
We will use variable symbols to construct our syntax in Definition~\ref{defn.predicate.syntax}.
\end{nttn}

\begin{defn}
\label{defn.predicate.syntax}
Suppose $\ns P$ is any set; we will always assume this is disjoint from $\tf{Var}$.
Define the syntax $\tf{Pred3}(\ns P)$ of \deffont{predicates over $\ns P$} by the grammar in Figure~\ref{fig.predicate.syntax}.

In this syntax:
\begin{enumerate*}
\item\label{item.propositional.constant}
We may call $p\in\ns P$ a \deffont{propositional atom} or \deffont{propositional constant}.
When we give our syntax a denotation, it will get assigned a truth-value in $\THREE$. 
\item
Variable symbols $x\in\tf{Var}$ range over the propositional atoms. 
\item
$\tall$ and $\texi$ are quantifiers.
They bind variable symbols.
We treat predicates up to $\alpha$-equivalence henceforth.
\item
We define the \deffont{free variables of a predicate} as usual:
$$
\begin{array}{r@{\ }l@{\quad}r@{\ }l@{\quad}r@{\ }l@{\quad}r@{\ }l}
\fv(v)=&\fv(p)=\varnothing
&
\fv(\tf Q \phi)=&\fv(\phi)
\\
\fv(\phi\tf Q\phi')=&\fv(\phi)\cup\fv(\phi')
\\
\fv(\Kmod{}\phi)=&\fv(\phi)
=\fv(\Emod{}\phi)
&
\fv(\tall x.\phi)=&\fv(\phi){\setminus}\{x\} =\fv(\texi x.\phi) \hspace{-4em}
\end{array}
$$
Note that $\fv(\phi)$ does not include the propositional constants in $\phi$; we only include the free variables.
\item
\label{item.closed.pred3}
We call a predicate \deffont[closed predicate]{closed} when it has no free variables. 
We write $\tf{ClosedPred3}(\ns P)$ for the set of closed predicates in $\tf{Pred3}(\ns P)$, in symbols:
$$
\tf{ClosedPred3}(\ns P) = \{\phi\in\tf{Pred3}(\ns P) \mid \fv(\phi)=\varnothing\} .
$$ 
Note that a closed predicate $\phi\in\tf{ClosedPred3}(\ns P)$ may still mention propositional constants from $\ns P$. 
\end{enumerate*}
\end{defn}

\begin{figure}
$$
\begin{array}{r@{\ }l}
\phi ::=& 
(v: v\in\THREE) 
\mid (x: x\in\tf{Var})
\mid (p : p\in\ns P) 
\\
\mid &(\tf Q \phi : \tf Q\in\unaryconnectives) 
\\
\mid &(\phi\tf Q\phi' :  \tf Q\in\binaryconnectives) 
\\
\mid &\Kmod{}\phi \mid \Emod{}\phi
\\
\mid &\tall x.\phi \mid \texi x.\phi
\end{array}
$$ 
\caption{Predicate syntax}
\label{fig.predicate.syntax}
\end{figure}

\begin{figure}
$$
\begin{array}{r@{\ }l@{\quad}l}
\model{v}_\avaluation =&v & v\in\THREE
\\
\model{p}_\avaluation =&\avaluation(p) & p\in\ns P
\\
\model{\tf Q\phi}_\avaluation =& Q\,\model{\phi}_\avaluation & 
\tf Q\in \unaryconnectives 
\\
\model{\phi\tf Q\phi'}_\avaluation =& \model{\phi}_\avaluation\,Q\,\model{\phi'}_\avaluation & 
\tf Q\in \binaryconnectives 
\\[2ex]
\model{\Kmod{}\phi}_\avaluation =& \bigwedge \{\model{\phi}_\avaluation \mid f:\ns P\to \THREE\}& 
\\
\model{\Emod{}\phi}_\avaluation =& \bigvee \{\model{\phi}_\avaluation \mid f:\ns P\to \THREE\}& 
\\
\model{\tall x.\phi}_\avaluation=& \bigwedge\{ \model{\phi[x\ssm p]}_\avaluation \mid p\in\ns P\}
\\
\model{\texi x.\phi}_\avaluation=& \bigvee\{ \model{\phi[x\ssm p]}_\avaluation \mid p\in\ns P\}
\end{array}
$$ 
\caption{Denotation of a predicate $\phi$ with respect to a valuation $f$}
\label{fig.3.phi.f}
\end{figure}

\begin{defn}[Denotation]
\label{defn.denotation}
Suppose $(\ns P,\opens)$ is a semitopology on a set $\ns P$ and $\avaluation:\ns P\to\THREE$ is a valuation (Definition~\ref{defn.3.top}(\ref{item.3.valuation})), and suppose $\phi\in\tf{ClosedPred3}(\ns P)$ is a closed predicate over $\ns P$ (Definition~\ref{defn.predicate.syntax}(\ref{item.closed.pred3})).

Then define $\model{\phi}_f$ inductively as in Figure~\ref{fig.3.phi.f}, where:
\begin{enumerate*}
\item
The unary and binary connectives (\unaryconnectives and \binaryconnectives) are interpreted using the tables in Figure~\ref{fig.3}, as discussed in Remark~\ref{rmrk.Q.vs.tfQ}.
\item
$\phi[x\ssm p]$ is obtained from $\phi$ by replacing (substituting) $x$ with $p$ throughout. 
\end{enumerate*} 
\end{defn}

\begin{rmrk}
\label{rmrk.Q.vs.tfQ}
For each unary and binary connective $Q$ considered in Figure~\ref{fig.3} there is a corresponding unary or binary syntax connective $\tf Q$ in the syntax of Figure~\ref{fig.predicate.syntax}.
(This is no different from how in the usual two-valued propositional logic, the symbol `$\land$' can represent both a function $\mathbb B\times\mathbb B\to\mathbb B$ and also a predicate connective.)

Each $\tf Q$ is interpreted as the corresponding $Q$ when we define the denotation in Figure~\ref{fig.3.phi.f}.
\end{rmrk}

\begin{rmrk}
\leavevmode
\begin{enumerate}
\item
The denotation of the connectives in \unaryconnectives and \binaryconnectives is straightforward: we look up the input(s) in the corresponding truth-table in Figure~\ref{fig.3}, and return the corresponding output; see Remark~\ref{rmrk.Q.vs.tfQ}.
\item
The modalities $\modT$, $\modTB$, and $\modB$ are not continuous as functions from $\THREE$ to $\THREE$; e.g. $\lambda v.\modT v$ is not continuous at $\tvB$.
Likewise $\lambda v.v\timp \tvF$ is not continuous. 
\item
$\Kmod{}\phi$ takes a conjunction (greatest lower bound) over the truth-values of $\phi$ for every possible assignment of truth-values from $\THREE$ to the propositional atoms in $\phi$.

Intuitively, $\Kmod{}$ universally quantifies over all possible valuations; if we view valuations as possible worlds, then $\Kmod{}$ is a box-style modality.
\item
$\Emod{}\phi$ is the de Morgan dual: it takes a disjunction (least upper bound) over the truth-values of $\phi$ for every possible assignment of truth-values from $\THREE$ to the propositional atoms in $\phi$. 

Intuitively, $\Emod{}$ existentially quantifies over all possible valuations; if we view valuations as possible worlds, then $\Emod{}$ is a diamond-style modality.
\item
$\tall x.\phi$ takes a conjunction over the truth-values of $\phi[x\ssm p]$ for every $p\in\ns P$.
So $\tall$ quantifies universally over propositional atoms.
\item
$\texi x.\phi$ takes a disjunction over the truth-values of $\phi[x\ssm p]$ for every $p\in\ns P$.
So $\texi$ quantifies existentially over propositional atoms.
\end{enumerate}
\end{rmrk}

\jamiesubsection{Validity}

\jamiesubsubsection{The definition}

\begin{defn}[Validity]
\label{defn.ment}
Suppose $(\ns P,\opens)$ is a semitopology and $f:\ns P\to\THREE$ is a valuation and $\phi\in\tf{ClosedPred3}(\ns P)$ is a closed predicate.
\begin{enumerate*}
\item\label{item.f.phi.valid}
Define $\avaluation\ment \phi$\index{$\avaluation\ment\phi$ (validity of a predicate in a valuation)} by
$$
\avaluation\ment\phi
\quad\text{when}\quad
\model{\phi}_\avaluation\in\{\tvT,\tvB\}.
$$
In this case we call $\phi$ \deffont[valid ($\phi$ in $\avaluation$)]{valid} in $\avaluation$. 
\item\label{item.designated.value}
We may call $\{\tvT,\tvB\}$ the set of \deffont[designated values $\{\tvT,\tvB\}$]{designated values}, following a terminology from paraconsistent logic --- these are the \emph{valid} truth-values. 

Note that $\tvT$ is valid, as we would expect, but so is $\tvB$.
\item
If $\avaluation$ does not matter to calculating $\model{\phi}_\avaluation$ (e.g. because $\phi$ mentions no variable symbols or predicate atoms) then we may write $\model{\phi}_\avaluation$ just as $\model{\phi}$, and $\avaluation\ment\phi$ just as $\ment\phi$.
\item
Suppose $\Phi\subseteq\tf{ClosedPred3}(\ns P)$ is a set of predicates.
Write 
$$
\avaluation\ment\Phi
\quad\text{when}\quad
\Forall{\phi{\in}\Phi}\avaluation\ment\phi,
$$
and say that $\Phi$ is \deffont[valid ($\Phi$ in $\avaluation$)]{valid} in $\avaluation$. 
\item\label{item.P.phi.valid}
Write $\ns P\ment \phi$ when $\avaluation\ment\phi$ for every valuation $\avaluation:\ns P\to\THREE$.
and similarly for $\ns P\ment\Phi$.
\item\label{item.P.O.phi.valid}
Write $\ns P,\opens\ment \phi$ when $\avaluation\ment\phi$ for every valuation $\avaluation:\ns P\to\THREE$ that is continuous on $(\ns P,\opens)$, and similarly for $\ns P\ment\Phi$ and $\ns P,\opens\ment\Phi$.
\end{enumerate*} 
\end{defn}

\begin{defn}
A \deffont[sequent $\Phi\cent\Psi$]{sequent}\index{$\Phi\cent\Psi$ (sequent)} is a pair of finite sets of predicates $\Phi\cent\Psi$. 

We define a notion of \deffont{sequent validity $\ns P\ment(\Phi\cent\Psi)$}\index{$\ns P\ment(\Phi\cent\Psi)$ (sequent validity)} by
$$
\ns P\ment (\Phi\cent\Psi)
\quad\text{when}\quad
\ns P\ment (\tand\Phi)\timp(\tor\Psi) .
$$
\end{defn}

\begin{rmrk}
\label{rmrk.commens.on.timp}
Recall from Remark~\ref{rmrk.comments.on.tt}(\ref{item.material.implication}) that $\tnotor$ is a material implication.
Yet $\timp$ can also claim to be an implication operator, and unlike $\tnotor$ it satisfies Modus Ponens.
We examine this in Propositions~\ref{prop.3.modus.ponens} and~\ref{prop.timp.MP}.
\end{rmrk}

In Definition~\ref{defn.ment}(\ref{item.designated.value}) we noted that $\tvT$ is valid, but so is $\tvB$.
We can make this formal with a small lemma:
\begin{lemm}
\label{lemm.diamond.valid.iff}
Suppose $\ns P$ is a set and $\avaluation:\ns P\to\THREE$ is a valuation and $\phi\in\tf{ClosedPred3}(\ns P)$.
Then 
$$
\avaluation\ment\phi
\quad\text{if and only if}\quad
\avaluation\ment\modTB\phi .
$$
\end{lemm}
\begin{proof}
Simply because $\modTB v\in \{\tvT,\tvB\}$ if and only if $v\in\{\tvT,\tvB\}$, for every $v\in\THREE$.
\end{proof}

\jamiesubsubsection{$\avaluation^\ment$ and $\avaluation^{\ment\tneg}$: the designated sets of a valuation}

In this easy Subsection we continue Remark~\ref{rmrk.looking.forward.to.f.ment} and follow a basic exercise in unpacking the definitions.
The only extra ingredient we have now, relative to the results surrounding Remark~\ref{rmrk.looking.forward.to.f.ment}, is that we can interpret an element of $\{\tvT,\tvB\}$ as `valid':
\begin{defn}
\label{defn.f^ment}
Suppose $\ns P$ is a set and $\avaluation:\ns P\to\THREE$ is a valuation.
Then define the \deffont{designated set $\avaluation^\ment\subseteq\ns P$}\index{$\avaluation^\ment$ (designated set of a valuation)} and the \deffont{neg-designated set $\avaluation^{\ment\tneg}\subseteq\ns P$}\index{$\avaluation^{\ment\tneg}$ (neg-designated set of a valuation)} of the valuation $\avaluation$ by: 
$$
\begin{array}{r@{\ }l}
\avaluation^\ment =& \avaluation^\mone\{\tvT,\tvB\}
\\ 
\avaluation^{\ment\tneg} =& \avaluation^\mone\{\tvF,\tvB\}
\end{array}
$$
\end{defn}

\begin{rmrk}
\label{rmrk.comments.on.f.ment}
\leavevmode
\begin{enumerate*}
\item
We call $\avaluation^\ment$ the designated set of $\avaluation$ because this is the set of $p\in\ns P$ such that $\avaluation(p)$ is a \emph{designated value} (i.e. is $\tvT$ or $\tvB$), as per Definition~\ref{defn.ment}(\ref{item.designated.value}).
\item
Similarly, $\avaluation^{\ment\tneg}$ is the set of $p\in\ns P$ such that $\avaluation(\tneg p)$ is a \emph{designated value} --- note that this is \emph{not the same thing} as the set of $p\in\ns P$ such that $\avaluation(p)$ is not a designated value, because $\{\tvF,\tvB\}$ is not the same thing as $\{\tvF\}$.
\item\label{item.f.ment.is.char}
The pair $(\avaluation^\ment,\avaluation^{\ment\tneg})$ characterises $\avaluation$ in the sense that 
$$
\f{char}_{CC}(\avaluation)=(\avaluation^\ment, \avaluation^{\ment\tneg})
\qquad\text{and}\qquad 
\avaluation=\indicator{C_{\tvT\tvB},C_{\tvF\tvB}}(\avaluation^\ment,\avaluation^{\ment\tneg}) ,
$$
where $\f{char}$ is from Definition~\ref{defn.char.function}, and $\indicator{}$ is from Definition~\ref{defn.indicator.functions}.
\end{enumerate*}
\end{rmrk}

\begin{lemm}
\label{lemm.f.ment.closed}
Suppose $(\ns P,\opens)$ is a semitopology and $\avaluation:\ns P\to\THREE$ is a continuous function on $(\ns P,\opens)$.
Then both $\avaluation^\ment$ and $\avaluation^{\ment\tneg}$ are closed sets.
\end{lemm}
\begin{proof}
By Definition~\ref{defn.3.top}(\ref{item.3.top.closed}) $\{\tvT,\tvB\}$ and $\{\tvF,\tvB\}$ are closed sets in $\THREE$, and by 
Corollary~\ref{corr.alternative.cont.closed} 
the inverse image of a closed set under a continuous function is closed.
\end{proof}

The notation in Definition~\ref{defn.f^ment} is justified by the following very easy lemma:
\begin{lemm}
\label{lemm.f.ment.iff.f.ment}
Suppose $\ns P$ is a set and $p\in\ns P$ and $\avaluation:\ns P\to\THREE$ is a valuation.
Then:
\begin{enumerate*} 
\item
$p\in \avaluation^\ment$ if and only if $\avaluation\ment p$.
\item
$p\in \avaluation^{\ment\tneg}$ if and only if $\avaluation\ment \tneg p$.
\end{enumerate*}
\end{lemm}
\begin{proof}
Direct from Figure~\ref{fig.3}, Definitions~\ref{defn.f^ment} and~\ref{defn.ment}(\ref{item.f.phi.valid}), and the case for $\avaluation\ment p$ in Figure~\ref{fig.3.phi.f}.
\end{proof}

\jamiesubsubsection{Validity of conjunction and quantification}

The $\tand$ and $\tall$ are `obviously' a conjunction and universal quantification respectively, but we still need to check that the truth-tables work and the notion of validity $\ment$ interacts correctly with them.
We do this and find that there are no surprises:

\begin{lemm}
\label{lemm.tand.valid.iff}
Suppose $\ns P$ is a set and $\avaluation:\ns P\to\THREE$ is a valuation and $\phi,\phi'\in\tf{ClosedPred3}(\ns P)$.
Then: 
\begin{enumerate*}
\item\label{item.tand.valid.iff.1}
$\avaluation\ment \phi\tand\phi'$ if and only if $\avaluation\ment\phi \land \avaluation\ment\phi'$.
\item\label{item.tand.valid.iff.2}
$\avaluation\ment \phi\tor\phi'$ if and only if $\avaluation\ment\phi \lor \avaluation\ment\phi'$.
\end{enumerate*}
\end{lemm}
\begin{proof}
We reason as follows for the case of $\tand$; the case of $\tor$ is exactly similar:
$$
\begin{array}[b]{r@{\ }l@{\quad}l}
\avaluation\ment\phi\tand\phi'
\liff&
\model{\phi\tand\phi'}_\avaluation\in\{\tvT,\tvB\}
&\text{Definition~\ref{defn.ment}(\ref{item.f.phi.valid})}
\\
\liff&
\model{\phi}_\avaluation\tand \model{\phi'}_\avaluation\in\{\tvT,\tvB\}
&\text{Figure~\ref{fig.3.phi.f}}
\\
\liff&
\model{\phi}_\avaluation\in\{\tvT,\tvB\}
\land
\model{\phi'}_\avaluation\in\{\tvT,\tvB\}
&\text{Lemma~\ref{lemm.tand.tt.iff} (or Figure~\ref{fig.3})}
\\
\liff&
\avaluation\ment \phi
\land
\avaluation\ment \phi'
&\text{Definition~\ref{defn.ment}(\ref{item.f.phi.valid})}
\end{array}
\qedhere$$
\end{proof}

\begin{lemm}
\label{lemm.tall.valid.iff}
Suppose $\ns P$ is a set and $\avaluation:\ns P\to\THREE$ is a valuation and $\tall x.\phi\in\tf{ClosedPred3}(\ns P)$.
Then:
\begin{enumerate*}
\item 
$\avaluation\ment \tall x.\phi$ if and only if 
$\Forall{p{\in}\ns P}\avaluation\ment\phi[x\ssm p]$.
\item
$\avaluation\ment \texi x.\phi$ if and only if 
$\Exists{p{\in}\ns P}\avaluation\ment\phi[x\ssm p]$.
\end{enumerate*}
\end{lemm}
\begin{proof}
For the case of $\tall$ we reason as follows:
$$
\begin{array}[b]{r@{\ }l@{\quad}l}
\avaluation\ment\tall x.\phi 
\liff&
\model{\tall x.\phi}_\avaluation\in\{\tvT,\tvB\}
&\text{Definition~\ref{defn.ment}(\ref{item.f.phi.valid})}
\\
\liff&
\bigwedge_{p\in\ns P} \model{\phi[x\ssm p]}_\avaluation\in\{\tvT,\tvB\}
&\text{Figure~\ref{fig.3.phi.f}}
\\
\liff&
\Forall{p{\in}\ns P}\model{\phi[x\ssm p]}_\avaluation\in\{\tvT,\tvB\}
&\text{Lemma~\ref{lemm.tand.tt.iff}}
\\
\liff&
\Forall{p{\in}\ns P}\avaluation\ment \phi[x\ssm p]
&\text{Definition~\ref{defn.ment}(\ref{item.f.phi.valid})}
\end{array}
$$
The case of $\texi$ is precisely similar.
\end{proof}

\jamiesubsubsection{Logical implications}
\label{subsect.implications}

\begin{rmrk}
Continuing Remarks~\ref{rmrk.comments.on.tt}(\ref{item.material.implication}) and~\ref{rmrk.commens.on.timp}, Figure~\ref{fig.3} has truth-tables for two binary connectives expressing notions of `implications': $\tnotor$ and $\timp$.
We will need both:
\begin{enumerate*}
\item
$\timp$ expresses logical implication (see Lemma~\ref{lemm.timp.MP}(\ref{item.timp.MP.is}) and Proposition~\ref{prop.timp.MP}(\ref{item.timp.is.timp})).
\item
$\tnotor$ helps us express the property of `being intertwined' in (see Lemma~\ref{lemm.char.intertwinedwith} and~\ref{prop.logical.intertwined}).
\end{enumerate*}
\end{rmrk}

Both implications interact nicely with $\tand$: 
\begin{prop}
\label{prop.3.modus.ponens}
Suppose $\ns P$ is any set and $p,q,r\in\ns P$ and $\avaluation:\ns P\to\THREE$ is a valuation. 
Then:
\begin{enumerate*}
\item
$\model{(p\tand q)\tnotor r}_\avaluation=\model{p\tnotor(q\tnotor r)}_\avaluation$.
\item 
$\model{(p\tand q)\timp r}_\avaluation=\model{p\timp(q\timp r)}_\avaluation$.
\end{enumerate*}
\end{prop}
\begin{proof}
We simplify using the equivalences in Figure~\ref{fig.3.equivalences} (we could also just check truth-tables, of course):
$$
\begin{array}[b]{r@{\ }l}
(p\tand q)\tnotor r =& \tneg(p\tand q)\tor r = \tneg p\tor \tneg q \tor r
\\
p\tnotor(q\tnotor r) =& \tneg p \tor \tneg q \tor r
\\[1.5ex]
(p\tand q)\timp r =& \tneg\modTB(p\tand q)\,\tor r = \tneg(\modTB p\tand \modTB q)\,\tor r = 
\modT\tneg p \,\tor\, \modT\tneg q \,\tor\, r
\\
p\timp(q\timp r) =& \tneg\modTB p \,\tor\, \tneg\modTB q\,\tor\, r = \modT\tneg p\,\tor\,\modT\tneg q\,\tor\, r .
\end{array}
\qedhere$$
\end{proof}

\begin{lemm}
\label{lemm.timp.MP}
Suppose $v,v'\in\THREE$.
Then:
\begin{enumerate*}
\item
If $v\in\{\tvT,\tvB\}$ implies $v'\in\{\tvT,\tvB\}$, then $v\tnotor v'\in\{\tvT,\tvB\}$.
\item
It is possible that $v\tnotor v'\in\{\tvT,\tvB\}$ and $v\in\{\tvT,\tvB\}$, but $v'\notin\{\tvT,\tvB\}$.
\item\label{item.timp.MP.is}
$v\timp v'\in\{\tvT,\tvB\}$ if and only if $v\in\{\tvT,\tvB\} \limp v'\in\{\tvT,\tvB\}$.
\end{enumerate*}
\end{lemm}
\begin{proof}
We consider each part in turn:
\begin{enumerate}
\item
We check the truth-table for $\tnotor$ in Figure~\ref{fig.3} and prove the contrapositive: if $v\tnotor v'=\tvF$, then $v=\tvT \land v'=\tvF$, and so $v\in\{\tvT,\tvB\}$ does not imply $v'\in\{\tvT,\tvB\}$.
\item
We set $v=\tvB$ and $v'=\tvF$ and note from the truth-table for $\tnotor$ in Figure~\ref{fig.3} that $\tvB\tnotor \tvF=\tvB$.
\item
We prove two implications:
\begin{itemize}
\item
For the right-to-left implication, we 
prove the contrapositive.
Suppose $v\timp v'=\tvF$.
We check the truth-table for $\timp$ in Figure~\ref{fig.3} and see that $v'=\tvF$ and $v\in\{\tvT,\tvB\}$.
Thus $v\in\{\tvT,\tvB\}$ implies $v'\notin\{\tvT,\tvB\}$, and it is \emph{not} the case that $v\in\{\tvT,\tvB\}$ implies $v'\in\{\tvT,\tvB\}$.
\item
For the left-to-right implication, there are three sub-cases:
\begin{itemize*}
\item
If $v'=\tvB$ or $v'=\tvT$ then $v'\in\{\tvT,\tvB\}$ and there is nothing to prove.
\item
If $v=\tvF$ then there is nothing to prove.
\item
Suppose $v'=\tvF$ and $v\in\{\tvB,\tvT\}$.
We check the truth-table for $\timp$ in Figure~\ref{fig.3} and see that the result holds.
\qedhere\end{itemize*}
\end{itemize}
\end{enumerate}
\end{proof}

\begin{prop}
\label{prop.timp.MP}
Suppose $\ns P$ is a set and $\avaluation:\ns P\to\THREE$ is a valuation and $\phi,\phi'\in\tf{ClosedPred3}(\ns P)$.
Then:
\begin{enumerate*}
\item
If $\avaluation\ment \phi$ implies $\avaluation\ment \phi'$, then $\avaluation\ment \phi\tnotor \phi'$.
\item
It is possible that $\avaluation\ment \phi\tnotor \phi'$ and $\avaluation\ment \phi$ but $\avaluation\not\ment \phi'$.
\item
\label{item.timp.is.timp}
$\avaluation\ment \phi\timp \phi'$ if and only if $\avaluation\ment \phi \limp \avaluation\ment \phi'$.
\end{enumerate*}
\end{prop}
\begin{proof}
By Definition~\ref{defn.ment}(\ref{item.f.phi.valid}), $\avaluation\ment\phi$ when $\model{\phi}_\avaluation\in\{\tvT,\tvB\}$, and similarly for $\avaluation\ment\phi\tnotor\phi'$, $\avaluation\ment\phi\timp\phi'$, and $\avaluation\ment\phi'$. 
By Figure~\ref{fig.3.phi.f} $\model{\phi\circ\phi'}_\avaluation=\model{\phi}_\avaluation\circ\model{\phi'}_\avaluation$ for $\circ\in\{\tnotor,\timp\}$. 
We use Lemma~\ref{lemm.timp.MP}.
\end{proof}

We can pull the results in this Subsection together as follows:
\begin{corr}
\label{corr.pull.results.together}
Suppose $\ns P$ is a set and $\avaluation:\ns P\to\THREE$ is a valuation and $\phi,\phi'\in\tf{ClosedPred3}(\ns P)$.
Then:
$$
\begin{array}{r@{\ }l}
\avaluation\ment \phi\timp\phi'
\liff&
\avaluation\not\ment\phi \lor \avaluation\ment \phi'
\\
\liff&
\model{\phi}_\avaluation\in\{\tvF\} \lor \model{\phi'}_\avaluation\in\{\tvT,\tvB\}
\\
\liff&
\model{\phi}_\avaluation\in\{\tvT,\tvB\} \limp \model{\phi'}_\avaluation\in\{\tvT,\tvB\}
\\[2ex]
\avaluation\ment \phi\tnotor\phi'
\liff&
\avaluation\ment\tneg\phi \lor \avaluation\ment \phi'
\\
\liff&
\model{\phi}_\avaluation\in\{\tvF,\tvB\} \lor \model{\phi'}_\avaluation\in\{\tvT,\tvB\}
\\
\liff&
\model{\phi}_\avaluation\in\{\tvT\} \limp \model{\phi'}_\avaluation\in\{\tvT,\tvB\}.
\end{array}
$$
\end{corr}
\begin{proof}
We reason as follows:
\begin{itemize}
\item
That $\avaluation\ment \phi\timp\phi'$ if and only if $\avaluation\not\ment\phi \lor \avaluation\ment \phi'$ just rephrases Proposition~\ref{prop.timp.MP}(\ref{item.timp.is.timp}).
\item
That $\avaluation\ment \phi\tnotor\phi'$ if and only if $\avaluation\ment\tneg\phi \lor \avaluation\ment \phi'$ follows from the characterisation $p\tnotor q = (\tneg p)\tor q$ of $\tnotor$ as material equivalence in Figure~\ref{fig.3.equivalences}, and from Lemma~\ref{lemm.tand.valid.iff}(\ref{item.tand.valid.iff.2}).
\item
The rest of the equivalences just unpack the definition of validity from Definition~\ref{defn.ment}(\ref{item.f.phi.valid}).
\qedhere\end{itemize}
\end{proof}

\jamiesubsection{Logical equivalence}

Lemma~\ref{lemm.MP.iff} is closely related to Proposition~\ref{prop.timp.MP}, though it is easiest to give a direct proof:
\begin{lemm}
\label{lemm.MP.iff}
Suppose $\ns P$ is a set and $\avaluation:\ns P\to\THREE$ is a valuation and $\phi,\phi'\in\tf{ClosedPred3}(\ns P)$ are closed predicates.
Then 
$$
\avaluation\ment \phi\tiff \phi'
\quad\text{if and only if}\quad
\avaluation\ment \phi\ \liff\ \avaluation\ment\phi'.
$$
\end{lemm}
\begin{proof}
A direct proof like that in Proposition~\ref{prop.timp.MP}(\ref{item.timp.is.timp}) is straightforward.
Or, we can note the equivalence $p\tiff q = (p\timp q)\tand (q\timp p)$ from Figure~\ref{fig.3.equivalences}, and use Proposition~\ref{prop.timp.MP}(\ref{item.timp.is.timp}) and Lemma~\ref{lemm.tand.valid.iff}.
\end{proof}

\begin{corr}
\label{corr.valuation.pp'.iff}
Suppose $(\ns P,\opens)$ is a semitopology and $\avaluation:\ns P\to\THREE$ is a valuation and $p,p'\in\ns P$.
Then the following are equivalent:
\begin{enumerate*}
\item\label{item.valuation.pp'.iff.1}
$\avaluation\ment p\tiff p'$.
\item\label{item.valuation.pp'.iff.2}
$p\in\avaluation^\ment\liff p'\in\avaluation^\ment$.
\item\label{item.valuation.pp'.iff.3}
$p\in\avaluation^\mone\{\tvF\}\liff p'\in\avaluation^\mone\{\tvF\}$.
\end{enumerate*}
\end{corr}
\begin{proof}
Equivalence of parts~\ref{item.valuation.pp'.iff.1} and~\ref{item.valuation.pp'.iff.2} follows from Lemmas~\ref{lemm.MP.iff} and~\ref{lemm.f.ment.iff.f.ment} (a direct calculation from the truth-table in Figure~\ref{fig.3} is also straightforward).

Equivalence of parts~\ref{item.valuation.pp'.iff.2} and~\ref{item.valuation.pp'.iff.3} follows noting from Definition~\ref{defn.f^ment} that $\avaluation^\ment=\avaluation^\mone\{\tvT,\tvB\}$ so that 
$p\in\avaluation^\ment\liff p\notin\avaluation^\mone\{\tvF\}$ 
and 
$p'\in\avaluation^\ment\liff p'\notin\avaluation^\mone\{\tvF\}$. 
\end{proof}

\begin{corr}
\label{corr.tall.MP.iff}
Suppose $\ns P$ is any set and $\avaluation:\ns P\to\THREE$ is a valuation, and suppose $\phi,\phi'\in\tf{Pred3}(\ns P)$ and $\fv(\phi)\cup\fv(\phi')\subseteq\{x\}$, so that $\tall x.\phi\tiff\phi'\in\tf{ClosedPred3}(\ns P)$.
Then the following are equivalent:
\begin{enumerate*}
\item\label{item.tall.MP.iff.base}
$\avaluation\ment \tall x.(\phi\tiff \phi')$.
\item\label{item.tall.iff.is.iff}
$\avaluation\ment \phi[x\ssm p]\liff \avaluation\ment\phi'[x\ssm p]$ for every $p\in\ns P$.
\item\label{item.tall.MP.iff.set}
$\{p \mid \avaluation\ment \phi[x\ssm p]\} = \{p \mid \avaluation\ment \phi'[x\ssm p]\}$. 
\item\label{item.tall.MP.iff.tb}
$\{p \mid \model{\phi[x\ssm p]}_\avaluation\in\{\tvT,\tvB\}\} = \{p \mid \model{\phi'[x\ssm p]}_\avaluation\in\{\tvT,\tvB\}\}$. 
\item\label{item.tall.MP.iff.f}
$\{p \mid \model{\phi[x\ssm p]}_\avaluation=\tvF\} = \{p \mid \model{\phi'[x\ssm p]}_\avaluation=\tvF\}$. 
\end{enumerate*}
\end{corr}
\begin{proof}
Equivalence of parts~\ref{item.tall.MP.iff.base} and~\ref{item.tall.iff.is.iff} is from Lemmas~\ref{lemm.MP.iff} and~\ref{lemm.tand.valid.iff}.
Equivalence of parts~\ref{item.tall.iff.is.iff} and~\ref{item.tall.MP.iff.set} is a fact of sets.
Equivalence of parts~\ref{item.tall.MP.iff.set} and~\ref{item.tall.MP.iff.tb} is immediate from Definition~\ref{defn.ment}(\ref{item.f.phi.valid}). 
Equivalence of parts~\ref{item.tall.MP.iff.tb} and~\ref{item.tall.MP.iff.f} follows since $\THREE=\{\tvT,\tvB,\tvF\}$.
\end{proof}

\jamiesubsection{A sequent system}
\label{subsect.three.sequent}

In this short subsection we develop a sequent system for our logic that is sound and complete with respect to validity.

The system here follows the style of the simple and elegant sequent system presented in~\cite{wintein:gencnt}.
There are important differences: they consider a four-valued purely propositional logic, whereas we have a three-valued logic but with far richer connectives including modalities and quantifiers.
So the sequent system here can be viewed as an elaboration of a simplification of the one in~\cite{wintein:gencnt}.

But in a way, the details do not matter. 
What matters is that a legitimate notion of derivation is shown to exist, so that we can talk meaningfully about derivability and proof-search.

\begin{defn}
For this Subsection, we fix some finite set $\ns P$ of propositional constant, as per Definition~\ref{defn.predicate.syntax}(\ref{item.propositional.constant}).
\end{defn}

\begin{defn}
\begin{enumerate*}
\item
Define \deffont{tags $\tvsTB$, $\tvsFF$, $\tvsFB$, and $\tvsTT$}\index{$\{\tvsTB,\tvsFF,\tvsFB,\tvsTT\}$ (tags)} by: 
$$
\tvsTB=\{\tvT,\tvB\}
\quad
\tvsFF=\{\tvF\}
\quad 
\tvsFB=\{\tvF,\tvB\}
\quad
\tvsTT=\{\tvT\}
$$
We will let $\atag$ and $\btag$ range over tags.
\item
A \deffont{tag-sequent $\Sigma$}\index{$\Sigma$ (tag-sequent)} is a finite set of pairs $\atag:\phi$ where $\atag$ is a tag and $\phi\in\tf{ClosedPred3}(\ns P)$ is a closed predicate (Definition~\ref{defn.predicate.syntax}(\ref{item.closed.pred3})).
\item
Define a $\f{neg}$ operation on tags such that $\f{neg}(t)=\{\tneg x\mid x\in t\}$.
Spelling this out:
$$
\f{neg}(\tvsTB) = \tvsFB
\quad
\f{neg}(\tvsFB) = \tvsTB
\quad
\f{neg}(\tvsFF) = \tvsTT
\quad
\f{neg}(\tvsTT) = \tvsFF
$$
\item
The \deffont{derivable tag-sequents} are defined inductively by the sequent rules in Figure~\ref{fig.3.sequents}.

Note that Figure~\ref{fig.3.sequents} presents a subsystem of the logic, but this is fine because we can derive rules for $\modTB$, $\modB$, $\texi x$, $\tor$, $\tnotor$, $\tlatticeiff$, $\timp$, $\tiff$, and $\Emod{}$, 
as per the following equivalences:
$$
\begin{array}{r@{\ }l@{\quad}r@{\ }l@{\quad}r@{\ }l}
\modTB\phi =& \tneg\modT\tneg\phi
&
\modB\phi =& \modTB\phi\tand\tneg\modT\phi 
&
\texi x.\phi=& \tneg\tall x.\tneg\phi
\\
\phi\tnotor\phi'=&(\tneg\phi)\tor\phi'
&
\phi\tlatticeiff\phi'=&(\phi\tnotor\phi')\tand(\phi'\tnotor\phi)
&
\phi\tor\phi'=&\tneg((\tneg\phi)\tand(\tneg\phi'))
\\
\phi\timp\phi' =&(\modTB\phi)\tnotor\phi'
&
\phi\tiff\phi' =& (\phi\timp\phi')\tand(\phi'\timp\phi)
&
\Emod{}\phi=& \tneg\Emod{}\tneg\phi
\end{array}
$$
\item
Write $\cent\Sigma$ for the judgement `$\Sigma$ is derivable'. 
\end{enumerate*}
\end{defn}

\begin{figure}
$$
\begin{array}{c@{}c}
\begin{prooftree}
\big((\atag,\btag)\in \{(\tvsTB,\tvsFF), (\tvsFB,\tvsTT)\big)
\justifies
\cent\Sigma, \atag:\phi, \btag:\phi
\using\rulefont{Ax}
\end{prooftree}
&
\begin{prooftree}
\cent\Sigma, \f{neg}(\atag):\phi
\justifies
\cent\Sigma, \atag:\tneg\phi
\using\rulefont{\tneg LR}
\end{prooftree}
\\[5ex]
\begin{prooftree}
\cent\Sigma,\atag:\phi 
\quad
\cent\Sigma,\atag:\psi 
\quad
(\atag{\in}\{\tvsTB,\tvsTT\})
\justifies
\cent\Sigma,\atag:\phi\tand\psi 
\using\rulefont{\tand R}
\end{prooftree}
&
\begin{prooftree}
\cent\Sigma,\atag:\phi,\atag:\psi 
\quad
(\atag{\in}\{\tvsFB,\tvsFF\})
\justifies
\cent\Sigma,\atag:\phi\tand\psi 
\using\rulefont{\tand L}
\end{prooftree}
\\[5ex]
\begin{prooftree}
\cent\Sigma,\tvsTT:\phi
\quad
(\atag{\in}\{\tvsTB,\tvsTT\})
\justifies
\cent\Sigma,\atag:\modT\phi
\using\rulefont{\modT R}
\end{prooftree}
&
\begin{prooftree}
\cent\Sigma,\tvsFF:\phi,\tvsFB:\phi
\quad
(\atag{\in}\{\tvsFB,\tvsFF\})
\justifies
\cent\Sigma,\atag:\modT\phi
\using\rulefont{\modT L}
\end{prooftree}
\\[5ex]
\begin{prooftree}
\cent\atag:\phi
\quad
(\atag{\in}\{\tvsTB,\tvsTT\})
\justifies
\cent\Sigma,\atag:\Kmod{}(\phi)
\using\rulefont{\Kmod{}\hspace{-1pt}R}
\end{prooftree}
&
\begin{prooftree}
\ncent\atag:\phi
\ \ 
((\atag,\btag){\in}\{(\tvsTB,\tvsFF),(\tvsTT,\tvsFB)\})
\justifies
\cent\Sigma,\btag:\Kmod{}(\phi)
\using\rulefont{\Kmod{}\hspace{-1pt}L}
\end{prooftree}
\\[5ex]
\begin{prooftree}
\cent\Sigma,\atag:\phi[x\ssm p]
\quad
(\atag{\in}\{\tvsTT,\tvsTB\}, \text{every $p$})
\justifies
\cent\Sigma,\atag:\tall x.\phi
\using\rulefont{\tall R}
\end{prooftree}
&
\begin{prooftree}
\cent\Sigma, \atag:\phi[x\ssm p]
\quad
(\atag{\in}\{\tvsFB,\tvsFF\})
\justifies
\cent\Sigma,\atag:\tall x.\phi
\using\rulefont{\tall L}
\end{prooftree}
\end{array}
$$
\caption{Derivable tag-sequents}
\label{fig.3.sequents}
\end{figure}

\begin{rmrk}
\label{rmrk.wd}
The derivation rule \rulefont{\Kmod{}\hspace{-1pt}L} is unusual because it has $\ncent$ above the line.
This is still a well-defined inductive definition, by induction on the syntax of the tag-sequent.
\end{rmrk}

\begin{lemm}
\label{lemm.sc.clauses}
Suppose $\phi,\psi\in\tf{ClosedPred3}(\ns P)$ and suppose $\avaluation:\ns P\to\THREE$ is any valuation.
We check properties corresponding to soundness and completeness of the rules in Figure~\ref{fig.3.sequents}:
\begin{enumerate*}
\item
\declaresoundness{Ax}{(\atag,\btag)=(\tvsTB,\tvsFF)} 

$\model{\phi}_\avaluation\in\tvsTB \lor \model{\phi}_\avaluation\in\tvsFF$ is a fact.
\item
\declaresoundness{Ax}{(\atag,\btag)=(\tvsFB,\tvsTT)} 

$\model{\phi}_\avaluation\in\tvsFB \lor \model{\phi}_\avaluation\in\tvsTT$ is a fact.
\item
\declaresoundnessshort{\tneg R}

$\model{\phi}_\avaluation\in x$ if and only if $\model{\tneg\phi}_\avaluation\in \f{neg}(x)$.
\item
\declaresoundness{\tand R}{\atag=\tvsTB} 

$\model{\phi}_\avaluation\in \tvsTB \land \model{\psi}_\avaluation\in\tvsTB$ if and only if $\model{\phi\tand\psi}_\avaluation\in\tvsTB$.
\item
\declaresoundness{\tand R}{\atag=\tvsTT}
 
$\model{\phi}_\avaluation\in \tvsTT \land \model{\psi}_\avaluation\in\tvsTT$ if and only if $\model{\phi\tand\psi}_\avaluation\in\tvsTT$.
\item
\declaresoundness{\tand L}{\atag=\tvsFB}
 
$\model{\phi}_\avaluation\in\tvsFB \lor \model{\psi}_\avaluation\in\tvsFB$ if and only if $\model{\phi\tand\psi}_\avaluation\in\tvsFB$.
\item
\declaresoundness{\tand L}{\atag=\tvsFF}
 
$\model{\phi}_\avaluation\in\tvsFF \lor \model{\psi}_\avaluation\in\tvsFF$ if and only if $\model{\phi\tand\psi}_\avaluation\in\tvsFF$.
\item
\declaresoundness{\modT R}{x\in\{\tvsTB,\tvsTT\}}

$\model{\phi}_\avaluation=\tvT$ if and only if $\model{\modT\phi}_\avaluation=\tvT$ if and only if $\model{\modT\phi}_\avaluation\in\tvsTB$ if and only if $\model{\modT\phi}_\avaluation\in\tvsTT$.
\item
\declaresoundness{\modT L}{x\in\{\tvsFB,\tvsFF\}}

$\model{\phi}_\avaluation\in\tvsFB \lor \model{\phi}_\avaluation\in\tvsFF$ if and only if $\model{\phi}_\avaluation\neq\tvT$ if and only if $\model{\modT\phi}_\avaluation=\tvF$ if and only if $\model{\modT\phi}_\avaluation\in\tvsFF$ if and only if $\model{\modT\phi}_\avaluation\in\tvsFB$.
\item
\declaresoundness{\Kmod{}\hspace{-1pt}R}{\atag}

If $\atag\in\{\tvsTB,\tvsTT\}$ then $\Forall{\avaluation'}\model{\phi}_{\avaluation'}\in\atag$ if and only if $\model{\Kmod{}\phi}_\avaluation\in\atag$.
\item
\declaresoundness{\Kmod{}\hspace{-1pt}L}{\atag}

If $\atag\in\{\tvsFB,\tvsFF\}$ then $\Exists{\avaluation'}\model{\phi}_{\avaluation'}\in\atag$ if and only if $\model{\Kmod{}\phi}_\avaluation\in\atag$.
\item
\declaresoundness{\tall R}{\atag=\tvsTB}

$\model{\phi[x\ssm p]}_\avaluation\in\tvsTB$ for every $p$, if and only if $\model{\tall x.\phi}_\avaluation\in\tvsTB$.
\item
\declaresoundness{\tall R}{\atag=\tvsTT}

$\model{\phi[x\ssm p]}_\avaluation=\tvsTT$ for every $p$, if and only if $\model{\tall x.\phi}_\avaluation=\tvsTT$.
\item
\declaresoundness{\tall L}{\atag=\tvsFB}

$\model{\phi[x\ssm p]}_\avaluation\in\tvsFB$ for some $p$, if and only if $\model{\tall x.\phi}_\avaluation\in\tvsFB$.
\item
\declaresoundness{\tall L}{\atag=\tvsFF}

$\model{\phi[x\ssm p]}_\avaluation=\tvF$ for some $p$, if and only if $\model{\tall x.\phi}_\avaluation=\tvF$.
\end{enumerate*}
\end{lemm}
\begin{proof}
These are all facts of the definition of $\model{\text{-}}_\avaluation$ from Figure~\ref{fig.3.phi.f} and the truth-tables from Figure~\ref{fig.3}.
\end{proof}

\begin{defn}
\label{defn.3.sequent.ment}
Suppose $\phi\in\tf{ClosedPred3}(\ns P)$ and $\atag$ is a tag and $\Sigma$ is a tag-sequent.
Then:
\begin{enumerate*}
\item
Write $\model{\atag:\phi}_\avaluation$ when $\model{\phi}_\avaluation\in \atag$.
\item
Write $\ment\Sigma$ when for every valuation $f:\ns P\to\THREE$, there exists some element $\atag:\phi$ in $\Sigma$ such that $\model{\atag:\phi}_\avaluation$.
\item
When $\ment\Sigma$ holds, we call $\Sigma$ a \deffont{valid tag-sequent}.
\end{enumerate*}
\end{defn}

\begin{prop}
Suppose $\Sigma$ is a tag-sequent.
Then we have:
\begin{enumerate*}
\item
\emph{Soundness:} \quad If $\cent\Sigma$ then $\ment\Sigma$.
\item
\emph{Completeness:}\quad If $\ment\Sigma$ then $\cent\Sigma$.
\end{enumerate*}
Thus, $\Sigma$ is derivable if and only if it is valid.
\end{prop}
\begin{proof}
We prove both by by a simultaneous induction on the syntax of tag-sequences:\footnote{An induction on derivations will not work, because of the $\ncent$ in \rulefont{\Kmod{}\hspace{-1pt}L} observed in Remark~\ref{rmrk.wd}.  However, the derivation rules are all syntax-directed, meaning that they reduce the size of a tag-sequent when read bottom-up, so we can work by induction on syntax.}
\begin{enumerate}
\item
By unpacking Definition~\ref{defn.3.sequent.ment} and using the relevant clause of Lemma~\ref{lemm.sc.clauses} from left-to-right.
\item
By unpacking Definition~\ref{defn.3.sequent.ment} and using the relevant clause of Lemma~\ref{lemm.sc.clauses} from right-to-left. 
\qedhere\end{enumerate}
\end{proof}

\jamiesection{Axiomatisations}
\label{sect.axiomatisations}

\jamiesubsection{Theory arising from a witness function}

\begin{defn}
\label{defn.Ax}
Suppose $\witness:\ns P\to\mathcal W(\ns P)$ is a witness function on a finite set $\ns P$ (Definition~\ref{defn.witnessed.set}(\ref{witness.function})).
\begin{enumerate*}
\item
Define the \deffont{axioms arising from $\witness$} to be sets of predicates in $\tf{ClosedPred3}(\ns P)$ as follows:
$$
\begin{array}{r@{\ }l@{\quad}l}
\tf{closedAx}_\witness(p) 
=&
\bigl( \tand_{w\in\witness(p)} \tor_{q\in w} q\bigr) \timp p
\\
\tf{closedAx}_\witness^\tneg(p) 
=&
\bigl( \tand_{w\in\witness(p)} \tor_{q\in w} \tneg q\bigr) \timp \tneg p
\\[2ex]
\tf{closedAx}_\witness=&\bigwedge\{\tf{closedAx}_\witness(p) \mid p\in\ns P\}
\\
\tf{closedAx}_\witness^\tneg=&\bigwedge\{\tf{closedAx}_\witness^\tneg(p) \mid p\in\ns P\}
\end{array}
$$
\item\label{item.Ax}
We can collect these axioms into a conjunction $\thyAxW$, which we call the \deffont{theory arising from $\witness$}:
$$
\thyAxW 
=
\tf{closedAx}_\witness\ \tand\ \tf{closedAx}_\witness^\tneg .
$$
\item
We may omit the $\witness$ annotation where this is unimportant or understood, writing (for example) $\tf{closedAx}_\witness^\tneg(p)$ just as $\tf{closedAx}^\tneg(p)$.
\end{enumerate*}
\end{defn}

\begin{rmrk}
We can read $\tf{closedAx}_\witness$ and $\tf{closedAx}_\witness^\tneg$ above as asserting that 
$$
\begin{array}{r@{\ }l}
\closure{\avaluation^\mone\{\tvT,\tvB\}}\subseteq&\avaluation^\mone\{\tvT,\tvB\}
\quad\text{and}
\\
\closure{\avaluation^\mone\{\tvF,\tvB\}}\subseteq&\avaluation^\mone\{\tvF,\tvB\}.
\end{array}
$$
The proofs below implicitly reflect this intuition, and make it formal.
\end{rmrk}

\begin{lemm}
\label{lemm.blocking.MP}
Suppose that:
\begin{itemize*}
\item
$\witness:\ns P\to \mathcal W(\ns P)$ is a witness function on a finite set $\ns P$.
\item
$p\in\ns P$ and $\blocking\subseteq\ns P$ blocks $p$ (Definition~\ref{defn.blocking.set}).
\item
$f:\ns P\to\THREE$ is a valuation and $f\ment\thyAxW$.
\end{itemize*}
Then:
\begin{enumerate*}
\item
$(\Forall{q{\in}\blocking}f\ment q)$ implies $f\ment p$.
\item
$(\Forall{q{\in}\blocking}f\ment \tneg q)$ implies $f\ment \tneg p$.
\end{enumerate*}
\end{lemm}
\begin{proof}
We consider each part in turn:
\begin{enumerate}
\item
Recall from Definition~\ref{defn.Ax} that 
$$
\tf{closedAx}_\witness(p)=\bigl( \tand_{w\in\witness(p)} \tor_{q\in w} q\bigr) \timp p,
$$
and $\tf{closedAx}_\witness(p)$ appears in $\thyAxW$ by Definition~\ref{defn.Ax}.

Choose one $w\in\witness(p)$ (by construction in Definition~\ref{defn.witnessed.set}(\ref{witness.function}) at least one such exists) and consider the disjunction $\tor_{q\in w}q$ for $w\in\witness(p)$ in $\tf{closedAx}_\witness$.
We assumed that $\blocking$ blocks $p$, so by assumption in Definition~\ref{defn.blocking.set} $\blocking\between w$.
We also assumed that $f\ment q$ for every $q\in\blocking$, so in particular $f\ment q$ for some $q\in w$, and thus by Lemma~\ref{lemm.tand.valid.iff} $f\ment\tor_{q\in w}q$.

Now this holds for \emph{every} $w\in\witness(p)$, so by Lemma~\ref{lemm.tand.valid.iff} $f\ment\tand_{w\in\witness(p)} \tor_{q\in w} q$.
We assumed $f\ment\tf{closedAx}_\witness(p)$, and by Proposition~\ref{prop.timp.MP}(\ref{item.timp.is.timp}) $f\ment p$ follows.
\item
The reasoning to show that $\Forall{q{\in}\blocking}f\ment \tneg q$ implies $f\ment \tneg p$ is precisely similar to the previous case, using $\tf{closedAx}_\witness^\tneg(p)=\bigl( \tand_{w\in\witness(p)} \tor_{q\in w} \tneg q\bigr) \timp \tneg p$.
\qedhere\end{enumerate}
\end{proof}

\begin{rmrk}
Note a curious thing:
\begin{itemize*}
\item
Definition~\ref{defn.trust.topology}(\ref{item.witness.semitopology}) characterises closed sets in the witness semitopology using a single property $p\blocks{\witness} P$.
\item
Definition~\ref{defn.Ax} contains \emph{two} axioms for each $p\in\ns P$: $\tf{closedAx}_\witness(p)$ and $\tf{closedAx}_\witness^\tneg(p)$.
\end{itemize*}
We can look at the definition of $p\blocks{\witness} P$ in Definition~\ref{defn.blocking.set}(\ref{item.p.blocks.P}), and we see that $\tf{closedAx}_\witness(p)$ visibly imitates it --- but what is $\tf{closedAx}_\witness^\tneg(p)$ there for?

A formal answer to this question is that the proof of Proposition~\ref{prop.Ax.iff.cont} --- which characterises those valuations $\avaluation:\ns P\to\THREE$ that are continuous on the witness semitopology $(\ns P,\opens(\witness))$ --- requires both axioms.

This formal answer raises a semi-formal question: why \emph{should} it require both axioms?
Because Definition~\ref{defn.Ax} (axiomatisation of continuity) has to work harder than Definition~\ref{defn.trust.topology}(\ref{item.witness.semitopology}) (definition of open/closed sets).
Defining an open (or a closed) set requires two truth-values (whether a point is in or not in that set) whereas the logic has three values and correspondingly a valuation $\avaluation:\ns P\to\THREE$ can return $\tvT$, $\tvF$, or $\tvB$.
This makes it a more complex entity than just an open (or closed) set.

In fact we saw in Remark~\ref{rmrk.comments.on.f.ment}(\ref{item.f.ment.is.char}) that a valuation $\avaluation$ can be thought of as a \emph{pair} of closed sets $\f{char}_{CC}(\avaluation)=(\avaluation^\ment,\avaluation^{\ment\tneg})=(\avaluation^\mone\{\tvT,\tvB\},\avaluation^\mone\{\tvF,\tvB\})$; see Definition~\ref{defn.char.function} and the surrounding discussion.
In this view, axiom $\tf{closedAx}_\witness(p)$ controls the behaviour of $\avaluation^\ment$, and $\tf{closedAx}_\witness^\tneg(p)$ controls the behaviour of $\avaluation^{\ment\tneg}$.
\end{rmrk}

\jamiesubsection{Axiomatisation of continuity}
\label{subsect.logical.continuity}

Recall from Definition~\ref{defn.Ax} $\thyAxW$ the theory arising from $\witness$.

\begin{lemm}
\label{lemm.Ax.to.cont}
Suppose $\witness:\ns P\to\mathcal W(\ns P)$ is a witness function on a finite set $\ns P$, and 
suppose $\avaluation:\ns P\to\THREE$ is a valuation.
Then 
\begin{itemize*}
\item
if $\avaluation\ment\thyAxW$ then 
\item
$\avaluation$ is continuous on the witness semitopology $(\ns P,\opens(\witness))$ from Definition~\ref{defn.trust.topology}(\ref{item.witness.semitopology}).
\end{itemize*}
\end{lemm}
\begin{proof}
Suppose $\avaluation\ment\thyAxW$.

To prove that $\avaluation$ is continuous, it suffices by Lemma~\ref{lemm.when.cont.3} to show that $\avaluation^\ment=\avaluation^\mone \{\tvT,\tvB\}$ and $\avaluation^{\ment\tneg}=\avaluation^\mone\{\tvF,\tvB\}$ are closed sets in $(\ns P,\opens(\witness))$. 
We consider each in turn:
\begin{itemize}
\item
\emph{We show that $\avaluation^\ment$ is closed.}\footnote{We cannot just use Lemma~\ref{lemm.f.ment.closed} to deduce this, because that Lemma requires us to know that $\avaluation$ is continuous, but this is what we are trying to prove.}

Following Definition~\ref{defn.trust.topology}(\ref{item.w.blocking}), $\avaluation^\ment$ is closed in $(\ns P,\opens(\witness))$ when for every $p\in\ns P$ and $\blocking\subseteq \avaluation^\ment$, if $\blocking$ blocks $p$ (Definition~\ref{defn.blocking.set}), then $p\in \avaluation^\ment$.

Suppose $p\in\ns P$ and suppose $\blocking\subseteq \avaluation^\ment$ for some blocking set for $p$, which by Lemma~\ref{lemm.f.ment.iff.f.ment} means precisely that $\avaluation\ment \blocking$. 
By assumption $\avaluation\ment \tf{Ax}(p)$, and it follows from Lemma~\ref{lemm.blocking.MP} that $\avaluation\ment p$, and so by Lemma~\ref{lemm.f.ment.iff.f.ment} $p\in \avaluation^\ment$ as required.
\item
\emph{We show that $\avaluation^\mone\{\tvF,\tvB\}$ is closed.}

This is just like the previous case, but using $\tf{closedAx}^\tneg(p)$.
\qedhere\end{itemize}
\end{proof}

\begin{lemm}
\label{lemm.cont.to.Ax}
Suppose $\witness:\ns P\to\mathcal W(\ns P)$ is a witness function on a finite set $\ns P$, and 
suppose $\avaluation:\ns P\to\THREE$ is a valuation.

Then if $\avaluation$ is continuous on the witness semitopology $(\ns P,\opens(\witness))$, then $\avaluation\ment\thyAxW$.
\end{lemm}
\begin{proof}
Suppose $\avaluation:\ns P\to \THREE$ is continuous.
By Definition~\ref{defn.\avaluation^ment} $\avaluation^\ment=\avaluation^\mone\{\tvT,\tvB\}$, and $\avaluation^{\ment\tneg}=\avaluation^\mone\{\tvF,\tvB\}$ are closed.

We consider $p\in\ns P$ and show that $\avaluation\ment \tf{closedAx}(p)$ and $\avaluation\ment \tf{closedAx}^\tneg(p)$.
By Proposition~\ref{prop.timp.MP}(\ref{item.timp.is.timp}) it would suffice to show that if $\avaluation\ment\tand_{w\in\witness(p)} \tor_{q\in w} q$ then $\avaluation\ment p$, and if $\avaluation\ment\tand_{w\in\witness(p)} \tor_{q\in w} \tneg q$ then $\avaluation\ment\tneg p$. 
\begin{itemize}
\item
\emph{Suppose $\avaluation\ment \tand_{w\in\witness(p)} \tor_{q\in w} q$.}\quad

Using Lemmas~\ref{lemm.tand.valid.iff} and~\ref{lemm.f.ment.iff.f.ment} this means precisely that there exists a blocking set $\blocking$ for $p$ such that $\blocking\subseteq \avaluation^\ment$.
We noted above that $f^\ment$ is closed, so that by Definition~\ref{defn.trust.topology}(\ref{item.w.blocking}) it follows from $B\subseteq \avaluation^\ment$ that $p\in\avaluation^\ment$, and so that $\avaluation\ment p$ as required.
\item
\emph{Suppose $\avaluation\ment \tand_{w\in\witness(p)} \tor_{q\in w} \tneg q$.}

By reasoning on $\avaluation^{\ment\tneg}$ precisely similar to the previous case, we deduce that $\avaluation\ment\tneg p$.
\qedhere\end{itemize}
\end{proof}

\begin{prop}
\label{prop.Ax.iff.cont}
Suppose $\witness:\ns P\to\mathcal W(\ns P)$ is a witness function on a finite set $\ns P$, and 
suppose $\avaluation:\ns P\to\THREE$ is a valuation.
Then the following are equivalent:
\begin{enumerate*}
\item
$\avaluation\ment\thyAxW$ (Definition~\ref{defn.Ax}). 
\item
$\avaluation$ is continuous on $(\ns P,\opens(\witness))$. 
\end{enumerate*}
In words we can write: 
\begin{quote}
$\thyAxW$ axiomatises continuity over the witness semitopology. 
\end{quote}
\end{prop}
\begin{proof}
The top-down implication is Lemma~\ref{lemm.Ax.to.cont}.
The bottom-up implication is Lemma~\ref{lemm.cont.to.Ax}.
\end{proof}

\begin{corr}
\label{corr.sc}
Suppose $\witness:\ns P\to\mathcal W(\ns P)$ is a witnessed function on a finite set $\ns P$.
Then 
$$
\ns P,\opens(\witness)\ment\phi
\quad\text{if and only if}\quad
\ns P\ment\thyAxW\timp \phi .
$$
\end{corr}
\begin{proof}
By Definition~\ref{defn.ment}(\ref{item.P.O.phi.valid}), $\ns P,\opens(\witness)\ment\phi$ holds when $\avaluation\ment\phi$ for every $\avaluation:\ns P\to\THREE$ that is continuous on $(\ns P,\opens(\witness))$.
By Propositions~\ref{prop.Ax.iff.cont} and~\ref{prop.timp.MP}(\ref{item.timp.is.timp}) and Lemma~\ref{lemm.tand.valid.iff}, this is precisely what $\ns P\ment\thyAxW\timp \phi$ expresses. 
\end{proof}

\begin{rmrk}[Alternative axiomatisation]
\label{rmrk.openAx}
An equivalent set of axioms to $\thyAxW$ from Definition~\ref{defn.Ax} is as follows:
$$
\begin{array}{r@{\ }l}
\tf{openAx}_\witness(p) =& \modT p \ \timp\  \tor_{w\in\witness(p)} \tand_{q\in w} \modT q
\\
\tf{openAx}_\witness^\tneg(p) =& \modT\tneg p \ \timp\  \tor_{w\in\witness(p)} \tand_{q\in w} \modT\tneg q
\\
\tf{OpenAx}(\witness) =& 
\bigwedge\bigl\{\tf{openAx}_\witness(p) \mid p\in\ns P\bigr\}
\ \tand\ 
\bigwedge\bigl\{\tf{openAx}_\witness^\tneg(p) \mid p\in\ns P\bigr\}
\end{array}
$$
These axioms characterise that $\avaluation^\mone\{\tvT\}$ and $\avaluation^\mone\{\tvF\}$ are open sets as per Definition~\ref{defn.blocking.set}(\ref{item.blocking.enables}), so by Lemma~\ref{lemm.when.cont.3} this characterises continuity of $\avaluation$, which is also what $\thyAxW$ does: thus we obtain a lemma that
$$
\avaluation\ment \tf{OpenAx}(\witness)
\quad\text{if and only if}\quad
\avaluation\ment\thyAxW.
$$
We leave filling in the details as an exercise to the reader.

Note a \emph{false} argument for \emph{false} axioms: it is not enough to claim that 
$p \timp \tor_{w\in\witness(p)} \tand_{q\in w} q$ is just the contrapositive of $\tf{closedAx}_\witness(p)$, and $\tf{closedAx}_\witness^\tneg(p)$ is just the contrapositive of $\tneg p \timp \tor_{w\in\witness(p)} \tand_{q\in w} \tneg q$, because by Lemma~\ref{lemm.no.contrapositive} we know that $\timp$ does not satisfy the contrapositive property. 

We use $\thyAxW$ because it is simpler than $\tf{OpenAx}(\witness)$, in the sense that the latter requires modalities.
This is because Definition~\ref{defn.ment}(\ref{item.f.phi.valid}) uses the set $\{\tvT,\tvB\}$ as designated values, and $\{\tvT,\tvB\}$ is a \emph{closed} set in the semitopology on $\THREE$ from Definition~\ref{defn.3.top}(\ref{item.3.top.closed}).
Thus, it is slightly more natural to use our logic to express topological properties in terms of closed sets.
\end{rmrk}

\jamiesubsection{Quantifying over continuous valuations, within the logic}

\jamiesubsubsection{Basic properties}

\begin{nttn}
\label{nttn.Kmod.witness}
Suppose $\witness:\ns P\to\mathcal W(\ns P)$ is a witness function on a finite set $\ns P$, and suppose $\phi\in\tf{Pred3}(\ns P)$.
Define $\Kmod{\witness}\phi$ and $\Emod{\witness}\phi$ by: 
$$
\begin{array}{r@{\ }l}
\Kmod{\witness}\phi
= &
\Kmod{}(\thyAxW\timp \phi)
\\
\Emod{\witness}\phi
= &
\Emod{}(\thyAxW\tand \phi) 
\end{array}
$$
\end{nttn}

\begin{prop}
\label{prop.Emod.char}
Suppose $\witness:\ns P\to\mathcal W(\ns P)$ is a witness function on a finite set $\ns P$, and 
suppose $\phi\in\tf{ClosedPred3}(\ns P)$ (Definition~\ref{defn.predicate.syntax}(\ref{item.closed.pred3})).\footnote{`Closed' means no free variables.  $\phi$ might still mention propositional atoms $p\in\ns P$.}
Suppose further that $\avaluation':\ns P\to\THREE$ is any valuation.
Then:
\begin{enumerate*}
\item\label{item.Kmod.char}
$\avaluation'\ment\Kmod{\witness}\phi$ if and only if $\avaluation\ment\phi$ for every continuous $\avaluation:(\ns P,\opens(\witness))\to\THREE$.
\item\label{item.Kmod.char.2}
$\avaluation'\ment\Kmod{\witness}\phi$ if and only if $\ns P,\opens(\witness)\ment \phi$.
\item\label{item.Emod.char}
$\avaluation'\ment\Emod{\witness}\phi$ if and only if $\avaluation\ment\phi$ for some continuous $\avaluation:(\ns P,\opens(\witness))\to\THREE$.
\end{enumerate*}
\end{prop}
\begin{proof}
We consider each part in turn:
\begin{enumerate}
\item
Following Definition~\ref{defn.ment}(\ref{item.f.phi.valid}), Figure~\ref{fig.3.phi.f}, and Notation~\ref{nttn.Kmod.witness}, it is routine to check that 
$\avaluation'\ment\Kmod{\witness}\phi$ precisely when 
$$
\Forall{\avaluation:\ns P\to\THREE}\avaluation\ment \thyAxW\timp\phi .
$$ 
By Propositions~\ref{prop.timp.MP}(\ref{item.timp.is.timp}) and~\ref{prop.Ax.iff.cont} this is equivalent to 
insisting that $\avaluation\ment \phi$ for every continuous $\avaluation:(\ns P,\opens(\witness))\to\THREE$, as required.
\item
From part~\ref{item.Kmod.char} of this result, using Definition~\ref{defn.ment}(\ref{item.P.O.phi.valid}).
\item 
By reasoning exactly similar to that used in part~\ref{item.Kmod.char}.
\qedhere\end{enumerate}
\end{proof}

Lemma~\ref{lemm.valuation.not.used} will be helpful later; see for instance Lemma~\ref{lemm.no.evaluation.intertwinedwithwitness}:
\begin{lemm}
\label{lemm.valuation.not.used}
Suppose $\witness:\ns P\to\mathcal W(\ns P)$ is a witness function on a finite set $\ns P$, and suppose $\phi\in\tf{ClosedPred3}(\ns P)$. 
Suppose $\avaluation,\avaluation':\ns P\to\THREE$ are valuations.
Then
$$
\begin{array}{r@{\ }l}
\avaluation\ment \Kmod{\witness}\phi
\liff&
\avaluation'\ment \Kmod{\witness}\phi 
,\quad\text{and}
\\
\avaluation\ment \Emod{\witness}\phi
\liff&
\avaluation'\ment \Emod{\witness}\phi . 
\end{array}
$$ 
In words: the valuation in which we evaluate $\Kmod{\witness}(\text{-})$ and $\Emod{\witness}(\text{-})$ is not relevant.
\end{lemm}
\begin{proof}
Routine from Proposition~\ref{prop.Emod.char}.
\end{proof}

\jamiesubsubsection{Valuation quantification treated as a modality}

Fix a witness function $\witness:\ns P\to\mathcal W(\ns P)$.
We saw in Proposition~\ref{prop.Ax.iff.cont} and Corollary~\ref{corr.sc} that continuity on the witness semitopology $(\ns P,\opens(\witness))$ can be captured in our logic as an axiom $\thyAxW$ (Definition~\ref{defn.logic}(\ref{item.logic.axiom})).

We can view $\Kmod{\witness}$ and $\Emod{\witness}$ as a pair of modalities, where 
\begin{itemize*}
\item
$\Kmod{\witness}$ behaves like a box modality, quantifying over all possible worlds, where possible worlds are valuations $\avaluation:\ns P\to\THREE$ that are continuous over $(\ns P,\opens(\witness))$, and 
\item
$\Emod{\witness}$ behaves like a diamond modality, quantifying over the existence of some possible world.
\end{itemize*}

\newcommand\wbox[0]{{\hspace{.5pt}\fboxsep=0pt\fbox{\color{lightgray}\rule{2mm}{2mm}}\hspace{1pt}}}
\newcommand\wdiamond[0]{\rotatebox[origin=c]{45}{$\wbox$}}

Just for this Subsection, we write
$$
\Kmod{\witness}
\ \text{as}\ \ \wbox
\qquad\text{and}\qquad
\Emod{\witness}
\ \text{as}\ \wdiamond
$$
and we fix a valuation $\avaluation:\ns P\to\THREE$ that is continuous on $(\ns P,\opens(\witness))$.

It is routine to check that $\wbox$ and $\wdiamond$ satisfy standard modal axioms, including:
\begin{enumerate*}
\item
Necessitation: If $\ns P,\opens(\witness)\ment\phi$ then $\ns P,\opens(\witness)\ment\wbox \phi$.
\item
Distribution / K: $\avaluation\ment\wbox(\phi\timp\phi')\timp\wbox\phi\timp\wbox\phi'$.
\item
Reflexivity / T: $\avaluation\ment(\wbox\phi)\timp\phi$.
\item
4: $\avaluation\ment (\wbox\phi)\timp\wbox\wbox\phi$.
\item
5: $\avaluation\ment\wdiamond\phi\timp\wbox\wdiamond\phi$.
\item
B: $\avaluation\ment \phi\timp\wbox\wdiamond\phi$
\end{enumerate*}
We recognise these as axioms for the well-known modal logic S5, which is equal to Necessitation + K + T + 5.\footnote{In the presence of the other axioms, axiom~5 is equivalent to axioms~4 and~B.} 

S5 is characteristic of Kripke structures in which the accessibility relation is an equivalence relation.
The proofs are easy, and it is nice to find S5 in our logic. 

What may be interesting to highlight here is that 
although our logical syntax from Figure~\ref{fig.predicate.syntax} assumes $\Kmod{}$ and $\Emod{}$ (without $\witness$), what actually interests us in here is to use this syntax to construct a family of modalities $\Kmod{\witness}$ and $\Emod{\witness}$; one for each witness function $\witness:\ns P\to\mathcal W(\ns P)$ and corresponding judgement-form $\ns P,\opens(\witness)\ment(\text{-})$.

So, it is the $\witness$-indexed modal structure $\Kmod{\witness}$ and $\Emod{\witness}$ obtained by quantifying over all continuous valuations on $(\ns P,\opens(\witness))$, rather than the raw syntax of $\Kmod{}$ and $\Emod{}$, that --- for now, at least --- we are using.

\jamiesection{Intertwined-ness \& regularity via three-valued logic}
\label{sect.logical.intertwinedwith}

We now set about characterising semitopological properties --- specifically, properties having to do with \emph{sets intersections}, like being intertwined --- inside our logic.
This is a first step to building a framework within which a participant can reason about network properties (or at least, about those parts that the participant can access information about).

\jamiesubsection{Logical characterisation of being intertwined}
\label{subsect.logical.intertwined}

\begin{rmrk}
In this Subsection we proceed in three steps:
\begin{enumerate*}
\item
Lemma~\ref{lemm.intertwinedwith.as.equiv} gives characterisations of points being interwined using closed sets, rather than the open sets used in Definition~\ref{defn.intertwined.points}(\ref{item.p.intertwinedwith.p'}).
\item
Lemma~\ref{lemm.char.intertwinedwith} uses this characterisation to express $p\intertwinedwith p'$ as a logical judgement $\ns P,\opens\ment p\tlatticeiff p'$.
\item
Notation~\ref{nttn.intertwined.pred} and Proposition~\ref{prop.logical.intertwined} use a $\Kmod{}$ binding to internalise this judgement as a predicate $\Kmod{\witness}(p\tlatticeiff p')$.
\end{enumerate*}
\end{rmrk}

\begin{lemm}
\label{lemm.intertwinedwith.as.equiv}
Suppose $(\ns P,\opens)$ is a semitopology and $p,p'\in\ns P$. 
Then the following are equivalent:
\begin{enumerate*}
\item\label{item.intertwinedwith.as.intertwined}\label{item.intertwinedwith.as.equiv.1}
$p\intertwinedwith p'$.
\item\label{item.intertwinedwith.as.equiv.2}
For every $O,O'\in\opens$, if $p\in O$ and $p'\in O'$ then $O\between O'$ (as per Definition~\ref{defn.intertwined.points}(\ref{item.p.intertwinedwith.p'})).
\item\label{item.intertwinedwith.as.equiv.3}
For every $C,C'\in\closed$, if $p\notin C$ and $p'\notin C'$ then $C\cup C'\neq\ns P$.
\item\label{item.intertwinedwith.as.equiv.4}
For every $C,C'\in\closed$, if $C\cup C'=\ns P$ then $p\in C$ or $p'\in C'$.
\item\label{item.intertwinedwith.as.equiv}\label{item.intertwinedwith.as.equiv.5}
For every $C,C'\in\closed$, if $C\cup C'=\ns P$ then $p,p'\in C$ or $p,p'\in C'$.
\end{enumerate*}
\end{lemm}
\begin{proof}
We check the assertions are equivalent by routine manipulations.

Equivalence of parts~\ref{item.intertwinedwith.as.equiv.1} and~\ref{item.intertwinedwith.as.equiv.2} was observed already in Definition~\ref{defn.intertwined.points}(\ref{item.p.intertwinedwith.p'}).

By Lemma~\ref{lemm.closed.complement.open} part~\ref{item.intertwinedwith.as.equiv.3} just restates part~\ref{item.intertwinedwith.as.equiv.2} using closed sets instead of open sets.
Part~\ref{item.intertwinedwith.as.equiv.4} is then just the contrapositive of part~\ref{item.intertwinedwith.as.equiv.3}.

We now prove equivalence of parts~\ref{item.intertwinedwith.as.equiv.4} and~\ref{item.intertwinedwith.as.equiv.5} (the more interesting part is to show that part~\ref{item.intertwinedwith.as.equiv.4} implies part~\ref{item.intertwinedwith.as.equiv.5}):
\begin{itemize}
\item
\emph{Assume that for every $C,C'\in\closed$, if $C\cup C'=\ns P$ then $p\in C\lor p'\in C'$.}

Suppose $C,C'\in\closed$ and $C\cup C'=\ns P$.
By assumption $p\in C\lor p'\in C'$; suppose $p\in C$ (the case of $p'\in C'$ is symmetric).

Now $C\cup C'=\ns P$ so $p'\in C$ or $p'\in C'$.
If $p'\in C$ then we have $p,p'\in C$ and we are done.
Otherwise, we have $p'\notin C$ and thus $p'\in C'$ --- and we have $p'\notin C$ and thus by our assumption (since $C'\cup C=\ns P$ and $p'\notin C$) also $p\in C'$.

Thus we have $p,p'\in C'$ and we are done. 
\item 
\emph{Assume for every $C,C'\in\closed$, if $C\cup C'=\ns P$ then $p,p'\in C\lor p,p'\in C'$.}

Suppose $C,C'\in\closed$ and $C\cup C'=\ns P$.
Clearly $p,p'\in C$ implies $p\in C$ and $p,p'\in C'$ implies $p'\in C'$, so $p\in C\lor p'\in C'$ is immediate.
\qedhere
\end{itemize}
\end{proof}

\begin{lemm}
\label{lemm.char.intertwinedwith}
Suppose $(\ns P,\opens)$ is a semitopology and $p,p'\in\ns P$.
Then the following are equivalent: 
\begin{enumerate*}
\item\label{item.char.intertwinedwith.1}
$p\intertwinedwith p'$ in $(\ns P,\opens)$.
\item\label{item.char.intertwinedwith.2}
$\ns P,\opens\ment p\tlatticeiff p'$.
\item\label{item.char.intertwinedwith.3}
$\ns P,\opens\ment (p\tand p') \tor (\tneg p\tand \tneg p')$.
\item\label{item.char.intertwinedwith.4}
$\ns P,\opens\ment (p\tor p') \tnotor (p\tand p')$.
\end{enumerate*}
\end{lemm}
\begin{proof}
Equivalence of parts~\ref{item.char.intertwinedwith.2}, \ref{item.char.intertwinedwith.3}, and~\ref{item.char.intertwinedwith.4} is routine from the equivalences in Figure~\ref{fig.3.equivalences} and Lemma~\ref{lemm.checking.truth.tables} (or just by checking truth-tables from Figure~\ref{fig.3}).
 
Equivalence of parts~\ref{item.char.intertwinedwith.1} and~\ref{item.char.intertwinedwith.3} is just by unfolding definitions and checking truth-tables, but we spell out the details.
We consider two implications:
\begin{itemize}
\item
\emph{Suppose $p\intertwinedwith p'$ in $(\ns P,\opens)$.}
We wish to show that $\ns P,\opens\ment (p\tand p') \tor (\tneg p\tand\tneg p')$.

Consider a valuation $\avaluation:\ns P\to\THREE$ that is continuous on $(\ns P,\opens)$, and consider $\avaluation^\ment=\avaluation^\mone\{\tvT,\tvB\}$ and $\avaluation^{\ment\tneg}=\avaluation^\mone\{\tvF,\tvB\}$ from Definition~\ref{defn.\avaluation^ment}.
These are closed sets in $(\ns P,\opens)$ by Lemma~\ref{lemm.f.ment.closed}.

We note that $\avaluation^\ment\cup \avaluation^{\ment\tneg}=\ns P$, and so by Lemma~\ref{lemm.intertwinedwith.as.equiv}(\ref{item.intertwinedwith.as.intertwined}\&\ref{item.intertwinedwith.as.equiv}) (since $p\intertwinedwith p'$) $p,p'\in \avaluation^\ment$ or $p,p'\in \avaluation^{\ment\tneg}$.
It follows using Lemma~\ref{lemm.tand.valid.iff} that $\avaluation\ment (p\tand p') \tor (\tneg p'\tand\tneg p)$.

Now $\avaluation$ was an arbitrary continuous valuation on $(\ns P,\opens)$, so the result follows by Definition~\ref{defn.ment}(\ref{item.P.O.phi.valid}).
\item
Suppose $\ns P,\opens\ment (p\tand p') \tor (\tneg p\tand \tneg p')$.

Consider $C,C'\in\closed$ such that $C\cup C'=\ns P$; by Lemma~\ref{lemm.intertwinedwith.as.equiv}(\ref{item.intertwinedwith.as.intertwined}\&\ref{item.intertwinedwith.as.equiv}) it would suffice to show that $p,p'\in C$ or $p,p'\in C'$.

We define $\avaluation=\indicator{C_{\tvT\tvB},C_{\tvF\tvB}}(C,C'):\ns P\to\THREE$ from Definition~\ref{defn.indicator.functions}, so $\avaluation(\ns P\setminus C)=\tvT$ and $\avaluation(\ns P\setminus C')=\tvF$ and $\avaluation(C\cap C')=\tvB$, and $C=\avaluation^\ment$ and $C'=\avaluation^{\ment\tneg}$.
This is well-defined since $C\cup C'=\ns P$, and continuous by Lemma~\ref{lemm.indicator.continuous}.
Thus, $\avaluation\ment (p\tand p') \tor (\tneg p\tand \tneg p')$.

By Lemma~\ref{lemm.tand.valid.iff}, we see that at least one of $f\ment p\land f\ment p'$ or $f\ment \tneg p\land f\ment \tneg p'$ must hold.
By Lemma~\ref{lemm.f.ment.iff.f.ment}, either $p,p'\in C$ or $p,p'\in C'$, as required.
\qedhere\end{itemize}
\end{proof}

\begin{nttn}
\label{nttn.intertwined.pred}
Suppose $\witness:\ns P\to\mathcal W(\ns P)$ is a witness function on a finite set $\ns P$, and 
suppose $p,p'\in\ns P$.
Then define a predicate $p\intertwinedwithwitness p'$\index{$p\intertwinedwithwitness p'$ (logical predicate)} in $\tf{ClosedPred3}(\ns P)$ by 
$$
p\intertwinedwithwitness p' \ =\ \Kmod{\witness}(p\tlatticeiff p') .
$$
\end{nttn}

We make an elementary observation:
\begin{lemm}
\label{lemm.no.evaluation.intertwinedwithwitness}
Suppose $\witness:\ns P\to\mathcal W(\ns P)$ is a witness function on a finite set $\ns P$ and $p,p'\in\ns P$, and suppose $\avaluation,\avaluation':\ns P\to\THREE$ are any valuations.
Then
$$
\avaluation\ment p\intertwinedwithwitness p'
\quad\text{if and only if}\quad
\avaluation'\ment p\intertwinedwithwitness p' .
$$ 
In words: $\avaluation\ment p\intertwinedwithwitness p'$ --- the validity of $p\intertwinedwithwitness p'$ --- depends on $p$, $p'$, and $\witness$.
It does not depend on $\avaluation$. 
\end{lemm}
\begin{proof}
Direct from Lemma~\ref{lemm.valuation.not.used}.
\end{proof}

\begin{prop}
\label{prop.logical.intertwined}
Suppose $\witness:\ns P\to\mathcal W(\ns P)$ is a witness function on a finite set $\ns P$, and suppose $\avaluation':\ns P\to\THREE$ is any valuation. 
Then the following are equivalent:
\begin{enumerate*}
\item\label{item.logical.intertwined.1}
$p\intertwinedwith p'$ in $(\ns P,\opens(\witness))$ (Definition~\ref{defn.intertwined.points}(\ref{item.p.intertwinedwith.p'})).
\item\label{item.logical.intertwined.2}
$\ns P,\opens(\witness)\ment p\tlatticeiff p'$.
\item\label{item.logical.intertwined.3}
$\avaluation'\ment p\intertwinedwithwitness p'$ ($\avaluation'$ does not matter, as per Lemma~\ref{lemm.no.evaluation.intertwinedwithwitness}).
\end{enumerate*}
\end{prop}
\begin{proof}
We prove two equivalences:
\begin{itemize}
\item
Equivalence of parts~\ref{item.logical.intertwined.1} and~\ref{item.logical.intertwined.2} is just Lemma~\ref{lemm.char.intertwinedwith}(\ref{item.char.intertwinedwith.1}\&\ref{item.char.intertwinedwith.2}), for the witness semitopology $(\ns P,\opens(\witness))$.
\item
To prove equivalence of parts~\ref{item.logical.intertwined.1} and~\ref{item.logical.intertwined.3}, we unpack Notation~\ref{nttn.intertwined.pred} and use Proposition~\ref{prop.Emod.char}(\ref{item.Kmod.char}) to see that 
$\avaluation'\ment p\intertwinedwithwitness p'$ when $\avaluation\ment p\tlatticeiff p'$ for every $\avaluation:\ns P\to\THREE$ that is continuous on $(\ns P,\opens(\witness))$.
We use Lemma~\ref{lemm.char.intertwinedwith}.
\qedhere\end{itemize}
\end{proof}

\begin{rmrk}
\label{rmrk.not.so.easy}
By Lemmas~\ref{lemm.valuation.not.used} and~\ref{lemm.no.evaluation.intertwinedwithwitness}, the valuation $\avaluation'$ is not used to evaluate $\avaluation'\ment p\intertwinedwithwitness p'$ in Proposition~\ref{prop.logical.intertwined}. 
The valuation $\avaluation'$ is there because the validity judgement 
`$\avaluation'\ment p\intertwinedwithwitness p'$' from Definition~\ref{defn.ment}(\ref{item.f.phi.valid}) requires us to provide a valuation.\footnote{This happens in first-order logic too, for example with $\avaluation\ment x\teq y \tiff y\teq x$.}

The reader might ask why in the statement of Proposition~\ref{prop.topindis}(\ref{item.topindis.3}) we do not replace 
\begin{itemize*}
\item
the judgement $\avaluation'\ment p\intertwinedwithwitness p'$ (Definition~\ref{defn.ment}(\ref{item.f.phi.valid})) with 
\item
the judgement $\ns P\ment  p\intertwinedwithwitness p'$ (Definition~\ref{defn.ment}(\ref{item.P.phi.valid})).
\end{itemize*}
Because this is a different result. 
It is the difference between proving 
$\Forall{\avaluation'}(A\tiff B)$ --- which permits us to replace $A$ with $B$, given some $\avaluation'$ --- and 
$A\tiff\Forall{\avaluation'}B$, which does not.
\end{rmrk}

Corollary~\ref{corr.logical.tall.intertwinedwith} will be useful later.
It shows how to characterise more sophisticated inclusions and equalities between $\avaluation^\ment$ and $\intertwined{p}$ 
($\avaluation^\ment$ is from Definition~\ref{defn.f^ment}; $p\intertwinedwithwitness x$ is from Definition~\ref{defn.intertwined.points}(\ref{intertwined.defn})):
\begin{corr}
\label{corr.logical.tall.intertwinedwith}
Suppose $\witness:\ns P\to\mathcal W(\ns P)$ is a witness function on a finite set $\ns P$, and 
suppose $p\in\ns P$ and $\avaluation:\ns P\to\THREE$ is a valuation.
Then: 
\begin{enumerate*}
\item\label{item.logical.tall.intertwinedwith.1}
$\avaluation\ment \tall x.x\tiff (p\intertwinedwithwitness x)$ if and only if $\avaluation^\ment=\intertwined{p}$.
\item\label{item.logical.tall.intertwinedwith.2}
$\avaluation\ment \tall x.x\timp (p\intertwinedwithwitness x)$ if and only if $\avaluation^\ment\subseteq\intertwined{p}$.
\item\label{item.logical.tall.intertwinedwith.3}
$\avaluation\ment \tall x.(p\intertwinedwithwitness x)\timp x$ if and only if $\intertwined{p}\subseteq\avaluation^\ment$.
\end{enumerate*}
\end{corr}
\begin{proof}
We consider each part in turn:
\begin{enumerate}
\item
Using Corollary~\ref{corr.tall.MP.iff}, $\avaluation\ment \tall x.x\tiff (p\intertwinedwithwitness x)$ holds when 
$$
\{p'\in\ns P \mid \avaluation\ment p'\} = \{p'\in\ns P \mid \avaluation\ment p\intertwinedwithwitness p'\} .
$$
By Definition~\ref{defn.f^ment}, the left-hand side of this equality is just $\avaluation^\ment$.

By Proposition~\ref{prop.logical.intertwined} $\avaluation\ment p\intertwinedwithwitness p'$ if and only if $p\intertwinedwith p'$, so 
the right-hand side of this equality is $\{p'\in\ns P \mid p\intertwinedwith p'\}=\intertwined{p}$ (Definition~\ref{defn.intertwined.points}(\ref{intertwined.defn})).

The result follows.
\item
Using Proposition~\ref{prop.timp.MP}(\ref{item.timp.is.timp}) and Definition~\ref{defn.f^ment}, $\avaluation\ment \tall x.x\timp (p\intertwinedwithwitness x)$ holds when
$$
\avaluation^\ment \subseteq \{p'\in\ns P \mid \avaluation\ment p\intertwinedwithwitness p'\} .
$$
We use Proposition~\ref{prop.logical.intertwined} just as we did in part~\ref{item.logical.tall.intertwinedwith.1} of this result.
\item
Precisely as for part~\ref{item.logical.tall.intertwinedwith.2} of this result.
\qedhere\end{enumerate}
\end{proof}

\jamiesubsection{Logical characterisation of topological indistinguishability}
\label{subsect.logical.topind}

By Proposition~\ref{prop.logical.intertwined}, $\Kmod{\witness}(p\tlatticeiff p')$ characterises being intertwined in $(\ns P,\opens(\witness))$.
But recall that Figure~\ref{fig.3} features two connectives for logical equivalence: $\tlatticeiff$ and $\tiff$.
In this Subsection we briefly consider the natural question: what does $\Kmod{\witness}(p\tiff p')$ characterise?
The answer is very simple: this is topological indistinguishability.

Recall from Definition~\ref{defn.topind} the notion of \emph{topological indistinguishability} $p\topind p'$.
This has a beautiful logical characterisation:
\begin{lemm}
\label{lemm.logical.topind.char}
Suppose $(\ns P,\opens)$ is a semitopology and $p,p'\in\ns P$.
Then the following are equivalent:
\begin{enumerate*}
\item
$p\topind p'$.
\item 
$\ns P,\opens\ment p\tiff p'$.
\end{enumerate*}
\end{lemm}
\begin{proof}
This result looks like it should have a trivial proof.
Not so: the proof is simple, but not trivial.
We prove two implications:
\begin{itemize}
\item
Suppose $p\topind p'$ and consider a continuous $\avaluation:\ns P\to\THREE$.
By assumption 
$p\in \avaluation^\mone\{\tvF\}\liff p'\in \avaluation^\mone\{\tvF\}$, so by Corollary~\ref{corr.valuation.pp'.iff}(1\&3) $\avaluation\ment p\tiff p'$ as required.
\item 
Suppose $\ns P,\opens\ment p\tiff p'$ and consider some $O\in\opens$.
We set $\avaluation=\indicator{O_\tvT,O_\tvF}(\varnothing,O)$ --- using $\indicator{O_\tvT,O_\tvF}(O,\varnothing)$ will not give us what we need --- so $\avaluation$ maps a point to $\tvF$ if it is in $O$, and to $\tvB$ otherwise.
By Corollary~\ref{corr.valuation.pp'.iff}(1\&3), $p\in O\liff p'\in O$ as required.
\qedhere\end{itemize}
\end{proof}

\begin{defn}
Suppose $\witness:\ns P\to\mathcal W(\ns P)$ is a witness function on a finite set $\ns P$, and suppose $p,p'\in\ns P$.
Then define a predicate $\tf{TopIndis}_\witness(p,p')$ in $\tf{ClosedPred3}(\ns P)$ by:
$$
\tf{TopIndis}_\witness(p,p') \ =\ \Kmod{\witness}(p\tiff p') .
$$
\end{defn}

\begin{prop}
\label{prop.topindis}
Suppose $\witness:\ns P\to\mathcal W(\ns P)$ is a witness function on a finite set $\ns P$ and $p,p'\in\ns P$, and suppose $\avaluation':\ns P\to\THREE$ is any valuation.
Then the following are equivalent:
\begin{enumerate*}
\item\label{item.topindis.1}
$p\topind p'$ in $(\ns P,\opens(\witness))$.
\item\label{item.topindis.2}
$\ns P,\opens(\witness)\ment p\tiff p'$.
\item\label{item.topindis.3}
$\avaluation'\ment \tf{TopIndis}_\witness(p,p')$ 

(By Lemma~\ref{lemm.valuation.not.used}, $\avaluation'$ is not used to evaluate $\avaluation'\ment\tf{TopIndis}_\witness(p,p')$.  See Remark~\ref{rmrk.not.so.easy}.)
\end{enumerate*}
\end{prop}
\begin{proof}
Equivalence of parts~\ref{item.topindis.1} and~\ref{item.topindis.2} is just Lemma~\ref{lemm.logical.topind.char}.
Equivalence of parts~\ref{item.topindis.2} and~\ref{item.topindis.3} is direct from Proposition~\ref{prop.Emod.char}(\ref{item.Kmod.char.2}).
\end{proof}

\jamiesubsection{Logical characterisation of $\interior(\avaluation^\ment)\neq\varnothing$}

To characterise (weak/quasi-)regularity, which we do below, we will need to logically characterise when $\avaluation$ takes value $\tvT$ at some point $p$, and when $\avaluation^\ment$ has a nonempty open interior and so is a closed neighbourhood.
These properties are closely related, and we study them in this Subsection:

\begin{lemm}
\label{lemm.logical.p.T}
Suppose $(\ns P,\opens)$ is a semitopology and $\avaluation:\ns P\to\THREE$ is continuous and $p\in\ns P$.
Then: 
\begin{enumerate*}
\item\label{item.logical.Box.char}
$\avaluation\ment \modT p$ if and only if $p\in \avaluation^\mone\{\tvT\}$.
\item\label{item.logical.Box.interior}
If $\avaluation\ment\modT p$ then $p\in\interior(\avaluation^\ment)$.
\item\label{item.logical.Box.closed.neighbourhood}
If $\avaluation\ment\modT p$ then $\avaluation^\ment$ is a closed neighbourhood of $p\in\ns P$ (a closed set with $p$ in its open interior; Definition~\ref{defn.cn}(\ref{item.closed.neighbourhood.of.p})). 
\item\label{item.logical.Box.no.converse}
The converse implication to part~\ref{item.logical.Box.closed.neighbourhood} need not hold: it is possible that $p\in\interior(\avaluation^\ment)$ yet $\avaluation\not\ment\modT p$ --- later on in Corollary~\ref{corr.precise.inverse.image} we will identify those valuations for which the reverse implication does hold, as an interesting class of \emph{extremal} valuations. 
\end{enumerate*}
\end{lemm}
\begin{proof}
We consider each part in turn:
\begin{enumerate}
\item
Following Definition~\ref{defn.ment}(\ref{item.f.phi.valid}) and the rules in Figures~\ref{fig.3.phi.f} and~\ref{fig.3}, $\avaluation\ment\modT p$ when $\avaluation(p)=\tvT$.
It is a fact that $\avaluation(p)=\tvT$ when $p\in\avaluation^\mone\{\tvT\}$.
\item
Suppose $\avaluation\ment\modT p$.
Then $p\in\avaluation^\mone\{\tvT\}$ by part~\ref{item.logical.Box.char} of this result and
$\avaluation^\mone\{\tvT\}\in\opens$  by Corollary~\ref{corr.valuation.open.char}(\ref{item.valuation.open.char}).
By construction
$\avaluation^\mone\{\tvT\}\subseteq \avaluation^\ment=\avaluation^\mone\{\tvT,\tvB\}$, so $\avaluation^\mone\{\tvT\}\subseteq \interior(\avaluation^\ment)$.
Thus we have
$$
p\in\avaluation^\mone\{\tvT\} \subseteq\interior(\avaluation^\ment)
$$
as required.
\item
By Lemma~\ref{lemm.f.ment.closed} $\avaluation^\ment$ is a closed set.
The result now just follows from part~\ref{item.logical.Box.interior} of this result. 
\item
Consider the \deffont{singleton semitopology $\tf{One} =\{\ast\}$}\index{$\tf{One}$ (singleton semitopology)} with $\opens=\{\varnothing,\tf{One}\}$.
Note that $\{\ast\}$ is a closed neighbourhood, with open interior $\{\ast\}$.
Define $\avaluation:\tf{One}\to\THREE$ by $\avaluation(\ast)=\tvB$. 
Then $\avaluation$ is continuous and $\avaluation^\ment=\{\ast\}$, but $\avaluation\not\ment\modT \ast$. 
\qedhere\end{enumerate}
\end{proof}

\begin{prop}
\label{prop.texi.modT.x.cn}
Suppose $(\ns P,\opens)$ is a semitopology and $\avaluation:\ns P\to\THREE$ is a valuation that is closed on $(\ns P,\opens)$.
Then:
\begin{enumerate*}
\item\label{item.texi.modT.iff}
$\avaluation\ment\texi x.\modT x$ if and only if $\avaluation^\mone\{\tvT\}\neq\varnothing$.
\item\label{item.texi.modT.cn}
If $\avaluation\ment\texi x.\modT x$ then $\avaluation^\ment$ is a closed neighbourhood.
\item
It is possible for $\avaluation^\ment$ to be a closed neighbourhood, yet $\avaluation\not\ment\texi x.\modT x$.
\end{enumerate*}
\end{prop}
\begin{proof}
We consider each part in turn:
\begin{enumerate}
\item
Suppose $\avaluation\ment\texi x.\modT x$.
By Lemma~\ref{lemm.tall.valid.iff} this is if and only if there exists some $p\in\ns P$ such that $\avaluation\ment \modT p$.
By Lemma~\ref{lemm.logical.p.T}(\ref{item.logical.Box.char}) $p\in \avaluation^\mone\{\tvT\}$, so that $\avaluation^\mone\{\tvT\}\neq\varnothing$.

The reverse implication follows just reversing this reasoning.
\item
Suppose $\avaluation\ment\texi x.\modT x$.
By Lemma~\ref{lemm.f.ment.closed} $\avaluation^\ment=\avaluation^\mone\{\tvT,\tvB\}$ is a closed set, and by part~\ref{item.texi.modT.iff} of this result it has a nonempty open interior. 
\item
See the counterexample in Lemma~\ref{lemm.logical.p.T}(\ref{item.logical.Box.no.converse}).
\qedhere\end{enumerate}
\end{proof}

\begin{corr}
\label{corr.char.closed.neighbourhood}
Suppose $(\ns P,\opens)$ is a semitopology and $C\subseteq\closed$ is a closed set.
Define
$$
\avaluation=\indicator{O_\tvT,C_{\tvT\tvB}}(\interior(C),C)
$$ 
and note from the construction of $\indicator{O_\tvT,C_{\tvT\tvB}}(\interior(C),C)$ in Definition~\ref{defn.indicator.functions}) that $\avaluation^\ment=C$. 
Then the following are equivalent:
\begin{enumerate*}
\item
$C$ is a closed neighbourhood (closed set with nonempty open interior; Definition~\ref{defn.cn}(\ref{item.closed.neighbourhood})).
\item\label{item.char.closed.neigbourhood.modT}
$\indicator{O_\tvT,C_{\tvT\tvB}}(O,C)\ment \texi x.\modT x$ . 
\end{enumerate*}
\end{corr}
\begin{proof}
By an easy argument from the results so far:
\begin{itemize}
\item
Suppose $C$ is a closed neighbourhood, meaning by Definition~\ref{defn.cn}(\ref{item.closed.neighbourhood}) that $\interior(C)\neq\varnothing$.
Choose $p\in\interior(C)$ and note that $\avaluation(p)=\tvT$.
By Proposition~\ref{prop.texi.modT.x.cn}(\ref{item.texi.modT.iff}), $\avaluation\ment \texi x.\modT x$ as required.
\item
Suppose $\avaluation\ment \texi x.\modT x$.
By Proposition~\ref{prop.texi.modT.x.cn}(\ref{item.texi.modT.cn}), $\avaluation^\ment$ is a closed neighbourhood. 
\qedhere\end{itemize}
\end{proof}

\jamiesubsection{Logical characterisation of being quasiregular}

By Definition~\ref{defn.tn}(\ref{item.quasiregular.point}), a point is quasiregular when its community is nonempty.
We express this as follows:
\begin{defn}
\label{defn.logical.quasiregular}
Suppose $\witness:\ns P\to\mathcal W(\ns P)$ is a witness function on a finite set $\ns P$, and suppose $p\in\ns P$.
We define a predicate $\tf{QuasiRegular}_\witness(p)$ by:
$$
\tf{QuasiRegular}_\witness(p) \ =\  \Emod{\witness}(\texi x.\modT x \land \tall x.x\tiff p\intertwinedwithwitness x) .
$$
\end{defn}

\begin{prop}
\label{prop.quasiregular.valuation}
Suppose $\witness:\ns P\to\mathcal W(\ns P)$ is a witness function on a finite set $\ns P$ and $p\in\ns P$, and suppose $\avaluation':\ns P\to\THREE$ is any valuation.
Then the following are equivalent:
\begin{enumerate*}
\item\label{item.quasiregular.valuation.1}
$p$ is quasiregular in $(\ns P,\opens(\witness))$.
\item\label{item.quasiregular.valuation.2}
There exists a valuation $\avaluation:\ns P\to\THREE$ such that 
$$
\avaluation\ment\thyAxW
\quad\text{and}\quad
\avaluation\ment \texi x.\modT x
\quad\text{and}\quad
\avaluation\ment \tall x.x\tiff p\intertwinedwithwitness x.
$$
\item\label{item.quasiregular.valuation.3}
$\avaluation'\ment \tf{Quasiregular}_\witness(p)$. 

(By Lemma~\ref{lemm.valuation.not.used}, $\avaluation'$ is not used to evaluate $\avaluation'\ment\tf{QuasiRegular}_\witness(p)$.  See Remark~\ref{rmrk.not.so.easy}.)
\end{enumerate*}
\end{prop}
\begin{proof}
Equivalence of parts~\ref{item.quasiregular.valuation.2} and~\ref{item.quasiregular.valuation.3} is just by unpacking Definition~\ref{defn.logical.quasiregular} and using Proposition~\ref{prop.Emod.char}(\ref{item.Emod.char}) and Lemma~\ref{lemm.tand.valid.iff}.

Note from Proposition~\ref{prop.views.of.quasiregularity} that $p$ is quasiregular when $\intertwined{p}$ is a closed neighbourhood (Definition~\ref{defn.cn}(\ref{item.closed.neighbourhood})). 
We prove equivalence of parts~\ref{item.quasiregular.valuation.1} and~\ref{item.quasiregular.valuation.2} by proving two implications:
\begin{itemize}
\item
\emph{Suppose $p$ is quasiregular, so that $\intertwined{p}$ is a closed neighbourhood and $\community(p)\neq\varnothing$.}

We set $\avaluation=\indicator{O_\tvT,C_{\tvT\tvB}}(\community(p),\intertwined{p}):\ns P\to\THREE$; so $\avaluation$ returns $\tvT$ on $\community(p)=\interior(\intertwined{p})$ --- which is nonempty since we assumed that $\intertwined{p}$ is a closed neighbourhood --- and $\tvB$ on $\boundary(\intertwined{p})=\intertwined{p}\setminus\community(p)$, and $\tvF$ on $\ns P\setminus\intertwined{p}$.
Then:
\begin{itemize*}
\item
By Lemma~\ref{lemm.indicator.continuous} $\avaluation$ is continuous on $(\ns P,\opens(\witness))$, so by Proposition~\ref{prop.Ax.iff.cont} $\avaluation\ment \thyAxW$.
\item
By Proposition~\ref{prop.texi.modT.x.cn}(\ref{item.texi.modT.iff}) (right-to-left direction, since $\avaluation^\mone\{\tvT\}=\community(p)\neq\varnothing$) $\avaluation\ment \texi x.\modT x$.
\item
By Corollary~\ref{corr.logical.tall.intertwinedwith}(\ref{item.logical.tall.intertwinedwith.1}) $\avaluation\ment \tall x.x\tiff p\intertwinedwithwitness x$ holds when $\avaluation^\ment=\intertwined{p}$. 
But we defined 
$\avaluation=\indicator{O_\tvT,C_{\tvT\tvB}}(\community(p),\intertwined{p})$, so this is a fact.
\end{itemize*}
\item
\emph{Suppose $\avaluation\ment\thyAxW$ and $\avaluation\ment \texi x.\modT x$ and $\avaluation\ment \tall x.x\tiff p\intertwinedwithwitness x$.}

By Proposition~\ref{prop.Ax.iff.cont} $\avaluation$ is continuous from $(\ns P,\opens(\witness))$ to $\THREE$.
Write $C=\avaluation^\ment=\avaluation^\mone\{\tvT,\tvB\}$.
By Proposition~\ref{prop.texi.modT.x.cn}(\ref{item.texi.modT.iff}) (left-to-right direction) $C$ is a closed neighbourhood.
By Corollary~\ref{corr.logical.tall.intertwinedwith} $C=\intertwined{p}$.
Thus $p$ is quasiregular.
\qedhere\end{itemize}
\end{proof}

\jamiesubsection{Logical characterisation of being weakly regular}

By Definition~\ref{defn.tn}(\ref{item.weakly.regular.point}), a point is weakly regular when it is an element of its own community. 
We express this as follows:
\begin{defn}
Suppose $\witness:\ns P\to\mathcal W(\ns P)$ is a witness function on a finite set $\ns P$, and suppose $p\in\ns P$.
We define a predicate $\tf{WeaklyRegular}_\witness(p)$ by:
$$
\tf{WeaklyRegular}_\witness(p) \ =\  \Emod{\witness}(\modT p \land \tall x.x\tiff (p\intertwinedwithwitness x)) .
$$
\end{defn}

\begin{prop}
\label{prop.logical.weakly.regular.char}
Suppose $\witness:\ns P\to\mathcal W(\ns P)$ is a witness function on a finite set $\ns P$ and $p\in\ns P$, and suppose $\avaluation':\ns P\to\THREE$ is any valuation.
Then the following are equivalent:
\begin{enumerate*}
\item
$p$ is weakly regular in $(\ns P,\opens(\witness))$.
\item
$\avaluation'\ment\tf{WeaklyRegular}_\witness(p)$. 

(By Lemma~\ref{lemm.valuation.not.used}, $\avaluation'$ is not used to evaluate $\avaluation'\ment\tf{WeaklyRegular}_\witness(p)$.  See Remark~\ref{rmrk.not.so.easy}.)
\end{enumerate*}
\end{prop}
\begin{proof}
We unpack what $\avaluation'\ment\tf{WeaklyRegular}_\witness(p)$ means. 
By Proposition~\ref{prop.Emod.char}(\ref{item.Emod.char}) this means that there exists a valuation $\avaluation:\ns P\to\witness$ that is continuous on $(\ns P,\opens(\witness))$ such that
$\avaluation^\ment=\avaluation^\mone\{\tvT,\tvB\}$ is a closed neighbourhood of $p$ (by Lemma~\ref{lemm.logical.p.T}(\ref{item.logical.Box.closed.neighbourhood})), and $\avaluation^\ment$ is equal to $\intertwined{p}$ (by Corollary~\ref{corr.logical.tall.intertwinedwith}).
But by Proposition~\ref{prop.views.of.regularity}(\ref{item.intertwined.p.closed.neighbourhood.of.p}) this precisely means that $p$ is weakly regular; the equivalence follows.
\end{proof}

\jamiesubsection{Logical characterisation of being unconflicted}

By Definition~\ref{defn.conflicted}(\ref{item.unconflicted}), a point is unconflicted when the `intertwined with' relation is transitive at that point.
We express this as follows:
\begin{defn}
Suppose $\witness:\ns P\to\mathcal W(\ns P)$ is a witness function on a finite set $\ns P$, and suppose $p\in\ns P$.
We define a predicate $\tf{Unconflicted}_\witness(p)$ by
$$
\tf{Unconflicted}_\witness(p) \ =\ \tall x',x''. (x\intertwinedwithwitness p \,\tand\, p\intertwinedwithwitness x'') \limp x'\intertwinedwithwitness x'' .
$$
\end{defn}

\begin{prop}
\label{prop.logical.unconflicted.char}
Suppose $\witness:\ns P\to\mathcal W(\ns P)$ is a witness function on a finite set $\ns P$ and $p\in\ns P$, and suppose $\avaluation':\ns P\to\THREE$ is any valuation.
Then the following are equivalent:
\begin{enumerate*}
\item
$p$ is unconflicted in the witness semitopology $(\ns P,\opens(\witness))$.
\item
$\avaluation'\ment \tf{Unconflicted}_\witness(p)$.

(By Lemma~\ref{lemm.valuation.not.used}, $\avaluation'$ is not used to evaluate $\avaluation'\ment\tf{Unconflicted}_\witness(p)$.  See Remark~\ref{rmrk.not.so.easy}.)
\end{enumerate*}
\end{prop}
\begin{proof}
We unpack what $\avaluation'\ment\tf{Unconflicted}_\witness(p)$ means. 
Using Lemma~\ref{lemm.tall.valid.iff}, Proposition~\ref{prop.timp.MP}(\ref{item.timp.is.timp}), and Lemma~\ref{lemm.tand.valid.iff}, 
$\avaluation'\ment\tf{Unconflicted}_\witness(p)$ if and only if 
$$
\Forall{p',p''{\in}\ns P}(\avaluation\ment p'\intertwinedwithwitness p \land \avaluation\ment p\intertwinedwithwitness p'') \limp \avaluation\ment p'\intertwinedwithwitness p'' .
$$
We simplify with Proposition~\ref{prop.logical.intertwined} to obtain 
$$
\Forall{p',p''{\in}\ns P}(p'\intertwinedwith p \land p\intertwinedwith p'') \limp p'\intertwinedwith p'' .
$$
But as per Definition~\ref{defn.conflicted}(\ref{item.unconflicted}), this is precisely what being unconflicted means.
\end{proof}

\jamiesubsection{Logical characterisation of being regular}

\begin{rmrk}
We will consider two characterisations of being regular in Definition~\ref{defn.logical.regular}:
\begin{enumerate*}
\item
By Theorem~\ref{thrm.r=wr+uc} a point is regular when it is weakly regular and unconflicted.
This is expressed by $\tf{Regular}_\witness$ in Definition~\ref{defn.logical.regular}.
\item
By Theorem~\ref{thrm.up.down.char} a point $p$ is regular when it is weakly regular and furthermore $\intertwined{p}$ is a minimal closed neighbourhood.
This is expressed by $\tf{Regular}'_\witness$ in Definition~\ref{defn.logical.regular}.
\end{enumerate*}
We translate both into our logic.
\end{rmrk}

\begin{defn}
\label{defn.logical.regular}
Suppose $\witness:\ns P\to\mathcal W(\ns P)$ is a witness function on a finite set $\ns P$, and suppose $p\in\ns P$.
We define predicates $\tf{Regular}_\witness(p)$ and $\tf{Regular}'_\witness(p)$ by:
$$
\begin{array}{r@{\ }l}
\tf{Regular}_\witness(p) =&
\tf{WeaklyRegular}_\witness(p) \tand \tf{Unconflicted}_\witness(p)
\\
\tf{Regular}'_\witness(p) =&
\tf{WeaklyRegular}_\witness(p) 
\tand 
\\
&\ \Kmod{\witness} (\texi x.\modT x)\timp (\tall x'.x'\timp (p\intertwinedwithwitness x'))\timp \tall x'.((p\intertwinedwithwitness x') \timp x')
\end{array}
$$
\end{defn}

\begin{prop}
Suppose $\witness:\ns P\to\mathcal W(\ns P)$ is a witness function on a finite set $\ns P$ and $p\in\ns P$, and suppose $\avaluation':\ns P\to\THREE$ is any valuation. 
Then the following are equivalent:
\begin{enumerate*}
\item
$p$ is regular in $(\ns P,\opens(\witness))$.
\item
$\avaluation'\ment \tf{Regular}_\witness(p)$.

(By Lemma~\ref{lemm.valuation.not.used}, $\avaluation'$ is not used to evaluate $\avaluation'\ment\tf{Regular}_\witness(p)$.  See Remark~\ref{rmrk.not.so.easy}.)
\end{enumerate*}
\end{prop}
\begin{proof}
The argument is routine from the machinery we have build:
\begin{enumerate*}
\item
By Theorem~\ref{thrm.r=wr+uc}, $p$ is regular when it is weakly regular and unconflicted.
\item
By Lemma~\ref{lemm.tand.valid.iff} and Propositions~\ref{prop.logical.weakly.regular.char} and~\ref{prop.logical.unconflicted.char}, $\avaluation'\ment\tf{Regular}_\witness(p)$ holds when $p$ is weakly regular and unconflicted. 
\end{enumerate*}
The result follows.
\end{proof}

\begin{prop}
\label{prop.witness.valuation.regular}
Suppose $\witness:\ns P\to\mathcal W(\ns P)$ is a witness function on a finite set $\ns P$ and $p\in\ns P$, and suppose $\avaluation':\ns P\to\THREE$ is any valuation.
Then the following are equivalent:
\begin{enumerate*}
\item\label{item.witness.valuation.regular.1}
$p$ is regular in $(\ns P,\opens(\witness))$.
\item\label{item.witness.valuation.regular.2}
$p\in\community(p)$ and $\intertwined{p}$ is a minimal closed neighbourhood.
\item\label{item.witness.valuation.regular.3}
$\avaluation'\ment \tf{Regular}'_\witness(p)$.

(By Lemma~\ref{lemm.valuation.not.used}, $\avaluation'$ is not used to evaluate $\avaluation'\ment\tf{Regular}'_\witness(p)$.  See Remark~\ref{rmrk.not.so.easy}.)
\end{enumerate*}
\end{prop}
\begin{proof}
Equivalence of parts~\ref{item.witness.valuation.regular.1} and~\ref{item.witness.valuation.regular.2} just repeats Theorem~\ref{thrm.up.down.char}.
We now consider equivalence of parts~\ref{item.witness.valuation.regular.2} and~\ref{item.witness.valuation.regular.3}.
We prove two implications:
\begin{itemize}
\item
\emph{Suppose $\avaluation'\ment\tf{Regular}'_\witness(p)$.}

Unpacking Definition~\ref{defn.logical.regular} and using Lemma~\ref{lemm.tand.valid.iff}, we have:
$$
\begin{array}{r@{\ }l}
\avaluation'\ment&\tf{WeaklyRegular}_\witness(p) 
\quad\text{and}
\\
\avaluation'\ment&\Kmod{\witness} (\texi x.\modT x)\timp (\tall x'.x'\timp (p\intertwinedwithwitness x'))\timp \tall x'.((p\intertwinedwithwitness x') \timp x')
\end{array}
$$
By Proposition~\ref{prop.logical.weakly.regular.char} $p$ is weakly regular, which by Proposition~\ref{prop.views.of.regularity}(\ref{item.views.of.regularity.cn}) means that $\intertwined{p}$ is a minimal closed neighbourhood of $p$.

We will now show that $\intertwined{p}$ is a minimal closed neighbourhood (not just of a minimal closed neighbourhood of $p$).
Suppose $C\subseteq\intertwined{p}$ is some closed neighbourhood, and write $O=\interior(C)\neq\varnothing$.
Let $\avaluation=\indicator{O_\tvT,C_{\tvT\tvB}}(O,C)$ from Definition~\ref{defn.indicator.functions}, so $\avaluation(p)=\tvT$ when $p\in O$, and $\avaluation(p)=\tvB$ when $p\in C\setminus O$, and $\avaluation(p)=\tvF$ when $p\in \ns P\setminus C$.
By Lemma~\ref{lemm.indicator.continuous} $\avaluation$ is continuous on $(\ns P,\opens(\witness))$.
It follows from Proposition~\ref{prop.Emod.char}(\ref{item.Kmod.char}) that
$$
\avaluation\ment (\texi x.\modT x)\timp (\tall x'.x'\timp (p\intertwinedwithwitness x'))\timp \tall x'.((p\intertwinedwithwitness x') \timp x') .
$$
Now $C$ is a closed neighbourhood so by Corollary~\ref{corr.char.closed.neighbourhood} (since $\avaluation=\indicator{O_\tvT,C_{\tvT\tvB}}(O,C)$) $\avaluation\ment\texi x.\modT x$.
By assumption $C\subseteq\intertwined{p}$ so by Corollary~\ref{corr.logical.tall.intertwinedwith}(\ref{item.logical.tall.intertwinedwith.2}) $\avaluation\ment\tall x'.x'\timp (p\intertwinedwithwitness x')$.
Then by Proposition~\ref{prop.timp.MP}(\ref{item.timp.is.timp}) it follows that $\avaluation\ment\tall x'.(p\intertwinedwithwitness x')\timp x'$, and by Corollary~\ref{corr.logical.tall.intertwinedwith}(\ref{item.logical.tall.intertwinedwith.3}) $\intertwined{p}\subseteq C$.

Thus $C\subseteq\intertwined{p}$ implies $\intertwined{p}\subseteq C$, and so $\intertwined{p}$ is a minimal closed neighbourhood as required.
\item
\emph{Suppose $p\in\community(p)$ and $\intertwined{p}$ is a minimal closed neighbourhood.}

The reasoning is essentially by reversing the argument of the previous case, but we spell out the details.

By Definition~\ref{defn.tn}(\ref{item.weakly.regular.point}) $p\in\community(p)$ means precisely that $p$ is weakly regular, and by Proposition~\ref{prop.logical.weakly.regular.char} it follows that $\avaluation'\ment\tf{WeaklyRegular}_\witness(p)$.

By Propositions~\ref{prop.Emod.char}(\ref{item.Kmod.char}) and~\ref{prop.timp.MP}(\ref{item.timp.is.timp}) and Corollary~\ref{corr.logical.tall.intertwinedwith}(\ref{item.logical.tall.intertwinedwith.2}), to prove
$$
\Kmod{\witness} (\texi x.\modT x)\timp (\tall x'.x'\timp (p\intertwinedwithwitness x'))\timp \tall x'.((p\intertwinedwithwitness x') \timp x')
$$
it suffices to show for every $\avaluation:\ns P\to\THREE$ that is continuous on $(\ns P,\opens(\witness))$, if $\avaluation\ment\texi x.\modT x$ and $\avaluation^\ment\subseteq\intertwined{p}$, then $\intertwined{p}\subseteq\avaluation^\ment$.

So suppose $\avaluation\ment\texi x.\modT x$, so that $\avaluation^\ment$ is a closed neighbourhood by Proposition~\ref{prop.texi.modT.x.cn}(\ref{item.texi.modT.cn}), and suppose $\avaluation^\ment\subseteq\intertwined{p}$.
By minimality of $\intertwined{p}$ it follows that $\intertwined{p}\subseteq\avaluation^\ment$.

We use Lemma~\ref{lemm.tand.valid.iff} to deduce 
$$
\begin{array}{r@{\ }l}
\avaluation'\ment&\tf{WeaklyRegular}_\witness(p)\ \tand\ 
\\
&\quad\Kmod{\witness} (\texi x.\modT x)\timp (\tall x'.x'\timp (p\intertwinedwithwitness x'))\timp \tall x'.((p\intertwinedwithwitness x') \timp x')
\end{array}
$$
as required.
\qedhere\end{itemize}
\end{proof}

\jamiesection{Computational complexity \& logic programming}
\label{sect.logic.programming}

We now consider some computational aspects of semitopologies, specifically the following two questions:
\begin{enumerate*}
\item
What is the computational complexity of deciding whether two points are intertwined?
\item
What is a suitable notion of logic programming over $\THREE$? 
\end{enumerate*}
Many excellent treatments of logic programming over two-valued logic are available~\cite{miller:unipfl,lifshitz:whaasp,lifshitz:anssp}.
There is some literature on the complexity of checking satisfiability of many-valued propositional logics; for example~\cite{hahnle:commvl,hanikova:comcpf} consider propositional many-valued logics, and~\cite{vidal:trammv} considers the modal case.
I am not aware of any comprehensive surveys.

\jamiesubsection{Translation of SAT to intertwinedness problem}
\label{subsect.translation.intertwined.sat}

\begin{defn}
\label{defn.sat}
Suppose $\psi$ is a Boolean logic proposition over some set of \deffont{propositional atoms} $\ns Q$.
So $\psi$ is built from propositional atoms $q\in\ns Q$, with connectives $\tbot$, $\ttop$, $\tneg,$ $\tand$, and $\tor$. 
Suppose without loss of generality that $\psi$ is in \deffont{conjunctive normal form}, so 
$$
\psi=\bigwedge_{i\in I} \bigvee_{j\in i} a_{ij}q_{ij} 
$$
where
\begin{enumerate*}
\item
$I$ is an indexing set and each $i\in I$ is itself an indexing set of $j\in i$,
\item
each $q_{ij}\in \ns Q$ is a propositional atom, and 
\item
$a_{ij}$ indicates an \emph{arity}, which is either \deffont{positive arity} (meaning that $q_{ij}$ is unnegated) or \deffont{negative arity} (meaning that $q_{ij}$ is negated).
\end{enumerate*}
\end{defn}

\begin{nttn}
\label{nttn.pn.literal}
We may call the combination $a_{ij}q_{ij}$ of an arity and a propositional atom a \deffont[literal (arity + propositional atom)]{literal}. 
We may call the literal 
\deffont[positive literal]{positive} when its arity is positive (so $a_{ij}q_{ij}$ is just a propositional atom $q\in\ns Q$), and 
\deffont[negative literal]{negative} when its arity is negative (so $a_{ij}q_{ij}$ is a negated proposition atom $\tneg q$).
\end{nttn}

\begin{rmrk}
Definition~\ref{defn.P.psi} below is a little fiddly to write out.
The reader might like to look at it together with Proposition~\ref{prop.sat.as.notintertwined}, since that Proposition motivates the design of this Definition.
Intuitively in Definition~\ref{defn.P.psi}:
\begin{enumerate*}
\item
Points on the right are intended to be assigned `true', and points on the left are intended to be assigned `false'.
\item
A conjunctive restriction is represented by a point with one witness-set, which may have several elements.
\item
A disjunctive restriction is represented by a point with several witness-sets, each of which contains just one element.
\end{enumerate*}
Details of how this works are in Proposition~\ref{prop.sat.as.notintertwined}.
\end{rmrk}

\begin{defn}
\label{defn.P.psi}
We fix some data to help us map a proposition $\psi$ from Definition~\ref{defn.sat} to a witness function in the sense of Definition~\ref{defn.witnessed.set}(\ref{witness.function}), as follows:
\begin{enumerate*}
\item
Fix symbols $\f{left}$ and $\f{right}$.
\item
For each $i\in I$, fix a symbol $\f{right}_i$.
\item
For each $q\in\ns Q$, fix symbols $\ttplus{q}$ and $\ttminus{q}$, and fix symbols $\f{left}_q$ and $\f{right}_q$.
\item
Define a set of \deffont[points (of NP-completeness construction)]{points}
$$
\ns P=\{\f{left},\f{right}\}\cup\{\f{right}_i\mid i\in I\}\cup\bigcup_{q{\in}\ns Q}\{\ttplus{q},\ttminus{q},\f{left}_q,\f{right}_q\}.
$$
\item
Define a function $m$ from literals (unnegated or negated propositional atoms, as per Notation~\ref{nttn.pn.literal}) to points $\ns P$, such that 
$$
m(q)=\ttplus{q}
\quad\text{and}\quad m(\tneg q)=\ttminus{q}.
$$
\item
Define a witness function $\witness$ on $\ns P$ as follows:
\begin{enumerate*}
\item\label{item.plus.and.minus.q}
For each $q\in\ns Q$, let $\ttplus{q}$ and $\ttminus{q}$ have witness-sets 
$$
\witness(\ttplus{q}) = \{\,\{\ttplus{q}\}\,\}
\quad\text{and}\quad
\witness(\ttminus{q}) = \{\,\{\ttminus{q}\}\,\} .
$$
So each of $\ttplus{q}$ and $\ttminus{q}$ has just one witness-set, which is singleton itself.
\item\label{item.rightq}
For each $q\in\ns Q$, let $\f{left}_q$ and $\f{right}_q$ have witness-sets
$$
\witness(\f{left}_q) = 
\{ \{\ttplus{q}\},\ \{\ttminus{q}\}\} = \witness(\f{right}_q) .
$$
So each $\f{left}_q$ has two singleton witness-sets, and so does $\f{right}_q$.
\item\label{item.left}
Let $\f{left}$ have just one witness-set, given by
$$
\witness(\f{left}) = 
\bigl\{\ \{\f{left}_q \mid q\in\ns Q\}\ \bigr\} .
$$
\item\label{item.right}
Let $\f{right}\in\ns P$ have just one witness-set, given by
$$
\witness(\f{right}) = 
\bigl\{\ \{\f{right}_q \mid q\in\ns Q\}\cup\{\f{right}_i \mid i\in I\}\ \bigr\} .
$$
\item\label{item.righti}
For each $i\in I$, let $\f{right}_i$ have witness-sets given by
$$
\witness(\f{right}_i) = 
\bigl\{\ \{m(a_{ij}q_{ij})\} \mid j\in i\bigr\} .
$$
So for each $j\in i$, the singleton set $\{m(a_{ij}q_{ij})\}$ is a witness-set of $\f{right}_i$.
\end{enumerate*}
\end{enumerate*}
\end{defn}

\begin{prop}
\label{prop.sat.as.notintertwined}
Suppose that:
\begin{itemize*}
\item
$\psi$ is a Boolean logic proposition over propositional atoms $\ns Q$, that is in conjunctive normal form in the sense of Definition~\ref{defn.sat}.
\item
$\witness:\ns P\to\mathcal W(\ns P)$ is the witness function derived from $\psi$ as per Definition~\ref{defn.P.psi}.
\end{itemize*}
Then the following are equivalent:
\begin{enumerate*}
\item
$\psi$ is 2-satisfiable in (the usual) two-valued Boolean logic, as per Definition~\ref{defn.2Sat}(\ref{item.2Sat.1}).
\item
$\f{left}\notintertwinedwith\f{right}$.
\end{enumerate*}
\end{prop}
\begin{proof}
We prove two implications.
\begin{itemize}
\item
Suppose $\f{left}\notintertwinedwith\f{right}$, so there exist a disjoint pair of open neighbourhoods 
$$
\f{left}\in L\in\opens
\quad\text{and}\quad 
\f{right}\in R\in\opens.
$$
Now by the construction of the witness semitopology in Definition~\ref{defn.trust.topology}(\ref{item.witness.semitopology}) 
and by the design of $\witness$ in Definition~\ref{defn.P.psi}, we can make observations as follows:
\begin{enumerate*}
\item
By Definition~\ref{defn.P.psi}(\ref{item.right}) $\f{right}\in R$ has only one witness-set, so by Definition~\ref{defn.trust.topology}(\ref{item.witness.semitopology}) that witness-set must be a subset of $R$.
It follows that $\f{right}_q\in R$ for every $q\in\ns Q$, and $\f{right}_i\in R$ for every $i\in I$.
\item
By Definition~\ref{defn.P.psi}(\ref{item.left}) $\f{left}\in L$ has only one witness-set, so by Definition~\ref{defn.trust.topology}(\ref{item.witness.semitopology}) that witness-set must be a subset of $L$.
It follows that $\f{left}_q\in L$ for every $q\in\ns Q$.
\item
By Definition~\ref{defn.P.psi}(\ref{item.rightq}) each $\f{right}_q\in R$ has just two witness-sets, so (as for the cases above) at least one of those witness-sets must be a subset of $R$.
Therefore 
$$
\ttplus{q}\in R \lor \ttminus{q}\in R.
$$
By similar reasoning on the $\f{left}_q$ we have 
$$
\ttplus{q}\in L \lor \ttminus{q}\in L.
$$
But we assumed $L$ and $R$ are disjoint, so we conclude that 
$$
(\ttplus{q}\in R\land \ttminus{q}\in L \land \ttplus{q}\notin L \land \ttminus{q}\notin R)
\ \lor\ 
(\ttplus{q}\in L\land \ttminus{q}\in R \land \ttplus{q}\notin R \land \ttminus{q}\notin L) .
$$
\item
Finally, by Definitions~\ref{defn.P.psi}(\ref{item.righti}) and~\ref{defn.trust.topology}(\ref{item.witness.semitopology}), for $i\in I$ it follows from $\f{right}_i\in R$ that there exists some $j\in i$ such that $\{m(a_{ij}q_{ij})\}\subseteq R$, thus that $m(a_{ij}q_{ij})\in R$.
\end{enumerate*}
We can now take our satisfying assignment $\avaluation$ to be such that $\avaluation(q)=\tvF$ when $\ttplus{q}\in L$, and $\avaluation(q)=\tvT$ when $\ttplus{q}\in R$.

Intuitively, this interprets $q$ as true when its positive literal $\ttplus{q}$ is in $R$, and it interprets $q$ as false when its positive literal $\ttplus{q}$ is in $L$.
\item
Suppose we have a satisfying assignment $\avaluation:\ns Q\to\{\tvF,\tvT\}$.
Then we set 
$$
\hspace{-2em}\begin{array}{r@{\ }l}
L=&
\{\ttplus{q} \mid q\compactin\avaluation^\mone\{\tvF\}\}\cup\{\ttminus{q} \mid q\compactin\avaluation^\mone\{\tvT\}\} 
\cup\{\f{left}_q \mid q\compactin\ns Q\}\cup\{\f{left}\}
\quad\text{and}
\\
R=&
\{\ttplus{q} \mid q\compactin\avaluation^\mone\{\tvT\}\}\cup\{\ttminus{q} \mid q\compactin\avaluation^\mone\{\tvF\}\} 
\cup\{\f{right}_q \mid q\compactin\ns Q\}\cup\{\f{right}_i \mid i\compactin I\}\cup\{\f{right}\} .
\end{array}
$$
Intuitively, this puts the positive and negative literals that are false on the left, and the positive and negative literals that are true on the right. 
By the same reasoning as above, it is routine to check that $L$ and $R$ are a disjoint pair of open sets.
\qedhere\end{itemize}
\end{proof}

\begin{thrm}
\label{thrm.NP.complete}
The problem of determining whether two points are intertwined in a finite semitopology, is NP complete in general. 
\end{thrm}
\begin{proof}
Proposition~\ref{prop.sat.as.notintertwined} exhibits a reduction of SAT into an intertwinedness problem, and the reduction clearly runs in polynomial time (indeed, it is linear).
\end{proof}

\begin{rmrk}
\label{rmrk.practical.intertwined}
Theorem~\ref{thrm.NP.complete} shows that checking intertwinedness is NP complete in general.
However, this is not the last word about how difficult intertwinedness is to check in practice, since there may be practical use cases where the problem is easier.

For example, suppose we have a semitopology $(\ns P,\opens)$ of which we know that it contains a strongly transitive open set $K$.  
This may (for example) arise as a kernel atom (Definition~\ref{defn.kernel}(\ref{item.kernel})) of core participants who for historical reasons are considered reliable, whose witness functions have been relatively stable, and of which we can prove (using our sequent system from Subsection~\ref{subsect.three.sequent}, or by model-checking the witness function restricted just to those participants) that they are intertwined, consensus-equivalent, and hypertwined with each other (Definition~\ref{defn.ht.ce}).\footnote{English has many terms for such a $K$: clique, cabal, eminences, founding committee, in-crowd, and so on.  This is a fairly typical situation in real life.  For example, a secret to getting along in an organisation is to find out which small clique of well-connected members actually makes the decisions and gets things done --- this may or may not line up with any formal hierarchy of the organisation.  And, such a group tends in practice to be stable over time (sometimes infuriatingly so).
This is the magic of mathematics: we may not be able to cure bureaucracy, but at least we can build a topological model of it!}

Then to check whether the space $\ns P$ is intertwined (meaning by Notation~\ref{nttn.intertwined.space} that \emph{all} points in $\ns P$ are pairwise intertwined), it suffices to compute $\closure{K}$.
By Lemma~\ref{lemm.intertwined.iff.closure}, $\closure{K}=\ns P$ if and only if $\ns P$ is intertwined.

By the algorithm given in Remark~\ref{rmrk.computing.closed.sets}, the closure $\closure{K}$ can be computed efficiently.
\end{rmrk}

\jamiesubsection{Horn satisfiability over $\tf{Bool}$ and $\THREE$}

\jamiesubsubsection{The (standard) HORNSAT algorithm for Boolean logic}

We start with a brief but precise introduction to Horn clause theories and an algorithm to check their satisfiability in Boolean (two-valued) logic:
\begin{nttn}
\label{nttn.Horn}
A \deffont{literal} is a possibly negated propositional atom: $p$ or $\tneg p$.
We call $p$ a \deffont{positive literal} and $\tneg p$ a \deffont{negative literal}.
A \deffont{Horn clause} is a possibly empty disjunct of positive and negative literals, that contains at most one positive literal.
For example: 
\begin{itemize*}
\item
$\tneg p\tor \tneg p'\tor \tneg p''\tor q$ and $\tneg p\tor\tneg p'\tor\tneg p''$ are Horn clauses. 
\item
$p\tor q$ is not a Horn clause, because it contains two positive literals.
\item
$\varnothing$ the empty disjunct is a Horn clause.
We may call this \deffont{empty Horn clause}, and (slightly abusing notation) we may write this as $\tbot$.
\item
$p$ and $\tneg p$ --- which are clauses consisting of a single literal --- are Horn clauses.
We may call a Horn clause consisting of a single literal a \deffont{unit clause}.
\end{itemize*}
Call a finite set of Horn clauses a \deffont{Horn clause theory}.
\end{nttn}

\begin{defn}
\label{defn.2Sat}
\leavevmode
\begin{enumerate*}
\item\label{item.2Sat.1}
Call a propositional theory (which need not consist only of Horn clauses) \deffont[satisfiable propositional theory (two-valued)]{2-satisfiable}\index{2-satisfiable (propositional theory)} when there is a valuation $\avaluation:\ns P\to\{\tvT,\tvF\}$ that makes all the clauses true in two-valued propositional logic.
\item\label{item.2Sat.2}
Call a propositional theory \deffont[satisfiable propositional theory (three-valued)]{3-satisfiable}\index{3-satisfiable (propositional theory)} when there is a valuation $\avaluation:\ns P\to\THREE$ that makes all the clauses either $\tvB$ or $\tvT$ in $\THREE$.
\end{enumerate*}
\end{defn}

HORNSAT is the computational problem of deciding whether a Horn clause theory over two-valued logic is 2-satisfiable. 
This can be solved in linear time~\cite{dowling:lintat}:
\begin{rmrk}[HORNSAT algorithm]
\label{rmrk.HORNSAT.2}
An overview of the algorithm is as follows:
\begin{enumerate*}
\item
If the set contains a positive unit clause $p$, then:
\begin{enumerate*}
\item
delete every clause that contains that literal in positive form, except for $p$ itself, and
\item
delete $\tneg p$ from any of the remaining clauses.
\end{enumerate*} 
\item
Repeat the previous step until no further changes to the theory occur.
\item
If the resulting theory contains an empty Horn clause then fail; otherwise succeed.
\end{enumerate*}
If we succeed, then we can read a satisfying assignment from the result by mapping $p$ to $\tvT$ for every unit clause $p$ that remains, and mapping every other $p'$ to $\tvF$. 
\end{rmrk}

\jamiesubsubsection{A proposal for HORNSAT over $\THREE$}

We now sketch a proposal for Horn clause programming in $\THREE$, adapting Notation~\ref{nttn.pn.literal} (Horn clause theories) and Remark~\ref{rmrk.HORNSAT.2} (the HORNSAT algorithm) to the three-valued case.

The syntax and algorithm in Definition~\ref{defn.3horn} and Remark~\ref{rmrk.HORNSAT.3} are just pencil-and-paper prototypes, but they are simple and elegant and uncover no obvious difficulties, which suggests that they could be implemented and that more advanced systems could also be practical, like those discussed in~\cite{miller:unipfl,lifshitz:whaasp,lifshitz:anssp} but based on $\THREE$ instead of $\tf{Bool}$.

\begin{defn}[Horn clause theories and HORNSAT over $\THREE$]
\label{defn.3horn}
Fix a set $\ns P$.
\begin{enumerate*}
\item
\begin{enumerate*}
\item
A \deffont[boxed positive literal $\modT p$]{boxed positive literal} is a proposition of the form $\modT p$, for $p\in\ns P$.
\item
An \deffont[unboxed positive literal $p$]{unboxed positive literal} is a proposition of the form $p$, for $p\in\ns P$.
\item
A \deffont[positive literal (boxed $\modT p$; unboxed $p$)]{positive literal} is a boxed or unboxed positive literal.
\end{enumerate*}
\item
\begin{enumerate*}
\item
A \deffont[boxed negative literal $\modT\tneg p$]{boxed negative literal} is a proposition of the form $\modT\tneg p$, for $p\in\ns P$.
\item
An \deffont[unboxed negative literal $\tneg p$]{unboxed negative literal} is a proposition of the form $\tneg p$, for $p\in\ns P$.
\item
A \deffont[negative literal (boxed $\modT\tneg p$; unboxed $\tneg p$)]{negative literal} is a boxed or unboxed negative literal.
\end{enumerate*}
\item
A \deffont[literal (three-valued logic)]{literal} is a positive or negative literal.
\item
A \deffont{3Horn clause} is a possibly empty disjunct of literals, at most one of which is positive:
\begin{itemize*}
\item
When a 3Horn clause contains a positive literal, we call this the \deffont[head (of a 3Horn clause)]{head} of the Horn clause.
\item
When a 3Horn clause is empty, we call it \deffont{unsatisfiable 3Horn clause}.
\item
When a 3Horn clause is a singleton, we call it a \deffont[unit 3Horn clause]{unit clause}.
\end{itemize*}
\item
Echoing Notation~\ref{nttn.Horn}, call a finite set of 3Horn clauses a \deffont{3Horn clause theory}.
\end{enumerate*}
\end{defn}

\begin{rmrk}
Useful classes of propositions fit into the 3Horn clause syntax of Definition~\ref{defn.3horn}.
In particular, it follows from the equivalences in Figure~\ref{fig.3.equivalences} that:
\begin{itemize*}
\item
$p \tnotor q$ is equivalent to $\tneg p\tor q$, which is a Horn clause.
\item
$p \tlatticeiff q$ is equivalent to $(\tneg p\tor q)\tand (\tneg q\tor p)$, which we can express as a Horn clause theory $\{\tneg p\tor q,\ p\tor\tneg q\}$.  
\item
We can express the invalidity $\nment p\tnotor q$ as $\{\modT p, \modT\tneg q\}$, because (checking Figure~\ref{fig.3}) $\model{p\tnotor q}_\avaluation=\tvF$ precisely when $\avaluation(p)=\tvT$ and $\avaluation(q)=\tvF$.
\item
$p \timp q$ is equivalent to $\modT\tneg p\tor q$.
\item
$(p_1\tand \dots\tand p_n) \timp p$ is equivalent to $\modT\tneg p_1\tor\dots\modT\tneg p_n\tor p$.
\end{itemize*}
Note from Definition~\ref{defn.ment}(\ref{item.f.phi.valid}) and Figure~\ref{fig.3.phi.f} that:
$\avaluation\ment \modT p$ precisely when $\avaluation(p)=\tvT$;
$\avaluation\ment p$ precisely when $\avaluation(p)\in\{\tvT,\tvB\}$;
$\avaluation\ment\modT\tneg p$ precisely when $\avaluation(p)=\tvF$; and 
$\avaluation\ment\tneg p$ precisely when $\avaluation(p)\in\{\tvF,\tvB\}$.
\end{rmrk}

\begin{rmrk}[Three-valued HORNSAT]
\label{rmrk.HORNSAT.3}
We now propose an algorithm to check satisfiability of Horn clause theories in $\THREE$ --- which as per Definition~\ref{defn.2Sat}(\ref{item.2Sat.2}) is the problem of deciding whether there exists a \emph{three-valued} valuation $\avaluation:\ns P\to\THREE$ that makes all the clauses valid (have truth-value equal to $\tvB$ or $\tvT$) in $\THREE$ as per the three-valued truth-tables in Figure~\ref{fig.3}.

It is based on the algorithm for two-valued logic from Remark~\ref{rmrk.HORNSAT.2}.
Rules are applied in decreasing order of priority, and are repeated as often as they continue to act to change the theory:
\begin{enumerate*}
\item
Suppose the theory contains a unit boxed positive literal $\modT p$ (so intuitively, a satisfying assignment must satisfy $\avaluation(p)=\tvT$).
Then:
\begin{enumerate*}
\item
We remove all other clauses that contain $p$ or $\modT p$, because they are now satisfied.
\item
We delete all literals of the form $\modT\tneg p$, 
because these literals cannot be satisfied.
\item
We delete all literals of the form $\tneg p$, 
because these literals cannot be satisfied. 
\end{enumerate*}
\item
Suppose the theory contains a unit positive literal $p$ (so for a satisfying assignment, $\avaluation(p)\neq\tvF$).
Then:
\begin{enumerate*}
\item
We delete all other clauses that contain a literal $p$ (but \emph{not} those that contain $\modT p$), because these clauses are now satisfied.
\item
We delete all literals of the form $\modT\tneg p$ from all clauses, 
because these literals cannot be satisfied. 
\end{enumerate*}
\item
If the theory contains an unsatisfiable (empty) Horn clause, then fail; otherwise succeed.
\end{enumerate*}
If we succeed, then we can read a satisfying assignment from the result by 
\begin{itemize*}
\item
mapping $p$ to $\tvT$ if a unit clause $\modT p$ remains, 
\item
mapping $p$ to $\tvB$ if a unit clause $p$ remains (in this case, from the form of the rules a unit clause $\modT p$ cannot remain), and 
\item
mapping $p$ to $\tvF$ otherwise.
\end{itemize*}
\end{rmrk}

\jamiesection{Extremal valuations}
\label{sect.extremal.valuations}

\jamiesubsection{Definition of an extremal valuation}

\begin{rmrk}
Intuitively, a valuation is \emph{extremal} when it is continuous and it returns as many $\tvT$ and $\tvF$ values as possible (is as definite as possible), and conversely when it returns as few $\tvB$ values as possible (is no more ambivalent than necessary); the precise definition is in Definition~\ref{defn.definite}.

For example, there are four extremal valuations from $\THREE$ to itself:
$$
\lambda v.\tvT,\quad \lambda v.\tvF, \quad \lambda v.v, \quad\text{and}\quad \lambda v.\tneg v.
$$
Conversely, $\lambda v.\modT v$ is not extremal because it is not continuous.
$\lambda v.\tvB$ is continuous but not extremal.
The function mapping $\tvT$ to $\tvT$, $\tvB$ to $\tvB$, and $\tvF$ to $\tvB$ is continuous but not extremal.

It turns out that extremal valuations are rather useful.
If we think in terms of agreement, an extremal valuation represents a system state where algorithms have run and succeeded as much as they can; if a participant is still returning the ambivalent truth-value $\tvB$ then this is because they must, and not just because they have not yet made up their mind.
\end{rmrk}

\begin{defn}
\label{defn.definite}
Suppose $(\ns P,\opens)$ is a semitopology and $p\in\ns P$ and $\avaluation,\avaluation':\ns P\to\THREE$ are continuous valuations.
\begin{enumerate*}
\item
Call the elements $\tvT,\tvF\in\THREE$ \deffont[definite elements ($\tvT,\tvB\in\THREE$)]{definite}, and call $\tvB\in\THREE$ \deffont[ambivalent element ($\tvB\in\THREE$)]{ambivalent} (because $\tvB$ is short for `both').
\item\label{item.definite.definite}
If $\avaluation(p)\in\THREE$ is ambivalent / definite then call $\avaluation$ \deffont[ambivalent (valuation at $p$)]{ambivalent} / \deffont[definite (valuation at $p$)]{definite} at $p$.

It is easy to check from Definition~\ref{defn.ment}(\ref{item.f.phi.valid}) that $\avaluation$ is definite at $p$ when $\avaluation(p)\nment\modB p$ and when $\avaluation\ment \modT p \tor \modT \tneg p$.
\item
Define $\f{definite}(\avaluation)\subseteq\ns P$ by
$$
\f{definite}(\avaluation)=\avaluation^\mone\{\tvT,\tvF\} .
$$
Thus, $\f{definite}(\avaluation)$ is the set of points at which $\avaluation$ is definite.
Note a nice characterisation of this using Definition~\ref{defn.f^ment} as
$$
\f{definite}(\avaluation)=\ns P\setminus(\avaluation^\ment\cap\avaluation^{\ment\tneg}) = (\ns P\setminus\avaluation^\ment)\cup(\ns P\setminus\avaluation^{\ment\tneg}).
$$ 
\item\label{item.definite.leq}
Define a partial order $\avaluation\leq \avaluation'$\index{$\avaluation\leq\avaluation'$ (partial order on valuations)} on valuations $\avaluation,\avaluation':\ns P\to\THREE$ by
$$
\avaluation\leq \avaluation'
\quad\text{when}\quad
\avaluation|_{\f{definite}(\avaluation)} = \avaluation'|_{\f{definite}(\avaluation)} . 
$$ 
In words: $\avaluation\leq\avaluation'$ when $\avaluation'$ agrees with $\avaluation$ whenever $\avaluation$ has a definite value (however, $\avaluation'$ may be definite at other values at which $\avaluation$ is ambivalent).
\item\label{item.extremal.valuation}
Call $\avaluation$ a \deffont{$(\ns P,\opens)$-extremal valuation} when
\begin{enumerate*}
\item
$\avaluation$ is continuous on $(\ns P,\opens)$ and 
\item
$\avaluation$ is $\leq$-maximal amongst valuations that are continuous on $(\ns P,\opens)$.
\end{enumerate*}
If the semitopology is clear from context, we may just call $\avaluation$ an \deffont{extremal valuation}.
\item\label{item.mentX}
We will write\index{$\mentX$} 
$$
\ns P,\opens\mentX \phi
$$ 
when $\avaluation\ment\phi$ for every extremal valuation.
\end{enumerate*}
\end{defn}

\begin{rmrk}
There is a slight wrinkle to the terminology here:
\begin{itemize*}
\item
In Definition~\ref{defn.3.top}(\ref{item.3.valuation}) we let a valuation on $\ns P$ be any function $\avaluation:\ns P\to\THREE$; so if we need it to be continuous then we have to say `$\avaluation$ is a \emph{continuous} valuation'.
\item
In Definition~\ref{defn.definite}(\ref{item.extremal.valuation}) we let an extremal valuation be a $(\ns P,\opens)$-continuous valuation that is $\leq$-maximal.
\end{itemize*}
We never need to say `$\avaluation$ is an extremal \emph{continuous} valuation', because if $\avaluation$ is extremal then it is assumed continuous.
The notion of being extremal assumes continuity; since without it being extremal just means being a function in $\ns P\to\{\tvT,\tvF\}$.
\end{rmrk}

\jamiesubsection{Topological characterisation of extremal valuations}

In this Subsection we characterise extremal valuations in terms of regular open/closed sets.
The key technical observation powering this Subsection is Lemma~\ref{lemm.extremal.closure.2}.

We note a standard fact from topology relating inverse images, closures, and interiors~\cite[Proposition~1.4.1(v'\&vi)]{engelking:gent}.
It is also valid in semitopologies:
\begin{lemm}
\label{lemm.closure.inv.cont}
Suppose $(\ns P,\opens)$ and $(\ns P',\opens')$ are semitopologies and $\avaluation:\ns P\to\ns P'$ is continuous and $P'\subseteq\ns P'$.
Then:
\begin{enumerate*}
\item\label{item.closure.inv.cont.1} 
$\closure{\avaluation^\mone(P')} \subseteq \avaluation^\mone\closure{P'}$.
\item\label{item.closure.inv.cont.2} 
$\interior(\avaluation^\mone(P')) \supseteq \avaluation^\mone(\interior(P'))$.
\end{enumerate*}
\end{lemm}
\begin{proof}
For part~\ref{item.closure.inv.cont.1}, write $P=\avaluation^\mone(P')$ and suppose $p\in\closure{P}$, so that by Lemma~\ref{lemm.closure.using.nbhd.intersections} $\nbhd(p)\between P$. 
Now consider $O'\in\nbhd(\avaluation(p))$; it would suffice to show that $O'\between P'$.
Write $O=\avaluation^\mone(O')$.
Note that $O\in\opens$ by continuity and $O\in\nbhd(p)$ by construction.
It follows that $O\between P$, and so that $O'\between P'$ as required.

For part~\ref{item.closure.inv.cont.2}, just use part~\ref{item.closure.inv.cont.1} and take complements, noting that $\avaluation^\mone(\ns P'\setminus P')=\ns P\setminus\avaluation^\mone(P')$ and by Lemma~\ref{lemm.closure.interior} that $\interior(\ns P'\setminus P')=\ns P'\setminus\closure{P'}$ and $\interior(\ns P'\setminus P')=\ns P'\setminus\closure{P'}$.
\end{proof}

\begin{corr}
\label{corr.closure.eval.inclusion}
Suppose $(\ns P,\opens)$ is a semitopology and $\avaluation:\ns P\to\THREE$ is a continuous valuation.
Then:
\begin{enumerate*}
\item
$\closure{\avaluation^\mone\{\tvT\}}\subseteq\avaluation^\mone\{\tvT,\tvB\}$ and
$\closure{\avaluation^\mone\{\tvF\}}\subseteq\avaluation^\mone\{\tvF,\tvB\}$.
\item
The inclusions may be strict.
\end{enumerate*}
\end{corr}
\begin{proof}
It is a fact that $\closure{\{\tvT\}}=\{\tvT,\tvB\}$ and $\closure{\{\tvF\}}=\{\tvF,\tvB\}$.
The inclusions follow from Lemma~\ref{lemm.closure.inv.cont}.

To show that the inclusions may be strict, it suffices to provide an example. 
Consider the semitopology in Figure~\ref{fig.012} (top-right example) with $\avaluation(0)=\tvF$, $\avaluation(1)=\tvB$, and $\avaluation(2)=\tvT$.
The reader can check that this is continuous and 
$\avaluation^\mone\{\tvF\}\subsetneq\avaluation^\mone\{\tvF,\tvB\}$  
and 
$\avaluation^\mone\{\tvT\}\subsetneq\avaluation^\mone\{\tvT,\tvB\}$.
\end{proof}

Remarkably, extremal valuations are characterised \emph{precisely} by the property that the inclusions of Corollary~\ref{corr.closure.eval.inclusion} are equalities, as we shall see in the next few results, and in Proposition~\ref{prop.extremal.closure} in particular.

\begin{lemm}
\label{lemm.extremal.closure.1}
Suppose $(\ns P,\opens)$ is a semitopology and $\avaluation:\ns P\to\THREE$ is a continuous extremal valuation.
Then 
$$
\closure{\avaluation^\mone\{\tvT\}}=\avaluation^\mone\{\tvT,\tvB\}
\quad\text{and}\quad 
\closure{\avaluation^\mone\{\tvF\}}=\avaluation^\mone\{\tvF,\tvB\}. 
$$
\end{lemm}
\begin{proof}
By Corollary~\ref{corr.closure.eval.inclusion}
$\closure{\avaluation^\mone\{\tvT\}}\subseteq\avaluation^\mone\{\tvT,\tvB\}$ 
and
$\closure{\avaluation^\mone\{\tvF\}}\subseteq\avaluation^\mone\{\tvF,\tvB\}$.
Now suppose $p\notin\closure{\avaluation^\mone\{\tvT\}}$ (the case of $\tvF$ is precisely similar).
So there exists $O\in\nbhd(p)$ such that $O\notbetween\avaluation^\mone\{\tvT\}$.
There are now two subcases:
\begin{itemize*}
\item
\emph{Suppose $O\subseteq\avaluation^\mone\{\tvF\}$.}\quad

Then $p\not\oldin\avaluation^\mone\{\tvT,\tvB\}$ and we are done.
\item
\emph{Suppose $O\not\subseteq\avaluation^\mone\{\tvF\}$.}\quad

Then we obtain a strictly more definite continuous valuation as $\indicator{O_\tvT,O_\tvF}(\avaluation^\mone\{\tvT\},\avaluation^\mone\{\tvF\}\cup O)$.
By extremality of $\avaluation$, this is impossible.
\qedhere\end{itemize*}
\end{proof}

The nontrivial technical content of this Subsection is here:
\begin{lemm}
\label{lemm.extremal.closure.2}
Suppose $(\ns P,\opens)$ is a semitopology and $\avaluation:\ns P\to\THREE$ is a continuous valuation.
Then 
$$
\closure{\avaluation^\mone\{\tvT\}}=\avaluation^\mone\{\tvT,\tvB\}
\ \land\ \closure{\avaluation^\mone\{\tvF\}}=\avaluation^\mone\{\tvF,\tvB\}
\quad\text{implies}\quad
\text{$\avaluation$ is extremal}.
$$
\end{lemm}
\begin{proof}
We prove the contrapositive.
Suppose $f$ is not extremal, so that there exists a continuous valuation $f':\ns P\to\THREE$ such that $f\lneq f'$.
So there is a $p'\in\ns P$ at which $f'$ is definite and $f$ is not; suppose $f'(p')=\tvT$ (the case that $f'(p')=\tvF$ is exactly similar).

Since $f'$ is continuous, by Definition~\ref{defn.nbhd} and Remark~\ref{rmrk.nbhd.concise} there exists an $O'\in\nbhd(p')$ such that $O'\subseteq (f')^\mone\{\tvT\}$.
From this it follows using Definition~\ref{defn.closure}(\ref{item.closure}) that $p'\notin\closure{\avaluation^\mone\{\tvF\}}$.
Now we assumed $f(p')=\tvB$, so that $p'\in\avaluation^\mone\{\tvF,\tvB\}$ is a fact.
Thus, $\closure{\avaluation^\mone\{\tvF\}}\neq\avaluation^\mone\{\tvF,\tvB\}$ as required.
\end{proof}

\begin{prop}
\label{prop.extremal.closure}
Suppose $(\ns P,\opens)$ is a semitopology and $\avaluation:\ns P\to\THREE$ is a continuous valuation.
Then the following are equivalent:
\begin{enumerate*}
\item\label{item.extremal.closure.1}
$\avaluation$ is extremal.
\item\label{item.extremal.closure.2}
$\closure{\avaluation^\mone\{\tvT\}}=\avaluation^\mone\{\tvT,\tvB\}$ and $\closure{\avaluation^\mone\{\tvF\}}=\avaluation^\mone\{\tvF,\tvB\}$. 
\item\label{item.extremal.closure.3}
$\avaluation^\mone\{\tvT,\tvB\}\subseteq\closure{\avaluation^\mone\{\tvT\}}$ and $\avaluation^\mone\{\tvF,\tvB\}\subseteq\closure{\avaluation^\mone\{\tvF\}}$. 
\end{enumerate*}
It is a fact that $\closure{\{\tvT\}}=\{\tvT,\tvB\}$ and $\closure{\{\tvF\}}=\{\tvF,\tvB\}$ in $\THREE$, so we can say:
\begin{quote}
A continuous valuation $\avaluation:\ns P\to\THREE$ is extremal precisely when inverse images commute with closures.
\end{quote}
\end{prop}
\begin{proof}
Equivalence of parts~\ref{item.extremal.closure.1} and~\ref{item.extremal.closure.2} is just Lemmas~\ref{lemm.extremal.closure.1} and~\ref{lemm.extremal.closure.2}.
Part~\ref{item.extremal.closure.2} certainly implies part~\ref{item.extremal.closure.3}, and part~\ref{item.extremal.closure.3} combined with Lemma~\ref{lemm.closure.inv.cont} implies part~\ref{item.extremal.closure.2}.
\end{proof}

It is easy and natural to dualise Proposition~\ref{prop.extremal.closure}:
\begin{corr}
\label{corr.extremal.interior}
Suppose $(\ns P,\opens)$ is a semitopology and $\avaluation:\ns P\to\THREE$ is a continuous valuation.
Then the following are equivalent:
\begin{enumerate*}
\item
$\avaluation$ is extremal.
\item
$\interior(\avaluation^\mone\{\tvF,\tvB\})=\avaluation^\mone\{\tvF\}$
and 
$\interior(\avaluation^\mone\{\tvT,\tvB\})=\avaluation^\mone\{\tvT\}$. 
\item
$\avaluation^\mone\{\tvF\}\supseteq\interior(\avaluation^\mone\{\tvT,\tvB\})$
and 
$\avaluation^\mone\{\tvT\}\supseteq\interior(\avaluation^\mone\{\tvF,\tvB\})$. 
\end{enumerate*}
It is a fact that $\interior(\{\tvF,\tvB\})=\{\tvF\}$ and $\interior(\{\tvT,\tvB\})=\{\tvT\}$ in $\THREE$, so we can say:
\begin{quote}
A continuous valuation $\avaluation:\ns P\to\THREE$ is extremal precisely when inverse images commute with interiors.
\end{quote}
\end{corr}
\begin{proof}
We take complements in Proposition~\ref{prop.extremal.closure} and dualise using Lemma~\ref{lemm.closure.interior} (just as we did to derive part~\ref{item.closure.inv.cont.2} of Lemma~\ref{lemm.closure.inv.cont} from part~\ref{item.closure.inv.cont.1}).
\end{proof}

In Lemma~\ref{lemm.logical.p.T}(\ref{item.logical.Box.no.converse}) we noted that $\avaluation\ment\modT p$ implies $p\in\interior(\avaluation^\ment)$, but the reverse implication need not hold.
Rather nicely, it turns out that extremal valuations are characterised \emph{precisely} as those valuations such that the reverse implications hold for $\modT p$ and $\modT\tneg p$:
\begin{corr}
\label{corr.precise.inverse.image}
Suppose $(\ns P,\opens)$ is a semitopology and $\avaluation:\ns P\to\THREE$ is a continuous valuation.
Then the following are equivalent:
\begin{enumerate*}
\item
$\avaluation$ is extremal.
\item
$p\in\interior(\avaluation^\ment)\liff \avaluation\ment\modT p$ 
and
$p\in\interior(\avaluation^{\ment\tneg})\liff \avaluation\ment\modT \tneg p$,\ 
for every $p\in\ns P$.
\item
$p\in\interior(\avaluation^\ment)\limp\avaluation\ment\modT p$ 
and
$p\in\interior(\avaluation^{\ment\tneg})\limp\avaluation\ment\modT \tneg p$,\ 
for every $p\in\ns P$.
\end{enumerate*}
\end{corr}
\begin{proof}
This just rephrases Corollary~\ref{corr.extremal.interior} using Lemma~\ref{lemm.logical.p.T}(\ref{item.logical.Box.char}) and Definition~\ref{defn.f^ment}.
\end{proof}

\jamiesubsection{Maximal disjoint pairs of open sets}

We now develop a different view of extremal valuations, based on maximal elements in the poset of disjoint pairs of open sets.
We start by introducing some notation and recalling an easy sets bijection:
\begin{nttn}
Suppose $\ns P$ and $\ns P'$ are sets and suppose $\mathcal P\subseteq\powerset(\ns P)$ and $\mathcal P'\subseteq\powerset(\ns P')$ are sets of subsets. 
Then write $\mathcal P\otimes\mathcal P'$ for the set of disjoint pairs in $\mathcal P\times\mathcal P'$.
In symbols:
$$
\mathcal P\otimes\mathcal P' = \{(P,P') \mid P\in\mathcal P,\ P'\in\mathcal P',\ P\notbetween P'\} .
$$
In particular, if $(\ns P,\opens)$ is a semitopology then $\opens\otimes\opens$ is the set of disjoint pairs of open sets.
\end{nttn}

\begin{rmrk}
\label{rmrk.char.correspondence}
Suppose $(\ns P,\opens)$ is a semitopology. 
Recall the \emph{indicator functions} $\indicator{O_\tvT,O_\tvF}$ from Definition~\ref{defn.indicator.functions} and the \emph{characteristic sets} $\f{char}_{OO}$ from Definition~\ref{defn.char.function}, and recall from Proposition~\ref{prop.what.is.an.indicator.function} that 
\begin{enumerate*}
\item 
$\f{char}_{OO}$ maps a continuous valuation $\avaluation:\ns P\to\THREE$ to a disjoint pair 
$$
\f{char}_{OO}(\avaluation)=(\avaluation^\mone\{\tvT\},\avaluation^\mone\{\tvF\})\oldin\opens\otimes\opens
$$ 
and 
\item
$\indicator{O_\tvT,O_\tvF}$ maps $(O,O')\oldin\opens\otimes\opens$ to the continuous valuation that maps $p\in\ns P$ to $\tvT$ if $p\in O$, and to $\tvF$ if $p\in O'$, and to $\tvB$ if $p\in \ns P\setminus(O\cup O')$, and
\item
$\f{char}_{OO}$ and $\indicator{O_\tvT,O_\tvF}$ are inverse and biject between valuations and pairs of disjoint open sets.  
\end{enumerate*}
We now note that $\f{char}$ and $\indicator{}$ can also be viewed as maps of posets:
\end{rmrk}

\begin{defn}
\label{defn.two.opens.ordering}
Suppose $(\ns P,\opens)$ is a semitopology.
Then:
\begin{enumerate*}
\item\label{item.componentwise.ordering}
Extend the subset inclusion ordering $\subseteq$ to disjoint pairs of open sets in $\opens\otimes\opens$ \deffont[componentwise inclusion ordering on pairs of sets]{componentwise} as follows:
$$
(O_1,O_1')\leq (O_2,O_2')
\quad\text{when}\quad
O_1\subseteq O_1' \ \land\  O_2\subseteq O_2'
$$
for $(O_1,O_1'),(O_2,O_2')\oldin\opens\otimes\opens$. 
\item
Call $(O,O')\oldin\opens\otimes\opens$ \deffont[maximal disjoint sets]{maximal disjoint} when it is $\leq$-maximal amongst pairs of disjoint open sets.
\end{enumerate*}
\end{defn}

\begin{lemm}
\label{lemm.two.leqs.across.bijection}
Suppose $(\ns P,\opens)$ is a semitopology.
Then:
\begin{enumerate*}
\item
If $\avaluation,\avaluation':\ns P\to\THREE$ are continuous valuations then the following are equivalent: 
\begin{itemize*}
\item
$\avaluation\leq\avaluation'$ in the sense of Definition~\ref{defn.definite}(\ref{item.definite.leq}). 
\item
$\f{char}_{OO}(\avaluation)\leq\f{char}_{OO}(\avaluation')$ in the sense of Definition~\ref{defn.two.opens.ordering}(\ref{item.componentwise.ordering}).
\end{itemize*}
\item
The following are equivalent for $(O_1,O'_1),(O_2,O'_2)\oldin\opens\otimes\opens$: 
\begin{itemize*}
\item
$(O_1,O'_1)\leq (O_2,O'_2)$ in the sense of Definition~\ref{defn.two.opens.ordering}(\ref{item.componentwise.ordering}).
\item 
$\indicator{O_\tvT,O_\tvF}(O_1,O'_1)\leq\indicator{O_\tvT,O_\tvF}(O_2,O'_2)$ in the sense of Definition~\ref{defn.definite}(\ref{item.definite.leq}).
\end{itemize*}
\end{enumerate*}
\end{lemm}
\begin{proof}
By routine calculations from the definitions.
\end{proof}

\begin{corr}
\label{corr.definite.disjoint}
Suppose $(\ns P,\opens)$ is a semitopology. 
Then $\indicator{O_\tvT,O_\tvF}$ and $\f{char}_{OO}$ determine a poset isomorphism between  
\begin{itemize*}
\item
continuous valuations ordered by $\leq$ (Definition~\ref{defn.definite}(\ref{item.definite.leq})) and 
\item
$\opens\otimes\opens$ ordered by $\leq$ (Definition~\ref{defn.two.opens.ordering}(\ref{item.componentwise.ordering})).
\end{itemize*}
\end{corr}
\begin{proof}
Direct from Lemma~\ref{lemm.two.leqs.across.bijection}.
\end{proof}

The maximal elements of $\opens\otimes\opens$ have some useful characterisations: 
\begin{prop}
\label{prop.max.disjoint.to.regular}
Suppose $(\ns P,\opens)$ is a semitopology and $(O,O')\oldin\opens\otimes\opens$.
Then the following are equivalent:
\begin{enumerate*}
\item\label{item.max.disjoint.to.regular.1}
$(O,O')$ is maximal disjoint.
\item\label{item.max.disjoint.to.regular.2}
$O=\interior(\ns P\setminus O')$ and $O'=\interior(\ns P\setminus O)$.
\item\label{item.max.disjoint.to.regular.3}
$O$ is regular (Definition~\ref{defn.regular.open.set}) and $O'=\interior(\ns P\setminus O)$.
\item\label{item.max.disjoint.to.regular.4}
$O'$ is regular and $O=\interior(\ns P\setminus O')$.
\item\label{item.max.disjoint.to.regular.5}
$O$ and $O'$ are regular and $O=\interior(\ns P\setminus O')$ and $O'=\interior(\ns P\setminus O)$.
\end{enumerate*}
\end{prop}
\begin{proof}
Equivalence of parts~\ref{item.max.disjoint.to.regular.1} and~\ref{item.max.disjoint.to.regular.2} is just an easy corollary of Lemma~\ref{lemm.interior.open}.
We spell out the details:
\begin{itemize}
\item
\emph{Suppose $(O,O')$ is maximal disjoint.}

Then $O'$ is a greatest open set that is disjoint from $O$, meaning equivalently that $O'$ is a greatest open set contained in $\ns P\setminus O$, so by Lemma~\ref{lemm.interior.open} $O'=\interior(\ns P\setminus O)$.
Similarly, $O=\interior(\ns P\setminus O')$. 
\item
\emph{Suppose $O'=\interior(\ns P\setminus O)$ and $O=\interior(\ns P\setminus O')$.}

Consider a disjoint pair $(O'',O''')\geq (O,O')$, meaning by Definition~\ref{defn.two.opens.ordering}(\ref{item.componentwise.ordering}) that $O''\supseteq O$ and $O'''\supseteq O'$.
Then $O''\subseteq\ns P\setminus O'''\subseteq\ns P\setminus O'$ so that $O''\subseteq\interior(\ns P\setminus O')=O$, so that $O''=O$.
Similarly, $O'''=O'$.
Since $(O'',O''')$ was arbitrary, $(O,O')$ is a maximal disjoint pair of open sets.
\end{itemize}
To prove equivalence of parts~\ref{item.max.disjoint.to.regular.2} and~\ref{item.max.disjoint.to.regular.3} we reason as follows:
\begin{itemize}
\item
\emph{Suppose $O=\interior(\ns P\setminus O')$ and $O'=\interior(\ns P\setminus O)$.}

We reason as follows using Lemma~\ref{lemm.closure.interior}(\ref{item.closure.interior.complement.closure}):
$$
\begin{array}{r@{\ }l@{\quad}l}
O
=&\interior(\ns P\setminus O')
&\text{Assumption}
\\
=&\interior(\ns P\setminus\interior(\ns P\setminus O))
&O'=\interior(\ns P\setminus O)
\\
=&\interior(\ns P\setminus(\ns P\setminus\closure{O}))
&\text{Lemma~\ref{lemm.closure.interior}(\ref{item.closure.interior.complement.closure})}
\\
=&\interior(\closure{O})
&\text{Fact of sets}
\end{array}
$$
\item
\emph{Suppose $O$ is regular (so $O=\interior(\closure{O})$) and $O'=\interior(\ns P\setminus O)$.}

We just reverse the reasoning of the previous case.
\end{itemize}
To prove equivalence of parts~\ref{item.max.disjoint.to.regular.3} and~\ref{item.max.disjoint.to.regular.4} follows by the same reasoning, on $O'$.
Equivalence of parts~\ref{item.max.disjoint.to.regular.1} and~\ref{item.max.disjoint.to.regular.5} then follows easily. 
\end{proof}

\begin{rmrk}
\label{rmrk.char.correspondence.extremal}
Putting Proposition~\ref{prop.max.disjoint.to.regular} and Corollary~\ref{corr.definite.disjoint} together, we see that the following items of data are in natural correspondence: 
\begin{enumerate*}
\item\label{item.char.bij}
\emph{Extremal valuations $\avaluation:\ns P\to\THREE$.}
An extremal $\avaluation$ corresponds to the by Corollary~\ref{corr.definite.disjoint} maximal disjoint pair $(\avaluation^\mone\{\tvT\},\avaluation^\mone\{\tvF\})$, and to the by Proposition~\ref{prop.max.disjoint.to.regular} regular open set $\avaluation^\mone\{\tvT\}$. 
\item 
\emph{Maximal pairs $(O,O')\oldin\opens\otimes\opens$.}
A maximal $(O,O')$ corresponds to the by Corollary~\ref{corr.definite.disjoint} extremal 
valuation $\indicator{O_\tvT,O_\tvF}(O,O'):\ns P\to\THREE$, and to the by Proposition~\ref{prop.max.disjoint.to.regular} regular open $O$. 
\item\label{item.char.extremal.regular}
\emph{Regular open sets $O\in\regularOpens$.}
A regular open set $O$ corresponds to the by Proposition~\ref{prop.max.disjoint.to.regular} maximal disjoint $(O,\interior(\ns P\setminus O))$ and to the by Corollary~\ref{corr.definite.disjoint} extremal $\indicator{O_{\tvT},O_{\tvF}}(O,\interior(\ns P\setminus O))$. 
\end{enumerate*} 
We sum this up in a small table in Figure~\ref{fig.max}.
\end{rmrk}

\begin{figure}
$$
\begin{array}{c@{\quad}c@{\quad}c}
\avaluation:(\ns P,\opens)\to\THREE
&
(\avaluation^\mone\{\tvT\},\avaluation^\mone\{\tvF\})
&
\avaluation^\mone\{\tvT\}
\\[1ex]
\indicator{O_\tvT,O_\tvF}(O,O')
&
(O,O')\oldin\opens\otimes\opens
&
O
\\[1ex]
\indicator{O_\tvT,O_\tvF}(O,\interior(\ns P\setminus O))
&
(O,\interior(\ns P\setminus O))
&
O\in\regularOpens
\end{array}
$$
\emph{In row~1 $\avaluation:(\ns P,\opens)\to\THREE$ is an extremal valuation; in row~2 $(O,O')$ is a maximal disjoint pair of open sets; in row~3 $O$ is a regular open set.}
\caption{Correspondence between extremal valuations, maximal disjoint pairs of open sets, and regular open sets}
\label{fig.max}
\end{figure}

We conclude with a small lemma noting that extremal valuations always exist.
A standard argument using Zorn's lemma is possible, but we give a more direct argument using closures and regular open sets: 
\begin{lemm}
Suppose $(\ns P,\opens)$ is a semitopology and $\avaluation:\ns P\to\THREE$ is a continuous valuation.
Then there exists an extremal valuation $\avaluation':\ns P\to\THREE$ such that $\avaluation\leq\avaluation'$.
\end{lemm}
\begin{proof}
Write $O=\avaluation^\mone\{\tvT\}$ and $O'=\avaluation^\mone\{\tvF\}$.
Note from Remark~\ref{rmrk.char.correspondence} that $O$ and $O'$ are open and disjoint.
Write $R=\interior(\closure{O})$.
Note by construction that $O\subseteq R$, and from Lemmas~\ref{lemm.ic.ci.regular} and~\ref{lemm.clint.between} that $R$ is open, regular, and disjoint from $O'$. 

Finally, write $R'=\interior(\ns P\setminus R)$.
By construction $O'\subseteq R'$ and using Proposition~\ref{prop.max.disjoint.to.regular}, $(R,R')$ is maximal disjoint above $(O,O')$ in $\opens\otimes\opens$.
We set $\avaluation'=\indicator{O_{\tvT}O_{\tvF}}(R,R')$.
By Corollary~\ref{corr.definite.disjoint} $\avaluation'$ is extremal and (since $O\subseteq R$ and $O'\subseteq R'$) $\avaluation\leq\avaluation'$.
\end{proof}

\jamiesubsection{Logical characterisation of extremal valuations}

\begin{rmrk}
Suppose $\witness:\ns P\to\mathcal W(\ns P)$ is a witness function on a finite set $\ns P$, and recall from Subsection~\ref{subsect.logical.continuity} and Proposition~\ref{prop.Ax.iff.cont} that a valuation $\avaluation:\ns P\to\THREE$ is continuous in the witness semitopology from Definition~\ref{defn.trust.topology}(\ref{item.witness.semitopology}) if and only if $\avaluation\ment\thyAxW$, for $\thyAxW$ the theory arising from $\witness$ in Definition~\ref{defn.Ax}.
Recall that $\thyAxW$ has for each $p\in\ns P$ the following pair of axioms: 
$$
\begin{array}{r@{\ }l}
\tf{closedAx}_\witness(p) 
=&
\bigl( \tand_{w\in\witness(p)} \tor_{q\in w} q\bigr) \timp p
\quad
\text{and}\quad
\\
\tf{closedAx}_\witness^\tneg(p) 
=&
\bigl( \tand_{w\in\witness(p)} \tor_{q\in w} \tneg q\bigr) \timp \tneg p .
\end{array}
$$
\end{rmrk}

\begin{rmrk}
Recall from Definition~\ref{defn.open.covers}(\ref{item.covers.p}) that $\f{Covers}(p)=\{O\in\nbhd(p)\mid p\lessdot O\}$ is the set of \emph{open covers} of $p$, which are the minimal open neighbourhoods of $p$.
Recall from Theorem~\ref{thrm.lim.O.open} that witness semitopologies are chain-complete, so that for the case of a witness semitopology we get Corollary~\ref{corr.closure.using.covers}, that $p\in\closure{O'}$ if and only if $O'\between\f{Covers}(p)$.
\end{rmrk}

\begin{defn}
\label{defn.AxEx}
Suppose $\witness:\ns P\to\mathcal W(\ns P)$ is a witness function on a finite set $\ns P$.
Continuing Definition~\ref{defn.Ax}: 
\begin{enumerate*}
\item
Define the \deffont[extremal axioms $\tf{ClosedPred3}(\ns P)$]{extremal axioms arising from $\witness$} to be sets of predicates in $\tf{ClosedPred3}(\ns P)$ as follows:
$$
\begin{array}{r@{\ }l@{\quad}l}
\tf{closedAxEx}_\witness(p) 
=&
p\timp \bigl( \tand_{Q\in\f{Covers}(p)} \tor_{q\in Q} \modT q\bigr) 
\\
\tf{closedAxEx}_\witness^\tneg(p) 
=&
\tneg p\timp \bigl( \tand_{Q\in\f{Covers}(p)} \tor_{q\in Q} \modT \tneg q\bigr) 
\\[2ex]
\tf{closedAxEx}_\witness=&\bigwedge\{\tf{closedAxEx}_\witness(p) \mid p\in\ns P\}
\\
\tf{closedAxEx}_\witness^\tneg=&\bigwedge\{\tf{closedAxEx}_\witness^\tneg(p) \mid p\in\ns P\}
\end{array}
$$
\item\label{item.AxX}
We collect these axioms into a large conjunction $\thyAxExW$, which we call the \deffont[extremal theory $\thyAxExW$]{extremal theory arising from $\witness$}:
$$
\thyAxExW 
= 
\thyAxW\ \tand\ \tf{closedAxEx}_\witness\ \tand \ \tf{closedAxEx}_\witness^\tneg .
$$
\end{enumerate*}
We may omit the $\witness$ annotation where this is unimportant or understood, writing (for example) $\tf{closedAxEx}_\witness^\tneg(p)$ just as $\tf{closedAxEx}^\tneg(p)$.
\end{defn}

\begin{prop}
Suppose $(\ns P,\opens)$ is a semitopology and $\avaluation:\ns P\to\THREE$ is a valuation.
Then the following conditions are equivalent:
\begin{enumerate*}
\item
$\avaluation$ is an extremal valuation.
\item
$\avaluation\ment\thyAxExW$.
\end{enumerate*}
\end{prop}
\begin{proof}
We prove two implications:
\begin{itemize}
\item
\emph{Suppose $\avaluation:\ns P\to\THREE$ is an extremal valuation.}

By assumption $\avaluation$ is continuous, so by Proposition~\ref{prop.Ax.iff.cont} $\avaluation\ment\thyAxW$.
It remains to show that $\avaluation\ment\thyAxExW$. 
From Proposition~\ref{prop.extremal.closure} we have that 
$\avaluation^\mone\{\tvT,\tvB\}\subseteq\closure{\avaluation^\mone\{\tvT\}}$ 
and
$\avaluation^\mone\{\tvF,\tvB\}\subseteq\closure{\avaluation^\mone\{\tvF\}}$.
But by Corollary~\ref{corr.closure.using.covers} this is precisely what $\tf{closedAxEx}_\witness$ and $\tf{closedAxEx}_\witness^\tneg$ express, so we are done.
\item
\emph{Suppose $\avaluation\ment\thyAxExW$.}

Then $\avaluation\ment\thyAxW$ so that by Proposition~\ref{prop.Ax.iff.cont} $\avaluation$ is continuous.
Also, by Corollary~\ref{corr.closure.using.covers} $\avaluation\ment\tf{closedAxEx}_\witness\land \avaluation\ment\tf{closedAxEx}_\witness^\tneg$ expresses precisely that $\avaluation^\mone\{\tvT,\tvB\}\subseteq\closure{\avaluation^\mone\{\tvT\}}$ and $\avaluation^\mone\{\tvF,\tvB\}\subseteq\closure{\avaluation^\mone\{\tvF\}}$, so by Proposition~\ref{prop.extremal.closure} $\avaluation$ is extremal as required.
\qedhere\end{itemize}
\end{proof}

\begin{rmrk}
We note in passing that:
\begin{enumerate}
\item
In Definition~\ref{defn.AxEx} we could equivalently set 
$$
\begin{array}{r@{\ }l}
\tf{closedAxEx}_\witness(p)=&p\timp \bigl( \tand_{Q\in\nbhd(p)} \tor_{q\in Q} \modT q\bigr)
\quad\text{and}
\\
\tf{closedAxEx}_\witness^\tneg(p)=&\tneg p\timp \bigl( \tand_{Q\in\nbhd(p)} \tor_{q\in Q} \modT \tneg q\bigr).
\end{array}
$$
This would not be wrong --- but in the case of a witness semitopology, which is chain-bounded, by Corollary~\ref{corr.closure.using.covers} we might as well restrict to $\f{Covers}(p)\subseteq\nbhd(p)$.
\item
We can use the equivalences in Figure~\ref{fig.3.equivalences} to rewrite the extremal axioms from Definition~\ref{defn.AxEx} into a more clausal form as follows:
\begin{itemize*}
\item
$p\timp \bigwedge_i \bigvee_j \modT q_{ij}$ 
becomes
$\modT \tneg p\, \tor (\bigwedge_i\bigvee_j \modT q_{ij})$.
\item
$\tneg p\timp \bigwedge_i \bigvee_j \tneg q_{ij}$ 
becomes 
$\modT p\, \tor (\bigwedge_i \bigvee_j \modT\tneg q_{ij})$.
\end{itemize*}
\end{enumerate}
\end{rmrk}

\jamiesubsection{Extremal valuations and regular points}

We can think of Proposition~\ref{prop.scp.extremal} (which considers continuous functions to $\THREE$) as generalising Theorem~\ref{thrm.correlated} (which regards partially continuous functions, but to a discrete space):
\begin{prop}
\label{prop.scp.extremal}
Suppose $(\ns P,\opens)$ is a semitopology and $\avaluation:\ns P\to\THREE$ is an extremal valuation and $\atopen\in\topens$ and $p,p'\in\atopen$.
Then:
\begin{enumerate*}
\item\label{item.scp.extremal.1}
$\avaluation(p)=\avaluation(p')$. 
\item\label{item.scp.extremal.2}
$\avaluation(p)\neq\tvB$.
\end{enumerate*}
\end{prop}
\begin{proof}
Write $(O,O')=\f{char}_{OO}(\avaluation)=(\avaluation^\mone\{\tvT\},\avaluation^\mone\{\tvF\})$, and note from Remark~\ref{rmrk.char.correspondence.extremal}(\ref{item.char.bij}) that this is a maximal pair of disjoint open sets.
We consider each part in turn:
\begin{enumerate}
\item
If $p\in O$ then $\atopen\between O$ and by Proposition~\ref{prop.open.consensus} $\atopen\subseteq O$, so that $p'\in O$.
By the same reasoning $p'\in O$ implies $p\in O$, and similarly $p\in O'$ if and only if $p'\in O'$.
Thus
$$
p\in O\liff p'\in O
\quad\text{and}\quad 
p\in O'\liff p'\in O'.
$$
It follows that $\avaluation(p)=\avaluation(p')$ as required.
\item
Suppose $\avaluation(p)=\tvB$; we will arrive at a contradiction.
By assumption $p$ has a topen neighbourhood $\atopen$ and it follows using Theorem~\ref{thrm.max.cc.char}(\ref{char.some.topen}\&\ref{char.Kp.greatest.topen}) that $p\in\community(p)\in\topens$.

Using part~\ref{item.scp.extremal.1} of this result (for the topen $\community(p)\in\topens$) we know that $\avaluation(p'')=\tvB$ for every $p''\in\community(p)$.
But then $\community(p)$ is an open set that is disjoint from both $O$ and $O'$, contradicting their maximality.
\qedhere\end{enumerate}
\end{proof}

\begin{lemm}
\label{lemm.regular.extremal}
Suppose $(\ns P,\opens)$ is a semitopology and $p\in\ns P$.
Then: 
\begin{enumerate*}
\item\label{item.regular.extremal.1}
If $p$ is regular then every extremal valuation is definite at $p$.
\item\label{item.regular.extremal.2}
The converse implication need not hold: it is possible for every extremal valuation to be definite at $p$, yet $p$ is not regular.
\end{enumerate*}
\end{lemm}
\begin{proof}
For part~\ref{item.regular.extremal.1}, suppose $p$ is regular, so that by Definition~\ref{defn.tn}(\ref{item.regular.point}) $p\in\community(p)\in\topens$, and suppose $\avaluation:\ns P\to\THREE$ is an extremal valuation.
By Proposition~\ref{prop.scp.extremal}(\ref{item.scp.extremal.2}) (taking $p'=p$) $\avaluation(p)\neq\tvB$, so by Definition~\ref{defn.definite}(\ref{item.definite.definite}) $\avaluation$ is definite at $p$.

For part~\ref{item.regular.extremal.2}, it suffices to provide a counterexample.
Consider the semitopology in Figure~\ref{fig.square.diagram}, so:
\begin{itemize*}
\item
$\ns P=\{0,1,2,3\}$ and
\item
$\opens$ is generated by $\bigl\{\{0,1\},\,\{1,2\},\,\{2,3\},\,\{3,4\}\bigr\}$.
\end{itemize*}
The reader can check that extremal valuations are constant and definite on $\{0,1\}$ and $\{2,3\}$, or on $\{0,3\}$ and $\{1,2\}$ --- for instance we can map $\{0,1\}$ to $\tvT$ and $\{2,3\}$ to $\tvF$.
However, by Lemma~\ref{lemm.square.diagram.not.qr} 
no points in this space are regular  
(see the discussion in Example~\ref{xmpl.not.intertwined}(\ref{item.square.diagram.not.regular})).
\end{proof}

\jamiesubsection{Characterisations of intertwinedness properties}

\jamiesubsubsection{Studying $p\intertwinedwith p'$ and $p\tlatticeiff p'$}

\begin{lemm}
\label{lemm.iw.reg}
For a semitopology $(\ns P,\opens)$ and $p,p'\in\ns P$,
the following are equivalent:
\begin{enumerate*}
\item\label{item.iw.reg.1}
$\Forall{O,O'{\in}\opens}(p\in O\land p'\in O')\limp O\between O'$ (meaning by Definition~\ref{defn.intertwined.points}(\ref{item.p.intertwinedwith.p'}) that $p\intertwinedwith p'$). 
\item\label{item.iw.reg.2}
$\Forall{O,O'{\in}\regularOpens}(p\in O\land p'\in O')\limp O\between O'$.
\end{enumerate*}
\end{lemm}
\begin{proof}
This just repackages Corollary~\ref{corr.nonintersect.nonintersect.regular}.
\end{proof}

Lemma~\ref{lemm.intertwinedwith.closures} gives yet another view of being intertwined; we use it in Definition~\ref{defn.ht.ce}:
\begin{lemm}
\label{lemm.intertwinedwith.closures}
For a semitopology $(\ns P,\opens)$ and $p,p'\in\ns P$,
the following are equivalent:
\begin{enumerate*}
\item\label{item.iw.closures.1}
$p\intertwinedwith p'$. 
\item\label{item.iw.closures.2}
$\Forall{O\in\opens} p,p'\in\closure{O} \lor p,p'\in \ns P\setminus O$.
\item\label{item.iw.closures.4}
$\Forall{O\in\regularOpens} p,p'\in \closure{O} \lor p,p'\in \ns P\setminus O$.
\item\label{item.iw.closures.5}
$\Forall{C\in\closed} p,p'\in C \lor p,p'\in \closure{\ns P\setminus C}$.
\item\label{item.iw.closures.6}
$\Forall{C\in\regularClosed} p,p'\in C \lor p,p'\in \closure{\ns P\setminus C}$.
\end{enumerate*}
\end{lemm}
\begin{proof}
For the equivalence of parts~\ref{item.iw.closures.1} and~\ref{item.iw.closures.2} we prove two implications:
\begin{itemize}
\item
\emph{Suppose $p\intertwinedwith p'$.}

Write $C=\closure{O}$ and $C'=\ns P\setminus O$.
By Lemma~\ref{lemm.closure.closed} $C$ is closed, and by Lemma~\ref{lemm.closed.complement.open} so is $\ns P\setminus O$. 
By construction and Lemma~\ref{lemm.closure.monotone}(\ref{closure.increasing}) $C\cup C'=\ns P$, so by Lemma~\ref{lemm.intertwinedwith.as.equiv}(\ref{item.intertwinedwith.as.equiv}) $p,p'\in C\lor p,p'\in C'$.
\item
\emph{Suppose $p\notintertwinedwith p'$,} so there exist $p\in O\in\opens$ and $p'\in O'\in\opens$ such that $O\notbetween O'$.

Because $p'\in O'\notbetween O$, we know from Definition~\ref{defn.closure}(\ref{item.closure}) that $p'\notin\closure{O}$. 
Also, $p\notin\ns P\setminus O$ (because $p\in O$). 
It follows that $p,p'\in\closure{O}$ is impossible, and $p,p'\in\ns P\setminus O$ is impossible. 
\end{itemize}
For the equivalence of parts~\ref{item.iw.closures.1} and~\ref{item.iw.closures.4} it suffices to show that part~\ref{item.iw.closures.2} implies part~\ref{item.iw.closures.4}, and that the negation of Lemma~\ref{lemm.iw.reg}(\ref{item.iw.reg.2}) implies the negation of part~\ref{item.iw.closures.4}:
\begin{itemize}
\item
\emph{Suppose $\Forall{O\in\opens} p,p'\in\closure{O} \lor p,p'\in\ns P\setminus O$.}

Then certainly $\Forall{O\in\regularOpens} p,p'\in \closure{O} \lor p,p'\in \ns P\setminus O$, because every regular open set is also an open set.
\item
\emph{Suppose we have $O,O'\in\regularOpens$ such that $p\in O\land p'\in O'$ and $O\notbetween O'$.}

Since $p\in O$, we have by Lemma~\ref{lemm.closure.monotone}(\ref{closure.increasing}) that $p\in\closure{O}$, and by construction that $p\notin \ns P\setminus O$.
Since $p'\in O'\notbetween O$, we have that $p'\in\ns P\setminus O$, and by Definition~\ref{defn.closure}(\ref{item.closure}) that $p'\notin \closure{O}$.
\end{itemize}
We prove equivalence of parts~\ref{item.iw.closures.1} and~\ref{item.iw.closures.5} as follows:
\begin{itemize}
\item
\emph{Suppose $p\intertwinedwith p'$.}

Suppose $C\in\closed$.
Write $C'=\closure{\ns P\setminus C}$.
By Lemma~\ref{lemm.closure.closed} $C'$ is closed, and 
by Lemma~\ref{lemm.closure.monotone}(\ref{closure.increasing}) $C\cup C'=\ns P$.
It follows by Lemma~\ref{lemm.intertwinedwith.as.equiv}(\ref{item.intertwinedwith.as.equiv}) that $p,p'\in C\lor p,p'\in C'$, so we are done.
\item
\emph{Suppose $p\notintertwinedwith p'$,} so there exist $p\in O\in\opens$ and $p'\in O'\in\opens$ such that $O\notbetween O'$.

Write $C=\closure{O}$ and $C'=\ns P\setminus O$.
Because $p'\in O'\notbetween O$, we know from Definition~\ref{defn.closure}(\ref{item.closure}) that $p'\notin C$. 
Also, $p\notin\ns P\setminus O$ (because $p\in O$).
Now using Lemma~\ref{lemm.closure.interior}(\ref{item.closure.interior.complement.closure}\&\ref{item.closure.interior.closed}) we have that
$$
\closure{\ns P\setminus\closure{O}}
\stackrel{L\ref{lemm.closure.interior}(\ref{item.closure.interior.complement.closure})}=
\closure{\interior(\ns P\setminus O)}
=
\closure{\interior(C')}
\stackrel{L\ref{lemm.closure.interior}(\ref{item.closure.interior.closed})}\subseteq 
C'.
$$
So if $p\notin C'$ then also $p\notin\closure{\ns P\setminus C}$.  
It follows that $p,p'\in C$ is impossible, and $p,p'\in\ns P\setminus C$ is impossible, so we are done. 
\end{itemize}
Equivalence of parts~\ref{item.iw.closures.4} and~\ref{item.iw.closures.6} follows by a routine argument from regularity (Definition~\ref{defn.regular.open.set}) using Corollary~\ref{corr.ro=rc} and Lemma~\ref{lemm.closure.interior}.
\end{proof}

\begin{prop}
\label{prop.intertwinedwith.ment.iff}
For a semitopology $(\ns P,\opens)$ and $p,p'\in\ns P$, the following are equivalent:
\begin{enumerate*}
\item\label{item.intertwinedwith.ment.iff.1}
$p\intertwinedwith p'$
\item\label{item.intertwinedwith.ment.iff.2}
$\ns P,\opens\ment p\tlatticeiff p'$
\item\label{item.intertwinedwith.ment.iff.3}
$\ns P,\opens\mentX p\tlatticeiff p'$ (Definition~\ref{defn.definite}(\ref{item.mentX}))
\end{enumerate*}
\end{prop}
\begin{proof}
Parts~\ref{item.intertwinedwith.ment.iff.1} and~\ref{item.intertwinedwith.ment.iff.2} just repeat Lemma~\ref{lemm.char.intertwinedwith}(\ref{item.char.intertwinedwith.1}\&\ref{item.char.intertwinedwith.2}).
For equivalence of parts~\ref{item.intertwinedwith.ment.iff.2} and~\ref{item.intertwinedwith.ment.iff.3} we prove two implications:
\begin{itemize}
\item
Suppose $\ns P,\opens\ment p\tlatticeiff p'$.
Then certainly $\ns P,\opens\mentX p\tlatticeiff p'$, since every extremal valuation is a continuous valuation.
\item
Suppose $\ns P,\opens\nment p\tlatticeiff p'$.
By the equivalence of parts~\ref{item.intertwinedwith.ment.iff.1} and~\ref{item.intertwinedwith.ment.iff.2} of this result and by Lemma~\ref{lemm.intertwinedwith.closures}, there exists an $O\in\regularOpens$ such that $p\notin\ns P\setminus O$ --- so $p\in O$ --- and $p'\notin \closure{O}$ --- so $p'\in\interior(\ns P\setminus O)$ --- or $p'\in O$ and $p\in\ns P\setminus O$.
We consider the former case; the proof for the latter is precisely similar.

We set $\avaluation=\indicator{O_\tvT O_\tvF}(O,\interior(\ns P\setminus O))$.
By Remark~\ref{rmrk.char.correspondence.extremal} this is extremal, and by construction $\avaluation(p)=\tvT$ and $\avaluation(p')=\tvF$ so that $\avaluation\nment p\tlatticeiff p'$. 
\qedhere\end{itemize}
\end{proof}

\begin{rmrk}
\label{rmrk.hyper.design.space}
Lemma~\ref{lemm.intertwinedwith.closures} and Proposition~\ref{prop.intertwinedwith.ment.iff} are interesting not just for what they are, but for the design space that they suggest:
\begin{enumerate*}
\item
Given that we have 
\begin{itemize*}
\item
two turnstiles ($\ment$ and $\mentX$), 
\item
two notions of logical equivalence ($\tlatticeiff$ and $\tiff$), and also 
\item
the $\modT$ modality, 
\end{itemize*}
when Proposition~\ref{prop.intertwinedwith.ment.iff} notes that these are equivalent
$$
p\intertwinedwith p'
\quad\liff\quad
\ns P,\opens\ment p\tlatticeiff p'
\quad\liff\quad
\ns P,\opens\mentX p\tlatticeiff p' ,
$$ 
this only draws attention to the rest of design space in which they are embedded (below, we omit $\ns P,\opens$ for compactness):\footnote{The design space is even larger than this, of course.  We have multiple modalities, and Remark~\ref{rmrk.two.implications} notes that there are actually sixteen different implications in three-valued logic.  But we have to start somewhere.}
\begin{multline*}
\ment p\tlatticeiff p'
\quad
\mentX p\tlatticeiff p'
\quad
\ment p\tiff p'
\quad
\mentX p\tiff p'
\quad
\\
\ment \modT(p\tlatticeiff p')
\quad
\mentX \modT(p\tlatticeiff p')
\quad
\ment \modT(p\tiff p')
\quad
\mentX \modT(p\tiff p')
\end{multline*}
\item
The equivalences in Lemma~\ref{lemm.intertwinedwith.closures} lack the tidy conciseness of the logical presentation, but they invite the same question: 
\begin{quote}
\emph{What is the design space of possible antiseparation properties between points?}
\end{quote}
\end{enumerate*}
We sum our answer up in a short table in Figure~\ref{fig.short.table},
and we can use Lemmas~\ref{lemm.how.hypertwined.fits.in} and~\ref{lemm.how.topind.fits.in} (proved below) to reformat Figure~\ref{fig.short.table} diagrammatically as per Figure~\ref{fig.corr.imp}, where arrows indicate implication / relation inclusion.
\end{rmrk} 

\begin{figure}
$$
\begin{array}{l@{\qquad}l@{\quad}l@{\quad}l}
\text{\emph{Intertwined} $\intertwinedwith$}
&
\ment p\tlatticeiff p'
&
\mentX p\tlatticeiff p'
&\text{See P.\ref{prop.intertwinedwith.ment.iff}}
\\[1ex]
\text{\emph{Consensus equivalent} $\intertwinedwithC$}
&
\mentX p\tiff p'
&
&\text{See L.\ref{lemm.ce.char}}
\\[1ex]
\text{\emph{Hypertwined} $\intertwinedwithX$}
&
\mentX \modT(p\tlatticeiff p')
&
\mentX \modT(p\tiff p')
&
\text{See L.\ref{lemm.ht.char}}
\\[1ex]
\text{\emph{Topologically indistinguishable} $\topind$}
&
\ment p\tiff p'
&
&\text{See L.\ref{lemm.logical.topind.char}}
\\[1ex]
\text{\emph{Always false} $\varnothing$}
&
\ment \modT(p\tlatticeiff p')
&
\ment \modT(p\tiff p')
&\text{See L.\ref{lemm.always.false}}
\end{array}
$$
\caption{Summary of antiseparation properties}
\label{fig.short.table}
\end{figure}

\begin{figure}
$$
\begin{tikzcd}
&\intertwinedwithX
\\
\varnothing & & \intertwinedwithC & \intertwinedwith  
\\
&\topind
\arrow[from=2-1, to=1-2]
\arrow[from=2-1, to=3-2]
\arrow[from=1-2, to=2-3]
\arrow[from=3-2, to=2-3]
\arrow[from=2-3, to=2-4]
\end{tikzcd}
$$
\caption{Summary of Lemmas~\ref{lemm.how.hypertwined.fits.in} \&~\ref{lemm.how.topind.fits.in}}
\label{fig.corr.imp}
\end{figure}

\jamiesubsubsection{Topological indistinguishability $\topind$, and the empty relation $\varnothing$}

In Lemma~\ref{lemm.logical.topind.char} we noted that in a semitopology $(\ns P,\opens)$,
$$
p\topind p'
\quad\text{holds precisely when}\quad 
\ns P,\opens\ment p\tiff p'.
$$
We now show that $\ns P,\opens\ment \modT(p\tlatticeiff p')$ and $\ns P,\opens\ment \modT(p\tiff p')$ yield the empty relation:
\begin{lemm}
\label{lemm.always.false}
Suppose $(\ns P,\opens)$ is a semitopology and $p,p'\in\ns P$.\footnote{If the semitopology is empty then there is no $p\in\ns P$ to choose.}
Then 
$$
\ns P,\opens\nment \modT(p\tlatticeiff p')
\quad\text{and}\quad
\ns P,\opens\nment \modT(p\tiff p')
$$ 
always (even if $p=p'$). 
\end{lemm}
\begin{proof}
We consider the valuation $\lambda x.\tvB:\ns P\to\THREE$.
The reader can check that this is continuous, and from Figures~\ref{fig.3} and~\ref{fig.3.phi.f} 
$$
\model{\modT(p\tlatticeiff p')}_\avaluation=\tvF
\quad\text{and}\quad
\model{\modT(p\tiff p')}_\avaluation=\tvF
. 
$$
The result follows from Definition~\ref{defn.ment}(\ref{item.P.O.phi.valid}\&\ref{item.f.phi.valid}).
\end{proof}

\jamiesubsubsection{Consensus equivalence $\intertwinedwithC$, hypertwined $\intertwinedwithX$, and hyperdefinite}

\begin{defn}
\label{defn.ht.ce}
Suppose $(\ns P,\opens)$ is a semitopology and $p,p'\in\ns P$.
Then: 
\begin{enumerate*}
\item
By Lemma~\ref{lemm.intertwinedwith.closures}(\ref{item.iw.closures.4}), $p$ and $p'$ are intertwined when for every $O\in\regularOpens$,
$$
p,p'\in\closure{O}\ \lor\ p,p'\in\ns P\setminus O .
$$ 
Then we can define:
\item\label{item.ce}
Call $p$ and $p'$ \deffont[consensus equivalent points $p\intertwinedwithC p'$]{consensus equivalent}\index{$p\intertwinedwithC p'$ (consensus equivalent points)} and write $p\intertwinedwithC p'$ when for every $O\in\regularOpens$, 
$$
p,p'\in O\ \lor\  p,p'\in\interior(\ns P\setminus O)\ \lor\ p,p'\notin O\cup\interior(\ns P\setminus O).
$$
By elementary propositional manipulations we can write this equivalently as 
$$
(p\in O\liff p'\in O)\ \land\ (p\in\interior(\ns P\setminus O)\liff p'\in \interior(\ns P\setminus O)).
$$
We may use these two forms synonymously without comment.
\item\label{item.ht}
Call $p$ and $p'$ \deffont[hypertwined points $p\intertwinedwithX p'$]{hypertwined}\index{$p\intertwinedwithX p'$ (hypertwined points)} and write $p\intertwinedwithX p'$ when for every $O\in\regularOpens$, 
$$
p,p'\in O \ \lor\  p,p'\in\interior(\ns P\setminus O).
$$
\item\label{item.hd}
Recall from Definition~\ref{defn.definite} the notion of \emph{definite} truth-values.
Expanding on this, call $p$ \deffont[hyperdefinite point]{hyperdefinite} when for every $O\in\regularOpens$,
$$
p\in O \ \lor\ p\in\interior(\ns P\setminus O) .
$$
\item
We may write $\intertwinedC{p}$ and $\intertwinedX{p}$ for the set of points that are consensus equivalent with and hypertwined with $p$ respectively:
$$
\intertwinedC{p} = \{p'\in\ns P \mid p\intertwinedwithC p'\}
\qquad
\intertwinedX{p} = \{p'\in\ns P \mid p\intertwinedwithX p'\} .
$$
\end{enumerate*}
\end{defn}

\begin{rmrk}
Definition~\ref{defn.ht.ce} uses quantifications over regular open sets, but recall from Corollary~\ref{corr.ro=rc} that taking closures and taking open interiors yield bijections between $\regularOpens$ and $\regularClosed$.
Thus, reasonable alternative forms of the Definition are easy to write out with regular closed sets.
\end{rmrk}

\begin{figure}
\vspace{-1em}
\centering
\includegraphics[width=0.4\columnwidth]{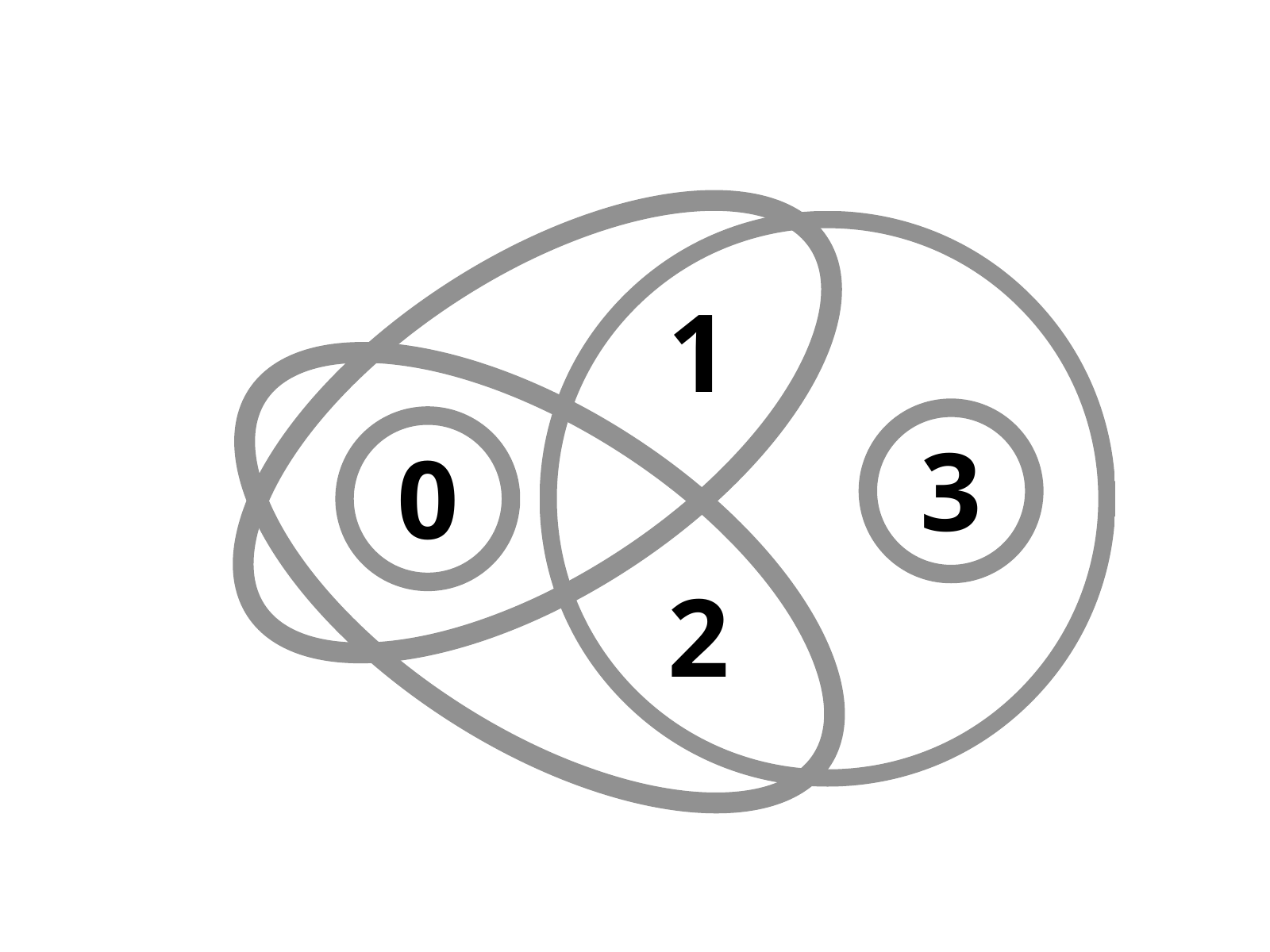}
\vspace{-1em}
\caption{$1$ and $2$ are hypertwined and hyperdefinite, but they are not even quasiregular (Example~\ref{xmpl.ht.not.wr})}
\label{fig.hypertwined12}
\end{figure}

\begin{xmpl}
\label{xmpl.ht.not.wr}
Figure~\ref{fig.hypertwined12} illustrates a space with two points, $1$ and $2$, that are unconflicted, hyperregular, and hypertwined with each other; but they are not even quasiregular.

This means that they have good consensus behaviour in the sense that they always agree on an unambivalent (non-$\tvB$) truth-value for any extremal valuation, but they are not intertwined with any topen set of points.
The `problem' (if we wish to call it a problem) is that $1$ and $2$ will agree with each other and will agree with at least one of $0$ or $3$, but (intuitively) $1$ and $2$ can choose \emph{which} of $0$ or $3$ to agree with when they decide their state. 

We make a few more small observations:
\begin{itemize}
\item
$\intertwinedX{p}$ and $\intertwinedC{p}$ are not necessarily open. 
Consider $p=0$ in the semitopology illustrated in Figure~\ref{fig.square.diagram}; then $\intertwinedX{p}=\{0\}=\intertwinedC{p}$.
\item
$\intertwinedX{p}$ and $\intertwinedC{p}$ are not necessarily a singleton.
Consider $p=1$ in the semitopology illustrated in Figure~\ref{fig.hypertwined12}; $\intertwinedX{p}=\{1,2\}=\intertwinedC{p}$.
\item
$\intertwinedX{p}$ and $\intertwinedC{p}$ do not necessarily have a nonempty open interior.
Again, consider $p=1$ in the semitopology illustrated in Figure~\ref{fig.hypertwined12}; $\interior(\intertwinedX{p})=\varnothing=\intertwinedC{p}$.
\end{itemize}
If $p$ is regular then things are simpler: see Proposition~\ref{prop.intertwined.implies.intertwinedX} and Remark~\ref{rmrk.simple.intertwinedX}.
\end{xmpl}

\begin{lemm}
\label{lemm.ht.ce.hd}
For a semitopology $(\ns P,\opens)$ and $p,p'\in\ns P$,
the following are equivalent:
\begin{enumerate*}
\item
$p\intertwinedwithX p'$ (Definition~\ref{defn.ht.ce}(\ref{item.ht})).
\item
$p\intertwinedwithC p'$ and at least one of $p$ and $p'$ is hyperdefinite.
\item
$p\intertwinedwithC p'$ and both $p$ and $p'$ are hyperdefinite.
\end{enumerate*}
We can write: 
\begin{quote}
\emph{hypertwined = consensus equivalent + hyperdefinite}.
\end{quote}
\end{lemm}
\begin{proof}
Routine from Definition~\ref{defn.ht.ce}(\ref{item.ce}\&\ref{item.ht}).
\end{proof}

By Definition~\ref{defn.ht.ce}(\ref{item.hd}) we call $p$ \emph{hyperdefinite} when for every regular $O\in\regularOpens$, $p\in O$ or $p\in\interior(\ns P\setminus O)$.
By Definition~\ref{defn.sc} we call $p$ \emph{hypertransitive} when for every $O,O'\in\opens$, if $p\in\closure{O}\cap\closure{O'}$ then $O\between O'$.
Remarkably, these two conditions are equivalent: 
\begin{prop}
\label{prop.hd=ht}
For a semitopology $(\ns P,\opens)$ and $p,p'\in\ns P$, the following are equivalent:
\begin{enumerate*}
\item
$p$ is hyperdefinite in the sense of Definition~\ref{defn.ht.ce}(\ref{item.hd}) ($p\in O$ or $p\in\interior(\ns P\setminus O)$ for every $O\in\regularOpens$).
\item
$p$ is hypertransitive in the sense of Definition~\ref{defn.sc}.
\end{enumerate*}
We can write:
\begin{quote}
\emph{Hyperdefinite = hypertransitive.}
\end{quote}
\end{prop}
\begin{proof}
We prove two implications:
\begin{itemize}
\item
Suppose $p$ is hypertransitive and consider $O\in\regularOpens$.
Write $O'=\interior(\ns P\setminus O)$; note by Lemma~\ref{lemm.regular.open.closed} that $\ns P\setminus O$ is also regular, so that $\closure{O'}=\ns P\setminus O$ and $O\cup O'=\ns P$.
By construction $O\notbetween O'$, and it follows by the contrapositive of Lemma~\ref{lemm.sc.op.reg.op}(\ref{item.sc.op.reg.op.3}) that $p\in O$ or $p\in O'$ as required.
\item
Suppose $p$ is hyperdefinite and consider $O,O'\in\regularOpens$ such that $p\in\closure{O}\cap\closure{O'}$.
Set $R'=\interior(\ns P\setminus O)$. 
By assumption (since $p$ is hyperdefinite) $p\in O$ or $p\in R'$.

If $p\in R'$ then by Definition~\ref{defn.closure}(\ref{item.closure}) (because $p\in\closure{O}$) also $R'\between O$, which is impossible by construction, so $p\notin R'$. 

Therefore it must be that $p\in O$.
Then by Definition~\ref{defn.closure}(\ref{item.closure}) (because $p\in\closure{O'}$) we have $O\between O'$ as required.
\qedhere\end{itemize}
\end{proof}

\begin{corr}
\label{corr.ht.hd}
For a semitopology $(\ns P,\opens)$ and $p\in\ns P$, the following are equivalent:
\begin{enumerate*}
\item
$p$ is hyperdefinite.
\item
$p$ is hypertransitive.
\item
$p\intertwinedwithX p$.
\item\label{item.ht.hd.any.point}
$p\intertwinedwithX p'$ for any $p'\in\ns P$.
\end{enumerate*}
\end{corr}
\begin{proof}
Routine from Definition~\ref{defn.ht.ce}(\ref{item.ce}\&\ref{item.ht}) and Proposition~\ref{prop.hd=ht}.
\end{proof}

\begin{lemm}
\label{lemm.how.hypertwined.fits.in}
Suppose $(\ns P,\opens)$ is a semitopology and $p,p'\in\ns P$.
Then:
\begin{enumerate*}
\item
If $p\intertwinedwithX p'$ then $p\intertwinedwithC p'$.
The converse implication need not hold.
\item
If $p\intertwinedwithC p'$ then $p\intertwinedwith p'$.
The converse implication need not hold.
\end{enumerate*}
In symbols, we can write:
$$
\intertwinedX{p}\subseteq\intertwinedC{p}\subseteq\intertwined{p} .
$$
\end{lemm}
\begin{proof}
The implications are routine from Definition~\ref{defn.ht.ce}, using Lemma~\ref{lemm.closure.monotone}(\ref{closure.increasing}) and arguments on sets.

For the non-implications it suffices to provide counterexamples.
\begin{enumerate*}
\item
Consider the semitopology $\THREE$ from Definition~\ref{defn.3.top} 
Set $p=p'=\tvB$.
The reader can check that $p\intertwinedwithC p'$ but $\neg(p\intertwinedwithX p')$.\footnote{If the reader finds it a little confusing to think about continuous valuations from $\THREE$ to itself, use the semitopology illustrated in Figure~\ref{fig.012}, top-left diagram, instead.
This is isomorphic, but instead of $\tvF$, $\tvB$, $\tvT$ we have $0$, $1$, and $2$.} 
\item
Consider the semitopology $\THREE$ again, but now set $p=\tvT$ and $p'=\tvB$. 
The reader can check that $p\intertwinedwith p'$ but $\neg(p\intertwinedwithC p')$.
\qedhere\end{enumerate*}
\end{proof}

Topological indistinguishability is stronger than consensus equivalence, but it is incomparable with being hypertwined: 
\begin{lemm}
\label{lemm.how.topind.fits.in}
Suppose $(\ns P,\opens)$ is a semitopology and $p,p'\in\ns P$.
Then:
\begin{enumerate*}
\item
If $p\topind p'$ then $p\intertwinedwithC p'$.
The converse implication need not hold.
\item
The properties $p\topind p'$ and $p\intertwinedwithX p'$ are incomparable.
That is:
\begin{enumerate*}
\item
$p$ and $p'$ can be topologically indistinguishable but not hypertwined.
\item
$p$ and $p'$ can be hypertwined but not topologically indistinguishable. 
\end{enumerate*}
\end{enumerate*}
\end{lemm}
\begin{proof}
Suppose $p$ and $p'$ are topologically indistinguishable.
Then it is routine to check Definition~\ref{defn.ht.ce}(\ref{item.ce}) and see that they are consensus equivalent.

For the non-implications, it suffices to provide counterexamples.
\begin{itemize*}
\item
Consider the semitopology $\THREE$ from Definition~\ref{defn.3.top}.
Set $p=p'=\tvB$.
Clearly, $p$ and $p'$ are topologically indistinguishable; however $\neg(p\intertwinedwithX p')$.
\item
Consider the Sierpi\'nski space $\ns P=\{0,1\}$ and $\opens=\{\varnothing,\{1\},\{0,1\}\}$ (cf. Example~\ref{xmpl.sk} and Remark~\ref{rmrk.three.scene}).

Set $p=0$ and $p'=1$; these are topologically distinguishable since $p'\in\{1\}$ and $p\notin\{1\}$.
But, there are only two extremal valuations --- $\lambda x.0$ and $\lambda x.1$ --- and it follows that $p\intertwinedwithC p'$ and also $p\intertwinedwithX p'$. 
\qedhere\end{itemize*}
\end{proof}

\begin{lemm}
\label{lemm.ce.char}
For a semitopology $(\ns P,\opens)$ and $p,p'\in\ns P$,
the following are equivalent:
\begin{enumerate*}
\item\label{item.ce.char.1}
$p\intertwinedwithC p'$ (Definition~\ref{defn.ht.ce}(\ref{item.ce})).
\item\label{item.ce.char.2}
$\ns P,\opens\mentX p\tiff p'$ (Definition~\ref{defn.definite}(\ref{item.mentX})).
\item\label{item.ce.char.3}
$\avaluation(p)=\avaluation(p')$ for every extremal $\avaluation:\ns P\to\THREE$.
\end{enumerate*}
\end{lemm}
\begin{proof}
For equivalence of parts~\ref{item.ce.char.1} and~\ref{item.ce.char.2} we prove two implications:
\begin{itemize}
\item
\emph{Suppose $p\intertwinedwithC p'$.}

Suppose $\avaluation:\ns P\to\THREE$ is an extremal valuation.
By Definition~\ref{defn.ht.ce}(\ref{item.ce}), $p\in\avaluation^\mone\{\tvF\} \liff p'\in\avaluation^\mone\{\tvF\}$. 
By Corollary~\ref{corr.valuation.pp'.iff} $\avaluation\ment p\tiff p'$.
Since $\avaluation$ was arbitrary, by Definition~\ref{defn.definite}(\ref{item.mentX}) $\ns P,\opens\mentX p\tiff p'$ as required.
\item
\emph{Suppose $\ns P,\opens\mentX p\tiff p'$}.

Consider some regular open set $O\in\regularOpens$.
Set 
$$
\avaluation=\indicator{O_\tvT,O_\tvF}(O,\interior(\ns P\setminus O))
\quad\text{and}\quad
\avaluation'=\indicator{O_\tvT,O_\tvF}(\interior(\ns P\setminus O),O).
$$
By Proposition~\ref{prop.max.disjoint.to.regular}(1\&3) and Corollary~\ref{corr.definite.disjoint} these are both extremal valuations, and from Corollary~\ref{corr.valuation.pp'.iff} it follows that 
$p\in O\liff p'\in O$ and $p\in\interior(\ns P\setminus O)\liff p'\in\interior(\ns P\setminus O)$. 
By Definition~\ref{defn.ht.ce}(\ref{item.ce}) $p\intertwinedwithC p'$ as required.
\end{itemize}
Equivalence of parts~\ref{item.ce.char.1} and~\ref{item.ce.char.3} is by routine calculations from Definition~\ref{defn.ht.ce}(\ref{item.ce}) and Remark~\ref{rmrk.char.correspondence}.
\end{proof}

\begin{lemm}
\label{lemm.hd.char}
For a semitopology $(\ns P,\opens)$ and $p\in\ns P$, the following are equivalent:
\begin{enumerate*}
\item
$p$ is hyperdefinite (Definition~\ref{defn.ht.ce}(\ref{item.hd})).
\item
$\ns P,\opens\mentX \modT p \tor \modT\tneg p$.
\item
$\ns P,\opens\nment^X \modB p$.
\item
$\avaluation(p)\in\{\tvT,\tvB\}$ for every extremal $\avaluation:\ns P\to\THREE$.
\end{enumerate*}
\end{lemm}
\begin{proof}
By routine calculations using Remark~\ref{rmrk.char.correspondence} and using the truth-tables in Figure~\ref{fig.3}.
\end{proof}

\begin{lemm}
\label{lemm.ht.char}
For a semitopology $(\ns P,\opens)$ and $p,p'\in\ns P$, the following are equivalent:
\begin{enumerate*}
\item
$p\intertwinedwithX p'$ (Definition~\ref{defn.ht.ce}(\ref{item.ht})).
\item
$\ns P,\opens\mentX \modT(p\tiff p')$.
\item
$\ns P,\opens\mentX \modT(p\tlatticeiff p')$.
\item\label{item.ht.char.truth-values}
$\avaluation(p)=\avaluation(p')\in\{\tvT,\tvB\}$ for every extremal $\avaluation:\ns P\to\THREE$.
\end{enumerate*}
\end{lemm}
\begin{proof}
By routine calculations combining Lemmas~\ref{lemm.hd.char}, \ref{lemm.ce.char}, and~\ref{lemm.ht.ce.hd} and using the truth-tables in Figure~\ref{fig.3}.
\end{proof}

\begin{rmrk}
\label{rmrk.sum.up.inter}
It may be helpful to sum up the high points of the results above (this summary is for intuition; references to the precise results are included):
\begin{enumerate*}
\item
Consensus equivalence $\intertwinedwithC$ is when two points return the same truth-value in $\THREE$ for all extremal valuations.
This is an equivalence relation, because equality is an equivalence relation.
\item
Being \emph{hypertwined} $\intertwinedwithX$ is the partial equivalence relation obtained by restricting consensus equivalence to the \emph{hyperdefinite points} (those that return $\tvT$ or $\tvF$ in all extremal valuations; never $\tvB$).
\item
Being hyperdefinite is the same as being hypertransitive, and also the same as being hypertwined with yourself or with any other point, and the same as validating $\ns P,\opens\ment \modT p \tor \modT \tneg p$ (Corollary~\ref{corr.ht.hd}).
\item
So in answer to our question of Remark~\ref{rmrk.hyper.design.space}, the design space is populated by four distinct entities: 
\begin{enumerate*}
\item
\emph{Topological indistinguishability} $\topind$.
By Lemma~\ref{lemm.logical.topind.char} this lives naturally in the world of possibly non-extremal valuations as $\ns P,\opens\ment p\tiff p'$; see Figure~\ref{fig.short.table}.

$p\topind p'$ when $\avaluation(p)=\avaluation(p')$ for every continuous valuation.
This relation is, of course, very familiar from topology; a space is $T_0$ precisely when $\topind$ coincides with $=$. 
\item
Being \emph{intertwined} $\intertwinedwith$.
This lives both in the world of extremal and possibly non-extremal valuations, as $\ns P,\opens\mentX p\tlatticeiff p'$ and $\ns P,\opens\ment p\tlatticeiff p'$ respectively, as per Figure~\ref{fig.short.table}.
We have studied $\intertwinedwith$ very closely, but there is also another one: 
\item
Being \emph{consensus equivalent} $\intertwinedwithC$.
This lives naturally in the world of extremal valuations as $\ns P,\opens\mentX p\tiff p'$, as per Figure~\ref{fig.short.table}.

$p\intertwinedwithC p'$ holds when $\avaluation(p)=\avaluation(p')$ for every \emph{extremal} valuation $\avaluation$ (i.e. like topological indistinguishability, but only for extremal valuations).
This is a very natural notion.
\item
Being \emph{hyperdefinite / hypertransitive}.
Remarkably, and like being intertwined, this lives naturally in both worlds, as per Proposition~\ref{prop.hd=ht}.
\end{enumerate*} 
\end{enumerate*} 
\end{rmrk}

\begin{rmrk}
\label{rmrk.mcn}
Remark~\ref{rmrk.sum.up.inter} invites the question of whether other relevant regularity or intertwinedness properties exist?\footnote{The list so far: (quasi/weak/indirect)-regularity; intertwined, consensus equivalent, hypertwined; and being unconflicted, hyperdefinite, hypertransitive, and strongly compatible.}
Yes --- and in fact we can note examples of both.

\emph{A regularity property.}\quad
In Theorem~\ref{thrm.up.down.char}(\ref{item.up.down.char.wr.mcn}) we showed that $p$ is regular when $p$ is weakly regular and $\intertwined{p}$ is a minimal closed neighbourhood.
If we write $\f{MCN}(p)$ for the property that $\intertwined{p}$ is a minimal closed neighbourhood, then we can write Theorem~\ref{thrm.up.down.char} as 
\begin{quote}
`regular = weakly regular + MCN' 
\end{quote}
following the style of other results of this genre such as Theorem~\ref{thrm.r=wr+uc} for `regular = weakly regular + unconflicted' and Theorem~\ref{thrm.regular=qr+sc} for `regular = quasiregular + hypertransitive'.
Now note that MCN does not coincide with being unconflicted (the points in Figure~\ref{fig.square.diagram} are unconflicted but not MCN).
And, MCN does not coincide with being hypertransitive (the points $1$ and $2$ in Figure~\ref{fig.hypertwined12} are hypertwined but not MCN).
So MCN is \dots another well-behavedness property.
It is an open problem how MCN fits in with the well-behavedness properties that we have considered so far.\footnote{Beyond a few simple comments, e.g. it can be proved that MCN implies indirect regularity (Definition~\ref{defn.indirectly.regular}).}

\emph{An intertwinedness property.}\quad
In Lemma~\ref{lemm.intertwined.not.transitive} we noted that `being intertwined' $\between$ is symmetric and reflexive, but not necessarily transitive, and it became clear that special cases where $\between$ \emph{is} transitive are of particular interest.
But, we can also directly study $\intertwinedwith^*$, the transitive closure of the intertwined-with relation $\intertwinedwith$: that is, call $p$ and $p'$ \deffont{transitively intertwined} and write $p\intertwinedwith^* p'$ when there exist points $p_1,\dots,p_n$ such that $p\between p_1\intertwinedwith \dots\intertwinedwith p_n\intertwinedwith p'$.\index{$p\intertwinedwith^* p'$ (transitively intertwined points)}
This transitive closure is another intertwinedness property with its own properties~\cite{gabbay:deccac}.

These are just some, of presumably many, such questions that remain to be considered. 
\end{rmrk}

We conclude with a brief investigation of the connections between $\intertwinedwithX$, regularity, and $\community(p)$.
This is just a routine application of the tools we have already built: 
\begin{prop}
\label{prop.intertwined.implies.intertwinedX}
Suppose $(\ns P,\opens)$ is a semitopology and $p,p'\in\ns P$.
Then: 
\begin{enumerate*}
\item\label{item.intertwined.implies.intertwinedX.1}
If there exists a topen $\atopen\in\topens$ such that $p,p'\in\atopen$, then $p\intertwinedwithX p'$.
\item\label{item.intertwined.implies.intertwinedX.2}
$p$ is regular if and only if $\community(p)=\intertwinedX{p}\neq\varnothing$.
\item\label{item.intertwined.implies.intertwinedX.3}
$p$ is regular if and only if $\community(p)\neq\varnothing$ and $\intertwinedX{p}\neq\varnothing$.
\end{enumerate*}
\end{prop}
\begin{proof}
We consider each part in turn:
\begin{enumerate}
\item
Suppose there exists a topen $\atopen\in\topens$ such that $p,p'\in\atopen$.
By Proposition~\ref{prop.scp.extremal}(\ref{item.scp.extremal.1}\&\ref{item.scp.extremal.2}) $\avaluation(p)=\avaluation(p')\in\{\tvT,\tvF\}$.
We use Lemma~\ref{lemm.ht.char}(\ref{item.ht.char.truth-values}).
\item
Suppose $p$ is regular, meaning by Definition~\ref{defn.tn}(\ref{item.regular.point}) that $p\in\community(p)\in\topens$.
We prove two subset inclusions:
\begin{itemize*}
\item
By part~\ref{item.intertwined.implies.intertwinedX.1} of this result $\community(p)\subseteq\intertwinedX{p}$.
\item
Set $\avaluation=\indicator{O_\tvT C_{\tvT\tvB}}(\community(p),\intertwined{p})$.
By Corollary~\ref{corr.community.regular.open} $\community(p)\in\regularOpens$, and from Remark~\ref{rmrk.char.correspondence.extremal}(\ref{item.char.extremal.regular}) this is an extremal valuation.
It follows that $\intertwinedX{p}\subseteq\community(p)$.
\end{itemize*}

Suppose $\community(p)=\intertwinedX{p}\neq\varnothing$.
By Definition~\ref{defn.tn}(\ref{item.quasiregular.point}) $p$ is quasiregular, and by Corollary~\ref{corr.ht.hd}(\ref{item.ht.hd.any.point}) 
$p$ is hypertransitive.
We use Theorem~\ref{thrm.regular=qr+sc}.
\item
Suppose $p$ is regular.
Then we use part~\ref{item.intertwined.implies.intertwinedX.2} of this result and we are done.

Suppose $\community(p)\neq\varnothing$ and $\intertwinedX{p}\neq\varnothing$.
By Definition~\ref{defn.tn}(\ref{item.quasiregular.point}) $p$ is quasiregular, and by Corollary~\ref{corr.ht.hd}(\ref{item.ht.hd.any.point}) 
$p$ is hypertransitive.
We just use Theorem~\ref{thrm.regular=qr+sc}. 
\qedhere\end{enumerate}
\end{proof}

\begin{rmrk}
\label{rmrk.simple.intertwinedX}
By Proposition~\ref{prop.intertwined.implies.intertwinedX}(\ref{item.intertwined.implies.intertwinedX.2}) and Lemma~\ref{lemm.how.hypertwined.fits.in}, if $p$ is regular then 
$$
\community(p)=\intertwinedC{p}=\intertwinedX{p}.
$$
If $p$ is not regular then the equalities need not hold.
The semitopology illustrated in Figure~\ref{fig.hypertwined12} gives an example, where $\community(1)=\varnothing$ but $\intertwinedX{1}=\{1,2\}$.
\end{rmrk}

\jamiepart{Conclusions}

\jamiesection{Conclusions}
\label{sect.conclusions}

We started by noticing that a notion of `actionable coalition' as discussed in the Introduction, suggests a topology-like structure which we call \emph{semitopologies}.

We simplified and purified our motivating examples to two mathematical questions: 
\begin{enumerate*}
\item
understand antiseparation properties, and 
\item
understand the implications of these for value assignments.\footnote{A value assignment is just a not-necessarily-continuous map from a semitopology to a discrete space (Definition~\ref{defn.value.assignment}(\ref{item.value.assignment})).}
\end{enumerate*}
We have surveyed the implications of these ideas and seen that they are mathematically rich and varied.
Point-set semitopologies have an interesting theory, and a family of results which resemble those of point-set topology, but which are different enough to have their own distinct character. 
We build and study a dual category of semiframes; and we provide a many-valued modal logic for describing them.

Generalising topology to drop the condition that intersections of open sets must be open, brings a wealth of new and interesting structure.
We have considered many results, but we also hope that putting this story together will serve as a stimulus to considering semitopologies as a new field of research.

\jamiesubsection{Topology vs. semitopology}
\label{subsect.vs}

We briefly compare and contrast topology and semitopology: 
\begin{enumerate}
\item
\emph{Topology:}\ 
Topology considers a wealth of separation properties, but we are not aware of a taxonomy of anti-separation properties in the topological literature.\footnote{The Wikipedia page on separation axioms~\cite{wiki:Separation_axiom} includes an excellent overview with over a dozen separation axioms; no anti-separation axioms are proposed.  Important non-Hausdorff spaces do exist; e.g. the \emph{Zariski topology}~\cite[Subsection~1.1.1]{hulek:eleag}.} 

\emph{Semitopology:}\ 
We consider a taxonomy of antiseparation properties, including: points being intertwined (see Definition~\ref{defn.intertwined.points} and Remark~\ref{rmrk.not.hausdorff}) and hypertwined and consensus equivalent (see the overviews in Figures~\ref{fig.short.table} and~\ref{fig.corr.imp}); and points being quasiregular, indirectly regular, weakly regular, and regular (Definitions~\ref{defn.tn} and~\ref{defn.indirectly.regular}), (un)conflicted (Definition~\ref{defn.conflicted}(\ref{item.unconflicted})), and hypertransitive (Definition~\ref{defn.sc}).\footnote{An extra word on the converse of this:  Our theory of semitopologies admits spaces whose points partition into distinct communities, as discussed in Theorem~\ref{thrm.topen.partition} and Remark~\ref{rmrk.partition}.  To a professional blockchain engineer it might seem terrible if two points points are \emph{not} intertwined, since this means they might not be in consensus in a final state. 
Should this not be excluded by the definition of semitopology, as is done in the literature on quorum systems, where it is typically definitionally assumed that all quorums in a quorum system intersect?  
No! 
Separation is a fact of life which we permit not only so that we can mathematically analyse it (and we do), but also because we may need it for certain \emph{normal situations}.
For example, most blockchains have a \emph{mainnet} and several \emph{testnets} and it is understood that each should be coherent within itself, but different nets \emph{need not} be in consensus with one another.  Indeed, if the mainnet had to agree with a testnet then this would likely be a bug, not a feature.  So the idea of having multiple partitions is nothing new \emph{per se}.  It is a familiar idea, which semitopologies put in a powerfully general mathematical context.}
\item
\emph{Topology:}\quad 
If a minimal open neighbourhood of a point exists then it is least, because we can intersect two minimal neighbourhoods to get a smaller one which by minimality is equal to both.

Yet, in topology the existence of a least open neighbourhood is not guaranteed (e.g. $0\in\mathbb R$ has no least open neighbourhood).

\emph{Semitopology:}\ 
A point may have multiple minimal open neighbourhoods --- examples are very easy to generate, see e.g. the top-right example in Figure~\ref{fig.012}.
Furthermore, in the useful special case of a chain-complete semitopology, every open neighbourhood of $p$ contains a(t least one) minimal open neighbourhood of $p$ (Corollary~\ref{corr.cover.exists}) so that existence of minimal open neighbourhoods is assured. 
\item
\emph{Topology:}\quad
Every finite $T_0$ topology is sober.
A topology is sober if and only if every nonempty irreducible closed set is the closure of a unique point.

\emph{Semitopology:}\quad
Neither property holds.  See Lemma~\ref{lemm.T0.not.sober}.
\item
\emph{Topology:}\quad 
We are typically interested in functions on topologies that are continuous (or mostly so, e.g. $f(x)=1/x$).
Thus for example, the definition of $\tf{Top}$ the category of topological spaces takes continuous functions as morphisms, essentially building in assumptions that continuous functions are of most interest and that finding them is enough of a solved problem that we can restrict to continuous functions in the definition.
 
\emph{Semitopology:}\quad 
For our intended application to consensus, 
we are explicitly interested in functions that may be discontinuous.
This models initial and intermediate states where local consensus has not yet been achieved, or final states on semitopologies that include disjoint topens and non-regular points (e.g. conflicted points), as well as adversarial or failing behaviour.
Thus, having continuity is neither a solved problem, nor even necessarily desirable.
\item
Sometimes, definitions from topology transfer to semitopology but split into multiple distinct notions when they do. 
For example: topology has one notion of \emph{dense subset of} and, as discussed in Remark~\ref{rmrk.top.ce}, when we transfer this to semitopologies it splits into two notions --- \emph{weakly} dense and \emph{strongly} dense (Definition~\ref{defn.dense}) --- both of which turn out to be important.

Sometimes, ideas that come from semitopology project down to topology but may lose impact in doing so; they make mathematical sense, but become less interesting, or at least lose finesse.
For example: our theory of semitopologies considers notions of \emph{topen set} and \emph{strongly topen set} (Definitions~\ref{defn.transitive} and~\ref{defn.strongly.transitive}).
In topology these are equivalent to one another, and to a known and simpler topological property of being \emph{hyperconnected} (Definition~\ref{defn.tangled}).\footnote{\dots but (strong) topens are their own thing.  Analogy: a projection from $\mathbb C$ to $\mathbb R$ maps $a+bi$ to $a$; this is not evidence that $i$ is equivalent to $0$!} 
Something similar happens with the semitopological notion of \emph{strongly dense for}; see the discussion in Remark~\ref{rmrk.strongly.dense.for}.
\item
A natural space of functions for describing topologies is continuous maps to the Sierpi\'nski space.
The natural space of functions for describing semitopologies seems to be possibly non-continuous maps to $\THREE$. 
\item
Semitopological questions such as \emph{`is this a topen set'} or \emph{`are these two points intertwined'} or \emph{`does this point have a topen neighbourhood'} --- and many other definitions, such as our taxonomy of points into \emph{regular}, \emph{weakly regular}, \emph{indirectly regular}, \emph{quasiregular}, \emph{unconflicted}, and \emph{hypertransitive}; or the notions of \emph{witnessed set}, and \emph{kernel} \dots and so on --- appear to be novel.

Also in the background 
is that we are particularly interested in properties and algorithms that work well using local and possibly incomplete or even partially incorrect information.

Thus semitopologies have their own distinct character: because they are mathematically distinct, and because modern applications having to do with actionable coalitions motivate us to ask questions that have not necessarily been considered before.
\end{enumerate}

\jamiesubsection{Related work}
\label{subsect.related.work}

\paragraph*{Union sets, closure spaces, and minimal structures}

There is a thread of research into \emph{union-closed families}; these are subsets of a finite powerset closed under unions, so that a union-closed family is precisely just a finite semitopology. 
The motivation is to study the combinatorics of finite subsemilattices of a powerset.
Some progress has been made in this~\cite{poonen:unicf}; the canonical reference for the relevant combinatorial conjectures is the `problem session' on page~525 (conjectures 1.9, 1.9', and 1.9") of~\cite{rival:grao}.
See also recent progress in a conjecture about union-closed families.\footnote{\url{https://web.archive.org/web/20230330170701/https://en.wikipedia.org/wiki/Union-closed_sets_conjecture\#Partial_results}.}
There is no direct connection to semitopologies, and certainly no consideration of duality results.
Perhaps the duality that we present may be of some interest in that community. 

A \emph{closure space} is a subset of a powerset that is closed under intersections~\cite[page~173]{erne:clo}, so that the set of complements of a closure space is just a semitopology (and likewise a finite closure space is, up to taking sets complements, just a union-closed family).
The motivation for closure spaces is, as the name suggests, to study closure operations in a topology-flavoured style, so closure spaces (unlike union-closed sets) share a topological flavour with semitopologies.

It is not clear that semitopologies are `just' closure spaces, union sets, or complete semilattices (see also Remark~\ref{rmrk.PtoP}).
This is because of how the theory of semitopologies (e.g. the basic notion of intertwined points) uses sets intersection $\between$.
Semiframes (the dual to semitopologies) make this explicit with the compatibility relation $\ast$ (Definitions~\ref{defn.compatibility.relation} and~\ref{defn.semiframe}), demonstrating that semitopologies are in fact \emph{compatible} complete sublattices of a powerset.
This seems to be a new idea.
And of course: the applications are new. 

A \emph{minimal structure} on a set $X$ is a subset of $\powerset(X)$ that contains $\varnothing$ and $X$.
Thus a semitopology is a minimal structure that is also closed under arbitrary unions.
There is a thread of research into minimal structures, studying how notions familiar from topology (such as continuity) fare in weak (minimal) settings~\cite{noiri:defsgf} and how this changes as axioms (such as closure under unions) are added or removed.
An accessible discussion is in~\cite{szaz:minsgt}, and see the brief but comprehensive references in Remark~3.7 of that paper.
Of course our focus is on properties of semitopologies 
which are not considered in that literature; but we share an observation with minimal structures that it is useful to study topology-like constructs, in the absence of closure under intersections.

\paragraph*{Gradecast converges on a topen}

Many consensus algorithms have the property that once consensus is established in a quorum $O$, it propagates to $\closure{O}$.
For example, in the Grade-Cast algorithm~\cite{feldman_optimal_1988}, participants assign a confidence grade of 0, 1 or 2 to their output and must ensure that if any participant outputs $v$ with grade 2 then all must output $v$ with grade at least 1.
If all the quorums of a participant intersect some set $S$ that unanimously supports value $v$, then the participant assigns grade at least 1 to $v$.

From the view of our paper, this is just taking a closure, which suggests that, to convince a topen to agree on a value, it would suffice to first convince an open neighbourhood that intersects the topen, and then use Grade-Cast to convince the whole topen.
More on this in Proposition~\ref{prop.open.strong-consensus} and Remark~\ref{rmrk.gradecast}.

\paragraph*{Algebraic topology as applied to distributed computing tasks}

The reader may know that solvability results about distributed computing tasks have been obtained from algebraic topology, starting with the impossibility of wait-free $k$-set consensus and the Asynchronous Computability Theorem~\cite{herlihy_asynchronous_1993,borowsky_generalized_1993,saks_wait-free_1993} in 1993.
See~\cite{herlihy_distributed_2013} for numerous such results.
 
The basic observation is that the set of states of a distributed algorithm forms a simplicial complex, called its \emph{protocol complex}, and topological properties of this complex, like connectivity, are constrained by the underlying communication and fault model.
These topological properties in turn can determine what tasks are solvable. 
For example: every algorithm in the wait-free model with atomic read-write registers has a connected protocol complex, and because the consensus task's output complex is disconnected, consensus in this model is not solvable~\cite[Chapter~4]{herlihy_distributed_2013}.

This work is also topological, but in a different way: we use (semi)topologies to study consensus in and of itself, rather than the solvability of consensus or other tasks in particular computation models.
Put another way: the papers cited above use topology to study the solvability of distributed tasks, but this work shows how the very idea of `distribution' can be viewed as having a semitopological foundation.

Of course we can imagine that these might be combined --- that in future work we may find interesting and useful things to say about the topologies of distributed algorithms when viewed as algorithms \emph{on} and \emph{in} a semitopology. 
See also the discussion of `algebraic semitopology' in Remark~\ref{rmrk.algebraic.semitopology}.

\paragraph*{Fail-prone systems and quorum systems}

Given a set of processes $\ns P$, a \emph{fail-prone} system~\cite{malkhi_byzantine_1998}  (or \emph{adversary structure}~\cite{hirt_player_2000}) is a set of \emph{fail-prone sets} $\mathcal{F}=\{F_1,...,F_n\}$ where, for every $1\leq i\leq n$, $F_i\subseteq \ns P$.
$\mathcal{F}$ denotes the assumptions that the set of processes that will fail (potentially maliciously) is a subset of one of the fail-prone sets.
A \emph{dissemination quorum system} for $\mathcal{F}$ is a set  $\{Q_1,..., Q_m\}$ of quorums where, for every $1\leq i\leq m$, $Q_i\subseteq \ns P$, and such that 
\begin{itemize*}
\item
for every two quorums $Q$ and $Q'$ and for every fail-prone set $F$, $\left(Q\cap Q'\right)\setminus F\neq\emptyset$ and 
\item
for every fail-prone set $F$, there exists a quorum disjoint from $F$.
\end{itemize*}
Several distributed algorithms, such as Bracha Broadcast~\cite{bracha_asynchronous_1987} and PBFT~\cite{castro_practical_2002}, rely on a quorum system for a fail-prone system $\mathcal{F}$ in order to solve problems such as reliable broadcast and consensus assuming (at least) that the assumptions denoted by $\mathcal{F}$ are satisfied.

Several recent works generalise the fail-prone system model to heterogeneous systems.
Under the failure assumptions of a traditional fail-prone system, Bezerra et al.~\cite{bezerra_relaxed_2022} study reliable broadcast when participants each have their own set of quorums.
Asymmetric Fail-Prone Systems~\cite{Alpos2024} generalise fail-prone systems to allow participants to make different failure assumptions and have different quorums.
In Permissionless Fail-Prone Systems~\cite{cachin_quorum_2023}, participants not only make assumptions about failures, but also make assumptions about the assumptions of other processes;
the resulting structure seems closely related to witness semitopologies, but the exact relationship still needs to be elucidated.

In Federated Byzantine Agreement Systems~\cite{mazieres2015stellar}, participants declare quorum slices and quorums emerge out of the collective quorum slices of their members.
Quorum slices are a special case of the notion of witness-set in Definition~\ref{defn.witnessed.set}(\ref{witness.witness}).
García-Pérez and Gotsman~\cite{garcia2018federated} rigorously prove the correctness of broadcast abstractions in Stellar's Federated Byzantine Agreement model and investigate the model's relationship to dissemination quorum systems.
The Personal Byzantine Quorum System model~\cite{losa:stecbi} is an abstraction of Stellar's Federated Byzantine Agreement System model and accounts for the existence of disjoint consensus clusters (in the terminology of the paper) which can each stay in agreement internally but may disagree between each other.
Consensus clusters are closely related to the notion of topen in Definition~\ref{defn.transitive}(\ref{transitive.cc}).

Sheff et al. study heterogeneous consensus in a model called Learner Graphs~\cite{sheff_heterogeneous_2021} and propose a consensus algorithm called Heterogeneous Paxos.

Cobalt, the Stellar Consensus Protocol, Heterogeneous Paxos, and the Ripple Consensus Algorithm~\cite{macbrough_cobalt_2018,mazieres2015stellar,sheff_heterogeneous_2021,schwartz_ripple_2014} are consensus algorithms that rely on heterogeneous quorums or variants thereof.
The Stellar network~\cite{lokhafa:fassgp} and the XRP Ledger~\cite{schwartz_ripple_2014} are two global payment networks that use heterogeneous quorums to achieve consensus among an open set of participants; the Stellar network is an instance of a witness semitopology.

Quorum systems and semitopologies are not the same thing.
Quorum systems are typically taken to be such that all quorums intersect (in our terminology: they are \emph{intertwined}), whereas semitopologies do not require this.
On the other hand, quorums are not always taken to be closed under arbitrary unions, whereas semitopologies are (see the discussion in Example~\ref{xmpl.semitopologies}(\ref{item.quorum.system})).

The literature on fail-prone systems and quorum systems is most interested in synchronisation algorithms for distributed systems and has been less concerned with their deeper mathematical structure.
Some work~\cite{losa:stecbi} gets as far as proving an analogue to Proposition~\ref{prop.cc.unions} (though it seems fair to say that the semitopological presentation is simpler and clearer), but it fails to notice the connection with topology and there is no consideration of algebra.

\paragraph*{Dualities}

We discussed duality results in detail in Remark~\ref{rmrk.categorical.duality}.
The reader may know that there are a great many such results, starting with Stone's classic duality between Boolean algebras and compact Hausdorff spaces with a basis of clopen sets~\cite{stone:therba,johnstone:stos}.
A nice representation result for semilattices (not the compatible semilattices we consider here) is in~\cite{bredhikin:repts}.
The duality between frames and topologies is described in \cite[page~479, Corollary~4]{maclane:sheglf}.
See also the encyclopedic treatment in~\cite{caramello:toptas}, with an overview in Example~2.9 on page~17.
See also a recent accessible text with clear exposition in~\cite{gherke:topddl}.

Our duality between semiframes and semitopologies fits into this canon.

\paragraph*{(Semi)lattices with extra structure}

We are not aware of semiframes having been studied in the literature, but they are in excellent company, in the sense that things have been studied that are structurally similar.
We mention two examples to give a flavour of this extensive literature:
\begin{enumerate}
\item
A \deffont{quantale} is a complete lattice $(\mathsf Q,\bigvee)$ with an associative \emph{multiplication} operation $\ast : (\ns Q \times \ns Q) \to \ns Q$ that distributes over $\bigvee$ in both arguments~\cite{rosenthal:quaata}.
A commutative quantale whose multiplication is restricted to map to either the top or bottom element in $\mathsf Q$ is close being a semiframe.\footnote{But not quite! We also need proper reflexivity (Definition~\ref{defn.compatibility.relation}(\ref{item.compatible.reflexive})), and quantale morphisms do not necessarily map the top element to the top element like semiframe morphisms should (Definitions~\ref{defn.complete.semilattice.morphism} and~\ref{defn.category.of.spatial.graphs}(\ref{item.category.spatial.morphism})).}
For reference, a pleasingly simple representation result for quantales is given in~\cite{brown:reptq}.
\item
An \deffont{overlap algebra} is a complete Heyting algebra $\ns X$ with an \emph{overlap relation} $\overlaps\subseteq\ns X\times\ns X$ whose intuition is that $x\overlaps y$ when $x\tand y$ is \emph{inhabited}.
The motivation for this comes from constructive logic, in which $\Exists{p}(p\in x\land p\in y)$ is a different and stronger statement than $\neg\Forall{p}\neg(p\in x\land p\in y)$.  
Accordingly, overlap algebras are described as `a constructive look at Boolean algebras'~\cite{ciraulo:oveacl}.

Overlap algebras are not semiframes, but they share an idea with semiframes in making a structural distinction between `intersect' and `have a non-empty join'.
\end{enumerate}

\jamiesubsection{Future work} 
\label{subsect.future.work}

The list of things that we have \emph{not} done is longer than the list of things that we have.
We see this as a feature, not a bug: it suggests that we may have tapped a rich vein of possible future research.

Here are a few comments and ideas.
This list is in no particular order and it is not exhaustive!

\begin{rmrk}[Theory of Byzantine behaviour]
Real networks are subject to Byzantine behaviour --- participants that don't follow the rules, e.g. through hostile intent, error, or communication difficulties.
Thus, a participant may fall silent due to a communications outage, or a deliberately hostile participant may misreport their view of the network in order to `invent' or sabotage action and so influence outcomes.

We have asked the question ``is $p$ intertwined with $p'$'' but not other important questions like:
\begin{itemize*}
\item
``\emph{How} intertwined are they; what is the minimal number of nodes to corrupt that would split $p$ apart from $p'$?'', or 
\item
``What conditions can we put on a witness function to guarantee that changing the witness function at one point $p$ will not change $\kernel(p')$ for any $p'\neq p$?''.
\end{itemize*}
Thus at a high level, given a semitopology $(\ns P,\opens)$ we are interested in asking how properties range over an `$\epsilon$-ball' of perturbed semitopologies --- as might be caused by various possible non-standard behaviours from a limited number of Byzantine points --- and in particular we are looking for criteria to guarantee that appropriately-chosen good properties be preserved under such perturbation.
This exciting and commercially relevant field of research remains to be explored. 
\end{rmrk}

\begin{rmrk}[Performance]
We have considered antiseparation properties such as two points being intertwined, or a space being regular, and we have provided logical specifications of these that could be checked by a SAT solver. 
It remains to develop optimised algorithms that are quicker on practically relevant use-cases.
We speculate on one such algorithm in Remark~\ref{rmrk.practical.intertwined}, but this is just a start.
Following up on this and on other algorithms is future work. 
\end{rmrk}

\begin{rmrk}[Other notions of morphism]
\label{rmrk.more.conditions}
In Definition~\ref{defn.morphism.semitopologies}(\ref{item.morphism.st}) we take a morphism of semitopologies $f:(\ns P,\opens)\to(\ns P',\opens')$ to be a continuous function $f:\ns P\to\ns P'$.\footnote{Correspondingly, in Definition~\ref{defn.category.of.spatial.graphs}(\ref{item.category.spatial.morphism}) we take a morphism of semiframes $g:(\ns X',\cti',\ast')\to(\ns X,\cti,\ast)$ to be a compatible morphism of complete semilattices.}

The reader may be familiar with conditions on maps between topologies other than continuity, such as being \emph{open} ($f$ maps open sets to open sets) and \emph{closed} ($f$ maps closed sets to closed sets).

These conditions also make sense in semitopologies, and furthermore semiframe and graph representations of semitopologies suggests a further design space, that includes conditions on sets intersections and strict inclusions.
We briefly list some of the conditions that we could impose on $f:\ns P\to\ns P'$:
\begin{enumerate*}
\item
If $O\between O'$ then $f^\mone(O)\between f^\mone(O')$.
(It is automatic that if $f^\mone(O)\between f^\mone(O')$ then $O\between O'$, but the reverse implication is a distinct condition.)
\item 
If $O\subsetneq O'$ then $f^\mone(O)\subsetneq f^\mone(O')$.
\item
$O\between O'$ and $f^\mone(O)\subseteq f^\mone(O')$ implies $O\subseteq O'$.

If we write this as a contrapositive --- $O\between O'$ and $O\not\subseteq O'$ implies $f^\mone(O)\not\subseteq f^\mone(O')$ --- then we see the connection to the subintersection relation from Definition~\ref{defn.ct}.
\end{enumerate*} 
See also related discussions in Remarks~\ref{rmrk.no.meet} and~\ref{rmrk.other.properties}, and in Remark~\ref{rmrk.different.notions.of.morphism}.
\end{rmrk}

\begin{rmrk}[Exponential spaces]
\label{rmrk.exponential.spaces}
It remains to check whether the category $\tf{SemiFrame}$ of semiframes is closed~\cite[page~180, Section~VII.7]{maclane:catwm}, or Cartesian.\footnote{It would be surprising if it were not Cartesian.}
We have checked that the category of semitopologies is Cartesian (it is), but it remains to check whether it is closed.

It also remains to look into the \emph{Vietoris} (also called the \emph{exponential}) semitopologies~\cite[Exercise~2.7.20, page~120]{engelking:gent}.
In view of our use of $\THREE$ to develop a logic for expressing properties of semitopologies, an exponential semitopology based on a many-valued domain might also be relevant.
More generally, it remains to consider functors of the from $\f{Hom}(\text{-},B)$ and $\f{Hom}(A,\text{-})$, for different values of $A$ and $B$.
\end{rmrk}

\begin{rmrk}[Computational/logical behaviour]
\label{rmrk.computation}
Semiframes stand as objects of mathematical interest in their own right (just as frames do) but the original motivation for them comes from semitopologies.
It might therefore be useful to think about `computable' semiframes.

What this would mean is not entirely clear at the moment, but of course this is what would make it research.
One possibility is to develop a theory of logic within semiframes.
On this topic, we can recall the discussion so far, and note that semiframes support a complementation operation $x^c =\bigvee \{x' \mid \neg(x'\ast x)\}$, so it is clearly possible to interpret propositional logic in a semiframe (implication would be $x\to y = x^c\tor y$).
\end{rmrk}

\begin{rmrk}[Finiteness and compactness]
\label{rmrk.finiteness.and.compactness}
The relation of semitopologies to finiteness is interesting.
On the one hand, our motivating examples %
are finite because they exist in the real world.
On the other hand %
participants cannot depend on an exhaustive search of the full system being practical (or even permitted --- this could be interpreted as a waste of resources or even as hostile or dangerous).

As touched on in Remark~\ref{rmrk.why.infinite}, this requires our models and algorithms to at least \emph{make sense} in a world of countably infinitely many points.\footnote{This is no different than a programming language including a datatype of arbitrary precision integers: the program must eventually terminate, but because we do not know when, we need the \emph{idea} of an infinity in the language.}
In fact, arguably even `countably large' is not quite right.
The natural cardinality for semitopologies may be \emph{uncountable}, since network latency means that we cannot even enumerate the network: no matter how carefully we count, we could always in principle discover new participants who have joined in the past (but we just had not heard of them yet).

This motivates us to consider algebraic conditions on a semiframe $(\ns X,\cti,\ast)$ that mimic some of the properties of open sets of finite point-set semitopologies, without necessarily insisting on finiteness itself.
We saw concrete examples of this in the \emph{chain-completeness} properties on point-set semitopologies from Definition~\ref{defn.chain-complete}, but for future work we could consider more, such as:
\begin{enumerate*}
\item
We could insist that a $\cti$-descending chain of non-$\tbot_{\ns X}$ elements in $\ns X$ have a non-$\tbot_{\ns X}$ greatest lower bound in $\ns X$.
\item 
We could insist that a $\cti$-descending chain of elements strictly $\cti$-greater than some $x\in \ns X$ have a greatest lower bound that is strictly $\cti$-greater than $x$.
\item
We could insist that if $(x_i\mid i\geq 0)$ and $(y_i\mid i\geq 0)$ are two $\cti$-descending chains of elements, and $x_i\ast y_i$ for every $i\geq 0$ --- in words: $x_i$ is compatible with $y_i$ --- then the greatest lower bounds of the two chains are compatible.
\end{enumerate*}
The reader may notice how these conditions are reminiscent of compactness conditions from topology: e.g. a metric space is compact if and only if every descending chain of open sets has a nonempty intersection.
This is no coincidence, since one of the uses of compactness in topology is precisely to recover some of the characteristics of finite topologies. 

Considering semiframes (and indeed semitopologies) with compactness/finiteness flavoured conditions, is future work.
\end{rmrk}

\begin{rmrk}[Generalising $\ast$]
\label{rmrk.generalising.ast}
In Remark~\ref{rmrk.compatibility.intuition} we mentioned that we can think of semitopologies not as \emph{`topologies without intersections'} so much as \emph{`topologies with a generalised intersection'}.
We have studied a relation called $\between$ (for point-set semitopologies) and $\ast$ (for semiframes), which intuitively measure whether two elements intersect.

But really, this is just a notion of generalised meet.
We would take $(\ns X,\cti)$ and $(\ns X',\cti')$ to be complete join-semilattices and the generalised meet ${\ast} : (\ns X\times\ns X)\to \ns X'$ is any commutative distributive map. 
Or, we could generalise in a different direction and consider (for example) cocomplete symmetric monoidal categories: $\ast$ becomes the (symmetric) monoid action.
These objects could be studied in their own right, or we could try to translate their structure back to sets, to see what point-set generalisations of semitopologies result. 
\end{rmrk}

\begin{rmrk}[Homotopy and convergence]
\label{rmrk.homotopy.and.convergence}
We have not looked in any detail at notions of \emph{path} and \emph{convergence} in semitopologies and semiframes.
We can give a flavour of why this might be new and different relative to the notions from topologies.

Let $(\ns P,\opens)$ be the semitopology defined as follows, and illustrated in Figure~\ref{fig.circle}:
\begin{itemize*}
\item
$\ns P=\mathbb Z\cup\{\top\}$ is thought of intuitively as a circle with $0$ at the bottom and $\top$ at the top.
\item 
For each $x\in\mathbb Z$ define  
$$
\f{left}(x)=\{\top\}\cup\{y\in\mathbb Z\mid y\leq x\}
\quad\text{and}\quad
\f{right}(x)=\{\top\}\cup\{y\in\mathbb Z\mid x\leq y\}
$$
and give $\ns P$ the semitopology $\opens$ generated by the sets $\f{left}(x)$ and $\f{right}(x)$ for all $x\in\mathbb Z$.
\end{itemize*}
Intuitively:
\begin{itemize*}
\item
$\f{left}(x)$ is a circle segment starting at $x$ ($x$ may be negative) and headed leftwards towards $\top$.
\item
$\f{right}(x)$ is a circle segment starting at $x$ ($x$ may be negative) and headed rightwards towards $\top$.
\end{itemize*}
We can converge on $\top$ from the left (via the negative numbers), and from the right (via the positive numbers) --- however, the descending sequences of open neighbourhoods intersect only at $\top$ and do not have a common open intersection.
This is not behaviour that would be possible in a topology.
This example is really just dressing up one of our earliest observations, from Lemma~\ref{lemm.two.min}: in semitopologies a point can have more than one minimal open neighbourhood, and the example illustrates that intuitively each of these minimal open neighbourhoods can be thought of as a distinct direction by which we can converge on the point.
Developing this part of the theory is future work.
\end{rmrk}

\ifgreyprint\newcommand\redorgrey{darkgray}\else\newcommand\redorgrey{red}\fi
\ifgreyprint\newcommand\cyanorgrey{gray}\else\newcommand\cyanorgrey{cyan}\fi
\newcommand{\curlybrace}[4]{%
\draw[thick, \redorgrey]  (#1:#3-0.1) -- (#1:#3) arc (#1:#2:#3)--(#2:#3-0.1) ; 
\draw[thick, \redorgrey] ({(#1+#2)/2}:#3) -- ({(#1+#2)/2}:#3+0.3);
\node at ({(#1+#2)/2}:#3+0.7) {\small #4}; 
}

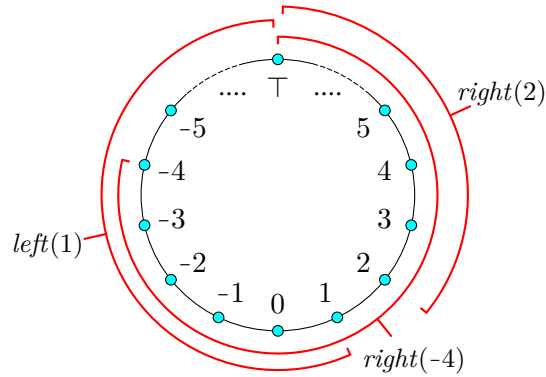
\begin{figure}[t]
\begin{center}
\begin{tikzpicture}
  \def\Radius{18mm}
  \draw %
    (0, 0) coordinate (M) circle[radius=\Radius]
  ;
  \begin{scope}  %
  \clip (-2cm,1.35cm) rectangle (2cm,1.725cm);
  \draw[white, dotted, thick]
    (0, 0) coordinate (M) circle[radius=\Radius]
  ;
  \end{scope}
  \filldraw[fill=\cyanorgrey, radius=2pt]
    \foreach \i in {0, 2, 3, 4, 5, 6, 7, 8, 9, 10, 11, 12} {
      (M) +(90 - \i * 360/14:\Radius) circle[]
    }
  ;
  \path
    \foreach \i/\a in {0/{\top}, 1/{\ddddot{}}, 2/5, 3/4, 4/3, 5/2, 6/1, 7/0, 13/{\ddddot{}}, 12/{\minus 5}, 11/{\minus 4}, 10/{\minus 3}, 9/{\minus 2}, 8/{\minus 1}} {
      (M) -- node[pos=.8] {$\a$} +(90 - \i * 360/14:\Radius)
    }
  ;
  \curlybrace{-90 + 2* 360/14}{88.5}{2.5}{$\ \f{right}(2)$}
  \curlybrace{-90 - 4 * 360/14}{90}{2.1}{$\ \f{right}(\minus 4)$}
  \curlybrace{91.5}{90 + 8* 360/14 - 1.5}{2.32}{$\f{left}(1)\ \ $}
\end{tikzpicture}
\end{center}
\caption{A point with two paths to it (Remark~\ref{rmrk.homotopy.and.convergence})}
\label{fig.circle}
\end{figure}

\begin{rmrk}[Constructive mathematics]
We have not considered what semiframes would look like in a constructive setting. 
Much of the interest in frames and locales (versus point-set topologies) comes from working in a constructive setting; e.g. in the topos of sheaves over a base space, locales give a good fibrewise topology of bundles.
To what extent similar structures might be built using semiframes, or what other structures might emerge instead, are currently entirely open questions.

It also remains to think about what a suitable Curry-Howard correspondence would be, based on residuated semilattices with a compatibility relation $\ast$ (instead of a conjunction $\tand$).
\end{rmrk}

\begin{rmrk}[Algebraic semitopology]
\label{rmrk.algebraic.semitopology}
We mentioned in Subsection~\ref{subsect.related.work} that our use of semitopology is not directly related to algebraic topology applied to solvability of distributed computing tasks.
These are distinct topics: they share a word in their name, but they are no more equal than a Great Dane and a Danish pastry.

But, it is an interesting question what algebraic \emph{semi}topology might look like.
Or to put this another way: 
\emph{What is the geometry of semitopological spaces?}
We would very much like to know.
\end{rmrk}

\begin{rmrk}[More values]
In Section~\ref{sect.three} we use a three-valued logic $\THREE$ with truth-values corresponding to `true', `false', and `both'.
However, in real systems a participant might also wish to return `neither', or `don't know', or `please wait', or even levels confidence of the above.
And why should we even restrict ourselves to that?
Why not consider an arbitrary lattice, representing all valid combinations of data and/or knowledge? 
$\THREE$ is a good place to start because it is minimal and still very expressive, but when we investigate practical applications it may well turn out to be the case that a richer domain of values is useful.
The maths suggests no fundamental obstacles to doing this.

In a related theme, in Definition~\ref{defn.value.assignment} we define a \emph{value assignment} $f:\ns P\to\tf{Val}$ to be a function from a semitopology to a codomain $\tf{Val}$ that is given the discrete semitopology.
This is a legitimate starting point, but of course we should consider more general codomains.
This could include an arbitrary semitopology on the right (for greatest generality), but even for our intended special case of consensus it would be interesting to try to endow $\tf{Val}$ with a semilattice structure (or something like it), at least, e.g. to model merging of distinct updates to a ledger.\footnote{We write `something like it' because we might also consider, or consider excluding, possibly conflicting updates.}
We can easily generate a (semi)topology from a semilattice by taking points to be elements of the lattice and open sets to be up-closed sets, and this would be a natural generalisation of the discrete semitopologies we have used so far.
\end{rmrk}

\begin{rmrk}
In Proposition~\ref{prop.open.strong-consensus} and Remark~\ref{rmrk.gradecast} we studied how consensus, once achieved on an open set $O$, propagates to its closure $\closure{O}$. 
But this is just half of the problem of consensus: it remains to consider (within our semitopological framework) what it is to attain consensus on some open set in the first place.

That is: suppose $(\ns P,\opens)$ is a semitopology and $f:\ns P\to\tf{Var}$ is a value assignment.
Then what does it mean, in maths and algorithms, to find a value assignment $f':\ns P\to\tf{Var}$ that is `close' to $f$ but is continuous on some open set $O$?
We have constructed a detailed theory of what it would then be to extend $f'$ to an $f''$ that continuously extends $f'$ to regular points; but we have not yet looked at how to build the $f'$. 
We speculate that unauthenticated Byzantine consensus algorithms (like Information-Theoretic HotStuff~\cite{abraham_information_2020}) can be understood in our setting; unlike authenticated algorithms, unauthenticated algorithms do not rely on one participant being able to prove to another, by exhibiting signed messages, that a quorum has acted in a certain way.
\end{rmrk}

\jamiesubsection{Open problems} 
\label{subsect.open.problems}

In addition to the future work mentioned above, we note some technical questions that have arisen in this text which we have not yet had time to answer:
\begin{enumerate}
\item
In view of Theorem~\ref{thrm.regular=qr+sc}, does there exist a space such that every point is quasiregular and unconflicted, but no point is hypertransitive (Definition~\ref{defn.sc})?
\item
As per Remark~\ref{rmrk.two.char.r}, it remains an open problem to check whether there is some natural property $X'$ such that regular = indirectly regular + $X'$ (Definition~\ref{defn.indirectly.regular}).
\item
In Remark~\ref{rmrk.chain-completeness.in.context} we draw an analogy between the chain-completeness condition on semitopologies and the Alexandrov condition (closure under arbitrary intersections) on topologies.
It is known that Alexandrov spaces are uniquely characterised by their specialisation preorder; it would be interesting to check whether the analogy extends and some characterisation in a similar spirit can be found for chain-complete semitopologies. 
\item
As per Remark~\ref{rmrk.two.open.problems}, we have no topological characterisation of witness semitopologies.
That is, it is an open problem to abstractly characterise the class of semitopologies that can be generated from a witness function.
\item
Also as per Remark~\ref{rmrk.two.open.problems}, it remains to investigate conditions on witness functions to guarantee good behaviour, such as quasiregularity, weak regularity, or regularity, of points --- or even the \emph{existence} of some (quasi/weakly) regular point.
\item\label{item.rounded.sober}
In Example~\ref{xmpl.sober.non-sober}(\ref{item.sober.R}) we prove that $\mathbb R$ is sober as a semitopological space.
This seems innocuous enough, since $\mathbb R$ is Hausdorff and it is known that Hausdorff topologies are sober --- however, Lemma~\ref{lemm.hausdorff.not.sober}(\ref{item.hausdorff.not.sober}) shows that Hausdorff \emph{semi}topologies are \emph{not} necessarily sober.
Thus $\mathbb R$ is not a sober semitopology just because it is Hausdorff; it has other properties that make this work.
Subsection~\ref{subsect.sober.semitopologies} contains some results, but developing a more rounded understanding of what makes a semitopology sober, is future work.
\item
As per the discussion in Remark~\ref{rmrk.mcn},
how does the property that $\community(p)$ is a minimal closed neighbourhood (though not necessarily of $p$) fit in to the other well-behavedness properties we have considered, such as (quasi/weak/indirect) regularity, being unconflicted, and being hypertransitive?
\end{enumerate}

\jamiesubsection{Final comments} 
\label{subsect.final.comments}

Distributed systems are an old idea; think: telephone exchanges, satellite systems --- and of course the generals in an army, as per the classic paper~\cite{lamport:byzgp}.
However, it is not hyperbole to note that the use and importance of distribution and decentralisation has expanded exponentially in the last twenty years.
This %
has provoked an explosion of new algorithms and new mathematics with which to understand them.  
This includes looking into generalisations of the notion of consensus and quorum~\cite{Alpos2024,sheff_heterogeneous_2021,cachin_quorum_2023,li_quorum_2023,bezerra_relaxed_2022,garcia2018federated,lokhafa:fassgp,losa:stecbi,florian_sum_2022,li_open_2023}, and new systems~\cite{macbrough_cobalt_2018,mazieres2015stellar,sheff_heterogeneous_2021,schwartz_ripple_2014}.

We have combined the research on consensus with a long mathematical tradition of studying topologies, algebras, and the dualities between them (references in Remark~\ref{rmrk.categorical.duality} and at the start of Subsection~\ref{subsect.related.work}).
We do this by applying a classic technique: \emph{topologise, then dualise}.
And, we think is fair to say that it works: we get new and interesting structures, a duality result, and a new logic.
As noted in Subsections~\ref{subsect.future.work} and~\ref{subsect.open.problems}, there is no shortage of scope for future research.

\cleartorecto\listoffigures
\ \\

\emph{A word on the figures:}\ \ 
As per a comment in Figure~\ref{fig.012}, we may omit open sets that are unions of open sets that are already illustrated.
In particular, we usually do not include the universal open set, where this is a union of open sets that are already illustrated.
Thus in Figure~\ref{fig.012} we \emph{do} draw the universal open set in the top left diagram, which is not a union of $\{0\}$ and $\{2\}$; but we \emph{do not} draw it in the top right diagram, since $\{0,1,2\}=\{0,1\}\cup\{1,2\}$.

\renewcommand\thispagestyle[1]{}

\newcommand{\etalchar}[1]{$^{#1}$}
\hyphenation{Mathe-ma-ti-sche}
\providecommand{\bysame}{\leavevmode\hbox to3em{\hrulefill}\thinspace}
\providecommand{\MR}{\relax\ifhmode\unskip\space\fi MR }
\providecommand{\MRhref}[2]{%
  \href{http://www.ams.org/mathscinet-getitem?mr=#1}{#2}
}
\providecommand{\href}[2]{#2}

\section*{Comments on citations and typesetting}

I have typeset URL pointers in full (even though this can be a bit long) so that readers of the print version can see them. 
Of course in the pdf, these links are clickable.
Where online sources are relevant, I provide links, and also permalinks to archived versions of the webpage. 

Where I need to reference standard definitions or lemmas, I cite standard textbooks where possible; but Wikipedia and the Stanford Encyclopedia of Philosophy are also very good, and being online they are easily accessible, so for the reader's convenience I cite them too (again, with permalinks).

\section*{Acknowledgements}

I would like to thank three anonymous referees for generously donating their time and careful attention to this material. 
Thanks to Giuliano Losa and the Stellar Development Foundation for their generous support and funding.
Giuliano was instrumental in explaining the issues, reviewing material, and guiding me to interesting problems; thank you for the many interesting discussions and insightful comments.
Thanks to Luca Zanolini and to Michael Gabbay for their editorial input. 
Thanks to the mathematics StackExchange community for helping to find citations for a simple but obscure lemma (see Remark~\ref{rmrk.pi-base}), and to Jobst Heitzig and Steve Vickers for their interest in these ideas and for asking good questions.
Thanks to Jane Spurr for editorial work.

This work would not have been possible in its current form without the support of these people and others. 
All remaining errors are mine.

\ \\
\vfill
\noindent\emph{This text is a revised version dated July 2025, which corrects some minor errors and omissions in the original version published August 2024.} 

\newpage
\printindex
\newpage


\begin{thebibliography}{SWRM21}

\bibitem[AAZ11]{arieli:idepl}
Ofer Arieli, Arnon Avron, and Anna Zamansky, \emph{Ideal paraconsistent
  logics}, Studia Logica \textbf{99} (2011), no.~1-3, 31--60.

\bibitem[ACTZ24]{Alpos2024}
Orestis Alpos, Christian Cachin, Bj{\"o}rn Tackmann, and Luca Zanolini,
  \emph{Asymmetric distributed trust}, Distributed Computing (2024), Available
  online at \url{https://doi.org/10.1007/s00446-024-00469-1}.

\bibitem[AS20]{abraham_information_2020}
Ittai Abraham and Gilad Stern, \emph{{Information} {Theoretic} {HotStuff}},
  24th International Conference on Principles of Distributed Systems, {OPODIS}
  2020, December 14-16, 2020, Strasbourg, France (Virtual Conference) (Quentin
  Bramas, Rotem Oshman, and Paolo Romano, eds.), LIPIcs, vol. 184, Schloss
  Dagstuhl - Leibniz-Zentrum f{\"{u}}r Informatik, 2020, pp.~11:1--11:16.

\bibitem[BG93a]{borowsky_generalized_1993}
Elizabeth Borowsky and Eli Gafni, \emph{Generalized {FLP} {Impossibility}
  {Result} for {T}-resilient {Asynchronous} {Computations}}, Proceedings of the
  {Twenty}-fifth {Annual} {ACM} {Symposium} on {Theory} of {Computing} (New
  York, NY, USA), {STOC} '93, ACM, 1993, pp.~91--100.

\bibitem[BG93b]{brown:reptq}
Carolyn Brown and Doug Gurr, \emph{A representation theorem for quantales},
  Journal of Pure and Applied Algebra \textbf{85} (1993), no.~1, 27--42.

\bibitem[BKK22]{bezerra_relaxed_2022}
João~Paulo Bezerra, Petr Kuznetsov, and Alice Koroleva, \emph{Relaxed reliable
  broadcast for decentralized trust}, Networked Systems (Mohammed-Amine
  Koulali and Mira Mezini, eds.), Lecture Notes in Computer Science, Springer
  International Publishing, 2022, pp.~104--118.

\bibitem[Bou98]{bourbaki:gent1}
Nicolas Bourbaki, \emph{General topology: Chapters 1-4}, Elements of
  mathematics, Springer, 1998.

\bibitem[Bra87]{bracha_asynchronous_1987}
Gabriel Bracha, \emph{Asynchronous {Byzantine} agreement protocols},
  Information and Computation \textbf{75} (1987), no.~2, 130--143.

\bibitem[Bre84]{bredhikin:repts}
D.~A. Bredhikin, \emph{A representation theorem for semilattices}, Proceedings
  of the American Mathematical Society \textbf{90} (1984), no.~2, 219--220.

\bibitem[BRV01]{blackburn:modl}
Patrick Blackburn, Maarten~de Rijke, and Yde Venema, \emph{Modal logic},
  Cambridge Tracts in Theoretical Computer Science, Cambridge University Press,
  2001.

\bibitem[Cam78]{campbell:orizl}
Paul~J Campbell, \emph{The origin of “{Z}orn's {L}emma”}, Historia
  Mathematica \textbf{5} (1978), no.~1, 77--89.

\bibitem[Car11]{caramello:toptas}
Olivia Caramello, \emph{A topos-theoretic approach to {S}tone-type dualities},
  2011, \url{https://doi.org/10.48550/arXiv.1103.3493}.

\bibitem[CC20]{ciraulo:oveacl}
Francesco Ciraulo and Michele Contente, \emph{{Overlap Algebras: a Constructive
  Look at Complete Boolean Algebras}}, {Logical Methods in Computer Science}
  \textbf{16} (2020).

\bibitem[Chu36]{church:unspen}
Alonzo Church, \emph{An unsolvable problem of elementary number theory},
  American Journal of Mathematics \textbf{58} (1936), no.~2, 345--363.

\bibitem[CL02]{castro_practical_2002}
Miguel Castro and Barbara Liskov, \emph{Practical {Byzantine} fault tolerance
  and proactive recovery}, ACM Transactions on Computer Systems (TOCS)
  \textbf{20} (2002), no.~4, 398--461.

\bibitem[CLZ23]{cachin_quorum_2023}
Christian Cachin, Giuliano Losa, and Luca Zanolini, \emph{Quorum systems in
  permissionless networks}, 26th International Conference on Principles of
  Distributed Systems ({OPODIS} 2022) (Eshcar Hillel, Roberto Palmieri, and
  Etienne Rivière, eds.), Leibniz International Proceedings in Informatics
  ({LIPIcs}), vol. 253, Schloss Dagstuhl – Leibniz-Zentrum für Informatik,
  2023, {ISSN}: 1868-8969, pp.~17:1--17:22.

\bibitem[DG84]{dowling:lintat}
William~F. Dowling and Jean~H. Gallier, \emph{Linear-time algorithms for
  testing the satisfiability of propositional horn formulae}, The Journal of
  Logic Programming \textbf{1} (1984), no.~3, 267--284.

\bibitem[DP02]{priestley:intlo}
B.~A. Davey and Hilary~A. Priestley, \emph{Introduction to lattices and order},
  2 ed., Cambridge University Press, 2002.

\bibitem[Edg20]{edgington:indc}
Dorothy Edgington, \emph{Indicative conditionals}, The Stanford Encyclopedia of
  Philosophy (Edward~N. Zalta, ed.), Metaphysics research lab, CSLI, Stanford
  University, fall 2020 ed., 2020, Available online at
  \url{https://plato.stanford.edu/archives/fall2020/entries/conditionals/}
  (permalink: \url{https://archive.ph/FzVOF}).

\bibitem[ELW]{emerson:trai}
Jonathan Emerson, Mark Lezama, and Eric~W. Weisstein, \emph{Transfinite
  induction}, Available online at
  \url{https://mathworld.wolfram.com/TransfiniteInduction.html}.

\bibitem[Eng89]{engelking:gent}
Ryszard Engelking, \emph{General topology}, Sigma Series in Pure Mathematics,
  Heldermann Verlag, Berlin, 1989.

\bibitem[Erd18]{erdman:protac}
John~M. Erdman, \emph{A problems based course in advanced calculus}, Pure and
  Applied Undergraduate Texts, no.~32, American Mathematical Society, 2018,
  Available online at
  \url{https://web.archive.org/web/20221128144749/https://web.pdx.edu/~erdman/PTAC/problemtext_pdf.pdf}.

\bibitem[Ern09]{erne:clo}
Marcel Erné, \emph{Closure}, Beyond topology, vol. 486, American Mathematical
  Society, 2009, pp.~163--238.

\bibitem[Fey14]{fey:strpat}
Mark Fey, \emph{{A straightforward proof of Arrow's theorem}}, Economics
  Bulletin \textbf{34} (2014), no.~3, 1792--1797.

\bibitem[FHNS22]{florian_sum_2022}
Martin Florian, Sebastian Henningsen, Charmaine Ndolo, and Björn Scheuermann,
  \emph{The {S}um of {I}ts {P}arts: {A}nalysis of {F}ederated {B}yzantine
  {A}greement {S}ystems}, Distributed Computing \textbf{35} (2022), no.~5,
  399--417 (en).

\bibitem[FM88]{feldman_optimal_1988}
Paul Feldman and Silvio Micali, \emph{Optimal algorithms for {B}yzantine
  agreement}, Proceedings of the twentieth annual {ACM} symposium on Theory of
  computing, {STOC} '88, Association for Computing Machinery, 1 1988,
  pp.~148--161.

\bibitem[Gab26]{gabbay:deccac}
Murdoch~J. Gabbay, \emph{Decentralised collaborative action: cryptoeconomics in
  space}, Cryptoeconomic Theory: Blockchain-AI Integration (M.~Swan, S.~Takagi,
  and F.~Witte, eds.), World Scientific, London, 2026, to appear; available at
  arXiv: 2504.12493 (\href{https://arxiv.org/abs/2504.12493}{permalink}).

\bibitem[Goo14]{tezos:whitepaper}
L.~M. Goodman, \emph{Tezos -- a self-amending crypto-ledger (white paper)},
  September 2014.

\bibitem[GPG18]{garcia2018federated}
{\'A}lvaro Garc{\'\i}a-P{\'e}rez and Alexey Gotsman, \emph{Federated byzantine
  quorum systems}, 22nd International Conference on Principles of Distributed
  Systems (OPODIS 2018), Schloss Dagstuhl-Leibniz-Zentrum fuer Informatik,
  2018.

\bibitem[Gut22]{gutev:simecu}
Valentin Gutev, \emph{Simultaneous extension of continuous and uniformly
  continuous functions}, Studia Mathematica \textbf{265} (2022), no.~2,
  121--139.

\bibitem[GvG24]{gherke:topddl}
Mai Gehrke and Sam van Gool, \emph{Topological duality for distributive
  lattices: Theory and applications}, Cambridge Tracts in Theoretical Computer
  Science, Cambridge University Press, 2024, Available online at
  \url{https://arxiv.org/abs/2203.03286}.

\bibitem[H\"03]{hahnle:commvl}
Reiner H\"ahnle, \emph{Complexity of many-valued logics}, Beyond two: theory
  and applications of multiple-valued logic (Melvin Fitting, Ewa Orłowska, and
  Janusz Kacprzyk, eds.), Studies in fuzziness and soft computing, Springer,
  January 2003.

\bibitem[Han11]{hanikova:comcpf}
Zuzana Hanikov\'a, \emph{Computational complexity of propositional fuzzy
  logics}, vol.~2, pp.~793--851, College Publications, 01 2011.

\bibitem[HKR13]{herlihy_distributed_2013}
Maurice Herlihy, Dmitry Kozlov, and Sergio Rajsbaum, \emph{Distributed
  computing through combinatorial topology}, Morgan Kaufmann, 2013.

\bibitem[HM00]{hirt_player_2000}
Martin Hirt and Ueli Maurer, \emph{Player {Simulation} and {General}
  {Adversary} {Structures} in {Perfect} {Multiparty} {Computation}}, Journal of
  Cryptology \textbf{13} (2000), no.~1, 31--60.

\bibitem[HS93]{herlihy_asynchronous_1993}
Maurice Herlihy and Nir Shavit, \emph{The asynchronous computability theorem
  for t-resilient tasks}, Proceedings of the twenty-fifth annual {ACM}
  symposium on {Theory} of computing, 1993, pp.~111--120.

\bibitem[Hul03]{hulek:eleag}
Klaus Hulek, \emph{Elementary algebraic geometry}, Student Mathematical
  Library, vol.~20, American Mathematical Society, 2003.

\bibitem[Jec73]{jech:axic}
Thomas Jech, \emph{The axiom of choice}, North-Holland, 1973, ISBN 0444104844.

\bibitem[Joh86]{johnstone:stos}
Peter~T. Johnstone, \emph{Stone spaces}, vol.~3, Cambridge University Press,
  1986.

\bibitem[Joh87]{johnstone:notlst}
\bysame, \emph{Notes on logic and set theory}, Cambridge University Press,
  1987.

\bibitem[JT04]{doi:10.1080/00029890.2004.11920120}
Tyler Jarvis and James Tanton, \emph{The hairy ball theorem via {S}perner's
  lemma}, The American Mathematical Monthly \textbf{111} (2004), no.~7,
  599--603.

\bibitem[Kop89]{koppelberg:hanba1}
Sabine Koppelberg, \emph{Handbook of boolean algebras, volume 1},
  North-Holland, 1989, Series editors Robert Bonnet and James Donald Monk.

\bibitem[Lam98]{lamport_part-time_1998}
Leslie Lamport, \emph{The part-time parliament}, {ACM} Transactions on Computer
  Systems \textbf{16} (1998), no.~2, 133--169.

\bibitem[LCL23]{li_quorum_2023}
Xiao Li, Eric Chan, and Mohsen Lesani, \emph{Quorum {Subsumption} for
  {Heterogeneous} {Quorum} {Systems}}, August 2023, arXiv:2304.04979 [cs].

\bibitem[LGM19]{losa:stecbi}
Giuliano Losa, Eli Gafni, and David Mazi{\`e}res, \emph{Stellar consensus by
  instantiation}, 33rd International Symposium on Distributed Computing (DISC
  2019) (Dagstuhl, Germany) (Jukka Suomela, ed.), Leibniz International
  Proceedings in Informatics (LIPIcs), vol. 146, Schloss
  Dagstuhl--Leibniz-Zentrum fuer Informatik, 2019, pp.~27:1--27:15.

\bibitem[Lif08]{lifshitz:whaasp}
Vladimir Lifschitz, \emph{What is answer set programming?}, Proceedings of the
  23rd National Conference on Artificial Intelligence - Volume 3, AAAI'08, AAAI
  Press, 2008, p.~1594–1597.

\bibitem[Lif19]{lifshitz:anssp}
Vladimir Lifschitz, \emph{Answer set programming}, Springer, 2019.

\bibitem[LL23]{li_open_2023}
Xiao Li and Mohsen Lesani, \emph{Open {Heterogeneous} {Quorum} {Systems}},
  April 2023, arXiv:2304.02156 [cs].

\bibitem[LLM{\etalchar{+}}19]{lokhafa:fassgp}
Marta Lokhava, Giuliano Losa, David Mazi\`eres, Graydon Hoare, Nicolas Barry,
  Eli Gafni, Jonathan Jove, Rafa\l{} Malinowsky, and Jed McCaleb, \emph{Fast
  and secure global payments with {S}tellar}, Proceedings of the 27th ACM
  Symposium on Operating Systems Principles (New York, NY, USA), SOSP '19,
  Association for Computing Machinery, 2019, p.~80–96.

\bibitem[LSP82]{lamport:byzgp}
Leslie Lamport, Robert Shostak, and Marshall Pease, \emph{{The Byzantine
  Generals Problem}}, ACM Trans. Program. Lang. Syst. \textbf{4} (1982), no.~3,
  382–401.

\bibitem[{Mac}71]{maclane:catwm}
Saunders {Mac Lane}, \emph{Categories for the working mathematician}, Graduate
  Texts in Mathematics, vol.~5, Springer, 1971.

\bibitem[Mac18]{macbrough_cobalt_2018}
Ethan MacBrough, \emph{Cobalt: {BFT} {Governance} in {Open} {Networks}} (en),
  Available online at \url{https://doi.org/10.48550/arXiv.1802.07240}.

\bibitem[Maz15]{mazieres2015stellar}
David Mazi{\`e}res, \emph{The {S}tellar consensus protocol: a federated model
  for {I}nternet-level consensus}, Tech. report, {Stellar Development
  Foundation}, 2015,
  \url{https://www.stellar.org/papers/stellar-consensus-protocol.pdf}
  (permalink:
  \url{https://web.archive.org/web/20240629063518/https://stellar.org/learn/stellar-consensus-protocol}).

\bibitem[MM92]{maclane:sheglf}
Saunders {Mac Lane} and Ieke Moerdijk, \emph{Sheaves in geometry and logic: A
  first introduction to topos theory}, Universitext, Springer, 1992.

\bibitem[MNPS91]{miller:unipfl}
Dale Miller, Gopalan Nadathur, Frank Pfenning, and Andre Scedrov, \emph{Uniform
  proofs as a foundation for logic programming}, Annals of Pure and Applied
  Logic \textbf{51} (1991), no.~1, 125--157.

\bibitem[MR98]{malkhi_byzantine_1998}
Dahlia Malkhi and Michael Reiter, \emph{Byzantine quorum systems}, Distributed
  computing \textbf{11} (1998), no.~4, 203--213.

\bibitem[NW94]{naor:loacaq}
Moni Naor and Avishai Wool, \emph{The load, capacity and availability of quorum
  systems}, Proceedings 35th Annual Symposium on Foundations of Computer
  Science, vol.~27, IEEE, 1994, pp.~214--225.

\bibitem[PN01]{noiri:defsgf}
Valeriu Popa and Takashi Noiri, \emph{On the definitions of some generalized
  forms of continuity under minimal conditions}, Memoirs of the Faculty of
  Science. Series A. Mathematics \textbf{22} (2001), 9--18.

\bibitem[Poo92]{poonen:unicf}
Bjorn Poonen, \emph{Union-closed families}, Journal of Combinatorial Theory,
  Series A \textbf{59} (1992), no.~2, 253--268.

\bibitem[PP12]{picado:fraltw}
Jorge Picado and Aleš Pultr, \emph{Frames and locales: Topology without
  points}, 1 ed., Frontiers in Mathematics, Birkhäuser, 2012.

\bibitem[PP21]{picado:seppft}
\bysame, \emph{Separation in point-free topology}, 1 ed., Birkhäuser, 2021.

\bibitem[Rik62]{riker:thepc}
William~H. Riker, \emph{The theory of political coalitions}, Yale University
  Press, 1962.

\bibitem[Riv85]{rival:grao}
Ivan Rival (ed.), \emph{Graphs and order. the role of graphs in the theory of
  ordered sets and its applications}, Proceedings of NATO ASI, Series C (ASIC),
  vol. 147, Banff/Canada, 1985.

\bibitem[Ros90]{rosenthal:quaata}
Kimmo~I. Rosenthal, \emph{Quantales and their applications}, Pitman Research
  Notes in Mathematics, no. 234, Longman Scientific \& Technical, UK, 1990.

\bibitem[Rus03]{russell:prim}
Bertrand Russell, \emph{The principles of mathematics}, Cambridge University
  Press, 1903,
  \onlineref{http://fair-use.org/bertrand-russell/the-principles-of-mathematics/s37}{https://web.archive.org/web/20231015061016/http://fair-use.org/bertrand-russell/the-principles-of-mathematics/s37}.

\bibitem[Rya10]{ryan:hisidf}
Johnny Ryan, \emph{A history of the internet and the digital future}, Reaktion
  Books, 2010.

\bibitem[Sto36]{stone:therba}
Marshall~H. Stone, \emph{The theory of representation for boolean algebras},
  Transactions of the American Mathematical Society \textbf{40} (1936), no.~1,
  37--111.

\bibitem[SWRM21]{sheff_heterogeneous_2021}
Isaac Sheff, Xinwen Wang, Robbert~van Renesse, and Andrew~C. Myers,
  \emph{Heterogeneous {Paxos}}, 24th {International} {Conference} on
  {Principles} of {Distributed} {Systems} ({OPODIS} 2020) (Dagstuhl, Germany)
  (Quentin Bramas, Rotem Oshman, and Paolo Romano, eds.), Leibniz
  {International} {Proceedings} in {Informatics} ({LIPIcs}), vol. 184, Schloss
  Dagstuhl–Leibniz-Zentrum für Informatik, 2021, ISSN: 1868-8969,
  pp.~5:1--5:17.

\bibitem[SYB14]{schwartz_ripple_2014}
David Schwartz, Noah Youngs, and Arthur Britto, \emph{The {R}ipple {P}rotocol
  {C}onsensus {A}lgorithm}, Ripple Labs Inc White Paper \textbf{5} (2014),
  no.~8, 151.

\bibitem[SZ93]{saks_wait-free_1993}
Michael Saks and Fotios Zaharoglou, \emph{Wait-free k-set agreement is
  impossible: {The} topology of public knowledge}, Proceedings of the
  twenty-fifth annual {ACM} symposium on {Theory} of computing, 1993,
  pp.~101--110.

\bibitem[Sz{\'a}07]{szaz:minsgt}
{\'A}rp{\'a}d Sz{\'a}z, \emph{Minimal structures, generalized topologies, and
  ascending systems should not be studied without generalized uniformities},
  Filomat (Nis) \textbf{21} (2007), 87--97.

\bibitem[Tec11]{ars-technica:howabg}
Ars Technica, \emph{How the atom bomb helped give birth to the internet},
  \url{https://arstechnica.com/tech-policy/2011/02/how-the-atom-bomb-gave-birth-to-the-internet/},
  2 2011, Permalink:
  \url{http://web.archive.org/web/20240622221756/https://arstechnica.com/tech-policy/2011/02/how-the-atom-bomb-gave-birth-to-the-internet/}.

\bibitem[Vic89]{vickers:topvl}
Steven Vickers, \emph{Topology via logic}, Cambridge University Press, USA,
  1989.

\bibitem[Vid21]{vidal:trammv}
Amanda Vidal, \emph{On transitive modal many-valued logics}, Fuzzy Sets and
  Systems \textbf{407} (2021), 97--114, Knowledge Representation and Logics.

\bibitem[Wes11]{sep-generalized-quantifiers}
Dag Westerst\r{a}hl, \emph{Generalized quantifiers}, The Stanford Encyclopedia
  of Philosophy (Edward~N. Zalta, ed.), Metaphysics research lab, CSLI,
  Stanford University, summer 2011 ed., 2011, Available online at
  \url{https://plato.stanford.edu/archives/sum2011/entries/generalized-quantifiers/}
  (permalink: \url{https://archive.ph/YlP09}).

\bibitem[Wik24a]{wiki:Equivalence_of_categories}
Wikipedia, \emph{{Equivalence of categories}},
  \url{http://en.wikipedia.org/w/index.php?title=Equivalence\%20of\%20categories&oldid=1227082771},
  2024, Permalink:
  \url{https://web.archive.org/web/20230316075107/https://en.wikipedia.org/wiki/Equivalence_of_categories}.

\bibitem[Wik24b]{wiki:Ideal_(order_theory)}
\bysame, \emph{{Ideal (order theory); maximal ideals}},
  \url{http://en.wikipedia.org/w/index.php?title=Ideal\%20(order\%20theory)&oldid=1200832357},
  2024, Permalink:
  \url{https://web.archive.org/web/20230724184908/https://en.wikipedia.org/wiki/Ideal\_(order\_theory)\#Maximal\_ideals}.

\bibitem[Wik24c]{wiki:Intersection_graph}
\bysame, \emph{{Intersection graph}},
  \url{http://en.wikipedia.org/w/index.php?title=Intersection\%20graph&oldid=1205563104},
  2024, Permalink:
  \url{https://web.archive.org/web/20230324152732/https://en.wikipedia.org/wiki/Intersection_graph}.

\bibitem[Wik24d]{wiki:Paraconsistent_logic}
\bysame, \emph{{Paraconsistent logic}; an ideal three-valued paraconsistent
  logic},
  \url{https://en.wikipedia.org/w/index.php?title=Paraconsistent\%20logic&oldid=1228606179#An_ideal_three-valued_paraconsistent_logic},
  2024, Permalink:
  \url{http://web.archive.org/web/20231004082704/https://en.wikipedia.org/wiki/Paraconsistent_logic\#An_ideal_three-valued_paraconsistent_logic}.

\bibitem[Wik24e]{wiki:Separation_axiom}
\bysame, \emph{{Separation axiom}; main definitions},
  \url{https://en.wikipedia.org/w/index.php?title=Separation\%20axiom&oldid=1230514922#Main_definitions},
  2024, Permalink:
  \url{https://web.archive.org/web/20221103233631/https://en.wikipedia.org/wiki/Separation_axiom#Main_definitions}.

\bibitem[Wik24f]{wiki:Sierpinski_space}
\bysame, \emph{{Sierpiński space}; categorical description},
  \url{http://en.wikipedia.org/w/index.php?title=Sierpi\%C5\%84ski\%20space&oldid=1228037897},
  2024, Permalink:
  \url{https://web.archive.org/web/20231015150722/https://en.wikipedia.org/wiki/Sierpi\%C5\%84ski_space\#Categorical_description}.

\bibitem[Wil70]{willard:gent}
Stephen Willard, \emph{General topology}, Addison-Wesley, 1970, Reprinted by
  Dover Publications.

\bibitem[WM16]{wintein:gencnt}
Stefan Wintein and Reinhard Muskens, \emph{{A Gentzen Calculus for Nothing but
  the Truth}}, Journal of Philosophical Logic \textbf{45} (2016), 451--465.

\bibitem[WR10]{whitehead:prim1}
Alfred~North Whitehead and Bertrand Russell, \emph{Principia mathematica},
  vol.~1, Cambridge University Press, 1910.

\end{thebibliography}
\end{document}